\documentclass[12pt, a4paper, twoside]{book}
\usepackage[utf8]{inputenc}
\usepackage[english]{babel}
\usepackage{amsmath, amssymb, amsthm, mathtools}
\allowdisplaybreaks[4]
\usepackage{graphicx}
\usepackage[most]{tcolorbox}
\usepackage{braket}
\usepackage{cite}
\usepackage{slashed}
\usepackage{booktabs, longtable, array, multirow}
\usepackage{dsfont}  
\usepackage[top=30mm,bottom=30mm,right=30mm,left=30mm]{geometry}
\usepackage{fancyhdr}
\usepackage[dvipsnames]{xcolor} 
\usepackage{microtype}
\usepackage{tikz}
\usetikzlibrary{arrows.meta}
\usepackage{simpler-wick} 
\usepackage{subcaption}
\usepackage[table]{xcolor}
\usepackage{makecell}
\usepackage{eso-pic}
\usepackage{float}
\makeatletter
\def\@makechapterhead#1{%
  \vspace*{0pt}%
  {\parindent \z@ \raggedright \normalfont
    \huge\bfseries \@chapapp\space \thechapter
    \par\nobreak
    \vskip 5pt
    \Huge \bfseries #1\par\nobreak
    \vskip 10pt
  }}
\makeatother

\newcommand{\GeV}{\mathrm{GeV}}

\definecolor{lightblue}{RGB}{65,105,225}

\definecolor{AM}{rgb}{0,0,1.0}

\usepackage[pdftex,colorlinks=true,
            linkcolor=RoyalBlue,
            citecolor=RoyalBlue,
            urlcolor=RoyalBlue,
            linktocpage=true,
            pdfstartpage=1,
            pdfstartview=FitV,
            breaklinks=true,
            pageanchor=true,
            pdfpagemode=UseNone,
            plainpages=false,
            bookmarksnumbered,
            bookmarksopen=true,
            bookmarksopenlevel=1,
            hypertexnames=true,
            pdfhighlight=/O,
            pdfdisplaydoctitle=true]{hyperref}

\usepackage[nameinlink,capitalize,noabbrev]{cleveref}

\hypersetup{
  pdftitle = {More than a Lifetime:
Towards Precision Predictions for
\texorpdfstring{$B$}{B}-Meson Decays in the Standard Model and Beyond},
  pdfauthor = {Ali Mohamed},
  pdfsubject = {Particle Physics, Heavy Quark Effective Theory},
  pdfkeywords = {B mesons, HQET, Lifetimes, Standard Model, BSM},
  pdfcreator = {pdfLaTeX},
  pdfproducer = {LaTeX with hyperref},
  pdfinfo = {
    CreationDate = {},
    ModDate = {}
  }
}

\crefname{equation}{Eq.}{Eqs.}
\Crefname{equation}{Equation}{Equations}

\crefname{appendix}{Appendix}{Appendices}
\Crefname{appendix}{Appendix}{Appendices}

\usepackage{csquotes}

\begin{document}

\begin{comment}
\thispagestyle{empty}
\AddToShipoutPictureBG*{%
  \includegraphics[width=\paperwidth,height=\paperheight]{PhD_Thesis_Cover.png}
}

\null
\clearpage
\end{comment}

% Minimal half-title page
\clearpage
\thispagestyle{empty}

\vspace*{0.36\textheight}

\begin{center}
    {\bfseries\Huge More than a Lifetime\par}

    \vspace{3.5cm}

    {\large\textsc{Dissertation}\par}

    \vspace{0.8cm}

    {\bfseries\Large Ali Mohamed\par}
\end{center}

\vfill
\clearpage

% Formal title page
\thispagestyle{empty}
\begin{center}

{\bfseries\huge
More than a Lifetime:\par
\vspace{3mm}
Towards Precision Predictions for $B$-Meson Decays in the
Standard Model and Beyond\par
}

\normalsize
\vspace{70pt}

{\textsc{Dissertation}}\\[1mm]
{\large
zur Erlangung des akademischen Grades eines Doktors\\
der Naturwissenschaften\par
}

\vspace{90pt}

{\large vorgelegt von}\\[2mm]
{\bfseries\Large M.Sc.\ Ali Mohamed}

\vfill

{\large
eingereicht bei der Naturwissenschaftlich-Technischen Fakultät\\
der Universität Siegen\par
}

\vspace{55pt}

{\large
Siegen\\[1mm]
August 2026
}

\end{center}

\newpage
\thispagestyle{empty}
\vspace*{4cm}
\begin{center}
Betreuer und erster Gutachter\\[2mm]
Prof.\ Dr.\ Alexander Lenz\\[2mm]
Universität Siegen\\[8mm]

Zweiter Gutachter\\[2mm]
Prof.\ Dr.\ Thomas Mannel\\[2mm]
Universität Siegen\\[8mm]

Weitere Mitglieder der Promotionskommission\\[4mm]

Prof.\ Dr.\ Simon Plätzer\\[2mm]
Universität Siegen\\[4mm]

Prof.\ Dr.\ Stefan Nimmrichter\\[2mm]
Universität Siegen\\[4mm]

und\\[4mm]

Prof.\ Dr.\ Florian Bernlochner\\[2mm]
Universität Bonn
\\[5cm]
Tag der mündlichen Prüfung\\[2mm]
9.\ Oktober 2026
\end{center}

\pagestyle{plain}

\chapter*{Abstract}
This thesis presents a systematic investigation of heavy hadron lifetimes, focusing on the $B$ mesons, as a probe of both Standard Model (SM) dynamics and possible effects of physics beyond the SM (BSM). 
Within the framework of the Heavy Quark Expansion (HQE), we compute, at leading order in QCD, the complete dimension-seven two-quark contributions, corresponding to the $1/m_b^4$ corrections, for the CKM-favoured non-leptonic decays $b\to c\bar{u}d$ and $b\to c\bar{c}s$, providing the full analytical expressions for the corresponding Wilson coefficients. These results, which reveal new matrix elements unique to non-leptonic decays, lay the foundation for future numerical studies and refined SM lifetime predictions. 
Having extended the SM calculation to the next order in the HQE, we turn to the dimension-six Darwin contribution, whose sizeable numerical impact in the SM motivates its generalization to a model-independent BSM framework. 
In particular, we derive, at leading order in QCD, the Wilson coefficients up to and including the dimension-six two-quark (Darwin) operators, 
corresponding to the $1/m_b^3$ corrections in the HQE, for non-leptonic decays induced by a general effective Hamiltonian. We further determine the previously missing dimension-six four-quark weak-annihilation contribution induced by the same general operator basis. 
Since the SM QCD-penguin operators are contained within this basis, our calculation also yields, as a special case, the previously missing dimension-five and dimension-six contributions induced by the SM QCD-penguin operators, thereby further improving the theoretical precision of HQE predictions.
Finally, we explore the phenomenological implications of the $B$-Mesogenesis model, which links the matter–antimatter asymmetry of the universe and dark matter to exotic $B$-meson decays. The model is tested through both inclusive and exclusive decay channels: the inclusive analysis, 
performed within the HQE, quantifies the impact of these new decay modes on the total decay rates and the lifetime ratios $\tau(B^+)/\tau(B_d)$ and $\tau(B_s)/\tau(B_d)$, 
while the exclusive analysis employs light-cone sum rules to calculate the
branching fraction for the exclusive decay $B^+\to p\psi$, involving a proton
and a dark-sector fermion in the final state. By comparing our predictions with experimental limits from BaBar, Belle, and Belle II, we constrain specific effective operators of the model. Furthermore, by examining power-suppressed terms, we identify regions in the dark-sector mass parameter space where subleading effects exceed the leading dimension-three contributions—offering new insight into the applicability of the operator product expansion within the HQE. Together, these studies advance the theoretical understanding of heavy-hadron lifetimes, strengthen the bridge between precision flavour physics and physics beyond the SM, and open new avenues for probing the interplay between heavy-quark dynamics and the dark sector.

\chapter*{Zusammenfassung}

Diese Dissertation präsentiert eine systematische Untersuchung der
Lebensdauern schwerer Hadronen mit besonderem Schwerpunkt auf
$B$-Mesonen als Sonden sowohl für die Dynamik des Standardmodells (SM)
als auch für mögliche Effekte von Physik jenseits des Standardmodells
(BSM). Im Rahmen der Heavy Quark Expansion (HQE) berechnen wir die
vollständigen führenden Zwei-Quark-Beiträge der Dimension sieben,
entsprechend den $1/m_b^4$-Korrekturen, für die CKM-begünstigten
nichtleptonischen Zerfälle $b\to c\bar{u}d$ und
$b\to c\bar{c}s$ und bestimmen die vollständigen analytischen
Ausdrücke der zugehörigen Wilson-Koeffizienten. Diese Ergebnisse
identifizieren neue Matrixelemente, die ausschließlich bei
nichtleptonischen Zerfällen auftreten, und bilden die Grundlage für
zukünftige numerische Untersuchungen sowie präzisere
SM-Vorhersagen der $B$-Mesonenlebensdauern.
\\
Nach der Erweiterung der SM-Berechnung auf die nächste Ordnung der HQE
untersuchen wir den Darwin-Beitrag der Dimension sechs, dessen
beträchtlicher numerischer Einfluss im Standardmodell eine
Verallgemeinerung auf einen modellunabhängigen BSM-Rahmen motiviert.
Insbesondere bestimmen wir auf führender Ordnung die
Wilson-Koeffizienten bis einschließlich der Zwei-Quark-Operatoren der
Dimension sechs (Darwin-Operatoren), entsprechend den
$1/m_b^3$-Korrekturen der HQE, für nichtleptonische Zerfälle, die
durch einen allgemeinen effektiven Hamiltonoperator beschrieben
werden. Darüber hinaus berechnen wir erstmals die bislang fehlenden
Beiträge der Dimension sechs aus der Vier-Quark-Schwachen-Annihilation
für dieselbe allgemeine Operatorbasis. Da die
QCD-Penguin-Operatoren des Standardmodells in dieser Basis enthalten
sind, umfasst unsere Berechnung als Spezialfall zugleich die bislang
fehlenden Beiträge der Dimension fünf und sechs dieser Operatoren und
verbessert damit die theoretische Präzision der HQE-Vorhersagen.
\\
Abschließend untersuchen wir die phänomenologischen Konsequenzen des
Modells $B$-Mesogenesis, das die Materie-Antimaterie-Asymmetrie des
Universums und Dunkle Materie mit exotischen $B$-Mesonen-Zerfällen
verknüpft. Das Modell wird anhand inklusiver und exklusiver
Zerfallskanäle untersucht. Die inklusive HQE-Analyse quantifiziert
den Einfluss dieser Zerfälle auf die Gesamtzerfallsraten sowie die
Lebensdauerverhältnisse $\tau(B^+)/\tau(B_d)$ und
$\tau(B_s)/\tau(B_d)$. Die exklusive Analyse basiert auf
Light-Cone Sum Rules (LCSR) und berechnet
Verzweigungsverhältnisse in baryonische Endzustände mit Fermionen des
dunklen Sektors. Durch den Vergleich unserer Vorhersagen mit den
experimentellen Grenzen von BaBar, Belle und Belle~II schränken wir
spezifische effektive Operatoren des Modells ein. Darüber hinaus
identifizieren wir durch die Untersuchung potenzunterdrückter Beiträge
Bereiche des Massenparameters des dunklen Sektors, in denen
subführende Beiträge die führenden Terme der Dimension drei
übertreffen. Dies liefert neue Erkenntnisse über den
Anwendungsbereich der Operatorproduktentwicklung innerhalb der HQE.
Insgesamt vertiefen diese Arbeiten das Verständnis der
Lebensdauern schwerer Hadronen, stärken die Verbindung zwischen
Präzisions-Flavourphysik und Physik jenseits des Standardmodells und
eröffnen neue Perspektiven für die Untersuchung des Zusammenspiels
von Schwerquarkdynamik und dem dunklen Sektor.
\cleardoublepage
\chapter*{Publications}

This thesis is based on the following
publications:

\begin{enumerate}

\item
A.~Lenz, A.~Mohamed, and Z.~W\"uthrich,
\emph{Constraining B-mesogenesis models with inclusive and exclusive decays},
JHEP \textbf{08} (2025) 141,
\href{https://arxiv.org/abs/2412.14947}{arXiv:2412.14947}.

\item
A.~Mohamed,
\emph{Are Subleading Effects Really Subleading? B-Meson Decays in Mesogenesis},
JHEP \textbf{06} (2026) 219,
\href{https://arxiv.org/abs/2511.04858}{arXiv:2511.04858}.

\item
M.~Lang, A.~Lenz, A.~Mohamed, M.~L.~Piscopo, and A.~V.~Rusov,
\emph{B-meson decay width up to $1/m_b^3$ corrections within and beyond the Standard Model},
JHEP \textbf{05} (2026) 100,
\href{https://arxiv.org/abs/2512.14635}{arXiv:2512.14635}.

\item
A.~Biswas, A.~Khodjamirian, and A.~Mohamed,
\emph{Light-cone sum rules with $B$-meson distribution amplitudes for the
$B\to p$ form factors in B-mesogenesis models},
Phys.\ Rev.\ D \textbf{113} (2026), no.~11, 116035,
\href{https://arxiv.org/abs/2603.01723}{arXiv:2603.01723}.

\end{enumerate}

\par\noindent
In addition, this thesis includes material from the following
manuscript currently in preparation:

\begin{itemize}
\item
A.~Mohamed, M.~Lang and A.~Lenz,
\emph{HQE at $\mathcal{O}(1/m_b^4)$: Two-Quark Operators Contribution to
Non-Leptonic $B$ Decays} (in preparation).
\end{itemize}

\par\noindent
The correspondence between the chapters of the thesis and the above
publications is as follows:

\begin{itemize}
\item Chapter~2 is based on the manuscript currently in preparation.
\item Chapter~3 is based primarily on Publication~3.
\item Chapter~4 is based primarily on Publications~1, 2, and~4.
\end{itemize}

\tableofcontents
\thispagestyle{empty}
\newpage
\null
\cleardoublepage
\thispagestyle{empty}
\vspace*{\fill}
\begin{center}
\textit{
For those who search without guarantee of discovery, and who persevere for the simple love of the work.
This thesis is dedicated to that spirit, and to the people who nurtured it in me.
}
\end{center}
\vspace*{\fill}
\thispagestyle{empty}
\newpage
\null
\cleardoublepage
\thispagestyle{empty}
\vspace*{\fill}
\begin{flushright}
\begin{minipage}{0.8\textwidth} 
\begin{quotation}
\textit{``It is thus the duty of the man who studies the writings of scientists, if learning the truth is his goal, to make himself an enemy of all that he reads, and, applying his mind to the core and margins of its content, attack it from every side.''}
\end{quotation}
\vspace{-0.5em}
\hfill --- Ibn al-Haytham
\end{minipage}
\end{flushright}
\vspace*{\fill}
\thispagestyle{empty}
\cleardoublepage
\chapter*{Introduction}
\addcontentsline{toc}{chapter}{Introduction}
The Standard Model (SM) of particle physics~\cite{Glashow1961,Weinberg1967,Salam1968} is a remarkably successful theory that describes the electromagnetic, weak, and strong interactions of elementary particles. It has withstood decades of experimental scrutiny, culminating in the discovery of the Higgs boson by the ATLAS~\cite{ATLAS:2012Higgs} and CMS~\cite{CMS:2012Higgs} collaborations in 2012, which confirmed the last missing element of the SM and provided direct evidence for the origin of particle masses via the Brout-Englert-Higgs mechanism~\cite{Englert:1964et, Higgs:1964ia, Guralnik:1964eu}.\\
Despite its success, the SM is widely regarded as incomplete. 
It does not incorporate gravity and fails to explain several key phenomena, such as the baryon asymmetry of the universe~\cite{Sakharov1967} and the nature of dark matter~\cite{Zwicky:1933gu,Rubin:1970zza} and dark energy~\cite{Riess1998,Perlmutter1999}.
Moreover, as experimental precision continues to improve, rare $B$-meson decays  provide some of the most sensitive tests of the SM. 
Examples include measurements of angular observables in $b\to s\ell^+\ell^-$ transitions and tests of lepton flavour universality in rare semileptonic $B$ decays, which have motivated extensive theoretical and experimental scrutiny in recent years, see e.g.~\cite{Altmannshofer:2021qrr,Bifani:2018zmi}.\\ 
To explore these unresolved mysteries, two complementary approaches can
be pursued. One is to refine SM predictions and compare them with increasingly precise experimental measurements. The other is
to construct and investigate extensions of the SM that could provide
solutions to these open questions and derive their phenomenological
consequences, see e.g.~\cite{Langacker:2008yv}. The former approach is
valuable in its own right, as it enables increasingly stringent tests of
the SM and may reveal deviations that point towards new physics.
Meanwhile, the latter provides concrete theoretical frameworks whose
predictions can be tested both indirectly through precision observables
and directly through the production of new particles at current and
future collider experiments. Furthermore, precise SM predictions play a
central role in interpreting indirect searches and constraining beyond the SM (BSM)
scenarios by allowing potential new-physics effects to be disentangled
from the SM contribution. One of the most powerful arenas for pursuing
both approaches is the field of flavour physics.\\
Flavour physics continues to be a highly active and central field in modern particle physics. It provides a unique window into testing the SM at high precision and probing potential new physics beyond it, see e.g. the review~\cite{Isidori:2010kg}. 
The field is driven by a broad experimental programme, with dedicated measurements performed at LHCb, Belle~II, and BESIII, complemented by the ATLAS and CMS experiments at the LHC. 
In particular, the study of $B$ mesons plays a crucial role due to the rich flavour structure of the quark sector and the wealth of precise experimental data available.
\\
While the SM successfully describes flavour-changing processes
through the Cabibbo--Kobayashi--Maskawa (CKM)
mechanism~\cite{Cabibbo:1963yz,Kobayashi:1973fv}, it does not explain
the origin of the fermion mass hierarchies, the pattern of quark
mixing~\cite{Froggatt:1978nt}, or the observed matter--antimatter
asymmetry in the Universe. Furthermore, flavour-changing neutral
currents are highly suppressed within the SM through the
Glashow--Iliopoulos--Maiani (GIM)
mechanism~\cite{Glashow:1970gm}, making flavour observables
particularly sensitive probes of new physics. Indeed, many BSM scenarios introduce additional sources of flavour
violation, which can manifest themselves as subtle deviations from SM
predictions, as discussed in, for example,
Ref.~\cite{DAmbrosio:2002vsn}.
Precision measurements of observables such as branching ratios and
$B$-meson lifetimes therefore provide complementary indirect probes of
physics beyond the SM and are sensitive to new particles that may lie
beyond the direct reach of current collider experiments. Their
interpretation relies on equally precise theoretical predictions,
motivating continued efforts to refine SM calculations and to explore
possible signatures of new physics in the flavour sector. \\
In this thesis, we pursue two complementary approaches. On the SM
side, we improve the theoretical description of inclusive
non-leptonic $B$-meson decays within the framework of the Heavy Quark
Expansion (HQE)~\cite{Shifman:1986mx,Chay:1990da,
Bigi:1993fe,Manohar:1993qn}. Specifically, we determine, at leading-order QCD, the matching coefficients of
the dimension-seven two-quark operators, corresponding to the
$1/m_b^4$ corrections, for the CKM-favoured transitions
$b\to c\bar{u}d$ and $b\to c\bar{c}s$. Furthermore, we calculate the
previously missing leading-order matching coefficients of the
dimension-five and dimension-six two-quark operators generated by the
QCD-penguin operators~\cite{Shifman:1977gc,Buras:1995db,
Buchalla:1995vs}, thereby further improving the theoretical precision
and completeness of HQE predictions.
On the BSM side, we extend the HQE for inclusive non-leptonic $B$ decays
to a generalised effective Hamiltonian by deriving the leading-order
Wilson coefficients of all two-quark contributions up to order
$1/m_b^3$. In addition, we determine the previously missing
dimension-six four-quark weak-annihilation contribution for the same
generalised effective Hamiltonian.
Complementing the BSM model-independent analysis, we also perform a
model-specific study of the
$B$-Mesogenesis scenario~\cite{Elor:2018twp,Alonso-Alvarez:2021qfd,
Elahi:2021jia,Nelson:2019fln}, which aims to explain the
matter--antimatter asymmetry of the Universe and the nature of dark
matter.
This model predicts exotic $B$-meson decays into a dark-sector fermion $\psi$, which would manifest as missing energy in detectors. 
We study the impact of these decays on phenomenological observables of the $B$ meson in both inclusive and exclusive channels. 
The inclusive analysis is carried out using the HQE, 
while exclusive decays are investigated within the framework of
Light-Cone Sum Rules (LCSR), discussed in
\cref{sec:LCSR}.\\
The calculations presented in this thesis rely on two robust theoretical frameworks. From a theoretical standpoint, the HQE has established itself as a powerful framework for studying inclusive decays and lifetimes of heavy hadrons. 
Over the past decades, the expansion in inverse powers of the $b$-quark mass has undergone significant development, leading to increasingly precise predictions at higher orders. 
As of the writing of this thesis, the state of the art includes the next-to-leading order (NLO) calculations of the dimension-five contributions $\Gamma_5$ in non-leptonic decays, both for the $b \to c \bar{c} s$~\cite{Mannel:2025fvj} and $b \to c \bar{u} d$~\cite{Mannel:2024uar} channels, 
the next-to-next-to-leading order (NNLO) determination of the leading dimension-three contribution $\Gamma_3$~\cite{Egner:2024azu}, 
and the NNLO calculation of the dimension-six four-quark spectator contribution $\tilde{\Gamma}_6$~\cite{Moretti:2026waq}.
Significant progress has also been made on the non-perturbative side. In particular, Heavy Quark Effective Theory (HQET) sum-rule determinations of the matrix elements of dimension-six $\Delta B=0$ four-quark operators, both within the SM and for a general BSM operator basis, have recently become available~\cite{Black:2024bus}. Furthermore, the first lattice-QCD determinations of dimension-six four-quark operator matrix elements using the gradient-flow method have been achieved~\cite{Black:2026rbz,Black:2026dzp}. 
Although currently restricted to charmed mesons, they represent an important step towards first-principles determinations of HQE matrix elements.
\\
The precision of HQE predictions has improved substantially in recent years and, for several observables, is now comparable to the experimental precision.
For many of these observables, the remaining theoretical uncertainties are dominated by the non-perturbative matrix elements entering the HQE, highlighting the importance of constraining them from experimental data and determining them using first-principles methods such as lattice QCD or QCD sum rules. 
A detailed discussion of the HQE formalism and its applications within the SM can be found in Refs.~\cite{Lenz:2014jha,Albrecht:2024oyn}.
\\
In the context of exclusive decays, the LCSR approach has proven to be a reliable and versatile tool,
see e.g. the review~\cite{Khodjamirian:2023wol}. 
It combines perturbative QCD with non-perturbative hadronic input and allows for systematic improvements in accuracy through the inclusion of higher-twist corrections and radiative effects~\cite{Braun:2003wx,Ball:2004rg}.
\\The thesis is divided as follows. \cref{Ch:1} introduces the theoretical framework used throughout this work. It begins with the formulation of the Effective Weak Hamiltonian, which provides the foundation for describing weak decays of heavy hadrons. The HQE is then presented as a systematic tool to compute inclusive observables in powers of $1/m_b$. The chapter further outlines the Background Field Method and the use of the Fock–Schwinger gauge, which are essential for generating the operator product expansion at the field-theoretical level. Finally, the framework of LCSR is introduced, providing the basis for the calculation of exclusive decay processes in later chapters. 
In \cref{Ch:2}, we investigate the dimension-seven sector of the HQE for
inclusive non-leptonic $B$-meson decays. We derive the complete leading order 
dimension-seven two-quark contributions, corresponding to the
$1/m_b^4$ corrections, for the CKM-favoured decay channels
$b\to c\bar{u}d$ and $b\to c\bar{c}s$, providing the full analytical
expressions for the corresponding Wilson coefficients.
In \cref{Ch:3}, we develop a model-independent extension of the HQE for
inclusive non-leptonic $B$ decays based on a generalised effective
Hamiltonian. Within this framework, we derive the leading-order Wilson
coefficients of all two-quark contributions up to order $1/m_b^3$ and
determine the previously missing dimension-six four-quark
weak-annihilation contribution. We further show that the SM
QCD-penguin operators arise as a special case of the general operator
basis, allowing us to determine their previously missing dimension-five
and dimension-six contributions.
In \cref{Ch:4}, we present a comprehensive phenomenological study
of the $B$-Mesogenesis scenario. On the inclusive side, we employ
the HQE to determine the impact of baryon-number-violating
operators on $B$-meson lifetimes and derive constraints from
lifetime ratios and collider searches. On the exclusive side, we
compute the hadronic matrix elements for
$B^+\to p\psi$ decays using LCSR. Combining both
approaches, we obtain stringent constraints on the viable
parameter space of the model.
Finally, \cref{Ch:summary} summarizes the main results and discusses directions
for future work.

\chapter{Theoretical Framework}
\label{Ch:1}
\vspace{3mm}
\begin{flushright}
\vspace{5mm}
\emph{
``Our job requires a great deal of resilience, and losing is part of the game.''
}
\\[0.3cm]
--- Alexander Lenz
\end{flushright}
\section{Effective Weak Hamiltonian}
One of the characteristic features of weak interactions is the
presence of widely separated energy scales. While weak decays occur at
energies of order a few GeV, they are mediated by much heavier
particles, such as the electroweak gauge bosons with masses of about
$80$--$90\,\mathrm{GeV}$ and the top quark with a mass of
approximately $173\,\mathrm{GeV}$.
Consequently, many weak-decay processes occur at energy scales far below
the masses of the particles that mediate the underlying interactions. A
complete description of such processes would require treating
simultaneously both the light degrees of freedom that actively
participate in the dynamics and much heavier particles. Moreover, the presence of these widely separated scales gives rise to large logarithms
of the form $\ln(M_W/\mu)$ in perturbative calculations, where $M_W$ denotes
the $W$-boson mass and $\mu$ the characteristic low-energy renormalisation
scale of the process.
Such logarithms can spoil the convergence of fixed-order
perturbation theory. 

Effective Field Theories (EFTs)~\cite{Weinberg:1978kz,Wilson:1969zs},
see also~\cite{Manohar:2018aog,Buras:1998raa}, provide a systematic
framework to deal with this situation. By ``integrating out'' the heavy
particles---i.e., removing them from the spectrum of active degrees of
freedom---we obtain an effective theory that describes the low-energy
physics in terms of local operators composed of light fields. The
effects of the heavy particles are not lost but are encoded in the
so-called Wilson coefficients, which multiply these operators and
capture the short-distance (high-energy) physics. Importantly, the
effective theory provides a systematic expansion in inverse powers of
the heavy scale, allowing higher-order corrections to be incorporated in
a controlled, order-by-order manner. Moreover, the separation of scales
permits the resummation of the large logarithms through the
renormalisation-group evolution of the Wilson coefficients, thereby
restoring the convergence of the perturbative expansion.

For weak decays of hadrons, the heavy particles that are integrated out
are primarily the electroweak gauge bosons and, depending on the process,
the top quark. As a result, the underlying electroweak interaction is
replaced by an effective weak Hamiltonian consisting of local 
operators multiplied by Wilson coefficients.

EFTs thus allow us to simplify complex interactions, perform consistent expansions in small parameters (such as $E/M_W$), where $E$ denotes the characteristic energy or momentum transfer of the process, and compute observables with controlled theoretical uncertainties. At the same time,
they offer a model-independent description of possible BSM effects by
parametrizing short-distance physics in terms of higher-dimensional
operators and their corresponding Wilson coefficients, without requiring
knowledge of the underlying ultraviolet (UV) completion. This framework is the foundational first step for all subsequent calculations of weak decay amplitudes in this thesis.

\begin{figure}[t]
\centering
\begin{tikzpicture}

\node[anchor=west] (SM) at (0,0)
{\includegraphics[width=0.34\textwidth]{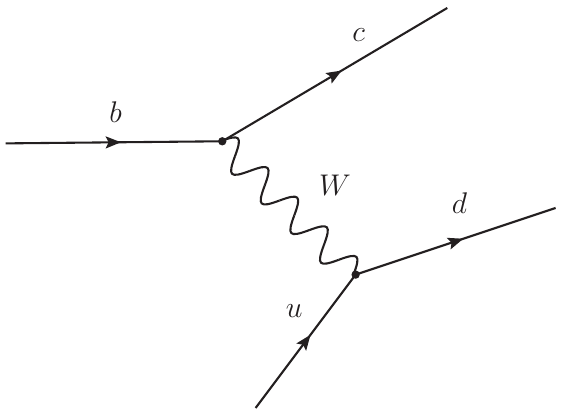}};

\node[anchor=east] (EFT) at (\textwidth,0)
{\includegraphics[width=0.30\textwidth]{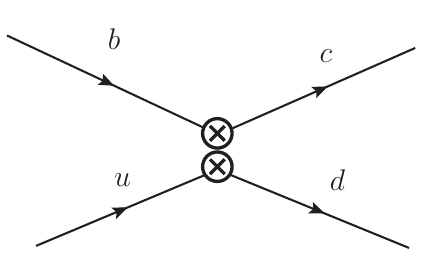}};

\draw[-{Latex[length=3.5mm,width=2.2mm]},
      line width=0.8pt,
      draw=black!70]
($(SM.east)+(1.3,0)$) --
($(EFT.west)+(-1.1,0)$);

\node[above,text=black!70]
at ($(SM.east)!0.5!(EFT.west)+(0,0.25)$)
{\small Integrating out the $W$ boson};

\end{tikzpicture}

\caption{
Illustration of the construction of the effective weak Hamiltonian for
the transition $b\to c\bar{u}d$. At energies well below the
electroweak scale, the $W$ boson is integrated out, replacing the full
SM interaction by a local four-fermion interaction.
}
\label{fig:SMeff}
\end{figure}

It will be instructive to illustrate the transition from the full Lagrangian to the effective Hamiltonian with a concrete example.   Consider the tree-level weak decay $b \to c \bar{u} d$, which proceeds via the exchange of a virtual $W$ boson in the SM, see~\cref{fig:SMeff}.
The second-order contribution to the $\hat{S}$-operator from the weak interaction is (see, \cref{app:1})
\begin{equation}
\label{eq:S_matrix_small}
\hat{S}^{(2)}
= -\frac{g^2}{2}V_{cb}\,V_{ud}^*
\int d^4x\,d^4y\;
T\Big\{ \big[\bar c^{i}(x)\gamma^\mu P_L b^{i}(x)\big]\,W^-_\mu(x)\;
\big[\bar d^{j}(y)\gamma^\nu P_L u^{j}(y)\big]\,W^+_\nu(y) \Big\} \,,
\end{equation}
where $i$ and $j$ are colour indices. After performing the time-ordered product to contract the charged gauge fields, one obtains the $W$-boson propagator in unitary gauge. Explicitly in momentum space \footnote{In the low-energy limit, the longitudinal term proportional to $q_\mu q_\nu / M_W^2$ gives a suppressed contribution and is therefore neglected.},
\begin{equation}
\langle 0 | T\{W^{-}_{\mu}(x) W^+_{\nu}(y)\} |0\rangle = \int \frac{d^4q}{(2\pi)^4}  \frac{-i}{q^2 - M_W^2 + i\varepsilon}\left( g_{\mu\nu} - \frac{q_\mu q_\nu}{M_W^2} \right) e^{-iq\cdot(x-y)} .
\end{equation}
Performing the $x$- and $y$-integrations yields the overall momentum-conserving delta function and the momentum-space amplitude 
\begin{equation}
iT_{fi} = \frac{g^2}{2} V_{cb}\,V_{ud}^* \frac{1}{q^2-M^2_W} \big[\bar c^{i}\gamma^\mu P_L b^{i} \big] \big[\bar d^{j}\gamma_\mu P_L u^{j} \big]
\end{equation}
At low energies, where $q^2 \ll M_W^2$, the propagator can be expanded as:
\begin{equation}
\frac{1}{q^2 - M_W^2} \approx -\frac{1}{M_W^2} \left( 1 + \frac{q^2}{M_W^2} + \cdots \right),
\end{equation}
keeping only the leading term, the amplitude reduces to
\begin{equation}
iT_{fi} = -i\,\frac{4G_F}{\sqrt{2}} V_{cb}\,V_{ud}^* \big[\bar c^{i}\gamma^\mu P_L b^{i} \big] \big[\bar d^{j}\gamma_\mu P_L u^{j} \big] + \mathcal{O}(q^2/M^2_W)
\label{eq:ampfull}
\end{equation}
where $G_F = \frac{\sqrt{2} g^2}{8 M_W^2}$ is the Fermi constant. 
To construct the corresponding low-energy description, one introduces an effective interaction Hamiltonian $\mathcal{H}_{\text{eff}}$ such that the $\hat{S}$-operator \footnote{
The minus sign in the exponent arises from the simple relation
$\mathcal{H}_{\text{eff}} = -\,\mathcal{L}_{\text{eff}}$. },
\begin{equation}
\hat{S} = T \exp \left[ -i \int d^4x \, \mathcal{H}_{\text{eff}}(x) \right],
\end{equation}
reproduces the same transition amplitude in \cref{eq:ampfull} as obtained from the full theory. In this framework, the effect of the heavy $W$-boson exchange is encoded in a local four-fermion operator, and the amplitude arises already at \textit{first order} in $\mathcal{H}_{\text{eff}}$, rather than from the second-order weak interaction in the full Lagrangian.
This transition from the full theory to the effective description is
illustrated schematically in~\cref{fig:SMeff}. 
From \cref{eq:ampfull}, the effective Hamiltonian then reads
\begin{equation}
\label{eq:Heff_tree}
\mathcal{H}_{\mathrm{eff}}^{\,b\to c\bar ud}(x)
=
\frac{4G_F}{\sqrt2}\,
V_{cb}V_{ud}^*\,
\big[\bar c^{i}(x)\gamma^\mu P_L b^{i}(x)\big]
\big[\bar d^{j}(x)\gamma_\mu P_L u^{j}(x)\big].
\end{equation}
This local four-fermion interaction encapsulates the short-distance effects of the $W$-boson exchange and constitutes the leading-order term in the weak effective theory. 

The operator in \cref{eq:Heff_tree} represents the first element in an infinite tower of higher-dimensional operators, each suppressed by powers of $1/M_W$ and multiplied by short-distance Wilson coefficients that encode the effects of heavy degrees of freedom. The general form of the Effective Weak Hamiltonian for a given decay process  can be written as
\begin{equation}
\mathcal{H}_{\text{eff}} = \frac{4G_F}{\sqrt{2}} \sum_i V^{\text{CKM}}  C_i(\mu)  O_i(\mu),
\end{equation}
where $V^{\text{CKM}}$ denotes the appropriate combination of CKM matrix elements associated with the process, $C_i(\mu)$ are the \textit{Wilson coefficients}, encoding the short-distance dynamics above the renormalisation scale $\mu$; they are process-independent and can be computed perturbatively, $O_i(\mu)$ are the \textit{local operators}, built from light (quark, gluon, photon) fields, which capture the long-distance physics. Their hadronic matrix elements must be evaluated using non-perturbative methods.

The separation of short- and long-distance dynamics in the effective
Hamiltonian introduces an explicit dependence on the renormalisation
scale~$\mu$. To ensure that physical observables remain independent of
this arbitrary scale, the Wilson coefficients and operator matrix
elements must evolve consistently under the Renormalisation Group.

As illustrated above for the $b\to c\bar{u}d$ transition, the
construction of the effective theory begins with a \emph{matching}
calculation, in which the Green's functions (or scattering amplitudes)
computed in the full SM are required to agree with those of the
effective theory at a matching scale $\mu_W\sim M_W$. In this way, the
effects of the heavy degrees of freedom, such as the $W$ boson and the
top quark, are absorbed into the Wilson coefficients, while the heavy
particles themselves are removed from the low-energy theory. The resulting Wilson
coefficients $C_i(\mu_W)$ are then evolved down to a characteristic
hadronic scale, $\mu_{\mathrm{had}}\sim m_b$ by solving the
Renormalisation Group Equations (RGEs). 
It is precisely this renormalisation-group evolution that resums the
large logarithms of the form $\ln(M_W^2/m_b^2)$ arising from the
hierarchy of scales discussed above, thereby restoring the convergence
of the perturbative expansion and ensuring a consistent connection
between the short-distance and low-energy descriptions. 
For detailed discussions of the renormalisation-group evolution and operator mixing in weak effective theories, see Refs.~\cite{Buchalla:1995vs,Chetyrkin:1996vx}.

Although the matching between the full SM and the effective theory can
be systematically improved by including higher-order perturbative QCD
corrections, the tree-level matching presented above is sufficient to
illustrate the underlying procedure. Since the calculations in this thesis are consistently
performed at leading order in QCD, we do not discuss the higher-order
matching in detail. Nevertheless, the numerical analyses throughout this
thesis employ Wilson coefficients evaluated at the appropriate
perturbative precision available in the literature.

Having introduced the general structure of the effective weak
Hamiltonian, we now present its explicit form in the SM. For
non-leptonic $b$-quark decays, it is given by~\cite{Buchalla:1995vs}
\begin{align}
{\cal H}_{\rm eff}^{\rm SM} = 
\frac{4 G_F}{\sqrt{2}} \sum_{q_3 = d, s} 
\Bigg\{ \sum_{q_1,q_2 = u,c } &
V_{q_1 b}\, V_{q_2 q_3}^{*}
\Bigl[
C_1(\mu_b)\, Q_1^{q_1 q_2 q_3}
+ C_2(\mu_b)\, Q_2^{q_1 q_2 q_3}
\Bigr]
\nonumber \\
&\quad
- V_{tb}\, V_{t q_3}^{*}
\sum_{k=3}^{6} C_k(\mu_b)
\sum_{q = u,d,s,c}
 Q_k^{qq_3}
\Bigg\}
+ {\rm h.c.}\, ,
\label{eq:Heff-SM}
\end{align}
where $G_F$ is the Fermi constant and $V_{qq^\prime}$ denote the elements
of the CKM matrix.
The operator basis relevant for the non-leptonic $|\Delta B|=1$
transitions in \cref{eq:Heff-SM} comprises two current-current operators,
$Q_{1,2}^{q_1q_2q_3}$, together with the QCD-penguin operators
$Q_{3,\ldots,6}^{qq_3}$. We adopt the following conventions:
\begin{equation}
Q_1^{q_1 q_2 q_3} 
 =   
(\bar q_1^i \, \gamma_\mu P_L \, b^i)
(\bar q_3^j \, \gamma^\mu P_L \, q_2^j)\,,
\qquad 
Q_2^{q_1 q_2 q_3} 
 =  
(\bar q_1^i  \, \gamma_\mu P_L \, b^j)
(\bar{q}_3^j \, \gamma^\mu P_L \, q_2^i)\,,
\label{eq:O12}
\end{equation}
for the current-current sector, while the QCD-penguin operators are
\begin{align}
Q_3^{qq_3} 
& 
= (\bar q_3^i \, \gamma_\mu P_L \, b^i) ( \bar q^j \, \gamma^\mu P_L \, q^j)
\,, \qquad 
Q_4^{qq_3} = (\bar q_3^i \, \gamma_\mu P_L \, b^j)  (\bar q^j \, \gamma^\mu P_L \, q^i)\,, 
\label{eq:O34}
\\[2mm]
Q_5^{qq_3} 
& 
=  (\bar q_3^i \, \gamma_\mu P_L \, b^i) (\bar q^j \, \gamma^\mu P_R \, q^j)\,, 
\qquad
Q_6^{qq_3} = (\bar q_3^i \, \gamma_\mu P_L \, b^j) 
(\bar q^j \, \gamma^\mu P_R \, q^i)\,.
\label{eq:O56}
\end{align}
The indices $i,j=1,2,3$ denote fundamental $SU(3)_c$ colour indices,
and the left- and right-handed chiral projectors are defined by
$P_{L,R}=(1\mp\gamma_5)/2$.

In \cref{eq:Heff-SM}, the Wilson coefficients $C_k(\mu_b)$,
with $k=1,\ldots,6$, are evaluated at the renormalisation scale
$\mu_b\sim m_b$ and are known up to NNLO in QCD~\cite{Gorbahn:2004my}.
In the SM, these coefficients are real-valued, whereas in generic
extensions with heavy new degrees of freedom integrated out at a scale
$\Lambda_{\mathrm{NP}}$, they can in general be complex-valued.
Additional operators,
such as the electroweak penguin operators $Q_7$--$Q_{10}$ and the
electromagnetic and chromomagnetic dipole operators $Q_{7\gamma}$ and
$Q_{8g}$, are omitted for brevity, as they only give subleading
contributions to the processes studied in this work.

While the discussion above introduces the SM $|\Delta B|=1$ weak effective Hamiltonian, this thesis will employ different effective Hamiltonians depending on the physical application. In the study of model-independent BSM contributions in \cref{Ch:3}, extended operator bases with different Dirac structures will be introduced.
In \cref{Ch:4}, the analysis of $B$-Mesogenesis  employs the effective Hamiltonian with charge-conjugated fields. Each Hamiltonian is presented explicitly in the corresponding chapter.

%\section{Heavy Quark Effective Theory}
\section{Heavy Quark Expansion}
\label{sec:HQE}
All inclusive $B$-meson decay widths considered in this work are
calculated within the framework of the HQE. This section introduces the HQE formalism, its application to $B$-meson decays, and the challenges associated with the determination of non-perturbative matrix elements.

The HQE is a systematic effective field theory approach for describing inclusive decay rates and lifetimes of hadrons containing a heavy quark, such as $B$ mesons. It is based on an operator product expansion (OPE)\cite{Wilson:1969zs, Wilson:1973jj, Novikov:1984rf} in inverse powers of the heavy-quark mass, $1/m_b$, which separates short-distance effects—encoded in perturbatively calculable Wilson coefficients—from long-distance contributions, contained in non-perturbative hadronic matrix elements.

Building upon the Effective Weak Hamiltonian introduced in the previous section, the HQE provides a framework in which the total decay rate of a heavy hadron can be expressed as a series of local operators of increasing dimension, suppressed by corresponding powers of $1/m_b$. This formalism allows for precise theoretical predictions of observables such as total $B$-meson decay widths and lifetime ratios, e.g., $\tau(B^+)/\tau(B_d)$ and $\tau(B_s)/\tau(B_d)$, by systematically incorporating both perturbative and non-perturbative QCD effects.

The HQE exploits the fact that the $b$-quark mass is much larger than the typical QCD scale ($m_b \gg \Lambda_{\text{QCD}}$), allowing the inclusive decay of a $B$~meson to be approximated as a free $b$-quark decay, corrected by systematic non-perturbative contributions from the light degrees of freedom (quarks and gluons) inside the hadron.

According to the optical theorem, the inclusive decay rate can be expressed in terms of the imaginary part of the forward scattering amplitude:
\begin{equation}
\Gamma(B \to X) 
= \frac{1}{2 m_B} 
\, \text{Im} \,
\langle B | 
\, i \! \int \! d^4x \;
\text{T} \{ \mathcal{H}_{\text{eff}}(x), \mathcal{H}_{\text{eff}}(0) \} 
| B \rangle,
\label{eq:GammaB}
\end{equation}
where $X$ denotes the sum over all hadronic final states accessible via
the effective Hamiltonian, and $\mathrm{T}$ denotes time ordering.  
Expanding the non-local time-ordered product into a series of local operators through the OPE yields a systematic expansion of the total decay rate in inverse powers of the heavy-quark mass:
\begin{equation}
\Gamma(B) =
\Gamma_3
+ \Gamma_5 \, \frac{\langle {\cal O}_5 \rangle}{m_b^2}
+ \Gamma_6 \, \frac{\langle {\cal O}_6 \rangle}{m_b^3}+ \Gamma_7 \, \frac{\langle {\cal O}_7 \rangle}{m_b^4}
+ \ldots
+ 16 \pi^2
\left[
  \tilde{\Gamma}_6 \, \frac{\langle \tilde{\mathcal{O}}_6 \rangle}{m_b^3}
+ \dots
\right].
\label{eq:HQE}
\end{equation}
\begin{figure}[t]
  \centering
  \setlength{\unitlength}{1pt}

  \begin{tabular}{ccccc}
    % First diagram
    \begin{picture}(150,120)
      \put(0,0){\includegraphics[width=0.25\textwidth]{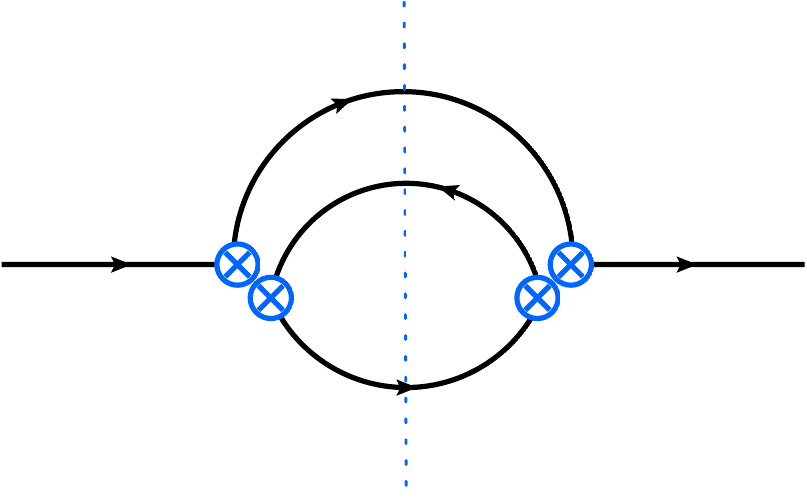}}
      \put(50,60){$f_1$}
      \put(50,30){$f_2$}
      \put(60,5){$f_3$}
      \put(10,34){$b$}
      \put(95,34){$b$}
       \put(120,28){$+$}
    \end{picture}
    % Second diagram
    \begin{picture}(130,120)
      \put(-10,0){\includegraphics[width=0.25\textwidth]{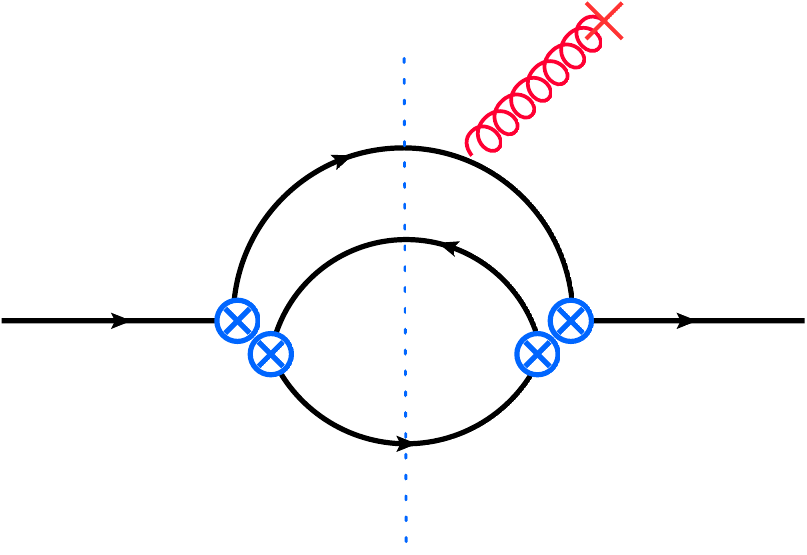}}
      \put(35,60){$f_1$}
      \put(45,30){$f_2$}
      \put(50,8){$f_3$}
      \put(2,34){$b$}
      \put(83,34){$b$}
      \put(110,28){$+ \cdots +$}
    \end{picture}
\hspace{0.3cm}
    % Third diagram
    \begin{picture}(150,120)
      \put(0,0){\includegraphics[width=0.25\textwidth]{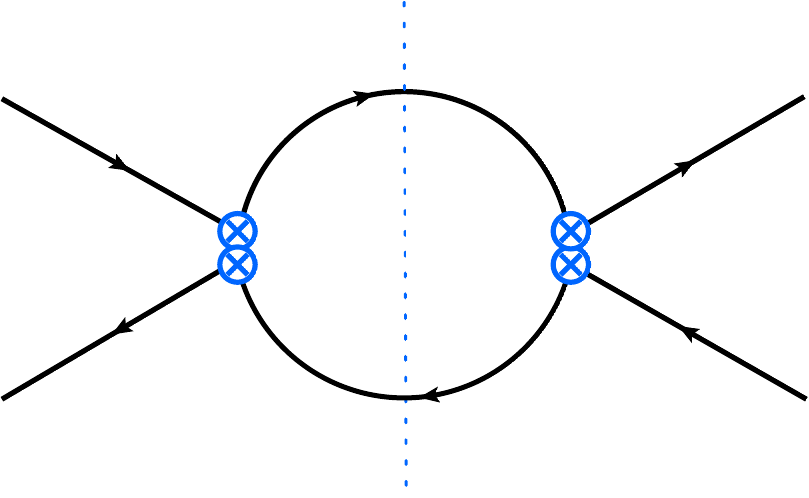}}
      \put(40,60){$f_1$}
      \put(65,7){$f_2$}
      \put(10,53){$b$}
      \put(100,55){$b$}
      \put(10,13){$q$}
      \put(95,10){$q$}
         \put(95,28){$+ \cdots $}
    \end{picture}
  \end{tabular}

\caption{Diagrammatic representation of the HQE for $B$-meson decays. The two-quark and four-quark contributions originate from two-loop and one-loop forward-scattering diagrams, respectively. Crossed vertices correspond to insertions of operators from the effective $|\Delta B|=1$ Hamiltonian. The vertical dashed line denotes the cut used to obtain the imaginary part of the forward amplitude. The particles $f_1$, $f_2$, and $f_3$ represent generic fermionic decay products of the $b$ quark.}
  \label{fig:HQEDiags}
\end{figure}

Here, the terms proportional to $\Gamma_3$, $\Gamma_5$,
$\Gamma_6$, $\tilde\Gamma_6$, and so on, correspond to contributions
from local $\Delta B=0$ operators of increasing mass dimension,
namely dimension-three, dimension-five, dimension-six two-quark,
dimension-six four-quark operators, and higher-dimensional operators,
whose expectation values encode non-perturbative information about the
internal structure of the $B$ meson.

The non-tilde operators, $\mathcal{O}_i$, are two-quark operators that describe corrections to the free heavy-quark decay, such as kinetic and chromomagnetic effects. In contrast, the tilde operators, $\tilde{\mathcal{O}}_i$, are four-quark operators that account for spectator interactions between the heavy $b$-quark and the light valence quarks within the meson. 
A diagrammatic illustration of the HQE is shown in \cref{fig:HQEDiags}.
The spectator contributions, represented by $\tilde{\Gamma}_i$, are enhanced by a factor of $16\pi^2$, reflecting the transition from the two-loop topology of the non-spectator diagrams to a one-loop topology in the corresponding forward-scattering diagrams.
This phase-space enhancement implies that, despite entering at order $1/m_b^3$, spectator effects can be numerically comparable to or even dominate over some lower-order non-spectator terms. 
Consequently, they play a crucial role in explaining the observed differences in lifetimes between $B^+$, $B_d$, and $B_s$ mesons.

At leading order, the HQE starts with the dimension-3 operator $\mathcal{O}_3 = \bar{b} b$, which corresponds to the free $b$-quark decay. Its matrix element is normalized to unity up to small corrections, so it does not induce any lifetime differences among $B$ mesons. 
The dimension-4 operator, $\mathcal{O}_4 = \bar{b} i \!\not\!\!{D} b$, does not contribute independently because it can be reduced to the leading operator by the equations of motion of the heavy quark, and is therefore absent in the final expansion. At order $1/m_b^2$, two dimension-5 operators appear: the kinetic operator $\mathcal{O}_{\pi} = \bar{b} (i D)^2 b$ and the chromomagnetic operator $\mathcal{O}_G =\frac{g_s}{2} \, \bar{b} \,\sigma_{\mu\nu} G^{\mu\nu} b$. Their matrix elements parameterize the motion of the heavy quark inside the hadron and its interaction with the gluon field, respectively.\\
At order $1/m_b^3$, one encounters the dimension-6 contributions. These include both two-quark operators, such as the Darwin and spin–orbit terms, and the four-quark operators that describe spectator effects. The Darwin operator encodes information about the spatial variation of the gluon field inside the hadron, while the spin–orbit term represents the interaction between the heavy-quark spin and the background chromomagnetic field. The four-quark operators, on the other hand, describe nonperturbative corrections sensitive to the flavour and colour of the spectator quark. These contributions dominate the lifetime differences between $B^+$, $B_d$, and $B_s$ mesons.

Going beyond this order, the HQE includes dimension-7 operators, which arise at $\mathcal{O}(1/m_b^4)$.
While numerically suppressed by an additional power of $1/m_b$, such contributions can become relevant in precision studies of inclusive observables and play a central role in the analyses presented later in this thesis. 

The HQE framework is general and not limited to SM calculations. In \cref{Ch:3}, it is applied to include BSM contributions, such as those affecting the Darwin term, in a model-independent way. 
\cref{Ch:4} applies the HQE to the study of $B$-Mesogenesis, where
decays into a dark-sector fermion $\psi$ contribute to the total
inclusive decay rate and modify the $B$-meson lifetime ratios. 
Therefore, in this work, we apply the HQE in both SM and BSM contexts, where all perturbative contributions, $\Gamma_i$ and $\tilde{\Gamma}_i$, associated with the operators shown in \cref{eq:HQE}, are explicitly calculated. 
In both cases, however, the operator expansion remains the same in structure, with additional operators\footnote{Here, additional operators
refer to operators in the effective weak Hamiltonian. New-physics
contributions are matched onto the same HQE operator basis defined in
\cref{sec:HQE-basis-appendix}.} or modified Wilson coefficients capturing the effects of new physics. 
All operators that contribute to the observables studied here are presented explicitly along with their Wilson coefficients in the sections where they are relevant.\\

\section{Quark Propagator in an External Gluon Field}
\label{sec:FSGauge}
The evaluation of many quantities in the HQE
and LCSR requires the quark propagator in the
presence of an external gluon field. A systematic expansion of this
propagator can be obtained by treating the gluon field as a slowly
varying classical background. This approach was pioneered in the
context of QCD sum rules by Shifman, Vainshtein, and
Zakharov~\cite{Shifman:1978bx}, where it provided the foundation for
the OPE in external gluonic fields.

A particularly convenient choice is the Fock--Schwinger (FS)
gauge~\cite{Fock:1937dy,Schwinger:1951nm}, in which the gauge potential
is expressed directly in terms of the gluon field-strength tensor and
its covariant derivatives. This establishes a transparent connection
between quark-level Feynman diagrams and the local operators appearing
in the HQE. The same formalism also forms the basis of the derivation
of the LCSR discussed in the following section.

The expansion of the quark propagator relies on the hierarchy of scales
between the soft external gluon field and the hard momentum carried by
the propagating quark.
The characteristic momentum carried by the external gluon field is of order $\Lambda_{\text{QCD}}$, which is small compared to the virtuality or external momentum of the quark line appearing in the short-distance expansion. 
Consequently, the gluon field can be treated as a slowly varying, or “soft” background, while the quark propagator is expanded perturbatively in powers of this background field. To implement the expansion, we begin by recalling the definition of the free quark propagator,
\begin{equation}
\langle 0 | {\rm T} \big\{\psi (x), \bar \psi (y) \big\} | 0 \rangle = i  S_0(x,y)\,,
\label{eq:corr-free-quark}
\end{equation}
with
\begin{equation}
S_0(x,y) = \int \frac{d^4 p}{(2 \pi)^4} \, \, {\cal S}_0(p)\, e^{- i p \cdot (x - y)}, \quad {\cal S}_0(p) =  \frac{\slashed p + m}{p^2 - m^2 + i \varepsilon} \,,
\label{eq:Sxy-free-prop}
\end{equation}
which describes the propagation of a quark in the absence of any gluonic background.
When a soft background gluon field is present, the full propagator can be expressed as an expansion around the free propagator,
\begin{equation}
i S (x, y) = i S_{0}(x,y) + i S_{1}(x,y) + i S_{2}(x,y)+  \ldots \,.
\label{eq:quark-propagator-definition1}
\end{equation}
$S_{1}(x,y)$ and $S_{2}(x,y)$ denote the first and second order corrections, describing the interaction of the quark with one and two gluon fields, respectively,  while the ellipsis stand for higher order terms with more than two gluons.
Explicitly,
\begin{equation}
i S_{1}(x,y) =\int d^4 z \,\,  i S_{0} (x - z)\,\, i \slashed A (z)\,\, i S_{0} (z - y).
\label{eq:quark-propagator-definition7}
\end{equation}
\begin{equation}
i S_{2}(x,y) =\int d^4 z d^4 z'\,\,  i S_{0} (x - z)\,\, i \slashed A (z)\,\, i S_{0} (z - z')\,\, i \slashed A (z')\,\, i S_{0} (z'-y).
\label{eq:quark-propagator-definition2}
\end{equation}
\begin{figure}[t]
\centering
\begin{tikzpicture}
  \node (fig) at (0,0)
    {\includegraphics[width=0.92\textwidth]{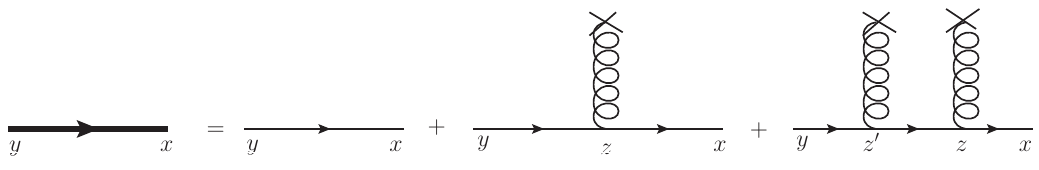}};

  \node[anchor=west] at ($(fig.east)+(-0.1cm,-0.65)$)
    {\footnotesize $+\cdots$};
\end{tikzpicture}
\caption{Diagrammatic representation of a quark propagating from $y$ to $x$ in an external gluon field, with zero, one, two, \ldots\ soft-gluon insertions.}
\label{fig:externalgluons}
\end{figure}
A schematic representation of \cref{eq:quark-propagator-definition1} is shown in \cref{fig:externalgluons}.

In the FS gauge, defined by the condition $x^\mu A_\mu(x) = 0$, the gauge potential at a point $x$ can be expressed in terms of the field–strength tensor and its covariant derivatives at the origin as
\begin{equation}
A_\mu^a (x) = 
\frac{1}{2} x^\rho G_{\rho\mu}^a(0) + \frac13 x^\rho x^\sigma D_\sigma G^a_{\rho \mu} (0)+ \frac18 x^\rho x^\sigma x^\eta D_\sigma D_\eta G^a_{\rho \mu} (0)+ \ldots \,,
\label{eq:A_mu}
\end{equation}
where $A_\mu^a$ denotes the gluon field, $G_{\mu\nu}^a$ the gluon
field-strength tensor, and $D_\mu$ the covariant derivative. Their
explicit definitions are collected in \cref{app:1}. The
ellipsis denotes terms involving higher covariant derivatives.
This representation ensures manifest gauge covariance of the expansion and makes explicit the connection between the propagator corrections and the local gluonic operators appearing in the OPE.
A notable feature of the FS gauge is that translational invariance is no longer manifest—the expansion is fixed around the reference point (commonly chosen as $x_0=0$) where the field–strength tensor is evaluated. 

A variant of the background field method in FS gauge was first introduced in Refs.~\cite{Dubovikov:1981bf, Shuryak:1981kj, Smilga:1982wx}. For comprehensive discussions of the technique and its applications, we refer the reader to the lecture notes~\cite{Shifman:1995dn}, the technical review~\cite{Novikov:1983gd}, and the references therein. For the purposes of this work, we present the explicit form of the quark propagator to the accuracy required, while a full derivation, following the same procedure without fundamentally new steps, is available in Ref.~\cite{Piscopo:2021ogu}.

In this work, the quark propagator will be expanded to generate operators up to dimension seven. This requires including terms with up to two covariant derivatives acting on the gluon field-strength tensor, as well as contributions containing two field-strength tensors. 

At this stage it is useful to note that, owing to the purely left-handed
$(V-A)$ structure of the SM current--current operators, only propagator
terms containing an odd number of Dirac matrices contribute to the
matching. Contributions involving an even number of gamma matrices
vanish after being contracted with the weak vertices and therefore do
not contribute to the two-quark matching coefficients. Consequently,
throughout this chapter only the terms with an odd number of Dirac
matrices are retained in the propagator expansion.

The quark propagator can be written in momentum space as
\begin{equation}
S(x,0)
=
\int\frac{d^4p}{(2\pi)^4}\,
{\cal S}(p)\,
e^{-ip\cdot x},
\label{eq:Sxy-prop}
\end{equation}
with
\begin{equation}
{\cal S}(p)
=
{\cal S}_0(p)
+
{\cal S}_1(p)
+
{\cal S}_2(p)
+\cdots,
\label{eq:Sp-expansion}
\end{equation}
where ${\cal S}_0(p)$ denotes the free propagator, while
${\cal S}_1(p)$ and ${\cal S}_2(p)$ contain one and two insertions of
the background gluon field, respectively. The expression for
${\cal S}_1(p)$ up to terms involving one covariant derivative on the
field-strength tensor is known e.g. from Refs.~\cite{Lenz:2020oce,Piscopo:2021ogu}. Here,
it is extended to include the contributions proportional to
$D_\sigma D_\eta G_{\rho\mu}$ required for the dimension-seven
matching, yielding
\begin{align}
{\cal S}_1(p)
={}&
\frac{\widetilde G_{\sigma\eta}\,p^\sigma\gamma^\eta\gamma^5}
     {(p^2-m^2+i\varepsilon)^2}
-
\frac{2}{3}
\frac{D_\alpha G_{\rho\mu}\,\gamma^\mu p^\alpha p^\rho}
     {(p^2-m^2+i\varepsilon)^3}
     +
2i\,
\frac{D_\alpha\widetilde G_{\tau\eta}\,
      p^\alpha p^\tau\gamma^\eta\gamma^5}
     {(p^2-m^2+i\varepsilon)^3}
\nonumber\\[2mm]
&
+\frac{2}{3}
\frac{D_\alpha G_{\alpha\mu}}
     {(p^2-m^2+i\varepsilon)^3}
\Big[
\gamma^\mu(p^2-m^2)
-p^\mu \slashed p
\Big]
-\frac{i}{8}\,
D_\sigma D_\eta G_{\rho\mu}\,
{\cal S}_{1}^{(7)\,\mu\rho\sigma\eta}(p)
+\ldots ,
\label{eq:S1p}
\end{align}
where
\(
\widetilde G_{\mu\nu}
=\frac12\epsilon_{\mu\nu\alpha\beta}G^{\alpha\beta}
\)
denotes the dual gluon field-strength tensor.
The last term in \cref{eq:S1p} contains the genuinely new
dimension-seven contribution. For compactness, the corresponding
Dirac--Lorentz structure has been collected in
${\cal S}_{1}^{(7)\,\mu\rho\sigma\eta}(p)$, which is given by
\begin{align}
{\cal S}_{1}^{(7)\,\mu\rho\sigma\eta}(p)
={}&
-2
\frac{\slashed p\,\gamma^\mu}
     {(p^2-m^2)^3}
\left(
\gamma^\sigma g^{\eta\rho}
+\gamma^\rho g^{\eta\sigma}
+\gamma^\eta g^{\rho\sigma}
\right)
\nonumber\\
&
+8
\frac{\slashed p\,\gamma^\mu}
     {(p^2-m^2)^4}
\left(
\gamma^\sigma p^\eta p^\rho
+\gamma^\rho p^\eta p^\sigma
+\gamma^\eta p^\rho p^\sigma
\right)
\nonumber\\
&
+8
\frac{2p^\mu\slashed p-\gamma^\mu(p^2-m^2)}
     {(p^2-m^2)^4}
\left(
p^\eta g^{\rho\sigma}
+p^\rho g^{\eta\sigma}
+p^\sigma g^{\eta\rho}
\right)
\nonumber\\
&
-48
\frac{2p^\mu\slashed p-\gamma^\mu(p^2-m^2)}
     {(p^2-m^2)^5}
p^\eta p^\rho p^\sigma .
\label{eq:S1p-dim7}
\end{align}
Similarly, for the propagator with reversed arguments we write
\begin{equation}
S(0,x)
=
\int \frac{d^4 p}{(2\pi)^4}\,
\widetilde{\cal S}(p)\,
e^{ip\cdot x},
\label{eq:S0x}
\end{equation}
with
\begin{equation}
\widetilde{\cal S}(p)
=
\widetilde{\cal S}_0(p)
+
\widetilde{\cal S}_1(p)
+
\widetilde{\cal S}_2(p)
+\ldots .
\label{eq:S0x-expansion}
\end{equation}
The one-gluon contribution is obtained analogously to \cref{eq:S1p}. It takes the form
\begin{align}
\widetilde{\cal S}_1(p)
={}&
\frac{\tilde G_{\sigma \eta}}{(p^2 -m^2  + i \varepsilon)^2} p^\sigma \gamma^\eta \gamma^5 - \frac23 \frac{p^\alpha D_\alpha  G_{\rho \mu}}{(p^2 -m^2 + i \varepsilon )^3} \gamma^\mu p^\rho
- 2 i \frac{D_\alpha \tilde G_{\tau \eta} \, p^\alpha p^\tau}{(p^2 - m^2 + i \varepsilon )^3}  \gamma^\eta \gamma^5
\nonumber \\[3mm]
&+ \frac23 \frac{D_\alpha G_{\alpha \mu}}{(p^2 - m^2 + i \varepsilon )^3} \, \Big[ \gamma^\mu \, (p^2 -m^2) - p^\mu \slashed p \Big] 
-\frac{i}{8}\,
D_\sigma D_\eta G_{\rho\mu}\,
\widetilde{\cal S}_{1}^{(7)\,\mu\rho\sigma\eta}(p)
+\ldots \,.
\label{eq:S1tildep}
\end{align}
As in the case of ${\cal S}_1(p)$, the genuinely new dimension-seven
contribution has been collected into the Dirac--Lorentz structure
$\widetilde{\cal S}_{1}^{(7)\,\mu\rho\sigma\eta}(p)$, defined by
\begin{align}
\widetilde{\cal S}_{1}^{(7)\,\mu\rho\sigma\eta}(p)
={}&
\frac{2}{(p^2-m^2)^3}
\left(
\gamma^\sigma g^{\eta\rho}
+\gamma^\rho g^{\eta\sigma}
+\gamma^\eta g^{\rho\sigma}
\right)
\slashed p\,\gamma^\mu
\nonumber\\
&-\frac{8}{(p^2-m^2)^4}
\left(
\gamma^\sigma p^\eta p^\rho
+\gamma^\rho p^\eta p^\sigma
+\gamma^\eta p^\rho p^\sigma
\right)
\slashed p\,\gamma^\mu
\nonumber\\
&-\frac{8}{(p^2-m^2)^4}
\left(
p^\eta g^{\rho\sigma}
+p^\rho g^{\eta\sigma}
+p^\sigma g^{\eta\rho}
\right)
\left(
2p^\mu\slashed p-\gamma^\mu(p^2-m^2)
\right)
\nonumber\\
&+\frac{48}{(p^2-m^2)^5}
\left(
2p^\mu\slashed p-\gamma^\mu(p^2-m^2)
\right)
p^\eta p^\rho p^\sigma .
\label{eq:S1tildep-dim7}
\end{align}
The contribution with two external-gluon insertions is obtained from
${\cal S}_2(p)$. Retaining only the leading term in the FS
expansion of the gauge field, namely the first term in
\cref{eq:A_mu}, and substituting it into
\cref{eq:quark-propagator-definition2} yields the propagator terms
containing two field-strength tensors. The corresponding momentum-space expression with finite quark mass is
\begin{align}
{\cal S}_2(p)
={}&
-\frac{\slashed{p}\gamma^\mu\gamma^\rho\gamma^\nu\gamma^\sigma}
{4(p^2-m^2)^3}
\,G_{\mu\rho}G_{\nu\sigma}
-\frac{\gamma^\mu g^{\nu\rho}}
{(p^2-m^2)^3}
\,p^\eta G_{\mu\rho}G_{\nu\eta}
-\frac{\gamma^\mu\gamma^\nu\gamma^\rho}
{(p^2-m^2)^3}
\,p^\eta G_{\nu\rho}G_{\mu\eta}
\nonumber\\
&
+\frac{1}{4}
\left(
\frac{2i \,\gamma^\rho \, \gamma^5 \epsilon^{\mu\nu\rho\eta} \, p^\eta-2\slashed{p}\,g^{\mu\nu}}
{(p^2-m^2)^3}
+\frac{4m^2\,\gamma^\nu p^\mu}
{(p^2-m^2)^4}
\right)
G_{\mu\lambda}G_{\nu\lambda}
\nonumber\\
&
+\frac{1}{4}
\left(
\frac{-8i \, \gamma^\rho \, \gamma^5 \epsilon^{\mu\nu\rho\eta} \, p^\eta+8\slashed{p}\,g^{\mu\nu}}
{(p^2-m^2)^4}
+\frac{24m^2 \,\gamma^\nu p^\mu}
{(p^2-m^2)^5}
\right)
p^\alpha p^\beta
G_{\mu\alpha}G_{\nu\beta}
\nonumber\\
&
+\frac{p^2\gamma^\nu}
{(p^2-m^2)^4}
\,p^\eta G_{\nu\rho}G_{\rho\eta}
-\frac{2\slashed{p}\,g^{\nu\rho}}
{(p^2-m^2)^4}
\,p^\alpha p^\beta
G_{\nu\alpha}G_{\rho\beta}
\nonumber\\
&
+\frac{2\slashed{p}}
{(p^2-m^2)^4}
\,p^\alpha p^\beta
G_{\sigma\alpha}G_{\sigma\beta}\,.
\label{eq:S2-prop-dim7}
\end{align}
Similarly, for the propagator with reversed arguments one obtains

\begin{samepage}
\begin{align}
\widetilde{\cal S}_2(p)
={}&
-\frac{\gamma^\rho\gamma^\mu\gamma^\sigma\gamma^\nu\slashed p}
{4(p^2-m^2)^3}
\,G_{\mu\rho}G_{\nu\sigma}
-\frac{\gamma^\nu g^{\mu\rho}}
{(p^2-m^2)^3}
\,p^\eta G_{\nu\rho}G_{\mu\eta}
-\frac{\gamma^\rho\gamma^\mu\gamma^\nu}
{(p^2-m^2)^3}
\,p^\eta G_{\mu\rho}G_{\nu\eta}
\nonumber\\
&
+\frac{1}{4}
\left(
\frac{2i \,\gamma^\rho \, \gamma^5 \epsilon^{\mu\nu\rho\eta} \, p^\eta-2\slashed{p}\,g^{\mu\nu}}
{(p^2-m^2)^3}
+\frac{4m^2\,\gamma^\nu p^\mu}
{(p^2-m^2)^4}
\right)
G_{\mu\lambda}G_{\nu\lambda}
\nonumber\\
&
+\frac{1}{4}
\left(
\frac{-8i \, \gamma^\rho \, \gamma^5 \epsilon^{\mu\nu\rho\eta} \, p^\eta+8\slashed{p}\,g^{\mu\nu}}
{(p^2-m^2)^4}
+\frac{24m^2 \,\gamma^\nu p^\mu}
{(p^2-m^2)^5}
\right)
p^\alpha p^\beta
G_{\mu\alpha}G_{\nu\beta}
\nonumber\\
&
+\frac{p^2\gamma^\nu}
{(p^2-m^2)^4}
\,p^\eta G_{\nu\rho}G_{\rho\eta}
-\frac{2\slashed p\,g^{\mu\sigma}}
{(p^2-m^2)^4}
\,p^\alpha p^\beta G_{\mu\alpha}G_{\sigma\beta}
\nonumber\\
&
+\frac{2\slashed p}
{(p^2-m^2)^4}
\,p^\alpha p^\beta G_{\sigma\alpha}G_{\sigma\beta}.
\label{eq:S20x-prop-dim7}
\end{align}
\end{samepage}

In the massless limit, $m\to0$, the propagator expressions in
\cref{eq:S1p-dim7,eq:S2-prop-dim7}, restricted throughout to terms
containing an odd number of Dirac matrices, reduce to the corresponding
results of Ref.~\cite{Novikov:1983gd}.
For finite quark masses, our results can be compared with the
background-field propagators derived in
Refs.~\cite{Rusov:2017chr,Chen:2011qu}.

Rusov~\cite{Rusov:2017chr} derived the massive background-field
propagator both in coordinate space, where the coefficients are
expressed in terms of the modified Bessel functions $K_n$, and in the
corresponding momentum-space representation. In both representations,
the gluon field strength is kept in the nonlocal form
$G_{\mu\nu}(ux)$, which is suitable for applications to LCSR.
For the HQE, however, a local operator expansion about a single
spacetime point is required. This is obtained by performing a
covariant Taylor expansion of the background field,
\begin{align}
G_{\mu\nu}(ux)
=
G_{\mu\nu}(0)
+u\,x^\alpha D_\alpha G_{\mu\nu}(0)
+\frac{u^2}{2}\,
x^\alpha x^\beta
D_\alpha D_\beta G_{\mu\nu}(0)
+\cdots ,
\end{align}
followed by replacing the coordinate factors with momentum derivatives
acting on the propagator and integrating by parts. Restricting to terms
linear in the background gluon field, the resulting local expression
is equivalent, after standard tensor and Dirac-algebra identities, to
\cref{eq:S1p-dim7}, including the terms proportional to
$D_\sigma D_\eta G_{\rho\mu}$.

An independent check of the terms containing two field-strength
tensors ($GG$) is provided by Ref.~\cite{Chen:2011qu}, where the
massive quark propagator is expanded to multiple insertions of the
background gluon field. The corresponding two-gluon contribution
agrees, after standard Dirac and tensor identities, with
\cref{eq:S2-prop-dim7}. Thus, the $DDG$ and $GG$ contributions to the
massive propagator are independently confirmed by
Refs.~\cite{Rusov:2017chr,Chen:2011qu}, respectively.

The reversed orientation $S(0,x)$ requires separate consideration.
Since the FS gauge singles out the origin, translational invariance is
not manifest and $S(0,x)$ cannot be obtained from $S(x,0)$ by a simple
translation. As both orientations appear naturally in the HQE, we
derive the corresponding local expansion for $S(0,x)$ explicitly
(see \cref{eq:S0x}).

%%%%%%%%%%%%%%%%%%%%%%%%%%%%
%%%%%%%%%%%%%%%%%%%%%%%%%%%%
%%%%%%%%%%%%%%%%%%%%%%%%%%%%

\section{Light-Cone Sum Rules}
\label{sec:LCSR}
In \cref{Ch:4}, we study exclusive $B$-meson decays within the
framework of LCSRs. This approach provides a
powerful nonperturbative tool for calculating hadronic matrix
elements in processes where both short-distance QCD dynamics and
long-distance hadronic effects play an important role, such as
exclusive heavy-meson decays, see e.g.~\cite{Khodjamirian:2020btr}.

The LCSR framework originated as an extension of the traditional QCD
sum rule approach developed by Shifman, Vainshtein, and
Zakharov~\cite{Shifman:1978bx,Shifman:1978by} in the late 1970s. In
that method, hadronic observables were related to QCD parameters and
vacuum condensates through a short-distance OPE. In the 1980s, Balitsky, Braun, and
Kolesnichenko~\cite{Balitsky:1986st,Balitsky:1989ry} reformulated
this framework near the light-cone, where vacuum-to-hadron matrix elements of nonlocal quark--gluon
operators are expanded in terms of hadronic distribution amplitudes
(DAs) instead of vacuum condensates. This light-cone expansion
enabled a systematic description of the longitudinal momentum
structure of hadrons and established a direct connection between
perturbative QCD and exclusive hadronic observables.

Since then, the LCSR approach has become an essential framework for
the study of exclusive processes involving heavy hadrons, combining
the advantages of the OPE with the flexibility of light-cone
factorization.
In the following, we describe the LCSR framework in a generic form in
order to outline the main conceptual steps and the general workflow of
the method before turning to the specific applications considered in
this thesis.

The starting point of the practical calculations in the LCSR approach is the analysis of suitably chosen correlation function involving an interpolating current for the hadron of interest and the relevant weak transition current. 
For a generic $B$-meson decay into a light hadron, one considers the correlator
\begin{equation}
\Pi_\mu(p,q) = i \int d^4x \, e^{iq \cdot x} \langle 0 | T\{ j_{\text{light}}(x), J_\mu^{\text{weak}}(0) \} | \bar{B}(p) \rangle ,
\label{eq:LCSR-correlator}
\end{equation}
where $J_\mu^{\text{weak}}$ denotes the relevant weak current and $j_{\text{light}}$ is an interpolating current that couples to the light hadronic final state. The external $B$ meson in this setup carries momentum $p$, while $q$ is the momentum transferred through the weak current.

The essential idea of the LCSR method is that the correlation
function in \cref{eq:LCSR-correlator} admits two complementary
representations, each valid in a different kinematical regime. For
large spacelike virtualities of the currents, the correlator is
dominated by short light-cone distances
($x^2\approx0$) and can be expanded using the light-cone OPE. In this
representation, the short-distance part of the correlation function
is calculable perturbatively in QCD, whereas the long-distance
dynamics are encoded in universal nonperturbative light-cone
distribution amplitudes (LCDAs). The nonlocal quark--gluon operators
appearing in the expansion are systematically organized according to
their twist, defined as the difference between the canonical
dimension and the Lorentz spin of the operator. Operators with lower
twist provide the leading contributions in the large-energy or heavy-mass limit,
while higher-twist terms are suppressed by additional powers of the scale. The corresponding LCDAs describe the longitudinal
momentum distribution of the partons inside the hadron and provide
the essential nonperturbative input of the LCSR framework.

On the other hand, for physical timelike momenta the same correlator
admits a hadronic representation. Using the unitarity relation, its
imaginary part in the variable $(p-q)^2$ can be written as
\begin{align}
\frac{1}{\pi}\,\mathrm{Im}\,
\Pi_\mu(p,q)
&=
\delta\!\left((p-q)^2-m_H^2\right)
\langle 0|j_{\text{light}}|H(p-q)\rangle
\langle H(p-q)|J_\mu^{\text{weak}}|\bar B(p)\rangle
\nonumber\\
&\quad
+
\rho^{\mathrm{had}}\!\left((p-q)^2,q^2\right),
\label{eq:LCSR-hadronic}
\end{align}
where $H$ denotes the lowest hadronic state carrying the quantum
numbers of the interpolating current $j_{\text{light}}$. The second
term, described by the spectral density
$\rho^{\mathrm{had}}((p-q)^2,q^2)$, contains the contributions of
all higher excited hadronic states and multiparticle continuum
configurations with the same quantum numbers. In general, the
hadronic representation is therefore not composed solely of isolated
pole contributions, but also contains continuous spectral
distributions associated with broad resonances and multiparticle
states.\footnote{
For example, if $j_{\text{light}}$ is an interpolating current with
the quantum numbers of a pion, the hadronic spectral density contains
not only the pion pole, but also excited pion resonances, such as
$\pi(1300)$ and $\pi(1800)$, together with multiparticle continuum
states (e.g.~$3\pi$, $\rho\pi$, etc.) carrying the same quantum
numbers.
}

A fundamental ingredient at the heart of the sum-rule method is the
dispersion relation, which follows from the analyticity properties of
the correlation function in the variable $(p-q)^2$. Exploiting this
analytic structure, the correlator can be represented as
\begin{equation}
\Pi_\mu(p,q)
=
\frac{1}{\pi}
\int ds\,
\frac{
\mathrm{Im}\,\Pi_\mu(s,q^2)
}{
s-(p-q)^2
}.
\label{eq:LCSR-dispersion}
\end{equation}

As discussed above, the same correlator can be evaluated in two
different kinematical regimes. In the physical timelike region, the
imaginary part of the correlator is expressed in terms of hadronic
intermediate states using the unitarity relation in
\cref{eq:LCSR-hadronic}. On the other hand, in the deep Euclidean
region, where $(p-q)^2\ll0$, the same correlation function is
calculated within QCD using the light-cone OPE. Its corresponding
spectral density is then obtained from the imaginary part of the OPE
result, yielding a perturbative spectral density convoluted with the
relevant LCDAs.

At this stage, an important conceptual difference between the two
representations becomes apparent. In the hadronic representation, the
spectral density is expressed in terms of physical hadronic states,
such as mesons, resonances, and multiparticle continuum
configurations. In contrast, in the deep Euclidean region, the
correlation function is calculated in terms of quark and gluon degrees
of freedom using the light-cone OPE, from which the corresponding
perturbative spectral density is obtained by taking its imaginary
part.

The central assumption allowing one to relate these two descriptions
is quark--hadron duality. More specifically, above an effective
continuum threshold $s_0$, the hadronic spectral density is assumed
to be sufficiently well approximated by its perturbative QCD
counterpart:
\begin{equation}
\mathrm{Im}\,\Pi_\mu^{\mathrm{had}}(s,q^2)
=
\mathrm{Im}\,\Pi_\mu^{\mathrm{OPE}}(s,q^2),
\qquad
s>s_0.
\label{eq:duality}
\end{equation}

Substituting the hadronic spectral representation from
\cref{eq:LCSR-hadronic} into the dispersion relation
\cref{eq:LCSR-dispersion}, one obtains the hadronic representation of
the correlator,
\begin{align}
\Pi_\mu^{\mathrm{had}}(p,q)
&=
\frac{
\langle 0|j_{\text{light}}|H(p-q)\rangle
\langle H(p-q)|J_\mu^{\text{weak}}|\bar B(p)\rangle
}{
m_H^2-(p-q)^2
}
+
\int\limits_{s_h}^{\infty} ds\,
\frac{
\rho^{\mathrm{had}}(s,q^2)
}{
s-(p-q)^2
}.
\label{eq:LCSR-hadronic-rep}
\end{align}
Here $s_h$ denotes the physical threshold at which
the higher-state contribution in the channel of
$j_{\text{light}}$ starts.\footnote{
For example, if $j_{\text{light}}$ carries the quantum numbers of the
pion, the lowest multiparticle continuum starts at the three-pion
threshold,
$s_h=(3m_\pi)^2=9m_\pi^2$.
}

In the Euclidean region, the same correlator is calculated from the
light-cone OPE. Its dispersion representation can be written as
\begin{equation}
\Pi_\mu^{\mathrm{OPE}}(p,q)
=
\frac{1}{\pi}
\int\limits_{s_{\rm OPE}}^{\infty} ds\,
\frac{
\mathrm{Im}\,\Pi_\mu^{\mathrm{OPE}}(s,q^2)
}{
s-(p-q)^2
},
\label{eq:LCSR-OPE-rep}
\end{equation}
where $s_{\rm OPE}$ is the partonic threshold determined by the
masses and approximations used in the OPE
calculation.\footnote{
 If $j_{\text{light}}$ carries the
$\pi^+$ quantum numbers, the perturbative spectral density corresponds to the
production threshold of the underlying $u\bar d$ quark pair,
$s_{\rm OPE}=(m_u+m_d)^2$. In the massless light-quark limit, this
reduces to $s_{\rm OPE}=0$.
}

Applying the quark--hadron duality approximation in
\cref{eq:duality} to the continuum parts of the dispersion relations,
one identifies the hadronic spectral density above the effective
threshold $s_0$ with the corresponding OPE spectral density. This
allows the higher-state contribution in
\cref{eq:LCSR-hadronic-rep} to be replaced by
\begin{equation}
\int\limits_{s_h}^{\infty} ds\,
\frac{\rho^{\mathrm{had}}(s,q^2)}
{s-(p-q)^2}
\simeq
\frac{1}{\pi}
\int\limits_{s_0}^{\infty} ds\,
\frac{
\mathrm{Im}\,\Pi_\mu^{\mathrm{OPE}}(s,q^2)
}
{s-(p-q)^2}.
\label{eq:LCSR-duality-continuum}
\end{equation}
As discussed above, the same correlation function admits both a
hadronic and an OPE description, corresponding to the physical
timelike and deep spacelike regions, respectively.
Writing the correlator in the form of the
dispersion relation in \cref{eq:LCSR-dispersion} allows one to connect
both kinematical domains through the analyticity of the correlation
function in the variable $s=(p-q)^2$. While the hadronic and OPE
calculations are each valid only in their corresponding kinematical
regions, the dispersion relation itself represents the analytic
continuation of the same correlator and is therefore valid throughout
the complex $s$-plane up to the physical singularities. In this way,
the hadronic and OPE representations correspond to the same
underlying analytic function and can therefore be matched:
\begin{equation}
\Pi_\mu^{\mathrm{had}}\!\left((p-q)^2,q^2\right)
=
\Pi_\mu^{\mathrm{OPE}}\!\left((p-q)^2,q^2\right),
\label{eq:LCSR-matching}
\end{equation}

Using the explicit hadronic and OPE 
representations in \cref{eq:LCSR-hadronic-rep,eq:LCSR-OPE-rep}, one obtains
\begin{align}
\frac{1}{\pi}
\int\limits_{s_{\rm OPE}}^{\infty} ds\,
\frac{
\mathrm{Im}\,\Pi_\mu^{\mathrm{OPE}}(s,q^2)
}{
s-(p-q)^2
}
&=
\frac{
\langle 0|j_{\text{light}}|H(p-q)\rangle
\langle H(p-q)|J_\mu^{\text{weak}}|\bar B(p)\rangle
}{
m_H^2-(p-q)^2
}
\nonumber\\
&\quad
+
\int\limits_{s_h}^{\infty} ds\,
\frac{
\rho^{\mathrm{had}}(s,q^2)
}{
s-(p-q)^2
}.
\label{eq:LCSR-before-duality}
\end{align}
Applying the quark--hadron duality approximation in
\cref{eq:LCSR-duality-continuum} and subtracting the continuum
contribution from both sides of
\cref{eq:LCSR-before-duality} isolates the ground-state pole
contribution.
This yields
\begin{align}
\frac{1}{\pi}
\int\limits_{s_{\rm OPE}}^{s_0} ds\,
\frac{\mathrm{Im}\,\Pi_\mu^{\mathrm{OPE}}(s,q^2)}{s-(p-q)^2}
=
\frac{
\langle 0|j_{\text{light}}|H(p-q)\rangle
\langle H(p-q)|J_\mu^{\text{weak}}|\bar B(p)\rangle
}{
m_H^2-(p-q)^2
}.
\label{eq:LCSR-preBorel}
\end{align}
It is important to emphasize that the duality relation
in \cref{eq:LCSR-duality-continuum} is weaker than the local duality
relation introduced in \cref{eq:duality}. The latter corresponds to a
pointwise equality between the hadronic and OPE spectral densities at
each value of $s$ above the effective threshold $s_0$. In practice,
however, such a local equality is not expected to hold because of the
resonance structure of the physical hadronic spectrum. Instead, the
duality assumption employed in QCD sum rules is understood in a
semi-local (or averaged) sense: after integration over a sufficiently
broad energy interval, the hadronic spectral density is approximated
by its perturbative QCD counterpart. It is this semi-local
duality that underlies the continuum subtraction in
\cref{eq:LCSR-duality-continuum}.

At this stage, the continuum contribution has been subtracted, but
the relation in \cref{eq:LCSR-preBorel} still contains possible
subtraction terms from the dispersion relation and depends explicitly
on the external variable $(p-q)^2$. To remove these subtraction
terms, improve the convergence of the OPE, and suppress the
sensitivity to the region close to the continuum threshold, one
performs a Borel transformation with respect to $(p-q)^2$.

The Borel transformation is defined by its action on the dispersion
kernel,
\begin{equation}
{\cal B}_{M^2}
\left[
\frac{1}{s-(p-q)^2}
\right]
=
\frac{1}{M^2}e^{-s/M^2},
\label{eq:BorelKernel}
\end{equation}
where $M^2$ is the Borel parameter. Applying this transformation to
both sides of \cref{eq:LCSR-preBorel} gives
\begin{align}
\frac{1}{\pi}
\int\limits_{s_{\rm OPE}}^{s_0} ds\,
e^{-s/M^2}\,
\mathrm{Im}\,\Pi_\mu^{\mathrm{OPE}}(s,q^2)
=
e^{-m_H^2/M^2}
\langle 0|j_{\text{light}}|H(p-q)\rangle
\langle H(p-q)|J_\mu^{\text{weak}}|\bar B(p)\rangle .
\label{eq:LCSR-Borel}
\end{align}
\cref{eq:LCSR-Borel} represents the generic structure of a
LCSR. The left-hand side contains the OPE
representation of the correlation function, calculated in QCD in
terms of perturbative kernels and LCDAs. The right-hand side is
expressed through hadronic quantities associated with the lowest
physical state, such as decay constants and form factors. In this
way, the sum rule connects the quark--gluon description of QCD with
observable hadronic parameters.\footnote{
The name ``sum rule'' originates from the unitarity relation in
\cref{eq:LCSR-hadronic}, where the correlation function is expressed
as a sum over hadronic states and their corresponding spectral
densities. After applying quark--hadron duality and the Borel
transformation, one obtains a relation (or ``rule'') that allows this
spectral sum to be evaluated in terms of QCD quantities.
}

To summarize, the LCSR method starts from a suitably chosen
correlation function involving the weak transition current and an
interpolating current for the hadronic state of interest. Using the
unitarity relation, the correlator is expressed in terms of hadronic
states, thereby relating it to the desired hadronic matrix element.
The same correlation function is then calculated in the deep
Euclidean region using the light-cone OPE within QCD. Assuming
quark--hadron duality, the contributions of higher hadronic states
and continuum configurations are approximated by the corresponding
OPE spectral density above an effective threshold. Finally, a Borel
transformation is applied to suppress the higher-state contributions
and improve the convergence of the expansion. The resulting sum rule
relates the perturbatively calculable OPE expression to the desired
hadronic parameters, such as decay constants and form factors. 
%The main steps entering the derivation of the LCSR
%are schematically summarized in \cref{fig:LCSR}. 

The main theoretical uncertainties in the LCSR approach typically
originate from the choice of the continuum threshold $s_0$, the
Borel parameter $M^2$, and the nonperturbative input entering the
LCDAs. The continuum threshold and Borel window are not arbitrary, but are
chosen according to the standard criteria of QCD sum rules. In
practice, the continuum threshold $s_0$ is typically determined from
the corresponding two-point sum rule and reflects the onset of excited
hadronic states and the continuum. The Borel parameter is then chosen
such that the light-cone OPE remains convergent while the ground-state
contribution is sufficiently enhanced relative to higher-state and
continuum contributions. Residual variations with respect to $s_0$ and
the Borel parameter are included in the uncertainty estimate of the
final results.

Having introduced the LCSR approach, it is instructive to briefly
compare it with another major first-principles non-perturbative method,
namely lattice QCD, originally introduced by
Wilson~\cite{Wilson:1974sk}. In this framework, spacetime is
discretized into a finite Euclidean lattice, allowing hadronic
observables to be computed numerically from first principles; for a
recent review, see~\cite{FlavourLatticeAveragingGroupFLAG:2024oxs}. In
contrast, LCSR is based on the light-cone OPE
and combines perturbatively calculable coefficient functions with
universal  LCDAs. Although
lattice QCD is systematically improvable and, for many observables,
provides the most precise theoretical predictions, it is
computationally demanding, requiring large-scale numerical
simulations on high-performance computing facilities. Consequently,
the calculation of a new hadronic quantity or the inclusion of
additional operator structures often represents a substantial
computational effort. By comparison, LCSR provides analytic
expressions that can be adapted more readily to new processes and
BSM operator structures. Moreover, in the context of hadronic form factors, lattice QCD and
LCSR are applicable in complementary kinematic regimes. Lattice QCD
provides the most precise results at low hadronic recoil (large
momentum transfer, large $q^2$), whereas LCSR is naturally suited to
the large-recoil region (small momentum transfer, small $q^2$), where
the light-cone expansion converges most effectively. Although lattice
QCD is applicable to a much broader class of non-perturbative
observables, including hadron masses, decay constants, bag
parameters, and matrix elements of local operators, the comparison
here is restricted to form-factor calculations. In this context,
LCSR and lattice QCD should be regarded as complementary rather than
competing approaches. Their predictions are frequently combined to
obtain reliable form-factor determinations over the full kinematic
range, while providing valuable cross-checks in the region where both
methods are applicable.

The explicit form of the correlation functions, the choice of
interpolating currents, and the relevant distribution amplitudes
employed in this work will be introduced and discussed in detail in
\cref{Ch:4}, where the general LCSR framework described above will be
applied to the study of the $B\to p$ transition form factors.
\begin{comment}
\begin{figure}[t]
    \centering
    \includegraphics[width=0.96\textwidth]{LCSR_graph.png}
    
\caption{
Schematic illustration of the main steps entering the derivation of
light-cone sum rules. The figure was generated with the assistance of
ChatGPT (OpenAI).
}
\label{fig:LCSR}
\end{figure}
\end{comment}
%%%%%%%%%%%%%%%%%%%%%%%%%%%%
%%%%%%%%%%%%%%%%%%%%%%%%%%%%
%%%%%%%%%%%%%%%%%%%%%%%%%%%%
\newpage
\newpage

\chapter{SM Dimension-Seven Two-Quark Contributions}
\label{Ch:2}
\vspace{3mm}
\begin{flushright}
\vspace{5mm}
\emph{
``I do not like that we get these new matrix elements.''\\
``Then complain.''\\
``To whom?''\\
``To God. Why did you make nature this way?''
}
\\[0.3cm]
--- A Discussion with Alexey Pivovarov
\end{flushright}
\vspace{5mm}
In this chapter, the complete two-quark contributions to the
dimension-seven term $\Gamma_7^{(0)}$ in the HQE are derived at leading order
in QCD, corresponding to corrections of order
$\mathcal{O}(1/m_b^4)$ to the inclusive non-leptonic decay width.
These contributions constitute the  $1/m_b$ corrections to the
dimension-six two-quark Darwin term $\Gamma_6$. The analysis is restricted to the
CKM-favoured transitions $b\to c\bar{u}d$ and $b\to c\bar{c}s$, which
provide the dominant contributions to the total decay width of
$B$ mesons. The calculation is performed within the framework
introduced in \cref{Ch:1}, and the complete analytical matching
coefficients for the dimension-seven two-quark operator basis are
presented.
\\
The chapter begins with a discussion of the HQE operator basis and the
matching procedure, followed by the derivation of the corresponding
Wilson coefficients. The resulting analytical expressions constitute the
first complete determination of the dimension-seven matching
coefficients for the CKM-favoured non-leptonic channels and provide an
important ingredient for future precision studies of heavy-hadron
lifetimes and related observables.
\section{Outline of the Calculation}
\label{sec:SMcalculation}

The starting point of the calculation is the transition operator
entering  \cref{eq:GammaB},
\begin{equation}
\mathcal{T}
=
\mathrm{Im}\, i\!\int d^4x\,
T\!\left\{
\mathcal{H}_{\rm eff}(x)
\mathcal{H}_{\rm eff}(0)
\right\},
\label{eq:trans_op}
\end{equation}
whose forward matrix element determines the inclusive decay width within
the HQE. The effective weak Hamiltonian is given by the SM Hamiltonian in
\cref{eq:Heff-SM}. The calculation presented in this chapter is restricted to the
current--current sector of the SM effective Hamiltonian for the generic
non-leptonic transition $b\to q_1\bar{q}_2q_3$. Accordingly, only the
operators $Q_1$ and $Q_2$ are considered. For completeness,
they are given by
\begin{equation}
\begin{aligned}
Q_1^{q_1q_2q_3}
&=
(\bar q_1^i\gamma_\mu P_L b^i)
(\bar q_3^j\gamma^\mu P_L q_2^j),
\qquad
Q_2^{q_1q_2q_3}
=
(\bar q_1^i\gamma_\mu P_L b^j)
(\bar q_3^j\gamma^\mu P_L q_2^i).
\end{aligned}
\end{equation}
Throughout this chapter, the operators are evaluated for the
$b\to c\bar{u}d$ and $b\to c\bar{c}s$ transitions.

The transition
operator can be decomposed according to the insertion of the
current--current operators as
\begin{equation}
{\cal T}^{(q_1\bar q_2q_3)}
=
C_1^2\,{\cal T}_{11}^{(q_1\bar q_2q_3)}
+
2\,C_1C_2\,{\cal T}_{12}^{(q_1\bar q_2q_3)}
+
C_2^2\,{\cal T}_{22}^{(q_1\bar q_2q_3)}\,,
\label{eq:Tsecond}
\end{equation}
where ${\cal T}_{ij}^{(q_1\bar q_2q_3)}$ denotes the contribution
originating from the insertion of the operator pair
$Q_i^{q_1q_2q_3}$ and $Q_j^{q_1q_2q_3}$.

Substituting the current--current operators into the time-ordered
product in \cref{eq:GammaB} and performing the Wick contractions yields
the following expressions for the two-quark contributions:
\begin{align}
{\cal T}_{11}^{ (2q)} & =  - \,  \frac{G_F^2}{2}  |V_{q_1b}|^2 |V_{q_3q_2}|^2\,  i \int d^4 x \,\, \bar b^j \gamma_\nu (1-\gamma_5) \, i S^{( q_1)}_{jk} (0, x) \gamma_\mu (1-\gamma_5) b^k (x)
\nonumber \\[3mm]
& \times   {\rm Tr} \Biggl[\gamma^\nu (1-\gamma_5) i S^{( q_3)}_{lm} (0, x) \gamma^\mu (1-\gamma_5) i S^{( q_2)}_{ml} (x,0) \Biggr]  + (x \leftrightarrow 0)
\, ,
\label{eq:T2q-11}
\end{align}\\
\begin{align}
{\cal T}_{12}^{ (2q)} & =  - \,  \frac{G_F^2}{2}  |V_{q_1b}|^2 |V_{q_3q_2}|^2 \, 
 i \int d^4 x \, \,  \bar b^j \gamma_\nu (1-\gamma_5)  \, i S^{( q_1)}_{lk} (0, x) \gamma_\mu (1-\gamma_5) b^k (x) 
\nonumber \\[3mm]
& \times    {\rm Tr} \Biggl[\gamma^\nu (1-\gamma_5) \, i S^{( q_3)}_{jr} (0, x) \gamma^\mu (1-\gamma_5) i S^{( q_2)}_{rl} (x,0) \Biggr]  + (x \leftrightarrow 0)
\, ,
\label{eq:T2q-12}
\end{align}\\
\begin{align}
{\cal T}_{22}^{ (2q)}& =  - \,  \frac{G_F^2}{2}  |V_{q_1b}|^2 |V_{q_3q_2}|^2 \, 
  i \int d^4 x \, \,  \bar b^j  \gamma_\nu (1-\gamma_5) \, i S^{( q_1)}_{lm} (0, x) \gamma_\mu (1-\gamma_5) b^k (x)
\nonumber \\[3mm]
& \times   {\rm Tr} \Biggl[\gamma^\nu (1-\gamma_5) i S^{( q_3)}_{jk} (0, x) \gamma^\mu (1-\gamma_5) i S^{( q_2)}_{lm} (x,0) \Biggr]  + (x \leftrightarrow 0)
\, ,
\label{eq:T2q-22}
\end{align}
The overall minus sign in
\cref{eq:T2q-11,eq:T2q-12,eq:T2q-22} originates from the closed
light-quark fermion loop.
The notation $(x\leftrightarrow0)$ denotes the contribution obtained by
interchanging the spacetime coordinates $x$ and $0$, corresponding to
the second ordering of the effective Hamiltonians in the time-ordered
product.

To obtain the two-quark contributions at dimension seven, the matching
must be performed up to operators containing four covariant derivatives
acting on the heavy-quark fields, schematically of the form
\[
\bar b \, iD_\mu\, iD_\nu\, iD_\rho\, iD_\sigma\, b\, .
\]
Using the commutation relations of the covariant derivatives, these
operators can be expressed in terms of structures containing either two
field-strength tensors,
\[
G_{\mu\nu}G_{\rho\sigma},
\qquad
G_{\mu\nu}=i[iD_\mu,iD_\nu],
\]
or a single field-strength tensor accompanied by two additional
covariant derivatives,
\[
D_\mu D_\nu G_{\rho\sigma}
=
i\bigl[iD_\mu,[iD_\nu,[iD_\rho,iD_\sigma]]\bigr].
\]

\begin{figure}[t]
    \centering
    \begin{subfigure}[b]{0.32\textwidth}
        \centering
        \includegraphics[width=\linewidth]{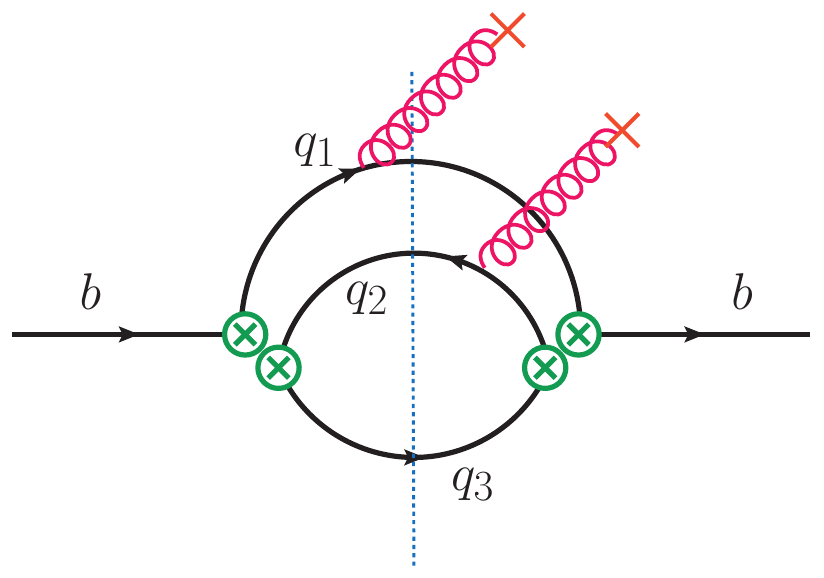}
        \caption{}
        \label{fig:2gluons-crossed}
    \end{subfigure}
    \hfill
    \begin{subfigure}[b]{0.32\textwidth}
        \centering
        \includegraphics[width=\linewidth]{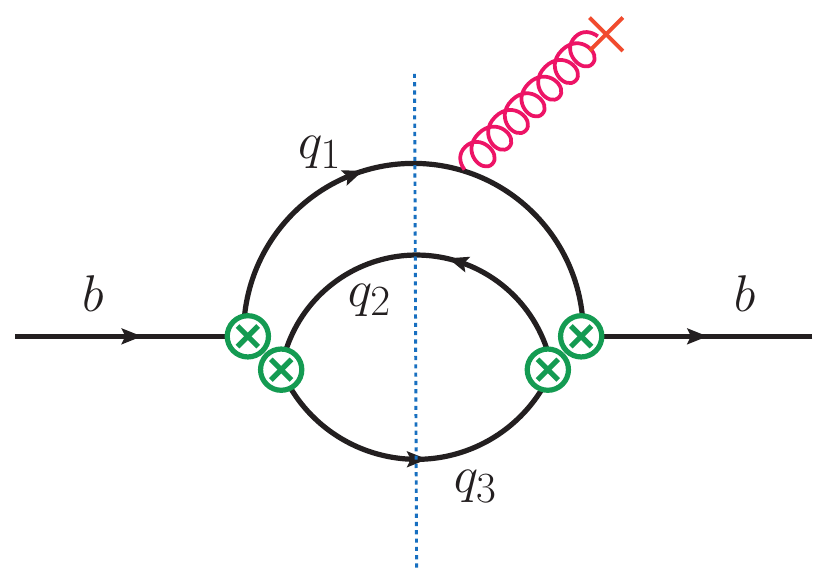}
        \caption{}
        \label{fig:1gluon-deriv}
    \end{subfigure}
    \hfill
    \begin{subfigure}[b]{0.32\textwidth}
        \centering
        \includegraphics[width=\linewidth]{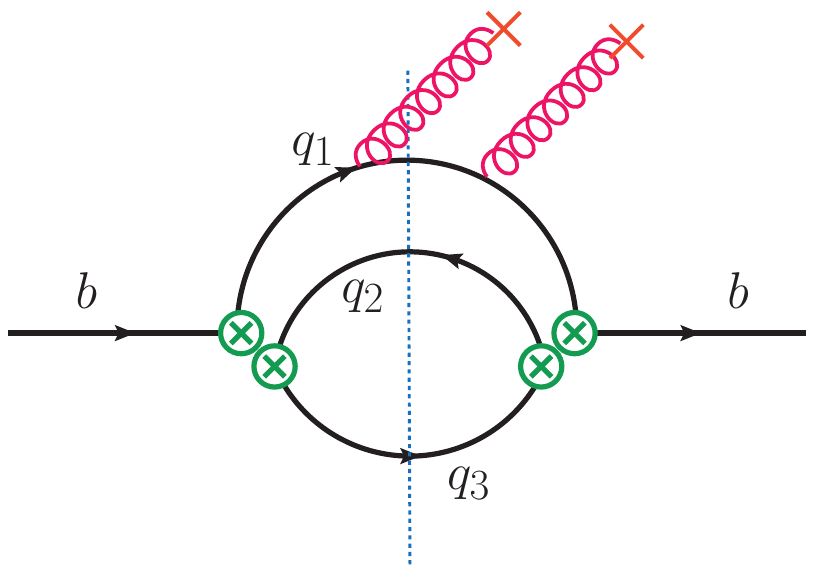}
        \caption{}
        \label{fig:2gluons-same}
    \end{subfigure}
\caption{Representative diagrams for dimension-7 two-quark contributions to 
         $b\to q_1 \bar{q}_2 q_3$ decays:
         (a) two gluons attached to different quark lines,
         (b) a single gluon with two covariant derivatives $D_{\mu}D_{\nu}G_{\rho\sigma}$,
         (c) two gluons attached to the same propagator.
         Shown here are representative topologies; the gluons can be attached to different 
         quark propagators depending on the colour structure (see text). Dashed blue lines indicate 
         the cut for forward matrix elements.}
    \label{fig:dim7-diagrams}
\end{figure}

Such structures are generated
by expanding the quark propagators in the external gluon field (see \cref{sec:FSGauge}). At
dimension seven, three distinct classes of propagator expansions
contribute:
\begin{enumerate}
\item \textbf{Different-propagator insertion (Fig.~\ref{fig:2gluons-crossed}):}
One background gluon is emitted from each of two different propagators,
giving rise to operators containing two field-strength tensors ($G_{\mu\nu}G_{\rho\sigma}$). These operators are generated by combining the first two terms of \cref{eq:S1p} (or
\cref{eq:S1tildep}), one from each propagator.

\item \textbf{Single-gluon insertion (Fig.~\ref{fig:1gluon-deriv}):}
A single background gluon is emitted from one propagator. The relevant
terms arise from the $D_\mu D_\nu G_{\rho\sigma}$ structures in the
one-gluon propagator expansion, see
\cref{eq:S1p,eq:S1p-dim7,eq:S1tildep}.

\item \textbf{Double-gluon insertion (Fig.~\ref{fig:2gluons-same}):}
Two background gluons are emitted from the same propagator through the
double-gluon term in the propagator expansion. The relevant
$G_{\mu\nu}G_{\rho\sigma}$ structures are contained in
\cref{eq:S2-prop-dim7,eq:S20x-prop-dim7}.
\end{enumerate}
Before proceeding with the explicit evaluation, we first analyse the
colour structure of the three classes of propagator expansions. The
colour algebra determines which external-gluon emissions give rise to
non-vanishing contributions and which diagrams vanish identically.
Consequently, only the allowed colour structures need to be considered
in the matching calculation. The momentum-space
propagators introduced above can be decomposed according to the number
of background-gluon insertions as
\begin{align}
{\cal S}^{(q_i)}_{jk}(p)
&=
{\cal S}^{(q_i)}_{0}(p)\,\delta_{jk}
+
{\cal S}^{(q_i)\,a}_{1}(p)\,t^a_{jk}
+
{\cal S}^{(q_i)\,ab}_{2}(p)\,(t^at^b)_{jk}
+\ldots ,
\label{eq:prop-colour-struct2}
\end{align}
where $i=1,2,3$ labels the quark flavour. The quantities
${\cal S}_0^{(q_i)}$, ${\cal S}_1^{(q_i)}$, and
${\cal S}_2^{(q_i)}$ denote the propagator contributions (see \cref{eq:Sxy-free-prop,eq:Sxy-prop,eq:Sp-expansion,eq:S0x,eq:S0x-expansion}), while the explicit colour
matrices have been factored out.
\begin{table}[t]
\centering
\setlength{\tabcolsep}{10pt}
\begin{tabular}{lccc}
\toprule
Contribution &
\makecell{Different\\propagators\\($GG$)} &
\makecell{Single\\insertion\\($DDG$)} &
\makecell{Double\\insertion\\($GG$)} \\
\midrule
$\mathcal{T}_{11}\;(Q_1\!\otimes\!Q_1)$
&
$(q_2,q_3)$
&
$q_1$
&
$q_1,\;q_2,\;q_3$
\\[1mm]

$\mathcal{T}_{22}\;(Q_2\!\otimes\!Q_2)$
&
$(q_1,q_2)$
&
$q_3$
&
$q_1,\;q_2,\;q_3$
\\[1mm]

$\mathcal{T}_{12}\;(Q_1\!\otimes\!Q_2)$
&
$(q_1,q_2),\;(q_1,q_3),\;(q_2,q_3)$
&
$q_1,\;q_2,\;q_3$
&
$q_1,\;q_2,\;q_3$
\\
\bottomrule
\end{tabular}

\caption{Allowed gluon-emission topologies for the different operator
products. The ordered pairs indicate the propagators from which the two
gluons are emitted.}
\label{tab:gluon-attachments}
\end{table}
To illustrate the role of the colour decomposition, consider the
contribution ${\cal T}_{11}^{(q_1\bar q_2 q_3)}$ in
\cref{eq:T2q-11}. The allowed propagator expansions are determined
entirely by the associated colour contractions:

\begin{itemize}
\item \textbf{Different-propagator insertion.}
When one background gluon is emitted from each of the $q_1$ and
$q_2$ propagators, the colour structure becomes
\begin{equation}
S^{(q_1)}_{jk}\,
S^{(q_3)}_{lm}\,
S^{(q_2)}_{ml}
\;\longrightarrow\;
t^a_{jk}\,
\delta_{lm}\,
t^b_{ml}
=
t^a_{jk}\,
\mathrm{Tr}(t^b)
=
0,
\end{equation}
and therefore this topology does not contribute.

\item \textbf{Single-gluon insertion.}
If the background gluon is emitted from the $q_1$ propagator while the
remaining propagators are taken at leading order, one finds
\begin{equation}
S^{(q_1)}_{jk}\,
S^{(q_3)}_{lm}\,
S^{(q_2)}_{ml}
\;\longrightarrow\;
t^a_{jk}\,
\delta_{lm}\,
\delta_{ml}
=
N_c\,t^a_{jk},
\end{equation}
which is non-vanishing.

\item \textbf{Double-gluon insertion.}
If both background gluons are emitted from the $q_1$ propagator, the
colour structure becomes
\begin{equation}
S^{(q_1)}_{jk}\,
S^{(q_3)}_{lm}\,
S^{(q_2)}_{ml}
\;\longrightarrow\;
(t^a t^b)_{jk}\,
\delta_{lm}\,
\delta_{ml}
=
N_c\,(t^a t^b)_{jk},
\end{equation}
which is likewise non-vanishing.
\end{itemize}
Applying the same procedure to the remaining gluon-emission topologies
and the other operator products yields the allowed gluon-emission
patterns summarized in \cref{tab:gluon-attachments}.

The calculation of the dimension-seven matching coefficients follows the
same general strategy as the determination of the Darwin-term
contribution at dimension six presented in Ref.~\cite{Piscopo:2021ogu}. The
individual steps can be summarized as follows:

\begin{enumerate}
\item \textbf{Insertion of the propagator expansion.} The appropriate
terms of the external-field propagator discussed in \cref{sec:FSGauge} are substituted into the
transition operators
in \cref{eq:T2q-11,eq:T2q-12,eq:T2q-22}, according to the gluon-emission
topology under consideration.

\item \textbf{Coordinate-space integrations.} The heavy-quark field is
written as
\begin{equation}
b(x)=e^{-ip_b\cdot x}\,b(0),
\end{equation}
where $p_b$ denotes the momentum of the external $b$ quark.\footnote{This
follows from translation invariance of the free field and reflects that
the external $b$ quark is described by a plane-wave state with momentum
$p_b$.}
The subsequent $x$-integration generates momentum-conserving
$\delta$-functions, allowing the loop momenta to be introduced.

\item \textbf{Loop-integral reduction.} After performing the
coordinate-space integration and using the resulting $\delta$-functions
to eliminate the constrained momenta, the remaining amplitudes can be
expressed in terms of scalar loop integrals of the generic form
\begin{equation}
\hspace*{-3.5mm}
{\cal I}_{n_1n_2n_3}
=
\int\frac{d^4l_1}{(2\pi)^4}
\int\frac{d^4l_2}{(2\pi)^4}
\frac{f(p_b,l_1,l_2)}
{\bigl[l_1^2-m_1^2\bigr]^{n_1}
 \bigl[l_2^2-m_2^2\bigr]^{n_2}
 \bigl[(l_1+l_2-p_b)^2-m_3^2\bigr]^{n_3}}\,,
\label{eq:MI-def}
\end{equation}
where $n_i\in\mathbb{N}_0$ and $f(p_b,l_1,l_2)$ denotes a polynomial in
the independent scalar products. The Feynman prescription
$+i\varepsilon$ is understood for all propagator denominators and is
suppressed throughout the following.

\item \textbf{IBP reduction.} Since the integrals are UV
finite, the calculation is performed directly in four dimensions. The
loop integrals are reduced to a small set of master integrals using
integration-by-parts (IBP) identities~\cite{Chetyrkin:1981qh,Laporta:2000dc}. The corresponding master integrals are defined in
\cref{sec:loop_int}; in particular, the sunrise master integrals are
introduced in \cref{eq:MI-def1,eq:MIs}. Their imaginary parts are known
analytically~\cite{Remiddi:2016gno} (see also
Refs.~\cite{Lenz:2020oce,Piscopo:2021ogu}).
The corresponding master integrals admit compact analytical
expressions whenever at most two internal masses are non-vanishing. For
the general case with three massive propagators, however, the analytic
structure becomes considerably more involved and is known to involve
elliptic functions~\cite{Pivovarov:1984ij,Groote:2000kz,Broedel:2017kkb,Broedel:2017siw}.
In the present calculation, finite light-quark masses are introduced
where required and are taken to zero only after the corresponding
expressions have been combined. Consequently, all master integrals can
be expressed analytically in terms of the known results with at most two
massive propagators, with the exception of the
$b\to c\bar cs$ transition. In this case, all three internal masses are
non-vanishing, such that the individual master integrals are not
available in closed analytical form. As discussed in
\cref{sec:loop_int}, the relevant combination entering the physical
matching coefficients becomes analytic only after combining the
two-quark contributions with the corresponding gluon corrections to the
four-quark operators.

\item \textbf{Matching onto the HQE operator basis.} After evaluating
the loop integrals and taking the imaginary part, the resulting
amplitudes assume the generic form
\begin{equation}
\mathcal{G}(p_b)\,
\langle B(p_B)|
\bar b_v(0)\,
\mathcal{M}^{\mu_1\cdots\mu_4}\,
(iD_{\mu_1})\cdots(iD_{\mu_4})\,
b_v(0)
|B(p_B)\rangle,
\label{eq:MEGeneral}
\end{equation}
where $\mathcal{G}(p_b)$ contains the kinematic dependence on the
internal quark masses and the external $b$-quark momentum $p_b$, while
$\mathcal{M}^{\mu_1\cdots\mu_4}$ denotes the corresponding Dirac-Lorentz
structure. The field 
$b_v$ is the HQET heavy-quark field introduced in \cref{eq:HQETfield}.
At order $\mathcal{O}(1/m_b^4)$, operators containing up to
four covariant derivatives are retained. Furthermore, employing the
decomposition
\begin{equation}
p_b^\mu = m_b v^\mu + iD^\mu,
\end{equation}
lower-dimensional operators generate additional contributions at
$\mathcal{O}(1/m_b^4)$ through the expansion in powers of the covariant
derivative. A detailed discussion of this procedure can be found in Appendix~A of
Ref.~\cite{Piscopo:2021ogu}. Finally,
the resulting matrix elements are projected onto the HQE operator basis
collected in \cref{eq:hadronic_parameters,eq:dim7MatrixElements,eq:color-decomp}
using the trace formulas of Ref.~\cite{Mannel:2023yqf}. 
In the
non-leptonic case, additional colour-singlet matrix elements arise,
requiring the supplementary trace relations derived in
\cref{sec:trace_formulas}.
\end{enumerate}

Following the procedure outlined above, the explicit evaluation reveals several general features of the
dimension-seven matching. For the contributions
${\cal T}_{11}^{(q_1\bar q_2 q_3)}$ and
${\cal T}_{22}^{(q_1\bar q_2 q_3)}$, all
different-propagator ($GG$) topologies and all single-gluon ($DDG$)
emissions vanish identically. For the mixed contribution
${\cal T}_{12}^{(q_1\bar q_2 q_3)}$, only a subset of the possible
topologies survives: among the different-propagator emissions, only the
$(q_1,q_3)$ configuration gives a non-vanishing result, while for the
single-gluon topology only emission from the $q_2$ propagator
contributes. In contrast, the double-gluon ($GG$) topology contributes
for all propagator attachments in
${\cal T}_{11}^{(q_1\bar q_2 q_3)}$,
${\cal T}_{12}^{(q_1\bar q_2 q_3)}$, and
${\cal T}_{22}^{(q_1\bar q_2 q_3)}$. These cancellations are not a
consequence of the colour structure alone, but arise from the interplay
between the Dirac algebra and the antisymmetry of the gluon
field-strength tensor. Similar patterns were already observed in the
dimension-five ($\mathcal{O}(1/m_b^2)$) HQE calculation in coordinate
space; see, for example, Ref.~\cite{Piscopo:2021ogu}. Furthermore,
although the FS gauge explicitly breaks translational
invariance, such that $S(x,0)\neq S(0,x)$ (see
\cref{sec:FSGauge}), the contributions involving $S(x,0)$ and
$S(0,x)$ always combine to produce an overall factor of two in the decay
rate. This behaviour is identical to that observed in previous
tree-level HQE calculations at lower dimensions.
\subsection{Mixing with Four-Quark Operators}
\label{sec:mixing}

\begin{figure}[t]
    \centering
    \raisebox{-0.09cm}{
    \begin{subfigure}[b]{0.3\textwidth}
        \centering
        \includegraphics[width=\linewidth]{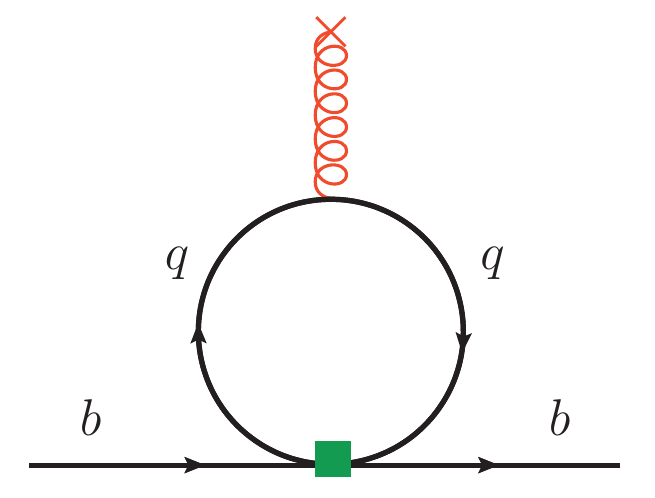}
        \caption{}
        \label{fig:mixonegluon}
    \end{subfigure}
    }
   \hspace{15mm}
    \begin{subfigure}[b]{0.3\textwidth}
        \centering
        \includegraphics[width=\linewidth]{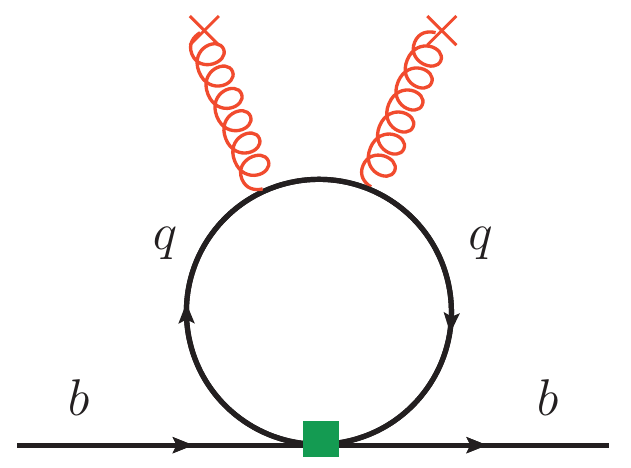}
        \caption{}
        \label{fig:mixtwogluon}
    \end{subfigure}
\caption{Diagrams for the mixing between two-quark and four-quark dimension-seven operators in the HQE.
         (a) Single gluon emission from a light quark propagator, corresponding to the $D^2G$ term.
         (b) Double gluon emission from a light quark propagator, corresponding to the $G^2$ term.
        }
    \label{fig:dim7-MixingDiagrams}
\end{figure}

Soft gluon emissions from massless light-quark propagators give rise to
infrared (IR) divergences in the matching calculation. In the present
work, these singularities are regulated by retaining finite light-quark
masses. Consequently, the IR divergences are represented by
logarithmic terms of the form
$\log\!\left({m_q}/{m_b}\right),$
where $m_q$ serves solely as an IR regulator.
These IR divergences cancel through the mixing of the gluon
corrections to the dimension-seven four-quark operators, illustrated in
\cref{fig:dim7-MixingDiagrams}, into the dimension-seven two-quark
operators. Consequently, the resulting
matching coefficients are IR finite.

In this work, we adopt the conventional normalized operator basis introduced in Ref.~\cite{Gabbiani:2004tp} to parameterize the matrix elements of the dimension-seven four-quark HQE operators. The corresponding operators are
\begin{align}
\nonumber
R_0^q &=
\frac{2}{m_b^2}\,
\bar b_i\gamma^{\mu}\gamma_5\vec{D}^{\alpha}b_i\,
\bar q_j\gamma_{\mu}P_L\vec{D}_{\alpha}q_j,
&
R_1^q &=
\frac{4}{m_b^2}\,
\bar b_i\gamma^{\mu}P_L\vec{D}^{\alpha}b_i\,
\bar q_j\gamma_{\mu}P_L\vec{D}_{\alpha}q_j,
\\[4pt]
R_2^q &=
\frac{4}{m_b^2}\,
\bar b_i\gamma^{\mu}P_L\vec{D}^{\nu}b_i\,
\bar q_j\gamma_{\nu}P_L\vec{D}_{\mu}q_j,
&
R_3^q &=
\frac{4m_q}{m_b}\,
\bar b_iP_Lb_i\,
\bar q_jP_Lq_j,
\label{eq:R-operators}
\end{align}
where $q=u,d,s$. The corresponding colour-rearranged operators,
denoted by $\widetilde R_i^q$, are obtained from $R_i^q$ by
interchanging the colour indices of the $b$- and $q$-quark fields.

To evaluate the mixing diagrams shown in
\cref{fig:dim7-MixingDiagrams}, we consider the one-loop matrix elements
of the four-quark operators in \cref{eq:R-operators} in the presence of a soft
background gluon field $A_\mu(x)$. These are obtained by computing the
time-ordered product of the operators $R_i^q$ (and analogously
$\widetilde R_i^q$) with the interaction part of the QCD Lagrangian in
\cref{eq:QCD_lag},
\begin{equation}
{\cal L}_{\rm QCD}^{\rm int}(x)
=
\sum_q
\bar q(x)\slashed A(x)q(x),
\end{equation}
which yields
\begin{align}
\langle R_i^q\rangle_{\rm 1-loop}
={}&
\left\langle
i\int d^4z\,
{\rm T}
\left\{
R_i^q,\,
{\cal L}_{\rm QCD}^{\rm int}(z)
\right\}
\right\rangle
\nonumber\\
&
+\frac{i^2}{2}
\left\langle
\int d^4z\,d^4z'\,
{\rm T}
\left\{
R_i^q,\,
{\cal L}_{\rm QCD}^{\rm int}(z)\,
{\cal L}_{\rm QCD}^{\rm int}(z')
\right\}
\right\rangle
+\ldots \,.
\label{eq:R-mixing-general}
\end{align}
The first term generates the single-gluon diagrams shown in
\cref{fig:mixonegluon}, whereas the second term gives rise to
the corresponding double-gluon diagrams in \cref{fig:mixtwogluon} . The ellipsis denotes
higher-order insertions of the interaction Lagrangian involving three or
more gluon fields, which do not contribute at the order considered in
this work.

The loop integrals generated by both the single- and double-gluon
insertions contain UV as well as IR
divergences. The UV divergences are regularized using dimensional
regularisation, working in $D=4-2\epsilon$ dimensions. Throughout the
calculation, the Dirac algebra is performed in the naive dimensional
regularisation (NDR) scheme, satisfying
\begin{equation}
\{\gamma_\mu,\gamma_\nu\}=2g_{\mu\nu},
\qquad
g_{\mu\nu}g^{\mu\nu}=D,
\qquad
\{\gamma_\mu,\gamma_5\}=0.
\label{eq:NDR}
\end{equation}
To ensure a consistent cancellation of the IR singularities with the
corresponding two-quark contributions, the IR divergences are
regularized by retaining finite masses for the light quarks, following
the same prescription adopted in the two-quark matching. Consequently,
the intermediate expressions contain logarithms of the form
$\log(m_q/m_b)$, which cancel in the final matching coefficients.

Before evaluating the one-loop matrix elements, it is convenient to
rewrite the heavy-quark field in terms of the HQET field according to
\begin{equation}
b(x)=e^{-im_bv\cdot x}b_v(x).
\end{equation}
Hence, a covariant derivative acting on the heavy-quark field
becomes
\begin{equation}
iD_\alpha b(x)
=
e^{-im_bv\cdot x}
\left(m_bv_\alpha+iD_\alpha\right)b_v(x).
\label{eq:DHQET}
\end{equation}
Moreover, in the FS gauge, the gauge potential vanishes at the origin,
$A_\mu(0)=0$ (see e.g. \cref{eq:A_mu}). Since all local operators are evaluated at the expansion
point $x=0$, the covariant derivative acting on the heavy-quark field
reduces to
\begin{equation}
iD_\alpha b(0)
=
\left(m_bv_\alpha+i\partial_\alpha\right)b_v(0).
\label{eq:DHQET-FS}
\end{equation}
This decomposition is applied throughout the evaluation of the
dimension-seven four-quark operator matrix elements. 
At the order considered in this work, the derivative acting on the
HQET field generates higher-order corrections and can therefore be
neglected, such that
\begin{equation}
D_\alpha b(0)\to -i\,m_b v_\alpha b_v(0).
\label{eq:field_dec}
\end{equation}

Using the decomposition in \cref{eq:field_dec} considerably simplifies the heavy-quark
structures appearing in the operators $R_i^q$. A further simplification
is achieved by performing a Fierz transformation of the four-quark
operators. This avoids the appearance of Dirac traces containing
$\gamma_5$, allowing the Dirac algebra to be carried out entirely within
the NDR scheme. Only the operators $R_0^q$ and $R_1^q$ are Fierz transformed. The
operators $R_2^q$ and $R_3^q$ are treated in their original form. Since
the Dirac indices in these operators are not part of the closed Dirac
traces arising in the loop calculation, the corresponding Dirac algebra
is identical in four and $D=4-2\epsilon$ dimensions. A Fierz
transformation is therefore unnecessary. The resulting operators read
\begin{align}
R_0^q
={}&
\frac{2i\,v^\alpha}{m_b}\,
\bar b_{v,i}\gamma_\mu P_L D_\alpha q_j\,
\bar q_j\gamma^\mu P_L\,b_{v,i}
+
\frac{4i\,v^\alpha}{m_b}\,
\bar b_{v,i}P_R D_\alpha q_j\,
\bar q_j P_L\,b_{v,i}
+\mathcal{O}\!\left(\frac{1}{m_b^2}\right),
\label{eq:R0-Fierz-HQET}
\\[3mm]
R_1^q
={}&
-\frac{4i\,v^\alpha}{m_b}\,
\bar b_{v,i}\gamma_\mu P_L D_\alpha q_j\,
\bar q_j\gamma^\mu P_L\,b_{v,i}
+\mathcal{O}\!\left(\frac{1}{m_b^2}\right),
\label{eq:R1-Fierz-HQET}
\\[3mm]
R_2^q
={}&
-\frac{4i\,v^\nu}{m_b}\,
\bar b_{v,i}\gamma^\mu P_L b_{v,i}\,
\bar q_j\gamma_\nu P_L D_\mu q_j
+\mathcal{O}\!\left(\frac{1}{m_b^2}\right),
\label{eq:R2-HQET}
\\[3mm]
R_3^q
={}&
4\,\frac{m_q}{m_b}\,
\bar b_{v,i}P_L b_{v,i}\,
\bar q_jP_L q_j .
\label{eq:R3-HQET}
\end{align}
The \(\mathcal O(1/m_b^2)\) terms shown for \(R_{0,1,2}\) 
originate from neglecting the residual derivative acting on the HQET field in \cref{eq:DHQET}. 
No analogous term appears for \(R_3\), since this operator contains no derivative acting on the heavy-quark field.

To illustrate the calculation of the mixing diagrams, we consider the
single-gluon contribution generated by the colour-rearranged operator
$\widetilde R_1^q$. The remaining operators and topologies are treated
analogously.
\begin{align}
\left\langle \widetilde R_1^q \right\rangle_{\rm 1-loop}
={}&
\Bigg\langle
i\int d^4z\,
{\rm T}
\Bigg\{
\frac{-4i\,v^\alpha}{m_b}\,
\bar b_{v,i}\gamma_\mu P_L D_\alpha q_i\,
\bar q_j\gamma^\mu P_L b_{v,j},
\bar q_l(z)\slashed A_{lm}(z)\,q_m(z)
\Bigg\}
\Bigg\rangle .
\label{eq:R1tilde-mixing}
\end{align}
Contracting the light-quark fields gives
\begin{align}
\left\langle \widetilde R_1^q \right\rangle_{\rm 1-loop}
={}&
\frac{4v^\alpha}{m_b}
\int d^4z\,
\left\langle
\bar b_{v,i}\gamma_\mu P_L
\left[D_\alpha S^{(q)}_{il}(0,z)\right]
\slashed A_{lm}(z)
S^{(q)}_{mj}(z,0)
\gamma^\mu P_L b_{v,j}
\right\rangle .
\label{eq:R1tilde-contracted}
\end{align}
Using the FS gauge, where $A_\alpha(0)=0$, the covariant derivative
acting on the endpoint of the propagator reduces to an ordinary
derivative,
\begin{equation}
D_\alpha S^{(q)}(0,z)
=
\partial_\alpha S^{(q)}(x,z)\big|_{x=0}.
\end{equation}
Using the momentum-space representation of the propagator, this becomes
\begin{equation}
D_\alpha S^{(q)}(0,z)
=
\int\frac{d^Dp}{(2\pi)^D}\,
(-ip_\alpha)\,
{\cal S}^{(q)}(p)\,
e^{ip\cdot z}.
\end{equation}
Therefore, at the level of the momentum-space integrand, the action
of the derivative amounts to inserting a factor $-ip_\alpha$ on the
corresponding propagator line,
\begin{equation}
\left[D_\alpha S^{(q)}_{il}(0,z)\right]
\longrightarrow
-i p_\alpha\,S^{(q)}_{il}(0,z).
\end{equation}
Using this replacement in \cref{eq:R1tilde-contracted}, one obtains
\begin{align}
\left\langle \widetilde R_1^q \right\rangle_{\rm 1-loop}
={}&
-\frac{4i\,v^\alpha}{m_b}
\int d^4z\,
\left\langle
\bar b_{v,i}\gamma_\mu P_L\,
p_\alpha\,
S^{(q)}_{il}(0,z)
\slashed A_{lm}(z)
S^{(q)}_{mj}(z,0)
\gamma^\mu P_L b_{v,j}
\right\rangle .
\label{eq:R1tilde-after-derivative}
\end{align}
We recognize the combination
\begin{equation}
\int d^4z\,
iS_0^{(q)}(0,z)\,
i\slashed A(z)\,
iS_0^{(q)}(z,0)
=
\lim_{x\to0}iS_1^{(q)}(x,0),
\label{eq:S1-local}
\end{equation}
as the limit of the one-gluon contribution to the quark
propagator, cf.~\cref{eq:quark-propagator-definition7}. In the
FS gauge, the latter admits the Fourier representation
\begin{equation}
\lim_{x\to0}iS_1^{(q)}(x,0)
=
\int\frac{d^Dl}{(2\pi)^D}\,
i{\cal S}_1^{(q)}(l),
\label{eq:S1-local-fourier}
\end{equation}
where ${\cal S}_1^{(q)}(l)$ is given in \cref{eq:S1p}. 
Inserting the explicit propagator expression in
\cref{eq:S1-local-fourier}, evaluating the resulting loop integrals
using the scalar tadpole formula of \cref{eq:scalar-tadpole}, and
simplifying the Dirac algebra in dimensional regularisation, the matrix
element is reduced to local two-quark operators. Finally, using the trace
formulas of Ref.~\cite{Mannel:2023yqf} to express these operators in
terms of the HQE basis ($s_B^4$, $s_{qB}^4$, $r_E^4$, $r_G^4$, $\ldots$) defined in
\cref{eq:hadronic_parameters,eq:dim7MatrixElements,eq:color-decomp}, one
obtains
\begin{equation}
\langle \widetilde R_1^q\rangle_{\rm 1-loop}
=\frac{m_B}{m_b}
\frac{\left(s_B^4-s_{qB}^4\right)}{48\pi^2}
\left[
-\frac{1}{\epsilon}
+\gamma_E
-1
-\log(4\pi)
+\log\!\left(\frac{m_q^2}{\mu^2}\right)
\right].
\label{eq:R1tilde-result}
\end{equation}

The UV divergence appearing in the matrix element
\cref{eq:R1tilde-result} is removed by the standard renormalisation
procedure in the $\overline{\rm MS}$ scheme. The corresponding
divergence is absorbed into the renormalisation constant of the
dimension-seven four-quark operator, from which the associated
anomalous dimension is obtained. The renormalised matrix element is then
evaluated at the matching scale $\mu=m_b$. For further details on the
renormalisation procedure and the derivation of the anomalous
dimensions, we refer the reader to Ref.~\cite{Piscopo:2021ogu}.
The same procedure applies to all remaining dimension-seven four-quark
operators, for both the single- and double-gluon   topologies.
The resulting renormalised matrix elements read
\begin{align}
\langle R^q_0 \rangle^{\mathrm{ren}} &= 0, \qquad
\langle R^q_1 \rangle^{\mathrm{ren}} = 
\frac{m_B}{24\pi^2 m_b} \left[ \log\!\left(\frac{m_b^2}{m_q^2}\right) - 1 \right] 
\big( 2r_E^{(1)} - 3r_G^{(1)} \big), \\[4pt]
\langle R^q_2 \rangle^{\mathrm{ren}} &= 
\frac{m_B}{24\pi^2 m_b} \log\!\left(\frac{m_b^2}{m_q^2}\right)\,
\big( 2r_E^{(1)} - 3r_G^{(1)} \big), \\[4pt]
\langle R^q_3 \rangle^{\mathrm{ren}} &= 
\frac{m_B}{12\pi^2 m_b} \big( 2r_E^{(1)} - r_G^{(1)} \big),
\end{align}
and for the colour-rearranged operators we obtain
\begin{align}
\langle \widetilde{R}^q_0 \rangle^{\mathrm{ren}}
&=
\frac{m_B}{96\pi^2 m_b}
\Big[
- 2\log\!\left(\frac{m_b^2}{m_q^2}\right)
\big(4s_B^4 - 4s_E^4 + s_{qB}^4\big)
\nonumber\\
&\qquad\qquad
+\,8r_E^4 - 2r_G^4 - 4s_B^4 + 4s_E^4 - s_{qB}^4
\Big], \\[4pt]
\langle \widetilde{R}^q_1 \rangle^{\mathrm{ren}}
&=
\frac{m_B}{48\pi^2 m_b}
\Big[
\log\!\left(\frac{m_b^2}{m_q^2}\right)
\big(8r_E^4 - 2r_G^4 + 4s_B^4 - 4s_E^4 + s_{qB}^4\big)
\nonumber\\
&\qquad\qquad
-\,8r_E^4 + 2r_G^4 + 4s_B^4 - 4s_E^4 + s_{qB}^4
\Big], \\[4pt]
\langle \widetilde{R}^q_2 \rangle^{\mathrm{ren}}
&=
\frac{m_B}{48\pi^2 m_b}
\log\!\left(\frac{m_b^2}{m_q^2}\right)
\Big[
8r_E^4 - 2r_G^4 - 4s_E^4 - s_{qB}^4
\Big], \\[4pt]
\langle \widetilde{R}^q_3 \rangle^{\mathrm{ren}}
&=
\frac{m_B}{12\pi^2 m_b}\,r_G^4.
\end{align}
Note that the operators $R_i$ receive contributions only from the colour-singlet two-quark operators defined in \cref{eq:color-decomp}, while the colour-rearranged operators $\widetilde{R}_i$ contain contributions from the combined colour-singlet and colour-octet structures given in \cref{eq:dim7MatrixElements}.

The cancellation of the IR logarithms is implemented at the level
of the physical dimension-seven contribution to the width. The
four-quark contributions at order $1/m_b$ are first written in terms of
the operators $R_i^q$ and $\widetilde R_i^q$, following
Ref.~\cite{Gabbiani:2004tp}. In a second step, the renormalised gluonic
matrix elements of these operators, derived above, are inserted into the
corresponding expressions for $\delta_{1/m}^q$. This generates
additional contributions to the same dimension-seven two-quark HQE
matrix elements that appear in the direct two-quark matching.

The finite coefficients quoted in the next section are obtained by adding these
induced four-quark contributions to the direct two-quark contributions
from ${\cal T}_{11}^{(q_1\bar q_2q_3)}$,
${\cal T}_{12}^{(q_1\bar q_2q_3)}$, and
${\cal T}_{22}^{(q_1\bar q_2q_3)}$. In this sum, the logarithms
depending on the auxiliary light-quark masses cancel, leaving
regulator-independent matching coefficients for the dimension-seven
two-quark operator basis.

In addition to the renormalised matrix elements of the dimension-seven
four-quark operators discussed above, the expansion of the
dimension-six Darwin contribution, parameterized by the HQE matrix
element $\rho_D^3$, to $\mathcal{O}(1/m_b^4)$ also induces mixing into
the dimension-seven HQE basis. Consequently, the renormalised matrix
elements of the corresponding dimension-six four-quark operators are
likewise required.
Following
Ref.~\cite{Lenz:2022pgw}, we employ the operator basis
\begin{align}
  \mathcal{O}^{q}_1 & =
  4\,(\bar b\,\gamma_\mu P_L\,q)
  (\bar q\,\gamma^\mu P_L\,b),
&
  \mathcal{O}^{q}_2 & =
  4\,(\bar b\,P_L\,q)
  (\bar q\,P_R\,b),
\label{eq:O1-O2}
\end{align}
whose renormalised matrix elements, expanded through the
dimension-seven HQE basis, are given by
\begin{align}
\langle \mathcal{O}^{q}_1 \rangle^{\mathrm{ren}}
&=
\frac{m_B}{3\pi^2}
\left[
\log\!\left(\frac{m_b^2}{m_q^2}\right)-1
\right]
\rho_D^3
\nonumber\\
&\quad
+\frac{m_B}{24\pi^2 m_b}
\left[
\log\!\left(\frac{m_b^2}{m_q^2}\right)-1
\right]
\left(
4r_G^4+4s_B^4-s_{qB}^4
\right)
+\mathcal{O}\!\left(
\frac{1}{m_b^2}
\right),
\\[2mm]
\langle \mathcal{O}^{q}_2 \rangle^{\mathrm{ren}}
&=
-\frac{m_B}{6\pi^2}
\log\!\left(\frac{m_b^2}{m_q^2}\right)
\rho_D^3
-\frac{m_B}{48\pi^2 m_b}
\log\!\left(\frac{m_b^2}{m_q^2}\right)
\left(
4r_G^4-4s_B^4+s_{qB}^4
\right)
+\mathcal{O}\!\left(
\frac{1}{m_b^2}
\right).
\end{align}
As for the dimension-seven four-quark operators, the renormalised matrix
elements given above are first inserted into the corresponding
expressions of Ref.~\cite{Lenz:2022pgw}. The contributions generated by
the expansion of the dimension-six Darwin term are then added to these
expressions. Finally, combining the resulting four-quark contributions
with the direct two-quark matching leads to the cancellation of all
logarithms depending on the auxiliary light-quark masses, yielding the
finite analytical results presented in the following section.

%%%%%%%%%%%%%%%%%%%%%%%%%%%%
%%%%%%%%%%%%%%%%%%%%%%%%%%%%
%%%%%%%%%%%%%%%%%%%%%%%%%%%%

\section{Analytical Results}
\label{sec:SMresults}

The dimension-seven two-quark contributions to the inclusive
non-leptonic decay width
$b\to q_1\bar q_2 q_3$ can be written as
\begin{align}
\Gamma^{(q_1\bar q_2 q_3)}_{1/m_b^4}(B)
=
\tilde{\Gamma}_0\,
\frac{1}{m_b^4}
\Big[
N_c C_1^2\,\mathcal W_{11}^{(q_1\bar q_2 q_3)}(\rho)
+
C_1C_2\,\mathcal W_{12}^{(q_1\bar q_2 q_3)}(\rho)
+
N_c C_2^2\,\mathcal W_{22}^{(q_1\bar q_2 q_3)}(\rho)
\Big],
\label{eq:Gamma-D7-main}
\end{align}
where
\begin{equation}
\tilde{\Gamma}_0
=
\frac{G_F^2m_b^5
|V_{q_1b}|^2
|V_{q_2q_3}|^2}
{192\pi^3},
\label{eq:Gamma0Tilde}
\end{equation}
$N_c=3$ denotes the number of colours and
$\rho=m_c^2/m_b^2$. The functions
$\mathcal W_{ij}^{(q_1\bar q_2q_3)}(\rho)$
encode the contributions associated with the operator products
$Q_i\otimes Q_j$. Each of them can be expressed as a linear combination
of the dimension-seven HQE matrix elements introduced in
\cref{eq:dim7MatrixElements,eq:color-decomp}. The explicit expressions
are given below.
\subsection
[\texorpdfstring{Results for the $b \to c \bar u d$ Transition}
{Results for the b -> c ubar d Transition}]
{Results for the $b \to c \bar u d$ Transition}
\label{sec:bcudSM}
{\footnotesize
\begin{align}
\mathcal{W}_{11}^{(c\bar ud)}
={}&
-\frac{8}{9}\,r_E^4\left(9\rho^4-20\rho^3+9\rho^2+6\log\rho+2\right)
+\frac{4}{9}\,r_G^4\left(3\rho^4-7\rho^3+9\rho^2-21\rho+12\log\rho+16\right)
\nonumber\\
&
-\frac{2}{9}\,s_E^4\left(9\rho^4-20\rho^3+36\rho-12\log\rho-25\right)
+\frac{2}{3}\,s_B^4(\rho-1)^3(5\rho+1)
\nonumber\\
&
-\frac{1}{36}\,s_{qB}^4\left(3\rho^4-16\rho^3+36\rho^2-48\rho+12\log\rho+25\right)
\nonumber\\
&
+\frac{2}{9N_c}\,r_E^{(1)}
\Bigl(
-26\rho^3-24\rho^3\log\rho
+48(\rho^3-1)\log(1-\rho)
+57\rho^2-12\rho-19
\Bigr)
\nonumber\\
&
-\frac{1}{N_c}\,r_G^{(1)}
\Bigl(
-8\rho^3\log\rho
+16(\rho^3-1)\log(1-\rho)
-(\rho-1)^2(6\rho+1)
\Bigr).
\label{eq:W11-cud}
\end{align}
}
{\footnotesize
\begin{align}
\mathcal{W}_{22}^{(c\bar{u}d)} &= 
-\frac{8}{9} r_E^4 \Bigl( 9\rho^4 - 62\rho^3 + 66\rho^2 - 6(4\rho-3)\rho^2\log\rho 
+ 12(4\rho^3-3\rho^2-1)\log(1-\rho) - 12\rho - 1 \Bigr) \nonumber \\
&\quad + \frac{2}{9} r_G^4 \Bigl( 6\rho^4 - 23\rho^3 - 6\rho^3\log\rho 
+ 12(\rho^3+3\rho-4)\log(1-\rho) + 3\rho^2 - 9\rho + 23 \Bigr) \nonumber \\
&\quad - \frac{2}{9} s_E^4 \Bigl( 9\rho^4 - 68\rho^3 + 102\rho^2 + 12(3-2\rho)\rho^2\log\rho 
- 12\rho \nonumber \\
&\qquad\qquad + 24(\rho-1)^2(2\rho+1)\log(1-\rho) - 31 \Bigr) \nonumber \\
&\quad + \frac{2}{3} s_B^4 \Bigl( 5\rho^4 - 23\rho^3 + 30\rho^2 - 6(\rho-2)\rho^2\log\rho 
- 5\rho + 12(\rho-1)^2\rho\log(1-\rho) - 7 \Bigr) \nonumber \\
&\quad - \frac{1}{36} s_{qB}^4 \Bigl( 3\rho^4 - 94\rho^3 + 96\rho^2 - 12(\rho-3)\rho^2\log\rho 
- 18\rho + 24(\rho-1)^3\log(1-\rho) + 13 \Bigr) \nonumber \\
&\quad + \frac{4}{9 N_c} r_{E}^{(1)}  \Bigl( 18(\rho^2-1)\log(1-\rho) + (3-9\rho^2)\log\rho 
+ 2(7\rho^3 - 18\rho^2 + 18\rho - 7) \Bigr) \nonumber \\
&\quad - \frac{2}{3 N_c} r_{G}^{(1)}  \Bigl( 18(\rho^2-1)\log(1-\rho) + (3-9\rho^2)\log\rho 
+ (5\rho+1)(\rho-1)^2 \Bigr).
\label{eq:F22-cud}
\end{align}
}

\begin{samepage}
{\footnotesize
\begin{align}
\mathcal{W}_{12}^{(c\bar{u}d)} &= 
-\frac{16}{9} r_E^4 \Bigl(
9\rho^4 - 76\rho^3 + 84\rho^2 + (-24\rho^3 + 36\rho^2 + 6)\log\rho
\nonumber\\
&\qquad\qquad
- 12\rho + 24(\rho-1)^2(2\rho+1)\log(1-\rho) - 5
\Bigr) \nonumber \\
&\quad + \frac{8}{9} r_G^4 \Bigl( 24(\rho^3 - 3\rho + 2)\log(1-\rho) + (3-12\rho^3)\log\rho 
+ (3\rho^2 - 20\rho - 10)(\rho-1)^2 \Bigr) \nonumber \\
&\quad - \frac{4}{9} s_E^4 \Bigl( 48(\rho^3-1)\log(1-\rho) - 12(2\rho^3+1)\log\rho 
+ (9\rho+11)(\rho-1)^3 \Bigr) \nonumber \\
&\quad + \frac{4}{9} s_B^4 \Bigl(
15\rho^4 - 92\rho^3 + 6\rho^2
+ 12(2\rho^3+9\rho^2-12\rho+1)\log(1-\rho)
\nonumber\\
&\qquad\qquad
- 6(2\rho^3+9\rho^2+1)\log\rho + 96\rho - 25
\Bigr) \nonumber \\
&\quad - \frac{1}{6} s_{qB}^4 (\rho-1) \Bigl( \rho^3 + \rho^2 + 31\rho - 96\log(1-\rho) + 87 \Bigr).
\label{eq:F12-cud}
\end{align}
}
\end{samepage}

\subsection
[\texorpdfstring{Results for the $b \to c \bar c s$ Transition}
{Results for the b -> c cbar s Transition}]
{Results for the $b \to c \bar c s$ Transition}
\label{sec:bccsSM}
For the $b\to c\bar cs$ channel, we define
$\xi_1=\sqrt{1-4\rho}$ and
$\xi_2=\log\!\left(\frac{1-\xi_1}{1+\xi_1}\right)$,
which will be used throughout the following expressions.
{\footnotesize
\begin{align}
\mathcal{W}_{11}^{(c\bar{c}s)} &= 
\frac{16}{9} \frac{r_E^4}{\xi_1^3} \Bigl[ 
\rho(80\rho^2 - 19\rho - 7)\xi_1 \xi_2 
- (54\rho^3 + 39\rho^2 - 4\rho + 1)\xi_1^4 
\nonumber \\
&\qquad\qquad
- (432\rho^5 + 132\rho^4 - 52\rho^3 - \rho^2 - 19\rho + 3)\xi_1 \xi_2
\Bigr] \nonumber \\
&\quad + \frac{2  r_G^4}{9 \xi_1} \Bigl[ 
-288\rho^4 - 144\rho^3 + 74\rho^2 - 133\rho + 32 
- 144\xi_1\rho^4 \xi_2 
+ 180\xi_1\rho^3 \log(1+\xi_1) \nonumber \\
&\qquad\qquad
+ 96\xi_1\rho^3 \xi_2 
- 18\xi_1\rho^2 \log(1+\xi_1) 
+ 14\xi_1\rho^2 \xi_2 
- 9\xi_1(20\rho^3 - 2\rho^2 + 2\rho + 1) \log(1-\xi_1) \nonumber \\
&\qquad\qquad
- 8\xi_1(4\rho+5)\rho \xi_2 
+ 18\xi_1\rho \log(1+\xi_1) 
+ 40\xi_1\rho \xi_2 
+ 9\xi_1 \log(1+\xi_1) 
+ 33\xi_1 \xi_2 \Bigr]
\nonumber \\
&\quad -\frac{2 s_E^4}{9 \xi_1} \Bigl[
-432\rho^4 - 348\rho^3 + 218\rho^2 + 74\rho - 25
\nonumber\\
&\qquad\qquad
- 4\xi_1(54\rho^4 + 48\rho^3 - 2\rho^2 + 2\rho + 3)\xi_2
+ 8\xi_1(8\rho + 1)\rho \xi_2
\Bigr] \nonumber \\
&\quad + \frac{2 s_B^4}{9\xi_1} \Bigl[
-720\rho^4 - 156\rho^3 + 264\rho^2 - 33\rho - 3
\nonumber\\
&\qquad\qquad
- 360\xi_1\rho^4 \xi_2
- 60\xi_1\rho^3 \log(1+\xi_1)
- 168\xi_1\rho^3 \xi_2
\nonumber\\
&\qquad\qquad
+ 18\xi_1\rho^2 \log(1+\xi_1)
+ 34\xi_1\rho^2 \xi_2
+ 3\xi_1(20\rho^3 - 6\rho^2 - 6\rho - 1)\log(1-\xi_1)
\nonumber\\
&\qquad\qquad
+ 4\xi_1(14\rho - 5)\rho \xi_2
+ 18\xi_1\rho \log(1+\xi_1)
+ 20\xi_1\rho \xi_2
+ 3\xi_1 \log(1+\xi_1)
+ 3\xi_1 \xi_2
\Bigr] \nonumber \\
&\quad -\frac{ s_{qB}^4}{36} \Bigl[ 
-72\rho^4 \xi_2 
+ 36\xi_1\rho^3 
- 120\rho^3 \log(1+\xi_1) 
- 144\rho^3 \xi_2 
+ 18\xi_1\rho^2 
+ 36\rho^2 \log(1+\xi_1) \nonumber \\
&\qquad\qquad
+ 60\rho^2 \xi_2 
+ 6(20\rho^3 - 6\rho^2 - 6\rho - 1) \log(1-\xi_1) 
- 52\xi_1\rho 
+ 25\xi_1 \nonumber \\
&\qquad\qquad
+ 48(\rho-1)\rho \xi_2 
+ 36\rho \log(1+\xi_1) 
+ 48\rho \xi_2 
+ 6\log(1+\xi_1) 
+ 18\xi_2 \Bigr]\nonumber \\
&\quad+ \frac{2 r_{E}^{(1)}}{9 N_c \, \xi_1^5} \Bigl[
-\xi_1^4 \Bigl(
48\rho^3 - 44\rho^2 + 40\rho + 1
+ 24(4\rho^2 - 2\rho + 1)\log\xi_1
\nonumber\\
&\qquad\qquad
- 12(4\rho^2 - 2\rho + 1)\log\rho
\Bigr)
\nonumber\\
&\qquad\qquad
+ 2\xi_1(200\rho^4 - 12\rho^3 + 90\rho^2 - 44\rho + 9)\xi_2
\nonumber\\
&\qquad\qquad
- 4\xi_1(96\rho^5 + 340\rho^4 - 144\rho^3 + 39\rho^2 - 10\rho + 3)\xi_2
\Bigr]\nonumber \\
&\quad -\frac{ r_{G}^{(1)}}{3 N_c \,\xi_1^5} \Bigl[
-3\xi_1^4 \Bigl(
16\rho^3 + 4\rho^2 + 28\rho - 5
+ 8(4\rho^2 - 2\rho + 1)\log\xi_1
\nonumber\\
&\qquad\qquad
- 4(4\rho^2 - 2\rho + 1)\log\rho
\Bigr)
\nonumber\\
&\qquad\qquad
+ 2\xi_1(176\rho^4 - 24\rho^3 + 138\rho^2 - 56\rho + 9)\xi_2
\nonumber\\
&\qquad\qquad
- 4\xi_1(96\rho^5 + 328\rho^4 - 150\rho^3 + 63\rho^2 - 16\rho + 3)\xi_2
\Bigr].
\label{eq:F11-ccs}
\end{align}
}

{\footnotesize
\begin{align}
\mathcal{W}_{22}^{(c\bar{c}s)} &= 
-\frac{8 r_E^4}{9 \xi_1^5} \Bigl[ 
2\rho(356\rho^3 - 66\rho^2 + 27\rho + 7)\xi_1 \xi_2 
- \xi_1^4 \bigl( 432\rho^4 + 156\rho^3 - 98\rho^2 + 22\rho + 1 \nonumber \\
&\qquad\qquad
+ 24(4\rho^2 - 2\rho + 1)\log\xi_1 
- 12(4\rho^2 - 2\rho + 1)\log\rho \bigr) \nonumber \\
&\qquad\qquad
- 2(1728\rho^6 - 96\rho^5 + 368\rho^4 - 162\rho^3 + 141\rho^2 - 41\rho + 6)\xi_1 \xi_2 \Bigr]\displaybreak[4]\nonumber\\
&\quad+\frac{2 r_G^4}{9 \xi_1^5} \Bigl[ 
8\rho(-16\rho^3 + 6\rho^2 + 9\rho + 1)\xi_1 \xi_2 
+ \xi_1^4 \bigl( -288\rho^4 - 24\rho^3 + 116\rho^2 - 112\rho + 29 \nonumber \\
&\qquad\qquad
+ 12\xi_1\xi_3\rho - 12\xi_1\xi_3
- 48(2\rho^2 - 7\rho + 2)\log\xi_1 
+ 24(2\rho^2 - 7\rho + 2)\log\rho \bigr) \nonumber \\
&\qquad\qquad
- 8(288\rho^6 - 96\rho^5 - 22\rho^4 - 135\rho^3 + 177\rho^2 - 56\rho + 6)\xi_1 \xi_2 \Bigr] \nonumber\\
&\quad -\frac{2 s_E^4}{9 \xi_1} \Bigl[ 
-432\rho^4 - 156\rho^3 + 338\rho^2 + 38\rho - 31 
+ 24(8\rho^2 + 2\rho - 1)\log\!\Bigl(\frac{\rho}{1-4\rho}\Bigr)\nonumber \\
&\qquad\qquad
- 8\xi_1(27\rho^4 + 12\rho^3 + 17\rho^2 + \rho - 3) \xi_2 
+ 8\xi_1(8\rho + 1)\rho \xi_2 \Bigr] \nonumber \\
&\quad +\frac{2 s_B^4}{3 \xi_1} \Bigl[ 
-240\rho^4 - 28\rho^3 + 110\rho^2 - 16\rho - 3 
+ 12\xi_1\rho^2 \xi_2 
+ 24\rho(1 - 4\rho)\log\!\Bigl(\frac{\rho}{1-4\rho}\Bigr) \nonumber \\
&\qquad\qquad
- 12\xi_1(10\rho^3 + 2\rho^2 + 3\rho - 2)\rho \xi_2 
+ 4\xi_1\xi_3(\rho - 1) \Bigr]\nonumber \\
&\quad -\frac{ s_{qB}^4}{36  \xi_1} \Bigl[ 
-144\rho^4 - 84\rho^3 + 310\rho^2 - 182\rho + 37 
+ 24(4\rho - 1)^2\log\!\Bigl(\frac{\rho}{1-4\rho}\Bigr)
+ 24\xi_1\xi_3(\rho - 1)\nonumber \\
&\qquad\qquad 
- 8\xi_1(9\rho^4 + 6\rho^3 + 10\rho^2 - 19\rho + 3) \xi_2 
+ 8\xi_1(\rho-1)\rho \xi_2 \Bigr]\nonumber \\
&\quad-\frac{4 r_{E}^{(1)}}{9 N_c \, \xi_1^3} \Bigl[ 
-(228\rho^2 - 16\rho - 5)\xi_1^4 
+ (608\rho^3 - 76\rho^2 - 10\rho - 9)\xi_1 \xi_2 \nonumber \\
&\qquad\qquad 
- (1824\rho^4 - 280\rho^3 + 32\rho^2 - 58\rho + 3)\xi_1 \xi_2 \Bigr]\nonumber \\
&\quad+ \frac{2 r_{G}^{(1)}}{3 N_c \, \xi_1^3} \Bigl[ 
-2(6\rho^2 + \rho + 5)\xi_1^4 
+ (32\rho^3 - 16\rho^2 + 20\rho - 9)\xi_1 \xi_2
\nonumber \\
&\qquad\qquad 
- (96\rho^4 + 8\rho^3 - 16\rho^2 - 28\rho + 3)\xi_1 \xi_2 \Bigr].
\label{eq:F22-ccs}
\end{align}
}

{\footnotesize
\begin{align}
\mathcal{W}_{12}^{(c\bar{c}s)} &=
-\frac{32 r_E^4}{9 \xi_1^5} \Bigl[
-\xi_1^4 \bigl(
216\rho^4 + 486\rho^3 - 155\rho^2 - 14\rho + 7
+ 12(4\rho^2 - 2\rho + 1)\log\xi_1
\nonumber\\
&\qquad\qquad
- 6(4\rho^2 - 2\rho + 1)\log\rho
\bigr)
\nonumber\\
&\qquad\qquad
+ (1444\rho^4 - 366\rho^3 - 90\rho^2 + 74\rho - 9)\xi_1 \xi_2
\nonumber\\
&\qquad\qquad
- (1728\rho^6 + 3168\rho^5 - 752\rho^4 + 30\rho^3 + 84\rho^2 - 22\rho + 3)\xi_1 \xi_2
\Bigr]
\nonumber\\
&\quad+ \frac{8 r_G^4}{9 \xi_1^5} \Bigl[ 
-2304\rho^6 - 5568\rho^5 + 5648\rho^4 - 164\rho^3 - 936\rho^2 + 268\rho - 22 
\nonumber \\
&\qquad\qquad
+ 6(64\rho^4 - 64\rho^3 + 36\rho^2 - 10\rho + 1)\log\!\Bigl(\frac{\rho}{1-4\rho}\Bigr)\nonumber \\
&\qquad\qquad
+ \xi_1(-576\rho^5 + 864\rho^4 - 612\rho^3 + 468\rho^2 - 162\rho + 18)\log(1+\xi_1)\nonumber \\
&\qquad\qquad
+ 18\xi_1^5(2\rho^3 - 2\rho^2 + \rho - 1)\log(1-\xi_1)\nonumber \\
&\qquad\qquad
+ \xi_1(-1152\rho^6 - 3456\rho^5 + 584\rho^4 - 516\rho^3 + 318\rho^2 - 20\rho - 6) \xi_2 \Bigr]\nonumber \\
&\quad + \frac{4 s_E^4}{9 \xi_1^3} \Bigl[ 
-1728\rho^5 + 1728\rho^4 + 868\rho^3 - 162\rho^2 - 174\rho + 35 
+ 24(32\rho^3 - 6\rho + 1)\log\!\Bigl(\frac{\rho}{1-4\rho}\Bigr)\nonumber \\
&\qquad\qquad
+ 144\xi_1\rho^3(4\rho - 1)\log(1+\xi_1)
+ 144\xi_1^3\rho^3 \log(1-\xi_1)\nonumber \\
&\qquad\qquad
+ \xi_1(-864\rho^5 + 1368\rho^4 - 576\rho^3 + 72\rho^2 + 192\rho - 48) \xi_2 \Bigr] \nonumber \\
&\quad+\frac{4 s_B^4}{9 \xi_1} \Bigl[ 
-720\rho^4 - 84\rho^3 + 26\rho^2 + 164\rho - 43 
+ 12(8\rho^2 + 2\rho - 1)\log\!\Bigl(\frac{\rho}{1-4\rho}\Bigr)\nonumber \\
&\qquad\qquad
+ \xi_1(-360\rho^4 - 72\rho^3 - 144\rho^2 + 36\rho + 6) \xi_2 \Bigr]\nonumber \\
&\quad-\frac{s_{qB}^4}{6 \xi_1} \Bigl[ 
-48\rho^4 + 292\rho^3 + 18\rho^2 + 258\rho - 67 
+ \xi_1(-24\rho^4 + 144\rho^3 + 24\rho - 48) \xi_2 \Bigr].
\label{eq:F12-ccs}
\end{align}
}
\section{Closing Remarks}

In this chapter, we have completed the leading-order calculation of the
dimension-seven two-quark contribution $\Gamma_7^{(0)}$ to inclusive
non-leptonic $b$-hadron decays for the CKM-favoured channels
$b\to c\bar ud$ and $b\to c\bar cs$. The resulting analytical matching
coefficients provide the complete $\mathcal{O}(1/m_b^4)$ two-quark
corrections for these channels and extend the non-leptonic HQE by one
order beyond the dimension-six Darwin contribution.

A distinctive feature of the non-leptonic calculation at leading order
in QCD is the separate
appearance of the colour-singlet matrix elements $r_G^{(1)}$ and
$r_E^{(1)}$ defined in \cref{eq:color-decomp}. In the semileptonic
case, the colour-singlet and colour-octet components enter only through
specific combinations, whereas the non-leptonic contributions probe
$r_G^{(1)}$ and $r_E^{(1)}$ individually. These matrix elements
contribute only to the terms proportional to $C_1^2$ and $C_2^2$ and
are absent from the interference term proportional to $C_1C_2$ at
$\mathcal{O}(1/m_b^4)$.

As an important consistency check, the corresponding semileptonic
result (see, e.g., Ref.~\cite{Mannel:2023yqf}) is recovered directly
from the $b\to c\bar ud$ expressions by taking
$N_c\to1$, $C_1\to1$, and $C_2\to0$, setting
$r_G^{(1)}=r_E^{(1)}=0$, and replacing the CKM factors accordingly.
In particular, the semileptonic contribution follows from
$\mathcal{W}_{11}^{(c\bar ud)}$ after omitting the terms proportional
to $r_G^{(1)}$ and $r_E^{(1)}$.

The results derived here provide the perturbative input required to
incorporate dimension-seven two-quark effects systematically into
precision predictions for $B$-meson decay widths and lifetime ratios.
Together with improved determinations of the corresponding
non-perturbative matrix elements, they will allow the numerical impact
of the $\mathcal{O}(1/m_b^4)$ corrections to be assessed and provide
a direct test of the convergence of the HQE for inclusive non-leptonic
decays at this order.

\chapter{Model-Independent BSM Contributions up to the Darwin Term} 
\label{Ch:3}
\vspace{3mm}
\begin{flushright}
\vspace{5mm}
\emph{
``These BSM calculations are getting rather involved.\\
After all, the Standard Model is not so bad, right?''
}
\\[0.3cm]
--- Thomas Mannel

\end{flushright}
The previous chapter focused on the SM description of
inclusive non-leptonic $b$-hadron decays within the HQE. We now turn to the complementary question of how the same
framework can be used to search for BSM physics.
In particular, new heavy degrees of freedom—too massive to be produced directly—can leave imprints at low energies by modifying the short-distance structure of the effective weak Hamiltonian.\\
This framework also provides a powerful and model-independent tool for BSM searches: new heavy particles manifest as either modified Wilson coefficients $C_i$ of existing SM operators or as entirely new local operators $O_i^{\text{BSM}}$ that respect the symmetries of the low-energy theory. These additional operators can contribute to observables such as the total inclusive width or lifetime ratios, allowing one to probe new-physics effects indirectly.\\
In this chapter, we extend the HQE to include the most general
dimension-six $|\Delta B|=1$ four-fermion operators contributing to non-leptonic
$b$-hadron decays. We determine their leading-order contributions to
the inclusive decay width, comprising the leading free-quark
contribution $\Gamma_3$, the dimension-five chromomagnetic correction
$\Gamma_5$, and the dimension-six two-quark Darwin term $\Gamma_6$.
This study is motivated by the sizable numerical contribution of the
Darwin term in the SM, suggesting that the corresponding contributions
arising from a generalised effective Hamiltonian may likewise have an
important phenomenological impact. We further derive the previously
uncalculated weak-annihilation (WA) topology of the dimension-six $\Delta B=0$ four-quark
contribution $\widetilde{\Gamma}_6$. As a direct consequence of this general analysis, we also
obtain new results for the contributions of the SM QCD-penguin operators at
dimension five and dimension six, which follow from the general BSM
operator basis after appropriate Fierz transformations. The resulting
expressions provide a model-independent parametrisation of
short-distance new-physics effects that can be readily applied to a
broad class of ultraviolet completions and used to constrain their
Wilson coefficients through precision measurements of inclusive
$b$-hadron decay widths and lifetime ratios.

\section{Generalised Effective Hamiltonian}
\label{subsec:Heff}
To parameterise possible BSM contributions in a model-independent
manner, we consider the most general effective Hamiltonian describing
non-leptonic $|\Delta B|=1$ transitions $b\to q_1\bar q_2 q_3$, with
$q_1,q_2=u,c$ and $q_3=d,s$. It is obtained by supplementing the SM
operator basis with all additional dimension-six four-fermion operators~\cite{Ciuchini:1997bw,Buras:2000if}:
\begin{equation}
{\cal H}_{\rm eff}
= 
\frac{4 G_F}{\sqrt 2} V_{q_1 b} V_{q_2 q_3}^*
\sum_{i = 1}^{10} 
\Bigl[
{\cal C}_i (\mu_b) {\cal Q}_i^{q_1q_2q_3} + {\cal C}_i^{\prime} (\mu_b) {\cal Q}_i^{q_1q_2q_3 \prime} 
\Bigr] 
+ {\rm h.c.}\,.
\label{eq:Heff-NP}    
\end{equation}
Here, the four-quark operators ${\cal Q}_i^{q_1q_2q_3}$ are defined as
\begin{align}
& {\cal Q}_1^{q_1q_2q_3} = (\bar q_1^i \gamma_\mu P_L \, b^i) (\bar q_3^j \gamma^\mu P_L \, q_2^j)\,,
\quad &
& {\cal Q}_2^{q_1q_2q_3} = (\bar q_1^i \gamma_\mu P_L \, b^j) (\bar q_3^j \gamma^\mu P_L \, q_2^i)\,,
\label{eq:Q1-Q2}
\\[2mm]
& {\cal Q}_3^{q_1q_2q_3} = (\bar q_1^i \gamma_\mu P_R \, b^i) (\bar q_3^j \gamma^\mu P_L \, q_2^j)\,,
\quad &
& {\cal Q}_4^{q_1q_2q_3} = (\bar q_1^i \gamma_\mu P_R \, b^j) (\bar q_3^j \gamma^\mu P_L \, q_2^i)\,,
\label{eq:Q3-Q4}
\\[2mm]
& {\cal Q}_5^{q_1q_2q_3} = (\bar q_1^i P_L \, b^i) (\bar q_3^j P_R \, q_2^j)\,,
\quad &
& {\cal Q}_6^{q_1q_2q_3} = (\bar q_1^i P_L \, b^j) (\bar q_3^j P_R \, q_2^i)\,,
\label{eq:Q5-Q6}
\\[2mm]
& {\cal Q}_7^{q_1q_2q_3} = (\bar q_1^i P_R \, b^i) (\bar q_3^j P_R \, q_2^j)\,, 
\quad &
& {\cal Q}_8^{q_1q_2q_3} = (\bar q_1^i P_R \, b^j) (\bar q_3^j P_R \, q_2^i)\,,
\label{eq:Q7-Q8}
\\[2mm]
& {\cal Q}_9^{q_1q_2q_3} = (\bar q_1^i \sigma_{\mu\nu} P_R\, b^i)(\bar q_3^j \sigma^{\mu\nu} P_R\, q_2^j)\,,
\quad &
& {\cal Q}_{10}^{q_1q_2q_3} = (\bar q_1^i \sigma_{\mu\nu} P_R\, b^j)(\bar q_3^j \sigma^{\mu\nu} P_R\, q_2^i)\,,
\label{eq:Q9-Q10}
\end{align}
where $i,j$ denote colour indices and
$\sigma^{\mu\nu}=(i/2)[\gamma^\mu,\gamma^\nu]$.
The corresponding primed operators,
${\cal Q}_i^{q_1q_2q_3\prime}$, are obtained by interchanging the
left- and right-handed chiral projectors,
\begin{equation}
{\cal Q}_i^{q_1q_2q_3\prime}
=
{\cal Q}_i^{q_1q_2q_3}\Big|_{P_L\leftrightarrow P_R}.
\end{equation}
The Wilson coefficients ${\cal C}_i^{(\prime)}$ in \cref{eq:Heff-NP} encode the
short-distance dynamics and are, in general, complex-valued. Their
dependence on the flavour combination $(q_1q_2q_3)$ is understood and
will be suppressed throughout this chapter whenever no ambiguity
arises.

The generalised effective Hamiltonian in \cref{eq:Heff-NP} contains the
SM Hamiltonian in \cref{eq:Heff-SM} as a special case. In particular,
the operators ${\cal Q}_1^{q_1q_2q_3}$ and
${\cal Q}_2^{q_1q_2q_3}$ coincide with the SM current--current
operators $Q_1^{q_1q_2q_3}$ and $Q_2^{q_1q_2q_3}$,
\begin{equation}
Q_1^{q_1q_2q_3}={\cal Q}_1^{q_1q_2q_3},
\qquad
Q_2^{q_1q_2q_3}={\cal Q}_2^{q_1q_2q_3}.
\label{eq:Q1_Q1cal}
\end{equation}

The SM QCD-penguin operators $Q_{3-6}^{qq_3}$ in \cref{eq:Heff-SM} with $q=u,c$ can likewise be mapped onto
the generalised operator basis. Using Fierz transformations, one finds
\begin{align}
Q_3^{qq_3} &= {\cal Q}_2^{qqq_3},
&
Q_4^{qq_3} &= {\cal Q}_1^{qqq_3},
\label{eq:O3_O4}
\\[2mm]
Q_5^{qq_3} &= -2\,{\cal Q}_6^{qqq_3},
&
Q_6^{qq_3} &= -2\,{\cal Q}_5^{qqq_3}.
\label{eq:O5_O6}
\end{align}
The remaining penguin contributions with $q=d,s$ are obtained from the
$q=u$ coefficient functions upon neglecting the light-quark masses, as
discussed in \cref{sec:SM_penguin}.

Correspondingly, the Wilson coefficients $\mathcal{C}_{i}^{(\prime)}$ can
be decomposed into their SM and BSM contributions,
\begin{equation}
\mathcal{C}_{i}^{(\prime)}
=
\mathcal{C}_{i,\mathrm{SM}}^{(\prime)}
+
\mathcal{C}_{i,\mathrm{BSM}}^{(\prime)}\,,
\end{equation}
where the SM part is fixed by the underlying theory, while the BSM
contribution is, in general, a complex quantity encoding the effects of
new short-distance physics. For example,
$\mathcal{C}_{1,\mathrm{SM}} = C_1$ and
$\mathcal{C}_{1,\mathrm{SM}}^{\prime}=0$.
With this decomposition, the double insertion of the generalised
effective Hamiltonian in \cref{eq:Heff-NP} into \cref{eq:GammaB}
naturally generates all contributions to the inclusive decay width:
the purely SM terms, the purely BSM terms, and the SM--BSM interference
terms. We therefore adopt the generalised effective Hamiltonian in
\cref{eq:Heff-NP} as the starting point of our analysis, and all results
presented in this chapter are expressed in terms of this operator basis.

Despite the considerable progress achieved in recent years, a complete HQE analysis based on the generalised effective Hamiltonian in \cref{eq:Heff-NP} has not yet been available. Existing studies have focused primarily on the leading-power contributions and on the dimension-six spectator effects. In particular, the matching coefficients of the four-quark operators, $\tilde{\Gamma}_6^{(0)}$, have been determined for the $b\to c\bar{c}s$ transition in Refs.~\cite{Jager:2017gal,Jager:2019bgk,Lenz:2022pgw} and for the $b\to c\bar{u}d$ transition in Ref.~\cite{Lenz:2022pgw}. The leading free-quark contribution to the generalised operator basis was subsequently derived for the $b\to c\bar{u}d$ channel at leading order in QCD in Ref.~\cite{Meiser:2024zea}, while the corresponding hadronic matrix elements of the dimension-six four-quark operators were obtained in the HQET limit using three-loop sum rules in Ref.~\cite{Black:2024bus}.

The results presented in this chapter complete the HQE matching for the generalised effective Hamiltonian through $\mathcal{O}(1/m_b^3)$, thereby providing the missing ingredients required for a complete model-independent description of inclusive non-leptonic $b$-hadron decays.

\section{Calculation Procedure}
\label{sec:BSMCal}
The matching calculation follows the same methodology presented in
detail in \cref{Ch:2}, and we only summarize the main steps here. We
start from the time-ordered product in \cref{eq:GammaB} and evaluate the
forward-scattering amplitude by expanding the internal quark propagators
in the external gluon field using the FS gauge.\footnote{In contrast to the SM calculation in \cref{Ch:2}, where only
terms containing an odd number of Dirac matrices in the propagator
expansion contribute, the general BSM operator basis involves different
Dirac structures for which both odd and even contributions are relevant.
We therefore employ the full background-field expansion of the quark
propagator, retaining terms with up to one covariant derivative acting
on the gluon field-strength tensor, as given, e.g., in
Refs.~\cite{Lenz:2020oce,Piscopo:2021ogu}.} Since the
present analysis is restricted to operators up to dimension six, the
background-field expansion is truncated accordingly. After evaluating
the resulting master loop integrals (see \cref{sec:loop_int}), the amplitudes are expanded in the
external momenta and matched onto the HQE operator basis defined in \cref{sec:HQE-basis-appendix}. The matching
onto the local HQE operators is performed using the trace formalism of
Refs.~\cite{Dassinger:2006md,Mannel:2023yqf}, yielding the Wilson
coefficients of the dimension-three, dimension-five, dimension-six
two-quark, and dimension-six four-quark operators.

As discussed in \cref{Ch:2}, the master loop integrals entering the
matching calculation are UV finite and can therefore be
evaluated directly in four dimensions. Starting at the
dimension-six two-quark level, however, IR singularities arise
from the massless light-quark propagators. We regulate these
singularities by retaining finite light-quark masses throughout the
calculation, which gives rise to logarithmic terms of the form
$\log(m_q^2/m_b^2)$ in the matching coefficients. The appearance of these IR logarithms reflects the mixing of the
dimension-six two-quark operators into the $\Delta B=0$ dimension-six
four-quark operators under renormalisation. To obtain IR-finite
Wilson coefficients, we explicitly compute the corresponding one-loop
renormalisation of the four-quark operators and combine it with the
bare matching coefficients. The logarithmic dependence on the
light-quark masses then cancels exactly, yielding finite matching
coefficients for the dimension-six HQE operators.

The $\Delta B=0$ four-quark operators required for the renormalisation
of the dimension-six two-quark matching coefficients were determined in
Ref.~\cite{Lenz:2022pgw}. Employing the same generalised effective
Hamiltonian in \cref{eq:Heff-NP}, they obtained the operator basis
\begin{align}
  \mathcal{O}_1
  & =  
  4\, (\bar b \, \gamma_\mu P_L \, q)\,(\bar{q} \, \gamma^\mu P_L \, b) \,,
  &
  \mathcal{O}_2  
  & =  
  4 \, (\bar b P_L \, q)\,(\bar{q} P_R \, b) \,,
  \label{eq:O1-O2-HQET} 
  \\[1mm]
  \mathcal{O}_3  
  & =  
  4 \,(\bar b \, \gamma_\mu P_L \, t^a q)\,(\bar{q}\,\gamma^\mu P_L\,t^a b) \,,
  &
  \mathcal{O}_4 
  & =  
  4\,(\bar b P_L \, t^a q)\,(\bar{q} P_R \, t^a b) \,,
  \label{eq:O3-O4-HQET} 
  \\[1mm]
  \mathcal{O}_5  
  & =  
  4\,(\bar b \, \gamma_\mu P_L \, q)\, (\bar{q} \, \gamma^\mu P_R \, b) \,,
  &
  \mathcal{O}_6  
  & =  
  4 \, (\bar b P_L \, q) \, (\bar{q} P_L \, b) \,,
  \label{eq:O5-O6-HQET} 
  \\[1mm]
  \mathcal{O}_7  
  & =  
  4\,(\bar b \, \gamma_\mu P_L \, t^a q) (\bar{q}\,\gamma^\mu P_R\,t^a b) \,,
  &
  \mathcal{O}_8 
  & =  
  4 \, (\bar b P_L \, t^a q) \,
  (\bar{q}P_L \, t^a b) \,,
  \label{eq:O7-O8-HQET}
\end{align}
Together with the corresponding chirality-flipped operators
$\mathcal{O}_i'$, obtained from $\mathcal{O}_i$ by the replacement
$P_L \leftrightarrow P_R$, these operators constitute the complete
$\Delta B=0$ operator basis.\footnote{In principle, tensor four-quark operators may also be included.
At dimension six, however, they are not independent: using the
$b$-quark equations of motion, they can be reduced to the remaining
Dirac structures up to $\mathcal{O}(1/m_b)$ corrections~\cite{Lenz:2022pgw}.
Moreover, their renormalised soft-gluon corrections vanish.}

\begin{figure}[t!]
\centering
\setlength{\unitlength}{1pt}

\begin{minipage}{0.32\textwidth}
\centering
\begin{picture}(150,120)
  \put(0,32){\includegraphics[width=0.95\linewidth]{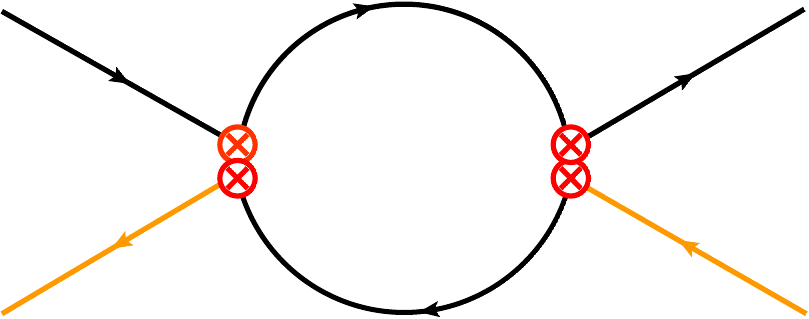}}
  \put(70,85){$q_1$}
  \put(70,25){$q_2$}
  \put(5,25){$q_3$}
  \put(120,25){$q_3$}
  \put(7,80){$b$}
  \put(115,80){$b$}
\end{picture}

\end{minipage}
\hfill
\begin{minipage}{0.32\textwidth}
\centering
\begin{picture}(150,120)
  \put(0,20){\includegraphics[width=0.95\linewidth]{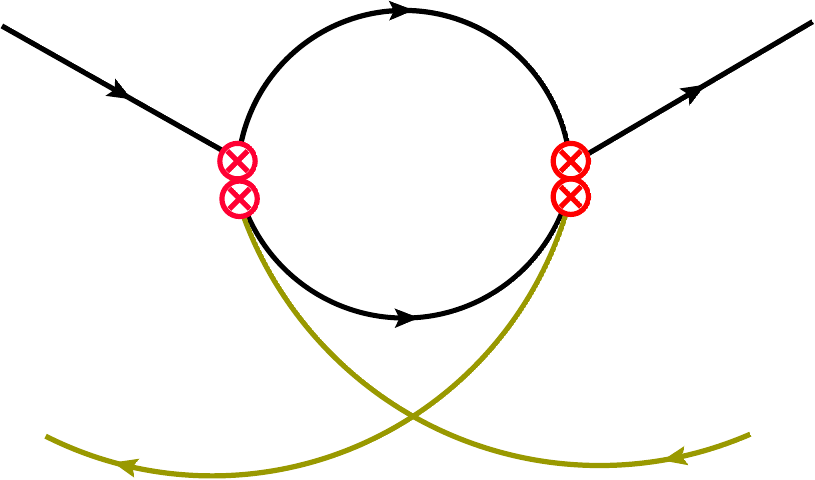}}
  \put(70,99){$q_1$}
  \put(70,55){$q_3$}
  \put(10,35){$q_2$}
  \put(115,35){$q_2$}
  \put(5,93){$b$}
  \put(115,90){$b$}
\end{picture}

\end{minipage}
\hfill
\begin{minipage}{0.32\textwidth}
\centering
\begin{picture}(150,120)
  \put(0,35){\includegraphics[width=0.95\linewidth]{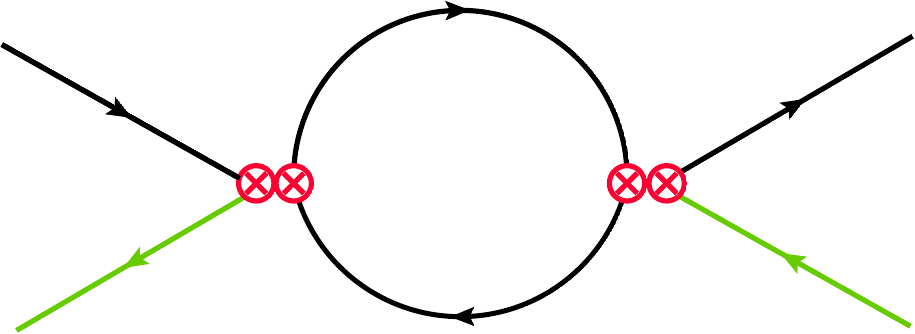}}
  \put(70,85){$q_3$}
  \put(70,30){$q_2$}
  \put(5,30){$q_1$}
  \put(120,30){$q_1$}
  \put(5,80){$b$}
  \put(120,80){$b$}
\end{picture}

\end{minipage}

\caption{Diagrammatic representation of the $\Delta B=0$ four-quark operator contributions in the HQE for the transition $b\to q_1\bar q_2 q_3$: weak exchange (WE), Pauli interference (PI), and weak annihilation (WA).}
\label{fig:4QDiags}
\end{figure}

The generalised effective Hamiltonian in \cref{eq:Heff-NP} contributes
to the four inclusive non-leptonic transitions
$b\to c\bar{u}d$, $b\to c\bar{c}s$, $b\to u\bar{c}s$, and
$b\to u\bar{u}d$. Although the matching procedure is identical for all
channels, the cancellation of IR singularities depends on the
underlying spectator topology.

For the $b\to c\bar{u}d$ and $b\to c\bar{c}s$ transitions, the IR
logarithms are completely cancelled by including the one-loop
renormalisation of the $\Delta B=0$ four-quark operators arising from
the Pauli interference (PI) and weak exchange (WE) topologies shown in
the left and middle panels of \cref{fig:4QDiags}. These contributions
were calculated in Ref.~\cite{Lenz:2022pgw} for the generalised
effective Hamiltonian in \cref{eq:Heff-NP}, and we make use of the
corresponding results.

The situation is different for the $b\to u\bar{c}s$ and
$b\to u\bar{u}d$ channels. In these cases, the PI and WE topologies
alone are insufficient to remove the IR logarithms. The
cancellation is completed only after including the
WA topology, shown in the right panel of
\cref{fig:4QDiags}. The corresponding WA contributions have not previously been determined
for the generalised effective Hamiltonian in \cref{eq:Heff-NP}. We
therefore compute these contributions in this work. Since the
calculation follows the same methodology as that presented in
\cref{sec:MesoIncl} (see also~\cite{Piscopo:2021ogu}), differing only in the underlying operator basis,
we do not repeat the technical details here and instead quote the final
matching coefficients.
\begin{figure}[t!]
  \centering
  \setlength{\unitlength}{1pt}

  \begin{tabular}{ccccc}
    \begin{picture}(150,120)
      \put(0,0){\includegraphics[width=0.27\textwidth]{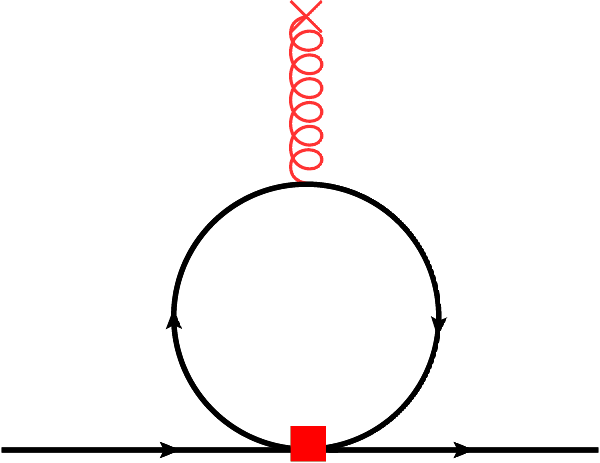}}
      \put(25,35){$q$}
      \put(90,35){$q$}
      \put(10,8){$b$}
      \put(110,8){$b$}
    \end{picture}
     \end{tabular}
  \caption{One-gluon corrections to the matrix elements of the dimension-six $\Delta B=0$ four-quark operators. Through mixing into the Darwin operator, these contributions cancel the infrared divergences arising in the two-quark matching calculation.}
\label{fig:MatrixMix}
    \end{figure}

As discussed in the \cref{Ch:2}, the one-loop diagram shown in
\cref{fig:MatrixMix} is both UV and IR
divergent. The UV divergences are regulated using dimensional regularisation and renormalised in the $\overline{\mathrm{MS}}$ scheme.
To regulate the IR divergences, we employ the same prescription as in
the matching calculation of the two-quark operators, namely by keeping
the light quark propagating in the loop massive. This gives rise to
logarithms of the form $\ln(m_q^2/m_b^2)$, which cancel exactly against
the corresponding logarithms in the bare two-quark matching
coefficients. 

The renormalisation of the generalised $\Delta B=0$ operator basis in \cref{eq:O1-O2-HQET,eq:O3-O4-HQET,eq:O5-O6-HQET,eq:O7-O8-HQET} is
particularly simple. Only ${\cal O}_{1-4}$ and the corresponding
chirality-flipped operators ${\cal O}_{1-4}'$ mix into the Darwin
operator at one loop. As these operators already occur in the SM, their
renormalised matrix elements are unchanged and can be taken directly
from the SM calculations. At the renormalisation scale $\mu=m_b$ they are given by
\begin{equation}
\begin{aligned}
\langle {{\cal O}}_{1,2} \rangle^{\mathrm{ren}}
&=
\langle {\cal O}'_{1,2}\rangle^{\mathrm{ren}}
=
 \frac{a_{1,2}}{12 \pi^2}
 \left[
\log\!\left(\frac{m_b^2}{m_q^2}\right)+b_{1,2}
 \right]
 \langle {\cal O}_{\rho_D}\rangle
 +{\cal O}\!\left(\frac{1}{m_b}\right),\\[2mm]
\langle {\cal O}^{(\prime)}_{3,4}\rangle^{\mathrm{ren}}
&=
-\frac{1}{2N_c}
\langle {\cal O}^{(\prime)}_{1,2}\rangle^{\mathrm{ren}}\,,
\end{aligned}
\label{eq:crD-finite}
\end{equation}
The numerical coefficients are given by
$a_1=2$, $a_2=-1$, $b_1=-1$, and $b_2=0$, in agreement with the
corresponding SM results~\cite{Lenz:2020oce,Piscopo:2021ogu}.\footnote{The numerical
values of these coefficients are basis dependent. In particular,
Ref.~\cite{Lenz:2020oce} employs a $\Delta B=0$ operator basis
constructed from colour-rearranged operators rather than the
colour-octet basis adopted here, leading to different values for these
constants.}
The remaining operators ${\cal O}_{5-8}$ and their chirality-flipped
counterparts ${\cal O}_{5-8}'$ do not contribute to the Darwin operator
through the one-loop diagrams of \cref{fig:MatrixMix}, as their
renormalised matrix elements vanish identically. Nevertheless, these
operators are not irrelevant at dimension six: they contribute at
tree level through the WA topology.

\section{Analytical Results}
\label{Sec:Results}

Having established the generalised effective Hamiltonian, the matching
procedure, and the renormalisation of the relevant operators, we are now
in a position to present the corresponding HQE matching coefficients.
Throughout this section, we consider the two-quark contributions to the
inclusive non-leptonic decay $b\to q_1\bar q_2 q_3$ up to and including
dimension six.

The decay width can be
written in the compact form
\begin{equation}
\Gamma(b\to q_1\bar q_2q_3)
=
\Gamma_0
|V_{q_1b}|^2
|V_{q_2q_3}|^2
\sum_{i,j=1}^{20}
{\cal C}_i{\cal C}_j^{*}
{\cal N}_{ij}
{\cal F}_{ij}^{(q_1q_2q_3)}(\rho)+\mathcal{O}(1/m_b^4),
\label{eq:Gamma_master_compact}
\end{equation}
where
\begin{equation}
\Gamma_0
\equiv
\frac{G_F^2m_b^5}{192\pi^3},
\label{eq:Gamma0}
\end{equation}
and, for convenience, we relabel the primed Wilson coefficients
according to
${\cal C}_{11}\equiv{\cal C}_1',\,\ldots,\,{\cal C}_{20}\equiv{\cal C}_{10}'$. 
The coefficient functions
${\cal F}_{ij}^{(q_1q_2q_3)}(\rho)$ encode the short-distance dynamics
associated with the operator product
${\cal Q}_i^{q_1q_2q_3}
{\cal Q}_j^{\dagger\,q_1q_2q_3}$.
Their HQE expansion is parameterised as
\begin{equation}
\begin{aligned}
{\cal F}_{ij}^{(q_1q_2q_3)}
&=
\left(1-\frac{\mu_\pi^2}{2m_b^2}\right)
P_{ij}^{(q_1q_2q_3)}(\rho)
+\frac{\mu_G^2}{m_b^2}
G_{ij}^{(q_1q_2q_3)}(\rho)
+\frac{\rho_D^3}{m_b^3}
D_{ij}^{(q_1q_2q_3)}(\rho),
\label{eq:Fij}
\end{aligned}
\end{equation}
where $\rho=m_c^2/m_b^2$. The coefficient of the kinetic-energy operator $\mu_\pi^2$ has
been fixed using reparametrisation invariance~\cite{Luke:1992cs}, implying that it is
identical to the leading-power coefficient up to the universal factor
$-1/2$. The colour factors $\mathcal{N}_{ij}$ entering
\cref{eq:Gamma_master_compact} are
\begin{equation}
\mathcal{N}_{ij}
=
\begin{cases}
    1, &\text{if } j - i \text{ odd}\, \\
    N_c, &\text{if } j - i \text{ even} \,.
\end{cases}
\label{eq:color_matrix}
\end{equation}
The coefficient functions ${\cal F}_{i,j}^{(q_1q_2q_3)}(\rho)$ satisfy the universal relations
\begin{align}
{\cal F}_{i,j}^{(q_1q_2q_3)}
&=
{\cal F}_{j,i}^{(q_1q_2q_3)},
\qquad
{\cal F}_{i+10,j+10}^{(q_1q_2q_3)}
=
{\cal F}_{i,j}^{(q_1q_2q_3)},
\\[2mm]
{\cal F}_{i,j}^{(q_1q_2q_3)}(\rho)
&=
0,
\qquad
1\leq i\leq10,\;
11\leq j\leq20,
\end{align}
which hold for all operator combinations. Consequently, it is sufficient
to present only the independent coefficient functions
${\cal F}_{i,j}^{(q_1q_2q_3)}(\rho)$ with
$1\leq i\leq j\leq10$.

The calculation is performed for all inclusive non-leptonic
$b$-quark decay channels by considering the distinct choices of massive
and massless final-state quarks listed in \cref{tab:massconfigs}.
\begin{table}[ht]
\centering
\begin{tabular}{cc}
\hline
Mass configuration & Corresponding transitions \\
\hline
$m_1\neq0,\;m_2=m_3=0$
&
$b\to c\bar ud,\; b\to c\bar us$
\\[1mm]
$m_1=m_2\neq0,\;m_3=0$
&
$b\to c\bar cs,\; b\to c\bar cd$
\\[1mm]
$m_2\neq0,\;m_1=m_3=0$
&
$b\to u\bar cs,\; b\to u\bar cd$
\\[1mm]
$m_1=m_2=m_3=0$
&
$b\to u\bar ud,\; b\to u\bar us$
\\
\hline
\end{tabular}
\caption{Mass configurations considered in the matching calculation.}
\label{tab:massconfigs}
\end{table}
To avoid interrupting the presentation with lengthy analytical
expressions, we restrict the discussion below to the CKM-favoured
transitions $b\to c\bar{u}d$ and $b\to c\bar{c}s$.
The analytical results for the additional
decay channels are collected in \cref{App:BSM Darwin}; the complete
set of results is also available in Ref.~\cite{Lang:2025ios} and the
accompanying ancillary files.

As discussed in \cref{subsec:Heff}, the operator identities in
\cref{eq:Q1_Q1cal,eq:O3_O4,eq:O5_O6} relate the generalised operator
basis to the SM QCD-penguin operators through Fierz transformations.
Consequently, once the matching coefficients
$P_{ij}^{(q_1q_2q_3)}(\rho)$,
$G_{ij}^{(q_1q_2q_3)}(\rho)$, and
$D_{ij}^{(q_1q_2q_3)}(\rho)$ have been determined, the corresponding
SM QCD-penguin contributions follow immediately. Since the light-quark
masses are neglected, only the mass configurations with
$m_1=m_2=m_c$ and $m_1=m_2=m_3=0$ are relevant. While the leading-power
penguin contributions were already known~\cite{Lenz:1997aa,Krinner:2013cja}, the
dimension-five and dimension-six corrections presented here constitute
new results and are discussed in \cref{sec:SM_penguin}.

\subsection
[\texorpdfstring{Results for the $b \to c \bar u d$ Transition}
{Results for the b -> c ubar d Transition}]
{Results for the $b \to c \bar u d$ Transition}
\label{sec:bcud}
\paragraph{Dimension-three contribution $(\Gamma_3)$:}

We begin with the leading-power contribution to the transition
$b\to c\bar{u}d$. Among the coefficient functions
$P_{ij}^{(cud)}(\rho)$ defined in \cref{eq:Fij}, only two are
independent. They are given by
\begin{align}
P_{1,1}^{(cud)}(\rho) &= 
1 - 8\rho - 12\rho^2 \log\rho + 8\rho^3 - \rho^4 \,, \label{eq:P11} \\[2mm]
P_{1,3}^{(cud)}(\rho) &=
- 2 \sqrt{\rho}\,\big(
1 + 9\rho + 6 \rho (1 + \rho)\log\rho - 9\rho^2 - \rho^3  
\big) \,. \label{eq:P13}
\end{align}
The remaining non-vanishing coefficient functions are obtained from the
following symmetry and proportionality relations:
\begin{equation}
\begin{aligned}
& P_{2,2}^{(cud)} =  P_{3,3}^{(cud)} = P_{4,4}^{(cud)} = P_{1,1}^{(cud)} \,,
\qquad
P_{5,5}^{(cud)} = P_{6,6}^{(cud)} = P_{7,7}^{(cud)} = P_{8,8}^{(cud)} = \frac{1}{4}P_{1,1}^{(cud)} \,,
\\[6pt]
&  P_{9,9}^{(cud)} = P_{10,10}^{(cud)} = 12\,P_{1,1}^{(cud)},
\qquad
P_{2,4}^{(cud)} = P_{1,3}^{(cud)} \,,
\qquad
P_{5,7}^{(cud)} = P_{6,8}^{(cud)} = -\frac{1}{2}P_{1,3}^{(cud)} \,,
\\[6pt]
&  
P_{i,j}^{(cud)} = P_{i,j-1}^{(cud)}
\quad \text{for odd } i \text{ and even } j \,,
\end{aligned}
\label{eq:Prellations}
\end{equation}
All other coefficient functions vanish identically.\footnote{After translating between the two operator bases and setting
$N_c=3$, we find complete agreement with the dimension-three results of
Ref.~\cite{Meiser:2024zea}.}
\paragraph{Dimension-five contribution ($\Gamma_5$):}
In view of the relation in \cref{eq:Fij}, only the chromomagnetic
matching coefficients need to be given explicitly. The independent
non-vanishing functions $G_{i,j}^{(cud)}(\rho)$ are
\begin{align}
G_{1,1}^{(cud)}(\rho) &= 
-\frac{1}{2}\big(5\rho^4 - 24\rho^3 + 24\rho^2 + 12\rho^2\log\rho - 8\rho + 3\big) \,, \label{eq:G11} \\[2pt]
G_{1,2}^{(cud)}(\rho) &= 
-\frac{1}{2}\big(5\rho^4 - 40\rho^3 + 72\rho^2 + 12\rho^2\log\rho - 56\rho + 19\big) \,, \label{eq:G12} \\[2pt]
G_{1,3}^{(cud)}(\rho) &= 
\frac{1}{3}\sqrt{\rho}\big(13\rho^3 - 27\rho^2 - 6(3\rho^2 - 3\rho + 2)\log\rho + 27\rho - 13\big) \,, \label{eq:G13} \\[2pt]
G_{2,4}^{(cud)}(\rho) &= 
\sqrt{\rho}\big(5\rho^3 - 3\rho^2 - 6(1 + \rho)\rho\log\rho + 3\rho - 5\big) \,, \label{eq:G24} \\[2pt]
G_{4,4}^{(cud)}(\rho) &= 
-\frac{1}{2}\big(5\rho^4 - 8\rho^3 - 24\rho^2 + 12\rho^2\log\rho + 40\rho - 13\big) \,, \label{eq:G44} \\[2pt]
G_{5,5}^{(cud)}(\rho) &= 
-\frac{1}{8}\big(5\rho^4 - 32\rho^3 + 72\rho^2 + 12(\rho-4)\rho\log\rho - 32\rho - 13\big) \,, \label{eq:G55} \\[2pt]
G_{5,7}^{(cud)}(\rho) &= 
-\frac{3}{2}\sqrt{\rho}\big(\rho^3 - 3\rho^2 - 2(\rho^2 - 5\rho - 2)\log\rho - 9\rho + 11\big) \,, \label{eq:G57} \\[2pt]
G_{6,8}^{(cud)}(\rho) &= 
\frac{3}{2}(1-\rho)\sqrt{\rho}\,(\rho^2 - 2\rho\log\rho - 1) \,, \label{eq:G68} \\[2pt]
G_{6,10}^{(cud)}(\rho) &= 
-4\sqrt{\rho}\,(\rho^3 - 6\rho\log\rho + 3\rho - 4) \,, \label{eq:G610} \\[2pt]
G_{8,10}^{(cud)}(\rho) &= 8(1-\rho)^3, \label{eq:G810} \\[2pt]
G_{9,9}^{(cud)}(\rho) &= 
-2\big(15\rho^4 - 64\rho^3 + 24\rho^2 + 12(3\rho+4)\rho\log\rho + 25\big) \,, \label{eq:G99} \\[2pt]
G_{10,10}^{(cud)}(\rho) &= 
-2\big(15\rho^4 - 40\rho^3 - 24\rho^2 + 36\rho^2\log\rho + 72\rho - 23\big) \,. \label{eq:G1010}
\end{align}
The remaining non-vanishing coefficient functions are obtained from the
relations
\begin{equation}
\begin{aligned}
& G_{3,3}^{(cud)} = G_{2,2}^{(cud)} = G_{1,1}^{(cud)},\qquad 
G_{7,7}^{(cud)} = G_{5,5}^{(cud)} \,,\qquad 
G_{8,8}^{(cud)} = G_{6,6}^{(cud)} = \frac{1}{4}G_{1,1}^{(cud)} \,, \\[6pt]
& 
G_{3,2}^{(cud)} = G_{1,4}^{(cud)} = G_{1,3}^{(cud)} \,, \qquad 
G_{3,4}^{(cud)} = G_{4,4}^{(cud)} \,,\qquad 
G_{5,6}^{(cud)} = G_{5,5}^{(cud)} \,, \\[6pt]
& 
 G_{5,8}^{(cud)} = G_{5,7}^{(cud)} \,,\qquad 
G_{5,10}^{(cud)} = 2G_{6,10}^{(cud)} \,,
\qquad
G_{9,6}^{(cud)} = 2G_{6,10}^{(cud)} \,,\\[6pt]
& 
G_{7,8}^{(cud)} = G_{7,7}^{(cud)} \,,\qquad 
G_{7,10}^{(cud)} = 2G_{8,10}^{(cud)} \,,\qquad 
G_{7,6}^{(cud)} = G_{5,7}^{(cud)} \,,
\\[6pt]
& G_{9,8}^{(cud)} = G_{7,10}^{(cud)} \,, \qquad 
G_{9,10}^{(cud)} = G_{9,9}^{(cud)} \, ,
\end{aligned}
\label{eq:Grelations}
\end{equation}
while all other coefficient functions vanish identically.
\paragraph{Dimension-six contribution ($\Gamma_6$):}
For the Darwin operator, the independent non-vanishing coefficient
functions $D_{i,j}^{(cud)}(\rho)$ in \cref{eq:Fij} are
\begin{flalign}
& D_{1,1}^{(cud)}(\rho)
= -\frac{2}{3}\big(5\rho^4 - 16\rho^3 + 12\rho^2
+ 16\rho - 17 - 12\log\rho\big) \,,
&& \label{eq:D11}
\\[2pt]
& D_{1,2}^{(cud)}(\rho)
= -\frac{2}{3}\big(5\rho^4 - 54\rho^3 + 90\rho^2
- 50\rho + 9\big)
- 16(\rho - 1)^3\log(1-\rho)
\notag\\
& \qquad\qquad
+ 8(\rho^3 - 3\rho^2 + 1)\log\rho \,,
&& \label{eq:D12}
\end{flalign}
\begin{align}
D_{1,3}^{(cud)}(\rho) &= \frac{8}{3} \sqrt{\rho} \big(2 \rho^3 - 9 \rho^2 + 18 \rho - 11 - 6 \log\rho\big) \,, \label{eq:D13} \\[2pt]
D_{1,4}^{(cud)}(\rho) &= \frac{16}{3} \sqrt{\rho} \big(\rho^3 - 6 \rho^2 + 12 \rho - 7 - 3 \log\rho\big) \,, \label{eq:D14} \\[2pt]
D_{2,2}^{(cud)}(\rho) &= \frac{2}{3} (1 - \rho) \big(5 \rho^3 - 25 \rho^2 + 11 \rho + 9 + 24 (\rho^2 - 1) \log(1 - \rho) - 12 \rho^2 \log\rho\big) \,, \label{eq:D22} \\[2pt]
D_{2,4}^{(cud)}(\rho) &= \frac{4}{3} \sqrt{\rho} \big(5 \rho^3 - 24 \rho^2 + 27 \rho - 8 + 12 (\rho - 1)^2 \log(1 - \rho) - 6 \rho (\rho - 2) \log\rho\big) \,, \label{eq:D24} \\[2pt]
D_{3,4}^{(cud)}(\rho) &= -\frac{2}{3} \big(5 \rho^4 + 18 \rho^3 - 78 \rho^2 + 118 \rho - 63 + 24 (1 - \rho)^3 \log(1 - \rho) \notag \\ 
& \qquad \qquad + 12 (\rho^3 - 3 \rho^2 - 1) \log\rho\big) \,, \label{eq:D34} \\[2pt]
D_{4,4}^{(cud)}(\rho) &= -\frac{2}{3} \big(5 \rho^4 + 8 \rho^3 - 66 \rho^2 + 112 \rho - 59 + 48 (1 - \rho)^2 \log(1 - \rho) - 24 \rho^2 \log\rho\big) \,, \label{eq:D44} \\[2pt]
D_{5,5}^{(cud)}(\rho) &= \frac{1}{6} \big(-5 \rho^4 + 22 \rho^3 - 54 \rho^2 - 22 \rho + 59 + 12 (5 \rho + 2) \log\rho\big) \,, \label{eq:D55} \\[2pt]
D_{5,6}^{(cud)}(\rho) &= \frac{1}{6} \big(-5 \rho^4 + 26 \rho^3 - 66 \rho^2 - 10 \rho + 55 + 12 (5 \rho + 2) \log\rho\big) \,, \label{eq:D56} \\[2pt]
D_{6,8}^{(cud)}(\rho) &= -\frac{4}{3} \sqrt{\rho} \big(\rho^3 - 6 \rho^2 + 6 \rho - 1 + 6 (\rho - 1)^2 \log(1 - \rho) - 3 \rho (\rho - 2) \log\rho\big) \,, \label{eq:D68} \\[2pt]
D_{6,10}^{(cud)}(\rho) &= -8 \sqrt{\rho} (\rho - 2) (\rho - 1)^2 \,, \label{eq:D610} \\[2pt]
D_{7,8}^{(cud)}(\rho) &= \frac{1}{6} \big(-5 \rho^4 + 18 \rho^3 - 42 \rho^2 - 34 \rho + 63\big) + 2 (5 \rho + 2) \log\rho \,, \label{eq:D78} \\[2pt]
D_{7,10}^{(cud)}(\rho) &= 16 \big(-3 \rho^3 + 7 \rho^2 - 7 \rho + 3 + 2 (\rho - 1)^3 \log(1 - \rho) - \rho^2 (\rho - 3) \log\rho\big) \,, \label{eq:D710} \\[2pt]
D_{8,8}^{(cud)}(\rho) &= \frac{1}{6} (1 - \rho) \big(5 \rho^3 - 21 \rho^2 + 3 \rho + 13 + 24 (\rho^2 - 1) \log(1 - \rho) - 12 \rho^2 \log\rho\big) \,, \label{eq:D88} \\[2pt]
D_{8,10}^{(cud)}(\rho) &= \frac{4}{3} \big(-17 \rho^3 + 45 \rho^2 - 51 \rho + 23 + 12 (\rho - 1)^3 \log(1 - \rho) - 6 \rho^2 (\rho - 3) \log\rho\big), \label{eq:D810} \\[2pt]
D_{9,9}^{(cud)}(\rho) &= -8 \big(5 \rho^4 - 14 \rho^3 - 2 \rho^2 + 14 \rho - 3 + 4 (5 \rho - 2) \log\rho\big) \,, \label{eq:D99} \\[2pt]
D_{9,10}^{(cud)}(\rho) &= -\frac{8}{3} \big(15 \rho^4 - 38 \rho^3 - 66 \rho^2 + 150 \rho - 61 + 12 (5 \rho - 2) \log\rho\big) \,, \label{eq:D910} \\[2pt]
D_{10,10}^{(cud)}(\rho) &= -\frac{8}{3} \big(15 \rho^4 - 10 \rho^3 - 108 \rho^2 + 234 \rho - 131 + 24 (\rho - 1)^2 (\rho + 5) \log(1 - \rho) \notag \\ 
& \qquad \qquad - 12 \rho^2 (\rho + 3) \log\rho\big) \,. \label{eq:D1010}
\end{align}
The remaining non-vanishing coefficient functions are obtained from
\begin{equation}
\begin{aligned}
& D_{3,3}^{(cud)} = D_{1,1}^{(cud)} \,, \qquad 
D_{7,7}^{(cud)} = D_{5,5}^{(cud)} \,, \qquad 
D_{6,6}^{(cud)} = \frac{1}{4}D_{2,2}^{(cud)} \,,\\[6pt]
&  
D_{5,7}^{(cud)} = -\frac{1}{4}D_{1,3}^{(cud)} \,, \qquad 
D_{5,8}^{(cud)} = -\frac{1}{4}D_{1,3}^{(cud)} \,, \qquad D_{5,10}^{(cud)} = \frac{2(\rho-1)}{\rho-2}D_{6,10}^{(cud)} \,,\\[6pt]
&  
D_{i,j}^{(cud)} = D_{i-1,j+1}^{(cud)} \qquad \text{for even } i \text{ and odd } j \,,
\end{aligned}
\label{eq:Drelations}
\end{equation}
while all other coefficient functions vanish identically.
\subsection
[\texorpdfstring{Results for the $b \to c \bar c s$ Transition}
{Results for the b -> c cbar s Transition}]
{Results for the $b \to c \bar c s$ Transition}
\label{sec:bccs}
\paragraph{Dimension-three contribution ($\Gamma_3$):}
For convenience, we define
\begin{equation}
\Delta_\rho \equiv \sqrt{1-4\rho},
\qquad
\Lambda_\rho \equiv
\log\!\left(\frac{1+\Delta_\rho}{1-\Delta_\rho}\right).
\end{equation}
The independent non-vanishing coefficient functions
$P_{i,j}^{(ccs)}(\rho)$ are
\begin{align}
P_{1,1}^{(ccs)}(\rho) &= \Delta_{\rho}\left(1 - 14\rho - 2\rho^2 - 12\rho^3 \right) + 24\rho^2 (1 - \rho^2) \Lambda_{\rho} \,, \label{eq:P11ccs} \\[6pt]
P_{1,3}^{(ccs)}(\rho) &= 2\sqrt{\rho} \left[ \Delta_{\rho}\left(-1 - 5\rho + 6\rho^2 \right) + 6\rho (1 - 2\rho + 2\rho^2) \Lambda_{\rho} \right] \,. \label{eq:P13ccs}
\end{align}
The remaining non-vanishing coefficient functions are obtained from
\begin{equation}
\begin{aligned}
&P_{2,2}^{(ccs)} = P_{3,3}^{(ccs)} = P_{4,4}^{(ccs)} = P_{1,1}^{(ccs)},
\\[4pt]
&P_{5,5}^{(ccs)} = P_{6,6}^{(ccs)} = P_{7,7}^{(ccs)} = P_{8,8}^{(ccs)}
= \tfrac14 P_{1,1}^{(ccs)},
\\[4pt]
&P_{9,9}^{(ccs)} = P_{10,10}^{(ccs)}
= 12\,P_{1,1}^{(ccs)},
\\[8pt]
&P_{2,4}^{(ccs)} = P_{1,3}^{(ccs)},
\qquad
P_{5,7}^{(ccs)} = P_{6,8}^{(ccs)}
= -\tfrac12 P_{1,3}^{(ccs)},
\\[8pt]
&P_{1,5}^{(ccs)} = P_{3,7}^{(ccs)} = P_{2,6}^{(ccs)} = P_{4,8}^{(ccs)}
= \sqrt{\rho}\,P_{1,3}^{(ccs)},
\\[4pt]
&P_{1,7}^{(ccs)} = P_{3,5}^{(ccs)} = P_{2,8}^{(ccs)} = P_{4,6}^{(ccs)}
= \tfrac12 P_{1,3}^{(ccs)},
\\[8pt]
&P_{1,9}^{(ccs)} = P_{2,10}^{(ccs)}
= -6\,P_{1,3}^{(ccs)},
\qquad
P_{3,9}^{(ccs)} = P_{4,10}^{(ccs)}
= 12\sqrt{\rho}\,P_{1,3}^{(ccs)},
\\[8pt]
&P_{i,j}^{(ccs)} = P_{i,j-1}^{(ccs)},
\qquad
\text{for odd } i \text{ and even } j.
\end{aligned}
\label{eq:Prelationsccs}
\end{equation}
while all other coefficient functions vanish identically.
\paragraph{Dimension-five contribution ($\Gamma_5$):}
The kinetic contribution is obtained from the leading-power result
according to \cref{eq:Fij}. The independent non-vanishing
chromomagnetic coefficient functions $G_{i,j}^{(ccs)}(\rho)$ are
\begin{align}
G_{1,1}^{(ccs)}(\rho) &= -\frac{1}{2\Delta_{\rho}} \Big[ 3 - 22\rho + 50\rho^2 + 20\rho^3 - 240\rho^4 + 24\Delta_{\rho}\rho^2(-1 + 5\rho^2)\Lambda_{\rho} \Big] \,, \label{eq:G11ccs} \\[2pt]
G_{1,2}^{(ccs)}(\rho)
&= \frac{1}{2\Delta_{\rho}} \Big[
-19 + 78\rho - 66\rho^2 + 172\rho^3 + 240\rho^4
\nonumber\\
&\qquad\qquad
- 24\Delta_{\rho}\rho(-2 - \rho + 4\rho^2 + 5\rho^3)\Lambda_{\rho}
\Big] \,,\label{eq:G12ccs} \\[2pt]
G_{1,3}^{(ccs)}(\rho) &= \frac{\sqrt{\rho}}{3\Delta_{\rho}} \Big[ -13 + 23\rho + 266\rho^2 - 600\rho^3 + 6\Delta_{\rho}(2 - 3\rho - 18\rho^2 + 50\rho^3)\Lambda_{\rho} \Big] \,, \label{eq:G13ccs} \\[2pt]
G_{1,4}^{(ccs)}(\rho) &= \frac{\sqrt{\rho}}{3\Delta_{\rho}} \Big[ -13 + 41\rho + 302\rho^2 - 1032\rho^3 + 6\Delta_{\rho}(2 - 9\rho - 18\rho^2 + 86\rho^3)\Lambda_{\rho} \Big] \,, \label{eq:G14ccs} \\[2pt]
G_{1,5}^{(ccs)}(\rho) &= \frac{\rho}{\Delta_{\rho}} \Big[ 10 - 71\rho + 154\rho^2 - 120\rho^3 + 6\Delta_{\rho}(-1 + 5\rho - 12\rho^2 + 10\rho^3)\Lambda_{\rho} \Big] \,, \label{eq:G15ccs} \\[2pt]
G_{1,6}^{(ccs)}(\rho) &= \frac{\rho}{\Delta_{\rho}} \Big[ 7 - 53\rho + 130\rho^2 - 120\rho^3 + 6\Delta_{\rho}\rho(1 - 10\rho + 10\rho^2)\Lambda_{\rho} \Big] \,, \label{eq:G16ccs} \\[2pt]
G_{1,7}^{(ccs)}(\rho) &= \frac{\sqrt{\rho}}{2\Delta_{\rho}} \Big[ -5 + 13\rho + 22\rho^2 + 24\rho^3 - 6\Delta_{\rho}\rho(-3 + 2\rho + 2\rho^2)\Lambda_{\rho} \Big] \,, \label{eq:G17ccs} \\[2pt]
G_{1,8}^{(ccs)}(\rho) &= -\frac{3\sqrt{\rho}}{2\Delta_{\rho}} \Big[ 9 - 37\rho + 10\rho^2 - 24\rho^3 + 2\Delta_{\rho}(-2 - \rho - 2\rho^2 + 6\rho^3)\Lambda_{\rho} \Big] \,, \label{eq:G18ccs}
\end{align}
\begin{align}
G_{1,9}^{(ccs)}(\rho) &= \frac{2\sqrt{\rho}}{\Delta_{\rho}} \Big[ -17 + 73\rho - 98\rho^2 + 312\rho^3 - 6\Delta_{\rho}\rho(-7 - 6\rho + 26\rho^2)\Lambda_{\rho} \Big] \,, \label{eq:G19ccs} \\[2pt]
G_{1,10}^{(ccs)}(\rho) &= -\frac{6\sqrt{\rho}}{\Delta_{\rho}} \Big[ -27 + 119\rho - 14\rho^2 - 120\rho^3 + 2\Delta_{\rho}(6 - 5\rho + 6\rho^2 + 30\rho^3)\Lambda_{\rho} \Big] \,, \label{eq:G110ccs}
\\[2pt]
G_{2,4}^{(ccs)}(\rho) &= \frac{\sqrt{\rho}}{\Delta_{\rho}} \Big[ -5 + 19\rho + 34\rho^2 - 120\rho^3 + 6\Delta_{\rho}\rho(1 - 2\rho + 10\rho^2)\Lambda_{\rho} \Big] \,, \label{eq:G24ccs} \\[2pt]
G_{2,8}^{(ccs)}(\rho) &= \frac{\sqrt{\rho}}{2\Delta_{\rho}} \Big[ 11 - 49\rho + 26\rho^2 - 24\rho^3 + 6\Delta_{\rho}\rho(-3 - 2\rho + 2\rho^2)\Lambda_{\rho} \Big] \,, \label{eq:G28ccs} \\[2pt]
G_{2,10}^{(ccs)}(\rho) &= \frac{2\sqrt{\rho}}{\Delta_{\rho}} \Big[ -1 + 11\rho - 94\rho^2 + 264\rho^3 + 6\Delta_{\rho}\rho(1 + 6\rho - 22\rho^2)\Lambda_{\rho} \Big] \,, \label{eq:G210ccs} \\[2pt]
G_{3,4}^{(ccs)}(\rho) &= \frac{1}{2\Delta_{\rho}} \Big[ 13 - 106\rho + 206\rho^2 - 20\rho^3 + 240\rho^4 - 24\Delta_{\rho}\rho^2(-1 + 5\rho^2)\Lambda_{\rho} \Big] \,, \label{eq:G34ccs} \\[2pt]
G_{3,5}^{(ccs)}(\rho) &= \frac{\sqrt{\rho}}{2\Delta_{\rho}} \Big[ -5 + 13\rho + 22\rho^2 + 24\rho^3 - 6\Delta_{\rho}\rho(-3 + 2\rho + 2\rho^2)\Lambda_{\rho} \Big] \,, \label{eq:G35ccs} \\[2pt]
G_{3,6}^{(ccs)}(\rho) &= \frac{\sqrt{\rho}}{6\Delta_{\rho}} \Big[ -17 + 73\rho - 98\rho^2 + 312\rho^3 - 6\Delta_{\rho}(2 - 15\rho - 6\rho^2 + 26\rho^3)\Lambda_{\rho} \Big] \,, \label{eq:G36ccs} \\[2pt]
G_{3,7}^{(ccs)}(\rho) &= \frac{\rho}{\Delta_{\rho}} \Big[ 10 - 71\rho + 154\rho^2 - 120\rho^3 + 6\Delta_{\rho}(-1 + 5\rho - 12\rho^2 + 10\rho^3)\Lambda_{\rho} \Big] \,, \label{eq:G37ccs} \\[2pt]
G_{3,8}^{(ccs)}(\rho) &= \frac{\rho}{\Delta_{\rho}} \Big[ 13 - 89\rho + 178\rho^2 - 120\rho^3 + 6\Delta_{\rho}(-2 + 9\rho - 14\rho^2 + 10\rho^3)\Lambda_{\rho} \Big] \,, \label{eq:G38ccs} \\[2pt]
G_{3,9}^{(ccs)}(\rho) &= \frac{12\rho}{\Delta_{\rho}} \Big[ -2 + \rho + 58\rho^2 - 120\rho^3 + 2\Delta_{\rho}(1 - \rho - 12\rho^2 + 30\rho^3)\Lambda_{\rho} \Big] \,, \label{eq:G39ccs} \\[2pt]
G_{3,10}^{(ccs)}(\rho) &= \frac{12\rho}{\Delta_{\rho}} \Big[ -11 + 55\rho - 14\rho^2 - 120\rho^3 + 2\Delta_{\rho}(2 - 5\rho + 6\rho^2 + 30\rho^3)\Lambda_{\rho} \Big] \,, \label{eq:G310ccs} \\[2pt]
G_{5,5}^{(ccs)}(\rho)
&= \frac{1}{8\Delta_{\rho}} \Big[
13 - 34\rho - 34\rho^2 - 212\rho^3 + 240\rho^4
\nonumber\\
&\qquad\qquad
- 24\Delta_{\rho}\rho(2 - \rho - 4\rho^2 + 5\rho^3)\Lambda_{\rho}
\Big] \,,
\label{eq:G55ccs} \\[2pt]
G_{5,7}^{(ccs)}(\rho) &= -\frac{3\sqrt{\rho}}{2\Delta_{\rho}} \Big[ 11 - 49\rho + 26\rho^2 - 24\rho^3 + 2\Delta_{\rho}(-2 - \rho - 6\rho^2 + 6\rho^3)\Lambda_{\rho} \Big] \,, \label{eq:G57ccs} \\[2pt]
G_{5,10}^{(ccs)}(\rho) &= 4\sqrt{\rho} \big[ \Delta_{\rho}(8 + \rho + 6\rho^2) + 6\rho(-3 + 2\rho^2)\Lambda_{\rho} \big] \,, \label{eq:G510ccs} \\[2pt]
G_{6,8}^{(ccs)}(\rho) &= \frac{3\sqrt{\rho}}{2\Delta_{\rho}} \Big[ -1 + 5\rho - 10\rho^2 + 24\rho^3 + 2\Delta_{\rho}\rho(1 + 2\rho - 6\rho^2)\Lambda_{\rho} \Big] \,, \label{eq:G68ccs} \\[2pt]
G_{6,10}^{(ccs)}(\rho) &= -4\sqrt{\rho} \big[ \Delta_{\rho}(-4 + \rho + 6\rho^2) + 6(\rho + 2\rho^3)\Lambda_{\rho} \big] \,, \label{eq:G610ccs} \\[2pt]
G_{7,10}^{(ccs)}(\rho) &= 4 \big[ \Delta_{\rho}(4 - 7\rho + 6\rho^2) + 6\rho(-1 + 2\rho^2)\Lambda_{\rho} \big] \,, \label{eq:G710ccs} \\[2pt]
G_{8,10}^{(ccs)}(\rho) &= 8\Delta_{\rho}^{3} \,, \label{eq:G810ccs} \\[2pt]
G_{9,9}^{(ccs)}(\rho)
&= \frac{2}{\Delta_{\rho}} \Big[
-25 + 122\rho - 166\rho^2 + 132\rho^3 + 720\rho^4
\nonumber\\
&\qquad\qquad
- 24\Delta_{\rho}\rho(-2 - 3\rho + 4\rho^2 + 15\rho^3)\Lambda_{\rho}
\Big] \,,
\label{eq:G99ccs} \\[2pt]
G_{9,10}^{(ccs)}(\rho)
&= \frac{2}{\Delta_{\rho}} \Big[
-25 - 22\rho + 314\rho^2 + 516\rho^3 + 720\rho^4
\nonumber\\
&\qquad\qquad
- 72\Delta_{\rho}\rho(-2 - \rho + 4\rho^2 + 5\rho^3)\Lambda_{\rho}
\Big] \,,
\label{eq:G910ccs} \\[2pt]
G_{10,10}^{(ccs)}(\rho) &= \frac{2}{\Delta_{\rho}} \Big[ 23 - 190\rho + 362\rho^2 - 60\rho^3 + 720\rho^4 - 72\Delta_{\rho}\rho^2(-1 + 5\rho^2)\Lambda_{\rho} \Big] \,. \label{eq:G1010ccs}
\end{align}
The remaining non-vanishing coefficient functions are obtained from
\begin{equation}
\begin{aligned}
& G_{2,2}^{(ccs)} = G_{1,1}^{(ccs)} \,,\quad
G_{2,6}^{(ccs)} = G_{1,6}^{(ccs)} \,,\quad
G_{3,3}^{(ccs)} = G_{1,1}^{(ccs)} \,,\quad
G_{4,4}^{(ccs)} = G_{3,4}^{(ccs)} \,,\\[6pt]
& G_{3,5}^{(ccs)} = G_{1,7}^{(ccs)} \,,\quad
G_{4,6}^{(ccs)} = \frac{1}{2}G_{2,4}^{(ccs)} \,,\quad
G_{4,8}^{(ccs)} = \sqrt{\rho}\,G_{2,4}^{(ccs)} \,,\quad
G_{4,10}^{(ccs)} = 12\sqrt{\rho}\,G_{2,4}^{(ccs)} \,,\\[6pt]
& G_{5,6}^{(ccs)} = G_{5,5}^{(ccs)} \,,\quad
G_{5,8}^{(ccs)} = G_{5,7}^{(ccs)} \,,\quad
G_{6,6}^{(ccs)} = \frac{1}{4}G_{3,3}^{(ccs)} \,,\quad
G_{7,8}^{(ccs)} = G_{7,7}^{(ccs)} = G_{5,5}^{(ccs)} \,,\\[6pt]
& G_{8,8}^{(ccs)} = G_{6,6}^{(ccs)} \,,\quad
G_{i,j}^{(ccs)} = G_{i-1,j+1}^{(ccs)}
\quad \text{for even } i \text{ and odd } j \,,
\end{aligned}
\label{eq:Grelationsccs}
\end{equation}
while all other coefficient functions vanish identically.
\paragraph{Dimension-six contribution ($\Gamma_6$):}
The independent non-vanishing Darwin coefficient functions
$D_{i,j}^{(ccs)}(\rho)$ in \cref{eq:Fij} are
\begin{align}
D_{1,1}^{(ccs)}(\rho) &= -\frac{2}{3} \Big[ \Delta_{\rho} (-17 - 8\rho + 22\rho^2 + 60\rho^3) + 12 (1 - \rho - 2\rho^2 + 2\rho^3 + 10\rho^4) \Lambda_{\rho} \Big] \,, \label{eq:D11ccs} \\[2pt]
D_{1,2}^{(ccs)}(\rho)& =
- \frac{2}{3} \Delta_\rho \left[33 - 46 \rho + 106 \rho^2 + 60 \rho^3 + 24 \log (1- 4 \rho) - 24 \log \rho \right]
\nonumber \\
&  \qquad \qquad + 8 \left[3 - 2 \rho + 4 \rho^2 - 16 \rho^3 - 10\rho^4 \right] \Lambda_\rho 
\label{eq:D12ccs} \\[2pt]
D_{1,3}^{(ccs)}(\rho) &= \frac{8}{3} \sqrt{\rho} \Big[ \Delta_{\rho} (-11 - 7\rho + 30\rho^2) + 6 (1 - 4\rho^2 + 10\rho^3) \Lambda_{\rho} \Big] \,, \label{eq:D13ccs} \\[2pt]
D_{1,4}^{(ccs)}(\rho) & = \frac{4}{3} \sqrt{\rho} \Big[\Delta_\rho (-22 - 23 \rho + 114 \rho^2 + 12 \log(1 - 4 \rho) - 
      12 \log(\rho))  \nonumber \\
& \qquad \qquad + 6 (-1 + 6 \rho - 14 \rho^2 + 38 \rho^3) \Lambda_\rho \Big]  \,,
\label{eq:D14ccs} \\[2pt]
D_{1,5}^{(ccs)}(\rho) &= \frac{4}{3} \rho \Big[ \Delta_{\rho} (7 - 31\rho + 30\rho^2) + 3 (-1 + 8\rho - 24\rho^2 + 20\rho^3) \Lambda_{\rho} \Big] \,, \label{eq:D15ccs} \\[2pt]
D_{1,7}^{(ccs)}(\rho) &= -\frac{4}{3} \sqrt{\rho} \Big[ 2\Delta_{\rho} (2 - 5\rho + 6\rho^2) + 3 (-1 + 4\rho - 8\rho^2 + 8\rho^3) \Lambda_{\rho} \Big] \,, \label{eq:D17ccs} \\[2pt]
D_{1,8}^{(ccs)} (\rho)  & = \frac{2}{3} \sqrt{\rho} \Big[ \Delta_\rho (7 + 29 \rho - 42 \rho^2 + 12 \log(1 - 4 \rho) - 12 \log \rho)
\notag \\[2pt]
& \qquad \qquad  - 12 (1 + \rho - 6 \rho^2 + 7 \rho^3) \Lambda_\rho \Big] \,, \label{eq:D18ccs} \\[2pt]
D_{1,9}^{(ccs)}(\rho) &= -16 \sqrt{\rho} \Big[ 2\Delta_{\rho} (-1 + \rho + 6\rho^2) + 3 (1 - 4\rho + 8\rho^3) \Lambda_{\rho} \Big] \,, \label{eq:D19ccs} \\[2pt]
D_{1,10}^{(ccs)} (\rho)  & = 8 \sqrt{\rho} \Big[-\Delta_\rho (3 + 17 \rho + 30 \rho^2 + 12 \log(1 - 4 \rho) - 
      12 \log \rho)  \nonumber \\ 
& \qquad \qquad + 12 (1 + \rho - 2 \rho^2 - 5 \rho^3) \Lambda_\rho \Big] \,, \label{eq:D110ccs} \\[2pt]
D_{2,2}^{(ccs)}(\rho) &= \frac{2}{3} \Delta_{\rho} \big[9 + 22\rho - 34\rho^2 - 60\rho^3 - 24\log(1 - 4\rho) + 24\log\rho\big] \nonumber \\
& \qquad \qquad - 16 \big[-1 + 2\rho + \rho^2 + 2\rho^3 + 5\rho^4\big] \Lambda_{\rho} \,, \label{eq:D22ccs} \\[2pt]
D_{2,4}^{(ccs)}(\rho) &= \frac{4}{3} \sqrt{\rho} \Big[ \Delta_{\rho} \big(-8 + 11\rho + 30\rho^2 + 12\log(1 - 4\rho) - 12\log\rho\big) \nonumber \\*
& \qquad \qquad + 12 \big(-1 + \rho + \rho^2 + 5\rho^3\big) \Lambda_{\rho} \Big] \,, \label{eq:D24ccs}
\\[2pt]
D_{2,6}^{(ccs)}(\rho) &= \frac{4}{3} \rho \Big[ \Delta_{\rho} \big(7 - 7\rho + 30\rho^2 + 12\log(1 - 4\rho) - 12\log\rho\big) \notag \\
& \qquad \qquad + 12 \big(-1 + \rho - 2\rho^2 + 5\rho^3\big) \Lambda_{\rho} \Big] \,, \label{eq:D26ccs} \\[2pt]
D_{2,8}^{(ccs)}(\rho)
&= \frac{2}{3}\sqrt{\rho}\Big[
\Delta_{\rho}\bigl(
4 + 5\rho - 6\rho^2 + 12\log(1 - 4\rho) - 12\log\rho
\bigr)
\nonumber\\
&\qquad\qquad
- 12(-1 + \rho)^2(1 + \rho)\Lambda_{\rho}
\Big] \,,
\label{eq:D28ccs} \\[2pt]
D_{2,10}^{(ccs)}(\rho) &= 8 \sqrt{\rho} \Big[ \Delta_{\rho} \big(4 - 9\rho - 18\rho^2 - 12\log(1 - 4\rho) + 12\log\rho\big) \nonumber \\
& \qquad \qquad - 12 \big(-1 + \rho + \rho^2 + 3\rho^3\big) \Lambda_{\rho} \Big] \,, \label{eq:D210ccs} \\[2pt]
D_{3,4}^{(ccs)}(\rho)
&= -\frac{2}{3}\Delta_{\rho}\Big[
-59 + 106\rho + 10\rho^2 + 60\rho^3
\nonumber\\
&\qquad\qquad
+ 48(1-2\rho)\log(1-4\rho)
+ 48(-1+2\rho)\log\rho
\Big]
\nonumber\\
&\qquad\qquad
+ 16\bigl(2-8\rho+2\rho^2-5\rho^4\bigr)\Lambda_{\rho}\,,
\label{eq:D34ccs} \\[2pt]
D_{3,6}^{(ccs)}(\rho) &= 2 \sqrt{\rho} \Big[ \Delta_{\rho} \big(2 + 15\rho - 18\rho^2 + 4\log(1 - 4\rho) - 4\log\rho\big) \notag \\
& \qquad \qquad - 2 \big(3 + 2\rho - 18\rho^2 + 18\rho^3\big) \Lambda_{\rho} \Big] \,, \label{eq:D36ccs} \\[2pt]
D_{3,8}^{(ccs)}(\rho) &= \frac{2}{3} \rho \Big[ \Delta_{\rho} \big(11 - 32\rho + 60\rho^2 + 24\log(1 - 4\rho) - 24\log\rho\big) \nonumber \\
& \qquad \qquad + 6 \big(-7 + 16\rho - 14\rho^2 + 20\rho^3\big) \Lambda_{\rho} \Big] \,, \label{eq:D38ccs} \\[2pt]
D_{3,9}^{(ccs)}(\rho) &= 16 \rho \Big[ \Delta_{\rho} (-2 - 13\rho + 30\rho^2) + (3 - 36\rho^2 + 60\rho^3) \Lambda_{\rho} \Big] \,, \label{eq:D39ccs} \\[2pt]
D_{3,10}^{(ccs)}(\rho) &= 8 \rho \Big[ \Delta_{\rho} \big(-25 + 40\rho + 60\rho^2 + 24\log(1 - 4\rho) - 24\log\rho\big) \nonumber \\
& \qquad \qquad + 6 \big(-3 + 10\rho^2 + 20\rho^3\big) \Lambda_{\rho} \Big] \,, \label{eq:D310ccs} \\[2pt]
D_{5,5}^{(ccs)}(\rho)
&= \frac{1}{6}\Big[
-\Delta_{\rho}(-59 + 16\rho - 50\rho^2 + 60\rho^3)
\nonumber\\
&\qquad\qquad
- 12(2 + \rho + 4\rho^2 - 10\rho^3 + 10\rho^4)\Lambda_{\rho}
\Big] \,,
\label{eq:D55ccs}\\[2pt]
D_{5,6}^{(ccs)}(\rho) &= \frac{1}{6} \Delta_{\rho} \big[51 - 2\rho + 38\rho^2 - 60\rho^3 - 24\log(1 - 4\rho) + 24\log\rho\big] \nonumber \\
& \qquad \qquad - 2 \big(-1 + 6\rho + 8\rho^2 - 8\rho^3 + 10\rho^4\big) \Lambda_{\rho} \,, \label{eq:D56ccs} \\[2pt]
D_{5,7}^{(ccs)}(\rho) &= -\frac{2}{3} \sqrt{\rho} \Big[ \Delta_{\rho} (-11 - 4\rho + 12\rho^2) + 6 (1 - 2\rho^2 + 4\rho^3) \Lambda_{\rho} \Big] \,, \label{eq:D57ccs} \\[2pt]
D_{5,8}^{(ccs)}(\rho)
&= \frac{8}{3}\sqrt{\rho}\Big[
\Delta_{\rho}\bigl(
2 + \rho - 3\rho^2 - 3\log(1-4\rho) + 3\log\rho
\bigr)
\nonumber\\
&\qquad\qquad
+ 3(1 - 3\rho + \rho^2 - 2\rho^3)\Lambda_{\rho}
\Big] \,,
\label{eq:D58ccs} \\[2pt]
D_{5,10}^{(ccs)}(\rho) &= 8 \sqrt{\rho} \Big[ \Delta_{\rho} (3 - 5\rho + 6\rho^2) + 12(-1 + \rho)\rho^2 \Lambda_{\rho} \Big] \,, \label{eq:D510ccs} \\[2pt]
D_{6,8}^{(ccs)}(\rho)
&= \frac{4}{3}\sqrt{\rho}\Big[
\Delta_{\rho}\bigl(
1 - 4\rho - 6\rho^2 - 6\log(1 - 4\rho) + 6\log\rho
\bigr)
\nonumber\\
&\qquad\qquad
- 6\bigl(-1 + \rho + \rho^2 + 2\rho^3\bigr)\Lambda_{\rho}
\Big] \,,
\label{eq:D68ccs} \\[2pt]
D_{6,10}^{(ccs)}(\rho) &= -8 \sqrt{\rho} \Big[ \Delta_{\rho} (-2 + \rho + 6\rho^2) + 12\rho^3 \Lambda_{\rho} \Big] \,, \label{eq:D610ccs} \\[2pt]
D_{7,8}^{(ccs)}(\rho) &= \frac{1}{6} \Delta_{\rho} \big[55 - 18\rho + 38\rho^2 - 60\rho^3 - 24\log(1 - 4\rho) + 24\log\rho\big] \nonumber \\
& \qquad \qquad - 2 \big(-1 + 6\rho + 8\rho^2 - 8\rho^3 + 10\rho^4\big) \Lambda_{\rho} \,, \label{eq:D78ccs}
\end{align}
\begin{align}
D_{7,10}^{(ccs)}(\rho) &= \frac{8}{3} \Big[ 2\Delta_{\rho} \big(11 - 17\rho + 12\rho^2 + 3(-1 + 4\rho)\log(1 - 4\rho) + 3(1-4\rho)\log\rho\big) \nonumber \\
& \qquad \qquad + (3 - 42\rho + 48\rho^3) \Lambda_{\rho} \Big] \,, \label{eq:D710ccs}
\\[2pt]
D_{8,8}^{(ccs)}(\rho) &= \frac{1}{6} \Delta_{\rho} \big[13 + 6\rho - 34\rho^2 - 60\rho^3 - 24\log(1 - 4\rho) + 24\log\rho\big] \nonumber \\
& \qquad \qquad - 4 \big[-1 + 2\rho + \rho^2 + 2\rho^3 + 5\rho^4\big] \Lambda_{\rho} \,, \label{eq:D88ccs} \\[2pt]
D_{8,10}^{(ccs)}(\rho) &= \frac{4}{3} \Delta_{\rho} \big[23 - 56\rho + 12\rho^2 + 12(-1 + 4\rho)\log(1 - 4\rho) + 12(1-4\rho)\log\rho\big] \nonumber \\
& \qquad \qquad + 16 \big(1 - 6\rho + 3\rho^2 + 2\rho^3\big) \Lambda_{\rho} \,, \label{eq:D810ccs} \\[2pt]
D_{9,9}^{(ccs)}(\rho) &= -8 \Big[ \Delta_{\rho} (-3 - 16\rho + 46\rho^2 + 60\rho^3) + 4 (2 - 5\rho - 12\rho^2 + 18\rho^3 + 30\rho^4) \Lambda_{\rho} \Big] \,, \label{eq:D99ccs} \\[2pt]
D_{9,10}^{(ccs)}(\rho) &= -\frac{8}{3} \Delta_{\rho} \big[-5 + 134\rho + 270\rho^2 + 180\rho^3 - 24(-5 + 8\rho)\log(1 - 4\rho) \nonumber \\
& \qquad \qquad + 24(-5 + 8\rho)\log\rho\big] - 32 \big(-13 + 30\rho - 24\rho^2 + 40\rho^3 + 30\rho^4\big) \Lambda_{\rho} \,, \label{eq:D910ccs} \\[2pt]
D_{10,10}^{(ccs)}(\rho) &= -\frac{8}{3} \Delta_{\rho} \big[-131 + 206\rho + 54\rho^2 + 180\rho^3 - 24(-5 + 8\rho)\log(1 - 4\rho) \nonumber \\
& \qquad \qquad + 24(-5 + 8\rho)\log\rho\big] - 64 \big[-5 + 18\rho - 3\rho^2 + 2\rho^3 + 15\rho^4\big] \Lambda_{\rho} \,. \label{eq:D1010ccs}
\end{align}
The remaining non-vanishing coefficient functions are obtained from the relations
\begin{equation}
\begin{aligned}
& D_{3,3}^{(ccs)} = D_{1,1}^{(ccs)} \,,\qquad 
D_{4,4}^{(ccs)} = D_{3,4}^{(ccs)} \,, \qquad
D_{3,5}^{(ccs)} = D_{1,7}^{(ccs)} \,,\qquad D_{3,7}^{(ccs)} = D_{1,5}^{(ccs)} \,, \\[6pt]
& D_{6,6}^{(ccs)} = \frac{1}{4}D_{2,2}^{(ccs)} \,, \qquad
D_{4,6}^{(ccs)} = \frac{1}{2}D_{2,4}^{(ccs)} \,,\qquad D_{4,8}^{(ccs)} = \sqrt{\rho}\,D_{2,4}^{(ccs)} \,,\\[6pt]
&  D_{4,10}^{(ccs)} = 12 \sqrt{\rho} \, D_{2,4}^{(ccs)} \,,\qquad D_{2,6}^{(ccs)} = D_{1,6}^{(ccs)} \,, \qquad D_{7,7}^{(ccs)} = D_{5,5}^{(ccs)}\,,\\[6pt]
& D_{i,j}^{(ccs)} = D_{i-1,j+1}^{(ccs)} \qquad \text{for even } i \text{ and odd } j \,,
\end{aligned}
\label{eq:Drelationsccs}
\end{equation}
while all other coefficient functions vanish identically.

Before concluding this chapter, we note that, although the calculations presented here
were performed predominantly in four dimensions, the complete set of
results has been independently reproduced using dimensional
regularisation in a separate calculation. The agreement between the
two calculations provides a highly non-trivial validation of the
analytical expressions obtained here. Details of the independent
calculation can be found in Ref.~\cite{Lang:2025ios}.

\section{Closing Remarks}

The results presented in this chapter provide the two-quark HQE
matching coefficients through dimension six for the most general
dimension-six $|\Delta B|=1$ four-fermion Hamiltonian considered in
\cref{eq:Heff-NP}. They can therefore be used to study generic
short-distance modifications of non-leptonic $b$-quark decays in a
model-independent way and to constrain the corresponding Wilson
coefficients using inclusive decay widths and lifetime ratios. Since
all relevant flavour combinations $b\to q_1\bar q_2q_3$, with
$q_1,q_2=u,c$ and $q_3=d,s$, have been considered, the results can be
applied simultaneously to CKM-favoured and CKM-suppressed decay
channels. In addition, the calculation of the previously missing
WA topology completes the dimension-six four-quark
matching contribution $\widetilde{\Gamma}_6$ relevant to the present
analysis.

An important consequence of the general analysis is the completion of
the dimension-six two-quark contribution in the SM at leading order in
QCD. In addition to the previously known current--current
contributions, the results of \cref{sec:SM_penguin} provide the
contributions induced by the QCD-penguin operators to the
chromomagnetic and Darwin coefficients at dimensions five and six,
respectively, while reproducing the known leading-power result.
Together with the current--current sector, these results therefore
complete the SM two-quark matching through dimension six at this
order.

The generalised operator basis also makes the results directly
applicable to phenomenological studies beyond the SM. They provide the
necessary short-distance input for global analyses combining
$B$-meson lifetimes and lifetime ratios with complementary flavour
observables, such as mixing parameters and CP asymmetries, thereby
allowing increasingly stringent constraints on BSM Wilson
coefficients. The potential of combining lifetime observables with complementary
flavour constraints has already been explored in
Refs.~\cite{Meiser:2024zea,Lenz:2022pgw}. In particular, lifetime
observables provide a complementary probe of possible new-physics
effects in non-leptonic $B$ decays and may help clarify the tensions
observed in several colour-allowed decay modes between experiment and
predictions based on QCD factorisation,
see e.g.~\cite{Bordone:2020gao,Huber:2016xod}.

Beyond a model-independent analysis, the matching coefficients derived
here can be mapped onto specific ultraviolet scenarios, including
models with additional charged gauge bosons, diquarks, or extended
Higgs sectors. The results can also be adapted to scenarios with
different colour structures after accounting for the corresponding
colour factors, as illustrated, for example, in
Ref.~\cite{Mohamed:2025zgx}. Further improvements, in particular
higher-order QCD corrections to the chromomagnetic and Darwin
coefficients and more precise determinations of the associated
non-perturbative matrix elements, will further enhance the sensitivity
of inclusive $b$-hadron observables to new physics.

\chapter{Phenomenology of \textit{B}-Mesogenesis Models} 
\label{Ch:4}
\vspace{3mm}

\begin{flushright}
\emph{
``Collaborating with someone is like entering a relationship with that person.''
}
\\[0.3cm]
--- Alexander Khodjamirian
\end{flushright}
The previous chapters were devoted to improving the HQE description of
inclusive non-leptonic $b$-hadron decays within the SM and extending it
to a model-independent effective Hamiltonian. We now shift our focus to
the phenomenology of a specific BSM scenario, namely
\textit{B}-Mesogenesis.\\
The chapter begins with an introduction to the \textit{B}-Mesogenesis
framework, including its effective Hamiltonian and the existing
constraints on the model parameter space. We then
investigate the impact of the operators introduced by the model on the
decay properties of $B$ mesons, with particular emphasis on the
lifetime ratios $\tau(B^+)/\tau(B_d)$ and
$\tau(B_s)/\tau(B_d)$, as well as on the total $B$-meson lifetimes.
The analysis is performed within the HQE, incorporating
power-suppressed corrections up to order $1/m_b^3$, corresponding to
dimension-six operators. In particular, we include both the
dimension-six two-quark (Darwin) contribution and the dimension-six
four-quark operators.\\
A distinctive feature of the \textit{B}-Mesogenesis scenario is the presence of a massive fermion in the final state, 
whose mass can approach that of the $b$-quark. 
In this regime, the convergence of the OPE becomes nontrivial, as higher-order power corrections may become comparable to—or even exceed—the leading term. 
We therefore examine in detail the range of dark-fermion masses for which the OPE remains valid, and identify the point at which the HQE description breaks down. 
This analysis provides valuable insight into the theoretical limitations of the inclusive approach in scenarios involving heavy final states.\\
Finally, we perform an exclusive calculation of the decay of a
$B^+$ meson into a proton and a dark-sector fermion,
$B^+ \to p\,\psi$, using the method of LCSR. To this end, we compute
the relevant $B\to p$ transition form factors based on $B$-meson
distribution amplitudes and use them to predict the corresponding
branching ratio (BR). By comparing the inclusive and exclusive
predictions, we establish a quantitative connection between the two
descriptions, enabling us to derive a lower bound on
$\mathrm{BR}(B^+\to p\,\psi)$. This connection provides a robust phenomenological test of the \textit{B}-Mesogenesis hypothesis and constrains its parameter space.

\section{The Model}
The $B$-Mesogenesis framework~\cite{Elor:2018twp,Alonso-Alvarez:2021qfd,Elahi:2021jia,Nelson:2019fln} offers an elegant and unified mechanism to explain both the observed baryon asymmetry of the Universe and the dark matter abundance through processes involving $B$-meson dynamics in the early Universe.\footnote{%
Several variations of this framework have been proposed in the literature, differing in their particle content—some even lacking a dark matter candidate~\cite{Nelson:2019fln}. 
In this work, however, we restrict ourselves to the specific realization presented in~\cite{Alonso-Alvarez:2021qfd}.} In this scenario, the early Universe is assumed to be dominated by radiation and a weakly coupled scalar field~$\Phi$, which decays predominantly into $b\bar{b}$ pairs. The produced $b$-quarks hadronize into $B$ and $\bar{B}$ mesons that subsequently decay not only through SM channels, but also via new, baryon-number–violating processes of the type 
$B \to \psi\,\mathcal{B}$, where $\mathcal{B}$ denotes a visible baryon and $\psi$ is a dark-sector fermion carrying baryon number\footnote{The dark fermion $\psi$ is not itself a dark matter candidate; it is required to decay rapidly into lighter, stable dark sector states (e.g., a scalar $\phi$ and a Majorana fermion $\xi$). These additional particles constitute the dark matter of the Universe but are irrelevant for the exotic $B$ decays studied in this work and are therefore not discussed further.}.

A baryon asymmetry arises from CP violation in neutral $B_d$--$\bar{B}_d$ and $B_s$--$\bar{B}_s$ meson mixing, which leads to unequal decay rates for $B_q$ and $\bar{B}_q$ into final states containing $\mathcal{B}$ and $\psi$. As a result, the matter–antimatter imbalance of the visible sector is directly linked to the CP-violating parameters of the $B$-meson system.

The new decay mode $B \to \psi\,\mathcal{B}$ proceeds through the exchange of a heavy colour-triplet scalar field~$Y$, which carries baryon number~$-2/3$ and hypercharge~$-1/3$ or~$+2/3$. The most general renormalizable Lagrangian governing the interactions of~$Y$ with the SM quarks and the gauge-singlet fermion~$\psi$ can be written as:
\begin{equation}
\begin{aligned}
\mathcal{L}_{-1/3} = -\sum_{\alpha,\beta}  y_{u_{\alpha}d_{\beta}} \epsilon_{{i}{j}{k}} \kern0.18emY^{*{i}} \kern0.18em \bar{u}^{{j}}_{\alpha R}\kern0.18em d^{c,{k}}_{\beta R} - \sum_{\gamma} y_{\psi d_{\gamma}}  Y_{{i}} \kern0.18em \bar{\psi}\kern0.18em d^{c,{i}}_{\gamma R} + \text{h.c.}\kern0.18em,\\
\mathcal{L}_{+2/3} = -\sum_{\alpha,\beta}  y_{d_{\alpha}d_{\beta}} \epsilon_{{i}{j}{k}} \kern0.18emY^{*{i}} \kern0.18em \bar{d}^{{j}}_{\alpha R}\kern0.18em d^{c,{k}}_{\beta R} - \sum_{\gamma} y_{\psi u_{\gamma}}  Y_{{i}} \kern0.18em \bar{\psi}\kern0.18em u^{c,{i}}_{\gamma R} + \text{h.c.}\kern0.18em,
\label{eq:ModelLagrangians}
\end{aligned}
\end{equation}
where $y$ denotes coupling constants, ${i},{j},{k}$ are colour indices, $\alpha,\beta,\gamma$ label quark flavours, and the superscript $c$ denotes charge conjugation. 
After integrating out the heavy colour-triplet $Y$ at tree level, the
resulting effective Hamiltonian mediating the $B$-meson decays can be
written as

\begin{align}
\mathcal H_{-1/3} \;=\; \sum_{u,d}\Bigg\{&
\underbrace{\epsilon_{{i}{j}{k}}\,
   C_{\psi}^{(ud)}\,(\bar u_R^{\,{i}} b_R^{c,{j}})\,(\bar\psi\, d_R^{c,{k}})
   + \epsilon_{{i}{j}{k}}\,
   \widetilde C_{\psi}^{(ud)}\,(\bar u_R^{\,{i}} d_R^{c,{j}})\,(\bar\psi\, b_R^{c,{k}})}_{\text{(i) $\psi$-type operators}}
\notag\\[4pt]
&+\,\underbrace{\epsilon_{{i}{j}{k}}\epsilon_{{i}{m}{n}}\,
   C_{uu}^{(d)}\,(\bar u_R^{\,{j}} b_R^{c,{k}})\,(\bar d_R^{c,{m}} u_R^{\,{n}})
   + C_{\psi\psi}^{(d)}\,(\bar\psi\, d_R^{c,{i}})\,(\bar b_R^{c,{i}}\,\psi)}_{\text{(ii) hadronic and $\psi\psi$ operators}}
\Bigg\} \;+\; \text{h.c.},
\label{eq:HeffMeso}
\end{align}
where $u \in \{u,c\},\ d \in \{d,s\},$ and the Wilson coefficients, including the heavy scale suppression, are
\begin{align}
C_{\psi}^{(ud)} &= -\frac{y_{ub}\,y_{\psi d}}{M_Y^2}, 
\ 
\widetilde C_{\psi}^{(ud)} = -\frac{y_{ud}\,y_{\psi b}}{M_Y^2}, \
C_{uu}^{(d)} = \frac{y_{ub}\,y_{ud}^{*}}{M_Y^2}, 
\ 
C_{\psi\psi}^{(d)} = -\frac{y_{\psi d}\,y_{\psi b}^{*}}{M_Y^2}.
\label{eq:Cpsipsi_map}
\end{align}
For concreteness, we focus the detailed analysis on the operators proportional to $C_{\psi}^{(ud)}$ and $\widetilde C_{\psi}^{(ud)}$.
We follow the convention in Ref.~\cite{Khodjamirian:2022vta} and refer to these operators as model $(d)$ and model $(b)$, respectively, and make the replacement $C_{\psi}^{(ud)} \to G_{(d)}$ and $\widetilde C_{\psi}^{(ud)} \to G_{(b)}$. 
The calculations for the remaining operators have been fully performed and their results are provided in \cref{app:BMesogenesisResutls}, with references to them given at the relevant points in the main text.

A comprehensive analysis of the existing experimental and SM
constraints on this model was performed in
Ref.~\cite{Alonso-Alvarez:2021qfd}. In the following, we briefly review
those constraints that are most relevant for the lifetime analysis
presented in this chapter.

\textbf{1. Lower limit on inclusive exotic $B$ decay}: 
A key feature of the $B$-Mesogenesis mechanism is the direct connection between the observed baryon asymmetry and two collider observables: the semileptonic CP asymmetries $A_{\text{SL}}^q$ in neutral $B_q$--$\bar{B}_q$ mixing, and the inclusive BR of the exotic decay $\text{BR}(B \to \psi X)$, where $X$ denotes any allowed final state containing at least one visible baryon $\mathcal{B}$ and possibly additional light mesons.
Given the experimental bound  $|A_{\text{SL}}^{d,s}| \lesssim 10^{-3}$ at 95\% CL~\cite{ParticleDataGroup:2026aaa}, this imposes a lower bound on the inclusive exotic BR. To reproduce the observed baryon asymmetry, one requires
\begin{equation}
\boxed{
\text{BR}(B \to \psi X) \gtrsim 10^{-4}.
\label{eq:LowerBoundBr}
}
\end{equation}
This bound is independent of the specific flavour structure of the effective operators in \cref{eq:HeffMeso}, and it holds for $B^+$, $B_d^0$, $B_s^0$, and $b$-flavoured baryons alike. This constraint will be used later in this work to derive corresponding lower bounds on the exclusive decay channel $B^+ \to p^+ \psi $.

\textbf{2. Effective couplings upper limits}: The colour-triplet scalar mediator $Y$, which generates the effective operators in \cref{eq:HeffMeso}, can be  produced at the LHC through its interactions with SM quarks. Searches for dijet resonances and jet plus missing transverse energy at ATLAS and CMS impose upper limits on the effective couplings. They will be used later in this work in the inclusive calculations of lifetimes and in deriving upper bounds on exclusive BR.

\textbf{3. Single operator dominance}:
flavour-changing neutral current and mixing observables in the $K$, $D$, and $B$ meson systems provide stringent constraints on the couplings of the colour-triplet scalar $Y$.
In particular, box diagrams involving $Y$ contribute to $\Delta F = 2$ processes — i.e., transitions that change a flavour quantum number (such as strangeness, charm, or bottomness) by two units — which govern neutral meson mixing.
The absence of large deviations from the SM predictions forces the coupling in \cref{eq:ModelLagrangians} to be highly hierarchical. Satisfying these flavour constraints while simultaneously achieving $\text{BR}(B \to \psi X) \gtrsim 10^{-4}$ requires that only one of the effective operators in \cref{eq:HeffMeso} dominates, with all others strongly suppressed.
Consequently, a single flavour combination—such as $(\psi b) (u d)$, $(\psi b) (u s)$, $(\psi d) (u b)$, or $(\psi s) (u b)$—is expected to be responsible for the entirety of the baryon asymmetry. 
This allows us to study the effect of each operator independently, without interference between different operator structures.

\textbf{4. Dark sector $\psi$ mass range}: The mass of the dark fermion $\psi$ is subject to several kinematic constraints. 
An upper bound follows from the requirement that the decay $B \to \psi X$ be kinematically allowed. The lightest visible baryon in the final state is the proton ($m_p \approx 938$~MeV), giving
$m_\psi < m_B - m_p \approx 4.34~\text{GeV}.$ A lower bound is imposed by proton stability. If $m_\psi < m_p - m_e \approx 937.8$~MeV, the decay $p^+ \to \psi e^+$ would become kinematically accessible, leading to unobserved proton decay. This requires $m_\psi > m_p - m_e \approx 937.8~\text{MeV}.$ Together, these constraints restrict the $\psi$ mass to the window
\begin{equation}
\boxed{
0.94~\text{GeV} < m_\psi < 4.34~\text{GeV}.
\label{eq:psiMassRange}
}
\end{equation}
All our results of the model analysis will be presented as functions of this dark sector mass range.

\textbf{5. Experimental searches for $B^+\to p+\text{invisible}$ decays:}
There have been dedicated efforts to search
for baryonic $B$-meson decays with missing energy, specifically the
channel $B^+\to p+\text{invisible}$. The most recent analysis was
performed by the Belle~II collaboration using the Belle data
sample~\cite{Belle-II:2026tyb}, yielding upper limits on the
branching fraction at the level of ${\cal O}(10^{-6})$. Earlier
searches for the same decay mode were carried out by the BaBar
collaboration~\cite{BaBar:2023dtq}, albeit with weaker sensitivity.
In the following, these experimental upper bounds will be combined
with our theoretical predictions for the lower limits on the
exclusive branching fraction in order to constrain the allowed
parameter space of the $B^+\to p\psi$ decay channel.

%%%%%%%%%%%%%%%%%%%%%%
\section{Inclusive Calculations: HQE}
\label{sec:MesoIncl}
The new decay modes introduced in the model necessarily modify $B$-meson observables. 
In particular, they contribute to the inclusive decay widths and, consequently, to the lifetime ratios. 
We study the effect of the new operators on these lifetime observables from an inclusive perspective, i.e., at the level of the quark transition $b \to \psi$. 
The analysis is performed within the framework of the HQE, outlined in \cref{sec:HQE}.
As presented in \cref{eq:HQE}, the total decay rate of the $B$ meson is expressed in terms of a leading dimension-three term representing the free quark decay, and $1/m_b$ power-suppressed terms of increasing dimension.
We perform the calculations of the coefficients of the two-quark operators up to the dimension-six Darwin term, as well as the contribution of the dimension-six four-quark operators.
Specifically, we compute the coefficients $\Gamma_3$, $\Gamma_5$, $\Gamma_6$, and $\tilde{\Gamma}_6$, all at leading order in $\alpha_s$.

We insert the operators of model $(d)$ and model $(b)$ from \cref{eq:HeffMeso} into the time-ordered product in \cref{eq:GammaB}. The calculations follow the same procedures outlined in \cref{Ch:2,Ch:3}.

For the two-quark operator contributions, we contract all quark fields, leaving the bilinear $\bar{b} b$. The quark propagators are then expanded in accordance with \cref{eq:quark-propagator-definition1} with the number of covariant derivatives included up to the desired accuracy. Since we perform the calculation up to dimension six, the quark propagator is expanded up to one covariant derivative acting on the gluon field strength tensor $G_{\mu\nu}$, yielding a total of three covariant derivatives sandwiched between the two $b$-quark fields — i.e., a dimension-six operator.
Since the calculation procedures have been described in detail in \cref{Ch:2,Ch:3}, we do not repeat them here. However, the new $B$-Mesogenesis operators contain a particular structure involving charge-conjugated fields and nontrivial colour contractions, to which we pay special attention.

In performing the Wick contraction to obtain the two-quark contribution, one encounters expressions of the form
\begin{equation}
:\epsilon_{ijk} \epsilon_{mnr}\,\Big(\bar{b}^{m,c} 
\wick{\c1 {u}^{n} \c1{\bar{u}}^{i}(x) } \,
b^{j,c}(x) \Big)\,
\Big(
\wick{\c2{\bar{\psi}}(x) 
\c1 {\bar{d}}^{r,c}(x) \c1{d}^{k,c} 
\c2{\psi}}
\Big):
\end{equation}
Using the definition of the charge-conjugated fields and the properties in \cref{conjugate_properties,conjugate_gamma_properties}, this can be reduced to
\begin{align}
\nonumber &\epsilon_{ijk} \epsilon_{mnr}\,\Bigl( \bar{b}^j(x) \, (S_{ni}^{(u)}(0,x))^{T} \, b^m \Bigr) \,
\mathrm{Tr}\Bigl( C P_L^{T} (S_{rk}^{(d)}(0,x))^{T} P_R^{T} C^{-1} S^{(\psi)}(0,x) \Bigr)  \\ 
&= \epsilon_{ijk} \epsilon_{mnr} \Bigl( \bar{b}^{j}(x) \, (S_{ni}^{(u)}(0,x))^{T} \, b^m \Bigr) \,
\mathrm{Tr}\Bigl( P_L \, S_{rk}^{(d)}(0,x) \, S^{(\psi)}(0,x) \Bigr),
\end{align}
\label{eq:yourlabel}
where $C$ is the charge conjugation operator and the following relation holds:
\begin{equation}
\wick{\c1 {\bar{q}}^{T}(x) \c1{q}^{T}(0)} = -(S^{(q)}(0,x))^{T}.
\end{equation}
Since the power-suppressed contributions arise from the emission of a soft gluon from the internal quark propagators, the associated colour structure becomes decisive. In the present case, the gluon can be emitted either from the $u$- or from the $d$-quark line. For emission from the $u$-quark propagator one has
$S^{(u)}_{ni} \to t^a_{ni}$ and $S^{(d)}_{rk} \to \delta_{rk}$,
leading to the colour factor
\begin{equation}
\epsilon_{ijk}\,\epsilon_{mnr}\,\delta_{kr}\,t^a_{ni}
= \epsilon_{ijk}\,\epsilon_{mnk}\,t^a_{ni}
= t^a_{jm}.
\end{equation}
An analogous contraction for emission from the $d$-quark line,
$S^{(d)}_{rk} \to t^a_{rk}$ and $S^{(u)}_{ni} \to \delta_{ni}$,
yields the same result. 
Thus, soft-gluon emission from both the $u$- and $d$-quark propagators contribute and must be included consistently in the power-suppressed analysis.

We now turn to the contribution of the dimension-six four-quark operators, corresponding to the coefficients $\tilde{\Gamma}_6$ in \cref{eq:HQE}. 
In contrast to the two-quark operator case discussed above, the structure of the Wick contractions is different. 
Here, only one quark line is contracted in addition to the dark-sector fermion $\psi$, while the remaining light quark fields remain uncontracted and give rise to local four-quark operators in the effective theory.

For model $(d)$, the relevant time-ordered product (suppressing colour indices) reads
\begin{align}
\nonumber
&\text{T}^{(d)} \Big\{
\bar{u}_R(x)\, b_R^{c}(x)\, \bar{\psi}(x)\, d_R^{c}(x),\,
\bar{b}_R^{c}\, u_R\, \bar{d}_R^{c}\, \psi\Big\} \\
\nonumber &= \,
:\wick{\bar{u}_R(x)\, b_R^{c}(x)\, \c1{\bar{\psi}}(x)\, \c2{d}_R^{c}(x)\,
\bar{b}_R^{c}\, u_R\, \c2{\bar{d}}_R^{c}\, \c1{\psi}}:
\\
&+ \,
:\wick{\c2{\bar{u}}_R(x)\, b_R^{c}(x)\, \c1{\bar{\psi}}(x)\, d_R^{c}(x)\,
\bar{b}_R^{c}\, \c2{u}_R\, \bar{d}_R^{c}\, \c1{\psi}}:
\label{eq:contract4q}
\end{align}

\begin{figure}[t]
\centering

\begin{subfigure}{0.31\textwidth}
\centering
\includegraphics[width=\linewidth]{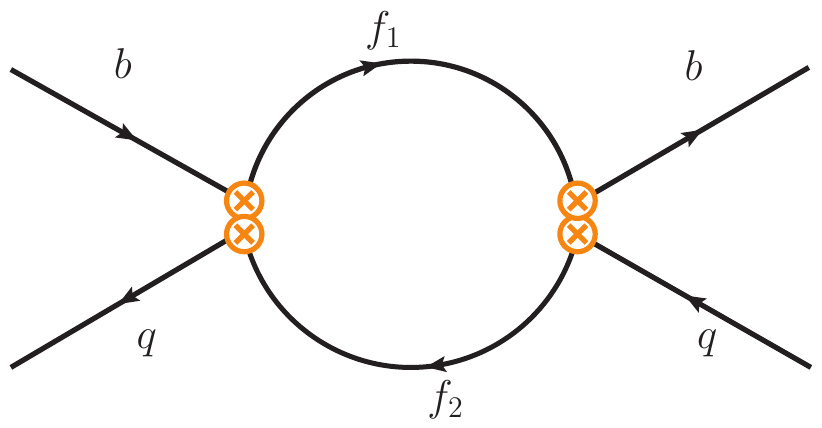}
\caption*{I}
\end{subfigure}
\hfill
\begin{subfigure}{0.31\textwidth}
\centering
\includegraphics[width=\linewidth]{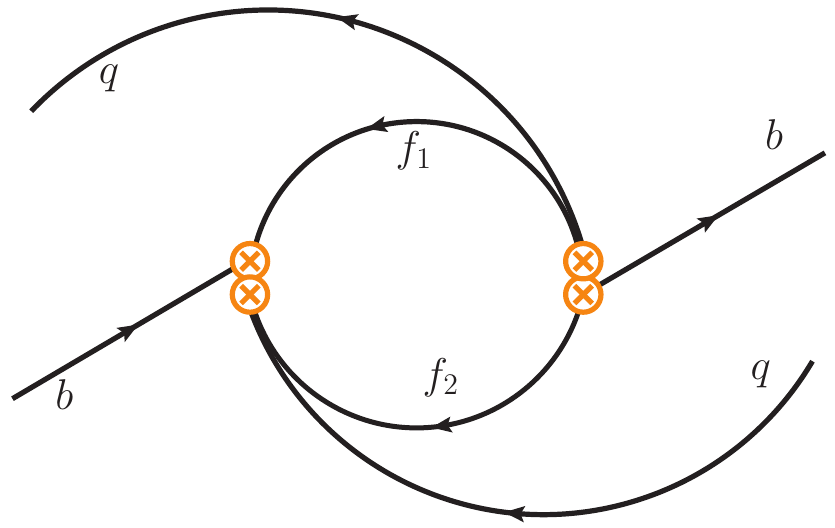}
\caption*{II}
\end{subfigure}
\hfill
\begin{subfigure}{0.31\textwidth}
\centering
\includegraphics[width=\linewidth]{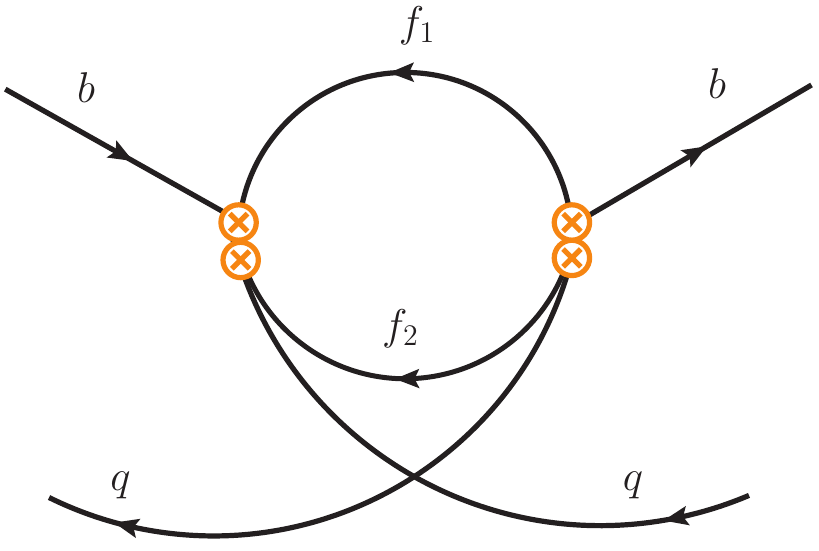}
\caption*{III}
\end{subfigure}

\vspace{3mm}

\begin{subfigure}{0.31\textwidth}
\centering
\includegraphics[width=\linewidth]{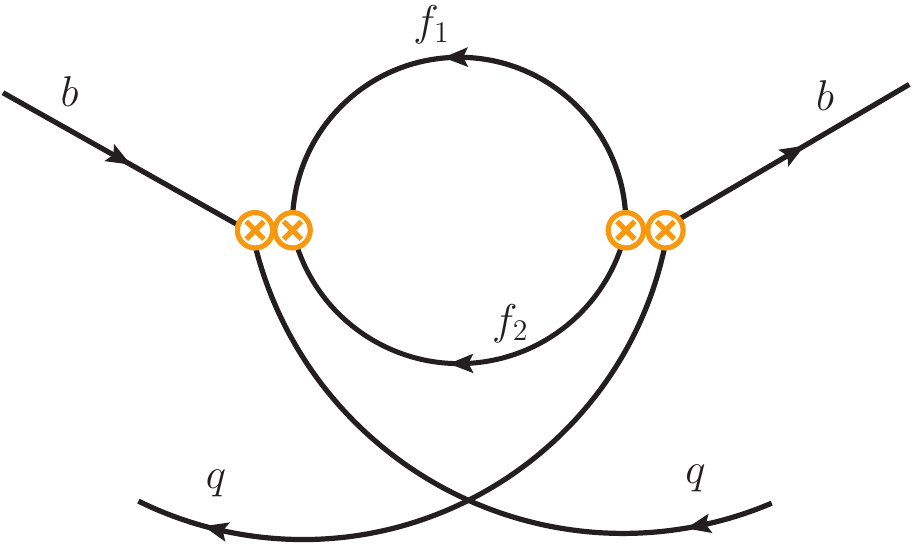}
\caption*{IV}
\end{subfigure}
\hfill
\begin{subfigure}{0.31\textwidth}
\centering
\includegraphics[width=\linewidth]{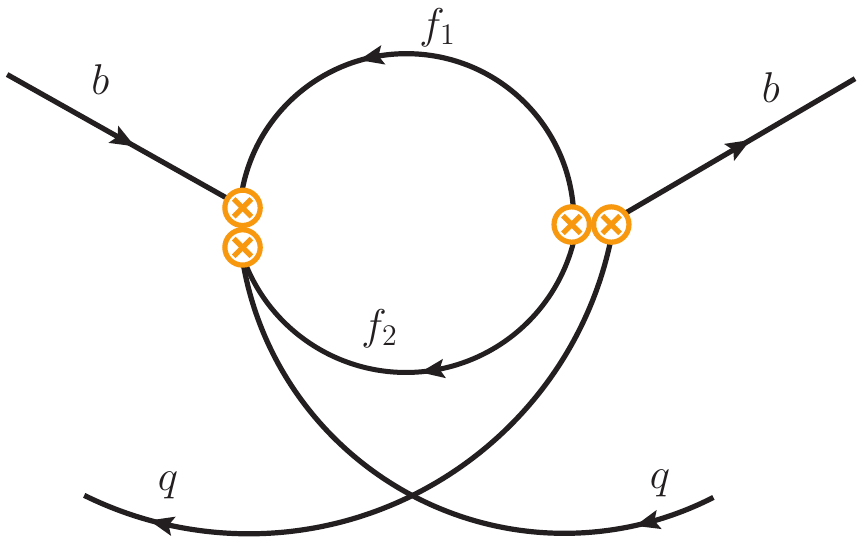}
\caption*{V}
\end{subfigure}
\hfill
\begin{subfigure}{0.31\textwidth}
\centering
\includegraphics[width=\linewidth]{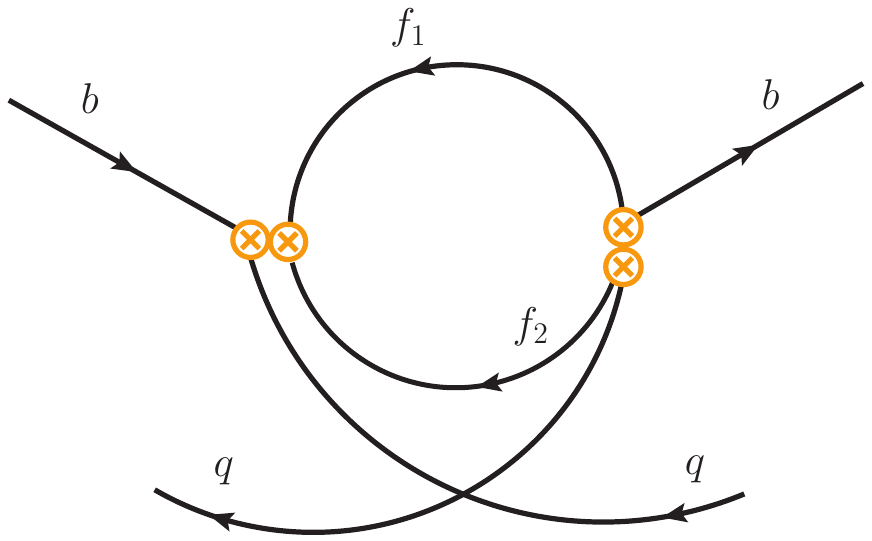}
\caption*{VI}
\end{subfigure}

\caption{Complete set of four-quark topologies (see \cref{eq:contract4q}) induced by the Mesogenesis operators in \cref{eq:ModelLagrangians}. The internal fermions $f_1$ and $f_2$ are generic and may correspond to any of the particles appearing in the Mesogenesis operators, namely $\psi$, $u$, $d$, $s$, or $c$.}
\label{fig:MesDiagrams}
\end{figure}
The contractions for model $(b)$ proceed analogously. The corresponding topologies are shown in \cref{fig:MesDiagrams}. 
As an illustrative example, we discuss in detail the evaluation of the first contraction in \cref{eq:contract4q}. Inserting the effective operator of model $(d)$ in \cref{eq:HeffMeso} into \cref{eq:GammaB}, the dimension-six four-quark operators contribution to the total decay rate of topology-IV reads
\begin{align}
\Gamma^{(4q)}_{(6,d),IV}(B)
&=
\frac{-\,|G_{(d)}|^2}{2m_B}\,
\text{Im} \,  \int d^4x\; 
\bra{B} i\, \mathcal{T}_{(d)}^{IV}(x)\ket{B}
+ (x \leftrightarrow 0),
\label{eq:Gamma4q_topI}
\end{align}
where we define for convenience
\begin{equation}
\mathcal{T}_{(d)}^{IV}(x)
=
\epsilon_{ijk}\,\epsilon_{mnr}\,
\text{T}\Big\{
\bar u_R^{i}(x)\, b_R^{c,j}(x)\, \bar\psi(x)\, d_R^{c,k}(x)\,
\bar b_R^{c,m}\, u_R^{n}\, \bar d_R^{c,r}\, \psi
\Big\}.
\label{eq:T_d_topI}
\end{equation}
Performing the Wick contractions and using
\begin{equation}
\wick{\c1 d_R^{c,k}(x)\,\c1{\bar d_R^{c,r}}(0)}
= C \, P_L^T \,\big(iS_{rk}^{(d)}(0,x)\big)^T\, P_R^T\, C^{-1},
\end{equation}
we obtain
\begin{equation}
\mathcal{T}_{(d)}^{IV}(x)
=
-\epsilon_{ijk}\,\epsilon_{mnr}\,
\bar u_R^{i}(x)\, b_R^{c,j}(x)\,\bar b_R^{c,m}\, u_R^{n}\, \text{Tr} \Big[ \, C \, P_L^T \, (iS^{(d)}_{rk}(0,x))^T  \, P_R^T \, C^{-1} \, iS^{(\psi)}(0,x) \Big].
\label{eq:T_d_topI2}
\end{equation}
The overall minus sign originates from the anticommuting nature of the fermion fields in the Wick contraction; in diagrammatic language, this is equivalently represented by the standard factor $(-1)$ associated with a closed fermion loop. Substituting the tree-level propagator in \cref{eq:propagators} (see also \cref{eq:Sxy-free-prop}), we arrive at 
\begin{align}
   \mathcal{T}_{(d)}^{IV}(x)
= -\int_{l_1 l_2} \, \frac{l_1^\alpha \, l_2^\beta \, e^{ix(l_1+l_2)}}{[l_1^2 - m_\psi^2] [l_2^2 - m_d^2]} \, \epsilon_{ijk}\,\epsilon_{mnk}\,
\bar u_R^{i}(x)\, b_R^{c,j}(x)\,\bar b_R^{c,m}\, u_R^{n} \, \text{Tr} \Big[P_L \gamma_\alpha \gamma_\beta \Big],
\label{eq:T_d_topI3}
\end{align}
where we define $\int_{l_1 l_2} \equiv \int \frac{d^4l_1}{(2\pi)^4} \,\frac{d^4l_2}{(2\pi)^4}$, $l_1$ and $l_2$ denote the four-momenta flowing in the $\psi$- and $d$-quark propagators, respectively.
The Dirac bilinear in front of the trace in \cref{eq:T_d_topI3} still involves charge-conjugated fields. 
To rewrite it in terms of ordinary quark bilinears, we apply a Fierz rearrangement for chiral spinors
(see, e.g., Ref.~\cite{Nieves:2003in}). In four dimensions one finds
\begin{equation}
\bar u_R^{i}(x)\, b_R^{c,j}(x)\,\bar b_R^{c,m}\, u_R^{n}
=
\frac{1}{2}\,
\big[\bar b^{j}(x)\gamma_\mu P_R\, b^{m}\big]\,
\big[\bar u^{i}(x)\gamma^\mu P_R\, u^{n}\big].
\label{eq:Fierz1}
\end{equation}
In order to match onto the forward matrix element (see \cref{eq:operators4q,eq:matrixelement4q}),
we perform a second Fierz rearrangement,
\begin{equation}
\big[\bar b^{j}(x)\gamma_\mu P_R\, b^{m}\big]\,
\big[\bar u^{i}(x)\gamma^\mu P_R\, u^{n}\big]
=
\big[\bar b^{j}(x)\gamma_\mu P_R\, u^{n}\big]\,
\big[\bar u^{i}(x)\gamma^\mu P_R\, b^{m}\big].
\label{eq:Fierz2}
\end{equation}
Since we evaluate the forward matrix element at parton level between free quark states,
we insert momentum eigenstates and use the plane-wave expansion of the fields,
\[
\bar b(x) \;\to\; \bar b(0)\, e^{i p_b x}, 
\qquad
\bar{u}(x) \;\to\; u(0)\, e^{i q x},
\]
where $p_b$ and $q$ denote the momenta of the corresponding quark lines.
Defining $p=p_b+q$ and using \cref{eq:Fierz1,eq:Fierz2}, and performing the trace,  \cref{eq:T_d_topI3} becomes
\begin{align}
\mathcal{T}_{(d)}^{IV}(x)
&= -g_{\alpha \beta} \int_{l_1 l_2} \, \frac{l_1^\alpha \, l_2^\beta \, e^{ix(l_1+l_2+p)}}{[l_1^2 - m_\psi^2] [l_2^2 - m_d^2]} \, \epsilon_{ijk}\,\epsilon_{mnk}\,
\big[\bar b^{j}\gamma_\mu P_R\, u^{n}\big]\,
\big[\bar u^{i}\gamma^\mu P_R\, b^{m}\big],
\label{eq:T_d_topI4}
\end{align}
The colour structure can now be simplified using the identity
\begin{equation}
\epsilon_{ijk}\,\epsilon_{mnk}
= \delta_{im}\,\delta_{jn}
- \delta_{in}\,\delta_{jm}.
\end{equation}
In addition, we make use of the standard relation for the
$SU(3)$ generators $t^a$ in the fundamental representation,
\begin{equation}
t^a_{ij}\, t^a_{kl}
= \frac{1}{2}\left(
\delta_{il}\,\delta_{kj}
- \frac{1}{N_c}\,\delta_{ij}\,\delta_{kl}
\right),
\end{equation}
with $N_c=3$. Applying these identities to \cref{eq:T_d_topI4} allows the
four-quark structure to be decomposed into colour-singlet and
colour-octet contributions
\begin{align}
\nonumber\epsilon_{ijk}\,\epsilon_{mnk}\,
\big[\bar b^{j}\gamma_\mu P_R\, u^{n}\big]\,
\big[\bar u^{i}\gamma^\mu P_R\, b^{m}\big] &= \left(  \frac{N_c-1}{N_c}\right)\,
\big[\bar b^{i}\gamma_\mu P_R\, u^{i}\big] \,
\big[\bar u^{j}\gamma^\mu P_R\, b^{j}\big] \\&- 2 \,
\big[\bar u^{i}\gamma^\mu P_R\, t^a_{ij}\,b^{j}\big]\,
\big[\bar u^{m}\gamma^\mu P_R\, t^a_{mn}\,b^{n}\big].
\label{eq:colorstructure}
\end{align}
Substituting \cref{eq:colorstructure} into \cref{eq:T_d_topI4} and using the four-quark operators definitions in \cref{eq:operators4q} 
\begin{align}
\mathcal{T}_{(d)}^{IV}(x)
&= \frac{-g_{\alpha \beta}}{4} \int_{l_1 l_2} \, \frac{l_1^\alpha \, l_2^\beta \, e^{ix(l_1+l_2+p)}}{[l_1^2 - m_\psi^2] [l_2^2 - m_d^2]} \, \Big[ \left( \frac{N_c-1}{N_c} \,  \right) \tilde{\mathcal{O}}^{(u)}_1 - 2 \, \tilde{\mathcal{O}}^{(u)}_3\Big].
\label{eq:T_d_topI5}
\end{align}
The operators
$\tilde{\mathcal{O}}^{(u)}_1$ and
$\tilde{\mathcal{O}}^{(u)}_3$
are the four-quark operators defined in
\cref{eq:operators4q}. 
Substituting \cref{eq:T_d_topI5}  back into \cref{eq:Gamma4q_topI}, we then perform the $x$-integration,
which yields $(2\pi)^4\delta^{(4)}(l_1+l_2+p)$, and use it to integrate over $l_2$.
Including the symmetry factor $2$ from $(x\leftrightarrow 0)$ and relabeling $l_1\to l$,
we arrive at
\begin{align}
\nonumber \Gamma^{(4q)}_{(6,d),IV}(B)
&=
\frac{\,|G_{(d)}|^2}{4m_B}\,g_{\alpha \beta} \,
\text{Im} \left(i \int_{l } \, \frac{l^\alpha \, (-l-p)^\beta \, }{[l^2 - m_\psi^2] [(-l-p)^2 - m_d^2]} \,\,  \right) \\& \qquad \times   
\bra{B}  \Big[ \left( \frac{N_c-1}{N_c} \,  \right) \tilde{\mathcal{O}}^{(u)}_1 - 2 \, \tilde{\mathcal{O}}^{(u)}_3\Big]  \ket{B}.
\end{align}
To match the convention of \cref{eq:integral4q}, we relabel the loop momentum by
$l\to -l$. Evaluating the imaginary part of the integration and contracting the Dirac indices, the dimension-six four-quark operator   
for topology-IV for model $(d)$ reads
\begin{align}
\Gamma^{(4q)}_{(6,d),IV}(B)
&=
\frac{\,-|G_{(d)}|^2}{128  \pi\,m_B} \, \left(  \frac{m_b^2-m_\psi^2}{m_b}\right)^2 \, \bra{B}  \Big[ \left( \frac{N_c-1}{N_c} \,  \right) \tilde{\mathcal{O}}^{(u)}_1 - 2 \, \tilde{\mathcal{O}}^{(u)}_3\Big]  \ket{B}.
\label{eq:modeld4qfinal}
\end{align}
Having outlined the calculation strategy for both the two-quark and the 
dimension-six four-quark contributions in the context of the Mesogenesis 
operators for representative examples, the remaining operator structures 
proceed in complete analogy. The steps — Wick contraction, treatment of 
charge-conjugated fields, Fierz rearrangements, colour decomposition, and 
evaluation of the loop integrals — follow the same pattern and introduce 
no qualitatively new features.
We are therefore now in a position to present the general results for the 
inclusive decay width induced by the Mesogenesis effective operators.
\subsubsection*{Inclusive Results from the HQE}
For $B$-meson decays mediated by the $B$-Mesogenesis  
operators in \cref{eq:HeffMeso}, the inclusive decay width corresponding 
to a given operator structure~$i$ can be expressed within the HQE as the 
sum of the two-quark and dimension-six four-quark contributions:
\begin{equation}
\Gamma^{(i)}_{\rm meso}
\;=\;
\Gamma^{(i)}_{\rm meso,\,2q}
\;+\;
\Gamma^{(i)}_{\rm meso,\,4q}
\;+\; \mathcal{O}\!\left(\frac{1}{m_b^4}\right).
\label{eq:Gamma_meso_decomp}
\end{equation}
The two-quark part encodes the free partonic decay and the power-suppressed corrections
from dimension-five and dimension-six two-quark operators and can be expressed in the compact form
\begin{equation}
\Gamma^{(i)}_{\rm meso,\,2q}
\;=\;
\frac{|C^{(i)}|^2\, m_b^5}{3072\,\pi^3}\;
\left[
   f_{0}^{(i)}(\rho_f)\left(1-\frac{\mu_\pi^2}{2m_b^2}\right)
 + f_{G}^{(i)}(\rho_f)\,\frac{\mu_G^2}{m_b^2}
 + f_{D}^{(i)}(\rho_f)\,\frac{\rho_D^3}{m_b^3}
\right],
\label{eq:MasterWidth_meso_2q}
\end{equation}
with $\rho_f \equiv (m_f/m_b)^2$ denoting the squared mass ratio of the massive final-state particle in the underlying transition $b \to f$.  The functions $f_{0}^{(i)}(\rho_f)$, $f_{G}^{(i)}(\rho_f)$, and $f_{D}^{(i)}(\rho_f)$ are dimensionless short-distance coefficient functions that depend on the Dirac and colour structure of the corresponding \textit{B}-Mesogenesis operator \footnote{ 
The spin–orbit parameter $\rho_{LS}^3$ does not contribute in the present case, as the associated operator yields a vanishing forward matrix element for the Dirac structures under consideration.}. 
As an example, we display the results for models $(d)$ and $(b)$, 
which correspond to the coefficients $C_\psi^{(ud)}$ and 
$\tilde{C}_\psi^{(ud)}$, respectively:
\begin{flalign}
&f_{0}^{(\psi ud)}(\rho_\psi) =
1 - 8\,\rho_\psi + 8\,\rho_\psi^{3} - \rho_\psi^{4}
- 12\,\rho_\psi^{2}\log\rho_\psi,
&&\label{eq:f0d}
\\[2pt]
&f_{G}^{(\psi ud)}(\rho_\psi) =
\frac{5}{2} - 8\,\rho_\psi + 8\,\rho_\psi^{3}
- \frac{5}{2}\,\rho_\psi^{4}
- 6\,\rho_\psi^{2}\log\rho_\psi,
&&\label{eq:fGd}
\\[2pt]
&f_{D}^{(\psi ud)}(\rho_\psi) =
\frac{2}{3}\,
\Biggl(
22 - 31\rho_\psi + 3\rho_\psi^{2}
+ 11\rho_\psi^{3} - 5\rho_\psi^{4}
\nonumber\\
&\qquad \qquad \qquad \quad 
- 12(1-\rho_\psi)^{2}(3+\rho_\psi)\log(1-\rho_\psi)
+ 6\rho_\psi^{2}(1+\rho_\psi)\log\rho_\psi
\Biggr),
&&\label{eq:fDd}
\\[2pt]
&f_{0}^{(\tilde\psi ud)}(\rho_\psi) =
1 - 8\rho_\psi + 8\rho_\psi^{3} - \rho_\psi^{4}
- 12\rho_\psi^{2}\log\rho_\psi,
&&\label{eq:f0b}
\\[2pt]
&f_{G}^{(\tilde\psi ud)}(\rho_\psi) =
-\frac{1}{2}\Big(
3 - 8\rho_\psi + 24\rho_\psi^{2} - 24\rho_\psi^{3}
+ 5\rho_\psi^{4} + 12\rho_\psi^{2}\log\rho_\psi
\Big),
&&\label{eq:fGb}
\\[2pt]
&f_{D}^{(\tilde\psi ud)}(\rho_\psi) =
\frac{2}{3}(1-\rho_\psi)\Big(
9 + 11\rho_\psi - 25\rho_\psi^{2} + 5\rho_\psi^{3}
\nonumber\\
&\qquad \qquad \qquad \qquad \qquad
- 24(1-\rho_\psi^{2})\log(1-\rho_\psi)
- 12\rho_\psi^{2}\log\rho_\psi
\Big).
&&\label{eq:fDb}
\end{flalign}
The results for the remaining operators are given in \cref{sec:MesoInclResults}.

For a fixed operator, the dimension-six four-quark
contribution depends on the external meson species.
Since the two possible Wick contractions correspond to different
spectator flavours, they contribute to different $B$ mesons and are not summed over.
Explicitly, we write
\begin{align}
\Gamma^{(i)}_{\rm meso,\,4q}(B^+)
&= \frac{|C^{(i)}|^2 \, m_b^2 }{192 \pi} \,
\sum_{a=1}^{4}
\tilde{\Gamma}^{(i,u)}_{X,a}\;
\frac{\langle B^+ | \tilde{\mathcal O}^{u}_{a} | B^+ \rangle}{2m_B},
\label{eq:Meso4qresultsB+}
\\[4pt]
\Gamma^{(i)}_{\rm meso,\,4q}(B_q)
&= \frac{|C^{(i)}|^2 \, m_b^2 }{192 \pi} \,
\sum_{a=1}^{4}
\tilde{\Gamma}^{(i,q)}_{X,a}\;
\frac{\langle B_q | \tilde{\mathcal O}^{q}_{a} | B_q \rangle}{2m_B},
\qquad q=d,s.
\label{eq:Meso4qresultsBq}
\end{align}
Here \(X\) labels the topology arising from the operator
structure (see \cref{fig:MesDiagrams}). The explicit expressions for 
$\tilde{\Gamma}^{(i,u)}_{X,a}$ and $\tilde{\Gamma}^{(i,q)}_{X,q}$ are given in \cref{sec:MesoInclResults}.

We stress that the analytical results presented in this section
have been subjected to several non-trivial cross-checks. The
four-quark contributions were independently reproduced in a separate
calculation, providing an important validation of the corresponding
Wilson coefficients. Furthermore, for the two-quark contributions, we
explicitly verified the cancellation of the IR divergences
against the corresponding four-quark contributions through the
operator-mixing procedure described in \cref{Ch:2,Ch:3}.
After accounting for the appropriate colour factors and Dirac
structures, we also reproduce the known SM
results~\cite{Mannel:2020fts,Lenz:2020oce}, with
$\rho_\psi$ in the present notation corresponding to $\rho_c$ in the
SM. A detailed comparison between the SM and
the present operator basis is presented in
\cref{sec:SMeffects}.

\subsection{Lifetimes and Lifetime Ratios}
\label{sec:Mesolifetimes}
Using the analytic results derived in the previous section, 
we now perform a numerical analysis of the $B$-meson lifetimes 
and lifetime ratios within the framework of the HQE. 
\subsubsection*{Total Lifetimes}
In general, the analysis of total lifetimes is not expected to provide 
stringent constraints on the couplings of the $B$-Mesogenesis model. 
The SM predictions for total lifetimes are affected by 
sizeable theoretical uncertainties, predominantly from non-perturbative 
inputs, see e.g.~\cite{Albrecht:2024oyn,Lenz:2022rbq, Egner:2024lay}. Moreover, collider searches already restrict the relevant 
BSM couplings to comparatively small values. Consequently, the primary purpose of the total lifetime analysis 
is not to derive competitive bounds, but rather to investigate the 
impact of the subleading power-suppressed contributions induced by 
the new operators. In particular, we study the hierarchy of the OPE, 
identifying the range of $m_\psi$ where subleading terms become relevant 
and determining which subleading term provide the dominant contributions. 
We emphasize that, in the following analysis of the total decay rate, we focus
exclusively on the new-physics (NP) contributions induced by the
Mesogenesis operators, omitting the SM part of the
width. This enables us to isolate and assess the relative importance
of the subleading HQE terms within the BSM framework.\footnote{For
coupling values consistent with current ATLAS and CMS bounds~\cite{Alonso-Alvarez:2021qfd},
the NP contribution amounts to at most $\sim 3\%$ of the SM total width
and is therefore negligible compared to the present experimental and
theoretical uncertainties.}

The total $B$-meson decay rate is given in \cref{eq:HQE}. 
In the previous section, we computed the two-quark contributions 
up to the dimension-six Darwin term, together with the four-quark operator 
contributions. These enter the HQE coefficients 
$\Gamma_3$, $\Gamma_5$, $\Gamma_6$, and $\tilde{\Gamma}_6$.

In the following, we analyse, for each operator in the effective Hamiltonian \cref{eq:HeffMeso}, 
the individual contributions of the HQE terms in \cref{eq:HQE} to the total decay width 
of the $B^+$ meson. We concentrate on the $B^+$ system since it exhibits distinctive 
topological structures and, as will be shown below, enables us to derive a relevant 
experimental lower bound for the exclusive decay $B^+ \to p \psi$ 
(see \cref{sec:MesoExcl}). In general, the HQE contributions from two-quark operators are spectator independent 
and therefore nearly identical for all $B$ mesons 
($B^+$, $B^0$, $B_s^0$), up to  $SU(3)_F$-breaking effects 
in the non-perturbative matrix elements. In contrast, the four-quark operator 
contributions are suppressed for the neutral mesons $B^0$ and $B_s^0$. 
Moreover, certain operators in \cref{eq:HeffMeso}—particularly those involving two additional massive 
fermions besides the $b$ quark—do not contribute to the $B^+$ decay width. 
For completeness, the corresponding contributions to the $B^0$ width 
are derived in \cref{app:BMesogenesisResutls}.

\begin{figure}[t]
    \centering
    \begin{subfigure}{0.49\textwidth}
        \centering
        \includegraphics[width=\textwidth]{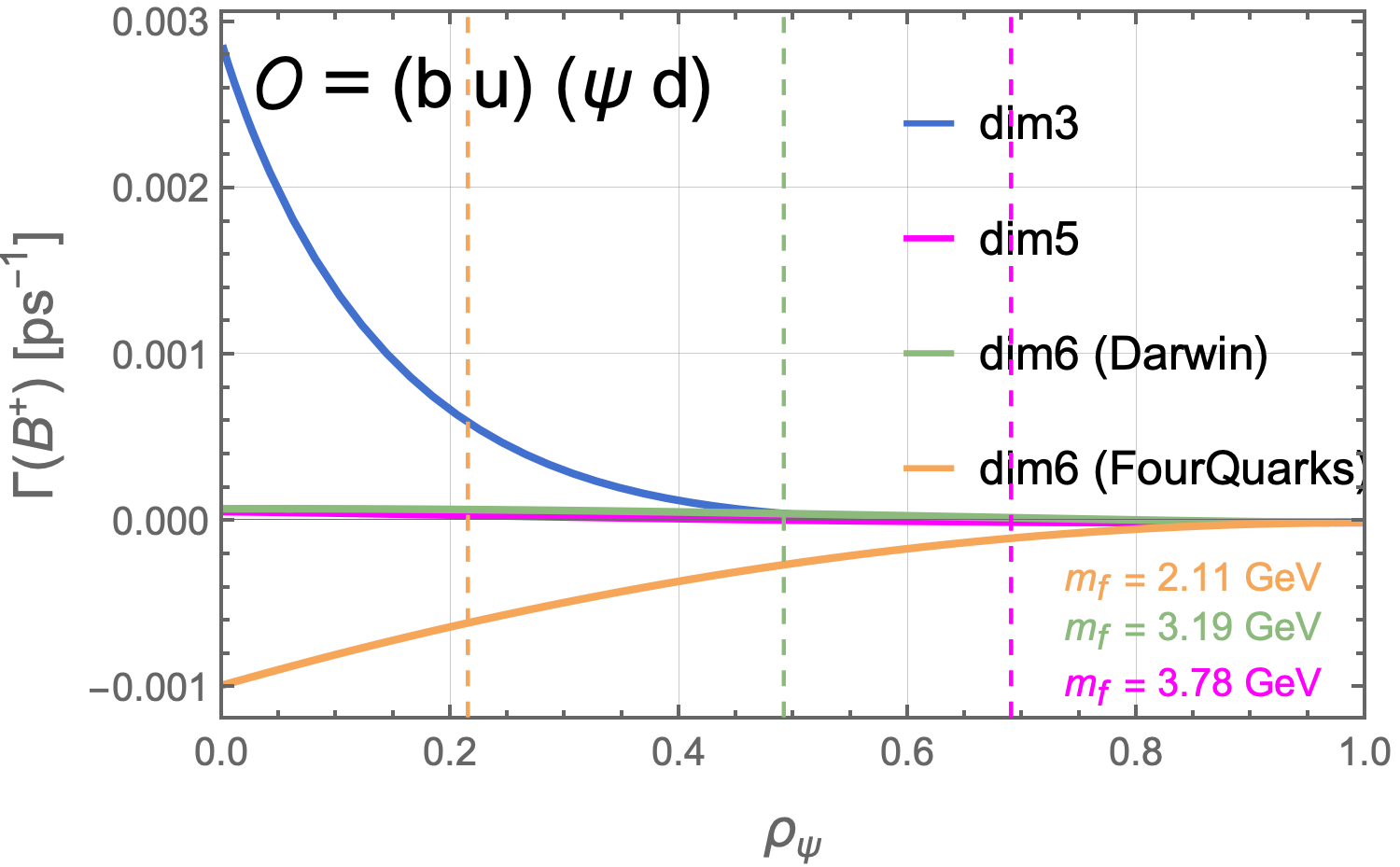}
    \end{subfigure}
    \hfill
    \begin{subfigure}{0.47\textwidth}
        \centering
        \includegraphics[width=\textwidth]{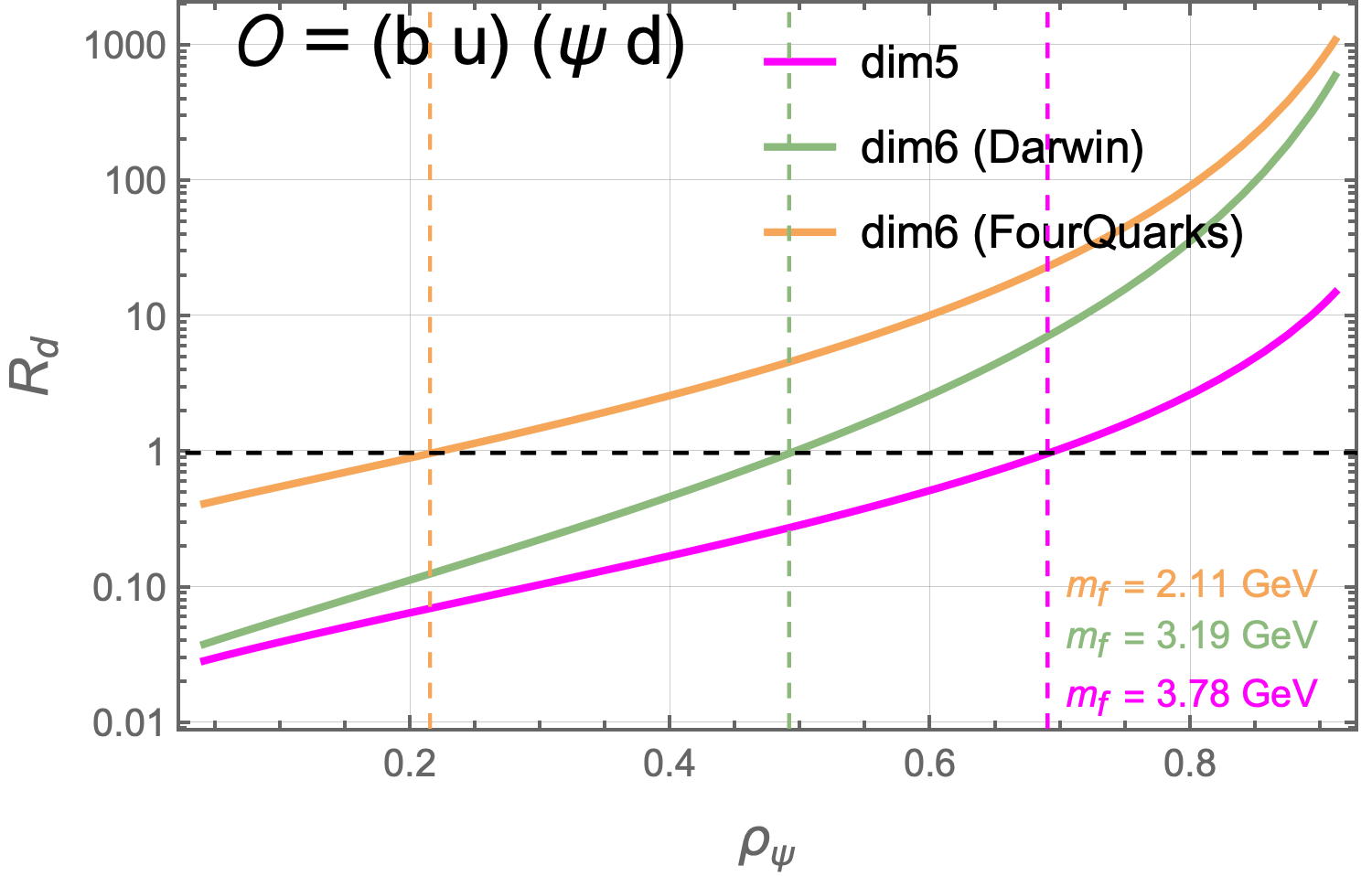}
    \end{subfigure}
    
    \medskip % Adds a small vertical space between the rows
    
    \begin{subfigure}{0.49\textwidth}
        \centering
        \includegraphics[width=\textwidth]{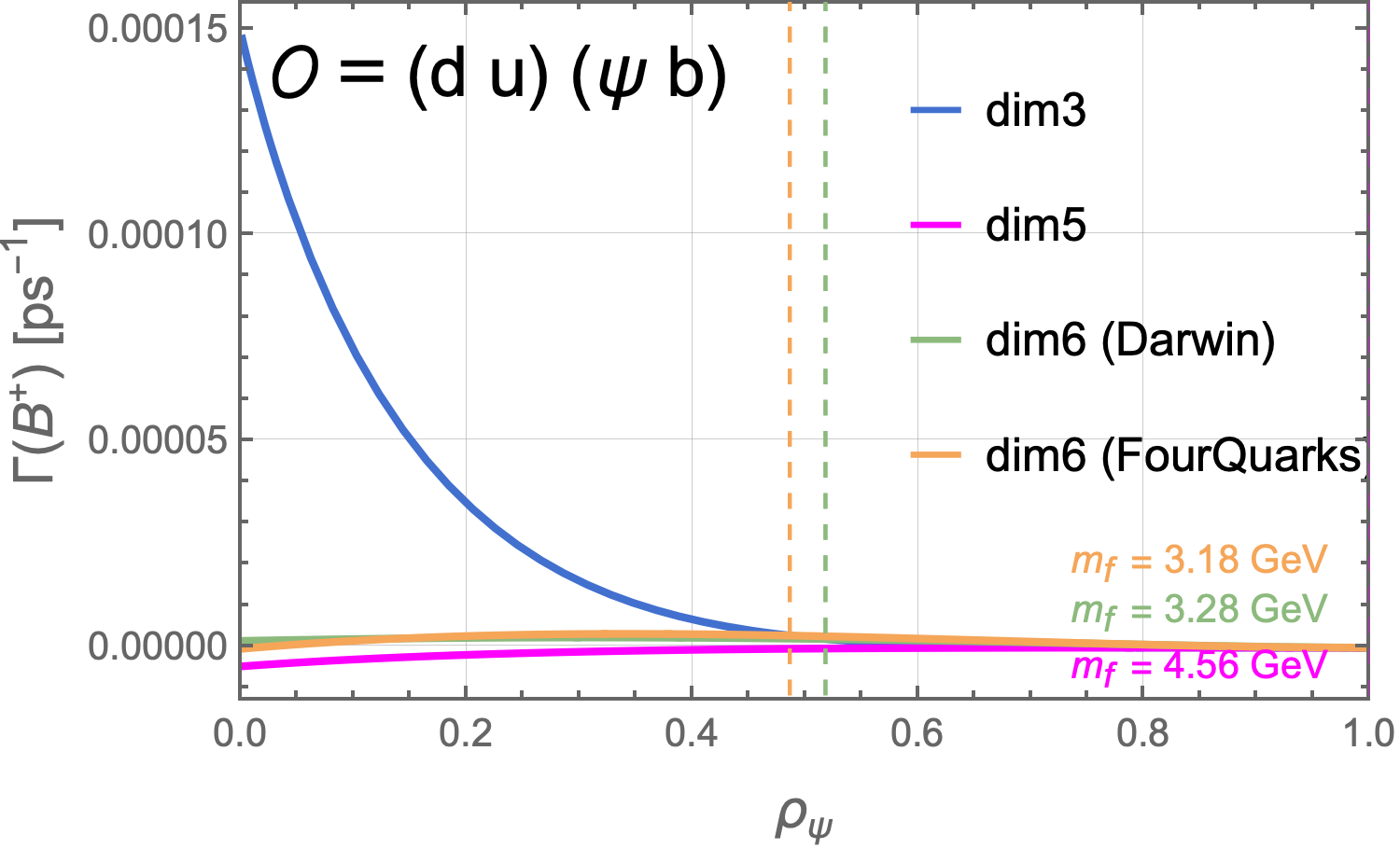}
    \end{subfigure}
    \hfill
    \begin{subfigure}{0.47\textwidth}
        \centering
        \includegraphics[width=\textwidth]{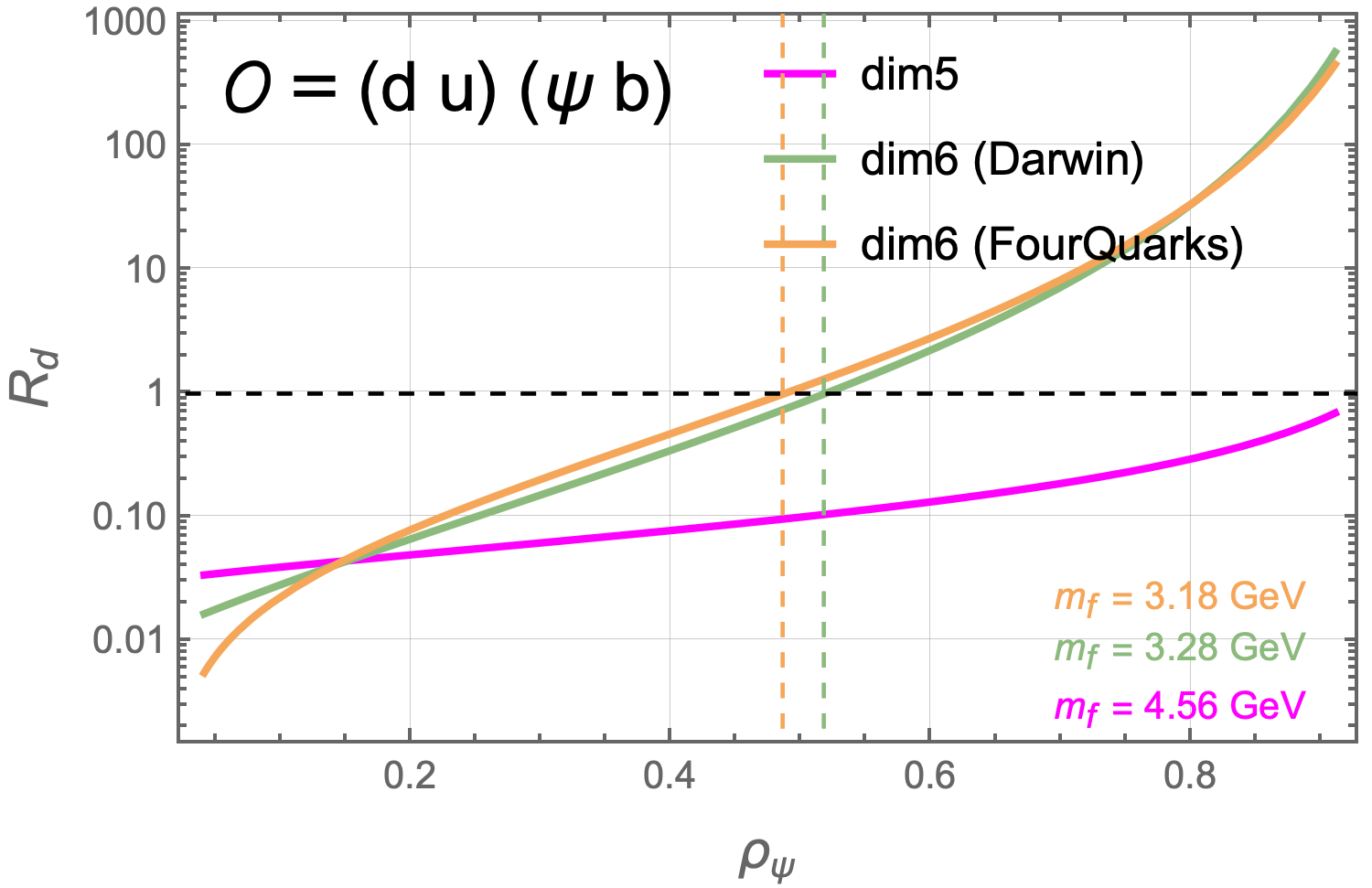}
    \end{subfigure}
\caption{Total decay width $\Gamma(B^+)$ induced solely by the Mesogenesis operators as a function of $\rho_f = (m_f/m_b)^2$. The left column shows the individual HQE contributions, while the right column displays the ratio $R_d = |\Gamma_d/\Gamma_3|$, where $\Gamma_d$ denotes the corresponding subleading term. The vertical lines indicate the values of $\rho_f$ at which the subleading contribution exceeds the leading dimension-three term $\Gamma_3$; the associated $m_f$ values are also shown.}
    \label{fig:DecayRates}
\end{figure}

\begin{figure}[ht]
    \centering
    \begin{subfigure}{0.49\textwidth}
        \centering
        \includegraphics[width=\textwidth]{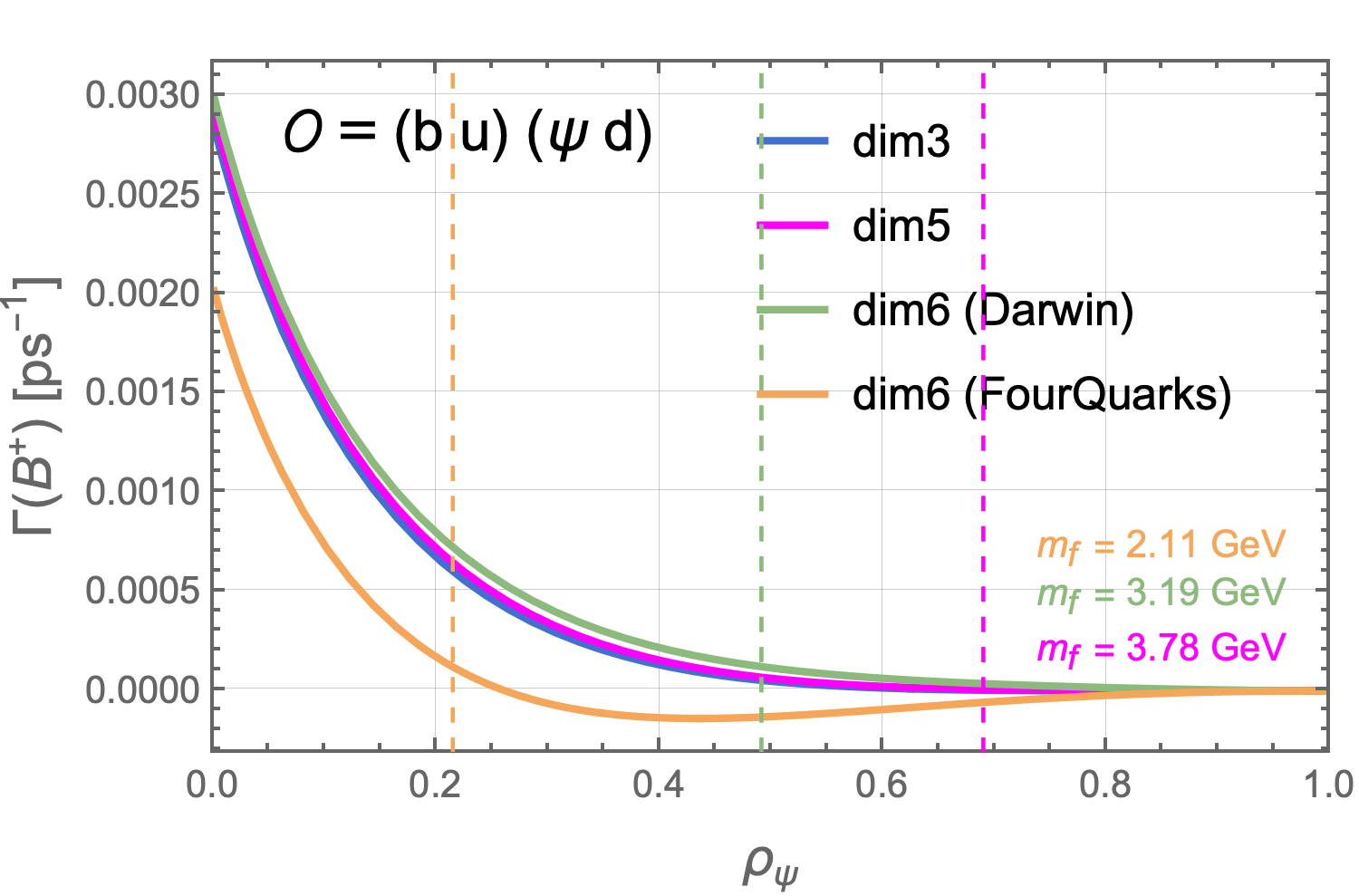}
    \end{subfigure}
    \hfill
    \begin{subfigure}{0.49\textwidth}
        \centering
        \includegraphics[width=\textwidth]{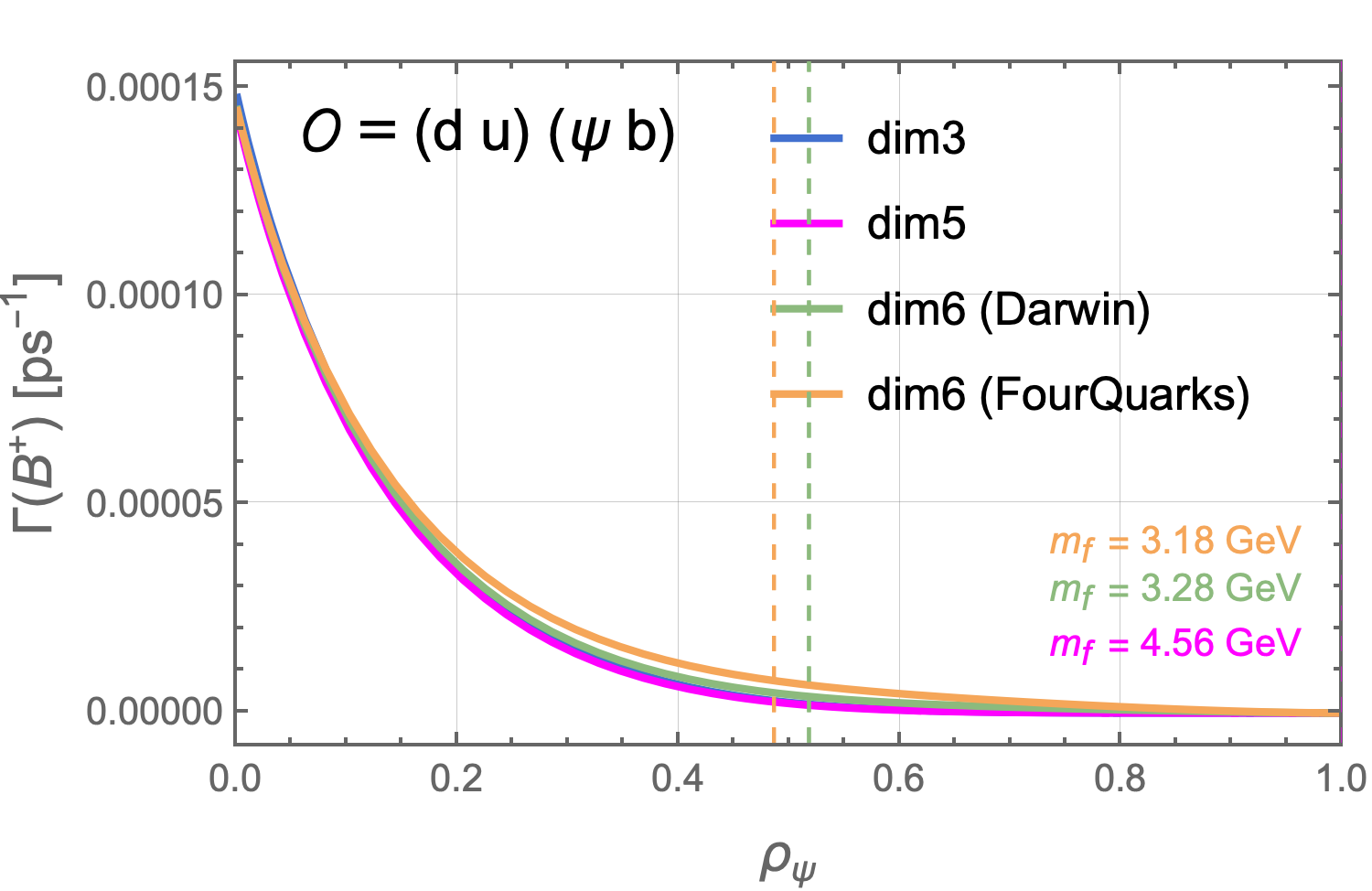}
    \end{subfigure}
        \caption{Total decay width of $\Gamma(B^+)$ arising purely from the Mesogenesis operators as a function of $\rho_f= (m_f/m_b)^2$. The figure displays the cumulative sum of all terms up to the indicated dimension. The vertical lines indicate the value of $\rho_f$, where the contribution exceeds the leading dimension-three $\Gamma_3$ term (the corresponding $m_f$ values are also indicated).}\label{fig:DecayRatesTotal}
\end{figure}

\begin{figure}[t]
  \centering
  % First diagram
  \begin{minipage}{0.42\textwidth}
    \centering
    \setlength{\unitlength}{1pt}
    \begin{picture}(150,120)
      \put(0,0){\includegraphics[width=0.9\textwidth]{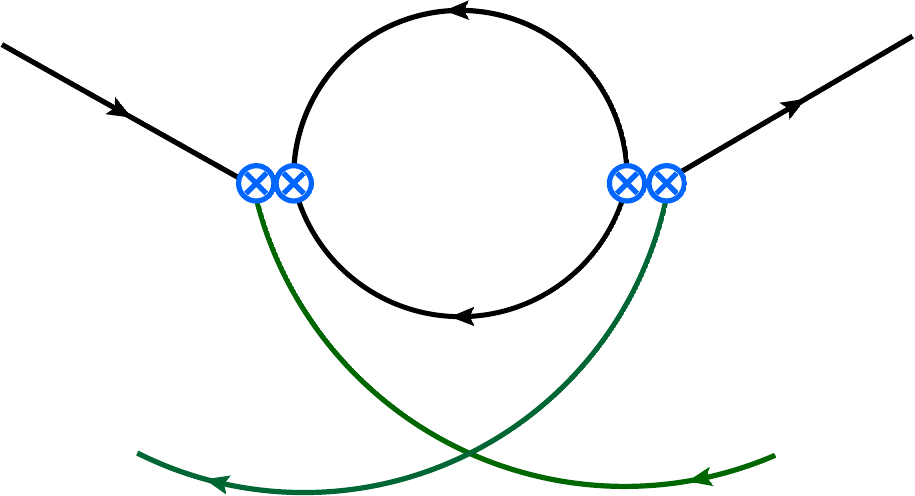}}
      \put(80,90){$d$}
      \put(80,40){$\psi$}
      \put(10,82){$b$}
      \put(30,10){$u$}
      \put(146,80){$b$}
      \put(130,10){$u$}
    \end{picture}
  \end{minipage}%
  \hfill
  % Second diagram
  \begin{minipage}{0.40\textwidth}
    \centering
    \setlength{\unitlength}{1pt}
    \begin{picture}(150,120)
      \put(0,0){\includegraphics[width=0.9\textwidth]{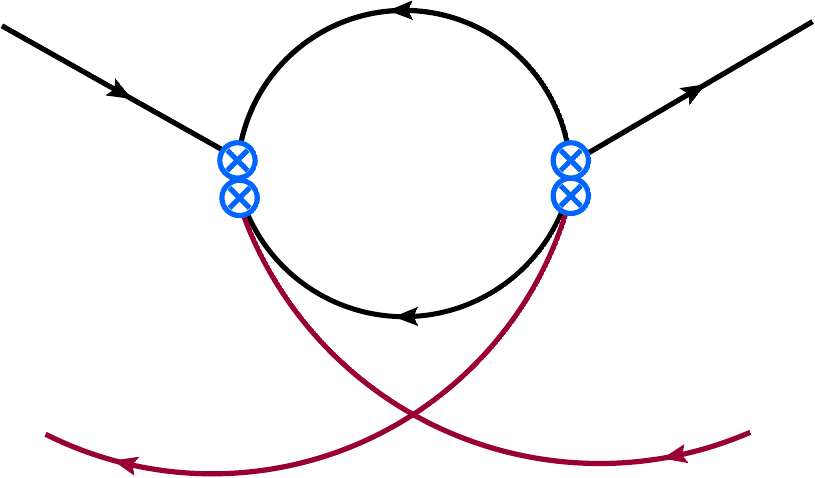}}
      \put(77,94){$\psi$}
      \put(77,37){$d$}
      \put(10,84){$b$}
      \put(13,10){$u$}
      \put(140,85){$b$}
      \put(135,11){$u$}
    \end{picture}
  \end{minipage}

  \caption{Topologies of the four-quark contributions to the total decay width $\Gamma(B^+)$ for the Mesogenesis models: model $(d)$ (left, $\overline{\text{PI}}$) and model $(b)$ (right, $\text{PI}$). The respective operators are
$\mathcal{O}=(b\,u)(\psi\,d)$ and
$\mathcal{O}=(d\,u)(\psi\,b)$. }
 \label{fig:Topologies}
\end{figure}

As in the previous section, we concentrate on models $(d)$ and $(b)$ for definiteness.
For each case, we present in \cref{fig:DecayRates} the individual contributions to $\Gamma(B^+)$ from the
leading dimension-three term $\Gamma_3$, the dimension-five correction $\Gamma_5$,
the Darwin term $\Gamma_6$, and the dimension-six four-quark contributions
$\tilde{\Gamma}_6$, as functions of $\rho_f=(m_f/m_b)^2$.
To quantify the relative size of subleading effects, we introduce $R_d\equiv|\Gamma_d/\Gamma_3|$ where $\Gamma_d$ denotes one of the subleading contributions listed above.
We also show in \cref{fig:DecayRatesTotal} the total decay width, obtained by summing all contributions,
as a function of $\rho_f$.
For convenience, the numerical inputs used in this analysis are collected in
\cref{sec:BMeso_input}.
The analysis leads to the following key observations:

\begin{itemize}
    \item \textbf{Hierarchy of contributions:}
    In general, we find $ \tilde{\Gamma}_6 > \Gamma_6 > \Gamma_5 ,$ i.e.\ the dimension-six four-quark contributions dominate over the
    Darwin term, which in turn exceeds the dimension-five corrections.
    This hierarchy closely mirrors the pattern observed in the SM,
    where the dominant dimension-six four-quark effects originate from
    PI topologies~\cite{Lenz:2020oce,Egner:2024lay}.
    The dominance of $\tilde{\Gamma}_6$ can be traced back to the
    phase-space and loop enhancement of four-quark spectator
    contributions (the familiar $16\pi^2$ factor, cf.~\cref{eq:HQE}),
    as well as to the specific colour and bag-parameter combinations
    entering the matrix elements (see \cref{sec:SMeffects}).
    The relevant dimension-six four-quark topologies for the two
    representative cases are illustrated in~\cref{fig:Topologies}.

    \item \textbf{Operator-dependent behaviour:}
For the operator $\mathcal{O} = (b\,u^c)(\psi\,d^c)$, the
dimension-six four-quark contribution exceeds the leading
dimension-three term over a significant portion of the
$\rho_\psi$ range. This behaviour is characteristic of operators
of the form $\mathcal{O} = (b\,q_1^c)(q_2\,q_3^c)$, where $q_1$
denotes the spectator quark and $q_2$, $q_3$ are arbitrary fermions.
Such structures generate a $\overline{\text{PI}}$ topology
(see \cref{fig:Topologies}), which can induce a sizeable
negative contribution to the decay width. A similar pattern appears in $\Gamma(B_d)$ for operators arising
from $\mathcal{L}_{2/3}$, for instance
$\mathcal{O} = (b\,d^c)(\psi\,u^c)$, where the spectator quark
couples directly to the $b$ quark, again leading to the
$\overline{\text{PI}}$ topology.

\item \textbf{Comparison to the SM:}
In the SM, operators such as $\mathcal{O}=(u\,b)(d\,u)$ also generate PI topologies. However, in that case the spectator
quark couples to a light quark rather than directly to the $b$ quark.
For operators of the form $\mathcal{O}=(u\,b)(d\,c)$, the $b$ quark
couples to the spectator, but the resulting contraction does not
produce a PI topology. In contrast, the Mesogenesis operators involve charge-conjugated
fields, leading to a modified $\overline{\rm PI}$ topology that
circumvents certain colour and contraction suppressions present in
the SM. This structural difference explains the enhanced behaviour
observed in the dimension-six contributions. A detailed operator-level
comparison is provided in \cref{sec:SMeffects}.

\end{itemize}
We therefore conclude that the hierarchy of HQE contributions can be
strongly operator dependent. In particular, models involving
non-SM operator structures should be examined carefully for such
effects before drawing phenomenological conclusions.

It is worth emphasizing, however, that the present analysis is performed at
leading order (LO). Experience from the SM indicates that next-to-leading
order (NLO) QCD corrections and higher-dimensional contributions can
alter the relative size of individual terms (see, e.g.,
Ref.~\cite{Egner:2024lay}). The hierarchy observed here should thus be
interpreted as an LO indication, subject to quantitative refinement
in more complete analyses.

\subsubsection*{Lifetime Ratios}

Experimentally, the lifetime ratios of $B$ mesons are measured with
remarkable precision (see e.g.\ the HFLAV review~\cite{HeavyFlavorAveragingGroupHFLAV:2024ctg}),
while the corresponding HQE predictions have reached a comparable
level of theoretical control, see e.g.  Ref.~\cite{Moretti:2026waq,Egner:2024lay}:
\begin{align}
\frac{\tau (B^+)}{\tau (B_d^0)}^{\rm Exp.} &=
1.076(4) \, , 
&
\frac{\tau (B_s^0)}{\tau (B_d^0)}^{\rm Exp.} &=
1.0021(34) \, , 
\\
\frac{\tau (B^+)}{\tau (B_d^0)}^{\rm HQE} &=
1.072 \pm 0.024 \, , 
&
\frac{\tau (B_s^0)}{\tau (B_d^0)}^{\rm HQE} &=
1.013(7) \, .
\label{eq:ratio_HQE}
\end{align}
We observe an excellent agreement between experiment and HQE
predictions. In contrast to the total lifetimes, the theoretical
uncertainties in the ratios are significantly reduced, since
dominant hadronic inputs largely cancel. The remaining difference
between experimental and theoretical precision amounts only to a
factor of a few, making lifetime ratios particularly sensitive probes
of subleading HQE contributions and potential new-physics effects.

In the HQE framework, the lifetime ratio can be expressed in terms of
the difference of total decay widths. In particular,
\begin{align}
\frac{\tau (B_s^0)}{\tau (B_d^0)}
&= \frac{\Gamma (B_d^0)}{\Gamma (B_s^0)}
= 1 + \tau (B_s^0)^{\rm Exp.}
\left[ \Gamma (B_d^0) - \Gamma (B_s^0) \right]^{\rm HQE}.
\label{eq:ratio_clean}
\end{align}
Expanding the width difference in powers of $1/m_b$ (see e.g. \cref{eq:HQE}), one obtains
\begin{align}
\frac{\tau (B_s^0)}{\tau (B_d^0)}
&= 1 + \tau (B_s^0)^{\rm Exp.}
\Bigg[
\Gamma_5 \frac{
\langle O_5 \rangle_{B_d^0} -
\langle O_5 \rangle_{B_s^0}}{m_b^2}
+ \Gamma_6 \frac{
\langle O_6 \rangle_{B_d^0} -
\langle O_6 \rangle_{B_s^0}}{m_b^3}
\nonumber\\
&\quad
+ 16\pi^2
\left(
\frac{
(\tilde{\Gamma}_6 \langle \tilde O_6 \rangle)_{B_d^0}
-
(\tilde{\Gamma}_6 \langle \tilde O_6 \rangle)_{B_s^0}
}{m_b^3}
\right.
\left.
+
\frac{
(\tilde{\Gamma}_7 \langle \tilde O_7 \rangle)_{B_d^0}
-
(\tilde{\Gamma}_7 \langle \tilde O_7 \rangle)_{B_s^0}
}{m_b^4}
+ \ldots
\right)
\Bigg] .
\end{align}
The leading dimension-three contribution,
$\Gamma_3$, cancels in the ratio
$\tau(B_s^0)/\tau(B_d^0)$, since it is universal for both
mesons.\footnote{This cancellation removes the strong $m_b^5$ dependence of the total decay width, substantially reducing the theoretical uncertainty.} As a consequence,
the ratio is governed by differences of subleading terms in the HQE,
starting at dimension five.

For the ratio $\tau(B^+)/\tau(B_d^0)$, an especially strong
cancellation occurs. In the isospin limit
($m_u = m_d$ and identical non-perturbative matrix elements),
all two-quark contributions cancel in the ratio.
As a result, the lifetime difference between charged and
neutral $B$ mesons is predominantly driven by spectator-dependent
effects arising from dimension-six four-quark operators. Unlike the two-quark terms,
both the short-distance Wilson coefficients
$\tilde{\Gamma}_{d=6,7,\ldots}$ and the associated hadronic
matrix elements depend explicitly on the flavour content of the
external hadron. Consequently, they differ for $B_d^0$, $B_s^0$,
and $B^+$ mesons and can generate sizeable lifetime splittings.

For the ratio $\tau(B_s^0)/\tau(B_d^0)$, however, spectator effects
are numerically small~\cite{Lenz:2022rbq}. This suppression originates
from the approximate $SU(3)_F$ symmetry between the $d$- and
$s$-spectator quarks, which reduces differences in the relevant
matrix elements.
In contrast, the ratio $\tau(B^+)/\tau(B_d^0)$ exhibits a sizeable
effect of order ${\cal O}(10\%)$~\cite{Lenz:2022rbq}. Here, PI between the spectator quark and the decay
products of the $b$ quark produces a substantial dimension-six
contribution. This mechanism distinguishes charged from neutral
$B$ mesons and dominates their lifetime difference.

For comparison, analogous spectator effects are even more pronounced
in the charm sector, where they exceed $100\%$ in the ratio
$\tau(D^+)/\tau(D^0)$~\cite{King:2021xqp}, reflecting the slower
convergence of the HQE for charm hadrons. For more details on the charm sector, see e.g. the review~\cite{Friday:2025gpj}.

The primary goal of the lifetime-ratio analysis is to determine whether
the measured $B$-meson lifetimes can provide additional constraints on
the couplings of the $B$-Mesogenesis model. Existing bounds on these
couplings have already been obtained from ATLAS and CMS searches for
colour-triplet scalars~\cite{Alonso-Alvarez:2021qfd}. Throughout this
analysis, we therefore adopt the corresponding upper limits on the
couplings.
Using these benchmark values, we compute the lifetime ratios as a
function of the dark-fermion mass $m_\psi$. The corresponding results
are shown in~\cref{sec:sub_plots}. In each case, a single coupling is assumed
to be non-zero, while all remaining couplings are set to zero.
We find that, within the current theoretical and experimental
uncertainties, the predicted lifetime ratios remain consistent with the
experimental measurements over the entire range of $m_\psi$.
Consequently, the lifetime ratios $\tau(B_s^0)/\tau(B_d^0)$ and
$\tau(B^+)/\tau(B_d)$ do not presently provide constraints on the
couplings beyond those already obtained from direct
collider searches.

It is nevertheless instructive to compare the implications of
constraining the Mesogenesis couplings using lifetime ratios with those
obtained from direct collider searches and flavour-mixing observables.
To this end, we determine the corresponding predictions for the
inclusive $B$-meson branching ratios using the couplings constrained by
each of these approaches. Besides the bounds from collider searches,
Ref.~\cite{Alonso-Alvarez:2021qfd} also derives  constraints
from flavour physics. These are obtained by evaluating the
contributions of the Mesogenesis operators to the neutral-meson mixing
observables $\epsilon_K$ and $\Delta M_{B,K}$ and comparing the
resulting predictions with the corresponding experimental measurements
and SM expectations.

To compute the inclusive BRs, we evaluate the leading HQE contribution,
$\Gamma_3$, and normalize it using the experimental lifetime of the
$B^+$ meson. In doing so, we implicitly assume that the SM prediction
for the total decay width agrees with the inverse of the measured
lifetime. This assumption is well justified, as the total $B$-meson
decay widths are currently predicted within the SM with a precision of
about $5\%$~\cite{Egner:2024lay} and are in good agreement with experiment.
Consequently, any additional decay channel contributing more than
approximately $5\%$ to the total width would be incompatible with the
measured lifetime. Such parameter points are therefore excluded and are
not shown in the branching-ratio plots beyond this limit.

This procedure allows us to compare the phenomenological implications of
the different constraints on the Mesogenesis couplings. In particular,
we determine the inclusive branching ratios obtained when the couplings
are constrained by lifetime ratios, direct collider searches, and
flavour-mixing observables, respectively. The corresponding allowed
parameter regions and branching-ratio predictions are presented in
\cref{fig:CouplingsConstraints23}--\ref{fig:model13-branching2} for the
$\mathcal{H}_{2/3}$ and $\mathcal{H}_{-1/3}$ scenarios.

 \cref{fig:CouplingsConstraints23,fig:CouplingsConstraints131,fig:CouplingsConstraints132}, show the regions of the
$\{G_X^{f_1f_2},m_\psi\}$ parameter space that are compatible with the
measured lifetime ratios. A parameter point is considered allowed if,
after varying all input parameters within their uncertainties, the
theoretical prediction overlaps with the experimentally measured
lifetime ratio at the $1\sigma$ level. The dominant theoretical
uncertainty originates from the SM HQE prediction for the lifetime
ratios, cf.~\cref{eq:ratio_HQE}. In addition, we vary the $b$-quark mass within a $10\%$ interval
to account for the ambiguity  of
different  mass schemes.

The corresponding predictions for the inclusive BRs are
shown in~\cref{fig:model23-branching,fig:model13-branching2}. For each value of $m_\psi$, the branching
ratio is evaluated using
$\mathrm{BR}^{\rm Th.}
=
\Gamma^{\text{meso}}_3(G_X^{f_1f_2},m_\psi)\,
\tau(B^+)^{\rm Exp.},$
where the coupling $G_X^{f_1f_2}$ is taken at the maximum value allowed
by the lifetime-ratio analysis. To obtain a conservative upper bound,
the experimental lifetime $\tau(B^+)^{\rm Exp.}$ and the $b$-quark mass
are both chosen at the upper edge of their respective allowed ranges.

\begin{table}[t]
\centering
\small
\setlength{\tabcolsep}{10pt} % Increase column spacing
\renewcommand{\arraystretch}{1.4} % Row height
\begin{tabular}{|p{2.2cm}|p{1.3cm}|p{1.3cm}|p{3.2cm}|p{3.2cm}|}
\hline
\textbf{Operator} & \textbf{Affects $B^+$?} & \textbf{Affects $B_d$?} & \textbf{Topology} & \textbf{BR ($\tau(B^+)/\tau(B_d)$)} \\
\hline
$(\bar{b} d^c)(\bar{\psi} u^c)$ & Yes & Yes & III (B$^+$), IV (B$_d$) & Figure~\ref{fig:model23-branching} (a)  \\
\hline
$(\bar{b} d^c)(\bar{\psi} c^c)$ & No & Yes & IV & Figure~\ref{fig:model23-branching} (b) \\
\hline
$(\bar{b} d^c)(\bar{d} s^c)$ & No & Yes & \{I, IV\} & Figure~\ref{fig:model23-branching} (c) \\
\hline
\rowcolor{red!15}
$(\bar{b} s^c)(\bar{d} s^c)$ & No & Yes & I & Figure~\ref{fig:model23-branching} (d) \\
\hline
\rowcolor{red!15}
$(\bar{b} s^c)(\bar{\psi} u^c)$ & Yes & No & III & Figure~\ref{fig:model23-branching} (e) \\
\hline
\end{tabular}
\caption{{$\mathcal{L}_{2/3}$ operator contributions to $B^+$ and $B_d$ meson decay rates from dimension-6 four-quark operators, showing topologies and inclusive BRs using couplings constrained by $\tau(B^+)/\tau(B_d)$. Rows highlighted in red correspond to the least-constrained operators.}}
\label{tab:operator-summary23}
\end{table}
\begin{table}[ht]
\centering
\small
\setlength{\tabcolsep}{10pt} % Increase column spacing
\renewcommand{\arraystretch}{1.4} % Row height
\begin{tabular}{|p{2.2cm}|p{1.3cm}|p{1.3cm}|p{3.2cm}|p{3.2cm}|}
\hline
\textbf{Operator} & \textbf{Affects $B^+$?} & \textbf{Affects $B_d$?} & \textbf{Topology} & \textbf{BR ($\tau(B^+)/\tau(B_d)$)} \\
\hline
$(\bar{b} u^c)(\bar{\psi} d^c)$ & Yes & Yes & IV$(B^+)$, III$(B_d)$ & Figure~\ref{fig:model13-branching1} (a)  \\
\hline
$(\bar{b} c^c)(\bar{\psi} d^c)$ & No & Yes & III & Figure~\ref{fig:model13-branching1} (b) \\
\hline
\rowcolor{yellow!10}
$(\bar{d} u^c)(\bar{\psi} b^c)$ & Yes & Yes & III both & unconstrained \\
\hline
$(\bar{d} c^c)(\bar{\psi} b^c)$ & No & Yes & III & Figure~\ref{fig:model13-branching1} (c) \\
\hline
$(\bar{\psi} d^c)(\bar{\psi} b^c)$ & No & Yes & I & Figure~\ref{fig:model13-branching1} (d) \\
\hline
$(\bar{b} u^c)(\bar{u} d^c)$ & Yes & Yes & \{IV, I\}$(B^+)$, I$(B_d)$ & Figure~\ref{fig:model13-branching1} (e) \\
\hline
$(\bar{b} c^c)(\bar{c} d^c)$ & No & Yes & I & Figure~\ref{fig:model13-branching1} (f) \\
\hline
\rowcolor{yellow!10}
$(\bar{b} c^c)(\bar{u} d^c)$ & Yes & Yes & I both & unconstrained \\
\hline
$(\bar{b} u^c)(\bar{c} d^c)$ & Yes & Yes & IV$(B^+)$, I$(B_d)$ & Figure~\ref{fig:model13-branching2} (a) \\
\hline
$(\bar{b} u^c)(\bar{\psi} s^c)$ & Yes & No & IV & Figure~\ref{fig:model13-branching2} (b) \\
\hline
\rowcolor{red!10}
$(\bar{\psi} b^c)(\bar{u} s^c)$ & Yes & No & III & Figure~\ref{fig:model13-branching2} (c) \\
\hline
$(\bar{b} u^c)(\bar{u} s^c)$ & Yes & No & \{IV, I\} & Figure~\ref{fig:model13-branching2} (d) \\
\hline
$(\bar{b} u^c)(\bar{c} s^c)$ & Yes & No & IV & Figure~\ref{fig:model13-branching2} (e) \\
\hline
$(\bar{b} c^c)(\bar{u} s^c)$ & Yes & No & I & Figure~\ref{fig:model13-branching2} (f) \\
\hline
\end{tabular}
\caption{{$\mathcal{L}_{-1/3}$ operator contributions to $B^+$ and $B_d$ meson decay rates from dimension-6 four-quark operators, showing topologies and inclusive BRs using couplings constrained by $\tau(B^+)/\tau(B_d)$. Rows highlighted in red correspond to the least-constrained operators. Rows highlighted in yellow indicate operators that contribute equally to both $B^+$ and $B_d$ mesons and thus cannot be constrained by the lifetime ratio.}}
\label{tab:operator-summary13}
\end{table}

Before discussing the results in detail, it is useful to highlight the
two main factors that determine the predicted inclusive BRs
when the Mesogenesis couplings are constrained by the lifetime ratio
$\tau(B^+)/\tau(B_d)$:

\begin{enumerate}
\item \textbf{Topology dependence.} The strength of the lifetime-ratio
constraint depends on the topology through which a given operator
contributes (see \cref{fig:MesDiagrams}). For example, operators that contribute only to one of the
two mesons, or through different topologies in the $B^+$ and $B_d$
channels, are generally subject to weaker constraints.

\item \textbf{Phase-space suppression.} The available phase space
directly determines the size of the partial decay width. Even if the
lifetime-ratio constraint on a coupling is relatively weak, the
corresponding branching ratio can remain small when the final state
contains heavy particles, such as a charm quark or the dark fermion
$\psi$.
\end{enumerate}

The interplay of these two effects governs the resulting BRs. In particular, weakly constrained operators may still predict
small branching ratios because of phase-space suppression, whereas
operators involving light final-state particles can give comparatively
large branching ratios even when their couplings are more tightly
constrained.

From the inclusive BR analysis, we identify a broad class of
Mesogenesis operators for which the predicted BR exceeds
the acceptable upper limit of approximately $5\%$ when the couplings are
constrained using the lifetime ratio
$\tau(B^+)/\tau(B_d)$ alone. This behaviour is most pronounced for
operators contributing exclusively to either the $B^+$ or the $B_d$
lifetime, in particular through topology~III with at most one massive
final-state particle or through topology~I involving only light quarks
($u$, $d$, and $s$). The corresponding operator classes, which are only
weakly constrained by the lifetime-ratio analysis, are highlighted in
red in~\cref{tab:operator-summary23,tab:operator-summary13} for the
$\mathcal{L}_{2/3}$ and $\mathcal{L}_{-1/3}$ scenarios, respectively.

We further find a small number of operators whose predicted BRs remain above the $5\%$ level even after imposing the collider
constraints. However, these deviations are well within the 
$10\%$ theoretical uncertainty assigned to the inclusive branching-ratio
calculation and therefore do not provide additional exclusions beyond
the existing collider bounds.

We also identify a class of operators whose contributions to the
$B^+$ and $B_d$ decay widths are identical, such that they cancel
exactly in the lifetime ratio $\tau(B^+)/\tau(B_d)$. Consequently,
the lifetime ratio does not constrain these operators. This occurs for
operators of the form $(d\,u^c)(f\,b^c)$, with
$f=\{c,\psi\}$, whose contributions arise through the same topology in
both decay channels. These operators are highlighted in yellow in~\cref{tab:operator-summary13}.

It is important to note that this cancellation relies on the presence
of the charge-conjugated fields. Operators with the same quark flavour
content but a different Lorentz structure, such as $(d\,u)(c\,b)$, do
not exhibit this behaviour. Instead, they contribute to the $B^+$ and
$B_d$ lifetimes through different topologies, so that no cancellation
occurs in the lifetime ratio (see, e.g., Ref.~\cite{Lenz:2022pgw}).

%%%%%%%%%%%%%%%%%%%%%%%%
\section{Exclusive Calculations: LCSR}
\label{sec:MesoExcl}

Having discussed the inclusive Mesogenesis analysis within the HQE,
we now turn to the corresponding exclusive decay channels.
In this section, we study the meson-to-baryon decay
$B^+ \to p\,\psi$, which constitutes the simplest baryonic transition
allowed in the Mesogenesis framework, using the method of LCSR introduced in \cref{sec:LCSR}.

The $B\to p$ transition has previously been studied
within the framework of LCSR using proton DAs up to twist-six~\cite{Khodjamirian:2022vta,Boushmelev:2023huu}.
In the present work, however, we employ the complementary formulation
based on $B$-meson DAs. One motivation for this
approach is that higher-twist corrections from the $B$-meson DAs can
be incorporated systematically and directly within the sum-rule
framework. In addition, the uncertainties associated with the
$B$-meson DAs are expected to be smaller than those arising from the
less well-constrained proton DAs. The present analysis therefore also
provides an important cross-check between the two different LCSR
formulations. A detailed comparison between the two approaches will
be presented later in this section.

For the exclusive analysis, it is convenient to rewrite the effective
Hamiltonian introduced in \cref{eq:HeffMeso} in the compact notation
used in \cite{Khodjamirian:2022vta}. Restricting ourselves to the
operators proportional to $G_{(d)}$ and $G_{(b)}$, relevant for the
$B^+\to p\Psi$ transition, we write
\begin{equation}
\begin{aligned}
{\cal H}_{(-1/3)}
=
-
G_{(d)}\overline{{\cal O}}_{(d)} \Psi^c
-
G^*_{(d)}\bar{\Psi}^c{\cal O}_{(d)}
+
\{d\leftrightarrow b \},
\end{aligned}
\label{eq:Heff2}
\end{equation}
where the shorthand $\{d\leftrightarrow b\}$ denotes the analogous
contributions proportional to $G_{(b)}$. The couplings
$G_{(d)}$ and $G_{(b)}$ correspond directly to the Wilson coefficients
$C_{\psi}^{(ud)}$ and $\widetilde C_{\psi}^{(ud)}$ introduced in
\cref{eq:Cpsipsi_map}, respectively. The local three-quark operators entering
\cref{eq:Heff2} are defined as
\begin{equation}
\overline{{\cal O}}_{(d)}
=
i \epsilon_{ijk}
\left(
\bar{u}^i_{R} b_R^{c\,j}
\right)
\bar{d}^{\,k}_R,
\qquad
{\cal O}_{(d)}
=
i \epsilon_{ijk}
d^{\,i}_R
\left(
\bar{b}^{c\,j}_R u^{\,k}_R
\right).
\label{eq:Od}
\end{equation}
where the superscript $c$ denotes charge-conjugated fields and
$q_R=\frac12(1+\gamma_5)q$. The indices
$i,j,k$ label the colour degrees of freedom. The corresponding operators for model $(b)$ are obtained through the
replacement $d\leftrightarrow b$:
\begin{equation}
\overline{{\cal O}}_{(b)}
=
i \epsilon_{ijk}
\left(
\bar{u}^i_{R} d_R^{c\,j}
\right)
\bar{b}^{\,k}_R,
\qquad
{\cal O}_{(b)}
=
i \epsilon_{ijk}
b^{\,i}_R
\left(
\bar{d}^{c\,j}_R u^{\,k}_R
\right).
\label{eq:Ob}
\end{equation}

The decay amplitude for the process
$B^+(p_B)\to p(P)\,\psi(q)$ is expressed as
\begin{align}
\mathcal{A}^{(X)}(B^+\to p\,\psi)
=
G^{(X)}
\,
\langle p(P)|\mathcal{O}^{(X)}|B^+(p_B)\rangle
\,u_\psi^c(q),
\label{eq:amp}
\end{align}
where $u_\psi^c(q)$ denotes the charge-conjugated spinor of the
outgoing dark-sector fermion $\psi$ with momentum $q$, and
$X=\{d,b\}$ labels the Mesogenesis model.
The hadronic matrix element is parameterised in terms of four
independent $B\to p$ form factors:
\begin{align}
\langle p(P)| \overline{{\cal  O}}_{(X)} |B^+(P+q)\rangle
&=
F^{(X)}_{B\to p_R}(q^2)\bar{u}_{pR}(P)
+
F^{(X)}_{B\to p_L}(q^2)\bar{u}_{pL}(P)
\nonumber\\
&\quad
+
\widetilde{F}^{(X)}_{B\to p_R}(q^2)\bar{u}_{pR}(P)
\frac{\slashed q}{m_p}
+
\widetilde{F}^{(X)}_{B\to p_L}(q^2)\bar{u}_{pL}(P)
\frac{\slashed q}{m_p}\,,
\label{eq:matrix_decomp}
\end{align}
where $q=p_B-P$ denotes the momentum transfer, and
$\bar u_{pR}(P)$ ($\bar u_{pL}(P)$) denotes the right-handed
(left-handed) proton Dirac spinor with momentum $P$. The above decomposition is dictated by Lorentz invariance together
with the quantum numbers of the external states and the operator.
The initial $B$ meson is a pseudoscalar state and therefore contributes
only through its momentum, while the final proton state provides the
Dirac spinor $u_p(P)$. Since the operator is chiral, the matrix
element can be expanded in terms of independent Dirac structures
constructed from the chiral projectors $P_{R,L}$ and the available
four-momenta. Using momentum conservation,
$p_B=P+q,$
one may equivalently choose either $\slashed p_B$ or $\slashed q$ as
the momentum-dependent Dirac structure. The use of $\slashed q$ is therefore merely a choice of basis. Note that structures proportional to $\slashed P$
are not independent and can be reduced using the proton equation
of motion,
$\slashed P\,u_p(P)=m_p\,u_p(P)$. The factor $1/m_p$ is introduced in the second line such that all form factors have the
same mass dimension.

The four form factors
$F^{(X)}_{B\to p_R}(q^2)$,
$F^{(X)}_{B\to p_L}(q^2)$,
$\widetilde F^{(X)}_{B\to p_R}(q^2)$, and
$\widetilde F^{(X)}_{B\to p_L}(q^2)$
encode the non-perturbative QCD dynamics of the baryonic transition
and constitute the main quantities to be determined from LCSR \footnote{Since the operators $\overline{{\cal  O}}_{(X)}$ contain the chiral structure
$\bar d\,P_L$, the matrix element inherits the same chirality.
Multiplying the decomposition from the right by $P_R$,
one finds that the structures proportional to
$\bar u_{pL}(P)$ and
$\bar u_{pR}(P)\slashed q$
vanish, implying
$F^{(X)}_{B\to p_L}(q^2)=0$ and
$\widetilde F^{(X)}_{B\to p_R}(q^2)=0$.
Nevertheless, for generality we retain the full decomposition in
\cref{eq:matrix_decomp}.}.

To derive the LCSR for the form factors introduced above, we consider
the following vacuum-to-$B$ correlation function:
\begin{equation}
{\cal F}^{(X)}(P,q)
=
i\int d^4x\, e^{iP\cdot x}
\bra{0}
T\left\{
\eta_p(x)\,
\overline{{\cal O}}_{(X)}(0)
\right\}
\ket{B^+(P+q)} \, ,
\label{eq:corr}
\end{equation}
where $\eta_p$ denotes the interpolating current with the quantum
numbers of the proton. The correlation function depends on the two
independent momenta $P$ and $q$, while the external $B$ meson carries
momentum $P+q$. 

The choice of the interpolating current is an important ingredient in
the construction of baryonic sum rules. In general, several local
three-quark currents with the quantum numbers of the proton can be
constructed, differing in their Dirac structure and in their overlap
with the proton ground state and excited baryonic resonances (see e.g.~\cite{Espriu:1983hu} for details). In the present work, we
adopt the Ioffe current~\cite{Ioffe:1981kw}, which corresponds to a
particular linear combination of two other three-quark
currents known to provide a strong coupling to
the proton ground state and stable sum rules in the nucleon channel. The proton interpolating current is therefore chosen as
\begin{equation}
 \eta_p(x)=
 \epsilon_{ijk}
 \big[
 u^{Ti}(x)\, C \gamma^{\mu} u^j(x)
 \big]
 \gamma_{5} \gamma_{\mu} d^k(x),
 \label{eq:IoffeCurrent}
\end{equation}
where the colour indices are shown explicitly and $C$ denotes the
charge-conjugation matrix, whose convention is specified in
\cref{conjugate_properties,conjugate_gamma_properties}.

The correlation function \cref{eq:corr} is a Dirac object depending
on the two independent four-momenta $P$ and $q$. In order to derive
separate sum rules for the form factors appearing in
\cref{eq:matrix_decomp}, it is convenient to decompose the correlation
function into a basis of independent Lorentz and Dirac structures:
\begin{align}
{\cal F}^{(X)}(P,q)
&=
\Big[
{\cal F}^{(X)}_{1}
+{\cal F}^{(X)}_{2}\frac{\slashed P}{m_B}
+{\cal F}^{(X)}_{3}\frac{\slashed q}{m_B}
+{\cal F}^{(X)}_{4}
\frac{\slashed P\slashed q}{m_B^2}
\Big]P_L
\nonumber\\
&\quad+
\Big[
{\cal F}^{(X)}_{5}
+{\cal F}^{(X)}_{6}\frac{\slashed P}{m_B}
+{\cal F}^{(X)}_{7}\frac{\slashed q}{m_B}
+{\cal F}^{(X)}_{8}
\frac{\slashed P\slashed q}{m_B^2}
\Big]P_R \, ,
\label{eq:corrdecomp}
\end{align}

The scalar coefficient functions
${\cal F}^{(X)}_{1,\ldots,8}$ are Lorentz-invariant amplitudes
depending only on the kinematic invariants $P^2$ and $q^2$; this
dependence is suppressed in \cref{eq:corrdecomp} for convenience.
Matching these amplitudes between the hadronic and QCD
representations yields the desired LCSR relations.

Employing the unitarity relation, we derive the hadronic
representation of the correlation function. In the timelike region
$P^2>0$, its imaginary part can be expressed as
\begin{align}
\frac{1}{\pi}\,\text{Im}\,{\cal F}^{(X)}(P,q)
&=
\delta(P^{2}-m_p^2)
\bra{0}\eta_p
\overline{
\ket{p(P)}\bra{p(P)}
}
\overline{{\cal O}}_{(X)}
\ket{B^+(P+q)}
\nonumber\\
&\quad+
\delta(P^{2}-m_{N^*}^2)
\bra{0}\eta_p
\overline{
\ket{N^{*}(P)}\bra{N^{*}(P)}
}
\overline{{\cal O}}_{(X)}
\ket{B^+(P+q)}
\nonumber\\
&\quad+
\rho_h^{(X)}(P,q)\, ,
\label{eq:unitary}
\end{align}
where the overlines indicate summation over baryon polarizations.
The inserted intermediate states are required to carry the same
conserved quantum numbers as the interpolating current
$\eta_p$, namely baryon number, isospin, electric charge, and spin.
The current in \cref{eq:IoffeCurrent} has non-vanishing overlap with
both positive- and negative-parity spin-$1/2$ baryons. Consequently,
in addition to the proton contribution, we explicitly isolate the
contribution of $N^*\equiv N(1535)$, the lowest nucleon resonance
with spin-parity $J^P=1/2^-$~\cite{ParticleDataGroup:2026aaa}.
Resonances and continuum states with quantum numbers
$J^P=1/2^+$ ($J^P=1/2^-$) and masses above the proton
(the $N^*$) are included in the spectral density
$\rho_h^{(X)}(P,q)$. 

Analogously to \cref{eq:matrix_decomp}, the matrix element involving
the negative-parity $N^*$ state is parameterized as 
\begin{align}
\langle N^*(P)| \overline{{\cal O}}_{(X)} |B^+(P+q)\rangle
&=
\Bigg[
F^{(X)}_{B\to N^*_R}(q^2)\,
\bar u_{N^*R}(P)
+
F^{(X)}_{B\to N^*_L}(q^2)\,
\bar u_{N^*L}(P)
\Bigg]\gamma_5
\nonumber\\
&+
\Bigg[
\widetilde F^{(X)}_{B\to N^*_R}(q^2)\,
\bar u_{N^*R}(P)
+
\widetilde F^{(X)}_{B\to N^*_L}(q^2)\,
\bar u_{N^*L}(P)
\Bigg]
\gamma_5\frac{\slashed q}{m_{N^*}}.
\label{eq:ffNst}
\end{align}
The couplings of the interpolating current to the proton and to the
lowest negative-parity nucleon resonance $N^*$ are defined as
\begin{equation}
 \bra{0} \eta_{p} \ket{p(P)}
 =
 \lambda_p m_p\, u_p(P)\,,
 \qquad
 \bra{0} \eta_{p} \ket{N^{*}(P)}
 =
 \lambda_{N^*} m_{N^*}\gamma_5 u_{N^*}(P)\,,
 \label{eq:deconst}
\end{equation}
where $u_p(P)$ and $u_{N^*}(P)$ denote the Dirac bispinors of the
proton and $N^*$ states with four-momentum $P$, respectively, while
$\lambda_p$ and $\lambda_{N^*}$ are the corresponding decay constants
describing the overlap of the interpolating current with the physical
baryon states. The additional $\gamma_5$ in the second matrix element
accounts for the negative parity of the $N^*$ state.

Substituting \cref{eq:matrix_decomp,eq:ffNst,eq:deconst} into
\cref{eq:unitary}, and using the spin-sum relations
\[
\sum_s u_p(P,s)\bar u_p(P,s)=\slashed P+m_p,
\qquad
\sum_s u_{N^*}(P,s)\bar u_{N^*}(P,s)=\slashed P+m_{N^*},
\]
we obtain the hadronic representation of the spectral density:
\begin{align}
\frac{1}{\pi}\,\mathrm{Im}\,{\cal F}^{(X)}(P,q)
&=
\delta(P^2-m_p^2)\,\lambda_p m_p
\Bigg\{
(\slashed P+m_p)
\Big[
F^{(X)}_{B\to p_R}(q^2)P_L
+
F^{(X)}_{B\to p_L}(q^2)P_R
\Big]
\nonumber\\
&\qquad
+
(\slashed P+m_p)\frac{\slashed q}{m_p}
\Big[
\widetilde F^{(X)}_{B\to p_R}(q^2)P_R
+
\widetilde F^{(X)}_{B\to p_L}(q^2)P_L
\Big]
\Bigg\}
\nonumber\\
&\hspace{-0.8cm}
-\delta(P^2-m_{N^*}^2)\,\lambda_{N^*}m_{N^*}
\Bigg\{
(\slashed P-m_{N^*})
\Big[
F^{(X)}_{B\to N^*_R}(q^2)P_L
+
F^{(X)}_{B\to N^*_L}(q^2)P_R
\Big]
\nonumber\\
&\qquad
+
(\slashed P-m_{N^*})\frac{\slashed q}{m_{N^*}}
\Big[
\widetilde F^{(X)}_{B\to N^*_R}(q^2)P_R
+
\widetilde F^{(X)}_{B\to N^*_L}(q^2)P_L
\Big]
\Bigg\}
\nonumber\\
&\quad
+\rho_h^{(X)}(P,q)\, .
\label{eq:hadronic_spectral}
\end{align}
The relative minus sign in the $N^*$ contribution originates from the
$\gamma_5$ factors in \cref{eq:ffNst,eq:deconst}, using
$\gamma_5(\slashed P+m_{N^*})\gamma_5=-\slashed P+m_{N^*}$. 

More generally, the hadronic spectral density can itself be decomposed
into the same basis of independent Lorentz and Dirac structures:
\begin{samepage}
\begin{align}
\rho^{(X)}(P,q)
&=
\Big[
\rho^{(X)}_{1}
+
\rho^{(X)}_{2}\frac{\slashed P}{m_B}
+
\rho^{(X)}_{3}\frac{\slashed q}{m_B}
+
\rho^{(X)}_{4}
\frac{\slashed P\,\slashed q}{m_B^2}
\Big]P_L
\nonumber\\
&
+
\Big[
\rho^{(X)}_{5}
+
\rho^{(X)}_{6}\frac{\slashed P}{m_B}
+
\rho^{(X)}_{7}\frac{\slashed q}{m_B}
+
\rho^{(X)}_{8}
\frac{\slashed P\,\slashed q}{m_B^2}
\Big]P_R \, ,
\label{eq:rhodecomp}
\end{align}
\end{samepage}
where the dependence of the invariant functions
$\rho_i^{(X)}$ on the kinematic variables $(P^2,q^2)$ has been suppressed for brevity.
As explained in \cref{sec:LCSR}, the imaginary part of the correlation
function enters a dispersion relation in the variable
$P^2\to s$, which is valid for all values of $P^2$ away from the
physical singularities at fixed $q^2$:
\begin{equation}
{\cal F}^{(X)}(P,q)
=
\frac{1}{\pi}
\int\limits_{s_{\rm th}}^\infty
ds\,
\frac{\mathrm{Im}\,{\cal F}^{(X)}(s,q)}
{s-P^2},
\label{eq:dispersion}
\end{equation}
where the lower integration limit $s_{\rm th}$ is determined by the
physical threshold in the channel interpolated by the current
$\eta_p$. In the present case, the lowest state carrying the proton
quantum numbers is the proton itself, such that
$s_{\rm th}=m_p^2$.

Using the dispersion relation in \cref{eq:dispersion} and comparing
the hadronic representation in \cref{eq:hadronic_spectral} with the
general decomposition in \cref{eq:corrdecomp}, one obtains dispersion
relations for the invariant amplitudes by matching the corresponding
Lorentz and Dirac structures. For the first four amplitudes, the
resulting relations read
\begin{align}
{\cal F}^{(X)}_{1}(P^2,q^2)
&=
\frac{\lambda_p m_p^2}{m_p^2-P^2}
F^{(X)}_{B\to p_R}(q^2)
+
\frac{\lambda_{N^*}m_{N^*}^2}{m_{N^*}^2-P^2}
F^{(X)}_{B\to N^*_R}(q^2)
+
\int_{s_h}^{\infty} ds\,
\frac{\rho_1^{(X)}(s,q^2)}{s-P^2},
\label{eq:F1disp}
\\[1mm]
{\cal F}^{(X)}_{2}(P^2,q^2)
&=
\frac{\lambda_p m_p m_B}{m_p^2-P^2}
F^{(X)}_{B\to p_R}(q^2)
-
\frac{\lambda_{N^*}m_{N^*}m_B}{m_{N^*}^2-P^2}
F^{(X)}_{B\to N^*_R}(q^2)
+
\int_{s_h}^{\infty} ds\,
\frac{\rho_2^{(X)}(s,q^2)}{s-P^2},
\label{eq:F2disp}
\\[1mm]
{\cal F}^{(X)}_{3}(P^2,q^2)
&=
\frac{\lambda_p m_p m_B}{m_p^2-P^2}
\widetilde F^{(X)}_{B\to p_L}(q^2)
+
\frac{\lambda_{N^*}m_{N^*}m_B}{m_{N^*}^2-P^2}
\widetilde F^{(X)}_{B\to N^*_L}(q^2)
+
\int_{s_h}^{\infty} ds\,
\frac{\rho_3^{(X)}(s,q^2)}{s-P^2},
\label{eq:F3disp}
\\[1mm]
{\cal F}^{(X)}_{4}(P^2,q^2)
&=
\frac{\lambda_p m_B^2}{m_p^2-P^2}
\widetilde F^{(X)}_{B\to p_L}(q^2)
-
\frac{\lambda_{N^*}m_B^2}{m_{N^*}^2-P^2}
\widetilde F^{(X)}_{B\to N^*_L}(q^2)
+
\int_{s_h}^{\infty} ds\,
\frac{\rho_4^{(X)}(s,q^2)}{s-P^2}.
\label{eq:F4disp}
\end{align}
The remaining four relations for
${\cal F}^{(X)}_{5,\ldots,8}$ are obtained from the above expressions
by interchanging the chiral labels,
$R \leftrightarrow L,$
together with the replacements
$1\to 5, 2\to 6, 3\to 7, 4\to 8.$
Here $s_h$ denotes the physical continuum threshold in the proton
channel. Since the lightest multi-particle state carrying the same
quantum numbers as the proton current is the $p\pi$ state, one has $s_h=(m_p+m_\pi)^2$ \footnote{In contrast to the proton and $N^*$ contributions, the
$p\pi$ state is a two-particle state whose invariant mass is not
fixed. It therefore generates a continuous spectral contribution
(branch cut) starting at $s=(m_p+m_\pi)^2$, rather than an isolated
pole in the dispersion relation.}.

The relations derived above constitute the hadronic representation of
the correlation function. The same correlation function can also be
evaluated in QCD using the light-cone OPE, which is valid in the spacelike region $P^2\ll 0,$
where the virtualities are sufficiently far away from the physical
hadronic poles and cuts. In this region, the correlation function
admits the dispersion representation
\begin{equation}
{\cal F}^{(X)}_{\rm OPE}(P,q)
=
\frac{1}{\pi}
\int_{0}^{\infty}
ds\,
\frac{\mathrm{Im}\,{\cal F}^{(X)}_{\rm OPE}(s,q)}
{s-P^2},
\label{eq:dispersionOPE}
\end{equation}
where the spectral density is calculated directly from quark and
gluon degrees of freedom. The lower integration limit reflects the approximation
$m_u=m_d=0$. We now
invoke the quark--hadron duality approximation to model the higher
resonance and continuum contributions appearing in
\cref{eq:F1disp,eq:F2disp,eq:F3disp,eq:F4disp}. Specifically, one
assumes that above a certain effective threshold $s_0$, the hadronic
spectral densities are well approximated by the corresponding OPE
spectral densities:
\begin{equation}
\int\limits_{s_h}^{\infty}
ds\,
\frac{\rho_n^{(X)}(s,q^2)}{s-P^2}
=
\frac{1}{\pi}
\int\limits_{s_{0}^{(n)}}^{\infty}
ds\,
\frac{\mathrm{Im}\,
{\cal F}_n^{(X)\,\mathrm{OPE}}(s,q^2)}
{s-P^2}\,,
\qquad
(n=1,\ldots,8),
\label{eq:dual}
\end{equation}
Here $s_{0}^{(n)}$ denotes the effective continuum-threshold parameter
associated with the invariant amplitude
${\cal F}^{(X)}_n$. It is important to distinguish the physical threshold $s_h$ from the
effective continuum thresholds $s_{0}^{(n)}$. The threshold $s_h$ is fixed
by the physical hadronic spectrum and is therefore universal for all
invariant amplitudes. In contrast, the parameters $s_{0}^{(n)}$ arise from
the quark--hadron duality approximation and effectively model the
transition between explicitly isolated low-lying states and the
higher-state continuum. Since different invariant amplitudes are
associated with different Dirac structures and may couple differently
to excited states, the corresponding effective thresholds need not be
identical.

Equating the hadronic and OPE representations of the invariant
amplitudes, implementing the quark--hadron duality approximation for
the higher-state contributions, and subsequently applying a Borel
transformation in the variable $P^2\to M^2$, one arrives at a system
of eight LCSR relations for the $B\to p$ and $B\to N^*$ transition
form factors. The first four sum rules are given by
\begin{align}
\lambda_p m_p^2\, e^{-m_p^2/M^2}
F^{(X)}_{B\to p_R}(q^2)
&+
\lambda_{N^*} m_{N^*}^2\,
e^{-m_{N^*}^2/M^2}
F^{(X)}_{B\to N^*_R}(q^2)
\nonumber\\
&=
\frac{1}{\pi}
\int_{0}^{s_{0}^{(1)}}
ds\,
e^{-s/M^2}
\,\mathrm{Im}\,
{\cal F}_1^{(X)\mathrm{OPE}}(s,q^2),
\label{eq:lcsr1}
\end{align}

\begin{align}
\lambda_p m_pm_B\, e^{-m_p^2/M^2}
F^{(X)}_{B\to p_R}(q^2)
&-
\lambda_{N^*} m_{N^*}m_B\,
e^{-m_{N^*}^2/M^2}
F^{(X)}_{B\to N^*_R}(q^2)
\nonumber\\
&=
\frac{1}{\pi}
\int_{0}^{s_{0}^{(2)}}
ds\,
e^{-s/M^2}
\,\mathrm{Im}\,
{\cal F}_2^{(X)\mathrm{OPE}}(s,q^2),
\label{eq:lcsr2}
\end{align}

\begin{align}
\lambda_p m_pm_B\, e^{-m_p^2/M^2}
\widetilde F^{(X)}_{B\to p_L}(q^2)
&+
\lambda_{N^*} m_{N^*}m_B\,
e^{-m_{N^*}^2/M^2}
\widetilde F^{(X)}_{B\to N^*_L}(q^2)
\nonumber\\
&=
\frac{1}{\pi}
\int_{0}^{s_{0}^{(4)}}
ds\,
e^{-s/M^2}
\,\mathrm{Im}\,
{\cal F}_3^{(X)\mathrm{OPE}}(s,q^2),
\label{eq:lcsr3}
\end{align}

\begin{align}
\lambda_p m_B^2\, e^{-m_p^2/M^2}
\widetilde F^{(X)}_{B\to p_L}(q^2)
&-
\lambda_{N^*} m_B^2\,
e^{-m_{N^*}^2/M^2}
\widetilde F^{(X)}_{B\to N^*_L}(q^2)
\nonumber\\
&=
\frac{1}{\pi}
\int_{0}^{s_{0}^{(4)}}
ds\,
e^{-s/M^2}
\,\mathrm{Im}\,
{\cal F}_4^{(X)\mathrm{OPE}}(s,q^2).
\label{eq:lcsr4}
\end{align}
The remaining four sum rules corresponding to
${\cal F}^{(X)}_{5,\ldots,8}$ are obtained analogously from the
relations below \cref{eq:F4disp}. The sum rules for the physical $B\to p$ form factors are obtained by
constructing suitable linear combinations of the LCSR relations such
that the unwanted $B\to N^*$ contributions cancel. One finds
\begin{equation}
F^{(X)}_{B\to p_R}(q^2)
=
\frac{e^{m_p^2/M^2}}
{\lambda_p m_p(m_p+m_{N^*})}
\left[
I_1^{(X)}(M^2,q^2)
+
\frac{m_{N^*}}{m_B}
I_2^{(X)}(M^2,q^2)
\right],
\label{eq:FpR_LCSR}
\end{equation}
and
\begin{equation}
\widetilde F^{(X)}_{B\to p_L}(q^2)
=
\frac{e^{m_p^2/M^2}}
{\lambda_p m_B(m_p+m_{N^*})}
\left[
I_3^{(X)}(M^2,q^2)
+
\frac{m_{N^*}}{m_B}
I_4^{(X)}(M^2,q^2)
\right],
\label{eq:FtpL_LCSR}
\end{equation}
where we introduced the shorthand notation
\begin{equation}
I_n^{(X)}(M^2,q^2)
\equiv
\frac{1}{\pi}
\int_0^{s_{0,n}}
ds\,
e^{-s/M^2}\,
\mathrm{Im}\,
{\cal F}_n^{(X)\mathrm{OPE}}(s,q^2)\,.
\label{eq:InDef}
\end{equation}
The corresponding sum rules for the $B\to N^*$ transition form
factors are obtained analogously by eliminating the proton
contributions instead.

Before turning to the explicit OPE calculation, it is important to
comment on an additional subtlety in the hadronic representation.
Between the proton ground state and the negative-parity resonance
$N^*(1535)$, there exists the broad positive-parity Roper resonance
$N(1440)$ with quantum numbers $J^P=1/2^+$\cite{ParticleDataGroup:2026aaa}. In principle, this
state should also be taken into account explicitly in the sum rules by
isolating its pole contribution in the same manner as for the proton
and $N^*$ states.

In practice, however, the treatment of the Roper resonance is
considerably more complicated. Owing to its large decay width and its
strong overlap with the nearby multi-particle continuum, the Roper
state does not manifest itself as a narrow isolated resonance.
Moreover, introducing an explicit Roper contribution would require
additional hadronic parameters, in particular its decay constant and
transition form factors, for which no reliable estimates currently
exist. The interpolating current furthermore couples simultaneously to
the proton, the Roper resonance, and other positive-parity
excitations, rendering a clean separation numerically difficult.

\begin{figure}[t]
\centering
\includegraphics[width=0.47\textwidth]{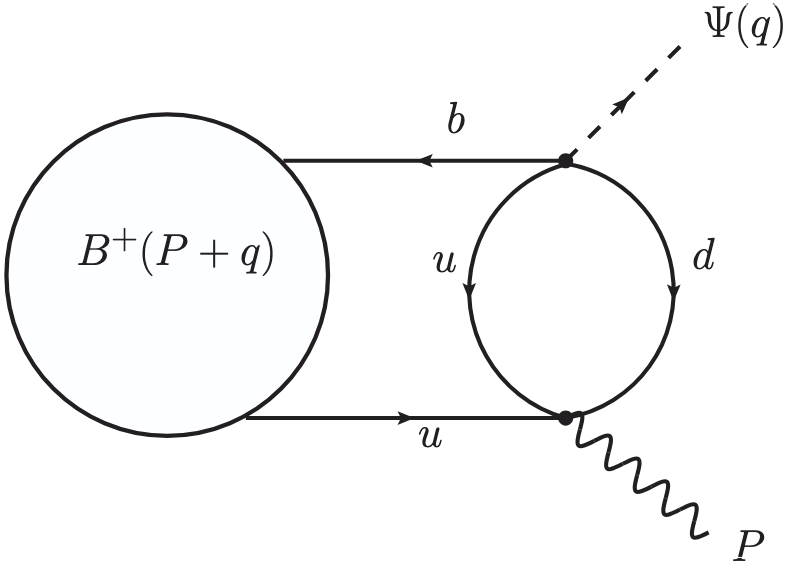}
\hfill
\includegraphics[width=0.47\textwidth]{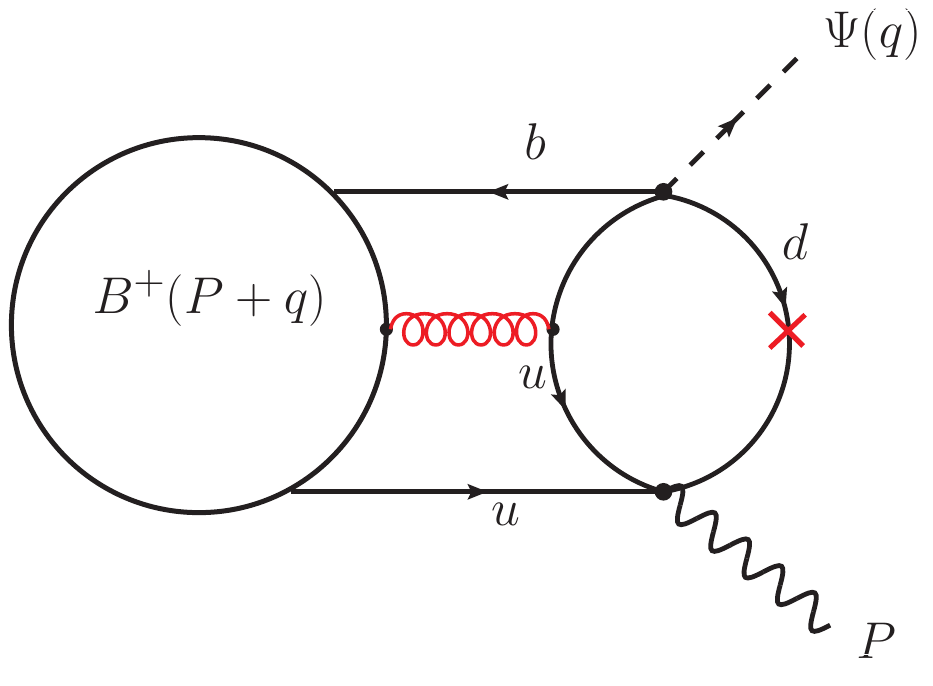}
\caption{Leading-order contributions to the correlation function in
the LCSR approach with $B$-meson DAs. Left: contribution from the
two-particle DAs. Right: diagram describing the contribution of the
three-particle (quark-antiquark-gluon) $B$-meson DAs. The curly line represents the
gluon component of the DA, while the cross indicates the second
diagram in which the gluon is emitted from the $d$-quark line.}
\label{fig:LCSRdiags}
\end{figure}

For these reasons, we adopt a simplified hadronic ansatz following
previous LCSR analyses of nucleon electromagnetic form factors
(see e.g.~\cite{Anikin:2013aka,Braun:2001tj}). In those studies, the
same proton interpolating current was employed, while the
$N(1440)$ and $N^*(1535)$ resonances were effectively included in the
spectral densities through the quark--hadron duality approximation in
\cref{eq:dual}. Following the same strategy, we absorb the Roper resonance contribution into the
higher-resonance and continuum part modeled through the quark--hadron
duality approximation. In addition, we adopt a common effective
continuum threshold for all invariant amplitudes,
$s_{0}^{(n)}\equiv s_{0}\,.$
These assumptions considerably simplify the resulting sum rules.
Furthermore, as will be shown below, all invariant amplitudes except
${\cal F}^{(X)}_1$ and ${\cal F}^{(X)}_4$ vanish in the adopted
approximation. Consequently, 
\begin{equation}
F^{(X)}_{B\to p_R}(q^2)
=
\frac{1}{\pi\lambda_p m_p^2}
\int\limits_{0}^{s_{0}} ds\,
e^{\frac{m_p^2-s}{M^2}}
\,\mathrm{Im}\,
{\cal F}_1^{(X)\mathrm{OPE}}(s,q^2)\,,
\label{eq:lcsrfin1}
\end{equation}
and
\begin{equation}
\widetilde F^{(X)}_{B\to p_L}(q^2)
=
\frac{1}{\pi\lambda_p m_B^2}
\int\limits_{0}^{s_{0}} ds\,
e^{\frac{m_p^2-s}{M^2}}
\,\mathrm{Im}\,
{\cal F}_4^{(X)\mathrm{OPE}}(s,q^2)\,.
\label{eq:lcsrfin4}
\end{equation}
Having established the hadronic representation and the corresponding
LCSR relations, we now turn to the explicit calculation of the OPE
side of the correlation function. For definiteness, we focus in the following on model $(d)$ and
consider the correlation function
\begin{equation}
{\cal F}^{(d)}(P,q)
=
i\int d^4x\, e^{iP\cdot x}
\bra{0}
T\left\{
\eta_p(x)\,
\overline{{\cal O}}_{(d)}(0)
\right\}
\ket{B^+(P+q)} \, .
\label{eq:corr_d}
\end{equation}
The extension to model $(b)$ proceeds analogously. 
The OPE receives contributions from both the two-particle and
three-particle $B$-meson DAs. The corresponding leading-order
diagrams are shown in \cref{fig:LCSRdiags}. The left diagram
corresponds to the two-particle DA contribution, while the right
diagram represents the three-particle DA contribution generated by
soft-gluon emission from the light-quark propagator.

The OPE calculation proceeds by contracting the time-ordered product
in \cref{eq:corr_d}. We first consider the contribution of the
two-particle $B$-meson DAs corresponding to the left diagram in
\cref{fig:LCSRdiags}. At leading order, the $u$- and $d$-quark fields
are contracted using the free light-quark propagator. We find it most
convenient to perform the calculation entirely in coordinate space.
After contracting the colour indices, simplifying the resulting Dirac
structures with the charge-conjugation identities in
\cref{conjugate_properties,conjugate_gamma_properties}, and taking into account the two possible contractions of
the $u$-quark fields, which yield identical contributions, we obtain
\begin{align}
{\cal F}^{(d)}(P,q)
&=
\int d^4x\,
e^{iP\cdot x}
\frac{-4 \,x_\alpha x_\beta}
{\big[(2\pi^2)(x^2)^2\big]^2}
\bra{0}
\bar b(0)\,
P_L\gamma^\alpha\gamma^\mu
u(x)
\ket{B^+(P+q)}
\,
\gamma_5\gamma_\mu\gamma^\beta P_L \, .
\label{eq:corr_coord}
\end{align}
The remaining nonlocal vacuum-to-$B$ matrix element is subsequently
expanded in terms of the $B$-meson light-cone DAs. To this end, we employ the HQET framework, in which the heavy $b$-quark field is decomposed as\footnote{At leading order in the heavy-quark expansion, this
decomposition is equivalent to \cref{eq:HQETfield}, with
$h_v\equiv b_v$. We retain the notation $h_v$ here to remain
consistent with the standard parametrisation of the $B$-meson
light-cone DAs; see \cref{app:BDA}.}
\[
b(x)=e^{-im_b v\cdot x}\, h_v(x),
\]
where $v^\mu$ denotes the four-velocity of the $B$ meson,
$v^\mu={(P+q)^\mu}/{m_B},$
and $h_v$ is the effective heavy-quark field in HQET. This decomposition factors out the dominant mechanical momentum
$m_b v^\mu$ carried by the heavy quark, while the residual soft
dynamics are encoded in the slowly varying field $h_v$. The resulting nonlocal matrix elements of the form
\[
\bra{0}\bar h_v(0)\,\Gamma\, u(x)\ket{\bar B(v)}
\]
are then parameterized in terms of the two-particle $B$-meson DAs,
whose definitions and conventions are collected in
\cref{app:BDA}.

Substituting the decomposition of the nonlocal matrix element in terms
of the $B$-meson DAs from \cref{app:BDA} into
\cref{eq:corr_coord}, we arrive at the following intermediate
expression for the OPE representation of the correlation function:
\begin{align}
&{\cal F}^{(d)\mathrm{OPE}}(P,q)
=
\left(
\frac{i f_B m_B}{4}
\right)
\int_0^\infty d\omega
\int d^4x\,
e^{i(P-\omega v)\cdot x}
\frac{x_\alpha x_\beta}
{\big[(2\pi^2)(x^2)^2\big]^2}
\nonumber\\
&\times
\Bigg\{
-
\mathrm{Tr}
\Big[
\gamma^\mu\gamma^\alpha(1+\slashed v)
\Big]
\phi_+^B(\omega)
+
\mathrm{Tr}
\Big[
\gamma^\mu\gamma^\alpha
(1+\slashed v)\slashed x
\Big]
\frac{i}{2}
\Phi_\pm^B(\omega)
\Bigg\}
\gamma_5\gamma_\mu\gamma_\beta
(1+\slashed v)
+\ldots ,
\label{eq:corrqq2}
\end{align}
where for brevity only the contributions of the twist-two and
twist-three DAs are shown explicitly. After performing the Dirac algebra and evaluating the coordinate-space
integrals, we obtain
\begin{align}
{\cal F}^{(d)\mathrm{OPE}}(P,q)
&=
\frac{f_B m_B}{16\pi^2}
\int_0^\infty d\omega
\Bigg(
(P-\omega v)^2
\ln\!\Big[-(P-\omega v)^2\Big]
\phi_+^B(\omega)
\nonumber\\
&\quad
-
\ln\!\Big[-(P-\omega v)^2\Big]
\big(
\slashed P\slashed v-\omega
\big)
\Phi_\pm^B(\omega)
-
8
\ln\!\Big[-(P-\omega v)^2\Big]
g_+^B(\omega)
\nonumber\\
&\quad
+
8
\frac{
\slashed P\slashed v-\omega
}
{(P-\omega v)^2}
G_\pm^B(\omega)
\Bigg)
P_L \, ,
\label{eq:corrfinald}
\end{align}
where the contributions of the twist-four and twist-five DAs are also
included. Note that in \cref{eq:corrfinald}, we retain only those terms that generate a
nonvanishing imaginary part in the variable $P^2$. Constant and
polynomial contributions in $P^2$ are omitted, since they disappear
after the Borel transformation. 

Repeating the same calculation for model $(b)$, we
find the simple relation
\begin{equation}
{\cal F}^{(b)\mathrm{OPE}}(P,q)
=
2\,
{\cal F}^{(d)\mathrm{OPE}}(P,q)\,.
\label{eq:formfactequal}
\end{equation}
Consequently, the corresponding sum rules and form factors of model
$(b)$ are directly obtained from those of model $(d)$ through an
overall factor of two.\footnote{Interestingly, this simple relation
does not arise in the alternative LCSR approach based on proton DAs~\cite{Khodjamirian:2022vta,Boushmelev:2023huu}.
This indicates that \cref{eq:formfactequal} is a consequence of the
specific Dirac and colour structure emerging in the present
$B$-meson-DAs formulation, rather than a fundamental property of the
underlying operators themselves.}

We now turn to the calculation of the three-particle DA
contributions. These arise from the soft-gluon emission off the light
$u$- and $d$-quark propagators, corresponding to the right diagram in
\cref{fig:LCSRdiags}. For massless light quarks, we employ the
light-cone expansion of the quark propagator in the external gluon
field~\cite{Balitsky:1987bk}:
\begin{equation}
\begin{aligned}
-i\langle 0|
T\{
q_\alpha^i(x)\,
\bar q_\beta^k(0)
\}
|0\rangle
&=
\frac{
\slashed x_{\alpha\beta}\delta^{ik}
}
{2\pi^2(-x^2)^2}
\\
&
+
\frac{
g_s(\lambda^a)^{ik}
}
{32\pi^2(-x^2)}
\int_0^1 du
\Big\{
u\,\sigma_{\mu\nu}
G^{a\mu\nu}(ux)\slashed x
+
\bar u\,
\slashed x
\sigma_{\mu\nu}
G^{a\mu\nu}(ux)
\Big\}_{\alpha\beta},
\end{aligned}
\label{eq:prop}
\end{equation}
where $\lambda^a$ denote the Gell-Mann matrices, and the Dirac and
colour indices are shown explicitly.\footnote{As explained at the end of  \cref{sec:FSGauge},
the expansion in \cref{eq:prop} is equivalent order by order to the
quark propagator expansion in the FS gauge discussed in
\cref{sec:FSGauge}. However, instead of expanding the gluon field
around a local point in terms of covariant derivatives, the
nonlocal dependence on $G_{\mu\nu}(ux)$ is kept explicitly. This form
is more convenient for LCSRs, since it directly matches onto the
nonlocal light-cone matrix elements defining the three-particle
$B$-meson DAs.
}

Using the propagator expansion in \cref{eq:prop}, following the same
colour and Dirac-algebra simplifications as in the two-particle case,
and accounting for the second identical contraction of the
$u$-quark fields, we obtain for the diagram in which the soft gluon is
emitted from the $u$-quark propagator the intermediate expression

\begin{samepage}
\begin{align}
{\cal F}^{(d)\mathrm{OPE}}(P,q)\big|_{u}
&=
\frac{1}{16\pi^4}
\int_0^1 du
\int d^4x\,
e^{i{\cal P}\cdot x}
\frac{x^{\alpha_1}x^{\alpha_2}}{(-x^2)^3}
\nonumber\\
&\quad\times
\Big[
u\,
\bra{0}
\bar b(0)\,
G_{\rho\beta}(ux)\,
P_L\gamma_{\alpha_1}\sigma^{\beta\rho}\gamma_\mu\,
u(x)
\ket{B^+(P+q)}
\nonumber\\
&\qquad
+\bar u\,
\bra{0}
\bar b(0)\,
G_{\rho\beta}(ux)\,
P_L\sigma^{\beta\rho}\gamma_{\alpha_1}\gamma_\mu\,
u(x)
\ket{B^+(P+q)}
\Big]
\gamma_5\gamma^\mu\gamma_{\alpha_2}P_L ,
\label{eq:3DA_u_intermediate}
\end{align}
\end{samepage}
where $\bar u\equiv 1-u$, and
${\cal P}=P-\sigma m_B v,$ with the function
$\sigma=\sigma(u,\omega_1,\omega_2)$
defined in \cref{eq:sigma}. Similarly, for the diagram in which the soft gluon is emitted from
the $d$-quark propagator, we obtain

\begin{samepage}
\begin{align}
{\cal F}^{(d)\mathrm{OPE}}(P,q)\big|_{d}
&=
-\frac{1}{16\pi^4}
\int_0^1 du
\int d^4x\,
e^{i{\cal P}\cdot x}
\frac{x^{\alpha_1}x^{\alpha_2}}{(-x^2)^3}
\Big[
\gamma_5\gamma_\mu
\big(
u\,\sigma^{\rho\beta}\gamma_{\alpha_1}
+
\bar u\,\gamma_{\alpha_1}\sigma^{\rho\beta}
\big)
P_L
\Big]
\nonumber\\
&\quad\times
\bra{0}
\bar b(0)\,
G_{\rho\beta}(ux)\,
P_L\gamma_{\alpha_2}\gamma^\mu
u(x)
\ket{B^+(P+q)} .
\label{eq:3DA_d_intermediate}
\end{align}
\end{samepage}
The resulting matrix elements containing two quark fields and one
gluon field-strength tensor, appearing in
\cref{eq:3DA_u_intermediate,eq:3DA_d_intermediate}, are then
expanded in terms of the three-particle $B$-meson DAs using the
decomposition in \cref{eq:3DA_compact}. Finally, for the diagram with gluon emission from the $u$-quark
propagator, we obtain
\begin{equation}
{\cal F}^{(d)\mathrm{OPE}}(P,q)\big|_{u}
=
-\frac{f_Bm_B}{64\pi^2}
\int_0^\infty d\omega_1
\int_0^\infty d\omega_2
\int_0^1 \frac{du}{{\cal P}^2}
\left[
u\,{\cal C}({\cal P},v)
+
\bar u\,\overline{\cal C}({\cal P},v)
\right]
P_L\,,
\label{eq:3partdiag}
\end{equation}
The expressions for the
coefficient functions ${\cal C}$ and $\overline{\cal C}$ are collected
in \cref{sec:MesoExclResults}. 

Evaluating the remaining diagrams, we
further obtain
\begin{equation}
\begin{aligned}
&{\cal F}^{(d)_{OPE}}(P,q)|_{d}= -{\cal F}^{(d)_{OPE}}(P,q)|_{u}\,,
\end{aligned}
\label{eq:3partrel1}
\end{equation}
\begin{equation}
\begin{aligned}
&{\cal F}^{(b)_{OPE}}(P,q)|_{d}= {\cal F}^{(b)_{OPE}}(P,q)|_{u} = 0.
\end{aligned}
\label{eq:3partrel2}
\end{equation}
Therefore, the net contribution of the three-particle $B$-meson DAs
to the correlation functions of models $(d)$ and $(b)$ vanishes
identically.

The vanishing of the three-particle DA contribution is highly
nontrivial and can be traced back to several ingredients.
1.~\textbf{Interpolating current:}
The specific form of the proton interpolating current plays a crucial
role. In particular, the Ioffe current is given by a fixed linear
combination of two independent three-quark currents (see
e.g.~\cite{Espriu:1983hu}). We verified explicitly that each of these
currents separately produces nonvanishing three-particle DA
contributions, whereas the cancellation occurs only after combining
them into the particular Ioffe-current structure. 2.~\textbf{Operator structure:}
The Lorentz structure of the effective interactions in
\cref{eq:Od,eq:Ob} is equally important. In the present setup, both
the diquark bilinear and the quark--$\Psi$ bilinear form Lorentz
scalars. Modifying this structure changes the cancellation pattern.
For example, considering an interaction of the form
$\left(
\bar u_R^i\gamma_\rho b_R^c
\right)
\left(
\bar d_R\gamma^\rho\Psi
\right),$
or the analogous replacement $d\leftrightarrow b$ in model $(b)$,
leads to nonvanishing contributions from the soft-gluon emission
diagrams in \cref{fig:LCSRdiags}, even in combination with the Ioffe
current. 3.~\textbf{Quark-mass effects:}
As an additional consistency check, we repeated the calculation in
momentum space using quark propagators in dimensional regularisation. For massless light quarks, we reproduced the same
relations in \cref{eq:3partrel1,eq:3partrel2}. On the other hand,
introducing a finite strange-quark mass, corresponding for instance
to a transition of the type $B\to\Lambda\Psi$, generates nonvanishing
three-particle DA contributions proportional to $m_s$.

We can now match the OPE result in \cref{eq:corrfinald} to the
general decomposition in \cref{eq:corrdecomp}. Using
$ \slashed P\slashed v
=
({P^2+\slashed P\slashed q})/{m_B},$
we see that the OPE contains only the scalar structure and the
$\slashed P\slashed q$ structure multiplying $P_L$. Therefore, only
the invariant amplitudes ${\cal F}^{(d)\mathrm{OPE}}_1$ and
${\cal F}^{(d)\mathrm{OPE}}_4$ are nonzero, while all remaining
amplitudes vanish:
${\cal F}^{(d)\mathrm{OPE}}_{2,3,5,6,7,8}=0.$ The nonvanishing invariant amplitudes are
\begin{align}
{\cal F}^{(d)\mathrm{OPE}}_1(P^2,q^2)
&=
\frac{f_B}{16\pi^2}
\int\limits_{0}^{\infty} d\omega
\Bigg[
m_B\Delta_\omega\ln(-\Delta_\omega)\,\phi_+^B(\omega)
-
\ln(-\Delta_\omega)
\left(
P^2-m_B\omega
\right)
\Phi_\pm^B(\omega)
\nonumber\\
&\quad
-
8m_B\ln(-\Delta_\omega)\,g_+^B(\omega)
+
8
\frac{\left(P^2-m_B\omega\right)}{\Delta_\omega}
G_\pm^B(\omega)
\Bigg],
\label{eq:OPEF1}
\\[2mm]
{\cal F}^{(d)\mathrm{OPE}}_4(P^2,q^2)
&=
\frac{f_Bm_B^2}{16\pi^2}
\int_0^\infty d\omega
\left[
-\ln(-\Delta_\omega)\,\Phi_\pm^B(\omega)
+
\frac{8}{\Delta_\omega}
G_\pm^B(\omega)
\right],
\label{eq:OPEF4}
\end{align}
where 
$\Delta_\omega \equiv (P-\omega v)^2$. These are the two invariant amplitudes that enter the final LCSR
relations in \cref{eq:lcsrfin1,eq:lcsrfin4}. 

The next step is to evaluate the imaginary parts of the invariant
amplitudes in the variable $s=P^2$. In the expressions obtained above,
the imaginary contributions originate from two different structures:
the logarithmic terms $\ln(-\Delta_\omega)$ and the pole terms
$1/\Delta_\omega$.
 We use the standard relation
\begin{equation}
\frac{1}{\pi}\,
\mathrm{Im}\,
\ln(-\Delta_\omega)
=
-\theta(\Delta_\omega),
\qquad
\frac{1}{\pi}\,
\mathrm{Im}\,
\frac{1}{\Delta_\omega+i0}
=
-\delta(\Delta_\omega).
\label{eq:ImRules}
\end{equation}
The step function generated by the logarithmic terms constrains the
integration region in the variable $\omega$ through the condition
$\Delta_\omega>0$. Solving the equation $(P-\omega v)^2=0$
for $\omega$, one obtains the physical root \footnote{
The second solution of the equation $\Delta_\omega=0$ corresponds to
$\omega_+
=
\frac{
m_B^2+s-q^2
+
\sqrt{
(m_B^2+s-q^2)^2-4m_B^2s
}
}{
2m_B
},$
which lies outside the physical integration region relevant for the
light-cone expansion. In particular, for the kinematics considered in
the LCSR derivation one has $\omega_+>m_B$, whereas the light-cone
DAs are supported in the region
$\omega={\cal O}(\Lambda_{\rm QCD})\ll m_B$. Therefore only the root
$\omega_-$ contributes.
}
\begin{equation}
\omega_-
=
\frac{
m_B^2+s-q^2
-
\sqrt{
(m_B^2+s-q^2)^2-4m_B^2s
}
}{
2m_B
}.
\label{eq:omegaMinus}
\end{equation}
Accordingly, the step function restricts the integration region to
$0\leq \omega \leq \omega_-,$
while the delta-function contribution is evaluated as
\begin{equation}
\delta(\Delta_\omega)
=
\frac{
m_B
}{
\sqrt{
(m_B^2+s-q^2)^2-4m_B^2s
}
} \,
\delta(\omega-\omega_-).
\label{eq:deltaDelta}
\end{equation}
This relation follows from the standard identity
\[
\delta(f(\omega))
=
\sum_i
\frac{\delta(\omega-\omega_i)}
{|f'(\omega_i)|},
\]
where $\omega_i$ are the roots of the equation
$f(\omega)=\Delta_\omega=0$. Evaluating the derivative
\[
f'(\omega)
=
2\omega-\frac{m_B^2+s-q^2}{m_B}
\]
at the physical root $\omega_-$ yields
\[
|f'(\omega_-)|
=
\frac{
\sqrt{
(m_B^2+s-q^2)^2-4m_B^2s
}
}{m_B},
\]
which directly leads to \cref{eq:deltaDelta}.

Using \cref{eq:ImRules,eq:omegaMinus,eq:deltaDelta}, we obtain the
spectral densities corresponding to the nonvanishing invariant
amplitudes in \cref{eq:OPEF1,eq:OPEF4}:
\begin{align}
\mathrm{Im}\,
{\cal F}^{(d)\mathrm{OPE}}_1(s,q^2)
&=
\frac{f_Bm_B}{16\pi}
\Bigg[
\int\limits_0^{\omega_-} d\omega\,
\left\{
-\Delta_\omega(s)\phi_+^B(\omega)
+
\left(
\frac{s}{m_B}-\omega
\right)
\Phi_\pm^B(\omega)
+
8g_+^B(\omega)
\right\}
\nonumber\\
&\qquad
-
8\,
\frac{m_B}{\sqrt{(m_B^2+s-q^2)^2-4m_B^2s}}\,
\left(
\frac{s}{m_B}-\omega_-
\right)
G_\pm^B(\omega_-)
\Bigg],
\label{eq:ImF1}
\\[2mm]
\mathrm{Im}\,
{\cal F}^{(d)\mathrm{OPE}}_4(s,q^2)
&=
\frac{f_Bm_B^2}{16\pi}
\Bigg[
\int\limits_0^{\omega_-} d\omega\,
\Phi_\pm^B(\omega)
-
8\,
\frac{m_B}{\sqrt{(m_B^2+s-q^2)^2-4m_B^2s}}\,
G_\pm^B(\omega_-)
\Bigg].
\label{eq:ImF4}
\end{align}
Here $\Delta_\omega(s)$ denotes $\Delta_\omega$ after the replacement
$P^2\to s$.

\subsection{Numerical Analysis}
Having derived the LCSR expressions for the $B\to p$ transition form
factors, we now turn to their numerical evaluation. We first specify
the input parameters entering the sum rules. 

\begin{table}[ht]
\centering
\small
\begin{tabular}{|l|l|c|}
\hline
Parameters  &  values/intervals & Source \\[1mm]
\hline 
\hline
&&\\[-3.0mm]
Hadron masses   & $m_B=5.279 $ GeV
 & \cite{ParticleDataGroup:2026aaa}
 \\[1mm]
& $m_p=938.272$  MeV &
\\[1mm]
& $m_{\Lambda_b}=5.620$  GeV &
\\[1mm]
& $m_{\pi^+}=139.57$  MeV &
\\[1mm]
 \hline
&&\\[-3.0mm]
Nucleon decay constant &  
$\lambda_p =-(27\pm 9)\times 10^{-3}$ 
GeV$^{\,2}$
 & \cite{Lenz:2009ar}\\[1mm]
\hline
$B$-meson decay constant & $f_B=190.0\pm 1.3$ MeV&\cite{FlavourLatticeAveragingGroupFLAG:2024oxs}
\\[1mm]
Inverse moment of tw2 $B$ DA  & 
 $\lambda_B =460 \pm 110$ MeV  
 &\cite{Braun:2003wx}\\[1mm]
 The ratio (\ref{eq:Rratio}) 
& $R=\lambda_E^2/\lambda_H^2=0.30^{+0.17}_{-0.12}$ & see \cref{app:BDA2}\\[1mm]
\hline
&&\\[-3.0mm]
Borel parameter squared & $M^2= 1.5$ - $2.0 $ GeV$^2$ &
\cite{Anikin:2013aka}
\\[1mm]
Duality threshold  & $ s_0 = 2.25 $ GeV$^2$ &  \\[1mm]
\hline
&&\\[-3.0mm]
$b$-quark $\overline{\text{MS}}$ mass & $\overline{m}_b(\mu=3~\mbox{GeV})= 4.47^{+0.04}_{-0.03} $ GeV  & 
\cite{ParticleDataGroup:2026aaa}\\[1mm]
\hline
&&\\[-3.0mm]
$B^+$-meson lifetime & $\tau_B= 1.638\pm 0.004$ ps  & 
\cite{ParticleDataGroup:2026aaa}\\[1mm]
\hline
\hline
\end{tabular}
\caption{The input parameters used in the LCSR numerical analysis.  }
\label{tab:input}
\end{table}

In~\cref{tab:input} we collect the values and intervals of all
input parameters used in the numerical analysis. Apart from the
hadron masses taken from \cite{ParticleDataGroup:2026aaa}, the LCSR
calculation requires the proton decay constant $\lambda_p$ defined in
\cref{eq:deconst}. For this parameter we adopt the interval obtained
in \cite{Lenz:2009ar} from a two-point QCD sum-rule analysis using
the same Ioffe interpolating current. The definitions and models of the $B$-meson DAs entering the LCSRs
are summarized in \cref{app:BDA}. The complete set of
two-particle and three-particle DAs depends on three nonperturbative
input parameters. The first one is the $B$-meson decay constant,
which is accurately determined from lattice QCD. We use the
$N_f=2+1+1$ FLAG average quoted in
\cite{FlavourLatticeAveragingGroupFLAG:2024oxs}. Considerably less
precise is the inverse moment of the twist-two DA, $\lambda_B$,
defined in \cref{eq:lambdaB}. For this quantity we adopt the QCD
sum-rule determination from \cite{Braun:2003wx}; see also
\cite{Khodjamirian:2020hob}. The third parameter is the ratio
$R=\lambda_E^2/\lambda_H^2$ introduced in \cref{eq:Rratio}. The interval
quoted in \cref{tab:input} is obtained by averaging the
available QCD sum-rule determinations of $\lambda_E$ and
$\lambda_H$ from
\cite{Grozin:1996pq,Nishikawa:2011qk,Rahimi:2020zzo}, which are based
on different two-point correlation functions.

Adopting the values in~\cref{tab:input}, we implicitly assume a
common renormalisation scale $\mu\sim1$ GeV for $\lambda_B$ and the
other scale-dependent quantities entering the DAs. This scale lies in
the same range as the typical Borel scale employed in the sum rules.
A more refined treatment would require the calculation of radiative
gluon corrections to the correlation function, which remains beyond
the scope of the present analysis.

The choice of the Borel parameter $M^2$ and the effective continuum
threshold $s_0$ listed in \cref{tab:input} follows the same
strategy adopted in previous LCSR analyses of nucleon electromagnetic
form factors \cite{Anikin:2013aka,Braun:2001tj}. Since in the present
work we employ the simplified version of the hadronic representation
described above (see \cref{eq:lcsrfin1,eq:lcsrfin4} and the
corresponding discussion), where the excited baryonic resonances are absorbed
into the continuum contribution through quark--hadron duality, it is
natural to use the same effective threshold
$s_0\simeq (1.5~\mbox{GeV})^2
=
2.25~\mbox{GeV}^2,$
together with the same Borel window
$M^2=1.5-2.0~\mbox{GeV}^2.$ Finally, for the estimate of the inclusive
width $B\to X_N\psi$ used later, we employ the
$\overline{\mathrm{MS}}$ $b$-quark mass at the scale $\mu=3$ GeV from
\cite{ParticleDataGroup:2026aaa}.

\begin{figure}[t]
    \centering
    \includegraphics[width=0.475\textwidth]{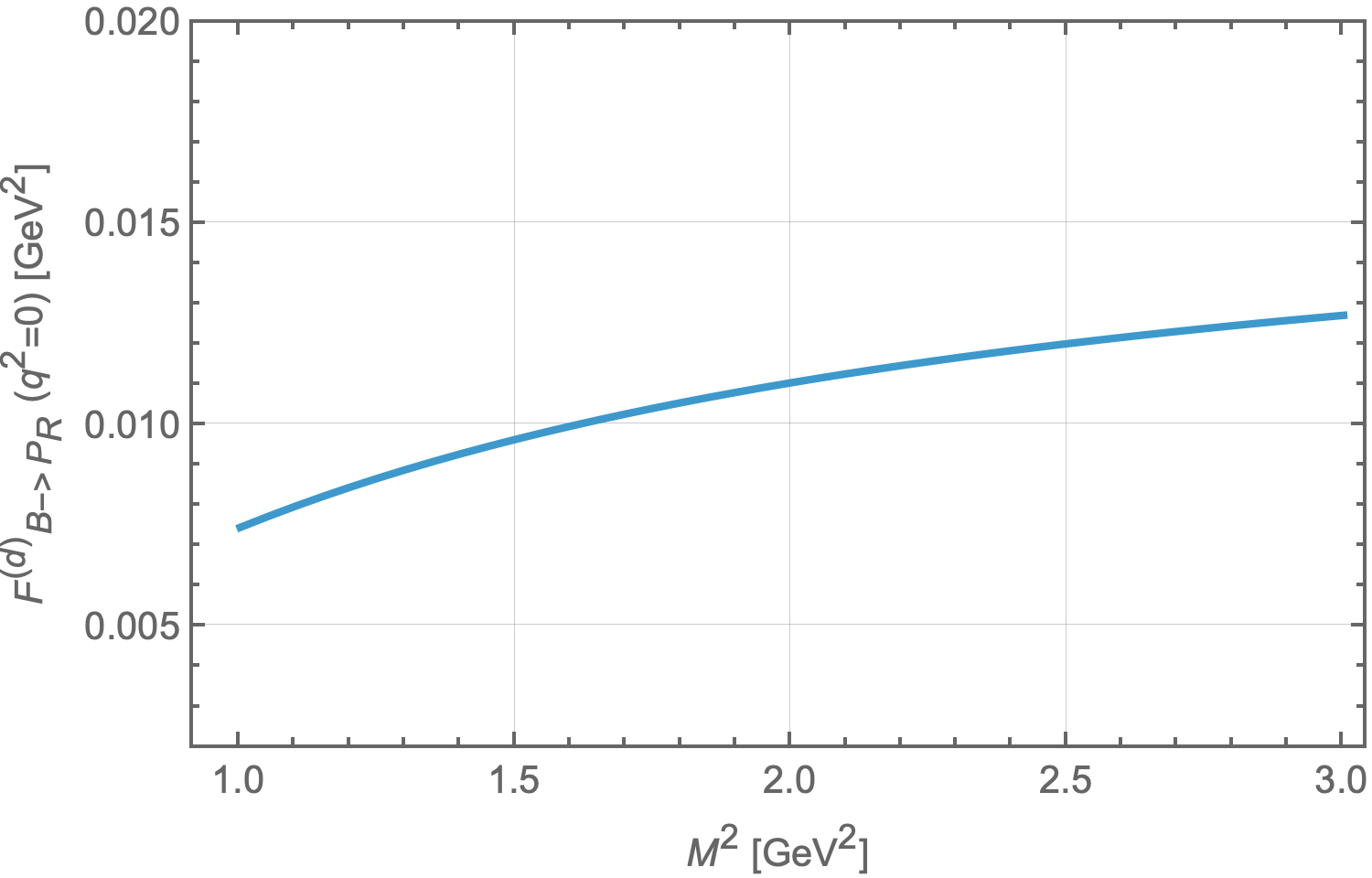}
    \hfill
    \includegraphics[width=0.475\textwidth]{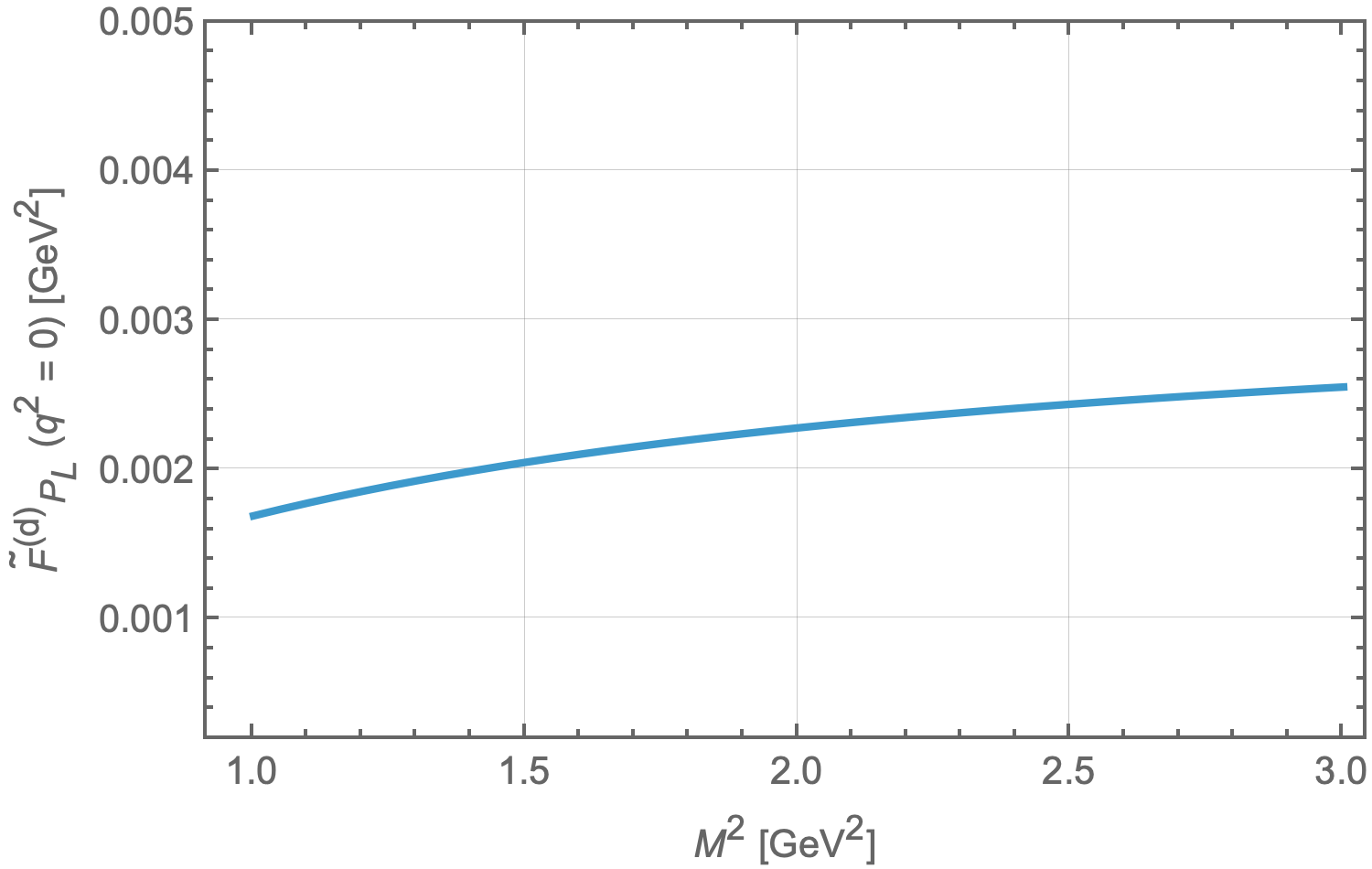}
\caption{Borel-mass dependence of the form factors $F^{(d)}_{B\to P_R}$ (left panel) and 
$\widetilde{F}^{(d)}_{B\to P_L}$ (right panel) at $q^2=0$, evaluated at the central values of the input parameters.
The interval $M^2=1.5 -2.0~\mbox{GeV}^2$ is used in LCSRs. 
}
\label{fig:FF_Borel}
\end{figure}

Having specified all input parameters, we first assess the stability
of the obtained form factors with respect to the Borel parameter
$M^2$. In \cref{fig:FF_Borel} we show the dependence of the
form factors on the Borel mass squared. One observes a good stability
behaviour throughout the adopted Borel window
$M^2=1.5-2.0~\mbox{GeV}^2$. In fact, the sum rules remain reasonably
stable even beyond this interval, indicating a mild sensitivity of
the results to the precise choice of the Borel parameter. 

In addition, by assessing the relative size of the proton-pole
contribution and the continuum contribution above the effective
threshold $s_0$ within the adopted Borel window, we find that the
simplified hadronic ansatz employed in the present analysis leads to
a numerically subdominant pole contribution in the LCSRs. This
behaviour is a direct consequence of using an effective threshold that
absorbs the nucleon resonances into the continuum through the
duality approximation. As a result, the extracted form factors show a
substantial sensitivity to the quark--hadron duality assumption
underlying the continuum subtraction procedure.
\begin{figure}[htbp]
    \centering
    \includegraphics[width=0.475\textwidth]{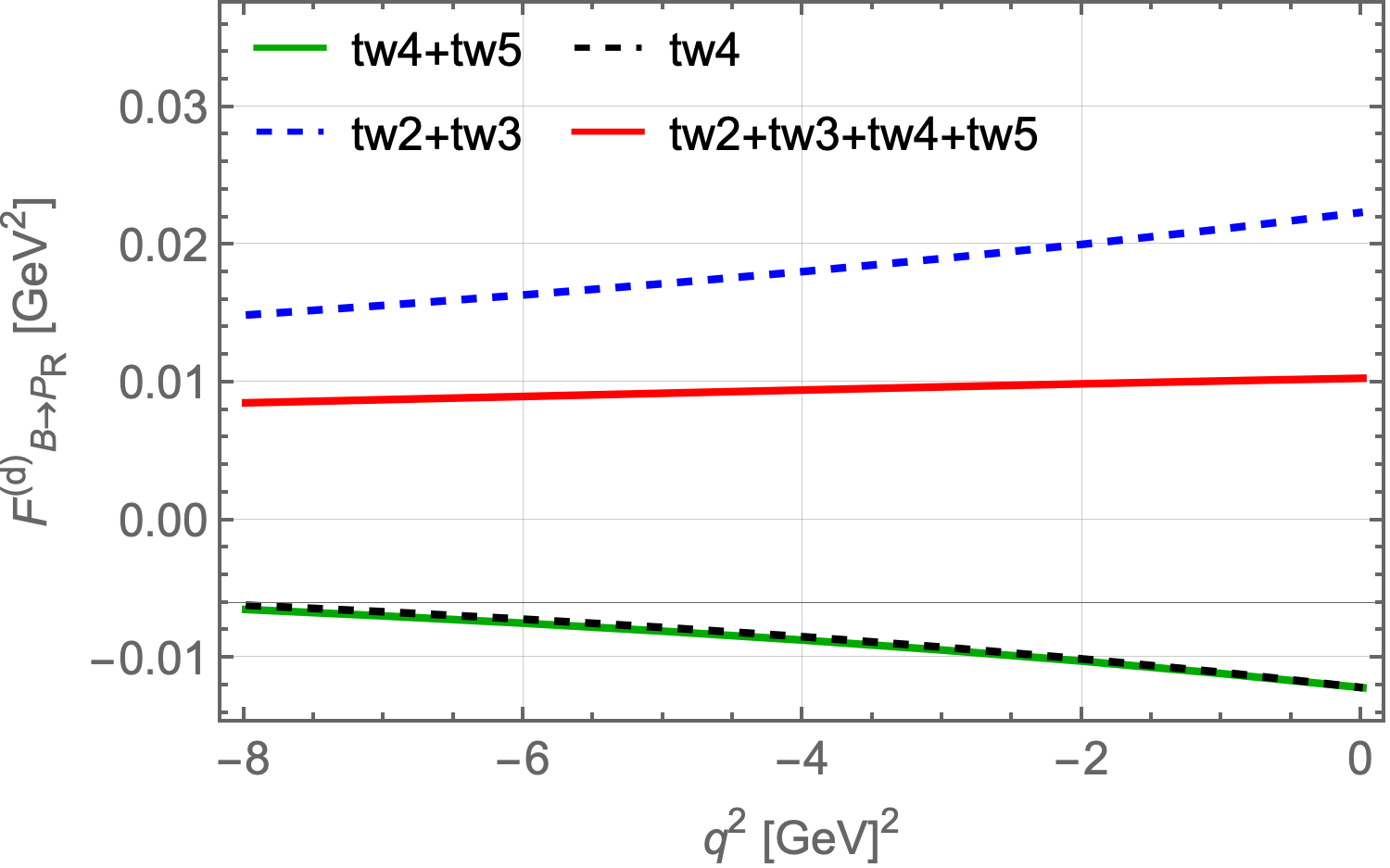}
    \hfill
    \includegraphics[width=0.49\textwidth]{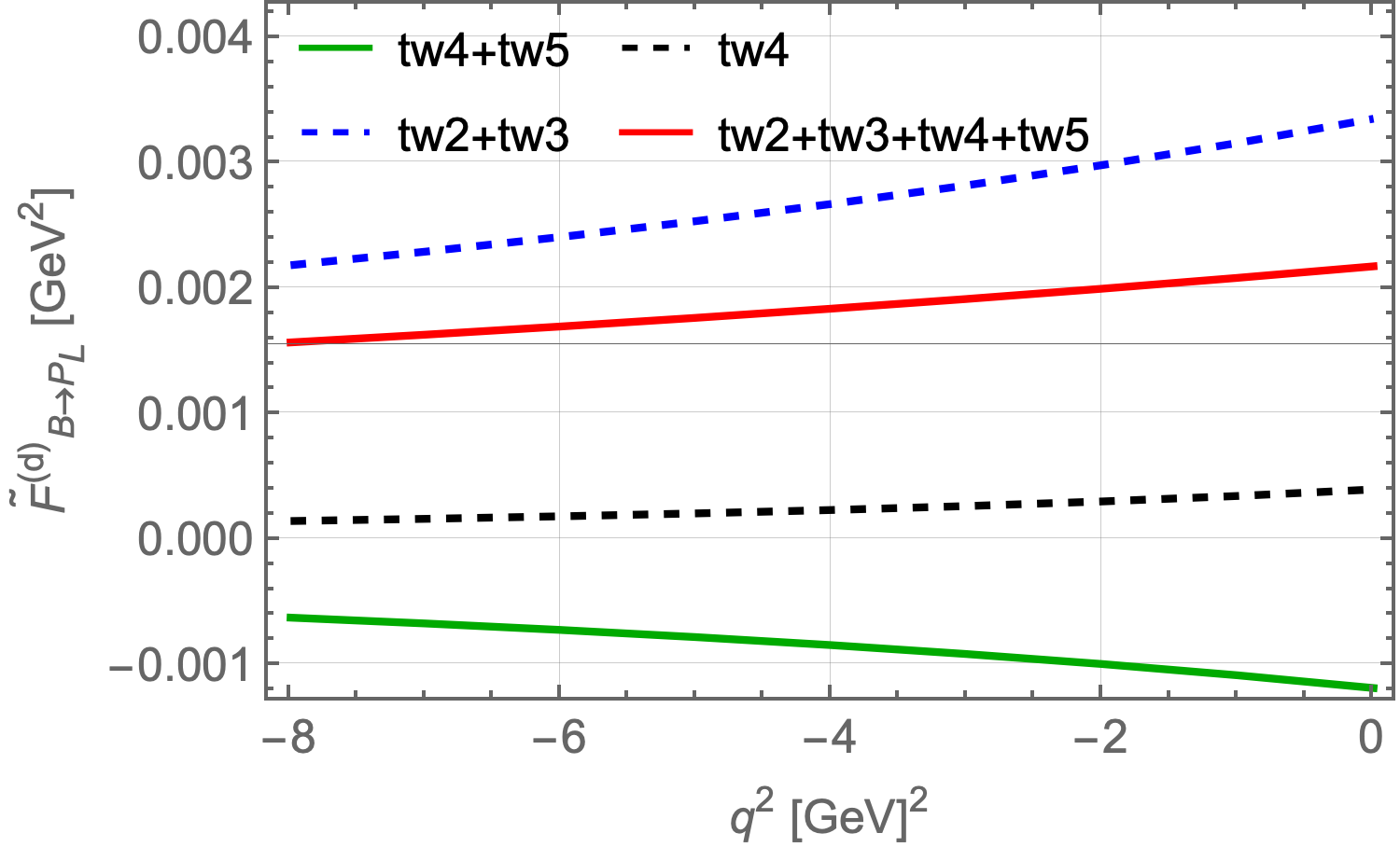}
\caption{The form factors  $F^{(d)}_{B\to P_R}$ (left panel) and $\widetilde{F}^{(d)}_{B\to P_L}$ (right panel) in the spacelike region and at central input,   shown  with the red solid lines.
 The blue dashed (green solid)
lines represent 
the sum of the twist-2 and twist-3 (twist-4 and twist-5) contributions, and the black dashed line represents the twist-4 
contribution.
}
\label{fig:DA_compar}
\end{figure}

We now evaluate the form factors at their central input values in the
region of validity of the LCSR, corresponding to spacelike momentum
transfer $q^2<0$. The resulting $q^2$ dependence of
$F^{(d)}_{B\to P_R}$ and
$\widetilde{F}^{(d)}_{B\to P_L}$ is displayed in
\cref{fig:DA_compar}. For both form factors, the dominant contribution originates from the
combined twist-two and twist-three terms.\footnote{
Since the twist-three DA $\phi_-^B$ enters the LCSRs only through the
integrated combination $\Phi_\pm^B$, involving also the twist-two DA
$\phi_+^B$, it is natural to combine the twist-two and twist-three
contributions together. Analogously, the twist-four and twist-five
terms appear through the combination $G_\pm^B$, and therefore their
contributions are also shown jointly.
} The higher-twist
contributions partially cancel the lower-twist result and reduce the
final values of the form factors. In the case of
$F^{(d)}_{B\to P_R}$, the twist-four contribution already provides
the dominant part of the higher-twist correction, indicating that the
genuine twist-five effects remain relatively small. In contrast, for
$\widetilde{F}^{(d)}_{B\to P_L}$ the difference between the isolated
twist-four term and the combined twist-four plus twist-five
contribution is substantial, showing that the twist-five DA
contribution is numerically important in this case.
Overall, the hierarchy between the different twist contributions
remains stable throughout the spacelike region considered.

We now turn to the discussion of the theoretical uncertainties of the
obtained LCSR predictions. In the numerical analysis, the individual
uncertainties associated with the input parameters listed in
\cref{tab:input} are combined in quadrature. As representative
results, we present in \cref{tab:compar} the values of the
$B\to p$ transition form factors at $q^2=0$ (denoted LCSR-$B$). The final predictions,
including all two-particle DA contributions up to twist five, are
shown in the column denoted by (tw2--tw5), whereas the column
(tw2--tw3) is displayed only for illustration of the impact of the
higher-twist corrections.
\begin{table}[ht]
\centering
\begin{tabular}{|c|c|c||c|c|}
\hline
\multirow{2}{*}{Form factor} & \multicolumn{2}{c||}{LCSR-$B$ (this work)} & \multicolumn{2}{c|}{LCSR-$N$} \\
\cline{2-5}
& (tw2-tw3) & (tw2-tw5) & (tw3) \cite{Khodjamirian:2022vta} & (tw3-tw6) \cite{Boushmelev:2023huu} \\
\hline
\hline
$F^{(d)}_{B\to p_R}(0)$ 
& $[1.4,\,3.9]$ & $[0.4,\,2.2]$ 
& $[2.4,\,2.8]$ & $[0.9,\,3.5]$ \\
\hline
$\widetilde{F}^{(d)}_{B\to p_L}(0)$ 
& $[0.24,\,0.52]$ & $[0.14,\,0.35]$ 
& $0$ & $[0.4,\,0.7]$ \\
\hline
\hline
$F^{(b)}_{B\to p_R}(0)$ 
& $[2.8,\,7.7]$ & $[0.8,\,4.4]$ 
& $[-0.9,\,-0.6]$ & $[-5.9,\,-2.2]$ \\
\hline
$\widetilde{F}^{(b)}_{B\to p_L}(0)$ 
& $[0.5,\,1.1]$ & $[0.28,\,0.70]$  
& $[0.07,\,0.10]$ & $[-0.9,\,-0.4]$ \\
\hline
\end{tabular}
\caption{Intervals of the $B \to p$ transition form factors at $q^2 = 0$ (in units of $10^{-2}$ GeV$^2$) for models $(d)$ and $(b)$, obtained from LCSRs with $B$-meson DAs (denoted as LCSR-$B$), are compared with the results from LCSRs with 
nucleon DAs (LCSR-$N$). Here the notation such as (tw$2$--tw$5$) indicates that contributions from twist-$2$ up to twist-$5$ are included.}
\label{tab:compar}
\end{table}

\begin{table}[t]
\centering
\begin{tabular}{|c|c|c||c|c|}
\hline
\multirow{2}{*}{Form factor} & \multicolumn{2}{c||}{Uncertainty from $\lambda_B$} & \multicolumn{2}{c|}{Uncertainty from $\lambda_p$} \\
\cline{2-5}
& (tw2-tw3) & (tw2-tw5) & {(tw2-tw3)} & (tw2-tw5)  \\
\hline
\hline
$F^{(d)}_{B\to p_R}(0)$ 
& {$[1.6,\,3.5]$} & {$[0.5,\,2.0]$} 
& {$[1.7,\,3.4]$} & {$[0.8,\,1.6]$} \\
\hline
{$\widetilde{F}^{(d)}_{B\to p_L}(0)$} 
& {$[0.29,\,0.39]$} & {$[0.16,\,0.29]$} 
& {$[0.25,\,0.49]$} & {$[0.16,\,0.33]$} \\
\hline
\hline
{$F^{(b)}_{B\to p_R}(0)$} 
& {$[3.2,\,6.9]$} & {$[1.0,\,4.1]$} 
& {$[3.3,\,6.7]$} & {$[1.6,\,3.1]$} \\
\hline
{$\widetilde{F}^{(b)}_{B\to p_L}(0)$} 
& {$[0.59,\,0.77]$} & {$[0.32,\,0.59]$}  
& {$[0.5,\,1.0]$} & {$[0.33,\,0.65]$} \\
\hline
\end{tabular}
\caption{The same form factors from LCSRs with $B$-meson DAs
as in~\cref{tab:compar}. Shown are the intervals stemming from  separate variations of the parameter $\lambda_B$ or $\lambda_p$ with all other input parameters fixed at their central values.}
\label{tab:lambdaBpdev}
\end{table}

The resulting uncertainties are rather sizeable. The dominant source
of the error budget originates from the broad allowed intervals of
the two most important nonperturbative input parameters: the inverse
moment $\lambda_B$ of the leading-twist $B$-meson DA and the proton
decay constant $\lambda_p$. To quantify their impact, we separately
vary $\lambda_B$ and $\lambda_p$ while keeping all remaining input
parameters fixed at their central values. The resulting variations of
the form factors at $q^2=0$ are collected in
\cref{tab:lambdaBpdev}. We find that the uncertainties induced by
$\lambda_B$ and $\lambda_p$ indeed account for a substantial fraction
of the total intervals quoted in \cref{tab:compar}. At the same time, both form factors in
\cref{eq:lcsrfin1,eq:lcsrfin4} depend on the proton decay constant
through the same overall normalisation factor $1/\lambda_p$.
Consequently, their ratio is independent of this parameter.

We also observe that the inclusion of the higher-twist contributions
reduces the overall uncertainty of the final predictions. This effect
originates from partial cancellations between the leading-twist and
higher-twist terms, which simultaneously decrease the central values
of the form factors and their sensitivity to the input parameters, in
agreement with the behaviour illustrated in
\cref{fig:DA_compar}.

Finally, it is important to stress that the quoted uncertainties
should not be interpreted statistically in the Gaussian sense. In
particular, both $\lambda_B$ and $\lambda_p$ are themselves extracted
from QCD sum rules, where the quoted intervals reflect systematic
variations of auxiliary sum-rule parameters rather than probability
distributions. For this reason, we prefer to present our final LCSR
predictions in the form of intervals instead of central values with
Gaussian errors.

In \cref{tab:compar} we also quote previous predictions for the
$B\to p$ form factors obtained from LCSRs based on nucleon DAs,
denoted by LCSR-$N$. The original analysis of
\cite{Khodjamirian:2022vta}, performed at twist-three accuracy, was
later extended in \cite{Boushmelev:2023huu} to include contributions
up to twist six. Comparing the LCSR-$B$ and LCSR-$N$ approaches, one immediately
notices that the simple relations between the form factors in models
$(d)$ and $(b)$ derived in \cref{eq:formfactequal} are absent in the
LCSR-$N$ framework at any twist order. In addition, within the
LCSR-$N$ approach the form factors in model $(b)$ acquire opposite
signs relative to those in model $(d)$.

At the same time, it should be emphasized that the notion of twist
for nucleon DAs differs fundamentally from the twist classification
used for $B$-meson DAs. Therefore, a direct term-by-term comparison
between twist contributions in the two approaches is not meaningful.
Nevertheless, the overall hierarchy of the twist expansion remains an
important criterion for the reliability of the OPE. In the present
LCSR-$B$ analysis this hierarchy is reasonably well respected, as
illustrated in \cref{fig:DA_compar}. In contrast, the LCSR-$N$
results of \cite{Boushmelev:2023huu} exhibit very large higher-twist
effects.

To investigate this point in more detail, we independently evaluated
the sum-rule expressions of \cite{Boushmelev:2023huu} in the
spacelike region using the same input parameters adopted there. In
particular, focusing on model $(b)$ at
$q^2=-4~\mathrm{GeV}^2$, where the twist expansion is expected to be
reliable, we find for the form factor
$F^{(b)}_{B\to p_R}$ that the twist-four contribution exceeds the
twist-three term by roughly a factor of four, while the twist-five
contribution amounts to about $60\%$ of the twist-three one. For the
second form factor,
$\widetilde F^{(b)}_{B\to p_L}$, the enhancement is even more
pronounced, with the twist-four contribution exceeding the
twist-three term by approximately one order of magnitude. This
behaviour explains the substantial difference between the twist-three
and twist-six predictions quoted in the first two columns of
\cref{tab:compar}, especially for model $(b)$.

For model $(d)$ the higher-twist effects in the LCSR-$N$ approach are
considerably less dramatic, and the agreement with our LCSR-$B$
results is correspondingly better. Moreover, once the rather large
uncertainties of both methods are taken into account, the predicted
intervals still overlap marginally. Nevertheless, the convergence
pattern of the OPE in the LCSR-$N$ framework, particularly for model
$(b)$, appears questionable and deserves further investigation.

The LCSR predictions derived above are valid only in the spacelike
region of momentum transfer $q^2 <0$, where the OPE is
expected to converge reliably. In contrast, the physical
decay region for the process $B\to p\psi$ corresponds to positive
timelike momentum transfer,
$0\le q^2 \le (m_B-m_p)^2.$
Therefore, in order to obtain phenomenological predictions for the
physical decay amplitudes and widths, the form factors calculated from
LCSRs in the spacelike domain must be extrapolated analytically to
the physical timelike region.

To extrapolate the LCSR predictions to the physical timelike region,
we employ the standard $z$-expansion method widely used for heavy-to-light
form factors. We introduce the conformal variable
\begin{equation}
z(q^2,t_0)=
\frac{
\sqrt{t_+-q^2}-\sqrt{t_+-t_0}
}{
\sqrt{t_+-q^2}+\sqrt{t_+-t_0}
},
\label{eq:zpar}
\end{equation}
with
$t_+=(m_{\Lambda_b}+2m_\pi)^2.$
For simplicity, we choose the second free parameter of the mapping to
be $t_0=0$, such that the point $q^2=0$ corresponds to $z=0$.

At this stage it is useful to recall that the correlation function in \cref{eq:corr}
depends on two independent kinematic variables, $P^2$ and $q^2$,
which correspond to two different hadronic channels. The variable
$P^2$ is associated with the baryonic channel interpolated by the
Ioffe current and is therefore saturated by hadronic states carrying
the proton quantum numbers. This is the channel in which the
dispersion relation and Borel transformation were constructed in the
derivation of the LCSRs. In contrast, the variable $q^2$ corresponds to the channel generated
by the local three-quark operators
$\overline{\mathcal O}_{(X)}$ in \cref{eq:Od,eq:Ob}. Consequently, its analytic structure
is determined by hadronic states carrying the same quantum numbers as
these operators. The lowest singularity in this channel is given by the $\Lambda_b$
pole, whereas the continuum contribution starts from the
$\Lambda_b\pi\pi$ threshold and receives additional contributions
from higher excited and multiparticle states carrying the same
quantum numbers. This is precisely the reason for choosing
$t_+=(m_{\Lambda_b}+2m_\pi)^2$
in the conformal mapping variable defined in \cref{eq:zpar}.\footnote{
In \cite{Khodjamirian:2022vta,Boushmelev:2023huu}, a similar
$z$-expansion was employed assuming instead the threshold
$t_+=(m_B+m_p)^2$. Here we use the physically lowest threshold
corresponding to the $\Lambda_b\pi\pi$ state. We verified that using
the higher threshold adopted in those works changes the extrapolated
form factors by less than $1\%$.
}

Under this transformation, the spacelike region where the LCSR is
evaluated and the physical decay region are mapped to
\begin{align}
-8.0~\mathrm{GeV}^2<q^2<0
&\quad\to\quad
0.083\ge z\ge0,
\label{eq:q2toz1}
\\
m_p^2<q^2<(m_B-m_p)^2
&\quad\to\quad
-0.067>z>-0.20.
\label{eq:q2toz2}
\end{align}

Choosing the lowest physical threshold in the definition of $t_+$ ensures
that only the $\Lambda_b$ pole remains inside the unit disk
$|z|<1$ to which the full complex $q^2$ plane is mapped.
To remove this pole contribution, we multiply the form factors by the
factor
$(1-{q^2}/{m_{\Lambda_b}^2}).$
The resulting functions become analytic inside the unit disk and can
therefore be expanded around $z=0$. Since both the spacelike region
used in the LCSR and the physical decay region correspond to rather
small values of $|z|$, the expansion converges rapidly and can be
truncated after the first few terms.

More specifically, we employ the standard Bourrely--Caprini--Lellouch (BCL) parametrisation
\cite{Bourrely:2008za} truncated at ${\cal O}(z^2)$:
\begin{align}
F^{(d)}_{B\to p_R}(q^2)
&=
\frac{
F^{(d)}_{B\to p_R}(0)
}{
1-q^2/m_{\Lambda_b}^2
}
\Bigg\{
1
+
\beta^{(d)}_{B\to p_R}
\Bigg[
z(q^2)-z(0)
+
\frac12
\Big(
z(q^2)^2-z(0)^2
\Big)
\Bigg]
\Bigg\}.
\label{eq:BCL}
\end{align}
This parametrisation is specified by two quantities: the form-factor
normalisation at $q^2=0$ and the slope parameter
$\beta^{(d)}_{B\to p_R}$. An analogous expression with the slope
parameter $\widetilde\beta^{(d)}_{B\to p_L}$ is used for the form
factor $\widetilde F^{(d)}_{B\to p_L}(q^2)$.

For the model-$(b)$ form factors we employ the relations in
\cref{eq:formfactequal}, which imply
\begin{equation}
F^{(b)}_{B\to p_R}(0)
=
2F^{(d)}_{B\to p_R}(0),
\qquad
\widetilde F^{(b)}_{B\to p_L}(0)
=
2\widetilde F^{(d)}_{B\to p_L}(0),
\label{eq:bF0}
\end{equation}
together with equal slope parameters:
\begin{equation}
\beta^{(b)}_{B\to p_R}
=
\beta^{(d)}_{B\to p_R},
\qquad
\widetilde\beta^{(b)}_{B\to p_L}
=
\widetilde\beta^{(d)}_{B\to p_L}.
\label{eq:bbeta0}
\end{equation}

We then fit the parametrisation in \cref{eq:BCL} to the
LCSR results obtained for the form factors in the spacelike region
$q^2\le0$. We first employ a simplified fitting procedure in which
the normalisation of the form factor at $q^2=0$ is fixed to the LCSR
prediction, while only the slope parameter is fitted. More
specifically, for the form factor
$F^{(d)}_{B\to p_R}(q^2)$ we determine the parameter
$\beta^{(d)}_{B\to p_R}$ using the LCSR results evaluated at eight
equidistant points in the interval defined in
\cref{eq:q2toz1}. The same procedure is repeated for the form factor
$\widetilde F^{(d)}_{B\to p_L}(q^2)$. To estimate the uncertainty of the fitted slope parameters, the input
value of the form factor at $q^2=0$ is varied within the parametric
uncertainty obtained from the LCSR analysis. The induced variation is
then propagated to the fitted slope parameters, yielding asymmetric
uncertainties for
$\beta^{(d)}_{B\to p_R}$ and
$\widetilde\beta^{(d)}_{B\to p_L}$. The resulting parameters of the
$z$-expansion are collected in
\cref{tab:FitParameters}. For the model-$(b)$ form factors, the
relations in \cref{eq:bF0,eq:bbeta0} are used.
\begin{table}[ht]
\centering
\renewcommand{\arraystretch}{1.8}
\begin{tabular}{|c|c|c|c|}
\hline
$F^{(d)}_{B\to p_R}(0)$ & $1.0^{+1.2}_{-0.6}$ & $\beta^{(d)}_{B\to p_R}$ & $0.99^{+3.4}_{-1.4}$ \\
\hline
$\widetilde{F}^{(d)}_{B\to p_L}(0)$ & $0.22^{+0.13}_{-0.08}$ & $\widetilde{\beta}^{(d)}_{B\to p_L}$ & $-1.78^{+0.76}_{-0.17}$ \\
\hline
\hline\hline
$F^{(b)}_{B\to p_R}(0)$ & $2.0^{+2.4}_{-1.2}$ & $\beta^{(b)}_{B\to p_R}$ & $0.99^{+3.4}_{-1.4}$ \\
\hline
$\widetilde{F}^{(b)}_{B\to p_L}(0)$ & $0.44^{+0.26}_{-0.16}$ & $\widetilde{\beta}^{(b)}_{B\to p_L}$ & $-1.78^{+0.76}_{-0.17}$ \\
\hline
\end{tabular}
\caption{Parameters of the $z$-expansion in \cref{eq:BCL} obtained from 
the fit of LCSRs. Presented are the
$B\to p$ form factors at $q^2=0$ (in units of $10^{-2}~ \mbox{GeV}^2$) and the dimensionless slope parameters.}
\label{tab:FitParameters}
\end{table}

The advantage of the fixed-normalisation fit is its simplicity and
its direct correspondence to the uncertainty intervals obtained from
the LCSR analysis. However, this procedure does not
explicitly account for possible correlations between the normalisation
and slope parameters entering the $z$-expansion. To estimate the size
of such effects, we also carried out an alternative fit in which the
normalisation $F^{(d)}_{B\to p_R}(0)$ is allowed to vary together
with the slope parameter. In this case, the covariance matrix derived
from the Hessian automatically incorporates the correlations between
both quantities. We find that the resulting predictions for the
branching fractions remain compatible, within uncertainties, with
those obtained in the fixed-normalisation approach. This indicates
that the effect of parameter correlations is numerically insignificant.

\begin{figure}[t]
    \centering
    \includegraphics[width=0.48\textwidth]{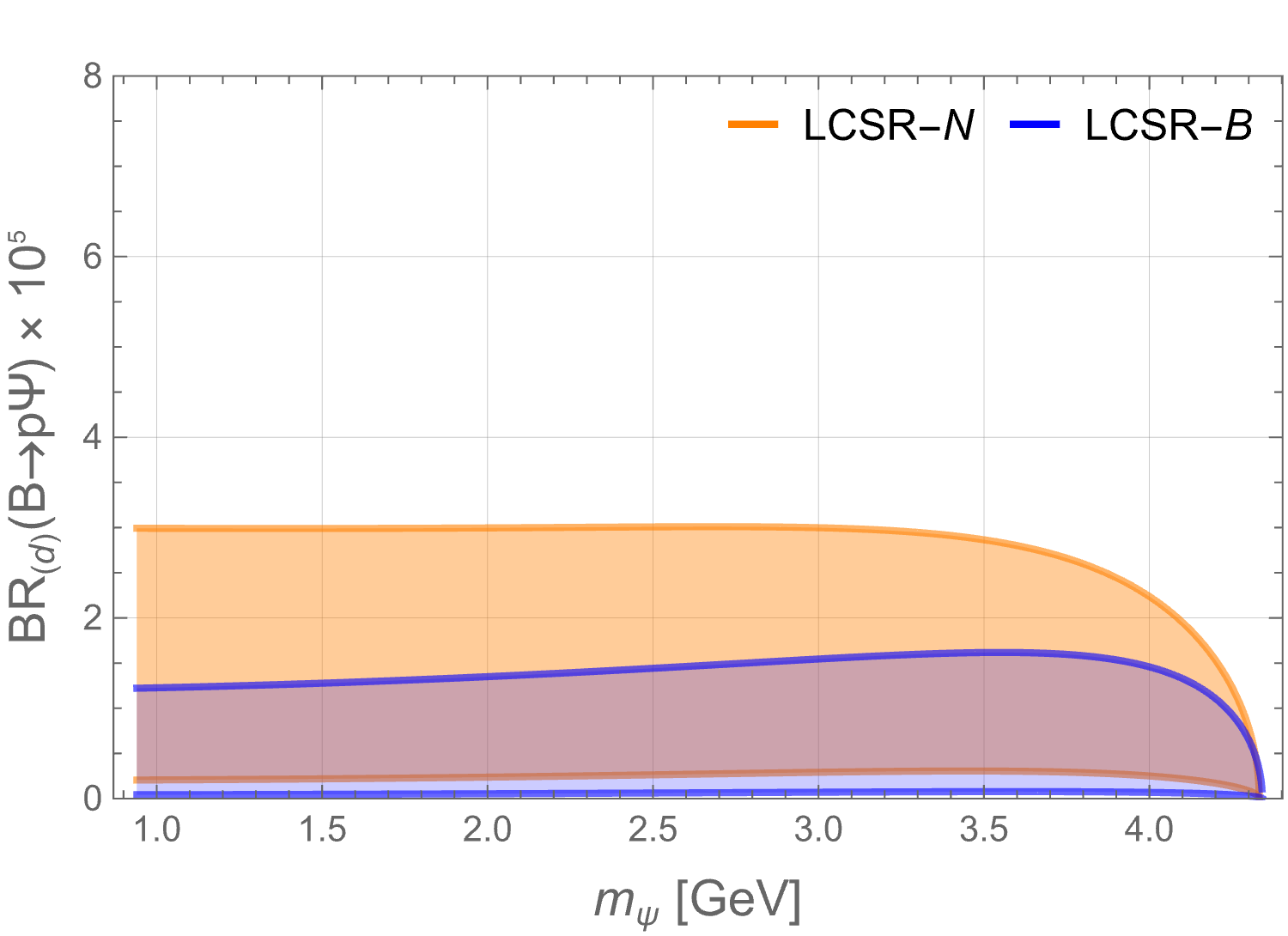}
    \hfill
    \includegraphics[width=0.48\textwidth]{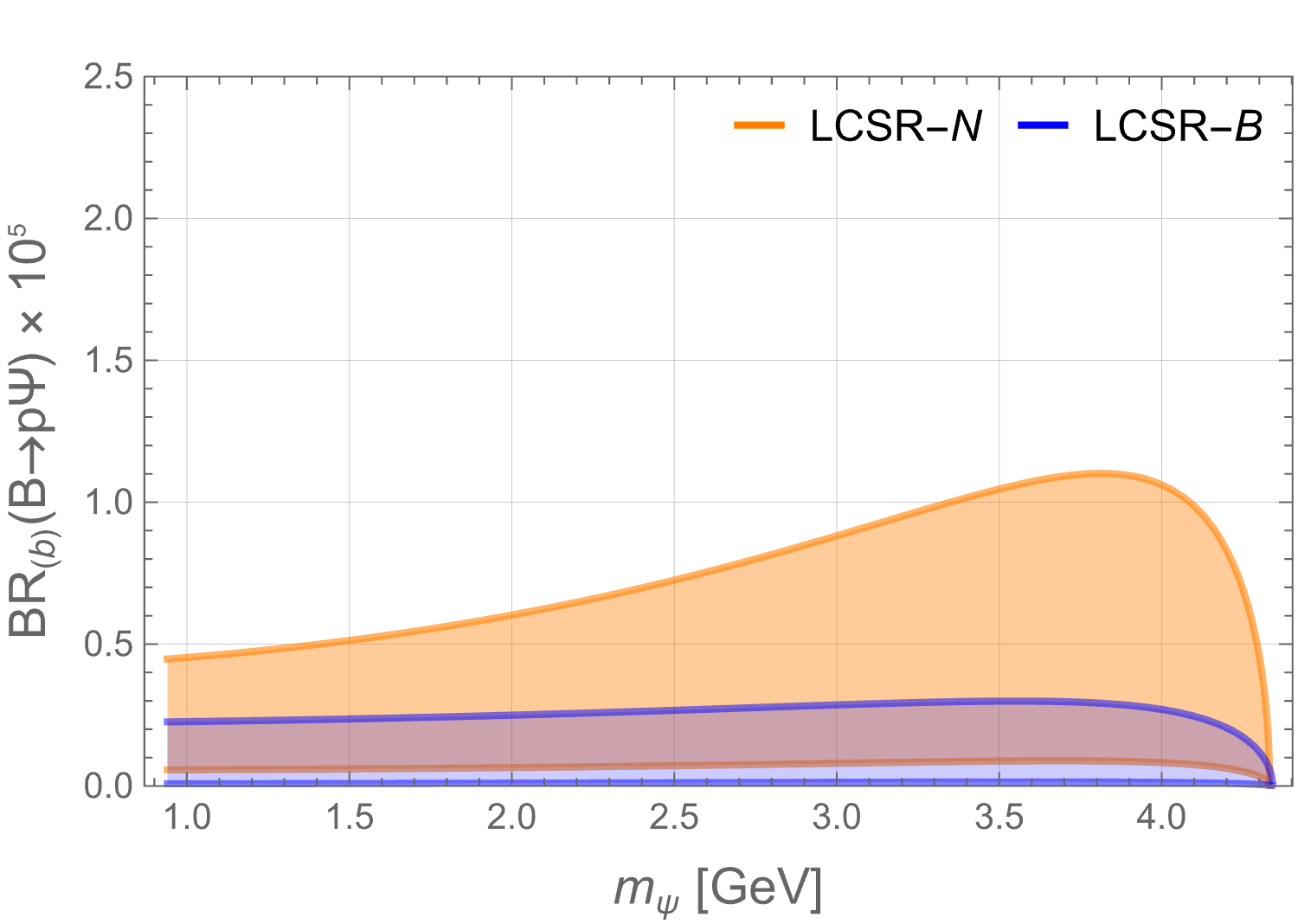}
    \caption{Branching fraction predicted for the $B^+ \to p \Psi$
    decay in model $(d)$ (left panel) and model $(b)$ (right panel).
    The blue bands represent the results of this work obtained with
    the LCSR-$B$ method. For comparison, the orange bands show the
    results obtained using the method with nucleon DAs (LCSR-$N$).}
    \label{fig:BR_comparisons}
\end{figure}

Having extrapolated the form factors to the physical timelike region,
we are now in a position to predict the branching fraction for the
exclusive decay $B^+\to p\Psi$. Using the decomposition in
\cref{eq:matrix_decomp}, the decay amplitude in \cref{eq:amp} can be written as
\begin{align}
{\cal A}^{(X)}(B^+\to p\Psi)
&=
G_{(X)}
\Bigg[
F^{(X)}_{B\to p_R}(q^2)\,
\bar u_{pR}(P)
+
F^{(X)}_{B\to p_L}(q^2)\,
\bar u_{pL}(P)
\nonumber\\
&\qquad
+
\widetilde F^{(X)}_{B\to p_R}(q^2)\,
\bar u_{pR}(P)\frac{\slashed q}{m_p}
+
\widetilde F^{(X)}_{B\to p_L}(q^2)\,
\bar u_{pL}(P)\frac{\slashed q}{m_p}
\Bigg]
u_\Psi^c(q),
\label{eq:AmpExclusive}
\end{align}
evaluated at $q^2=m_\Psi^2$.
The total decay width is obtained in the standard way from the
spin-summed squared amplitude integrated over the two-body phase
space:
\begin{equation}
\Gamma(B^+\to p\Psi)_{(X)}
=
\frac{1}{2m_B}
\int d\Phi_2\,
\overline{|{\cal A}^{(X)}|^2},
\label{eq:GammaDef}
\end{equation}
where
\begin{equation}
d\Phi_2
=
(2\pi)^4
\delta^{(4)}(p_B-p_p-p_\Psi)
\frac{d^3p_p}{(2\pi)^3\,2E_p}
\frac{d^3p_\Psi}{(2\pi)^3\,2E_\Psi}
\end{equation}
denotes the Lorentz-invariant two-body phase-space measure. In the
present case, the momenta are identified as
$p_B=P+q$, $p_p=P$, and $p_\Psi=q$.
For a two-body decay, the phase-space integration can be carried out
analytically and yields the familiar kinematic factor
$\lambda^{1/2}(m_B^2,m_p^2,m_\Psi^2)$,
where $\lambda(x,y,z)$ is the Källén function defined in
\cref{eq:kaellen}. Retaining only the nonvanishing form factors, the decay width for real-valued form factors reads
\begin{align}
\Gamma_{(X)}(B^+\to p\psi)
&=
\frac{|G_{(X)}|^2\,
\lambda^{1/2}(m_B^2,m_p^2,m_\psi^2)}
{16\pi m_B^3}
\Bigg\{
\Bigg[
\Big(F^{(X)}_{B\to p_R}(m_\psi^2)\Big)^2
\nonumber\\
&\qquad
+
\frac{m_\psi^2}{m_p^2}
\Big(\widetilde F^{(X)}_{B\to p_L}(m_\psi^2)\Big)^2
\Bigg]
\big(m_B^2-m_p^2-m_\psi^2\big)
\nonumber\\
&\qquad
+
2m_\psi^2\,
F^{(X)}_{B\to p_R}(m_\psi^2)\,
\widetilde F^{(X)}_{B\to p_L}(m_\psi^2)
\Bigg\}.
\label{eq:widthb}
\end{align}
The branching fraction is then obtained from
\begin{equation}
{\text{BR}}_{(X)}(B^+\to p\Psi)
=
\tau_{B^+}\,
\Gamma_{(X)}(B^+\to p\Psi).
\label{eq:BRExclusive}
\end{equation}
The resulting branching fractions are shown in
\cref{fig:BR_comparisons} as functions of the dark-fermion mass
$m_\Psi$.

The uncertainty bands are obtained by varying independently the
$z$-expansion parameters entering the decay rate in
\cref{eq:widthb}. For each form factor, both the normalisation and
the slope parameter are varied over the ranges given in
\cref{tab:FitParameters}. We use three representative values for each
parameter, corresponding to its central, maximal, and minimal values.
For every fixed value of $m_\Psi$, the branching fraction is then
evaluated over the full set of parameter combinations, and the
resulting uncertainty band is defined by the global minimum and
maximum obtained from this scan. This procedure provides a conservative estimate of the theoretical uncertainty, since it explores the full allowed parameter ranges
without assuming any particular statistical distribution or
correlation pattern among the fit parameters. Consequently, the
resulting bands are expected to encompass the maximal variation of
the branching-fraction predictions compatible with the adopted LCSR
input.

Moreover, in \cref{fig:BR_comparisons} we compare our predictions with the
results obtained in the LCSR-$N$ approach of
\cite{Boushmelev:2023huu}. In both approaches the
effective couplings in \cref{eq:Heff2} are fixed to their 
maximal allowed values (see \cref{Constants1/3}),
\[
|G_{(d)}|^2
=
1.0\times10^{-13}\,\mathrm{GeV}^{-4},
\qquad
|G_{(b)}|^2
=
4.7\times10^{-15}\,\mathrm{GeV}^{-4},
\]
which correspond to the upper bounds inferred from the ATLAS and CMS
searches for heavy coloured particles, as discussed in
\cite{Alonso-Alvarez:2021qfd}.\footnote{
In \cite{Khodjamirian:2022vta,Boushmelev:2023huu}, equal values
$|G_{(d)}|^2=|G_{(b)}|^2=10^{-13}\,\mathrm{GeV}^{-4}$ were used.
} We find that the predictions obtained from the two LCSR methods are
compatible within the rather large theoretical uncertainties.
Nevertheless, throughout the full kinematic range our LCSR-$B$
results tend to lie below the corresponding LCSR-$N$ predictions.
Moreover, the uncertainty bands obtained in the present analysis are
noticeably narrower.

Now that we have obtained the final predictions for the form factors
and the corresponding branching fractions, it is worthwhile to
perform a more detailed comparison between the two complementary
approaches based on $B$-meson DAs (LCSR-$B$) and nucleon DAs
(LCSR-$N$). Such a comparison is particularly instructive since both
methods rely on different realisations of the light-cone expansion
and organize the nonperturbative dynamics in substantially different
ways.
\begin{figure}[t]
    \centering
    \includegraphics[width=0.475\textwidth]{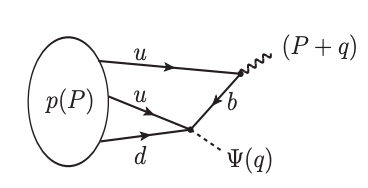}
\caption{Representative diagram entering the LCSR-$N$ approach based
on proton DAs. In this case only the $b$-quark field is contracted
perturbatively, while the three light-quark fields are absorbed into
the proton DAs.}
\label{fig:DA_proton}
\end{figure}

\begin{itemize}

\item
{\bf Structure of the correlation function and OPE diagrams:}
In the LCSR-$B$ framework, the nonperturbative $B$-meson matrix element
contains only two light-cone partons, such that two quark fields in
the correlation function are contracted perturbatively, generating a
closed quark loop in the OPE diagrams (see \cref{fig:LCSRdiags}). In
contrast, the LCSR-$N$ approach is based on nucleon DAs involving
three quark fields, so that only the $b$-quark propagator is
contracted perturbatively (see \cref{fig:DA_proton}). As a result, the analytical structure of
the resulting form factors becomes more complicated in
the LCSR-$N$ approach, in particular when extracting the imaginary
parts required for the dispersion relations.

\item
{\bf Structure of the light-cone expansion:}
Both approaches are based on a twist expansion on the light cone, but
the organisation and implementation of higher-twist effects differ
substantially. In the LCSR-$B$ framework, higher-twist two-particle
$B$-meson DAs can be incorporated rather directly through the
vacuum-to-$B$ matrix elements, as done in the present work up to
twist five. In contrast, extending the LCSR-$N$ approach to higher
twists is technically more involved, since it requires the full set
of higher-twist nucleon DAs together with their corresponding
contributions to the correlation function. This leads to a much more
complicated analytical structure of the resulting sum rules.\\
At the same time, the concept of twist for $B$-meson DAs is not
identical to the twist classification employed for nucleon DAs.
Therefore, a direct term-by-term comparison of individual twist
contributions between the two approaches is not meaningful. 

\item
{\bf Interpolating currents and hadronic representation:}
An important conceptual difference between the two approaches concerns
the choice of interpolating current and the resulting structure of the
hadronic representation. In the present LCSR-$B$ framework, the proton
is interpolated by a baryonic current, which leads to a considerably
more involved hadronic spectrum in the proton channel. In particular,
besides the proton pole itself, one encounters negative-parity baryon
states and, in the intermediate region, the broad Roper resonance,
which cannot be represented by a simple narrow-pole contribution.
\\
A fully realistic treatment of this hadronic structure would require a
multi-resonance parametrisation together with additional hadronic
inputs, such as the decay constants of the
excited baryons. In particular, no reliable determination of the
Roper-resonance decay constant is currently available. For this
reason, we adopted a simplified hadronic ansatz in which only the proton
pole is separated explicitly, while all excited and continuum states
are absorbed into the duality continuum. As a consequence, the
resulting sum rules exhibit an enhanced sensitivity to the continuum
contribution and rely more strongly on the quark--hadron duality
approximation.
\\
In contrast, the LCSR-$N$ approach employs an interpolating current
for the $B$ meson. The hadronic structure in this channel is much
simpler, since no nearby opposite-parity resonance analogous to the
nucleon case appears below the continuum threshold. Consequently, the
ground-state pole remains parametrically more dominant and the
continuum contribution is relatively more suppressed compared to the
LCSR-$B$ framework.

\item
{\bf Three-particle contributions:}
In the LCSR-$B$ framework, the contributions of the three-particle
$B$-meson DAs vanish identically for both models $(d)$ and $(b)$.
For model $(d)$, the cancellation occurs because the soft-gluon
emission from the $d$-quark propagator is equal in magnitude and
opposite in sign to the corresponding $u$-quark emission contribution,
as shown in \cref{eq:3partrel1}. In contrast, for model $(b)$ the
three-particle DA contributions vanish term by term already at the
level of the individual diagrams, see
\cref{eq:3partrel2}. These cancellations originate from the interplay
between the Dirac structure of the effective operators and the
specific linear combination defining the Ioffe current. In addition,
the massless limit adopted for the light quarks also plays a role in
these cancellations; retaining finite light-quark masses would
generate nonvanishing three-particle contributions proportional to the
light-quark masses. No analogous cancellation mechanism is present in
the LCSR-$N$ approach.

\item
{\bf Relations between models $(d)$ and $(b)$:}
A characteristic feature of the present LCSR-$B$ calculation is the
simple relation in \cref{eq:formfactequal}.
These relations hold for the complete two-particle DA contributions. The three-particle DA contributions, however, do not
satisfy such relations individually. In the LCSR-$N$ approach, no analogous
relations between models $(d)$ and $(b)$ emerge at any twist order,
and the model-$(b)$ form factors even acquire opposite signs relative
to those in model $(d)$ (see \cref{tab:compar}).
\item
{\bf Numerical size of the form factors:}
Although the predictions of both methods overlap within uncertainties,
the form factors obtained in the present LCSR-$B$ analysis are
systematically smaller in magnitude than the corresponding
LCSR-$N$ results. This difference propagates directly to the predicted
branching fractions.

\item
{\bf Uncertainty pattern:}
The dominant uncertainties in the present work originate from the
inverse moment $\lambda_B$ and the proton decay constant
$\lambda_p$. Even with these sizeable uncertainties, the resulting
bands remain narrower than those obtained in the LCSR-$N$ approach.
Moreover, in the present analysis the inclusion of higher-twist
contributions tends to stabilize the final predictions due to partial
cancellations between different twist components.

\item
{\bf Hierarchy of twist contributions and convergence of the OPE:}
The overall hierarchy of the twist expansion provides an important
criterion for assessing the convergence and reliability of the OPE in
both sum-rule approaches. In the present LCSR-$B$ analysis, the
lowest-twist contributions dominate the form factors, while the
higher-twist corrections remain systematically suppressed throughout
the spacelike region considered. This hierarchical pattern, together
with the observed stability under variation of the Borel parameter,
indicates a reasonably controlled OPE. 
\\
In contrast, the situation is substantially different in the
LCSR-$N$ framework, especially for model $(b)$. Using the sum-rule
expressions and input parameters of Ref.~\cite{Boushmelev:2023huu}, we
evaluated the separate twist contributions at
$q^2=-4~\mathrm{GeV}^2$. For the form factor
$F^{(b)}_{B\to p_R}$, the twist-four contribution is approximately
four times larger than the twist-three term, while the twist-five
correction still amounts to roughly $60\%$ of the twist-three
contribution. An even stronger enhancement is observed for
$\widetilde F^{(b)}_{B\to p_L}$, where the twist-four contribution
exceeds the twist-three one by nearly an order of magnitude.
\\
Tracing the origin of this enhancement in more detail, one finds that
the dominant effect arises from the twist-four
${\cal{T}}_{2,4}$-contributions in the decomposition of the nucleon DAs~\cite{Boushmelev:2023huu}. These
terms introduce the parameter $f_2^d$, directly related to the
next-to-conformal coefficient $\xi_{10}$, whose numerical value is
known rather poorly and is typically assigned a conservative
uncertainty of about $50\%$~\cite{Braun:2006hz}. As a result, the twist-four correction
becomes anomalously large in model $(b)$ and significantly enhances
the predicted branching fraction. In contrast, for model $(d)$ these
specific contributions cancel identically in the course of the
calculation, explaining why the higher-twist effects remain much more
moderate in that case.
\\
Overall, this pattern implies that higher-twist terms dominate the
final result in the LCSR-$N$ framework for model $(b)$, raising
questions about the convergence of the corresponding OPE and
indicating that the higher-twist structure deserves further
investigation.

\item
{\bf Overall consistency:}
Despite these conceptual and numerical differences, both approaches
lead to branching-fraction predictions that remain compatible within
their relatively large theoretical uncertainties. This agreement
provides a nontrivial cross-check of the general magnitude of the
exclusive $B\to p\Psi$ transition amplitudes.
\end{itemize}
\section{Ratio of Exclusive to Inclusive Decay Widths}
\label{sect:excl-incl}
The branching-fraction predictions shown in \cref{fig:BR_comparisons} were obtained by fixing
the effective couplings $G_{(d,b)}$ to their maximal values allowed
by the current LHC constraints from ATLAS and CMS searches for heavy
coloured particles. On the other hand, the inclusive contribution generated by the
corresponding quark-level transition
$b\to\psi ud$ was calculated within the HQE framework in
\cref{sec:MesoIncl}. Having both the exclusive and inclusive decay
rates at hand, we can now relate them within the Mesogenesis
framework and derive a lower bound on the exclusive branching
fraction.

Using the condition on the inclusive decay rate in
\cref{eq:LowerBoundBr}, one can obtain a corresponding lower bound on
the exclusive decay by considering the ratio of the exclusive and
inclusive decay widths. Multiplying this ratio by the lower limit on
the inclusive branching fraction directly yields a lower bound for
the exclusive branching fraction. To this end, we introduce the exclusive-to-inclusive ratio
\begin{equation}
R_{[\mathrm{excl/incl}]}(B^+\to p\Psi)
\equiv
\frac{
\Gamma(B^+\to p\Psi)
}{
\Gamma(B^+\to X_N\Psi)
}
\times10^{-4}
=
\frac{
\mathrm{BR}(B^+\to p\Psi)
}{
\mathrm{BR}(B^+\to X_N\Psi)
}
\times10^{-4}.
\label{eq:Rdef}
\end{equation}
Here $X_N$ denotes all possible kinematically accessible baryonic
hadronic states produced after the underlying quark-level transition
$\bar b\to u d \Psi$. Together with the spectator $u$-quark of the
$B^+$ meson, the produced quarks form a visible $uud$ system carrying
the quantum numbers of a proton. The lowest such state is the proton
itself, corresponding to the exclusive channel $B^+\to p\Psi$,
whereas the inclusive state $X_N$ also contains excited nucleon and
baryonic resonance states, as well as multiparticle hadronic
configurations such as $N^\ast$, $\Delta$, $N^\ast\pi$, $p\pi\pi$,
and similar states, whenever kinematically allowed.

For the inclusive decay width entering the ratio
\cref{eq:Rdef}, we use the HQE prediction including all contributions
up to dimension six. More specifically, we include the dimension-three
free-quark contribution, the dimension-five chromomagnetic and
kinetic corrections, and the dimension-six Darwin-term contribution
for the two-quark operators, together with the dimension-six
four-quark operator contributions. The corresponding expressions can be 
found in
\cref{eq:MasterWidth_meso_2q,eq:f0d,eq:fGd,eq:fDd,eq:f0b,eq:fGb,eq:fDb,eq:Meso4qresultsB+}. 

\begin{figure}[t]
    \centering
    \includegraphics[width=0.49\textwidth]{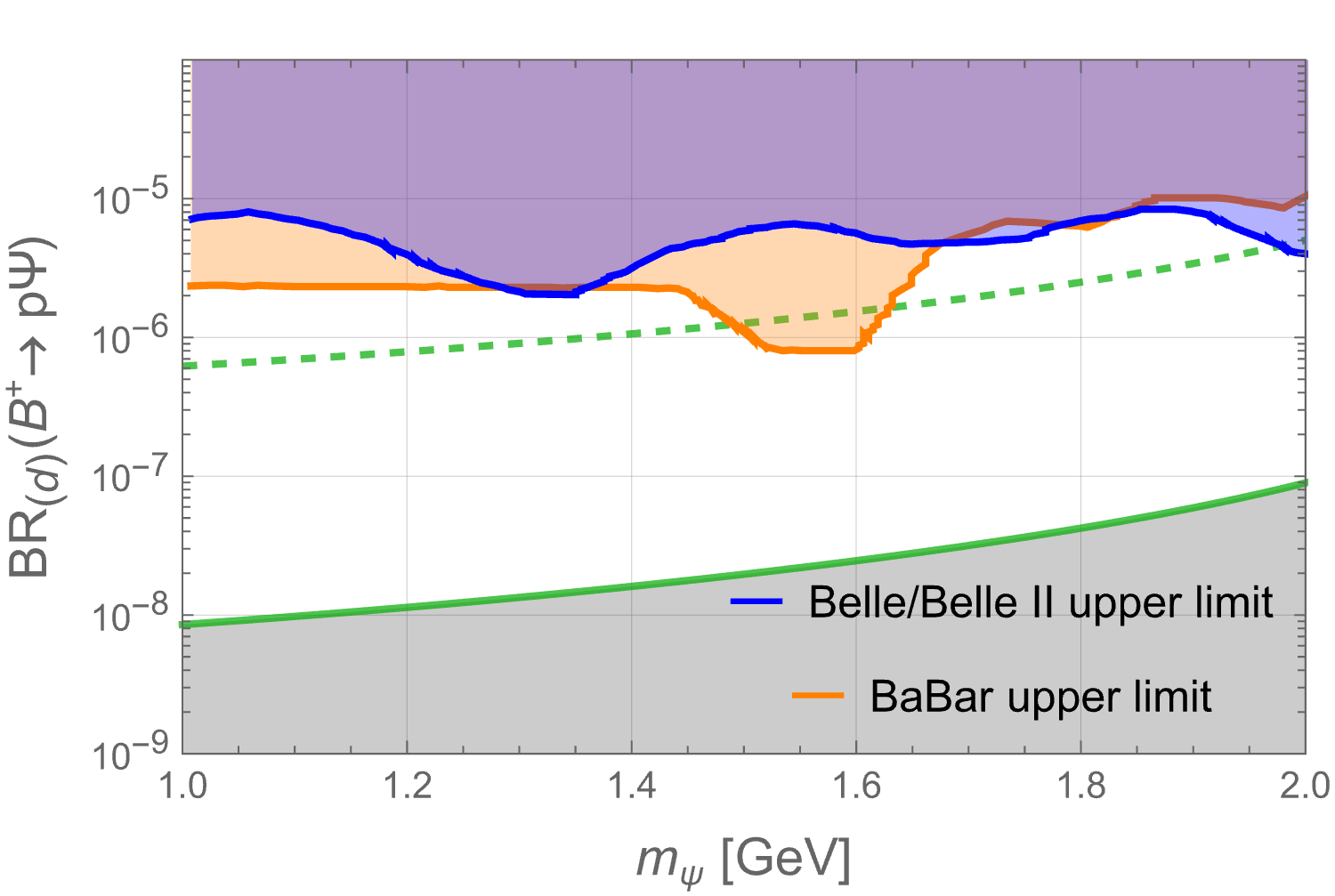}
    \hfill
    \includegraphics[width=0.49\textwidth]{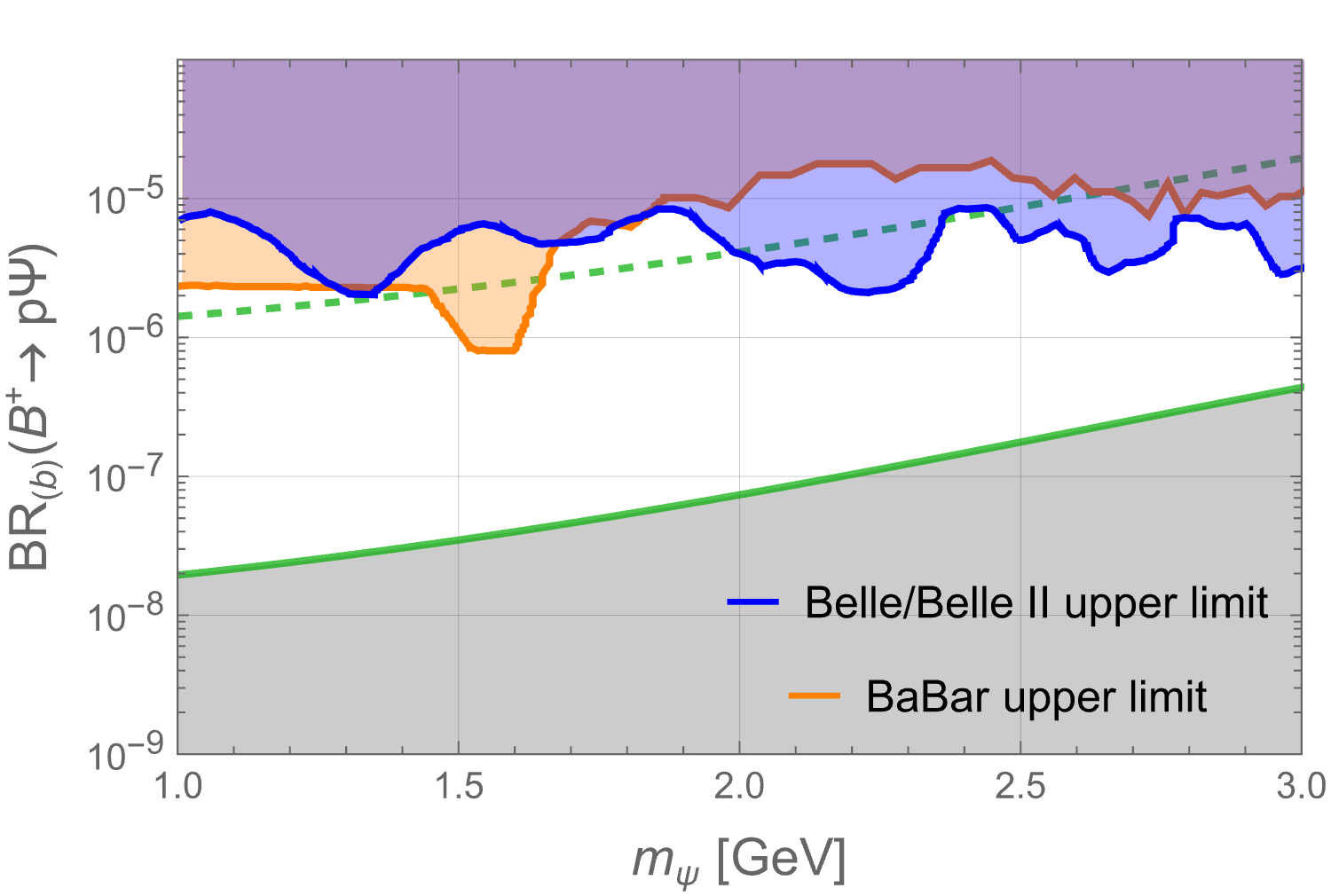}
    
\caption{Branching fraction $\mathrm{BR}(B^+ \to p \Psi)$ as a function of $m_\Psi$ for model $(d)$ (left panel) and model $(b)$ (right panel). The solid green lines present the predicted lower limit in \cref{eq:Rdef}, and the dashed green lines 
indicate the maximal value of this limit within estimated uncertainties; orange and blue shaded regions display  experimental upper limits set by, respectively,
BaBar~\cite{BaBar:2023dtq} and Belle/Belle II~\cite{Belle-II:2026tyb} measurements of 
the $B\to p+invisible$  decays.
The white areas indicate the parameter space allowed by current experimental data. 
}
\label{fig:Rkh}
\end{figure}

The resulting lower limits on the exclusive branching fraction are
shown in \cref{fig:Rkh} for models $(d)$ and $(b)$. The solid green
curves represent the central predictions obtained from the
exclusive-to-inclusive ratio in \cref{eq:Rdef}, while the dashed
green curves indicate the maximal values of the lower limits after
including the estimated theoretical uncertainties. For comparison,
the shaded orange and blue regions display the current experimental
upper bounds from BaBar~\cite{BaBar:2023dtq} and
Belle/Belle~II~\cite{Belle-II:2026tyb} searches for
$B\to p+\text{invisible}$ decays. The remaining white regions
therefore correspond to the parameter space currently allowed by
experiment.
As discussed in \cref{sec:MesoIncl}, the inclusion of subleading
terms in the HQE reveals specific regions of the dark-fermion mass
$m_\psi$ where the OPE description ceases to be reliable. More
precisely, for sufficiently large values of $m_\psi$, the subleading
power corrections become larger than the leading dimension-three
contribution, signaling a breakdown of the convergence of the
expansion. In model $(d)$ this behaviour occurs for
$m_\psi\sim2~\mathrm{GeV}$, whereas in model $(b)$ it appears around
$m_\psi\sim3~\mathrm{GeV}$. For this reason, when evaluating the
exclusive-to-inclusive ratio and the corresponding lower bounds in \cref{fig:Rkh}, we
restrict the plots to the regions below these values of $m_\psi$,
where the HQE remains under theoretical control.

The uncertainty band shown in \cref{fig:Rkh}, enclosed between the
solid and dashed green curves, is obtained following the same
conservative procedure adopted for the exclusive branching fractions
in \cref{fig:BR_comparisons}. More specifically, the parameters entering the
exclusive-to-inclusive ratio are varied independently over their
allowed ranges, and the extremal values obtained from the full scan
are used to define the uncertainty interval.
In addition, we assign an extra $10\%$ uncertainty to the inclusive
width in order to account for residual theoretical effects not
included explicitly in the calculation, such as renormalisation-scale
dependence, scheme ambiguities, and neglected higher-order
perturbative and power corrections within the HQE. Numerically, the total uncertainty is dominated by the form factors
$F^{(d)}_{B\to p_R}(0)$ and
$F^{(b)}_{B\to p_R}(0)$ determined in
\cref{tab:compar}. To remain conservative in deriving
phenomenological constraints, we interpret the lower edge of the
uncertainty band as our final prediction for the lower bound on
$\mathrm{BR}(B^+\to p\Psi)$.
\\\\
\textbf{Important note:}
Shortly after completing our work on
$B$-Mesogenesis, the study~\cite{Hiller:2026osz} appeared, employing
LHC searches for jets accompanied by missing energy to constrain
baryon-number violating operators. A subset of these
operators corresponds to the Mesogenesis scenarios considered in the
present work. Their analysis indicates that the flavour structure
associated with the quark combination $bud$ is strongly constrained by
current data, whereas scenarios involving heavier flavours, such as
$bcs$, remain phenomenologically viable. In the latter case, the
corresponding exclusive decay channels are expected to involve charmed
baryons, including
$B^{0,+}\to \Xi_c^{0,+}\Psi$ and
$B_s\to \Omega_c^0\Psi$.
This observation further highlights the importance of a flavour-flexible
framework for exclusive form-factor calculations. In particular, the
LCSR approach developed in the present work can be generalised to a
broad class of baryonic final states beyond the proton channel
considered here. At the same time, our inclusive HQE analysis already
incorporates the complete set of Mesogenesis operators, independently
of the flavour assignment, making the framework suitable for studying
a wide variety of exclusive-to-inclusive decay patterns in future
investigations.

\chapter{Summary and Outlook}  
\label{Ch:summary}

In this thesis, we have explored several aspects of heavy-hadron lifetimes within and beyond the SM, employing both the HQE and the method of LCSR. The results presented here collectively contribute to a more complete theoretical understanding of inclusive and exclusive $B$-meson decays and provide a model-independent framework for testing new-physics effects in heavy-flavour observables.\\
In the SM part of this work, we extended the theoretical description of
inclusive non-leptonic $B$-hadron decays within the HQE by deriving
previously unknown higher-order contributions. These include the
complete leading-order dimension-seven two-quark Wilson coefficients
for the CKM-favoured decay channels $b\to c\bar u d$ and $b\to c\bar c s$, as well as the previously missing
dimension-five and dimension-six contributions of the SM QCD-penguin
operators. Together, these results provide essential ingredients for
future high-precision predictions of inclusive $B$-hadron decay widths
and lifetime ratios.\\
In the second part of this thesis, we extended the HQE to a
model-independent effective Hamiltonian describing $|\Delta B|=1$ non-leptonic
$b$-hadron decays beyond the SM. We derived the leading-order Wilson
coefficients of the HQE up to and including the dimension-six
contributions, comprising the free-quark term, the chromomagnetic
correction, the two-quark Darwin contribution, and the previously
uncalculated WA topology of the dimension-six four-quark
operators.  The resulting expressions provide a model-independent
framework for studying short-distance new physics in inclusive
$b$-hadron decays and can be readily matched onto a broad class of
ultraviolet completions.\\
Finally, we investigated the phenomenological implications of
\textit{B}-Mesogenesis, in which the origin of the baryon asymmetry of
the Universe is linked to the dark-matter sector through exotic
$b$-hadron decays. We analysed the impact of the new interactions on
inclusive $B$-meson decay widths and lifetime ratios within the HQE,
including power corrections up to the Darwin term. We further studied
the convergence of the HQE in the presence of a massive dark-sector
fermion in the final state, identifying the region of parameter space
where the operator product expansion remains reliable.
\\
Complementing the inclusive analysis, we performed an exclusive calculation of the baryon-number-violating decay $B^+\to p\psi$ using the method of LCSR. By computing the relevant $B\to p$ transition form factors from $B$-meson distribution amplitudes, we predicted the corresponding branching fraction and established a quantitative connection between the inclusive and exclusive descriptions. This allowed us to derive a lower bound on $\mathrm{BR}(B^+\to p\psi)$. 
A comparison with the experimental upper bounds from BaBar and Belle/Belle~II then enabled us to determine the regions of parameter space that are currently allowed by experiment.\\
Several directions for future research naturally follow from the work presented in this thesis. On the theoretical side, an important next step is the computation of next-to-leading-order (NLO) QCD corrections to the Darwin term.
These corrections are crucial for further reducing the theoretical uncertainty in lifetime predictions and for enabling a more precise comparison with experimental data.\\
Another promising direction is the detailed matching of specific BSM scenarios---such as extended Higgs sectors, leptoquarks, or heavy neutral fermions---onto the general operator basis developed here. This would allow for direct phenomenological applications of the formalism and provide quantitative constraints on the corresponding Wilson coefficients from lifetime measurements.\\
Furthermore, a natural next step is the synthesis of these calculations: integrating the new dimension-seven SM corrections with the full set of dimension-six BSM operators to perform a global phenomenological analysis. Such a study would sharpen the SM prediction for lifetimes and lifetime ratios while providing a consistent framework to constrain general New Physics.\\
The LCSR framework can also be extended to a broader class of exclusive channels, including baryonic and multi-body decays, where hadronic uncertainties remain significant. Such analyses would improve our understanding of the transition between inclusive and exclusive regimes and could play an important role in testing quark-hadron duality.\\
Moreover, future experimental progress in flavour physics---at Belle~II,
LHCb Upgrade~II, and future colliders---will provide unprecedented
precision in heavy-hadron lifetime measurements. Confronting these data
with the refined theoretical predictions developed in this thesis will
offer a powerful probe of both SM dynamics and possible new-physics
contributions.
\\
Finally, the questions addressed in this thesis represent only a small
part of a much broader programme in heavy-flavour physics. The theoretical tools developed and extended here---including effective
field theories, the HQE, and LCSRs---are sufficiently versatile to be
applied to a wide range of precision studies within and beyond the SM. As experimental measurements continue to
improve, these methods will remain indispensable for interpreting the
data and for uncovering possible signs of new physics. In this sense,
this thesis is not only about heavy-hadron lifetimes, but about
developing a broader framework for precision flavour physics.
It is, indeed, \emph{more than a lifetime.}

\appendix
\crefalias{chapter}{appendix} 

\chapter{Collection of Useful Formulas}
\label{app:1}
The $S$-matrix element $S_{fi}$ connects the asymptotic initial and final states in scattering processes. It is commonly written as
\begin{equation}
S_{fi} = \langle f \,|\, \hat{S} \,|\, i \rangle \,,
\end{equation}
where $|i\rangle$ and $|f\rangle$ denote the initial and final states, respectively. The $S$-operator is the time-ordered exponential of the interaction Lagrangian $\mathcal{L}_{\rm int}$,
\begin{equation}
\hat{S} = T \exp \Biggl[ i \int d^4 x \, \mathcal{L}_{\rm int}(x) \Biggr] \,,
\end{equation}
which can be expanded to second order as
\begin{equation}
S_{fi} = \langle f | 1 + i \int d^4 x \, T\{\mathcal{L}_{\rm int}(x)\} 
- \frac{1}{2} \int d^4 x \, d^4 y \, T\{\mathcal{L}_{\rm int}(x) \mathcal{L}_{\rm int}(y)\} + \cdots | i \rangle \,.
\end{equation}
Here, the first-order term gives the leading transition amplitude, while the second-order term accounts for processes with two interaction vertices or loop corrections. \\
The transition amplitude $T_{fi}$ is then defined through
\begin{equation}
S_{fi} = \delta_{fi} + i (2\pi)^4 \delta^{(4)}(p_f - p_i) \, T_{fi} \,.
\end{equation}
The interaction Lagrangian consists of QCD and electroweak contributions:
\begin{equation}
\mathcal{L}_{\rm int} = \mathcal{L}_{\rm QCD} + \mathcal{L}_{\rm EW} \,.
\end{equation}
In this thesis, we focus on weak transitions of quarks. The relevant part of the electroweak Lagrangian for quark flavour-changing processes is
\begin{equation}
\mathcal{L}_{\rm weak} = - \frac{g}{\sqrt{2}}\, V_{ij} \, [\bar{u}_i \gamma^\mu P_L  \, d_j ] \, W_\mu^+ + \text{h.c.} \,,
\end{equation}
where $P_L = (1-\gamma_5)/2$, $V_{ij}$ are CKM matrix elements, $u_i$ and $d_j$ denote up- and down-type quark fields, and $W_\mu^+$ is the charged weak boson.

The QCD Lagrangian for quarks and gluons is
\begin{equation}
\mathcal{L}_{\rm QCD}
=
\bar{q} (i \gamma^\mu D_\mu - m_q) q
-
\frac{1}{4} G^a_{\mu\nu} G^{a\,\mu\nu},
\qquad
G_{\mu\nu}^a
=
\partial_\mu A_\nu^a
-
\partial_\nu A_\mu^a
+
g_s f^{abc} A_\mu^b A_\nu^c .
\label{eq:QCD_lag}
\end{equation}
where $q$ denotes the quark fields, $m_q$ their masses, $G^a_{\mu\nu}$ is the gluon field strength tensor, $g_s$ the strong coupling, $T^a$ the $SU(3)_c$ generators, and $A^a_\mu$ the gluon fields.

The covariant derivative acting on a field in the fundamental
representation is defined by
\begin{equation}
iD_\mu
=
i\partial_\mu
+
g_sA_\mu^at^a,
\label{eq:D-fund}
\end{equation}
where $t^a=\lambda^a/2$ are the generators of
${SU}(3)_c$. For fields transforming in the adjoint
representation, the covariant derivative takes the analogous form
\begin{equation}
iD_\mu
=
i\partial_\mu
+
g_sA_\mu^aT^a,
\label{eq:D-adj}
\end{equation}
where the generators are given by
\begin{equation}
(T^a)_{bc}=-if^{abc},
\label{eq:T-def}
\end{equation}
with $f^{abc}$ the totally antisymmetric structure constants of
${SU}(3)_c$. The generators satisfy the Lie algebra
\begin{equation}
[t^a,t^b]=if^{abc}t^c,
\qquad
[T^a,T^b]=if^{abc}T^c.
\label{eq:SU3-comm}
\end{equation}
\\
The gluon field-strength tensor is defined by
\begin{equation}
G_{\mu\nu}
=
G_{\mu\nu}^a\,t^a,
\qquad
G_{\mu\nu}^a
=
\partial_\mu A_\nu^a
-
\partial_\nu A_\mu^a
+
g_sf^{abc}A_\mu^bA_\nu^c.
\label{eq:Gmunu}
\end{equation}
Equivalently, in the fundamental representation it can be expressed
through the commutator of two covariant derivatives (see e.g.\  Refs.~\cite{Shifman:1980ui,Piscopo:2021ogu}),
\begin{equation}
G_{\mu\nu}
=
i[iD_\mu,iD_\nu].
\label{eq:G-comm}
\end{equation}
Successive covariant derivatives acting on the field-strength tensor are
therefore given by
\begin{align}
D_\alpha G_{\mu\nu}
&=
i[iD_\alpha,[iD_\mu,iD_\nu]],
\label{eq:DG}
\\
D_\beta D_\alpha G_{\mu\nu}
&=
i[iD_\beta,[iD_\alpha,[iD_\mu,iD_\nu]]].
\label{eq:DDG}
\end{align}
The following are the properties of the charge conjugation operator which are used in this work:
\begin{equation}\label{conjugate_properties}
\begin{aligned}
\bar{\psi}_{R} &= \bar{\psi}P_{L},\kern5.2em \psi_{R}=P_{R}\kern0.14em \psi,\\
\bar{\psi}^{c} &= -\psi^{T}\kern0.14em C^{-1},\kern3.3em \psi^{c}= C (\bar{\psi})^{T},\\
\bar{\psi}_{R}^{c} &= - \psi^{T}(P_{R})^{T}\kern0.14em C^{-1},\ \psi_{R}^{c} = C (P_{L})^{T}(\bar{\psi})^{T},
\end{aligned}
\end{equation}
\begin{equation}\label{conjugate_gamma_properties}
\begin{aligned}
C \gamma^{\mu} C^{-1} &= -(\gamma^{\mu})^T,\kern4.8em C\gamma_{5} C^{-1} = (\gamma_{5})^T,\\C \gamma^{\mu} \gamma^{\nu} C^{-1} &= (\gamma^{\mu})^T (\gamma^{\nu})^T,\kern2em
C P_{R}\gamma^{\mu} C^{-1} = - (\gamma^{\mu})^{T} (P_{L})^{T},\\\kern0.14em C (P_{L})^{T}(\gamma^{\mu})^{T} (P_{R})^{T} C^{-1} &= -P_{L}\gamma^{\mu},\\
P_{L}&\equiv \left(\dfrac{1-\gamma_{5}}{2}\right),\kern5em P_{R} \equiv \left(\dfrac{1+\gamma_{5}}{2}\right).
\end{aligned}
\end{equation}

\chapter{Supplementary Material for the Dimension-Seven Calculations}
\label{App:SM_Dim7}

\section{HQE Operator Basis}
\label{sec:HQE-basis-appendix}

Within the HQE, the transition operator in \cref{eq:trans_op} is expanded in terms of local
operators of increasing dimension. This appendix summarizes the
complete basis of local two-quark HQE operators up to dimension seven
employed throughout this thesis. The operators are grouped according to
their mass dimension and follow the conventions of
\cref{app:1}.

Following the conventions of
Refs.~\cite{Mannel:2018mqv,Fael:2018vsp}, the hadronic matrix elements
are parameterised as follows.

The dimension-five and dimension-six HQE parameters are defined by
\begin{equation}
\begin{aligned}
2m_B\,\mu_\pi^2(B)
&=
-\langle B(p_B)|
\bar b_v (iD_\mu)(iD^\mu)b_v
|B(p_B)\rangle,
\\[3pt]
2m_B\,\mu_G^2(B)
&=
\langle B(p_B)|
\bar b_v (iD_\mu)(iD_\nu)(-i\sigma^{\mu\nu})b_v
|B(p_B)\rangle,
\\[6pt]
2m_B\,\rho_D^3(B)
&=
\langle B(p_B)|
\bar b_v (iD_\mu)(iv\!\cdot\!D)(iD^\mu)b_v
|B(p_B)\rangle,
\\[3pt]
2m_B\,\rho_{LS}^3(B)
&=
\langle B(p_B)|
\bar b_v (iD_\mu)(iv\!\cdot\!D)(iD_\nu)
(-i\sigma^{\mu\nu})b_v
|B(p_B)\rangle,
\end{aligned}
\label{eq:hadronic_parameters}
\end{equation}
Here, $\mu_\pi^2$ and $\mu_G^2$ parameterise the kinetic and
chromomagnetic operators at dimension five, respectively, whereas
$\rho_D^3$ and $\rho_{LS}^3$ correspond to the Darwin and spin--orbit
operators at dimension six.
Moreover,
$\sigma^{\mu\nu}=i[\gamma^\mu,\gamma^\nu]/2$,
$v^\mu=p_B^\mu/m_B$, and $b_v$ denotes the rephased full-QCD
$b$-quark field, defined through
\begin{equation}
b(x)=e^{-im_b v\cdot x}\,b_v(x),
\label{eq:HQETfield}
\end{equation}
where the exponential factor removes the large momentum
$m_b v^\mu$ carried by the heavy quark. 
The independent dimension-seven matrix elements are defined as
\begin{equation}
\begin{aligned}
2m_B\,r_G^4
&=
\langle B(p_B)|
\bar b_v
[(iD_\mu),(iD_\nu)]
[(iD^\mu),(iD^\nu)]
b_v
|B(p_B)\rangle,
\\[3pt]
2m_B\,r_E^4
&=
\langle B(p_B)|
\bar b_v
[(iv\!\cdot\!D),(iD_\mu)]
[(iv\!\cdot\!D),(iD^\mu)]
b_v
|B(p_B)\rangle,
\\[3pt]
2m_B\,s_B^4
&=
\langle B(p_B)|
\bar b_v
[(iD_\mu),(iD_\alpha)]
[(iD^\mu),(iD_\beta)]
(-i\sigma^{\alpha\beta})
b_v
|B(p_B)\rangle,
\\[3pt]
2m_B\,s_E^4
&=
\langle B(p_B)|
\bar b_v
[(iv\!\cdot\!D),(iD_\alpha)]
[(iv\!\cdot\!D),(iD_\beta)]
(-i\sigma^{\alpha\beta})
b_v
|B(p_B)\rangle,
\\[3pt]
2m_B\,s_{qB}^4
&=
\langle B(p_B)|
\bar b_v
[(iD_\mu),[(iD^\mu),[(iD_\alpha),(iD_\beta)]]]
(-i\sigma^{\alpha\beta})
b_v
|B(p_B)\rangle.
\end{aligned}
\label{eq:dim7MatrixElements}
\end{equation}
Here, $r_G^4$ and $r_E^4$ parameterise the spin-independent
field-strength and chromoelectric structures, respectively, while
$s_B^4$ and $s_E^4$ describe spin-dependent structures involving two
field strengths. The parameter $s_{qB}^4$ instead corresponds to a
spin-dependent structure involving covariant derivatives of the
field-strength tensor.

Furthermore, the matrix elements $r_G^4$ and $r_E^4$ can be decomposed
into colour-singlet and colour-octet components using the
${SU}(3)_c$ identity
$\{t^a,t^b\}=\delta^{ab}/N_c+d^{abc}t^c$. The corresponding matrix
elements are defined by
\begin{equation}
\begin{aligned}
2m_B\,r_G^{(1)}
&=
\langle B(p_B)|
\bar b_v b_v\,
G_{\mu\nu}^aG^{\mu\nu\,a}
|B(p_B)\rangle,
\\[3pt]
2m_B\,r_G^{(8)}
&=
\langle B(p_B)|
\bar b_v t^c b_v\,
d^{abc}G_{\mu\nu}^aG^{\mu\nu\,b}
|B(p_B)\rangle,
\\[6pt]
2m_B\,r_E^{(1)}
&=
\langle B(p_B)|
\bar b_v b_v\,
v_\alpha v_\beta
G_{\alpha\nu}^aG^{\beta\nu\,a}
|B(p_B)\rangle,
\\[3pt]
2m_B\,r_E^{(8)}
&=
\langle B(p_B)|
\bar b_v t^c b_v\,
v_\alpha v_\beta
d^{abc}G_{\alpha\nu}^aG^{\beta\nu\,b}
|B(p_B)\rangle.
\end{aligned}
\label{eq:color-decomp}
\end{equation}
The colour decomposition is particularly relevant for distinguishing
semileptonic from non-leptonic decays at leading order QCD. In the semileptonic case, the
colour structure involves a product of two colour generators, such that
the colour-singlet and colour-octet matrix elements enter only through
the combination implied by the identity above. In non-leptonic decays,
additional colour contractions allow the colour-singlet matrix elements
$r_G^{(1)}$ and $r_E^{(1)}$ to appear separately.

\section{Loop Integrals}
\label{sec:loop_int}
The two-loop integrals arising in the matching of the dimension-seven and dimension six
two-quark operators can be reduced, by means of IBP identities, to a
small set of master integrals. We define the generic scalar integral as
\begin{equation}
{\cal I}_{n_1n_2n_3}
=
\int\frac{d^4l_1}{(2\pi)^4}
\int\frac{d^4l_2}{(2\pi)^4}
\frac{1}
{\left[l_1^2-m_1^2\right]^{n_1}
 \left[l_2^2-m_2^2\right]^{n_2}
 \left[(l_1+l_2-p_b)^2-m_3^2\right]^{n_3}},
\label{eq:MI-def1}
\end{equation}
where $n_i\in\mathbb N_0$ and the Feynman $+i\varepsilon$ prescription is understood for all propagators. The complete set of master integrals required
for the present calculation is
\begin{equation}
\left\{
{\cal I}_{111},
{\cal I}_{211},
{\cal I}_{121},
{\cal I}_{112},
{\cal I}_{011},
{\cal I}_{101},
{\cal I}_{110}
\right\}.
\label{eq:MIs}
\end{equation}

The first four integrals correspond to the scalar sunrise topology with
three propagators of masses $m_1$, $m_2$, and $m_3$
\cite{Tarasov:1997kx,Caffo:1998du}. Since increasing the power of a
propagator is equivalent to differentiating with respect to the
corresponding mass squared, they satisfy
\begin{align}
{\cal I}_{211}
&=
\frac{\partial}{\partial m_1^2}\,
{\cal I}_{111},
&
{\cal I}_{121}
&=
\frac{\partial}{\partial m_2^2}\,
{\cal I}_{111},
&
{\cal I}_{112}
&=
\frac{\partial}{\partial m_3^2}\,
{\cal I}_{111}.
\label{eq:Im_MI112}
\end{align}

The remaining three master integrals factorize into products of scalar
tadpole integrals. For example,
\begin{equation}
{\cal I}_{101}
=
\int\frac{d^4l_1}{(2\pi)^4}
\frac{1}{l_1^2-m_1^2+i\varepsilon}
\int\frac{d^4l_2}{(2\pi)^4}
\frac{1}{l_2^2-m_3^2+i\varepsilon},
\label{eq:prod-scal-tadopole}
\end{equation}
which follows after the change of variables
$l_1+l_2-p_b\rightarrow l_2$. Analogous relations hold for
${\cal I}_{011}$ and ${\cal I}_{110}$. Since the imaginary part of a
product of tadpole integrals vanishes,
\begin{equation}
{\rm Im}\,{\cal I}_{101}
=
{\rm Im}\,{\cal I}_{110}
=
{\rm Im}\,{\cal I}_{011}
=
0,
\label{eq:Im-prod-tadpole}
\end{equation}
these integrals do not contribute to the absorptive part of the
transition operator.

Consequently, it is sufficient to determine the discontinuity of the
single master integral ${\cal I}_{111}$ in the physical region
$p_b^2\ge(m_1+m_2+m_3)^2$, from which the remaining master integrals
follow through differentiation. Using the result of
Ref.~\cite{Remiddi:2016gno}, evaluated in four dimensions, one finds
\begin{equation}
{\rm Im}\,{\cal I}_{111}
=
\frac{1}{256\pi^3}
\int_{(m_2+m_3)^2}^{(\sqrt{s}-m_1)^2}
dt\,
\frac{\sqrt{\lambda(t,m_2^2,m_3^2)\,
\lambda(s,t,m_1^2)}}{ts},
\label{eq:Im_sunset}
\end{equation}
where $s=p_b^2$ and
\begin{equation}
\lambda(x,y,z)
=
x^2+y^2+z^2-2xy-2xz-2yz
\label{eq:kaellen}
\end{equation}
denotes the Källén function.
The master integral ${\cal I}_{111}$ admits a closed analytical
representation only when at most two internal masses are non-vanishing.
This is sufficient for all decay channels considered in this work except
for the $b\to c\bar cs$ transition, where all three internal propagators
carry finite masses. In this case, the analytical expression is obtained
only after combining the two-quark contribution with the corresponding
four-quark mixing terms, leading to the cancellation of the IR
logarithms. Using the result of Ref.~\cite{Mannel:2020fts}, one finds
\begin{align}
\left[
{\cal M}_{112}(\rho,\eta)
-
\sqrt{1-4\rho}\,\log(\eta)
\right]_{\eta\to0}
={}&
2(1-\rho)
\log\!\left(
\frac{1+\sqrt{1-4\rho}}
{1-\sqrt{1-4\rho}}
\right)
\nonumber\\
&
+\sqrt{1-4\rho}
\left[
1+2\log(\rho)
-4\log\!\left(\sqrt{1-4\rho}\right)
\right],
\end{align}
where
\begin{equation}
{\cal M}_{112}(\rho,\eta)
=
256\pi^3\,{\rm Im}\,{\cal I}_{112}
\bigg|_{m_1^2=m_2^2=m_c^2,\;m_3^2=m_q^2,\;p_b^2=m_b^2},
\end{equation}
with
$\rho={m_c^2}/{m_b^2}$ and $\eta={m_q^2}/{m_b^2}.$

Finally, the gluon corrections to the four-quark operators reduce to 
one-loop tadpole integrals. Throughout the calculations we employ the
standard dimensional-regularisation result
\begin{equation}
\mu^{2\epsilon}
\int\frac{d^Dq}{(2\pi)^D}
\frac{1}{(q^2-m^2+i0)^b}
=
i(-1)^{-b}
\frac{\mu^{2\epsilon}}{(4\pi)^{D/2}}
(m^2)^{D/2-b}
\frac{\Gamma\!\left(b-\frac D2\right)}
{\Gamma(b)},
\label{eq:scalar-tadpole}
\end{equation}
where $D=4-2\epsilon$ and $\mu$ denotes the
renormalisation scale introduced to keep the strong coupling
dimensionless in dimensional regularisation.

\section{Trace Formulas for Colour-Decomposed Matrix Elements}
\label{sec:trace_formulas}
The trace formulas provide the systematic translation between general hadronic matrix elements 
and the independent HQE parameters defined in \cref{eq:hadronic_parameters,eq:dim7MatrixElements,eq:color-decomp}~\cite{Mannel:2010wj,Mannel:1994kv,Dassinger:2006md}. 
While the complete set of trace formulas at $\mathcal{O}(1/m_b^4)$ for the combined 
matrix elements $r_G^4$ and $r_E^4$  was derived in~\cite{Mannel:2023yqf,Mannel:2010wj}, 
the colour decomposition of these matrix elements requires corresponding 
formulas for the separate colour-singlet and colour-octet components $r_G^{(1)}$, $r_G^{(8)}$, 
$r_E^{(1)}$, and $r_E^{(8)}$ presented in \cref{eq:color-decomp}. 
In this appendix, we derive these relations.

We start with the colour singlet case and write the most general matrix element
\begin{equation}
\langle B(v)|\bar{b}_{v,j}b_{v,i}\left(G^{a}_{\mu\nu}G^{a}_{\alpha\beta}\right)|B(v)\rangle=\mathcal{M}_{ij,\mu\nu\alpha\beta},
\label{eq:gen-mat-singlet}
\end{equation}
where $i,j$ are Dirac indices and $\mu,\nu,\alpha,\beta$ are Lorentz indices. 
Note that no colour indices appear, since everything is colour singlet. 
Concerning the Lorentz indices, we have to construct this from the metric tensor $g$ 
and the velocity vector $v$, and possibly also from Lorentz indices related to the Dirac matrices.

For the trivial case of Dirac matrices, we have the four indices generated by the two structures
\begin{align}
R^{(G)}_{\mu\nu\alpha\beta}&=g^{\perp}_{\mu\alpha}g^{\perp}_{\nu\beta}-g^{\perp}_{\nu\alpha}g^{\perp}_{\mu\beta},
\label{eq:R-G-structure} \\
R^{(E)}_{\mu\nu\alpha\beta}&=g^{\perp}_{\mu\alpha}v_{\nu}v_{\beta}-g^{\perp}_{\nu\alpha}v_{\mu}v_{\beta}
-g^{\perp}_{\mu\beta}v_{\alpha}v_{\nu}+g^{\perp}_{\nu\beta}v_{\alpha}v_{\mu},
\label{eq:R-E-structure}
\end{align}
with $g^{\perp}_{\mu\nu}=g_{\mu\nu}-v_{\mu}v_{\nu}$, which are antisymmetric under 
$\mu\leftrightarrow\nu$ and $\alpha\leftrightarrow\beta$ and symmetric under 
$(\mu\nu)\leftrightarrow(\alpha\beta)$. Dirac matrices that carry Lorentz indices are $\gamma_{\mu}$ and $\sigma_{\mu\nu}$. 
Since we are at highest order, $\gamma_{\mu}$ is equivalent to $v_{\mu}$, while 
$\sigma_{\mu\nu}$ is antisymmetric in $\mu$ and $\nu$ and no scalar operator can be 
formed with a single $\sigma_{\mu\nu}$. In this way we obtain
\begin{equation}
\mathcal{M}_{ij,\mu\nu\alpha\beta}=\left(\frac{\not{v}+1}{2}\right)_{ij}
\left[a^{(1)}_{G}R^{(G)}_{\mu\nu\alpha\beta}+a^{(1)}_{E}R^{(E)}_{\mu\nu\alpha\beta}\right],
\label{eq:M-singlet-decomp}
\end{equation}
where $a_{G}^{(1)}$ and $a_{E}^{(1)}$ are determined from \cref{eq:R-G-structure,eq:R-E-structure} 
in terms of $r_{G}^{(1)}$ and $r_{E}^{(1)}$, and we get
\begin{align}
2M_{B}r_{G}^{(1)} &= 12a_{G}^{(1)},
\label{eq:rG1-relation} \\
2M_{B}r_{E}^{(1)} &= 2\left[6a_{G}^{(1)}+3a_{E}^{(1)}\right].
\label{eq:rE1-relation}
\end{align}
The formulae for the octet part look completely the same, with the  replacement 
superscript $(1)\to(8)$.

\chapter{Supplementary Analytical Results for the BSM Darwin Term}
\label{App:BSM Darwin}
This appendix collects the supplementary analytical results for the
generalised effective Hamiltonian discussed in \cref{Sec:Results}. Owing to
the length of the complete expressions, machine-readable versions of
all matching coefficients are provided in the ancillary files
accompanying Ref.~\cite{Lang:2025ios}, to which the interested reader
is referred. The notation and conventions adopted throughout this
appendix are the same as those introduced in \cref{Ch:3}.
\section{\texorpdfstring{\boldmath Results for the $b \to u \bar c s$ Transition}{Results for the b -> u cbar s Transition}}
\label{sec:resultsucs}

For completeness, we collect here the matching coefficients for the
CKM-suppressed transition $b\to u\bar{c}s$. As discussed in
\cref{sec:BSMCal}, the cancellation of the IR divergences in this
channel requires the inclusion of the WA contribution in
addition to the PI and WE topologies (see \cref{fig:4QDiags}). The
corresponding WA matching coefficients are given in
\cref{sec:WA}.
\paragraph{Dimension-three contribution ($\Gamma_3$):}
The independent non-vanishing coefficient functions
$P_{i,j}^{(ucs)}(\rho)$ are
\begin{align}
P_{1,1}^{(ucs)}(\rho) &= 
1 - 8\rho - 12\rho^2 \log\rho + 8\rho^3 - \rho^4 \,, \label{eq:P11ucs} \\[2mm]
P_{1,7}^{(ucs)}(\rho) &= \sqrt{\rho} \big(-1 - 9\rho + 9\rho^2 + \rho^3 - 6\rho(1 + \rho) \log\rho \big) \,. \label{eq:P17ucs}
\end{align}
The remaining non-vanishing coefficient functions are obtained from
\begin{equation}
\begin{aligned}
&P_{2,2}^{(ucs)} = P_{3,3}^{(ucs)} = P_{4,4}^{(ucs)}
= P_{1,1}^{(ucs)},
\\[4pt]
&4P_{5,5}^{(ucs)} = 4P_{6,6}^{(ucs)}
= 4P_{7,7}^{(ucs)} = 4P_{8,8}^{(ucs)}
= P_{1,1}^{(ucs)},
\\[4pt]
&\frac{1}{12}P_{9,9}^{(ucs)}
= \frac{1}{12}P_{10,10}^{(ucs)}
= P_{1,1}^{(ucs)},
\\[8pt]
&P_{3,4}^{(ucs)}
= 4P_{5,6}^{(ucs)}
= 4P_{7,8}^{(ucs)}
= \frac{1}{12}P_{9,10}^{(ucs)}
= P_{1,1}^{(ucs)},
\\[8pt]
&P_{1,8}^{(ucs)} = P_{2,8}^{(ucs)}
= P_{3,5}^{(ucs)}
= P_{4,5}^{(ucs)}
= P_{1,7}^{(ucs)},
\\[4pt]
&-\frac{1}{12}P_{1,9}^{(ucs)}
= -\frac{1}{12}P_{1,10}^{(ucs)}
= -\frac{1}{12}P_{2,10}^{(ucs)}
= P_{1,7}^{(ucs)},
\\[8pt]
&P_{i,j}^{(ucs)} = P_{i-1,j+1}^{(ucs)},
\qquad
\text{for even } i \text{ and odd } j\,,
\end{aligned}
\label{eq:relations_ucsdim3}
\end{equation}
while all other coefficient functions vanish identically.
\paragraph{Dimension-five contribution ($\Gamma_5$):}
The independent non-vanishing chromomagnetic coefficient functions
$G_{i,j}^{(ucs)}(\rho)$ are
\begin{align}
G_{1,1}^{(ucs)}(\rho) &= 
-\frac{1}{2}\big(5\rho^4 - 24\rho^3 + 24\rho^2 + 12\rho^2\log\rho - 8\rho + 3\big) \,, \label{eq:G11ucs} \\[2mm]
G_{1,2}^{(ucs)}(\rho) &= -\frac{1}{2} \big(19 + 16\rho - 24\rho^2 - 16\rho^3 + 5\rho^4 + 12\rho(4 + \rho)\log\rho\big) \,, \label{eq:G12ucs} \\[2mm]
G_{1,7}^{(ucs)}(\rho) &= \frac{1}{2} \sqrt{\rho} \big(-5 + 3\rho - 3\rho^2 + 5\rho^3 - 6\rho(1+\rho)\log\rho\big) \,, \label{eq:G17ucs} \\[2mm]
G_{1,8}^{(ucs)}(\rho) &= \frac{3}{2} \sqrt{\rho} \big(-9 + 7\rho + \rho^2 + \rho^3 - 2(2 + 3\rho + \rho^2)\log\rho\big) \,, \label{eq:G18ucs} \\[2mm]
G_{1,9}^{(ucs)}(\rho) &= 2 \sqrt{\rho} \big(-17 + 15\rho + 9\rho^2 - 7\rho^3 + 6\rho(-5 + 3\rho)\log\rho\big) \,, \label{eq:G19ucs} \\[2mm]
G_{2,8}^{(ucs)}(\rho) &= \frac{1}{2} \sqrt{\rho} \big(11 - 9\rho - 3\rho^2 + \rho^3 - 6(-3 + \rho)\rho\log\rho\big) \,, \label{eq:G28ucs} \\[2mm]
G_{2,10}^{(ucs)}(\rho) &= -2 \sqrt{\rho} \big(1 - 3\rho - 9\rho^2 + 11\rho^3 + 6(1 - 3\rho)\rho\log\rho\big) \,, \label{eq:G210ucs} \\[2mm]
G_{3,4}^{(ucs)}(\rho) &= \frac{1}{2} \big(13 - 40\rho + 24\rho^2 + 8\rho^3 - 5\rho^4 - 12\rho^2\log\rho\big) \,, \label{eq:G34ucs} \\[2mm]
G_{3,6}^{(ucs)}(\rho) &= \frac{1}{6} \sqrt{\rho} \big(-17 - 9\rho + 9\rho^2 + 17\rho^3 - 6(-2 + 9\rho + 3\rho^2)\log\rho\big) \,, \label{eq:G36ucs} \\[2mm]
G_{7,10}^{(ucs)}(\rho) &= -4 \big(-4 + 3\rho + \rho^3 - 6\rho\log\rho\big) \,, \label{eq:G710ucs} \\[2mm]
G_{8,10}^{(ucs)}(\rho) &= 8 (1 - \rho)^3 \,, \label{eq:G810ucs} \\[2mm]
G_{9,9}^{(ucs)}(\rho) &= -2 \big(25 - 72\rho + 120\rho^2 - 88\rho^3 + 15\rho^4 + 36\rho^2\log\rho\big) \,, \label{eq:G99ucs} \\[2mm]
G_{9,10}^{(ucs)}(\rho) &= -2 \big(25 + 72\rho - 72\rho^2 - 40\rho^3 + 15\rho^4 + 12\rho(8 + 3\rho)\log\rho\big) \,, \label{eq:G910ucs} \\[2mm]
G_{10,10}^{(ucs)}(\rho) &= 2 \big(23 - 72\rho + 24\rho^2 + 40\rho^3 - 15\rho^4 - 36\rho^2\log\rho\big) \,. \label{eq:G1010ucs}
\end{align}
The remaining non-vanishing coefficient functions are obtained from the relations
\begin{equation}
\begin{aligned}
& G_{2,2}^{(ucs)} = G_{3,3}^{(ucs)} = 4G_{6,6}^{(ucs)} = 4G_{8,8}^{(ucs)} = G_{1,1}^{(ucs)} \,, \quad G_{1,10}^{(ucs)} = -12G_{1,8}^{(ucs)} \,, \\[6pt]
& G_{4,4}^{(ucs)} = 4G_{5,5}^{(ucs)} = 4G_{5,6}^{(ucs)} = 4G_{7,7}^{(ucs)} = 4G_{7,8}^{(ucs)} = G_{3,4}^{(ucs)} \,, \quad G_{3,5}^{(ucs)} = G_{4,6}^{(ucs)} = G_{1,7}^{(ucs)} \,, \\[6pt]
& G_{i,j}^{(ucs)} = G_{i-1,j+1}^{(ucs)} \quad \text{for even } i \text{ and odd } j \,,
\end{aligned}
\label{eq:relations_ucsdim5}
\end{equation}
while all other coefficient functions vanish identically.
\paragraph{Dimension-six contribution ($\Gamma_6$):}
The independent non-vanishing Darwin coefficient functions
$D_{i,j}^{(ucs)}(\rho)$ are
\begin{align}
D_{1,1}^{(ucs)}(\rho) &= \frac{2}{3}(1-\rho) \big(9 + 11\rho - 25\rho^2 + 5\rho^3 - 24(1 - \rho^2)\log(1-\rho) - 12\rho^2\log\rho\big) \,, \label{eq:D11ucs} \\[2mm]
D_{1,2}^{(ucs)}(\rho) &= -\frac{2}{3} \big(41 - 26\rho + 18\rho^2 - 38\rho^3 + 5\rho^4 + 48(1-\rho)^2(1 + \rho)\log(1-\rho) \nonumber \\
&\quad + 12(2 + 5\rho + 2\rho^2 - 2\rho^3)\log\rho\big) \,, \label{eq:D12ucs} \\[2mm]
D_{1,7}^{(ucs)}(\rho)
&= \frac{2}{3}\sqrt{\rho}\Big(
-8 + 27\rho - 24\rho^2 + 5\rho^3
\nonumber\\
&\qquad\qquad
+ 12(1-\rho)^2\log(1-\rho)
- 6\rho(-2+\rho)\log\rho
\Big)\,,
\label{eq:D17ucs} \\[2mm]
D_{1,8}^{(ucs)}(\rho)
&= \frac{2}{3}\sqrt{\rho}\Big(
7 + 6\rho - 15\rho^2 + 2\rho^3
\nonumber\\
&\qquad\qquad
+ 24(1-\rho)^2\log(1-\rho)
+ 6(1 + 4\rho - 2\rho^2)\log\rho
\Big)\,,
\label{eq:D18ucs} \\[2mm]
D_{1,9}^{(ucs)}(\rho) &= -8 \sqrt{\rho} \big(\rho(7 - 8\rho + \rho^2 - 6(-2 + \rho)\log\rho) + 12(1-\rho)^2\log(1-\rho)\big) \,, \label{eq:D19ucs} \\[2mm]
D_{2,8}^{(ucs)}(\rho) &= -\frac{2}{3} \sqrt{\rho} \big(-4 + 3\rho + \rho^3 - 12(1 - \rho)^2\log(1-\rho) + 6\rho(-2 + \rho)\log\rho\big) \,, \label{eq:D28ucs} \\[2mm]
D_{2,10}^{(ucs)}(\rho)
&= -8\sqrt{\rho}\Big(
-4 + 17\rho - 16\rho^2 + 3\rho^3
\nonumber\\
&\qquad\qquad
+ 12(1-\rho)^2\log(1-\rho)
- 6\rho(-2+\rho)\log\rho
\Big)\,,
\label{eq:D210ucs} \\[2mm]
D_{3,4}^{(ucs)}(\rho) &= -\frac{2}{3} \big(-51 + 94\rho - 42\rho^2 - 6\rho^3 + 5\rho^4 + 24(1 - \rho)^2(3 + \rho)\log(1-\rho) \nonumber \\
&\quad - 12(-1 + \rho^2 + \rho^3)\log\rho\big) \,, \label{eq:D34ucs} \\[2mm]
D_{3,6}^{(ucs)}(\rho) &= 4 \sqrt{\rho} \big(1 + 3\rho - 5\rho^2 + \rho^3 + 4(1-\rho)^2\log(1-\rho) + 2(1 + 2\rho - \rho^2)\log\rho\big) \,, \label{eq:D36ucs} \\[2mm]
D_{4,4}^{(ucs)}(\rho)
&= -\frac{2}{3}\Big(
-59 + 112\rho - 66\rho^2 + 8\rho^3 + 5\rho^4
\nonumber\\
&\qquad\qquad
+ 48(1-\rho)^2\log(1-\rho)
- 24\rho^2\log\rho
\Big)\,,
\label{eq:D44ucs} \\[2mm]
D_{7,8}^{(ucs)}(\rho) &= -\frac{1}{6} \big(-55 + 106\rho - 54\rho^2 - 2\rho^3 + 5\rho^4 + 24(1 - \rho)^2(3+\rho)\log(1-\rho) \nonumber \\
&\quad - 12(-1 + \rho^2 + \rho^3)\log\rho\big) \,, \label{eq:D78ucs} \\[2mm]
D_{7,10}^{(ucs)}(\rho)
&= \frac{8}{3}\Big(
22 - 27\rho + 12\rho^2 - 7\rho^3
\nonumber\\
&\qquad\qquad
- 6(1-\rho)^3\log(1-\rho)
+ 3(1 + 5\rho + 3\rho^2 - \rho^3)\log\rho
\Big)\,,
\label{eq:D710ucs} \\[2mm]
D_{8,8}^{(ucs)}(\rho) &= \frac{1}{6} (1-\rho) \big(13 + 3\rho - 21\rho^2 + 5\rho^3 - 24(1-\rho^2)\log(1-\rho) - 12\rho^2\log\rho\big) \,, \label{eq:D88ucs} \\[2mm]
D_{8,10}^{(ucs)}(\rho) &= \frac{4}{3} \big(23 - 51\rho + 45\rho^2 - 17\rho^3 - 12(1-\rho)^3\log(1-\rho) + 6(3 - \rho)\rho^2\log\rho\big) \,, \label{eq:D810ucs} 
\end{align}
\begin{align}
D_{9,9}^{(ucs)}(\rho) &= -\frac{8}{3} \big(7 - 72\rho + 162\rho^2 - 112\rho^3 + 15\rho^4 + 48(1-\rho)^2(1+2\rho)\log(1-\rho) \nonumber \\
&\quad + 24(3 - 2\rho)\rho^2\log\rho\big) \,, \label{eq:D99ucs} \\[2mm]
D_{9,10}^{(ucs)}(\rho) &= -\frac{8}{3} \big(11 + 102\rho - 66\rho^2 - 62\rho^3 + 15\rho^4 + 24(1-\rho)^2(7+5\rho)\log(1-\rho) \nonumber \\
&\quad + 12(5 + 10\rho + 3\rho^2 - 5\rho^3)\log\rho\big) \,, \label{eq:D910ucs} \\[2mm]
D_{10,10}^{(ucs)}(\rho) &= -\frac{8}{3} \big(-131 + 234\rho - 108\rho^2 - 10\rho^3 + 15\rho^4 + 24(1-\rho)^2(5+\rho)\log(1-\rho) \nonumber \\
&\quad - 12\rho^2(3 + \rho)\log\rho\big) \,. \label{eq:D1010ucs}
\end{align}
The remaining non-vanishing coefficient functions are obtained from
\begin{equation}
\begin{aligned}
& D_{2,2}^{(ucs)} = D_{3,3}^{(ucs)} = 4D_{6,6}^{(ucs)} = D_{1,1}^{(ucs)}\,,
\qquad
D_{1,10}^{(ucs)} = -12D_{1,8}^{(ucs)}\,,
\\[6pt]
& D_{3,5}^{(ucs)} = D_{4,6}^{(ucs)} = D_{1,7}^{(ucs)}\,,
\qquad
D_{5,5}^{(ucs)} = D_{7,7}^{(ucs)} = \frac14 D_{4,4}^{(ucs)}\,,
\qquad
D_{5,6}^{(ucs)} = \frac14 D_{3,4}^{(ucs)}\,,
\\[6pt]
& D_{i,j}^{(ucs)} = D_{i-1,j+1}^{(ucs)},
\qquad
\text{for even } i \text{ and odd } j\,,
\end{aligned}
\label{eq:relations_ucsdim6}
\end{equation}
while all other coefficient functions vanish identically.
\section{\texorpdfstring{\boldmath Results for the $b \to u \bar u d$ Transition}{Results for the b -> u ubar d Transition}}
\label{sec:resultsuud}
The dimension-three and dimension-five matching coefficients for the
$b\to u\bar{u}d$ transition are obtained directly from the corresponding
$b\to c\bar{u}d$ results by taking the massless limit $\rho\to0$.
The Darwin contribution is an exception, as the IR divergences arising
from the expansion of the $q_1$-quark propagator require a dedicated
matching calculation in the massless case. The corresponding coefficient
functions are presented below.
\paragraph{Dimension-six contribution ($\Gamma_6$):}
The independent non-vanishing Darwin coefficient functions
$D_{i,j}^{(uud)}$ are
\begin{align}
D_{1,1}^{(uud)}(\rho) &= 6 \,, & D_{1,2}^{(uud)}(\rho) &= -\frac{34}{3} \,, & D_{3,4}^{(uud)}(\rho) &= \frac{110}{3} \,, & D_{4,4}^{(uud)}(\rho) &= \frac{118}{3} \,, \label{eq:D11uud} \\[2pt]
D_{5,5}^{(uud)}(\rho) &= \frac{59}{6} \,, & D_{5,6}^{(uud)}(\rho) &= \frac{55}{6} \,, & D_{6,6}^{(uud)}(\rho) &= \frac{3}{2} \,, & D_{7,8}^{(uud)}(\rho) &= \frac{21}{2} \,, \label{eq:D55uud} \\[2pt]
D_{7,10}^{(uud)}(\rho) &= 48 \,, & D_{8,8}^{(uud)}(\rho) &= \frac{13}{6} \,, & D_{8,10}^{(uud)}(\rho) &= \frac{92}{3} \,, & D_{9,9}^{(uud)}(\rho) &= -\frac{56}{3} \,, \label{eq:D710uud} \\[2pt]
D_{9,10}^{(uud)}(\rho) &= 120 \,, & D_{10,10}^{(uud)}(\rho) &= \frac{1048}{3} \,. & & & & \label{eq:D910uud}
\end{align}
The remaining non-vanishing coefficients follow from the relations

\begin{samepage}
\begin{align}
D_{3,3}^{(uud)} &= D_{1,1}^{(uud)} \,, \qquad D_{7,7}^{(uud)} = D_{5,5}^{(uud)} \,, \qquad D_{2,2}^{(uud)} = D_{1,1}^{(uud)} \,, \nonumber
\\[6pt]
D_{i,j}^{(uud)} & = D_{i-1,j+1}^{(uud)} \qquad \quad \text{for even } i \text{ and odd } j, 
\label{eq:relations_uud}
\end{align}
\end{samepage}
while all other coefficient functions vanish identically.

\section{\boldmath Results for the WA Topology}
\label{sec:WA}

In this appendix we collect the matching coefficients for the WA topology induced by the generalised effective Hamiltonian in \cref{eq:Heff-NP}. As discussed in \cref{sec:BSMCal}, these contributions are required to complete the dimension-six four-quark matching and to ensure the cancellation of the IR divergences in the $b\to u\bar{c}s$ and $b\to u\bar{u}d$ channels.

Following the notation of Ref.~\cite{Lenz:2022pgw}, the WA contribution to the transition operator for internal quark flavours $q_2=c,u$ and $q_3=d,s$ is parametrised as
\begin{equation}
\mathrm{Im}\,\mathcal{T}_{\mathrm{WA}}^{q_2 q_3}
=
\frac{G_F^2 m_b^2}{6\pi}\,|V_{ub}|^2 |V_{q_{2}q_{3}}|^2\, (1-\rho)^2\,
\bigg[
\sum_{m,n = 1}^{20}
{\cal{C}}_m\, {\cal{C}}_n^{*}\, {\mathcal{N}}^{\prime}_{mn}\, A^{\mathrm{WA},\, q_2 q_3}_{mn}
\bigg] \,,
\label{eq:T-NP}
\end{equation}
where the colour factor ${\cal N}_{mn}'$ is given by
${\cal N}_{mn}'=N_c$ if both indices $m$ and $n$ are odd, and
${\cal N}_{mn}'=1$ otherwise. For convenience, the primed Wilson
coefficients are incorporated into the same index according to
\begin{equation}
{\cal{C}}_m \equiv {\cal{C}}^{\prime\,}_{\,m-10}\,, \qquad m>10 \,.
\end{equation}
The independent non-vanishing coefficient functions
$A_{mn}^{{\rm WA},\,q_2q_3}$ are given below for the
$q_2q_3=cs$ channel, retaining the full charm-mass dependence. The
corresponding expressions for $q_2q_3=ud$ follow from the massless
limit, $\rho\to0$.
\begin{align}
A^{\rm WA, c s}_{1,1}
&=
-\Bigl[
\mathcal{O}_1\!\left(1+\frac{\rho}{2}\right)
-\mathcal{O}_2(1+2\rho)
\Bigr] , \quad
A^{\rm WA, c s}_{1,3}
=
-\Bigl[
\mathcal{O}_5'\!\left(1+\frac{\rho}{2}\right)
-\mathcal{O}_6'(1+2\rho)
\Bigr] \, , \\
A^{\rm WA, c s}_{1,5} &= -\frac{3}{2} \, \mathcal{O}_6' \,\sqrt{\rho}\, , \quad 
A^{\rm WA, c s}_{1,7} = -\frac{3}{2}  \,\mathcal{O}_2 \,\sqrt{\rho}\, , \quad 
A^{\rm WA, c s}_{1,9} = -6 \sqrt{\rho} \bigl[\mathcal{O}_1 - \mathcal{O}_2 \bigr] \,, \\
A^{\rm WA, c s}_{3,5} &= -\frac{3}{2}\,  \mathcal{O}_2' \, \sqrt{\rho} \, ,\quad 
A^{\rm WA, c s}_{3,7} = -\frac{3}{2}  \, \mathcal{O}_6 \, \sqrt{\rho}\, , \quad 
A^{\rm WA, c s}_{3,9} = -6 \sqrt{\rho} \bigl[ \mathcal{O}_5 - \mathcal{O}_6 \bigr]\,,\\ 
A^{\rm WA, c s}_{5,5} &= \frac{3}{2}\, \mathcal{O}_2' \, ,\quad 
A^{\rm WA, c s}_{5,7} = \frac{3}{2}\, \mathcal{O}_6 \, ,\quad 
A^{\rm WA, c s}_{9,9} = -8 \, (\mathcal{O}_1 - \mathcal{O}_2) \, (1 + 2 \rho) \,,
\end{align}
where the $\Delta B=0$ operators ${\cal O}_i^{(\prime)}$ are defined in
\cref{eq:O1-O2-HQET,eq:O3-O4-HQET,eq:O5-O6-HQET,eq:O7-O8-HQET}. The
remaining non-vanishing coefficient functions follow from the symmetry
relations
\begin{align}
&A^{{\rm WA}, cs}_{3,3}
=
A^{{\rm WA}, cs}_{1,1}\Big|_{\mathcal O_i \leftrightarrow \mathcal O'_i},
\qquad
A^{{\rm WA}, cs}_{7,7}
=
A^{{\rm WA}, cs}_{5,5}\Big|_{\mathcal O_i \leftrightarrow \mathcal O'_i},
\\[6pt]
&A^{{\rm WA}, cs}_{m,n}
=
\begin{cases}
A^{{\rm WA}, cs}_{m,n-1},
& m \text{ odd},\; n \text{ even},
\\[4pt]
A^{{\rm WA}, cs}_{m-1,n+1},
& m \text{ even},\; n \text{ odd},
\end{cases}
\\[6pt]
&A^{{\rm WA},cs}_{m,n}
=
A^{{\rm WA},cs}_{m-1,n-1}
\Bigg|_{{\cal O}_i^{(\prime)}
\rightarrow
\left(
\dfrac{{\cal O}_i^{(\prime)}}{N_c}
+
2\,{\cal O}_{i+2}^{(\prime)}
\right)},
\qquad
m,n \text{ even},
\\[6pt]
&A^{{\rm WA},cs}_{m,n}
=
A^{{\rm WA},cs}_{n,m}
\Big|_{\mathcal O_i \leftrightarrow \mathcal O'_i},
\qquad
m\neq n,
\\[6pt]
&A^{{\rm WA},cs}_{m,n}
=
A^{{\rm WA},cs}_{m+10,n+10}
\Big|_{\mathcal O_i \leftrightarrow \mathcal O'_i},
\qquad
m,n\in\{1,\ldots,10\},
\end{align}
while all other coefficient functions vanish identically.

\section{Contribution of SM QCD-Penguin Operators up to Dimension-Six}
\label{sec:SM_penguin}
As discussed in \cref{subsec:Heff}, the SM QCD-penguin operators are related
to the generalised operator basis through the Fierz identities in
\cref{eq:Q1_Q1cal,eq:O3_O4,eq:O5_O6}. Consequently, once the matching
coefficients ${\cal F}_{i,j}^{(q_1q_2q_3)}$ have been determined (see \cref{eq:Gamma_master_compact,eq:Fij}), the
corresponding SM penguin contributions follow directly without requiring
an independent calculation.

The interference between the current-current operators
$Q_{1,2}^{qqq_3}$ and the QCD-penguin operators
$Q_{3,\ldots,6}^{qq_3}$ is given by
\begin{align}
\frac{\Gamma_{\rm PO}^{\rm int}(b\to q\bar q q_3)}{\Gamma_0}
&=
2\,\mathrm{Re}\!\left(\xi_{qq_3}\xi_{tq_3}^*\right)
\Bigg[
\nonumber\\
&\qquad
C_1\Big(
C_3{\cal F}_{1,2}^{(qqq_3)}
+C_4{\cal F}_{1,1}^{(qqq_3)}
-2C_5{\cal F}_{1,6}^{(qqq_3)}
-2C_6{\cal F}_{1,5}^{(qqq_3)}
\Big)
\nonumber\\
&\qquad
+
C_2\Big(
C_3{\cal F}_{2,2}^{(qqq_3)}
+C_4{\cal F}_{2,1}^{(qqq_3)}
-2C_5{\cal F}_{2,6}^{(qqq_3)}
-2C_6{\cal F}_{2,5}^{(qqq_3)}
\Big)
\Bigg].
\label{eq:Peng_mixed}
\end{align}
where $q=u,c$, $q_3=d,s$, and we have introduced the shorthand
$\xi_{xy}\equiv V_{xb}V_{xy}^{*}$.

The contribution arising from the double insertion of the QCD-penguin
operators takes the form
\begin{align}
\frac{\Gamma_{\rm PO}^{\rm quad}(b\to q\bar q q_3)}{\Gamma_0}
&=
|\xi_{tq_3}|^2
\Bigg[
\Big(
C_3^2{\cal F}_{2,2}^{(qqq_3)}
+2C_3C_4{\cal F}_{2,1}^{(qqq_3)}
+C_4^2{\cal F}_{1,1}^{(qqq_3)}
\Big)
\nonumber\\
&\hspace{0cm}
+4\Big(
C_5^2{\cal F}_{6,6}^{(qqq_3)}
+2C_5C_6{\cal F}_{6,5}^{(qqq_3)}
+C_6^2{\cal F}_{5,5}^{(qqq_3)}
\Big)
\nonumber\\
&\hspace{0cm}
-4\Big(
C_3C_5{\cal F}_{2,6}^{(qqq_3)}
+C_3C_6{\cal F}_{2,5}^{(qqq_3)}
+C_4C_5{\cal F}_{1,6}^{(qqq_3)}
+C_4C_6{\cal F}_{1,5}^{(qqq_3)}
\Big)
\Bigg].
\label{eq:Peng_quadratic}
\end{align}
The quadratic contribution is valid for
$q=u,d,s,c$. Since the masses of the light quarks are neglected, the
corresponding coefficient functions satisfy
\begin{equation}
    {\cal F}_{i,j}^{ddq_3} =  {\cal F}_{i,j}^{ssq_3} =  {\cal F}_{i,j}^{uuq_3}\,.
\end{equation}
As a consistency check, we have verified that the leading-power
contributions obtained by inserting the functions
$P_{i,j}^{(ccs)}(\rho)$ into
\cref{eq:Peng_mixed,eq:Peng_quadratic} reproduce the known SM results
of Ref.~\cite{Krinner:2013cja}. Throughout this section the Wilson
coefficients are understood to be the real-valued SM coefficients.

\chapter{Supplementary Material for the \textit{B}-Mesogenesis Analysis}
The free quark propagator in coordinate space is given by
\begin{equation}
\begin{aligned}
S^{(q)}_{jn}(x,y)
&=
\delta_{jn}\,S^{(q)}_0(x-y),\\
S^{(q)}_0(x-y)
&=
\int\frac{d^4p}{(2\pi)^4}
\frac{\slashed p+m}
{p^2-m^2+i\epsilon}
e^{-ip\cdot(x-y)},\\
\left(S^{(q)}_0\right)^T(x-y)
&=
\int\frac{d^4p}{(2\pi)^4}
\frac{p_\mu(\gamma^\mu)^T+m}
{p^2-m^2+i\epsilon}
e^{-ip\cdot(x-y)}.
\end{aligned}
\label{eq:propagators}
\end{equation}
Furthermore, the following one-loop tensor integral frequently appears
in the evaluation of the dimension-six four-quark contributions. Its
absorptive part is given by (see, e.g.,
Appendices~B and~C of Ref.~\cite{Piscopo:2021ogu})
\begin{equation}
\begin{aligned}
\text{Im} &\left(i \int \dfrac{d^{4}k}{(2\pi)^{4}}\kern0.16em \dfrac{k_{\mu}k_{\nu}-k_{\mu}p_{\nu}}{[k^{2}-m^{2}_{1}][ (p-k)^{2}-m^{2}_{2}]} \right) = \\
\dfrac{p^{2}\sqrt{\lambda(1,r_{1},r_{2})}}{192\pi}&\left(g_{\mu\nu}\lambda(1,r_{1},r_{2})-2\dfrac{p_{\mu}p_{\nu}}{p^{2}}(2(r_{1}-r_{2})^{2}-(1+r_{1}+r_{2}))\right),
\label{eq:integral4q}
\end{aligned}
\end{equation}
where \(r_i=m_i^2/p^2\) and \(\lambda(a,b,c)=(a-b-c)^2-4bc\) denotes the
Källén function.

The matrix elements of the four-quark operators arising in the $B$-Mesogenesis 
calculations are expressed in terms
of the operator basis\footnote{The operator basis differs from the corresponding SM basis
only by the replacement $P_L \leftrightarrow P_R$ (see \cref{eq:O7-O8-HQET}). Owing to parity
invariance of QCD, the parametrisation of the hadronic matrix elements
is identical to that of the SM operators.}
\begin{equation}
\begin{aligned}
\tilde{\mathcal{O}}^{q}_{1}
&=
4\,(\bar q^{i}\gamma^{\mu}P_{R}b^{i})
(\bar b^{k}\gamma_{\mu}P_{R}q^{k}),
\\
\tilde{\mathcal{O}}^{q}_{2}
&=
4\,(\bar q^{i}P_{L}b^{i})
(\bar b^{k}P_{R}q^{k}),
\\
\tilde{\mathcal{O}}^{q}_{3}
&=
4\,(\bar q^{i}\gamma^{\mu}P_{R}t^{a}_{ik}b^{k})
(\bar b^{r}\gamma_{\mu}P_{R}t^{a}_{rm}q^{m}),
\\
\tilde{\mathcal{O}}^{q}_{4}
&=
4\,(\bar q^{i}P_{L}t^{a}_{ik}b^{k})
(\bar b^{r}P_{R}t^{a}_{rm}q^{m}),
\end{aligned}
\label{eq:operators4q}
\end{equation}
where $q=u,d$. Their hadronic matrix elements are parameterized as~\cite{Kirk:2017juj,Black:2024bus}
\begin{equation}
\begin{aligned}
\langle B|\tilde{\mathcal O}_{1,2}^{q}|B\rangle
&=
f_B^2M_B^2\,B_{1,2}^{q},
\\
\langle B|\tilde{\mathcal O}_{3,4}^{q}|B\rangle
&=
f_B^2M_B^2\,B_{3,4}^{q},
\end{aligned}
\label{eq:matrixelement4q}
\end{equation}
where $B_{1,2}^{q}$ and $B_{3,4}^{q}$ denote the corresponding bag
parameters. Throughout this work, isospin symmetry is assumed,
\begin{equation}
B_i^u=B_i^d,
\qquad
i=1,\ldots,4.
\label{eq:isospin-bag}
\end{equation}
\section{Input Parameters}
\label{sec:BMeso_input}
For the bottom and charm quark masses we use the values obtained from
the global fit of Ref.~\cite{Finauri:2023kte},
\begin{equation}
m_b^{\rm kin}(1\,\GeV)=(4.573\pm0.012)\,\GeV,
\qquad
\overline{m}_c(2\,\GeV)=(1.090\pm0.010)\,\GeV.
\label{eq:mb-mc}
\end{equation}
The $B$-meson decay constants are taken from the lattice-QCD averages of
Ref.~\cite{Aoki:2019cca},
\begin{equation}
f_B=(0.1900\pm0.0013)\,\GeV,
\qquad
f_{B_s}=(0.2303\pm0.0013)\,\GeV.
\label{eq:decay-constants}
\end{equation}
The $B$-meson masses are taken from the
Particle Data Group~\cite{ParticleDataGroup:2026aaa},
\begin{equation}
m_{B^+}=5.27934\,\GeV,
\qquad
m_{B_d}=5.27965\,\GeV,
\qquad
m_{B_s}=5.36688\,\GeV.
\end{equation}
The CKM matrix elements are taken from the global fit of the
CKMfitter collaboration~\cite{Charles:2004jd} (online update). The HQE matrix elements entering the dimension-three, -five, and
-six contributions are summarised in
\cref{tab:Non-perturbative-input}. The corresponding Bag
parameters parameterising the forward matrix elements of the
dimension-six $\Delta B=0$ four-quark operators are collected in~\cref{tab:Bag-parameters}.%
\footnote{The parametrisations of the SM and Mesogenesis
four-quark operators are identical, since the two operator bases differ
only by their chiral structure; see Ref.~\cite{Lenz:2024rwi}.}
\begin{table}[h!]
\centering
\small
\renewcommand{\arraystretch}{1.4}
    \begin{tabular}{|c||c|c||c|c|}
    \hline 
    Parameter 
    & $B^{+}, B_{d}$ 
    & Source
    & $B_s$
    & Source \\
    \hline
    \hline
    $\mu_\pi^2 (B_q)$ [GeV$^2$]
    & $0.454 \pm  0.043 $
    & Exp. fit \cite{Finauri:2023kte}
    & $ 0.534 \pm  0.074 $
    & Exp. fit + Ref.~\cite{Egner:2024lay}
    \\
    \hline
    $\mu_G^2 (B_q)$ [GeV$^2$]
    & $0.274 \pm  0.053 $
    & Exp. fit \cite{Finauri:2023kte} 
    & $ 0.321 \pm  0.072$ 
    & Exp. fit + Ref.~\cite{Egner:2024lay}
    \\
    \hline
    $\rho_D^3 (B_q)$ [GeV$^3$]
    & $0.176 \pm  0.019 $
    & Exp. fit \cite{Finauri:2023kte}
    & $0.210 \pm  0.034$ 
    & Exp. fit +  Ref.~\cite{Egner:2024lay} \\
    \hline
    \end{tabular}
    \caption{Input values of the two-quark non-perturbative parameters.
    }
    \label{tab:Non-perturbative-input}
\end{table}

\begin{table}[th]
\centering
\small
\renewcommand{\arraystretch}{1.5} 
\begin{tabular}{|c||c|c|c|c|}
\hline
$\mu_0 = 1.5\GeV$    
& $ \tilde B_1^q$ 
& $ \tilde B_2^q$ 
& $ \tilde B_3^q$ 
& $ \tilde B_4^q$ 
\\
\hline
\hline
    $\langle B_{u,d} | \tilde O^{u,d}_i | B_{u,d} \rangle$ 
     & $1.0026^{+0.0246}_{-0.0221}$ 
     & $0.9982^{+0.0206}_{-0.0214}$ 
     & $-0.0057^{+0.0221}_{-0.0225}$ 
     & $-0.0014^{+0.0216}_{-0.0221}$
\\
\hline
     $\langle B_{s} | \tilde O^{s}_i | B_{s} \rangle$  
     & $1.0022^{+0.0246}_{-0.0221}$  
     & $0.9983^{+0.0246}_{-0.0221}$ 
     & $-0.0036^{+0.0265}_{-0.0270}$  
     & $-0.0009^{+0.0259}_{-0.0265}$
\\
\hline
\end{tabular}
\caption{Input values of the dimension-six Bag parameters at the renormalisation scale 
$\mu_0 = 1.5 \,\GeV$, collected from Ref.~\cite{Egner:2024lay} and references therein.}
\label{tab:Bag-parameters}
\end{table}
\cref{Constants1/3,Constants2/3} define the effective $B$-Mesogenesis 
couplings $G_X^{f_1f_2}$ and list their corresponding upper bounds from
Ref.~\cite{Alonso-Alvarez:2021qfd}. Here,
$X=\mathrm{I},\ldots,\mathrm{VI}$ denotes the topologies shown in~\cref{fig:MesDiagrams}, whereas $f_1$ and $f_2$ identify the
internal fermions propagating in the loop of the corresponding
diagram.
\begin{longtable}{|>{\centering\arraybackslash}p{4.5cm} | >{\centering\arraybackslash}p{4cm} | >{\centering\arraybackslash}p{4cm}|}
    \caption{$\mathcal{H}_{-1/3}$ Constants and their Max values, taken from Ref.~\cite{Alonso-Alvarez:2021qfd}.}
    \label{Constants1/3}\\
    \hline
    Parameter & Max Value $\text{(GeV)}^{-4}$& Source \\
    \hline\hline
    \endfirsthead
    \multicolumn{3}{c}%
    {{\bfseries \tablename\ \thetable{} -- continued from previous page}} \\
    \hline
    Parameter & Max Value $\text{(GeV)}^{-4}$& Source \\
    \hline\hline
    \endhead
    \hline
    \multicolumn{3}{|r|}{{Continued on the next page}} \\
    \hline
    \endfoot
    \hline
    \endlastfoot
    $G^{\psi u}_{III}= \dfrac{y_{ub}y_{\psi d}}{M_{Y}^{2}}\dfrac{y^{*}_{ub}y^{*}_{\psi d}}{M_{Y}^{2}}$& $1\times 10^{-13}$ & Table II \\
    \hline
    $G^{\psi c}_{III}= \dfrac{y_{cb}y_{\psi d}}{M_{Y}^{2}}\dfrac{y^{*}_{cb}y^{*}_{\psi d}}{M_{Y}^{2}}$ & $1.5\times 10^{-12}$  &  Table II \\
    \hline
    $G^{u \psi}_{III}= \dfrac{y_{ud}y_{\psi b}}{M_{Y}^{2}}\dfrac{y^{*}_{ud}y^{*}_{\psi b}}{M_{Y}^{2}}$ & $4.7 \times 10^{-15}$ & Table II\\
    \hline
    $G^{c \psi}_{III}= \dfrac{y_{cd}y_{\psi b}}{M_{Y}^{2}}\dfrac{y^{*}_{cd}y^{*}_{\psi b}}{M_{Y}^{2}}$ &  $1.6 \times 10^{-13}$ &Table II\\
    \hline
    $G^{\psi u}_{II}= \dfrac{y_{ub}y_{\psi d}}{M_{Y}^{2}}\dfrac{y^{*}_{ud}y^{*}_{\psi b}}{M_{Y}^{2}}$ & $2.2 \times 10^{-14}$ & Table II \\
    \hline
    % Additional rows (total 21 rows)
    $G^{\psi c}_{II}= \dfrac{y_{cb}y_{\psi d}}{M_{Y}^{2}}\dfrac{y^{*}_{cd}y^{*}_{\psi b}}{M_{Y}^{2}}$& $4.9 \times 10^{-13}$ & Table II \\
    \hline
    $G^{u \psi}_{II}= \dfrac{y_{ud}y_{\psi b}}{M_{Y}^{2}}\dfrac{y^{*}_{ub}y^{*}_{\psi d}}{M_{Y}^{2}}$ & $2.2 \times 10^{-14}$ & Table II\\
    \hline
    $G^{c \psi}_{II}= \dfrac{y_{cd}y_{\psi b}}{M_{Y}^{2}}\dfrac{y^{*}_{cb}y^{*}_{\psi d}}{M_{Y}^{2}}$ &  $4.9 \times 10^{-13}$ & Table II \\
    \hline
    $G^{\psi \psi}_{I}= \dfrac{y_{\psi d}y^{*}_{\psi b}}{M_{Y}^{2}}\dfrac{y_{\psi b}y^{*}_{\psi d}}{M_{Y}^{2}}$  & $1.7 \times 10^{-15}$ & Eq. (43j), Eq. (41a) \\
    \hline
    $G^{u u}_{I}= \dfrac{y_{u d}y^{*}_{u b}}{M_{Y}^{2}}\dfrac{y_{u b}y^{*}_{u d}}{M_{Y}^{2}}$ &$1.7 \times 10^{-15}$ & Eq. (43f), Eq. (41a) \\
    \hline
    $G^{c c}_{I}= \dfrac{y_{c d}y^{*}_{c b}}{M_{Y}^{2}}\dfrac{y_{c b}y^{*}_{c d}}{M_{Y}^{2}}$  &$1.7 \times 10^{-15}$& Eq. (43f), Eq. (41a) \\
    \hline
    $G^{c u}_{I}= \dfrac{y_{u d}y^{*}_{c b}}{M_{Y}^{2}}\dfrac{y_{c b}y^{*}_{u d}}{M_{Y}^{2}}$  & $8.5 \times 10^{-16}$ & Eq. (43e), Eq. (41a) \\
    \hline
    % The rest of the rows
    $G^{uc}_{I}= \dfrac{y_{c d}y^{*}_{u b}}{M_{Y}^{2}}\dfrac{y_{u b}y^{*}_{c d}}{M_{Y}^{2}}$& $8.5 \times 10^{-16}$ & Eq. (43e), Eq. (41a) \\
    \hline
    $G^{s \psi}_{IV}= \dfrac{y_{\psi s}y_{u b}}{M_{Y}^{2}}\dfrac{y^{*}_{u b}y^{*}_{\psi s}}{M_{Y}^{2}}$ & $1 \times 10^{-13}$ & Table II \\
     \hline
    $G^{d \psi}_{IV}= \dfrac{y_{\psi d}y_{u b}}{M_{Y}^{2}}\dfrac{y^{*}_{u b}y^{*}_{\psi d}}{M_{Y}^{2}}$ & $1 \times 10^{-13}$ & Table II \\
    \hline
    $G^{\psi s}_{III}= \dfrac{y_{\psi b}y_{u s}}{M_{Y}^{2}}\dfrac{y^{*}_{u s}y^{*}_{\psi b}}{M_{Y}^{2}}$  & $3.7 \times 10^{-14}$ & Table II \\
    \hline
    $G^{\psi d}_{III}= \dfrac{y_{\psi b}y_{u d}}{M_{Y}^{2}}\dfrac{y^{*}_{u d}y^{*}_{\psi b}}{M_{Y}^{2}}$  & $4.7 \times 10^{-15}$ & Table II \\
    \hline
    $G^{\psi s}_{V}= \dfrac{y_{\psi s}y_{u b}}{M_{Y}^{2}}\dfrac{y^{*}_{u s}y^{*}_{\psi b}}{M_{Y}^{2}}$  & $6.3 \times 10^{-14}$ & Table II \\
    \hline
    $G^{\psi d}_{V}= \dfrac{y_{\psi d}y_{u b}}{M_{Y}^{2}}\dfrac{y^{*}_{u d}y^{*}_{\psi b}}{M_{Y}^{2}}$  & $2.2 \times 10^{-14}$ & Table II \\
    \hline
    $G^{\psi s}_{VI}= \dfrac{y_{\psi b}y_{u s}}{M_{Y}^{2}}\dfrac{y^{*}_{u b}y^{*}_{\psi s}}{M_{Y}^{2}}$  & $6.3 \times 10^{-14}$ & Table II \\
    \hline
    $G^{\psi d}_{VI}= \dfrac{y_{\psi b}y_{u d}}{M_{Y}^{2}}\dfrac{y^{*}_{u b}y^{*}_{\psi d}}{M_{Y}^{2}}$  & $2.2 \times 10^{-14}$ & Table II \\
    \hline
    $G^{cs}_{IV}= \dfrac{y_{c s}y^{*}_{u b}}{M_{Y}^{2}}\dfrac{y_{u b}y^{*}_{c s}}{M_{Y}^{2}}$ & $8.5 \times 10^{-15}$ & Eq. (43g), Eq. (41a) \\
    \hline
    $G^{us}_{IV}= \dfrac{y_{u s}y^{*}_{u b}}{M_{Y}^{2}}\dfrac{y_{u b}y^{*}_{u s}}{M_{Y}^{2}}$  &$8.5 \times 10^{-15}$& Eq. (43g), Eq. (41a) \\
     \hline
    $G^{cd}_{IV}= \dfrac{y_{c d}y^{*}_{u b}}{M_{Y}^{2}}\dfrac{y_{u b}y^{*}_{c d}}{M_{Y}^{2}}$ & $8.5 \times 10^{-16}$ & Eq. (43e), Eq. (41a) \\
    \hline
    $G^{ud}_{IV}= \dfrac{y_{u d}y^{*}_{u b}}{M_{Y}^{2}}\dfrac{y_{u b}y^{*}_{u d}}{M_{Y}^{2}}$ & $1.7 \times 10^{-15}$ & Eq. (43f), Eq. (41a) \\
     \hline
    $G^{cs}_{I}= \dfrac{y_{u s}y^{*}_{c b}}{M_{Y}^{2}}\dfrac{y_{c b}y^{*}_{u s}}{M_{Y}^{2}}$  & $8.5\times 10^{-15}$ & Eq. (43g), Eq. (41a)\\
     \hline
    $G^{us}_{I}= \dfrac{y_{u s}y^{*}_{u b}}{M_{Y}^{2}}\dfrac{y_{u b}y^{*}_{u s}}{M_{Y}^{2}}$  & $8.5\times 10^{-15}$ & Eq. (43g), Eq. (41a)\\
    \hline
    $G^{cd}_{I}= \dfrac{y_{u d}y^{*}_{c b}}{M_{Y}^{2}}\dfrac{y_{c b}y^{*}_{u d}}{M_{Y}^{2}}$  & $8.5\times 10^{-16}$ & Eq. (43e), Eq. (41a)\\
    \hline
    $G^{ud}_{I}= \dfrac{y_{u d}y^{*}_{u b}}{M_{Y}^{2}}\dfrac{y_{u b}y^{*}_{u d}}{M_{Y}^{2}}$  & $1.7\times 10^{-15}$ & Eq. (43f), Eq. (41a)\\
    
\end{longtable}
\begin{table}[h]
    \centering
    \renewcommand{\arraystretch}{2}
    \caption{ $\mathcal{H}_{2/3}$ Constants and their Max values, taken from Ref.~\cite{Alonso-Alvarez:2021qfd}.}
    \label{Constants2/3}
    \begin{tabular} { |>{\centering\arraybackslash}p{4.5cm} | >{\centering\arraybackslash}p{4cm} | >{\centering\arraybackslash}p{4cm}|}
        \hline
        Parameter & Max Value $\text{(GeV)}^{-4}$& Source \\
        \hline\hline
        $G^{u \psi}_{IV}= \dfrac{y_{bd}y_{\psi u}}{M_{Y}^{2}}\dfrac{y^{*}_{bd}y^{*}_{\psi u}}{M_{Y}^{2}}$& $2.5\times 10^{-13}$ & Table II \\
        \hline
        $G^{c \psi}_{IV}= \dfrac{y_{bd}y_{\psi c}}{M_{Y}^{2}}\dfrac{y^{*}_{bd}y^{*}_{\psi c}}{M_{Y}^{2}}$ & $2.5\times 10^{-13}$  &  Table II \\
        \hline
        $G^{d s}_{IV}= \dfrac{y_{ds}y^{*}_{b d}}{M_{Y}^{2}}\dfrac{y_{bd}y^{*}_{d s}}{M_{Y}^{2}}$ &  $8.5 \times 10^{-15}$  & Eq. (44d), Eq. (41a)\\
       \hline
        $G^{d s}_{I}= \dfrac{y_{ds}y^{*}_{b d}}{M_{Y}^{2}}\dfrac{y_{bd}y^{*}_{d s}}{M_{Y}^{2}}$ &  $8.5 \times 10^{-15}$  & Eq. (44d), Eq. (41a)\\
        \hline
        $G^{s s}_{I}= \dfrac{y_{ds}y^{*}_{b s}}{M_{Y}^{2}}\dfrac{y_{bs}y^{*}_{d s}}{M_{Y}^{2}}$ &  $1.7 \times 10^{-15}$  & Eq. (44c), Eq. (41a)\\
         \hline
        $G^{d \psi}_{III}= \dfrac{y_{bd}y_{\psi u}}{M_{Y}^{2}}\dfrac{y^{*}_{bd}y^{*}_{\psi u}}{M_{Y}^{2}}$ & $2.5\times 10^{-13}$ & Table II \\
        \hline
        $G^{s \psi}_{III}= \dfrac{y_{bs}y_{\psi u}}{M_{Y}^{2}}\dfrac{y^{*}_{bs}y^{*}_{\psi u}}{M_{Y}^{2}}$ & $1.2\times 10^{-12}$ & Table II \\
        \hline
    \end{tabular}
\end{table}

\section{\texorpdfstring{\boldmath Power-Suppressed Effects in the SM: A Comparison}{Power-suppressed Effects in the SM: A Comparison}}
\label{sec:SMeffects}
In \cref{sec:Mesolifetimes}, we showed that within the
Mesogenesis framework, power-suppressed contributions can exceed the
leading dimension-three term in the HQE in certain regions of the dark-sector
fermion mass $m_\psi$. It is instructive to compare this behaviour with
the corresponding situation in the SM. Since the
Mesogenesis analysis is performed under the assumption that a single
operator dominates at a time, we consider the analogous SM operator
${\cal O}=(u\,d)(c\,b)$, which induces the familiar
PI topology in the $B^+$ meson. We therefore study the HQE
contributions to the inclusive decay
$b\to c\bar{u}d$ as a function of the scaled charm-quark mass
$\rho_c=m_c^2/m_b^2$. The corresponding SM effective Hamiltonian reads
\begin{equation}
{\cal H}_{\rm eff}^{\rm SM}
=
\frac{4G_F}{\sqrt{2}}\,
V_{cb}V_{ud}^{*}
\left[
C_1(\mu_b)\,Q_1^{cud}
+
C_2(\mu_b)\,Q_2^{cud}
\right]
+{\rm h.c.}\, ,
\end{equation}
where $Q_{1,2}^{cud}$ are the current–current operators defined in~\cref{eq:Heff-SM}.
\begin{figure}[t]
    \centering
    \begin{subfigure}{0.45\textwidth}
        \centering
        \includegraphics[width=\textwidth]{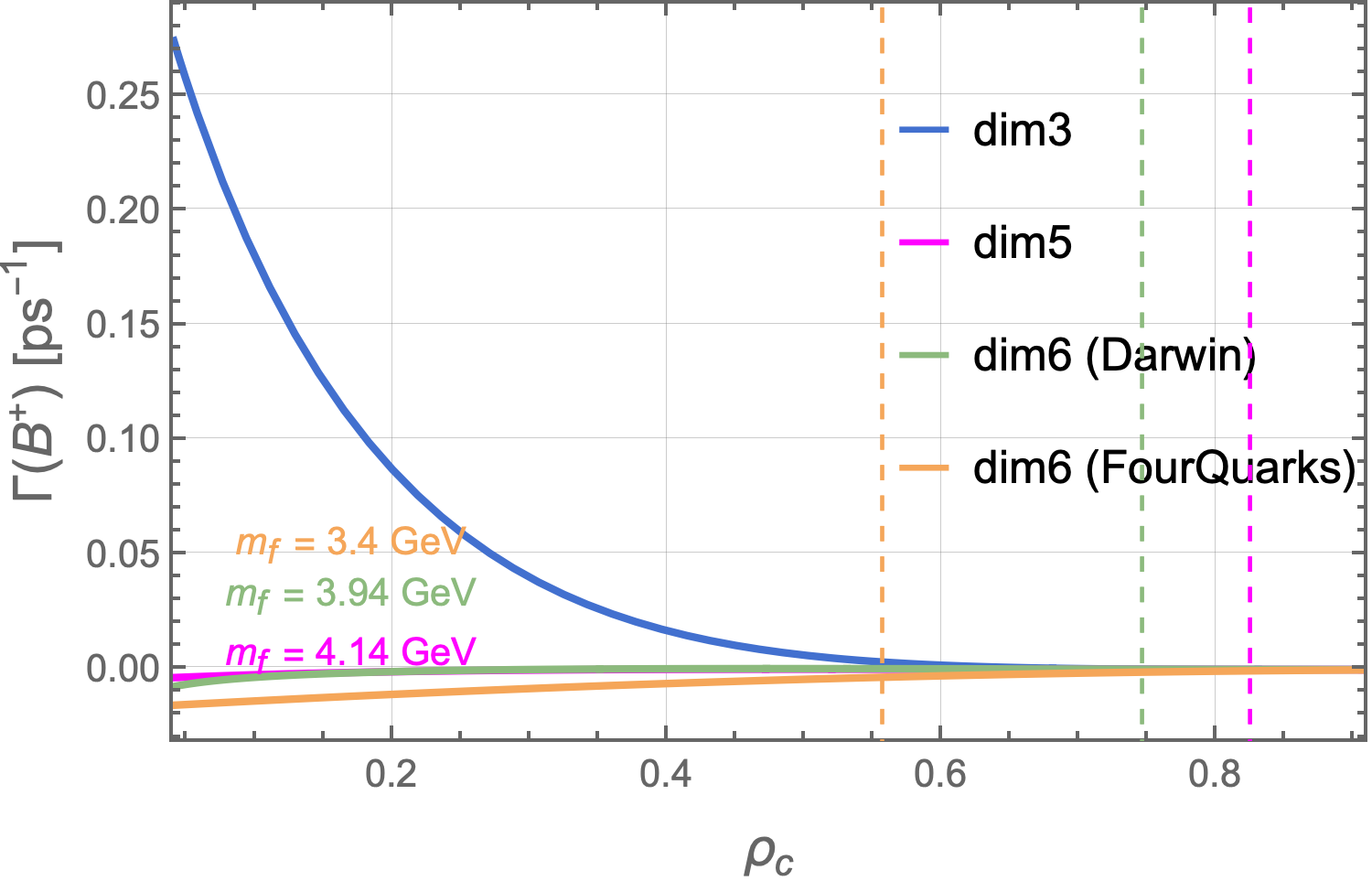}
    \end{subfigure}
    \hfill
    \begin{subfigure}{0.45\textwidth}
        \centering
        \includegraphics[width=\textwidth]{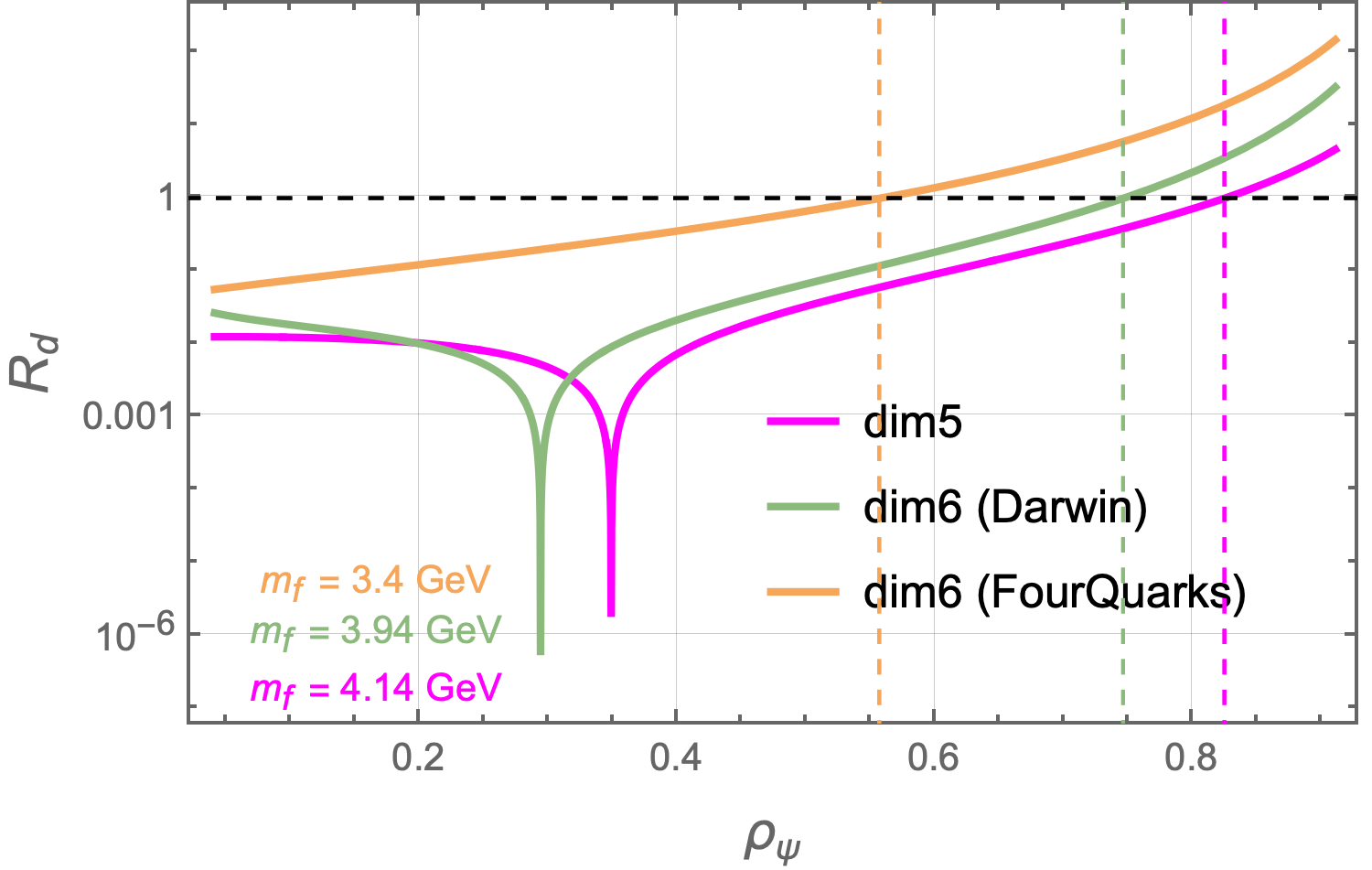}
    \end{subfigure}
    
   % \medskip % Adds a small vertical space between the rows
\caption{Total decay width $\Gamma(B^+)$ arising from the SM operator $(c b)( d u)$ , shown as a function of $\rho_f = (m_f/m_b)^2$. 
The plots display the individual HQE contributions (left) and the corresponding ratios $R_d = |\Gamma_d/\Gamma_3|$ (right).
The vertical lines indicate the values of $\rho_f$ where subleading terms exceed the leading contribution (the corresponding $m_f$ values are also indicated). 
For the dimension-six four-quark contribution, which originates from the PI topology and is negative across the full range, the line marks the point where its \textit{absolute value} exceeds $\Gamma_3$. 
Both $C_1$ and $C_2$ are included. The sharp spikes in the $R_d$ plot occur where the numerator, $\Gamma_d$, changes sign. }
\label{fig:DecayRatesSM}
\end{figure}
\begin{figure}[ht]
    \centering
    \begin{subfigure}{0.47\textwidth}
        \centering
        \includegraphics[width=\textwidth]{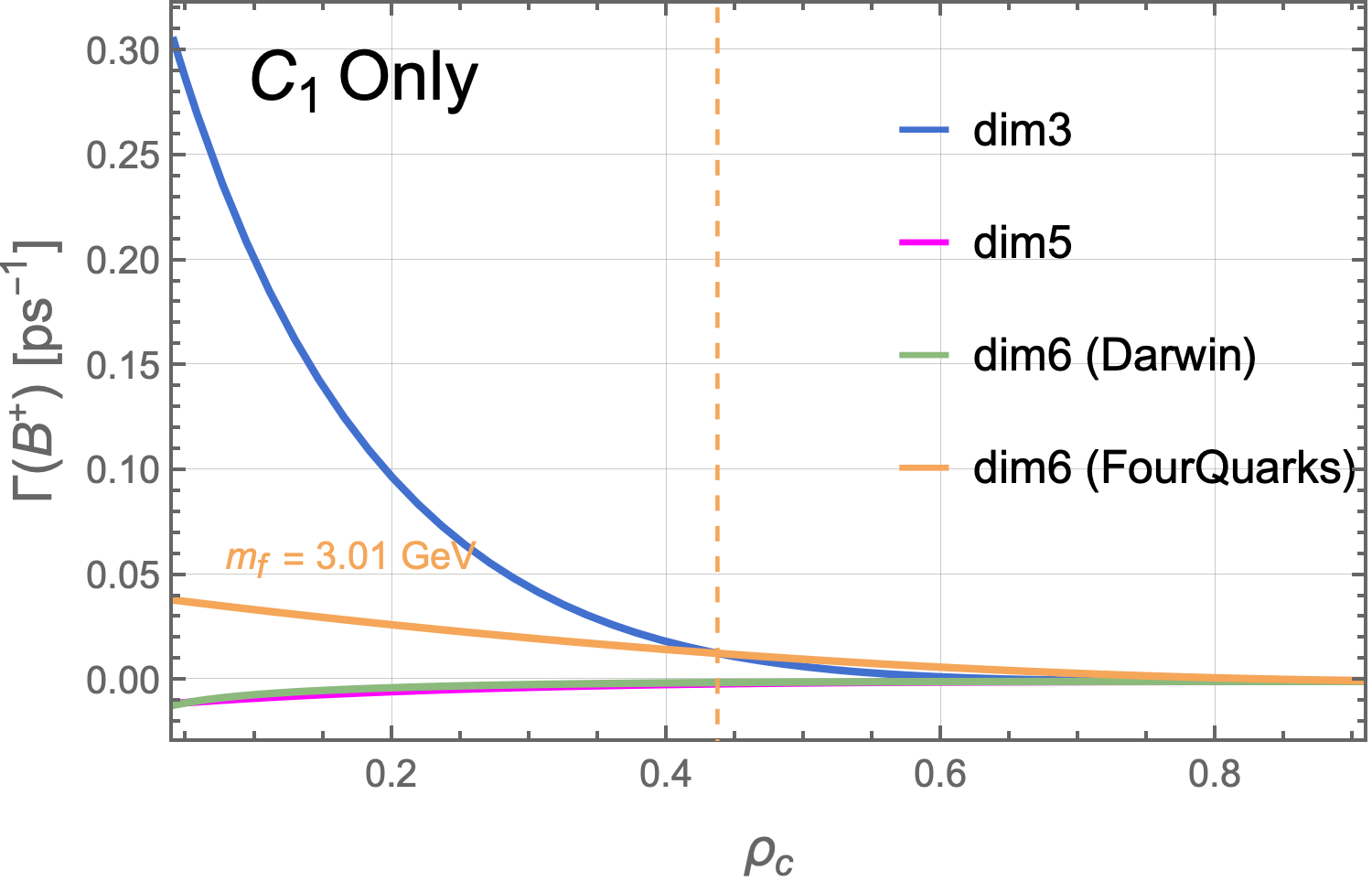}
    \end{subfigure}
    \hfill
    \begin{subfigure}{0.47\textwidth}
        \centering
        \includegraphics[width=\textwidth]{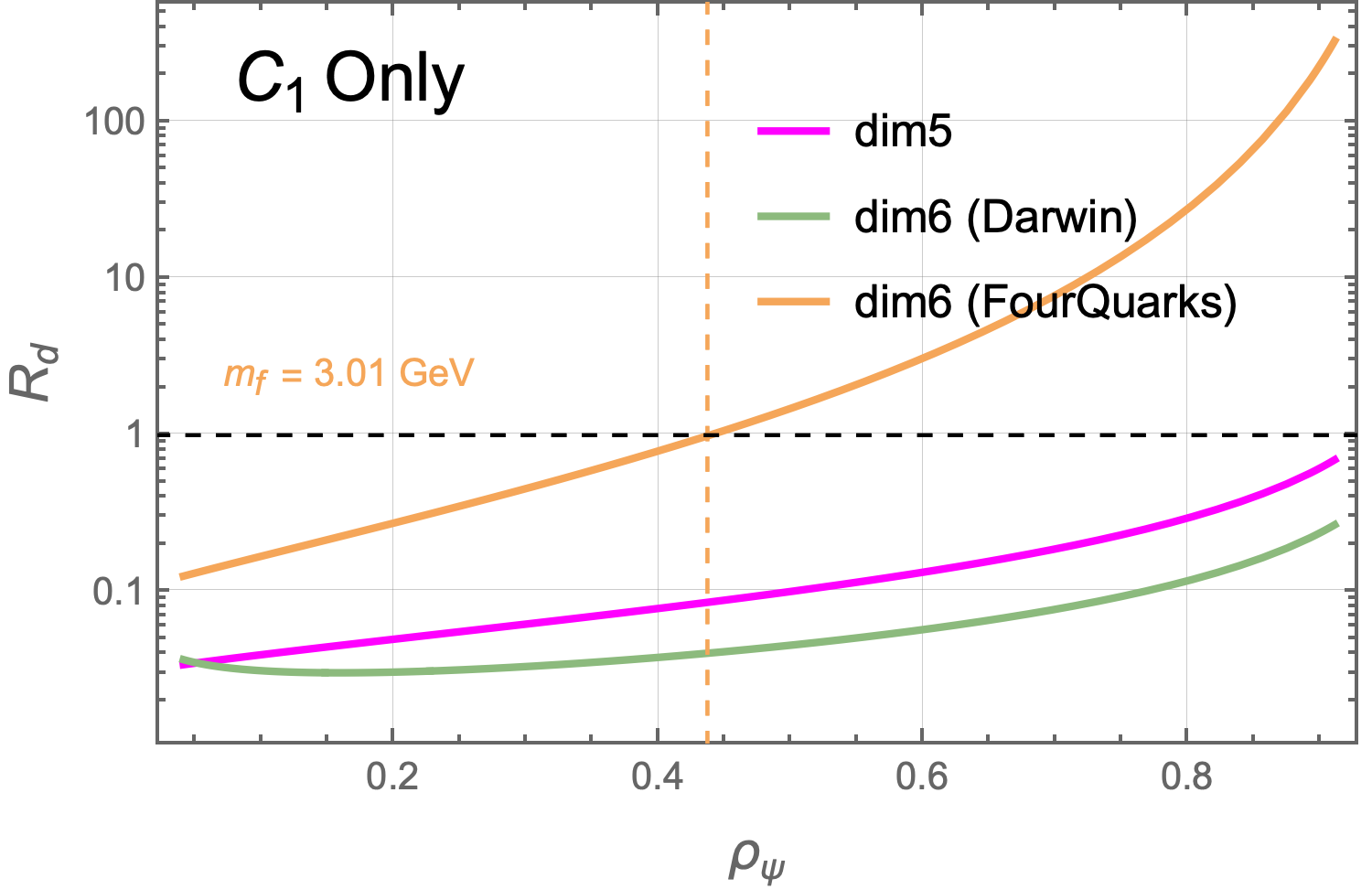}
    \end{subfigure}
    
   % \medskip % Adds a small vertical space between the rows

   \begin{subfigure}{0.47\textwidth}
        \centering
        \includegraphics[width=\textwidth]{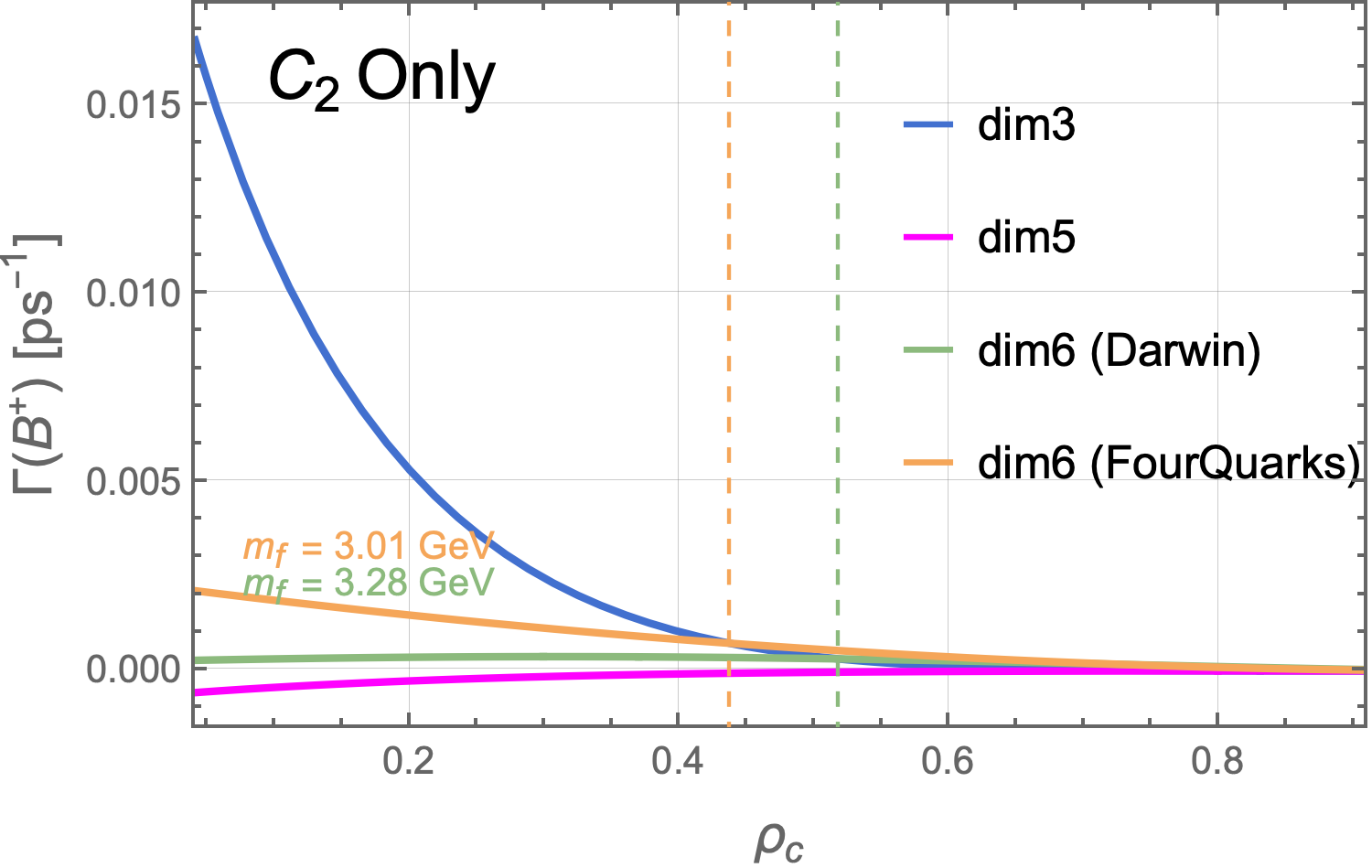}
    \end{subfigure}
    \hfill
    \begin{subfigure}{0.47\textwidth}
        \centering
        \includegraphics[width=\textwidth]{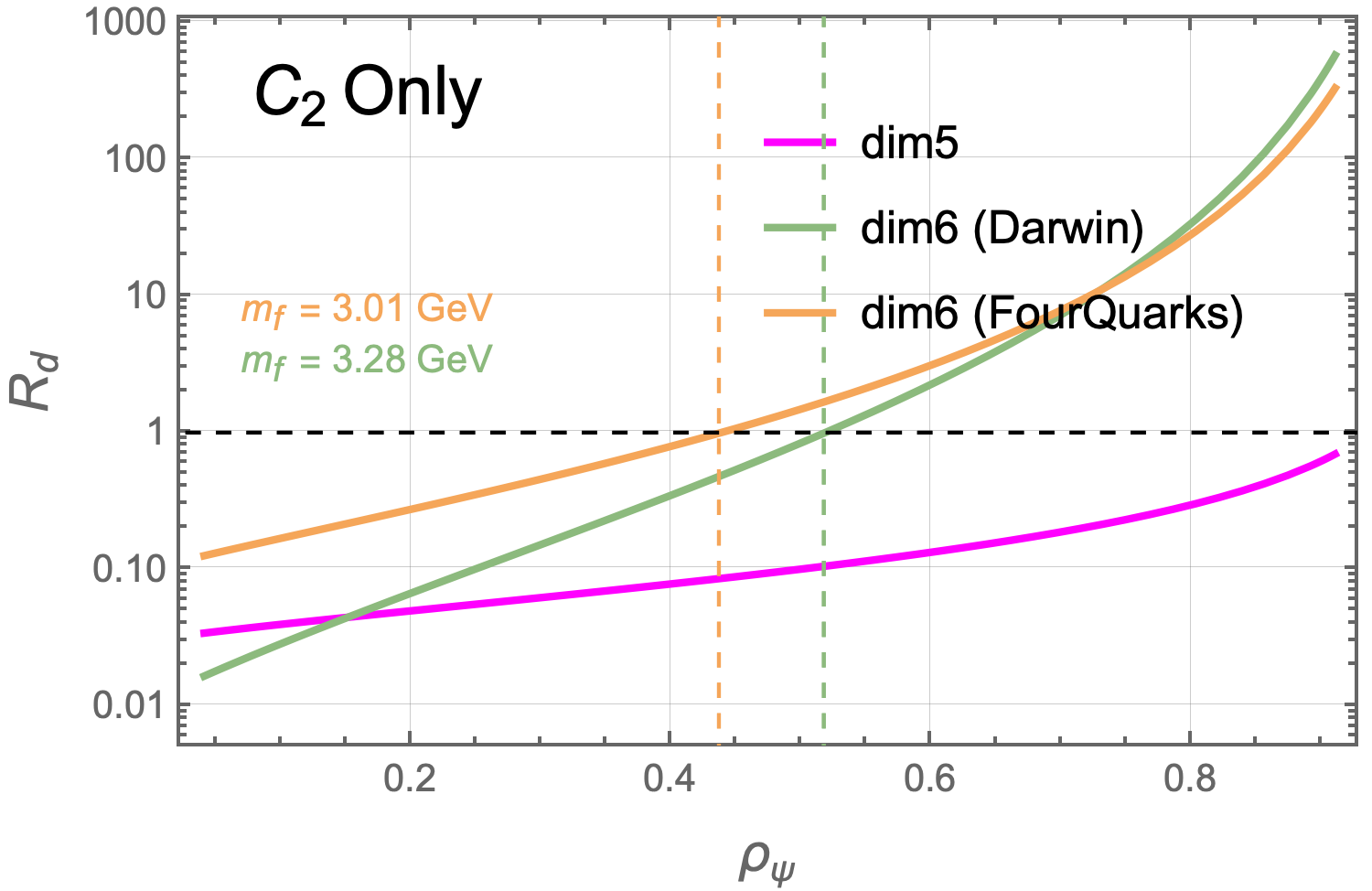}
    \end{subfigure}
\caption{Total decay width $\Gamma(B^+)$ arising from the SM operator $(c b)(d u)$, as a function of $\rho_f = (m_f/m_b)^2$. 
The plots show the individual HQE contributions (left) and the ratios $R_d = |\Gamma_d/\Gamma_3|$ (right). 
The upper panels display the case with only $C_1$ switched on, while the lower panels correspond to only $C_2$. 
Vertical lines mark the values of $\rho_f$ where subleading terms exceed the leading dimension-three width (the corresponding $m_f$ values are also indicated). }
    \label{fig:DecayRatesSMC1C2}
\end{figure}

Using the LO Wilson coefficients
$C_1(4.5\,\GeV)=1.11$ and
$C_2(4.5\,\GeV)=-0.26$ from Ref.~\cite{Egner:2024lay},
together with the numerical inputs collected in \cref{sec:BMeso_input}, we show in~\cref{fig:DecayRatesSM} the individual HQE contributions and their
ratios to the leading dimension-three term.
By varying the charm-quark mass over the kinematic range analogous to
that spanned by the dark-fermion mass $m_\psi$ in the Mesogenesis
analysis, we observe a qualitatively similar hierarchy: in certain
regions of parameter space, power-suppressed contributions become larger
than the leading term. This demonstrates that the breakdown of the
HQE hierarchy observed in the Mesogenesis scenario is not a
peculiarity of that model, but rather a generic feature of the HQE when
the relevant mass scales approach the kinematic boundaries of the decay.

Despite this qualitative similarity, the size of the power corrections differs between the two scenarios. 
In the SM, the dimension-six four-quark contributions generated by PI topology remain modest compared with the leading dimension-three term throughout the physical parameter space. 
In the Mesogenesis framework, however, the corresponding $\overline{\rm PI}$ topology can produce much larger corrections, causing the power-suppressed terms to compete with—or even exceed—the leading contribution, as demonstrated in~\cref{fig:DecayRates,fig:DecayRatesTotal}.

To isolate the differences between the SM PI topology and
the corresponding $\overline{\rm PI}$ topology in Mesogenesis, we first
eliminate the interference between the two current--current operators by
setting $C_2=0$. The SM Hamiltonian then reduces to
\begin{equation}
 {\cal H}_{b \to c u d}^{\rm SM}(x)
=
\widetilde{G}_F\,
(\bar c^i \gamma_\mu P_L b^i)
(\bar d^j \gamma^\mu P_L u^j)(x)
+{\rm h.c.},
\end{equation}
where
\(
\widetilde{G}_F
=
\tfrac{4G_F}{\sqrt2}V_{cb}V_{ud}^*C_1.
\)
The corresponding Mesogenesis operator is
\begin{equation}
\mathcal H^{-1/3}_{b\to \psi ud}(x)
=
C_{\psi}^{(ud)}
\,\epsilon_{ijk}\,
(\bar u_R^{\,i} b_R^{c,j})
(\bar\psi\, d_R^{c,k})(x)
+{\rm h.c.}.
\end{equation}

The two operators differ only in their Dirac and colour structures.
These differences can be identified separately already at leading order.
Replacing
\(
|C_\psi^{(ud)}|^2\rightarrow |\widetilde G_F|^2
\),
the SM dimension-three contribution is related to the corresponding
Mesogenesis result by
\begin{equation}
\Gamma_3^{\rm SM}(B^+)
=
\Gamma_3^{-1/3}(B^+)
\times
\underbrace{4}_{\rm Dirac}
\times
\underbrace{\frac32}_{\rm color}.
\end{equation}

The same Dirac and colour structures also determine the dimension-six
four-quark contributions. In the SM, the PI contribution is
\begin{equation}
\Gamma_{\rm PI}^{\rm SM}
\propto
4
\left(
\frac{1}{N_c}\tilde Q_1
+
2\tilde Q_3
\right),
\end{equation}
where the operators $\tilde Q_{1,3}$ are defined in \cref{eq:operators4q}. By contrast, the corresponding
$\overline{\rm PI}$ contribution in Mesogenesis reads
\begin{equation}
\Gamma_{\overline{\rm PI}}^{-1/3}
\propto
\left(
\frac{N_c-1}{N_c}\tilde Q_1
-
2\tilde Q_3
\right)\Big|_{P_L\to P_R}.
\end{equation}

The SM contribution retains the overall factor of $4$ originating from
the Dirac structure. The colour structure, however, differs from that of
the Mesogenesis operator, leading to a different linear combination of
the four-quark operators. 

Employing the parametrisation of the forward
matrix elements in terms of the bag parameters listed in~\cref{tab:Bag-parameters}, we find that the combination
$\left(\frac{N_c-1}{N_c}\right)\tilde{B}_1^u-2\tilde{B}_3^u$
appearing in Mesogenesis is approximately twice as large in magnitude as
the corresponding SM combination
$\left(\frac{1}{N_c}\right)\tilde{B}_1^u+2\tilde{B}_3^u.$
This modified colour structure therefore enhances the dimension-six
four-quark contribution in Mesogenesis relative to the SM.

Finally, reinstating the full SM operator basis by including the
interference with the operator proportional to $C_2$ further suppresses
the dimension-six four-quark contribution. Numerically, the magnitude of
the PI term is reduced by approximately a factor of two, shifting the
point at which it overtakes the leading dimension-three contribution
from $\rho_c \simeq 0.4$ (for $C_2=0$) to
$\rho_c \simeq 0.6$ (for the physical values of $C_1$ and $C_2$), as
shown in~\cref{fig:DecayRatesSMC1C2}. No analogous cancellation
occurs in the Mesogenesis framework, where a single operator is assumed
to dominate, allowing the enhanced bag-parameter combination to remain
fully effective.

Taken together, these observations explain why the relative size of the
$\overline{\rm PI}$ contribution is substantially larger in Mesogenesis
than the corresponding PI contribution in the SM. Two effects are
responsible for this enhancement. First, the different colour structure
of the Mesogenesis operator selects a more favorable combination of bag
parameters. Second, under the assumption of single-operator dominance,
the Mesogenesis contribution is not subject to the destructive
interference between current--current operators that further suppresses
the PI contribution in the SM. Therefore, the ratio
$\Gamma_{\rm dim6}/\Gamma_{\rm dim3}$ is systematically enhanced in the
Mesogenesis framework, allowing the dimension-six contribution to exceed
the leading dimension-three term over sizable regions of parameter
space. For calculations involving BSM operators with the same colour structure as the SM but
different Dirac structures, see Ref.~\cite{Lenz:2022pgw}.

\section{\texorpdfstring{$B$-meson Distribution Amplitudes}{B-Meson Distribution Amplitudes}}
\label{app:BDA}

\subsection{\texorpdfstring{Two-Particle $B$-meson DAs}{Two-particle B-meson DAs}}
\label{app:BDA2}
The nonlocal vacuum-to-$B$ matrix element appearing in
\cref{eq:corr_coord} is parameterised in terms of the two-particle
$B$-meson light-cone distribution amplitudes (DAs). Employing the
HQET decomposition of the heavy-quark field, we use the standard
parametrisation~\cite{Grozin:1996pq} (for its derivation, see \cref{sec:DA_derive}) \footnote{The parametrisation given here differs by an overall minus sign from that used in Ref.~\cite{Biswas:2026oxq}. This originates from a different convention for the charge-conjugation matrix, where Ref.~\cite{Biswas:2026oxq} adopts $C=\gamma^2\gamma^0$. While such an overall phase does not affect physical observables, since the form factors enter quadratically in the branching fraction (see \cref{eq:widthb}), the same convention also enters the perturbative derivation of the correlation function in \cref{eq:corr} through the charge-conjugation identities. Consequently, the overall sign is compensated, and the final results are identical.}
\begin{align}
i\langle 0|
\bar h_{v\alpha}(0)[0,x]\,u_\beta(x)
|\bar B(v)\rangle
&=
\frac{f_B m_B}{4}
\int\limits_0^\infty\!\!d\omega\,
e^{-i\omega v\cdot x}
\Bigg[
\Big\{
\phi_+^B(\omega)
+
x^2 g_+^B(\omega)
\Big\}
\Big[
(1+\slashed v)\gamma_5
\Big]_{\beta\alpha}
\nonumber\\
&\hspace{-1.8cm}
-\frac{1}{2v\cdot x}
\Big\{
\phi_+^B(\omega)
-
\phi_-^B(\omega)
+
x^2\big(
g_+^B(\omega)-g_-^B(\omega)
\big)
\Big\}
\Big[
\slashed x(1+\slashed v)\gamma_5
\Big]_{\beta\alpha}
\Bigg] ,
\label{eq:BDA}
\end{align}
where $[0,x]$ denotes the gauge link ensuring gauge invariance, and
$\omega$ corresponds to the light-cone projection of the spectator
quark momentum in the $B$-meson rest frame.

The functions
$\phi_+^B(\omega)$,
$\phi_-^B(\omega)$,
$g_+^B(\omega)$, and
$g_-^B(\omega)$
represent the two-particle $B$-meson DAs of twist two, three, four,
and five, respectively, following the classification of
\cite{Braun:2017liq}. The twist-two and twist-three DAs satisfy the
normalisation conditions
\begin{equation}
\int_0^\infty d\omega\,
\phi_\pm^B(\omega)
=
1\,.
\label{eq:DAnorm}
\end{equation}

To eliminate the explicit factors $1/(v\cdot x)$ appearing in
\cref{eq:BDA}, it is convenient to introduce the integrated
combination~\cite{Khodjamirian:2006st}
\begin{equation}
\Phi_\pm^B(\omega)
\equiv
\int_0^\omega d\tau\,
\big(
\phi_+^B(\tau)-\phi_-^B(\tau)
\big),
\label{eq:PhiDef}
\end{equation}
such that
\[
\frac{d}{d\omega}\Phi_\pm^B(\omega)
=
\phi_+^B(\omega)-\phi_-^B(\omega),
\qquad
\Phi_\pm^B(0)=\Phi_\pm^B(\infty)=0.
\]
Analogously, we define
\begin{equation}
G_\pm^B(\omega)
\equiv
\int_0^\omega d\tau\,
\big(
g_+^B(\tau)-g_-^B(\tau)
\big).
\label{eq:GDef}
\end{equation}

For the numerical analysis we adopt the standard exponential model
for the twist-two DA~\cite{Grozin:1996pq},
\begin{equation}
\phi_+^B(\omega)
=
\frac{\omega}{\lambda_B^2}
e^{-\omega/\lambda_B}\,,
\label{eq:phiBplus}
\end{equation}
while for the twist-three and twist-four DAs we employ the
corresponding exponential ansatz proposed in
\cite{Braun:2017liq}
\begin{align}
\phi_-^B(\omega)
&=
\frac{1}{\lambda_B}
e^{-\omega/\lambda_B}
+
\frac{1}{2\lambda_B}
\left[
1-\frac{2\omega}{\lambda_B}
+\frac{\omega^2}{2\lambda_B^2}
\right]
\frac{1-R}{1+2R}
e^{-\omega/\lambda_B},
\label{eq:phiBminus}
\\
g_+^B(\omega)
&=
\frac{3\omega^2}{16\lambda_B}
\frac{3+4R}{1+2R}
e^{-\omega/\lambda_B}.
\label{eq:gBplus}
\end{align}

For the twist-five DA we employ the Wandzura--Wilczek inspired model
used in \cite{Gubernari:2018wyi},
\begin{equation}
g_-^B(\omega)
=
\frac{3\omega}{4}
e^{-\omega/\lambda_B}.
\label{eq:gBminus}
\end{equation}

To assess the numerical impact of the twist-five contribution, we also
consider the approximation in which the term proportional to
$g_-^B$ is neglected. In this case, one simply replaces
\begin{equation}
G_\pm^B(\omega)
\rightarrow
G_+^B(\omega)
=
\int_0^\omega d\tau\,
g_+^B(\tau)
\label{eq:GplusOnly}
\end{equation}
in the final expressions for the invariant amplitudes.

The set of $B$-meson DAs employed in this work depends on three
nonperturbative parameters: the decay constant $f_B$, the inverse
moment of the twist-two DA,
\begin{equation}
\frac{1}{\lambda_B}
=
\int_0^\infty d\omega\,
\frac{\phi_+^B(\omega)}{\omega},
\label{eq:lambdaB}
\end{equation}
and the ratio
\begin{equation}
R\equiv\frac{\lambda_E^2}{\lambda_H^2},
\label{eq:Rratio}
\end{equation}
where $\lambda_E$ and $\lambda_H$ parameterize the matrix elements of
local heavy-light quark-antiquark-gluon operators in HQET.
\subsection{\texorpdfstring{Three-Particle $B$-meson DAs}{Three-particle B-meson DAs}}
\label{app:BDA3}

In addition to the two-particle contributions discussed above, the OPE
also receives corrections from three-particle quark-antiquark-gluon
$B$-meson DAs. For completeness, we summarise here the corresponding
definitions and conventions used in the present work.

We employ the HQET parametrisation of the vacuum-to-$B$ matrix
element with one gluon-field insertion derived in
\cite{Braun:2017liq}. More specifically, we use the form presented in
\cite{Descotes-Genon:2019bud} for the $\bar B$ meson, rewritten with
explicit Dirac indices and charge-conjugated to the $B^+$ matrix
element, adopting the same charge-conjugation convention as in the
two-particle DA parametrisation in \cref{eq:BDA}:
\begin{align}
&
\langle 0|
\bar h_{v\alpha}(0)\,
g_s\frac{\lambda^a}{2}
G^a_{\mu\nu}(ux)\,
u_\beta(x)
|B^+(v)\rangle
=
\frac{f_Bm_B}{4}
\int_0^\infty d\omega_1\,d\omega_2\,
e^{-i(\omega_1+u\omega_2)v\cdot x}
\nonumber\\
&\times
\Bigg\{
\gamma_5
\Bigg[
-(v_\mu\gamma_\nu-v_\nu\gamma_\mu)\phi_3
-\frac{i}{2}\sigma_{\mu\nu}(\phi_3-\phi_4)
+\frac{x_\mu v_\nu-x_\nu v_\mu}{2v\cdot x}
(\phi_3+\phi_4-2\psi_4)
\nonumber\\
&\qquad
+\frac{x_\mu\gamma_\nu-x_\nu\gamma_\mu}{2v\cdot x}
(\phi_3+\widetilde\psi_4)
+\frac{i\epsilon_{\mu\nu\alpha\beta}x^\alpha v^\beta}
{2v\cdot x}\gamma_5
(\phi_3-\phi_4+2\widetilde\psi_4)
\nonumber\\
&\qquad
-\frac{i\epsilon_{\mu\nu\alpha\beta}x^\alpha}
{2v\cdot x}\gamma^\beta\gamma_5
(\phi_3-\phi_4+\widetilde\psi_4)
+\frac{(x_\mu v_\nu-x_\nu v_\mu)\slashed x}
{2(v\cdot x)^2}
(\phi_4-\psi_4-\widetilde\psi_4)
\nonumber\\
&\qquad
-\frac{\slashed x(x_\mu\gamma_\nu-x_\nu\gamma_\mu)}
{4(v\cdot x)^2}
(\phi_3-\phi_4+2\widetilde\psi_4)
\Bigg]
(1-\slashed v)
\Bigg\}_{\beta\alpha}.
\label{eq:def3DAsBpl}
\end{align}
Here $\phi_3$ denotes the twist-three DA, while
$\phi_4$, $\psi_4$, and $\widetilde\psi_4$ are twist-four DAs.
For brevity, their dependence on
$(\omega_1,\omega_2)$ is suppressed, and the Wilson lines on the
left-hand side are not shown explicitly.

Similarly to the two-particle case, it is convenient to eliminate the
explicit factors of $1/(v\cdot x)$ by introducing integrated DAs:
\begin{equation}
\overline\Phi(\omega_1,\omega_2)
\equiv
\int_0^{\omega_1}
d\tau\,
\Phi(\tau,\omega_2),
\label{eq:barPhi}
\end{equation}
where
\[
\Phi=
\{
\phi_3,\phi_4,\psi_4,\widetilde\psi_4
\}.
\]
To remove the terms proportional to $1/(v\cdot x)^2$, we further
define the doubly integrated DAs
\begin{equation}
\overline{\overline\Phi}(\omega_1,\omega_2)
\equiv
\int_0^{\omega_1}
d\tau\,
\overline\Phi(\tau,\omega_2).
\label{eq:doublebarPhi}
\end{equation}

Using these definitions, the matrix element
\cref{eq:def3DAsBpl} can be rewritten in a more compact form free of
inverse powers of $(v\cdot x)$:
\begin{align}
&
\langle 0|
\bar h_{v\alpha}(0)\,
g_s\frac{\lambda^a}{2}
G^a_{\mu\nu}(ux)\,
u_\beta(x)
|B^+(v)\rangle
=
-\frac{f_Bm_B}{4}
\int_0^\infty d\omega_1\,d\omega_2\,
e^{-i\sigma m_B v\cdot x}
\nonumber\\
&\times
\Bigg\{
\gamma_5
\Bigg[
-(v_\mu\gamma_\nu-v_\nu\gamma_\mu)\phi_3
-\frac{i}{2}\sigma_{\mu\nu}(\phi_3-\phi_4)
+\frac{i}{2}(x_\mu v_\nu-x_\nu v_\mu)
(\overline\phi_3+\overline\phi_4-2\overline\psi_4)
\nonumber\\
&\qquad
+\frac{i}{2}(x_\mu\gamma_\nu-x_\nu\gamma_\mu)
(\overline\phi_3+\overline{\widetilde\psi}_4)
-\frac12
\epsilon_{\mu\nu\alpha\beta}
x^\alpha v^\beta\gamma_5
(\overline\phi_3-\overline\phi_4+2\overline{\widetilde\psi}_4)
\nonumber\\
&\qquad
+\frac12
\epsilon_{\mu\nu\alpha\beta}
x^\alpha\gamma^\beta\gamma_5
(\overline\phi_3-\overline\phi_4+\overline{\widetilde\psi}_4)
-\frac12
(x_\mu v_\nu-x_\nu v_\mu)\slashed x
(
\overline{\overline\phi}_4
-
\overline{\overline\psi}_4
-
\overline{\overline{\widetilde\psi}}_4
)
\nonumber\\
&\qquad
+\frac14
\slashed x(x_\mu\gamma_\nu-x_\nu\gamma_\mu)
(
\overline{\overline\phi}_3
-
\overline{\overline\phi}_4
+
2\overline{\overline{\widetilde\psi}}_4
)
\Bigg]
(1-\slashed v)
\Bigg\}_{\beta\alpha}.
\label{eq:3DA_compact}
\end{align}
where we introduced
\begin{equation}
\sigma(u,\omega_1,\omega_2)
\equiv
\frac{\omega_1+u\omega_2}{m_B}.
\label{eq:sigma}
\end{equation}

For completeness, we also quote the exponential model for the
three-particle DAs proposed in \cite{Braun:2017liq}:
\begin{align}
\phi_3(\omega_1,\omega_2)
&=
-\frac{3\omega_1\omega_2^2}
{4\lambda_B^3}
\frac{1-R}{1+2R}
e^{-(\omega_1+\omega_2)/\lambda_B},
\nonumber\\
\phi_4(\omega_1,\omega_2)
&=
\frac{3\omega_2^2}
{4\lambda_B^2}
\frac{1+R}{1+2R}
e^{-(\omega_1+\omega_2)/\lambda_B},
\nonumber\\
\psi_4(\omega_1,\omega_2)
&=
R\,\widetilde\psi_4(\omega_1,\omega_2)
=
\frac{3\omega_1\omega_2}
{2\lambda_B^2}
\frac{R}{1+2R}
e^{-(\omega_1+\omega_2)/\lambda_B}.
\label{eq:3DAsexpmod}
\end{align}

All three-particle DAs in this model are therefore determined by the
same nonperturbative parameters $\lambda_B$ and $R$ already entering
the two-particle DAs.

\subsection{\texorpdfstring{$B^+$}{B+}-meson DAs from the \texorpdfstring{$B^-$}{B-}-meson Parametrisation}
\label{sec:DA_derive}
In this section, we obtain the parametrisation of the $B^+$-meson
light-cone DAs from the corresponding
$B^-$-meson parametrisation using charge-conjugation symmetry. 

We start from the parametrisation in Ref.~\cite{Beneke:2000wa}
\begin{align}
\langle0|
\bar q_\beta(z)\,b_\alpha(0)
|B^{-}(v)\rangle
&=
-\frac{i f_B M_B}{4}
\int_0^\infty d\omega\,
e^{-i\omega v\cdot z}
\nonumber\\
&\qquad\times
\Biggl[
(1+\slashed v)
\left(
\phi_+^B(\omega)
-
\frac{\phi_+^B(\omega)-\phi_-^B(\omega)}
{2\,v\!\cdot\! z}
\,\slashed z
\right)
\gamma_5
\Biggr]_{\alpha\beta}.
\end{align}
For simplicity, the gauge link between the quark fields is suppressed throughout this derivation.
We write this in the form
\begin{align}
\langle0|
\bar q_\beta(z)\,b_\alpha(0)
|B^{-}(v)\rangle
=
A\,
\bigl[(1+\slashed v)\gamma_5\bigr]_{\alpha\beta}
+
B\,
\bigl[(1+\slashed v)\slashed z\gamma_5\bigr]_{\alpha\beta},
\end{align}
with
\begin{align}
A
&=
-\frac{i f_B M_B}{4}
\int_0^\infty d\omega\,
e^{-i\omega v\cdot z}\,
\phi_+^B(\omega),
\\
B
&=
\frac{i f_B M_B}{8\,v\!\cdot\! z}
\int_0^\infty d\omega\,
e^{-i\omega v\cdot z}
\left(
\phi_+^B(\omega)-\phi_-^B(\omega)
\right).
\end{align}
We now obtain the corresponding $B^+$ matrix element by applying
charge conjugation with the operator $\widehat C$:
\begin{align}
\langle0|
\bar q_\beta(z)\,b_\alpha(0)
|B^{-}\rangle
&=
\langle0|
\widehat C^{-1}
\underbrace{\widehat C\,\bar q_\beta(z)\,\widehat C^{-1}}
\underbrace{\widehat C\,b_\alpha(0)\,\widehat C^{-1}}
\widehat C
|B^{-}\rangle .
\end{align}
Using (see the relations in \cref{conjugate_properties})\footnote{Note that $\widehat C$ denotes the charge-conjugation \emph{operator}, whereas $C$ denotes the corresponding charge-conjugation \emph{matrix} acting on Dirac indices.}
\begin{align}
\widehat C |B^{-}\rangle &= |B^{+}\rangle,
\\
\widehat C\,\bar q_\beta(z)\,\widehat C^{-1}
&=
- C_{\beta\beta'}\,q_{\beta'}(z),
\\
\widehat C\,b_\alpha(0)\,\widehat C^{-1}
&=
C^{-1}_{\alpha'\alpha}\,\bar b_{\alpha'}(0),
\end{align}
one obtains
\begin{align}
\langle0|
\bar b_{\alpha'}(0)\,q_{\beta'}(z)
|B^{+}\rangle
C^{-1}_{\alpha'\alpha}
C_{\beta\beta'}
=
A\,
\bigl[(1+\slashed v)\gamma_5\bigr]_{\alpha\beta}
+
B\,
\bigl[(1+\slashed v)\slashed z\gamma_5\bigr]_{\alpha\beta}.
\end{align}
Multiplying by $C_{\alpha\alpha_2}$ and $C^{-1}_{\beta_2\beta}$ gives
\begin{align}
\langle0|
\bar b_{\alpha_2}(0)\,q_{\beta_2}(z)
|B^{+}\rangle
=
A\,
\Bigl[
C(1+\slashed v)\gamma_5 C^{-1}
\Bigr]_{\alpha_2\beta_2}
+
B\,
\Bigl[
C(1+\slashed v)\slashed z\gamma_5 C^{-1}
\Bigr]_{\alpha_2\beta_2}.
\end{align}

\begin{align}
C(1+\slashed v)\gamma_5 C^{-1}
&=
(1-\slashed v^{T})\gamma_5^{T}
=
\bigl[\gamma_5(1-\slashed v)\bigr]^T
=
\bigl[(1+\slashed v)\gamma_5\bigr]^T,
\\
C(1+\slashed v)\slashed z\gamma_5 C^{-1}
&=
(1-\slashed v^{T})(-\slashed z^{T})\gamma_5^{T}
=
-\bigl[\gamma_5\slashed z(1-\slashed v)\bigr]^T
\nonumber\\
&=
\bigl[\slashed z(1+\slashed v)\gamma_5\bigr]^T .
\end{align}
Therefore,
\begin{align}
\langle0|
\bar b_\alpha(0)\,q_\beta(z)
|B^{+}\rangle
=
A\,
\bigl[(1+\slashed v)\gamma_5\bigr]_{\beta\alpha}
+
B\,
\bigl[\slashed z(1+\slashed v)\gamma_5\bigr]_{\beta\alpha}.
\end{align}

Finally, substituting the expressions for $A$ and $B$ yields
\begin{align}
\langle0|
\bar b_\alpha(0)\,q_\beta(z)
|B^{+}(v)\rangle
&=
-\frac{i f_B M_B}{4}
\int_0^\infty d\omega\,
e^{-i\omega v\cdot z}
\nonumber\\
&\qquad\times
\Biggl[
\phi_+^B(\omega)\,(1+\slashed v)\gamma_5
-
\frac{\phi_+^B(\omega)-\phi_-^B(\omega)}
{2\,v\!\cdot\! z}
\,\slashed z(1+\slashed v)\gamma_5
\Biggr]_{\beta\alpha},
\end{align}
which coincides with the expression in Ref.~\cite{Grozin:1996pq}.
\section{\textit{B}-Mesogenesis Additional Results}
\label{app:BMesogenesisResutls}
\subsection{Inclusive Results from HQE}
\label{sec:MesoInclResults}
In this section we collect the explicit expressions for the coefficients
entering
\cref{eq:MasterWidth_meso_2q,eq:Meso4qresultsB+,eq:Meso4qresultsBq}.

Throughout this appendix, we employ the abbreviations
\begin{equation}
\Delta_{\rho_\psi}\equiv\sqrt{1-4\rho_\psi},
\qquad
\Lambda_{\rho_\psi}\equiv
\log\!\left(
\frac{1+\Delta_{\rho_\psi}}
     {1-\Delta_{\rho_\psi}}
\right).
\end{equation}
\\
We start with the two-quark operators. \\\\
For $C_{\psi\psi}^{(d)}$:
\begin{flalign}
f^{(\psi\psi d)}_0
&=
\frac{1}{2}\Big[
\Delta_{\rho_\psi}
\bigl(
1-14\rho_\psi-2\rho_\psi^2-12\rho_\psi^3
\bigr)
+
24\rho_\psi^2(1-\rho_\psi^2)\,
\Lambda_{\rho_\psi}
\Big] \, & \\[4pt] 
  f^{(\psi\psi d)}_G
&=
\frac{1}{4}\Big[
24\rho_\psi^2(1-5\rho_\psi^2)\,
\Lambda_{\rho_\psi}
-
\Delta_{\rho_\psi}
\bigl(
3-10\rho_\psi+10\rho_\psi^2+60\rho_\psi^3
\bigr)
\Big] \, & \\[4pt]  \nonumber f^{(\psi\psi d)}_D
&=
\frac{1}{3}\Bigg[
\Delta_{\rho_\psi}
\bigl(
9 + 22\rho_\psi - 34\rho_\psi^2 - 60\rho_\psi^3
\bigr)
+ 24\Delta_{\rho_\psi}
\log\!\left(\frac{\rho_\psi}{1-4\rho_\psi}\right)
\\
&\qquad
+ 24\bigl(
1 - 2\rho_\psi - \rho_\psi^2
- 2\rho_\psi^3 - 5\rho_\psi^4
\bigr)
\Lambda_{\rho_\psi}
\Bigg] \, .
\end{flalign}\\
For $C_{cu}^{(d)}$:
\begin{flalign}
f^{(cu d)}_0
&= -2\Big[
\rho_c^4 - 8\rho_c^3 + 8\rho_c
+ 12\rho_c^2\log\rho_c - 1
\Big],
&&\\[2pt]
f^{(cu d)}_G
&= -5\rho_c^4 + 28\rho_c^3 - 48\rho_c^2 + 20\rho_c
- 12(\rho_c-2)\rho_c\log\rho_c + 5,
&&\\[2pt]
f^{(cu d)}_D
&= \frac{4}{3}\Big[
-5\rho_c^4 + 26\rho_c^3 - 45\rho_c^2 - 10\rho_c
- 12(\rho_c-1)^2(\rho_c+1)\log(1-\rho_c)
\nonumber\\
&\qquad
+ 6(\rho_c^3-\rho_c^2+5\rho_c+2)\log\rho_c
+ 34
\Big].
&&
\end{flalign}
For $C_{uc}^{(d)}$:
\begin{flalign}
&f^{(uc d)}_0 =
f^{(cu d)}_0,
&&\\[2pt]
&f^{(uc d)}_G =
-5\rho_c^4 + 16\rho_c^3 - 16\rho_c
- 12\rho_c^2\log\rho_c + 5,
&&\\[2pt]
&f^{(uc d)}_D =
-\frac{2}{3}\Big[
10\rho_c^4 - 15\rho_c^3 - 18\rho_c^2 + 71\rho_c
+ 12(\rho_c-1)^2(\rho_c+5)\log(1-\rho_c)
\nonumber\\
&\qquad \qquad \qquad
- 6(\rho_c^3+3\rho_c^2+1)\log\rho_c
- 48
\Big]
&&
\end{flalign}
\\
For $C_{cc}^{(d)}$:
\begin{flalign}
f^{(cc d)}_0 &= \frac{1}{4} f^{(\psi\psi d)}_0 |_{\rho_\psi \to \rho_c}, \\  f^{(cc d)}_G
&=
\Delta_{\rho_c}
\bigl(
5 + 14\rho_c + 14\rho_c^2 - 60\rho_c^3
\bigr)
-24\rho_c
\bigl(
1-\rho_c-2\rho_c^2+5\rho_c^3
\bigr)
\Lambda_{\rho_c}\,,& \\[4pt]  \nonumber f^{(cc d)}_D
&=
\frac{2}{3}\Bigg[
\Delta_{\rho_c}
\left(
72-\rho_c+22\rho_c^2-120\rho_c^3
+12\log\!\frac{\rho_c}{1-4\rho_c}
\right)
\nonumber\\
&\qquad
-6\left(
40\rho_c^4-14\rho_c^3+8\rho_c^2+5\rho_c+3
\right)\Lambda_{\rho_c}
\Bigg] \, . &
\end{flalign}
For operators involving two distinct
massive final states (e.g.\ $c$ and $\psi$), the Darwin term requires
two-scale master integrals that are not available in closed form.
Since these contributions are expected to be numerically subleading,
with the dominant subleading effects arising from the dimension-6 four-quark operators,
we restrict the analysis in such cases to the dimension-5 level.
The complete expressions for the dimension-3 and dimension-5 terms
are lengthy and can be provided upon request. 

The above results also apply to the Wilson coefficients obtained by replacing $d \to s$. For operators with only light quarks $(u, d, s)$ in the final state, the corresponding expressions are obtained in the massless limit of the massive cases.

We now turn to the four-quark contributions. 
The result for topology~IV in model $(d)$,
corresponding to the coefficients
$\Gamma^{(\psi u d),u}_{\text{IV},a}$,
was already derived in \cref{eq:modeld4qfinal};
for completeness, we repeat it here in the unified notation.
\\\\
\textbf{Contributions to  $\Gamma(B^+)$:}
\begin{align}
 &\tilde{\Gamma}^{(\psi u d),u}_{\text{IV},1} = -2 (1-\rho_\psi)^2, \qquad 
 \tilde{\Gamma}^{(\psi u d),u}_{\text{IV},2} = \tilde{\Gamma}^{(\psi u d),u}_{\text{IV},4} = 0, \qquad 
 \tilde{\Gamma}^{(\psi u d),u}_{\text{IV},3} = 6 (1-\rho_\psi)^2,
\end{align}
%\vspace{-2em}
\begin{align}
\tilde{\Gamma}^{(\tilde{\psi} u d),u}_{\text{IV},1}
&=
-\frac{1}{3}(1-\rho_\psi)^2(2+\rho_\psi),
\qquad
\tilde{\Gamma}^{(\tilde{\psi} u d),u}_{\text{IV},2}
=
\frac{2}{3}(1-\rho_\psi)^2(2\rho_\psi+1),
\\
\tilde{\Gamma}^{(\tilde{\psi} u d),u}_{\text{IV},3}
&=
(1-\rho_\psi)^2(2+\rho_\psi),
\qquad
\tilde{\Gamma}^{(\tilde{\psi} u d),u}_{\text{IV},4}
=
-2(1-\rho_\psi)^2(2\rho_\psi+1).
\end{align}
%\vspace{-2em}
\begin{align}
\nonumber 
& \tilde{\Gamma}^{( c u d),u}_{\text{I},1} 
= -\frac{2}{3} (1-\rho_c)^2 (2+\rho_c), \qquad 
\tilde{\Gamma}^{( c u d),u}_{\text{I},2} 
= \frac{4}{3} (1-\rho_c)^2 (2\rho_c+1), \\
& \tilde{\Gamma}^{( c u d),u}_{\text{I},3} 
= - (1-\rho_c)^2 (2+\rho_c), \qquad 
\tilde{\Gamma}^{( c u d),u}_{\text{I},4} 
= 2 (1-\rho_c)^2 (2\rho_c+1),
\end{align}
%\vspace{-2em}
\begin{align}
 &\tilde{\Gamma}^{( u c d),u}_{\text{IV},1} = -4 (1-\rho_c)^2, \qquad 
 \tilde{\Gamma}^{( u c d),u}_{\text{IV},2} = \tilde{\Gamma}^{( u c d),u}_{\text{IV},4} = 0, \qquad 
 \tilde{\Gamma}^{( u c d),u}_{\text{IV},3} = 12 (1-\rho_c)^2,
\end{align}
where $\rho_f = m_f^2/m_b^2$. Finally, the massless limit follows from
\begin{align}
 &\tilde{\Gamma}^{( u u d),u}_{\text{IV},a} 
 = \tilde{\Gamma}^{( u c d),u}_{\text{IV},a}\Big|_{\rho_c \to 0}, \qquad  
 \tilde{\Gamma}^{( u u d),u}_{\text{I},a} 
 = \tilde{\Gamma}^{( u c d),u}_{\text{I},a}\Big|_{\rho_c \to 0}.
\end{align}
\textbf{Contributions to  $\Gamma(B_q)$:}
  \begin{align}
  \nonumber &\tilde{\Gamma}^{(\psi c d),q}_{\text{III},1} = \frac{1}{3}\sqrt{\lambda(1,\rho_c,\rho_\psi)} \, (\rho_\psi^2+\rho_c^2+\rho_\psi+\rho_c-2\rho_\psi \rho_c -2), \\ \nonumber & 
  \tilde{\Gamma}^{(\psi c d),q}_{\text{III},2} = \frac{-2}{3}\sqrt{\lambda(1,\rho_c,\rho_\psi)} \, (2\rho_\psi^2+2\rho_c^2-\rho_\psi-\rho_c-4\rho_\psi \rho_c -1), \\ & 
  \tilde{\Gamma}^{(\psi c d),q}_{\text{III},3} = -3\,\tilde{\Gamma}^{(\psi c d),q}_{\text{III},1}, \qquad  \tilde{\Gamma}^{(\psi c d),q}_{\text{III},4} = -3\,\tilde{\Gamma}^{(\psi c d),q}_{\text{III},2}, \end{align}
  \vspace{-2em}
\begin{align}
  \nonumber  &\tilde{\Gamma}^{(\psi \psi d),q}_{\text{I},1} = - \sqrt{1-4\rho_\psi} \, (1-\rho_\psi), \qquad \tilde{\Gamma}^{(\psi \psi d),q}_{\text{I},2} = \sqrt{1-4\rho_\psi} \, (2\rho_\psi+1), \\ & \qquad \qquad \qquad \qquad \qquad
  \tilde{\Gamma}^{(\psi \psi d),q}_{\text{I},3} = \tilde{\Gamma}^{(\psi \psi d),q}_{\text{I},4} = 0, 
\end{align}
\begin{align}
  \nonumber  &\tilde{\Gamma}^{(cc d),q}_{\text{I},1} = - \frac{4}{3}\sqrt{1-4\rho_c} \, (1-\rho_c), \qquad \tilde{\Gamma}^{(cc d),q}_{\text{I},2} = \frac{4}{3}\sqrt{1-4\rho_c} \, (2\rho_c+1), \\ & 
  \tilde{\Gamma}^{(cc d),q}_{\text{I},3} = -2 \sqrt{1-4\rho_c} \, (1-\rho_c) , \qquad \tilde{\Gamma}^{(cc d),q}_{\text{I},4} = 2\sqrt{1-4\rho_c} \, (2\rho_c+1), 
\end{align}
\begin{align}
  \nonumber  &\tilde{\Gamma}^{(cu d),q}_{\text{I},1} = - \frac{2}{3}(1-\rho_c)^2 \, (2+\rho_c) , \qquad \tilde{\Gamma}^{(cu d),q}_{\text{I},2} = \frac{4}{3} (1-\rho_c)^2 \, (2\rho_c+1), \\ & 
  \tilde{\Gamma}^{(cu d),q}_{\text{I},3} = - (1-\rho_c)^2 \, (2+\rho_c) , \qquad \tilde{\Gamma}^{(cu d),q}_{\text{I},4} = 2(1-\rho_c)^2 \, (2\rho_c+1).
\end{align}
The following relations hold 
\begin{align}
  \nonumber  &\tilde{\Gamma}^{(\tilde{\psi}c d),q}_{\text{III},a} = \tilde{\Gamma}^{({\psi}c d),q}_{\text{III},a},
  \qquad 
  \tilde{\Gamma}^{( \psi u d),q}_{\text{III},a} =\tilde{\Gamma}^{( \tilde{\psi} u d),q}_{\text{III},a} 
 = \tilde{\Gamma}^{( \psi cd),q}_{\text{III},a}\Big|_{\rho_c \to 0},
  \qquad \tilde{\Gamma}^{(uc d),q}_{\text{I},a} = \tilde{\Gamma}^{(cu d),q}_{\text{I},a}, \\& 
\qquad \qquad \qquad \qquad \qquad \qquad 
\tilde{\Gamma}^{( uud),q}_{\text{I},a} = \tilde{\Gamma}^{( cud),q}_{\text{I},a}\Big|_{\rho_c \to 0}.
\end{align}
Moreover, since in our analysis we work in the massless limit for the light quarks,
$m_d = m_s = 0$, the corresponding 
coefficients above are identical under the interchange $d \leftrightarrow s$.

\subsection{Exclusive Results from LCSR}
\label{sec:MesoExclResults}

In this section we collect the final expressions for the LCSR results
derived in \cref{eq:3partdiag}.

\begin{align}
{\cal C}({\cal P},v)
&=
\Big[
8({\cal P}\!\cdot\! v)\slashed{\cal P}\slashed v
-
16({\cal P}\!\cdot\! v)^2
\Big]\phi_3
+
8({\cal P}\!\cdot\! v-\slashed{\cal P}\slashed v)
(\overline\phi_3-\overline\phi_4)
\nonumber\\
&\quad
+
8(4{\cal P}\!\cdot\! v-\slashed{\cal P}\slashed v)
\overline{\widetilde\psi}_4
-
\frac{4}{{\cal P}^2}
\Big[
{\cal P}^2
+
2({\cal P}\!\cdot\! v)\slashed{\cal P}\slashed v
\Big]
(
\overline{\overline\phi}_4
-
\overline{\overline\psi}_4
-
\overline{\overline{\widetilde\psi}}_4
)
\nonumber\\
&\quad
-
12
(
\overline{\overline\phi}_3
-
\overline{\overline\phi}_4
+
2\overline{\overline{\widetilde\psi}}_4
),
\nonumber\\[2mm]
\overline{\cal C}({\cal P},v)
&=
12{\cal P}^2
\text{Log}(-{\cal P}^2)\phi_4
-
4({\cal P}\!\cdot\! v-\slashed{\cal P}\slashed v)
(\overline\phi_3-\overline\phi_4)
\nonumber\\
&\quad
-
4(4{\cal P}\!\cdot\! v-\slashed{\cal P}\slashed v)
\overline{\widetilde\psi}_4
+
\frac{4}{{\cal P}^2}
\Big[
{\cal P}^2
+
2({\cal P}\!\cdot\! v)\slashed{\cal P}\slashed v
\Big]
(
\overline{\overline\phi}_4
-
\overline{\overline\psi}_4
-
\overline{\overline{\widetilde\psi}}_4
)
\nonumber\\
&\quad
+
12
(
\overline{\overline\phi}_3
-
\overline{\overline\phi}_4
+
2\overline{\overline{\widetilde\psi}}_4
),
\label{eq:3partdiag1}
\end{align}
where
${\cal P}=P-\sigma m_B v.$ In the above expressions, terms that do not generate an imaginary
part in the variable $P^2$ are omitted, since they do not contribute
to the final LCSR expressions.
\subsection{\boldmath Supplementary Plots}
\label{sec:sub_plots}
This appendix collects the results for the remaining operators considered in our analysis. While the main text focuses on the representative cases ($\mathcal{O} = (b u)(\psi d)$ and $\mathcal{O} = (d u)(\psi b)$), the following plots provide a systematic overview and illustrate to what extent the observed features generalize across the operator basis.
\label{app:D}
\begin{figure}[ht]
    \centering
    \begin{subfigure}{0.43\textwidth}
        \centering
        \includegraphics[width=\textwidth]{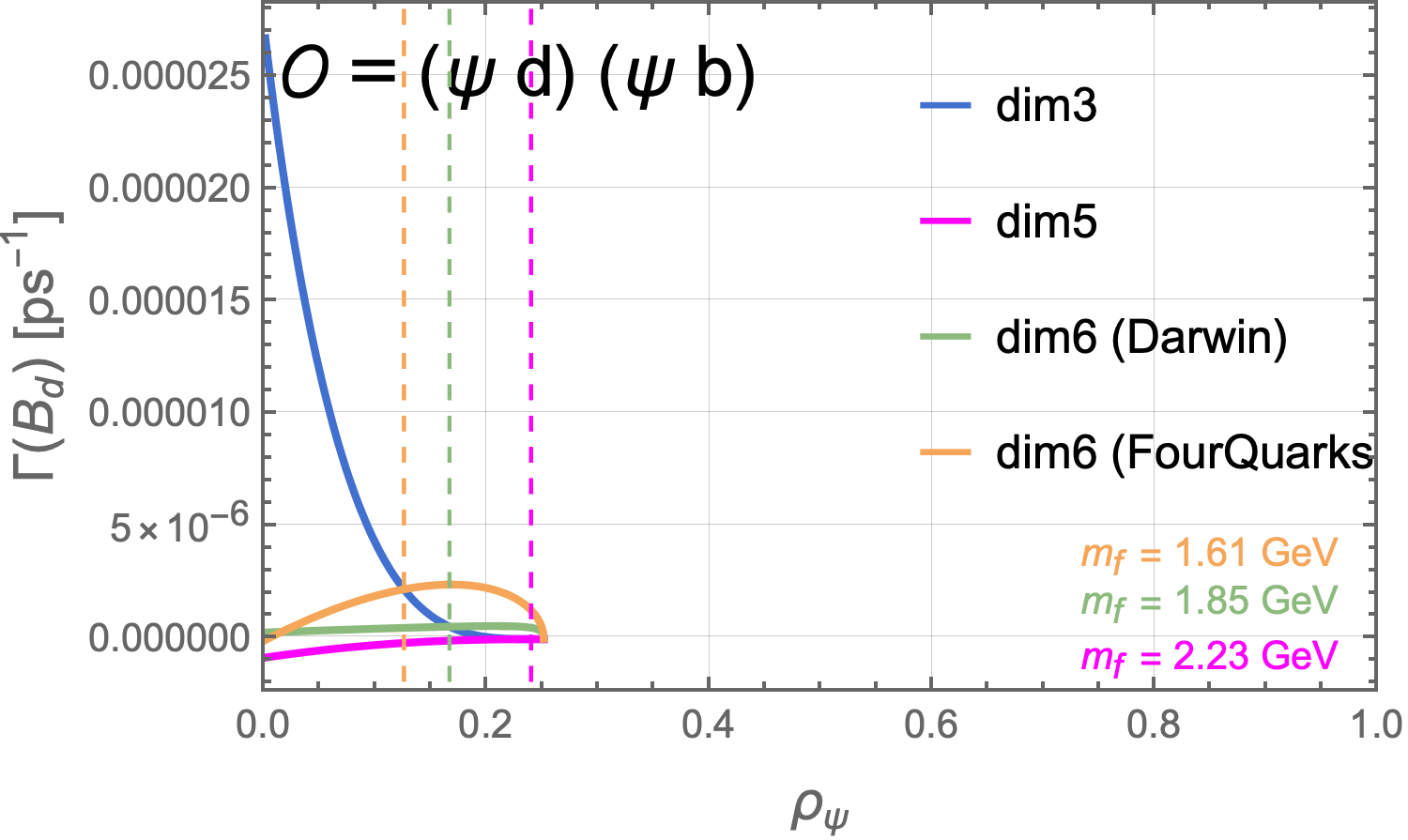}
    \end{subfigure}
    \hfill
    \begin{subfigure}{0.40\textwidth}
        \centering
        \includegraphics[width=\textwidth]{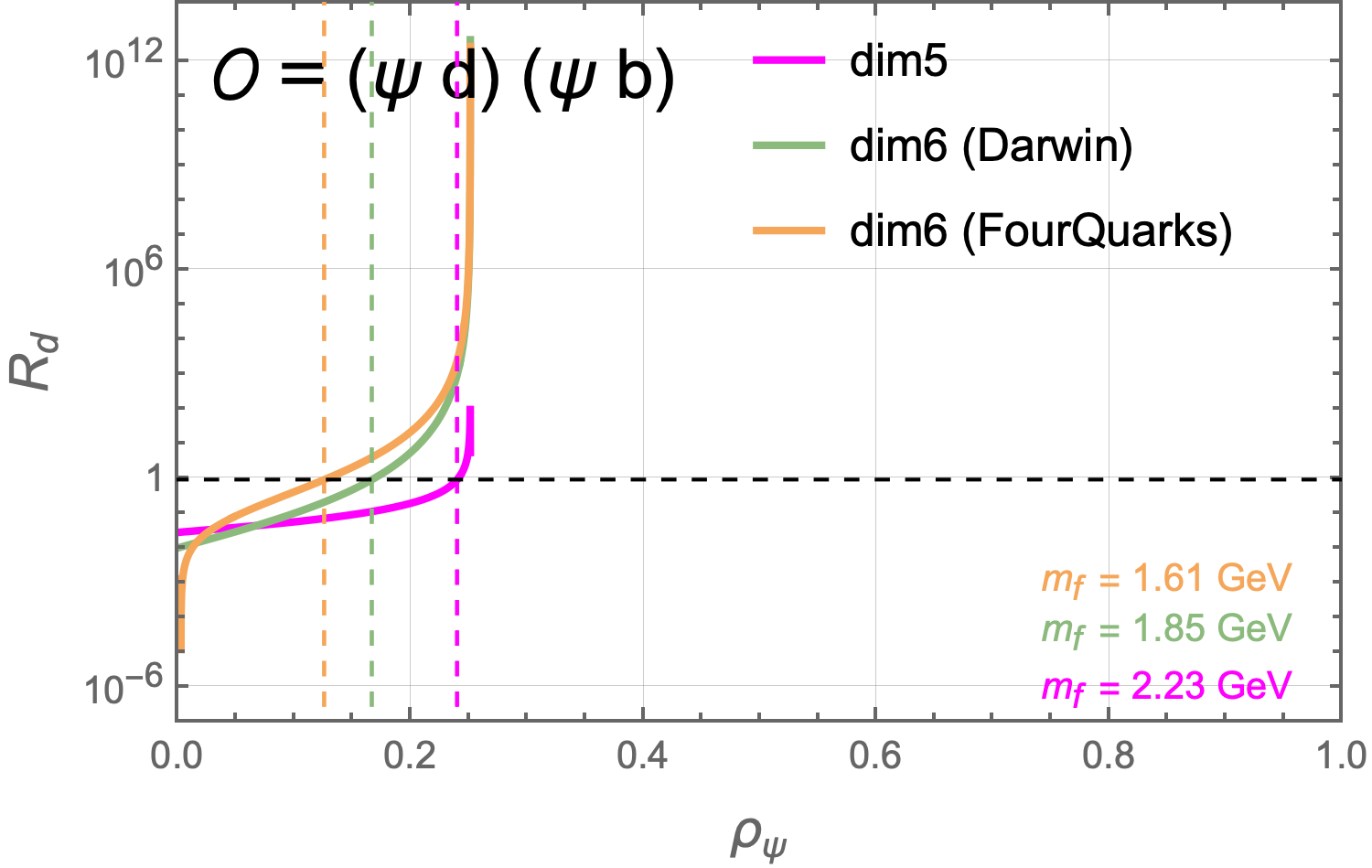}
    \end{subfigure}
    
    \medskip % Adds a small vertical space between the rows
    
    \begin{subfigure}{0.43\textwidth}
        \centering
        \includegraphics[width=\textwidth]{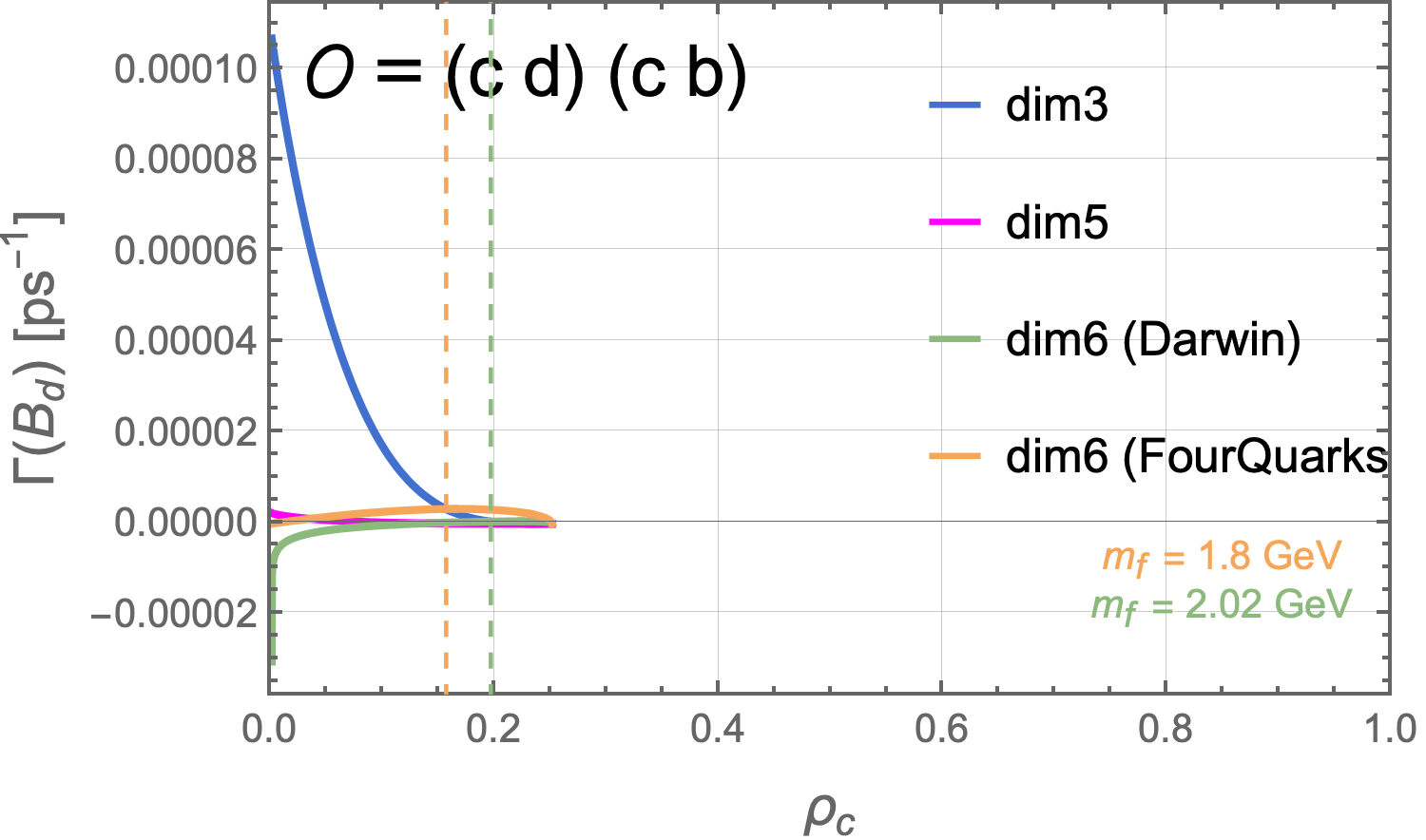}
    \end{subfigure}
    \hfill
    \begin{subfigure}{0.40\textwidth}
        \centering
        \includegraphics[width=\textwidth]{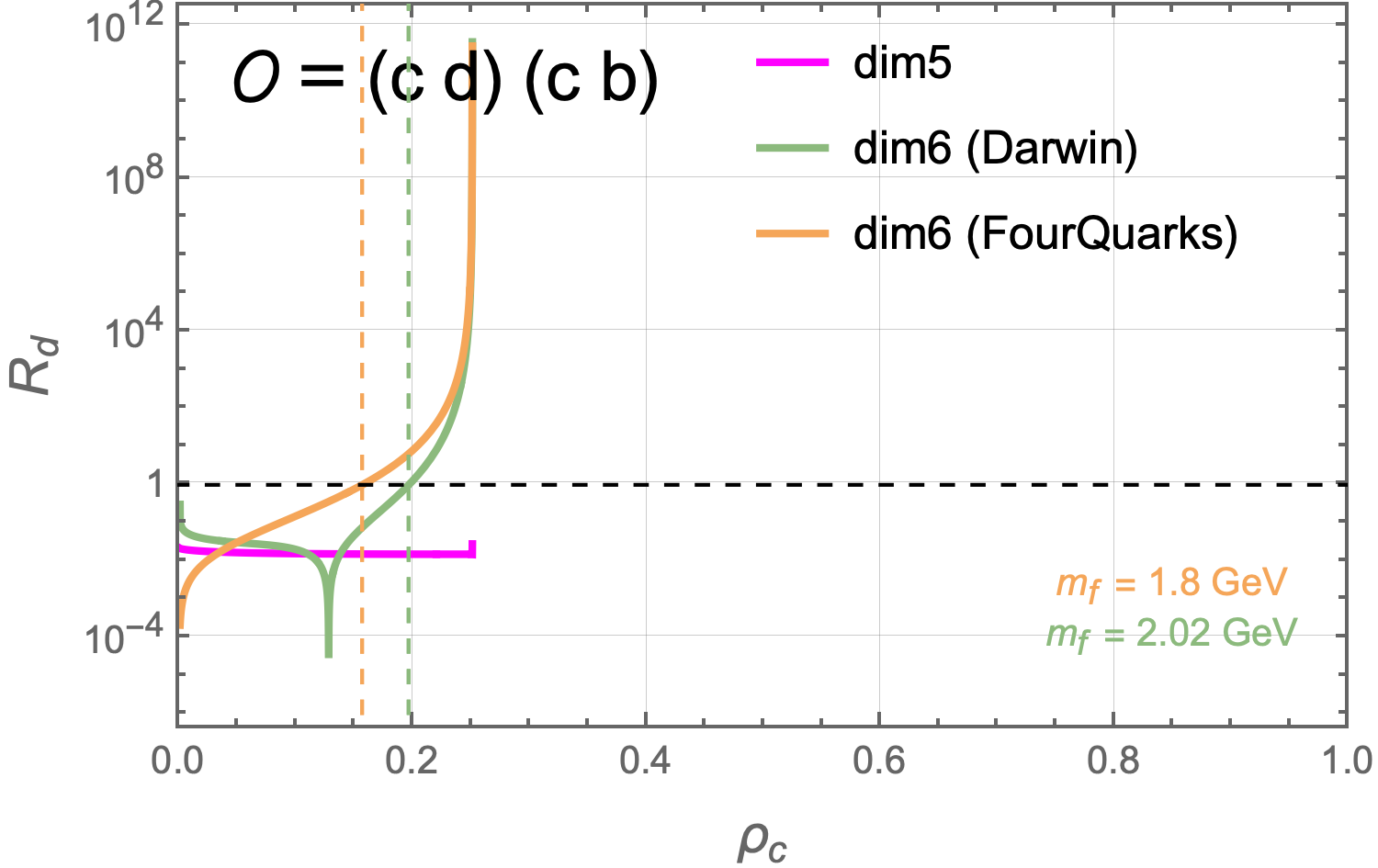}
    \end{subfigure}

     \medskip % Adds a small vertical space between the rows
    
    \begin{subfigure}{0.43\textwidth}
        \centering
        \includegraphics[width=\textwidth]{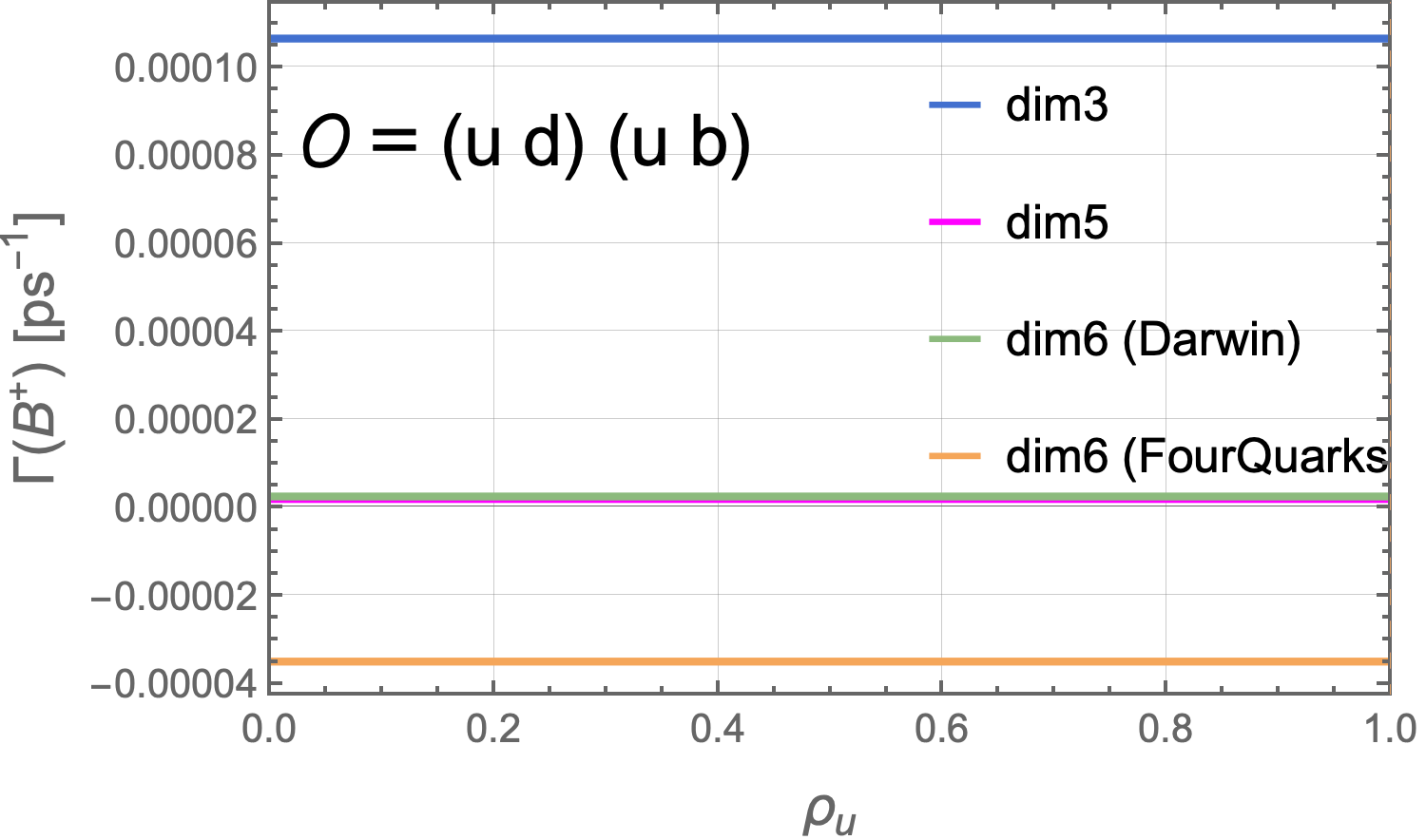}
    \end{subfigure}
    \hfill
    \begin{subfigure}{0.40\textwidth}
        \centering
        \includegraphics[width=\textwidth]{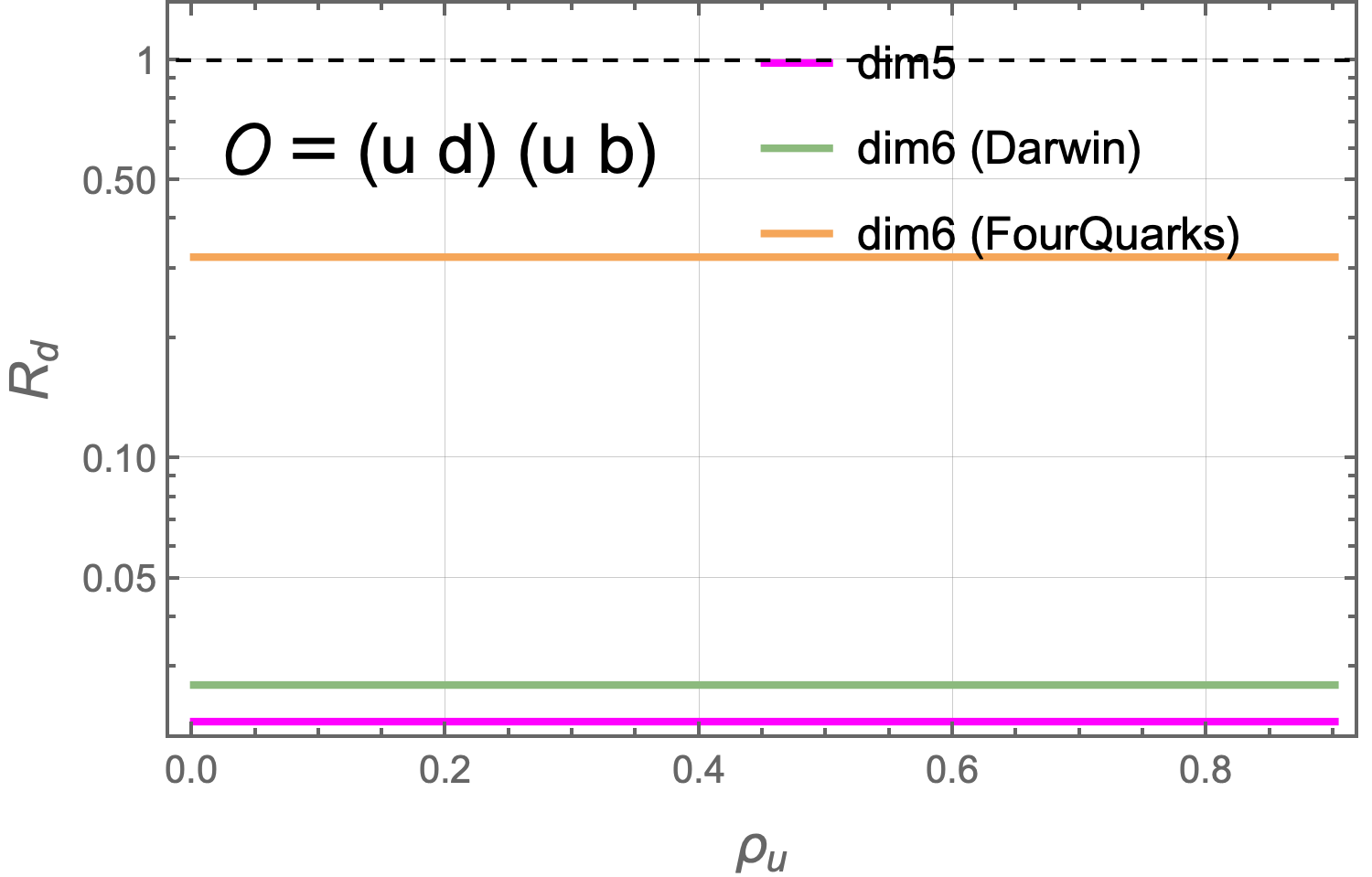}
    \end{subfigure}
    \caption{(Part 1) Total decay width $\Gamma(B^+)$ induced by the Mesogenesis (BSM) operators as a function of $\rho_f=(m_f/m_b)^2$. The left and right columns show the individual HQE contributions and the corresponding ratios $R_d=|\Gamma_d/\Gamma_3|$, respectively. The vertical lines indicate the values of $\rho_f$ (and the corresponding $m_f$) where a subleading contribution exceeds the leading dimension-three term $\Gamma_3$. The curves terminate at the kinematic threshold where the loop particles can no longer be produced on shell, while the spikes in $R_d$ arise from zeros of the corresponding subleading contribution.}
    \label{fig:DecayRates2}
\end{figure}

\begin{figure}[ht]
    \centering
    \begin{subfigure}{0.43\textwidth}
        \centering
        \includegraphics[width=\textwidth]{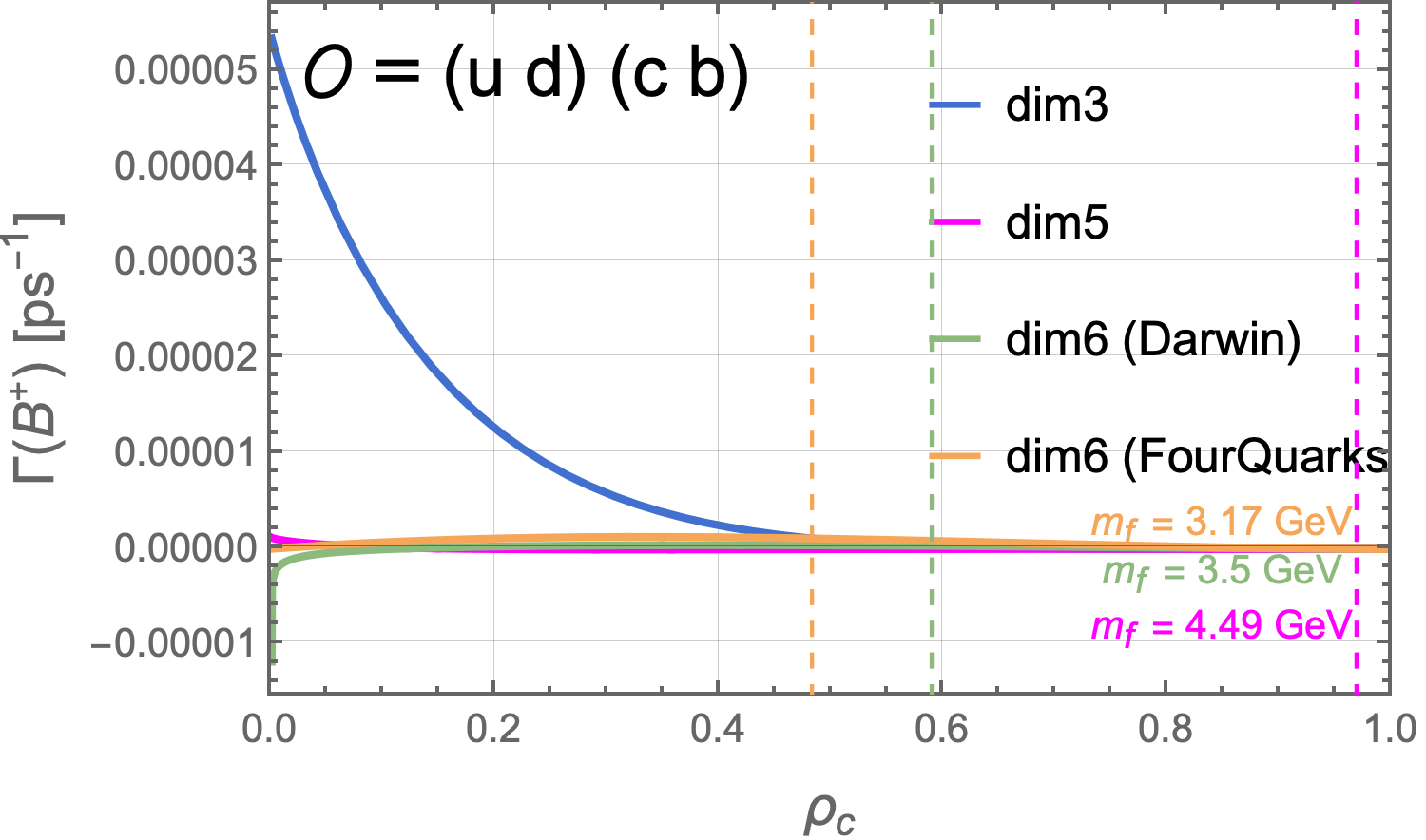}
    \end{subfigure}
    \hfill
    \begin{subfigure}{0.40\textwidth}
        \centering
        \includegraphics[width=\textwidth]{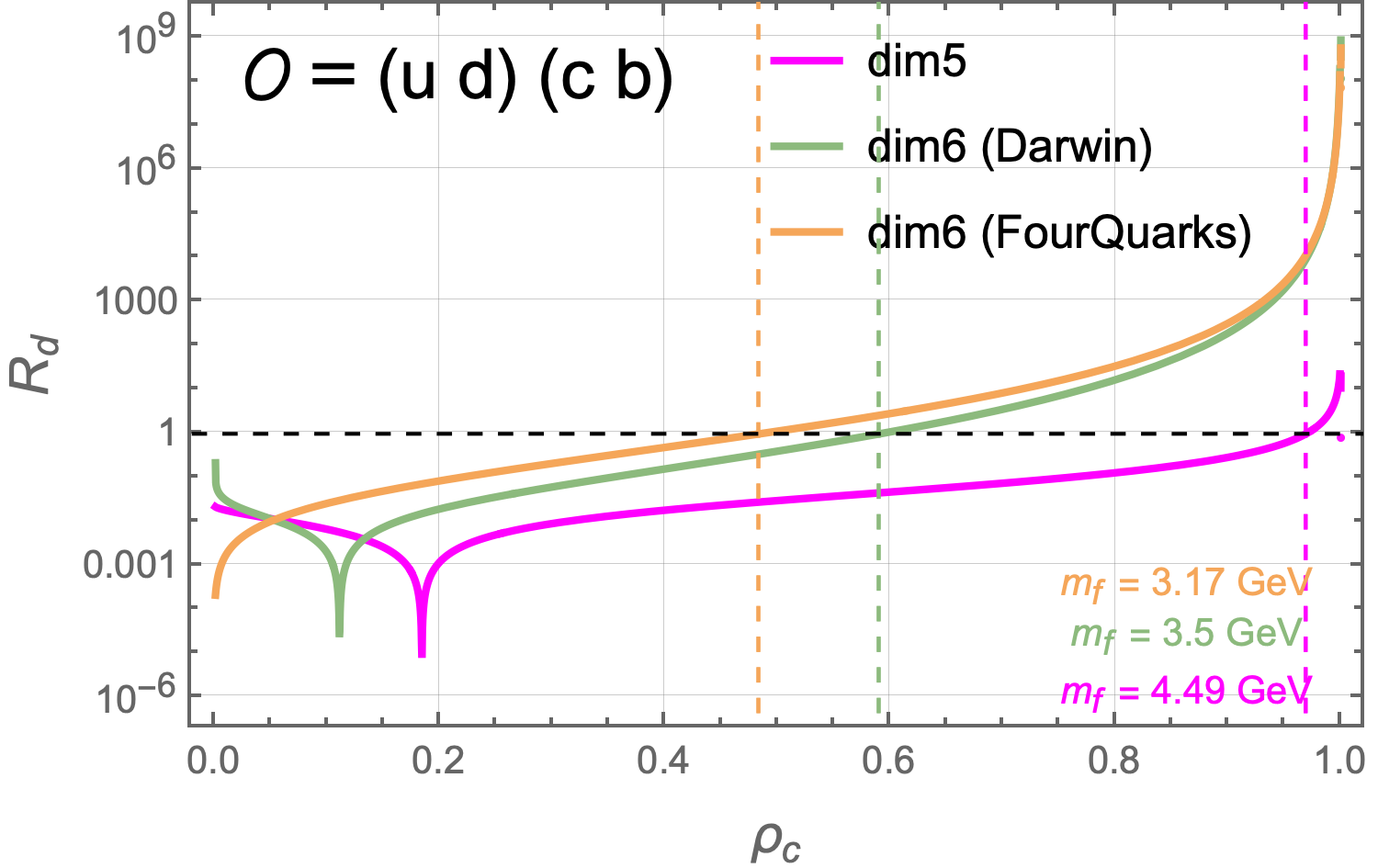}
    \end{subfigure}

     \medskip % Adds a small vertical space between the rows
    
    \begin{subfigure}{0.43\textwidth}
        \centering
        \includegraphics[width=\textwidth]{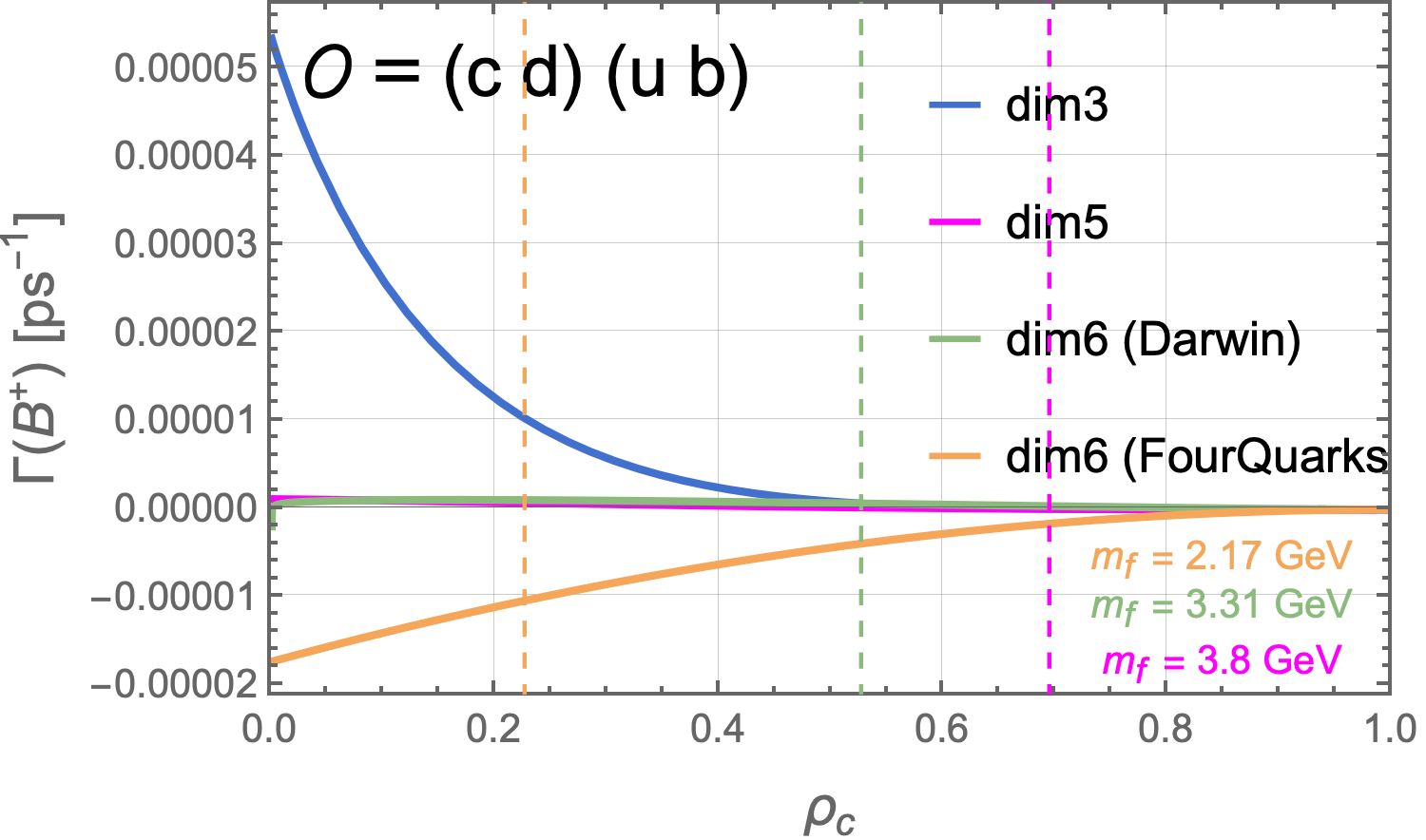}
    \end{subfigure}
    \hfill
    \begin{subfigure}{0.40\textwidth}
        \centering
        \includegraphics[width=\textwidth]{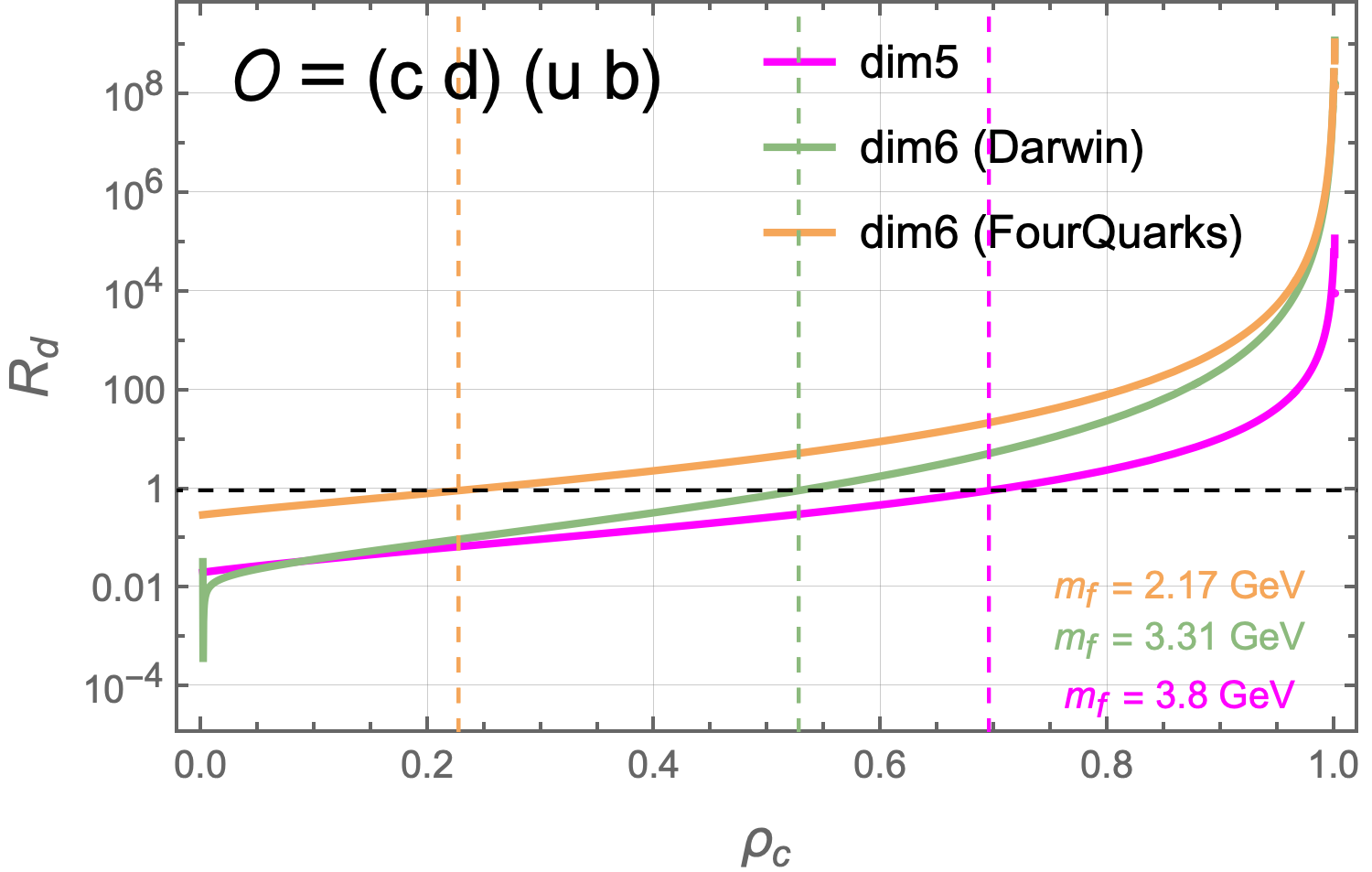}
    \end{subfigure}

    \medskip % Adds a small vertical space between the rows
    
    \begin{subfigure}{0.43\textwidth}
        \centering
        \includegraphics[width=\textwidth]{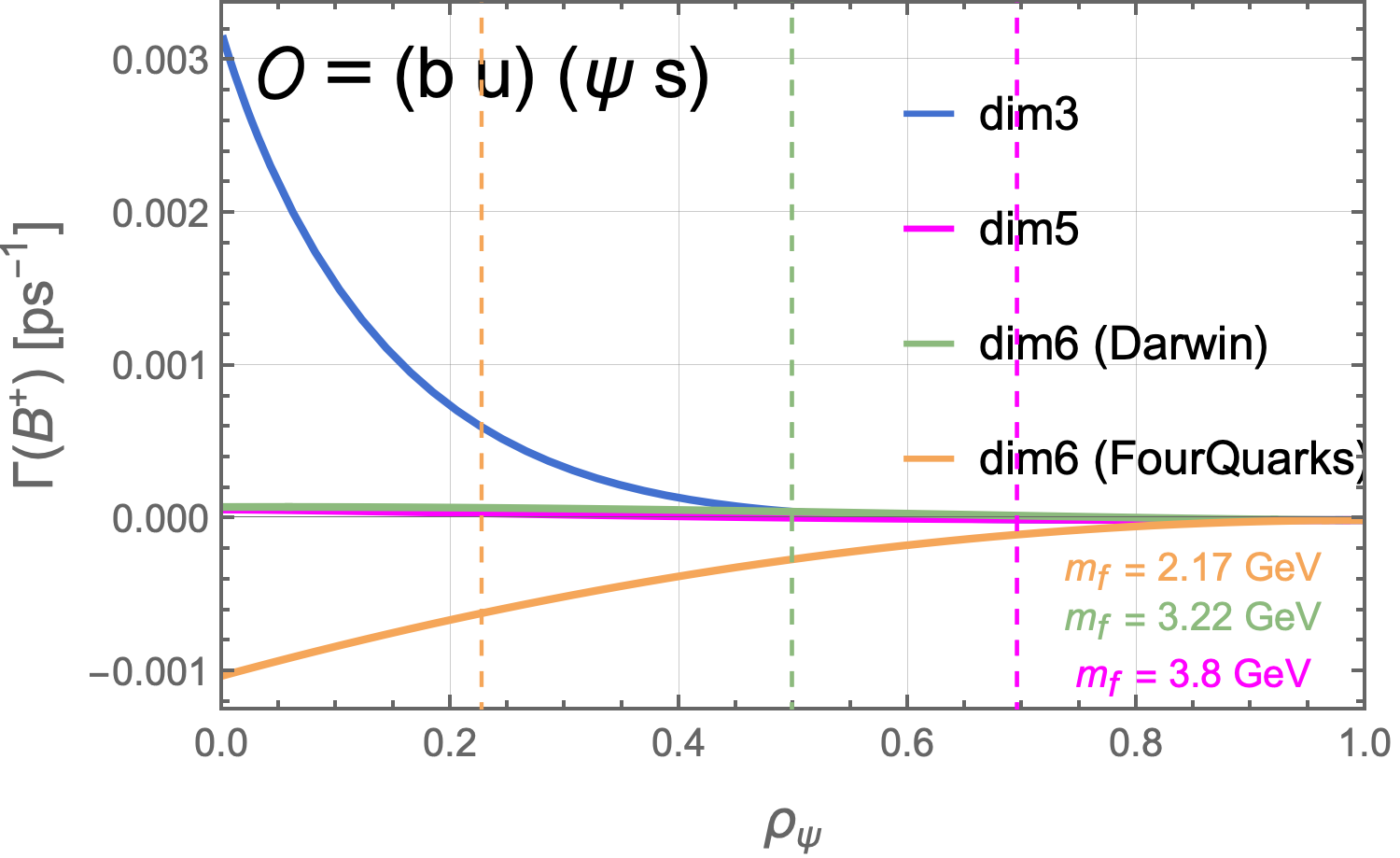}
    \end{subfigure}
    \hfill
    \begin{subfigure}{0.40\textwidth}
        \centering
        \includegraphics[width=\textwidth]{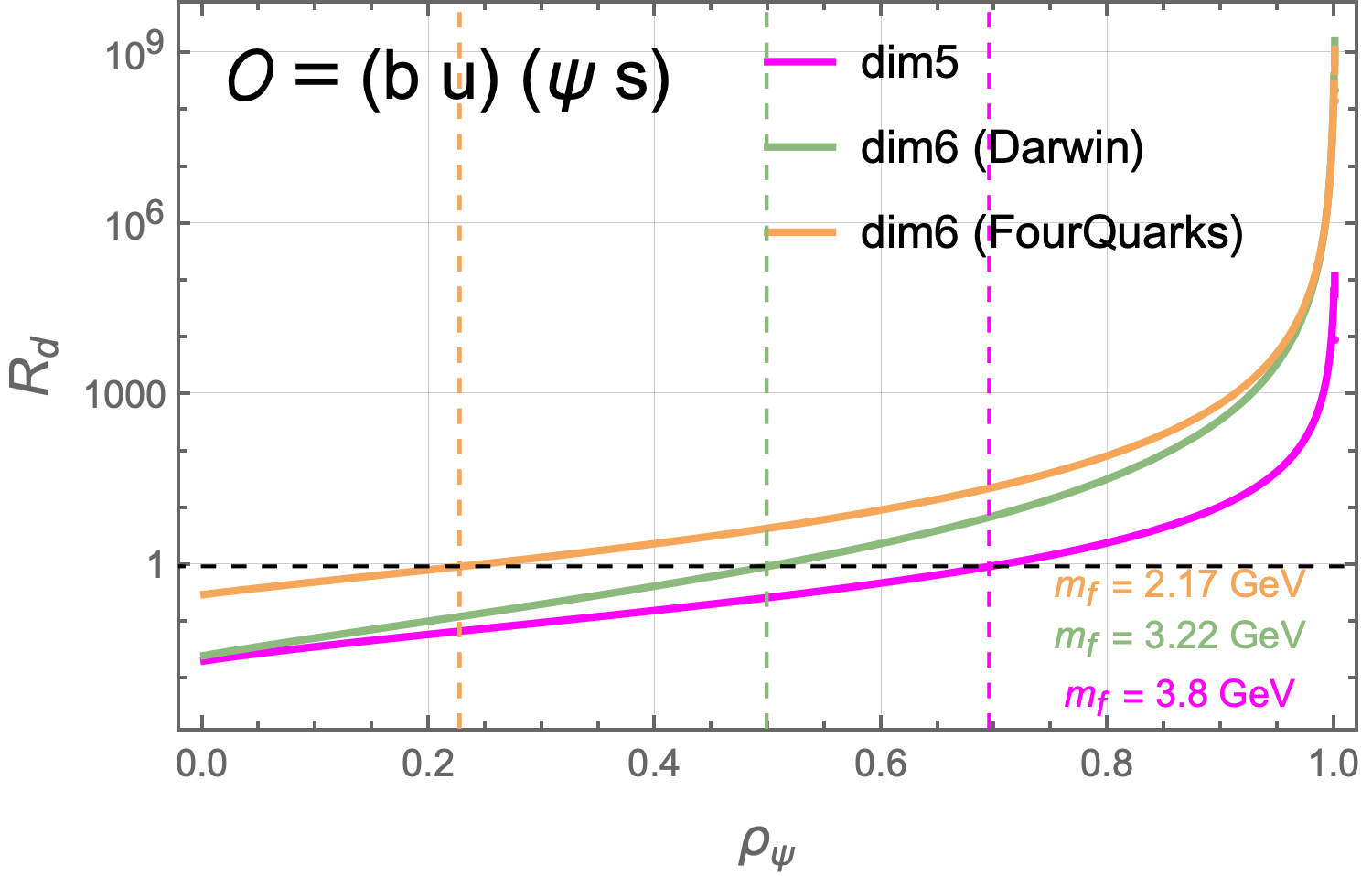}
    \end{subfigure}

    \medskip % Adds a small vertical space between the rows
    
    \begin{subfigure}{0.43\textwidth}
        \centering
        \includegraphics[width=\textwidth]{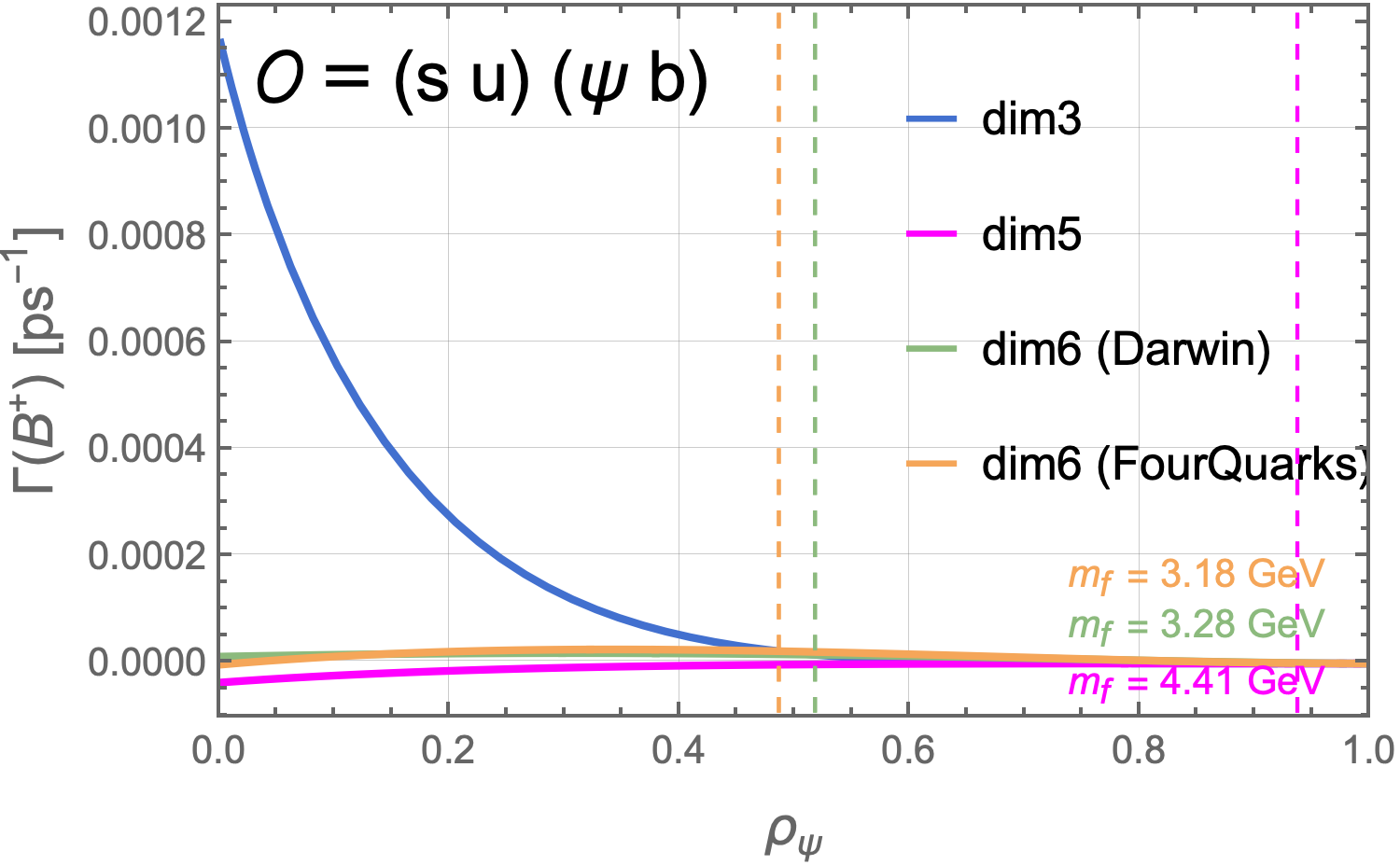}
    \end{subfigure}
    \hfill
    \begin{subfigure}{0.40\textwidth}
        \centering
        \includegraphics[width=\textwidth]{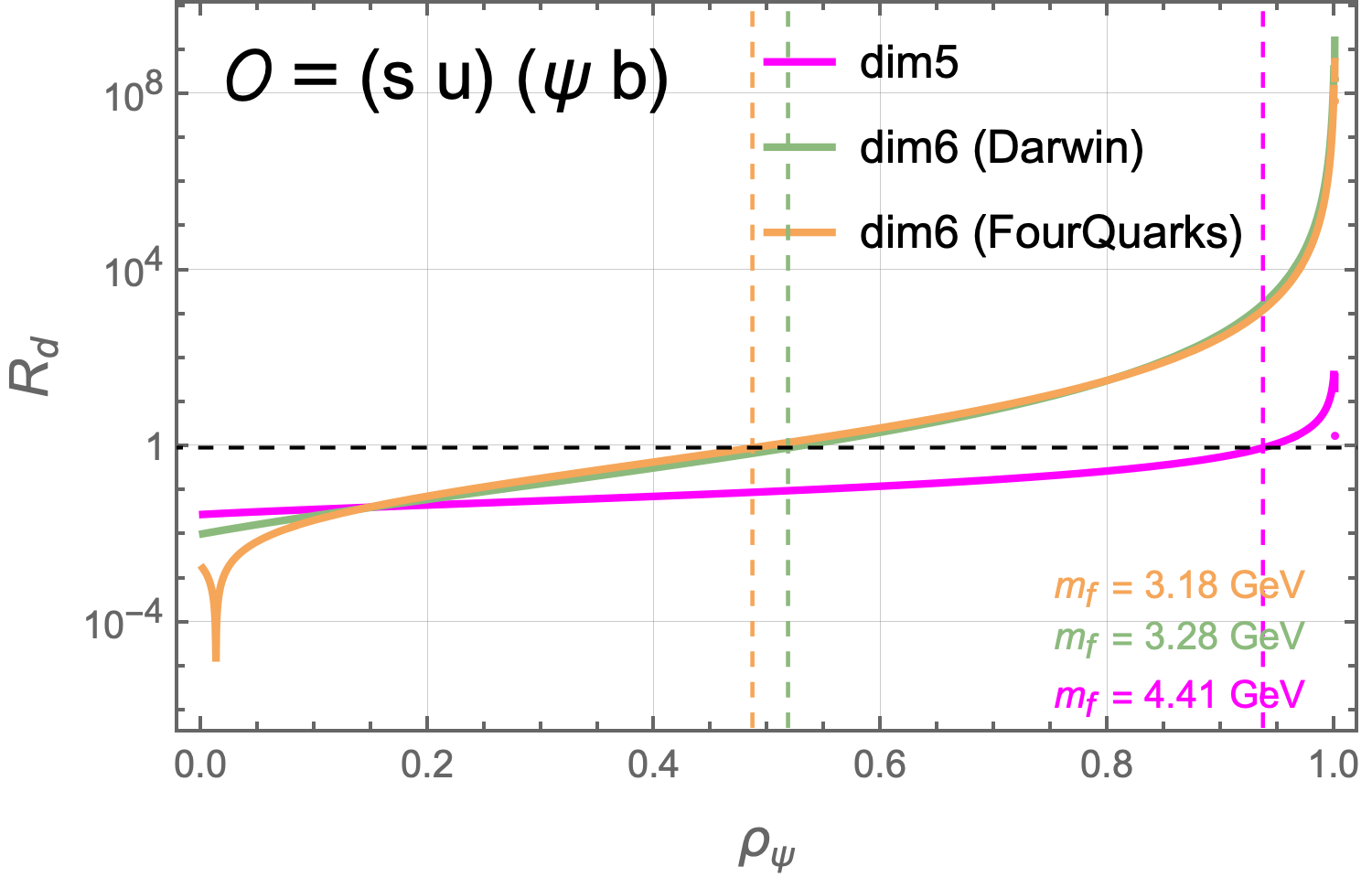}
    \end{subfigure}

    \caption{(Part 2): Additional plots for different operators with the same colour coding as in the previous Figure.}
    \label{fig:DecayRates3}
\end{figure}

\begin{figure}[ht]
    \centering
    \begin{subfigure}{0.43\textwidth}
        \centering
        \includegraphics[width=\textwidth]{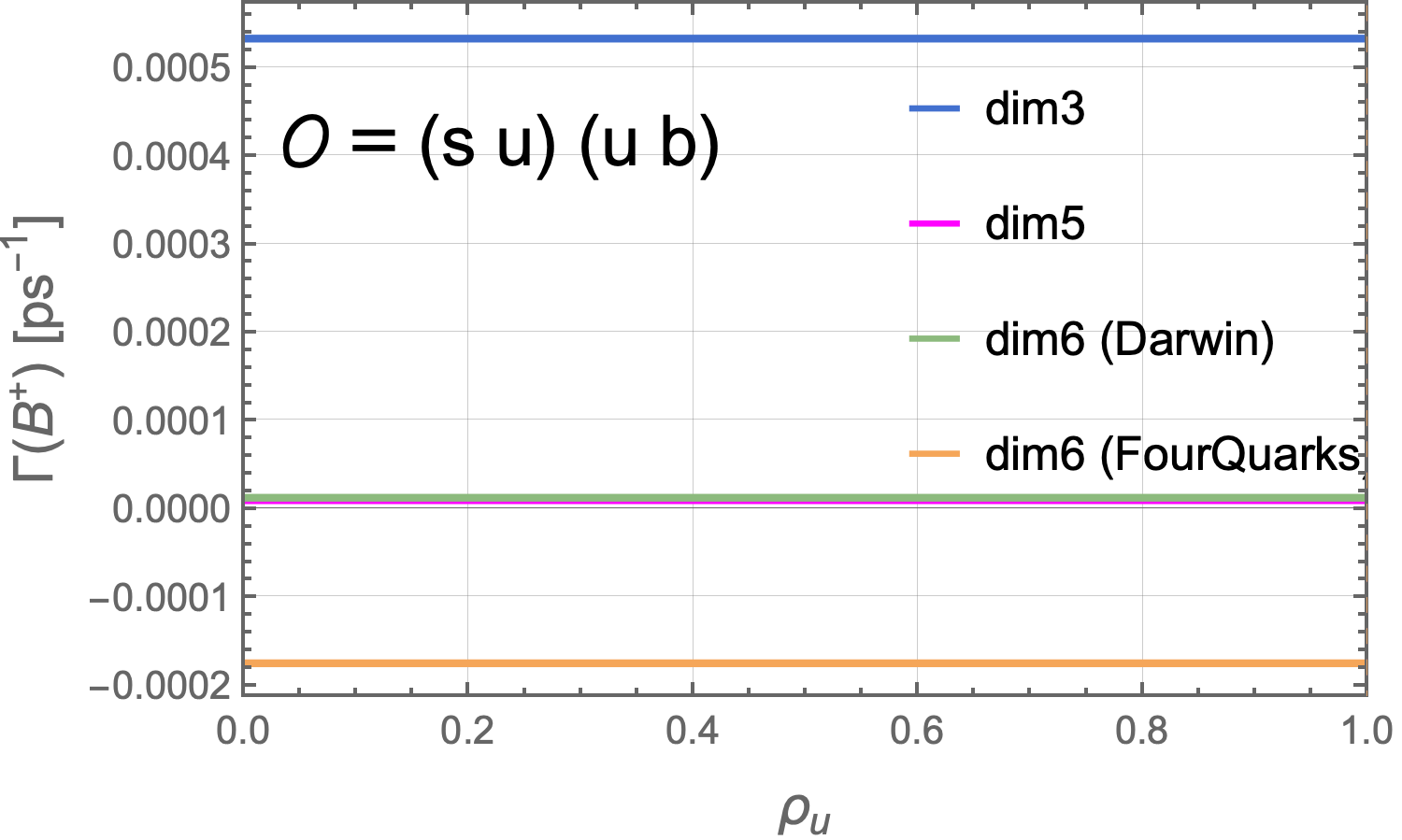}
    \end{subfigure}
    \hfill
    \begin{subfigure}{0.40\textwidth}
        \centering
        \includegraphics[width=\textwidth]{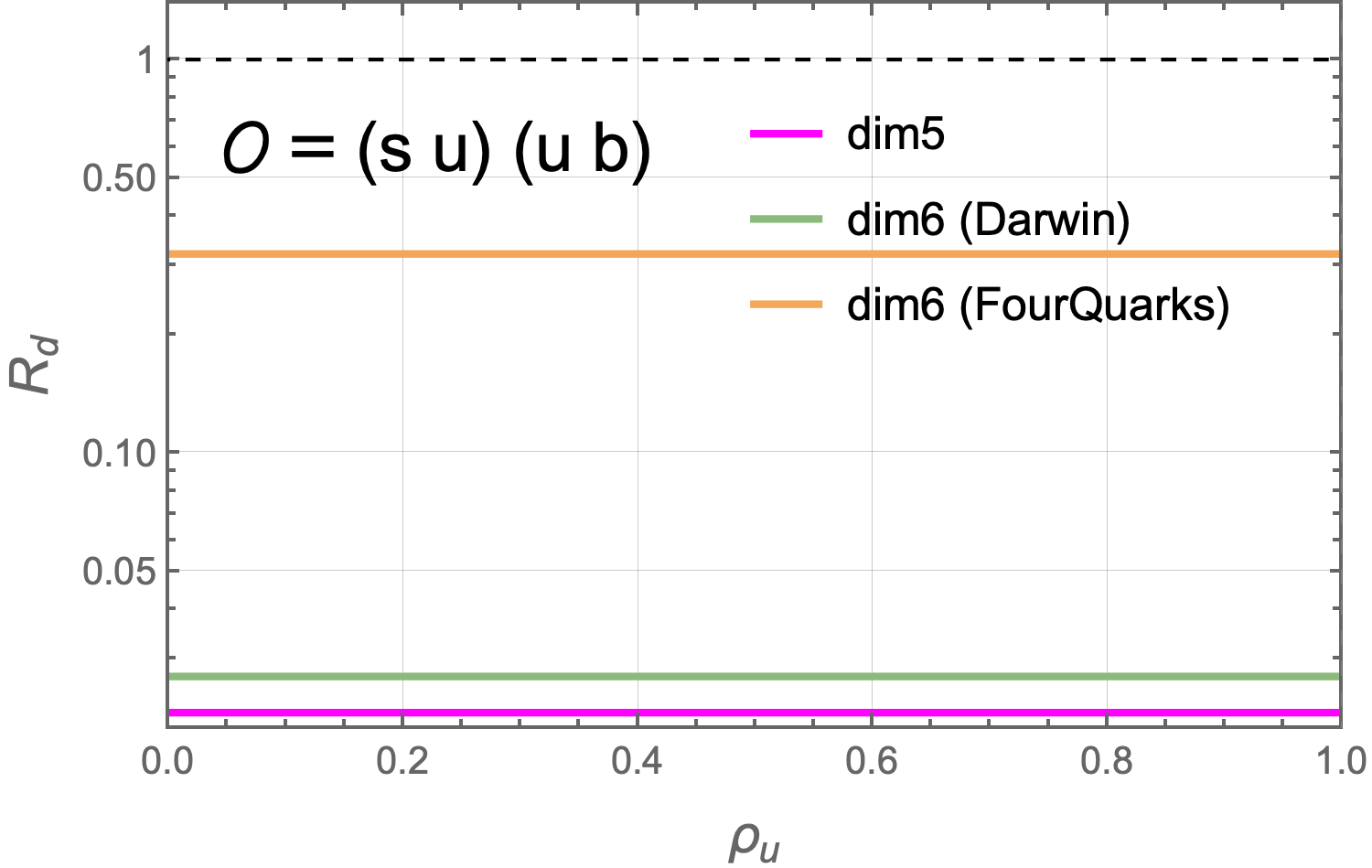}
    \end{subfigure}

     \medskip % Adds a small vertical space between the rows
    
    \begin{subfigure}{0.43\textwidth}
        \centering
        \includegraphics[width=\textwidth]{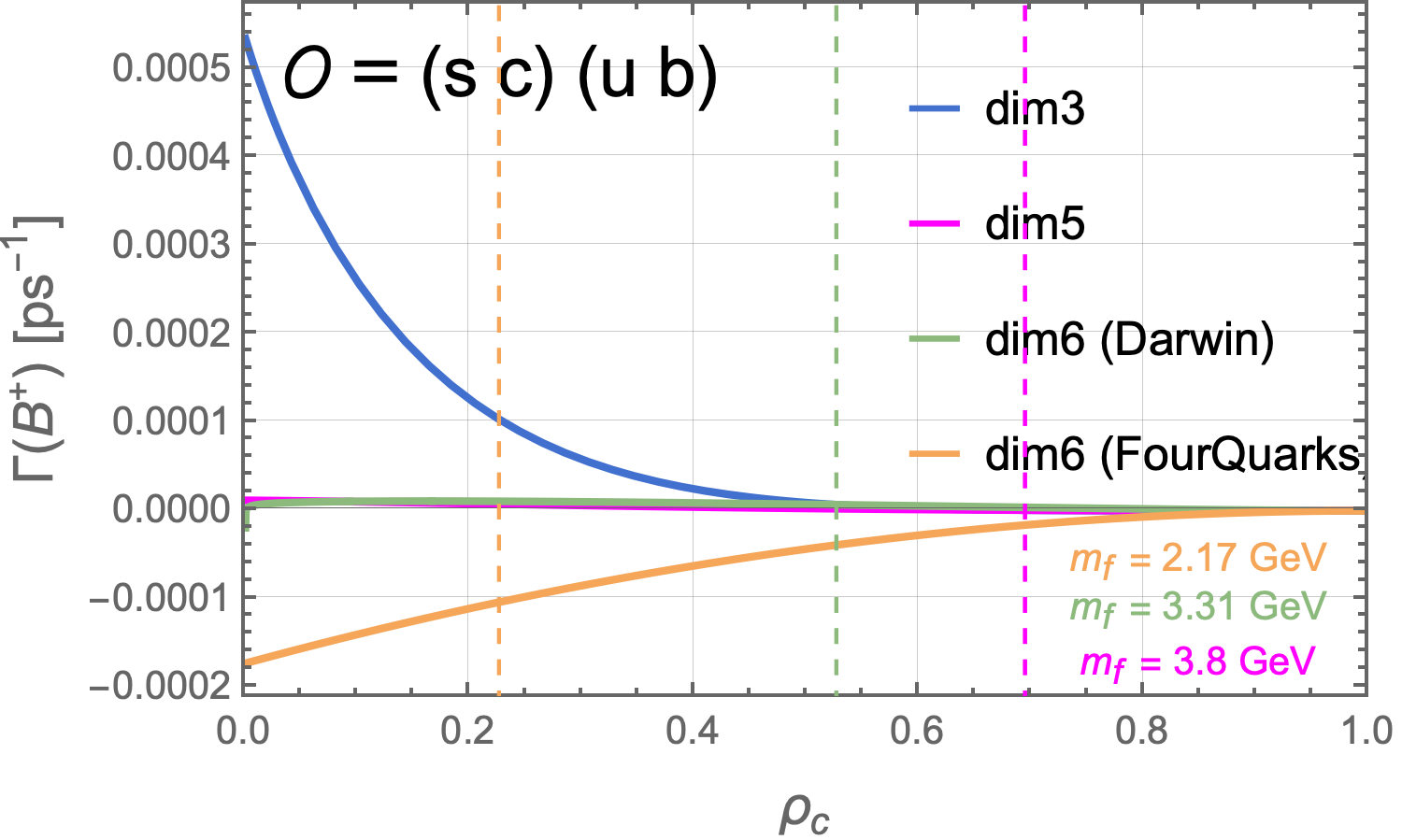}
    \end{subfigure}
    \hfill
    \begin{subfigure}{0.40\textwidth}
        \centering
        \includegraphics[width=\textwidth]{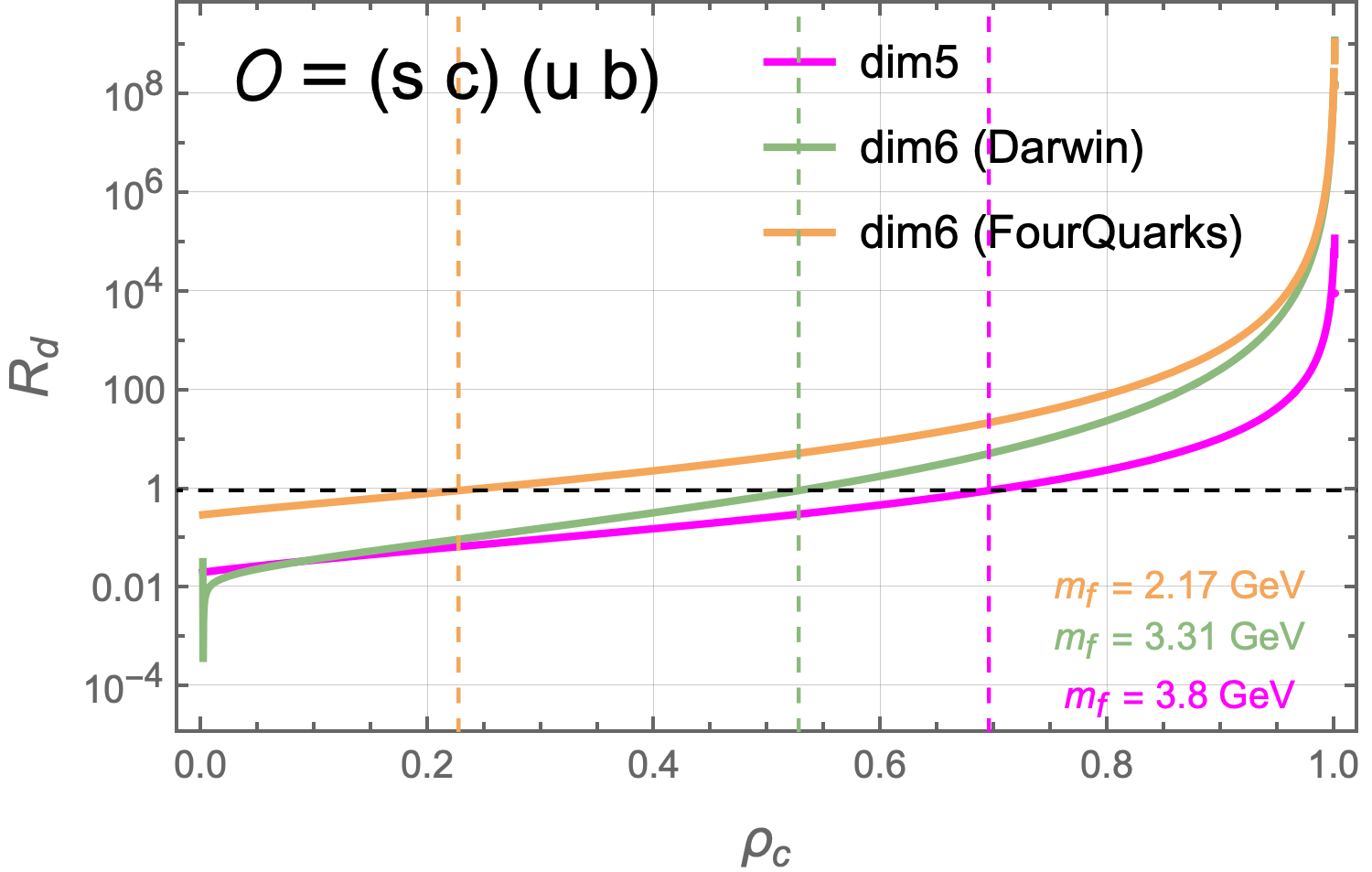}
    \end{subfigure}

    \medskip % Adds a small vertical space between the rows
    
    \begin{subfigure}{0.43\textwidth}
        \centering
        \includegraphics[width=\textwidth]{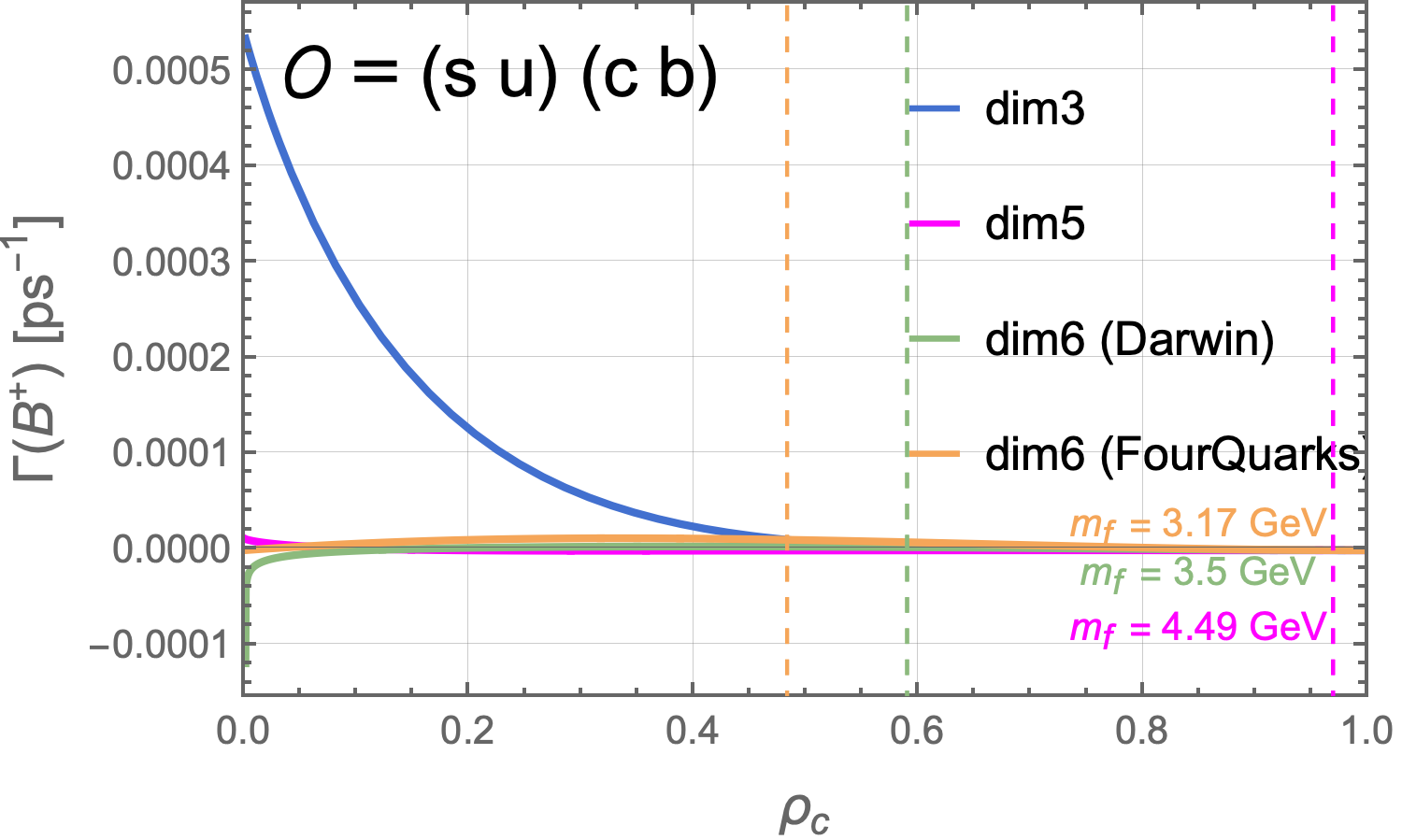}
    \end{subfigure}
    \hfill
    \begin{subfigure}{0.40\textwidth}
        \centering
        \includegraphics[width=\textwidth]{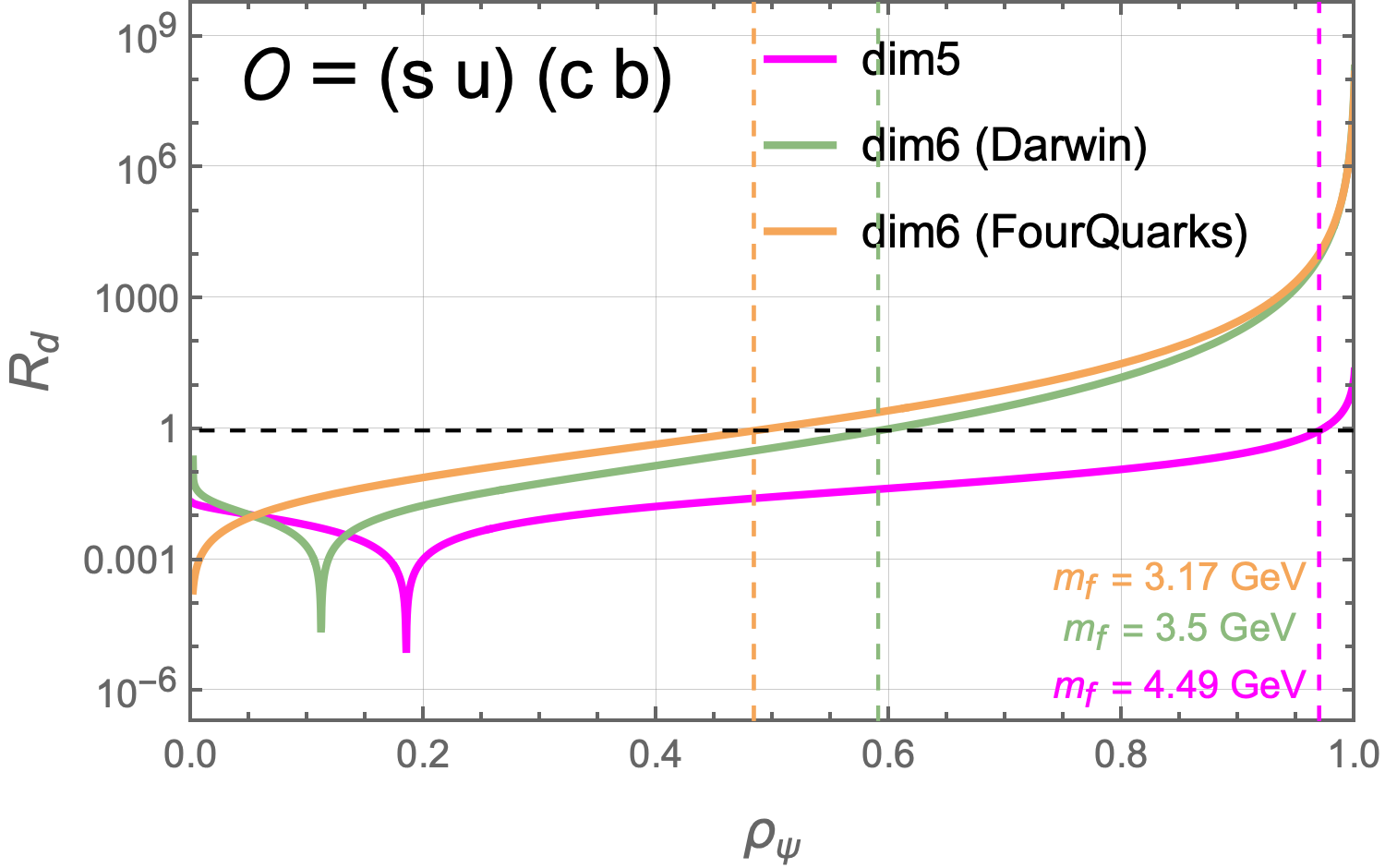}
    \end{subfigure}

      \medskip % Adds a small vertical space between the rows
    
    \begin{subfigure}{0.43\textwidth}
        \centering
        \includegraphics[width=\textwidth]{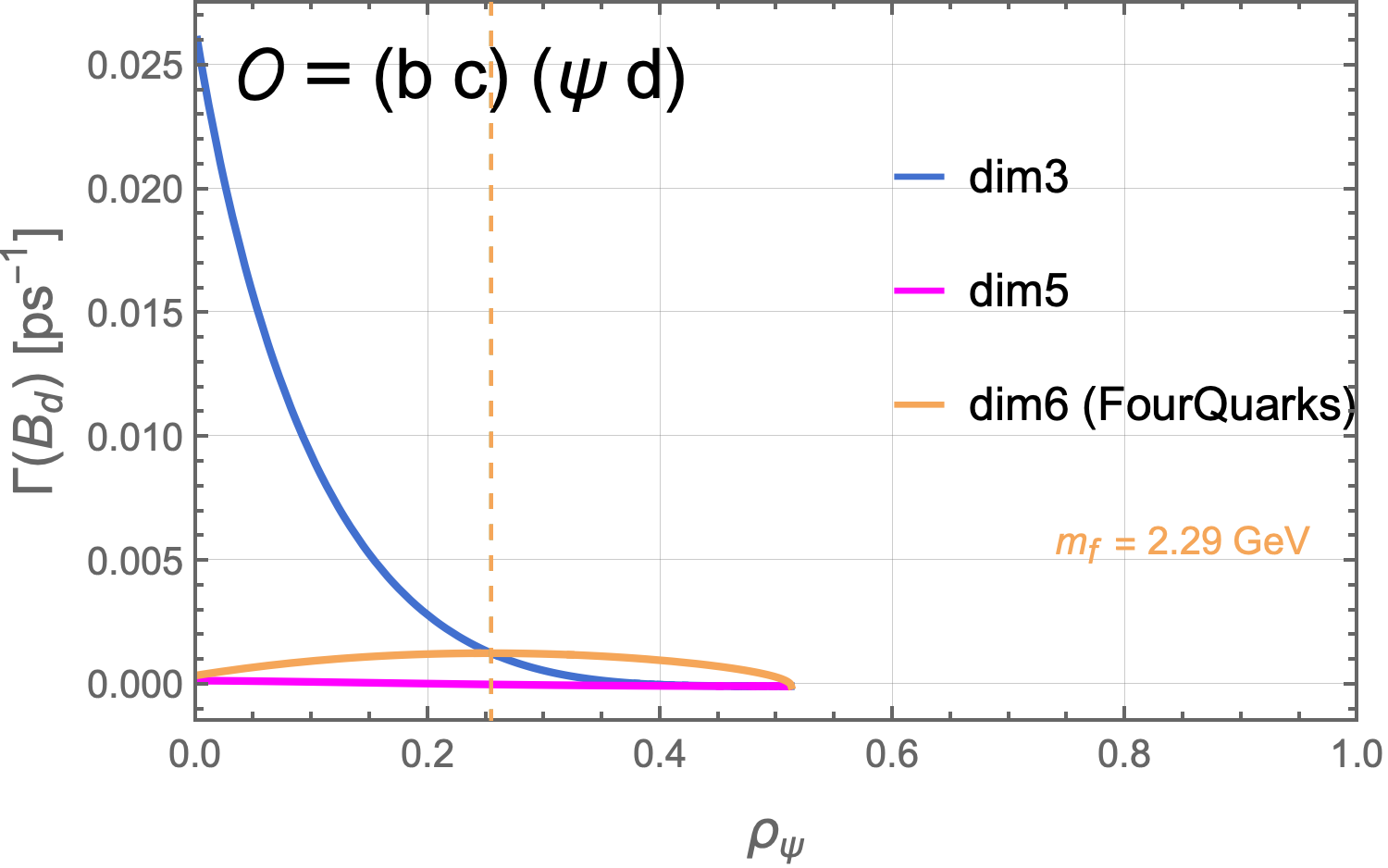}
    \end{subfigure}
    \hfill
    \begin{subfigure}{0.40\textwidth}
        \centering
        \includegraphics[width=\textwidth]{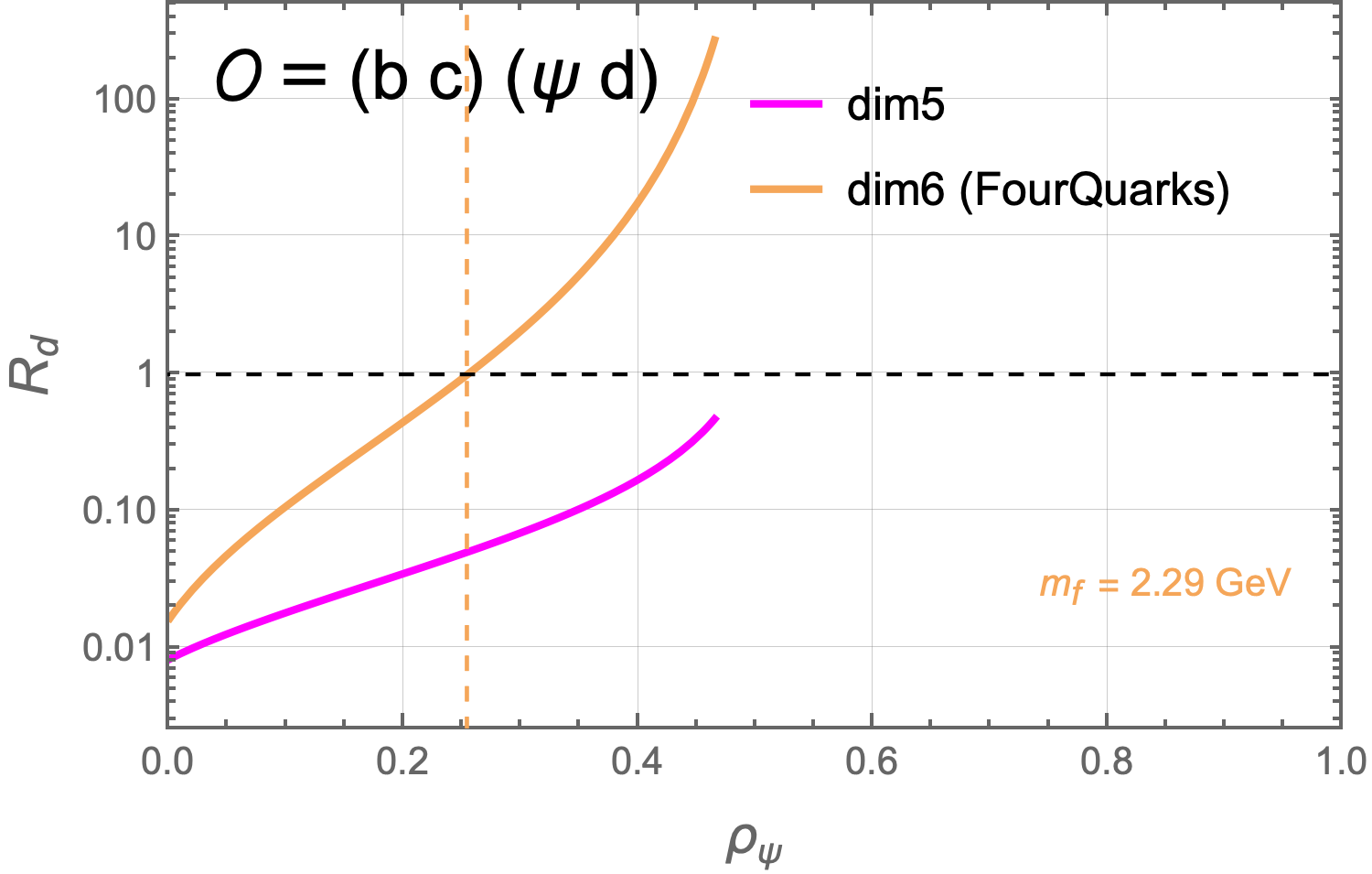}
    \end{subfigure}

    \caption{(Part 3): Additional plots for different operators with the same colour coding as in the previous Figure.}
    \label{fig:DecayRates4}
\end{figure}
%%%%%%%%%%%%%%%%%%%%%%%%%%%%%%%%%%%%%%%%%%%%%%%
\begin{figure}[ht]
    \centering
    \begin{subfigure}{0.43\textwidth}
        \centering
        \includegraphics[width=\textwidth]{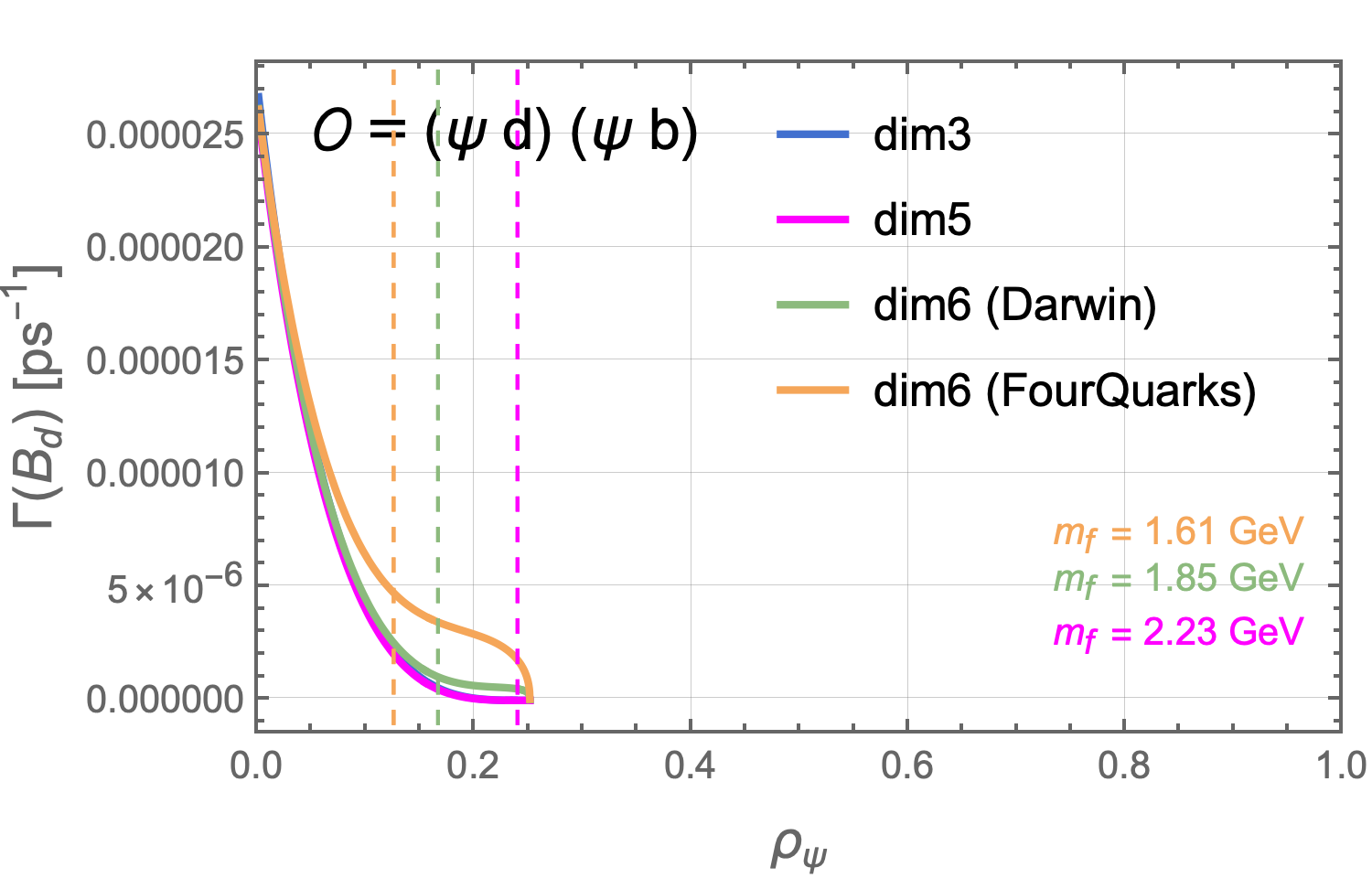}
    \end{subfigure}
    \hfill
    \begin{subfigure}{0.40\textwidth}
        \centering
        \includegraphics[width=\textwidth]{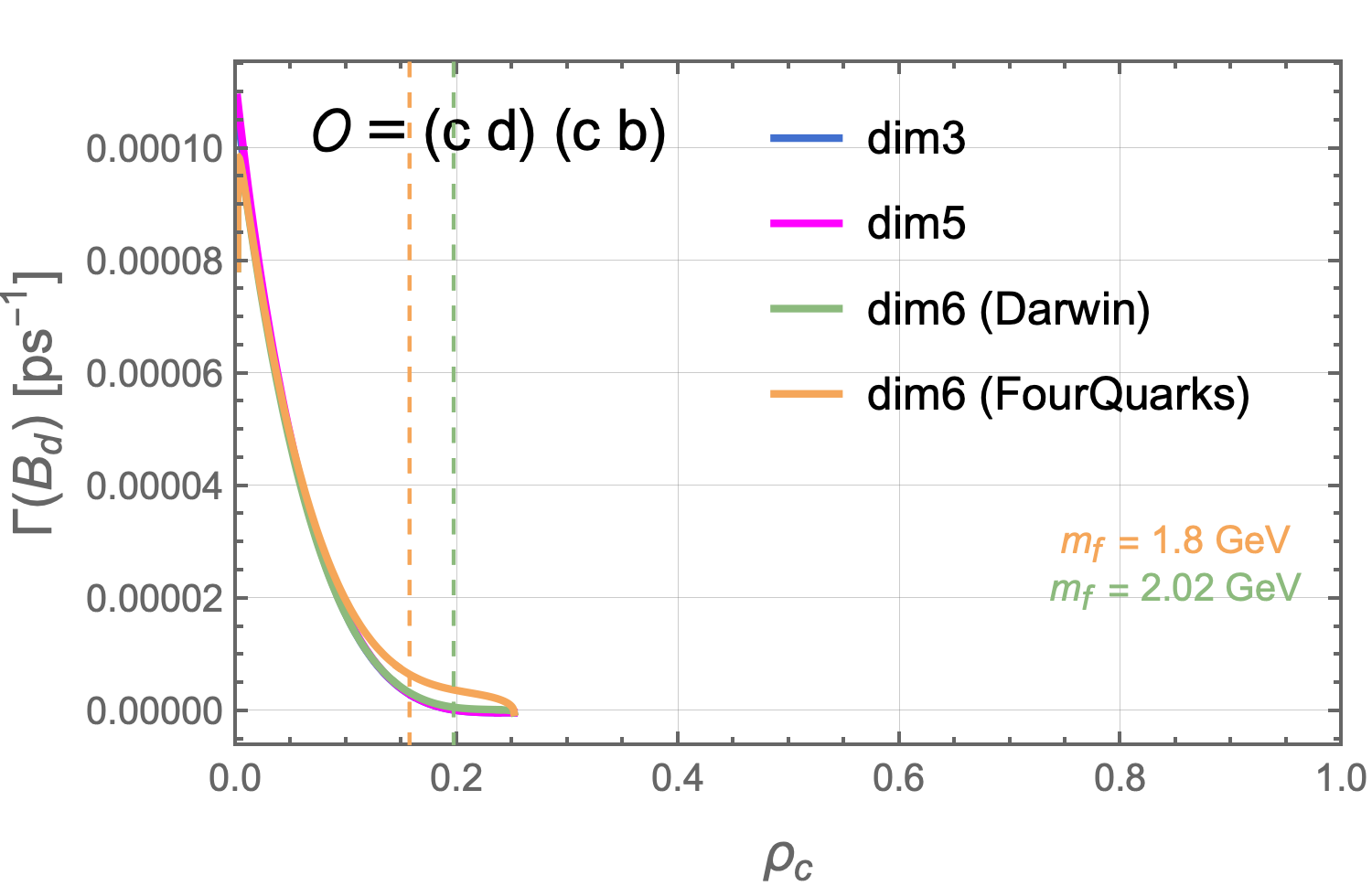}
    \end{subfigure}
     \hfill
    \begin{subfigure}{0.43\textwidth}
        \centering
        \includegraphics[width=\textwidth]{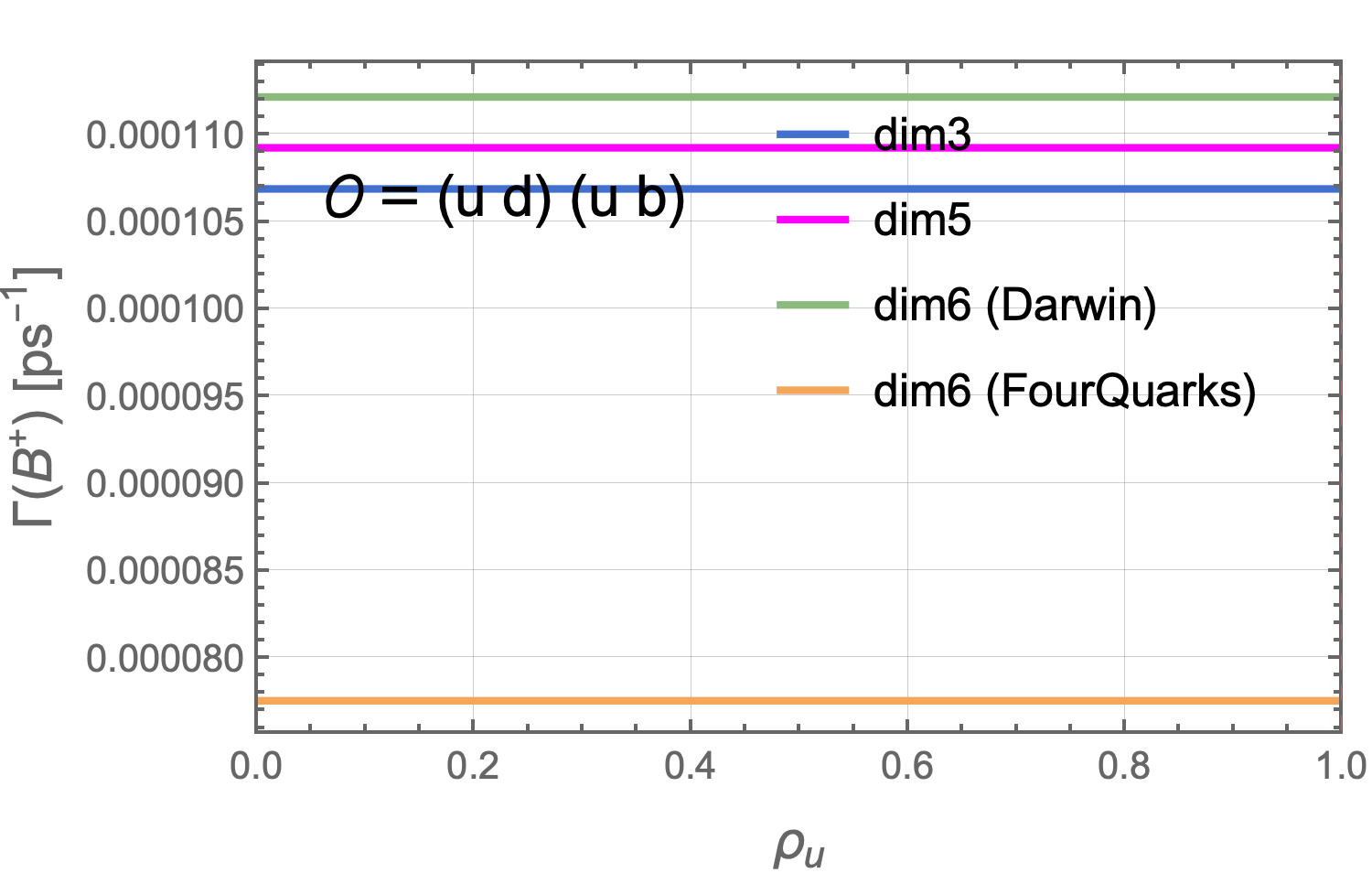}
    \end{subfigure}
     \hfill
    \begin{subfigure}{0.40\textwidth}
        \centering
        \includegraphics[width=\textwidth]{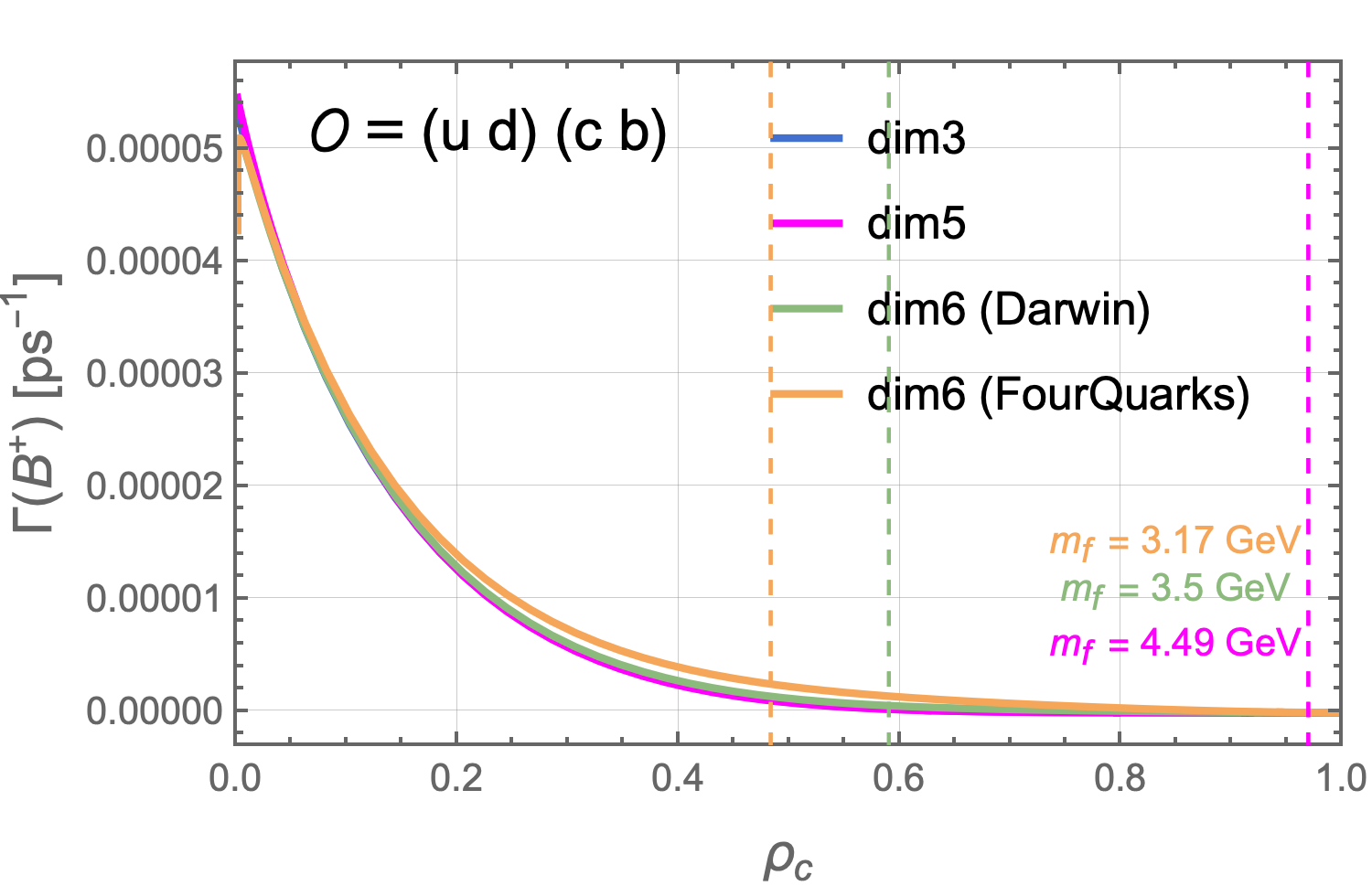}
    \end{subfigure}
     \hfill
    \begin{subfigure}{0.43\textwidth}
        \centering
        \includegraphics[width=\textwidth]{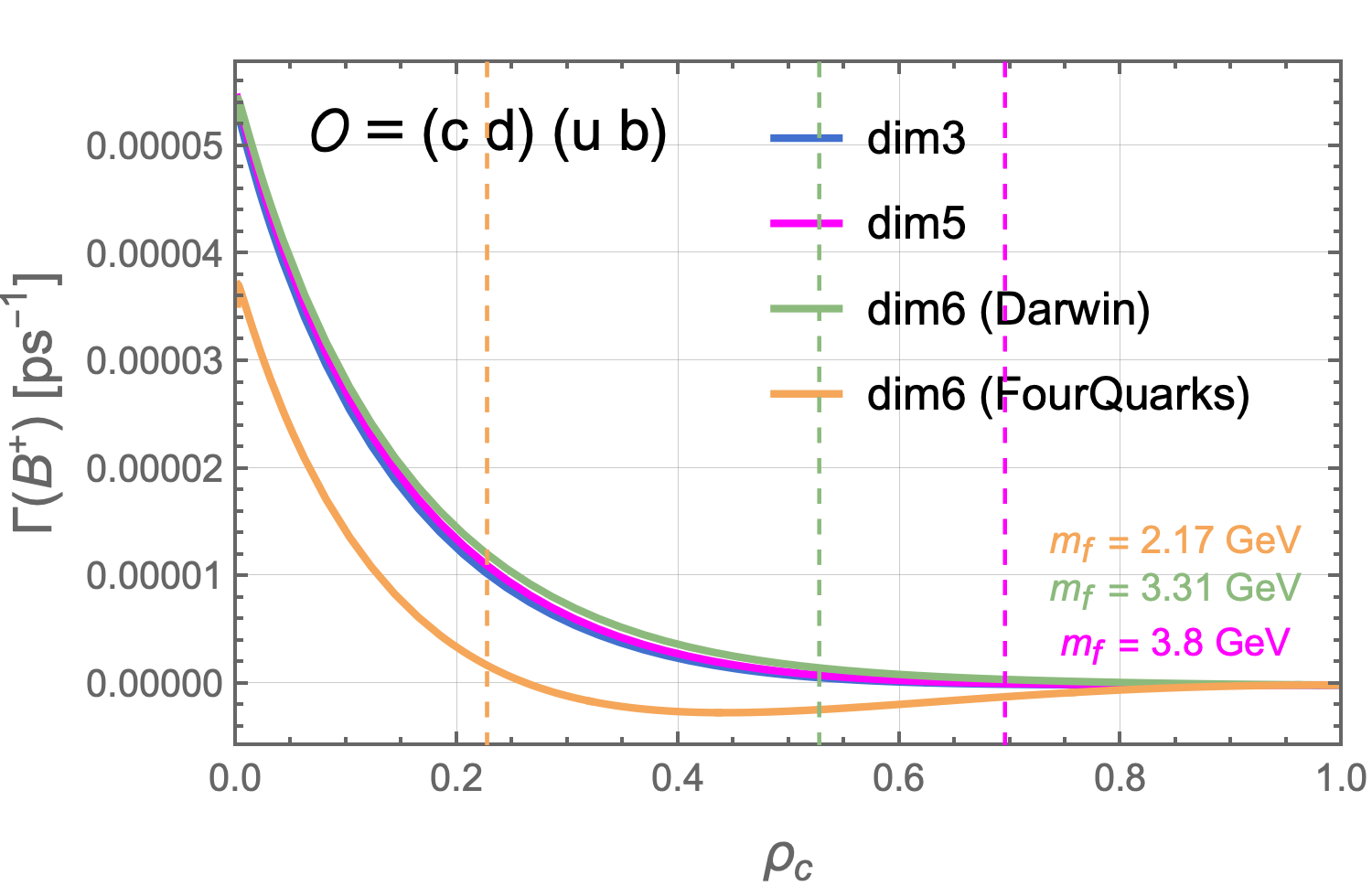}
    \end{subfigure}
     \hfill
    \begin{subfigure}{0.40\textwidth}
        \centering
        \includegraphics[width=\textwidth]{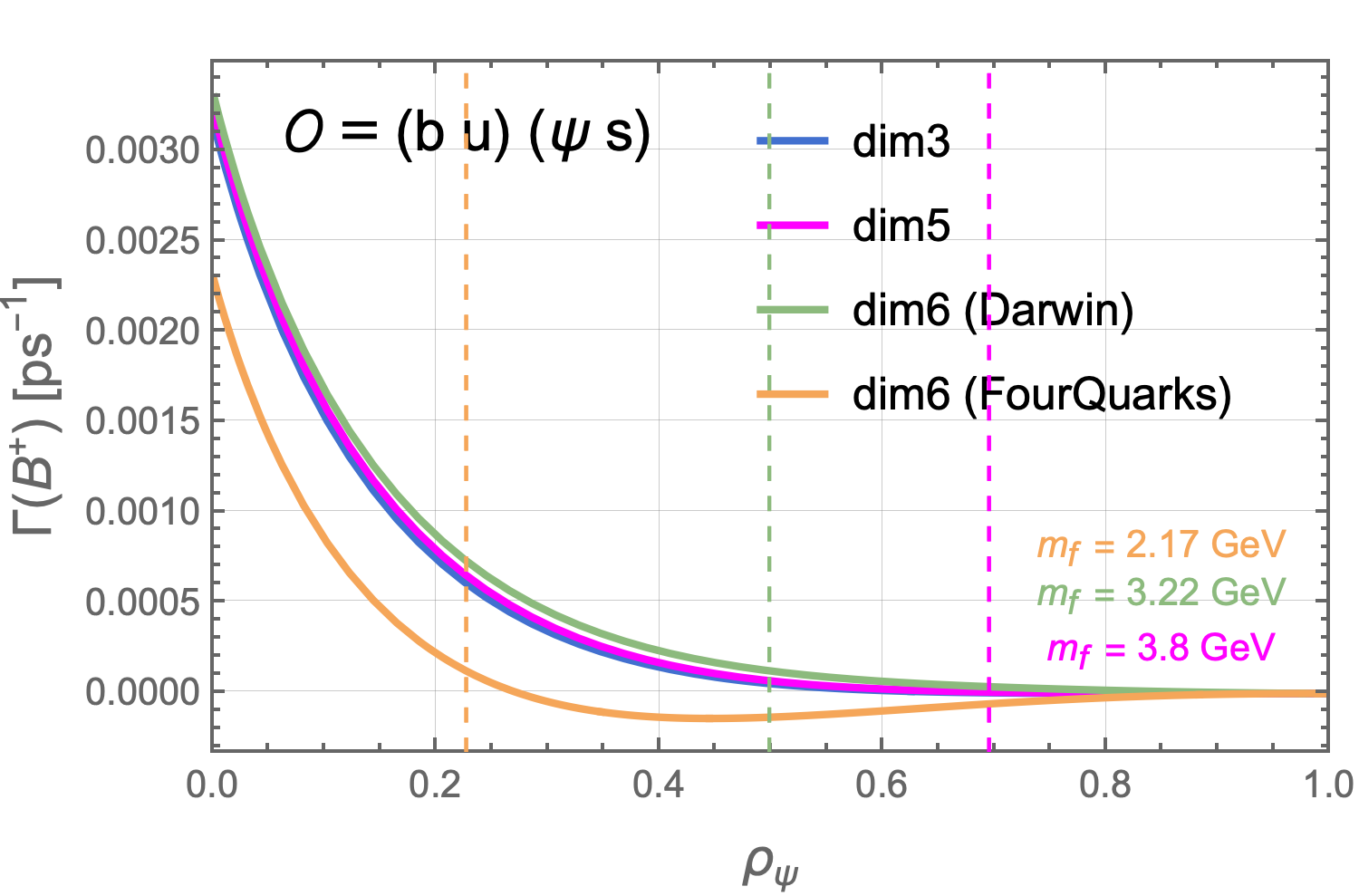}
    \end{subfigure}
     \hfill
    \begin{subfigure}{0.43\textwidth}
        \centering
        \includegraphics[width=\textwidth]{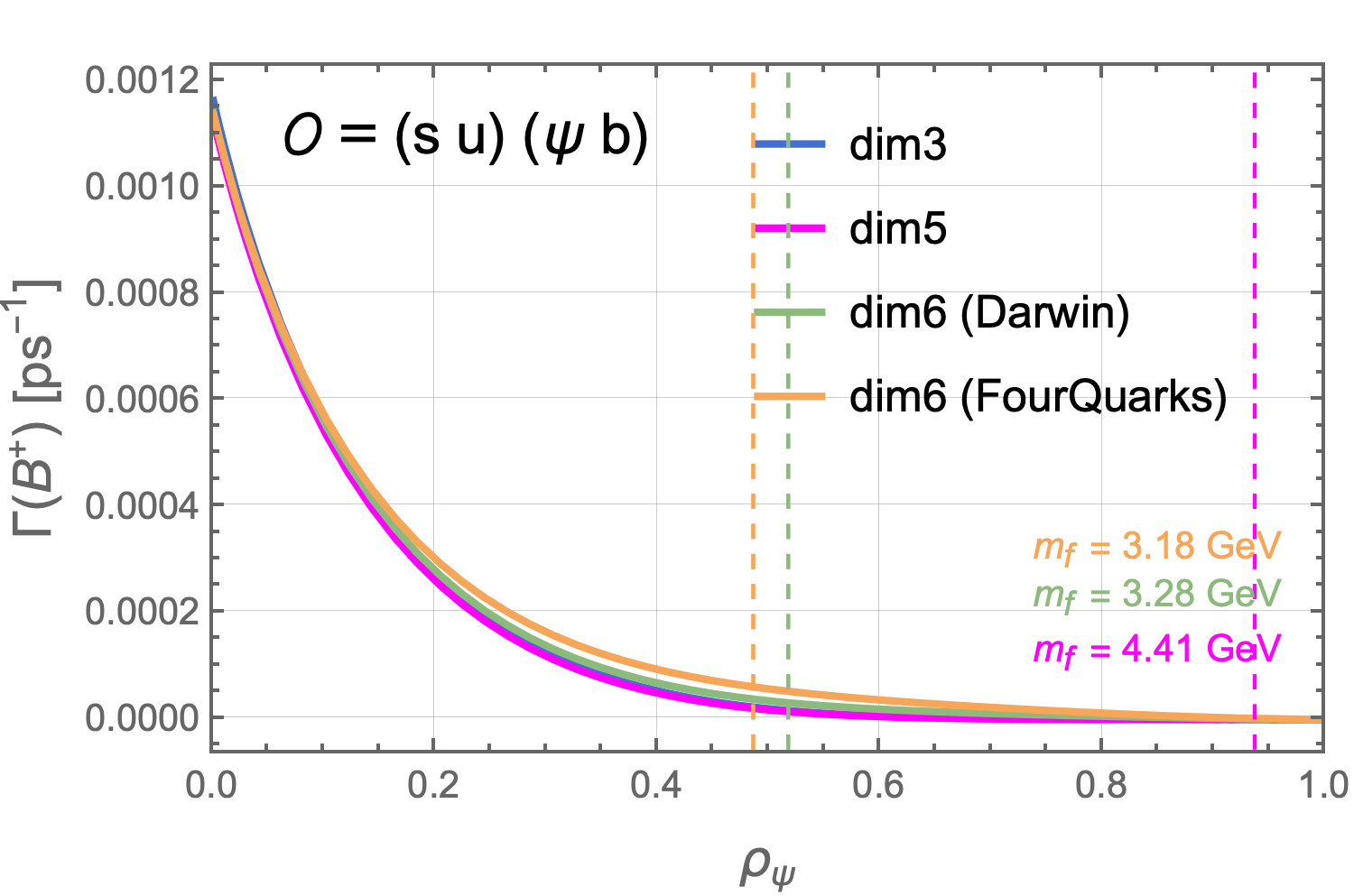}
    \end{subfigure}
    \hfill
    \begin{subfigure}{0.40\textwidth}
        \centering
        \includegraphics[width=\textwidth]{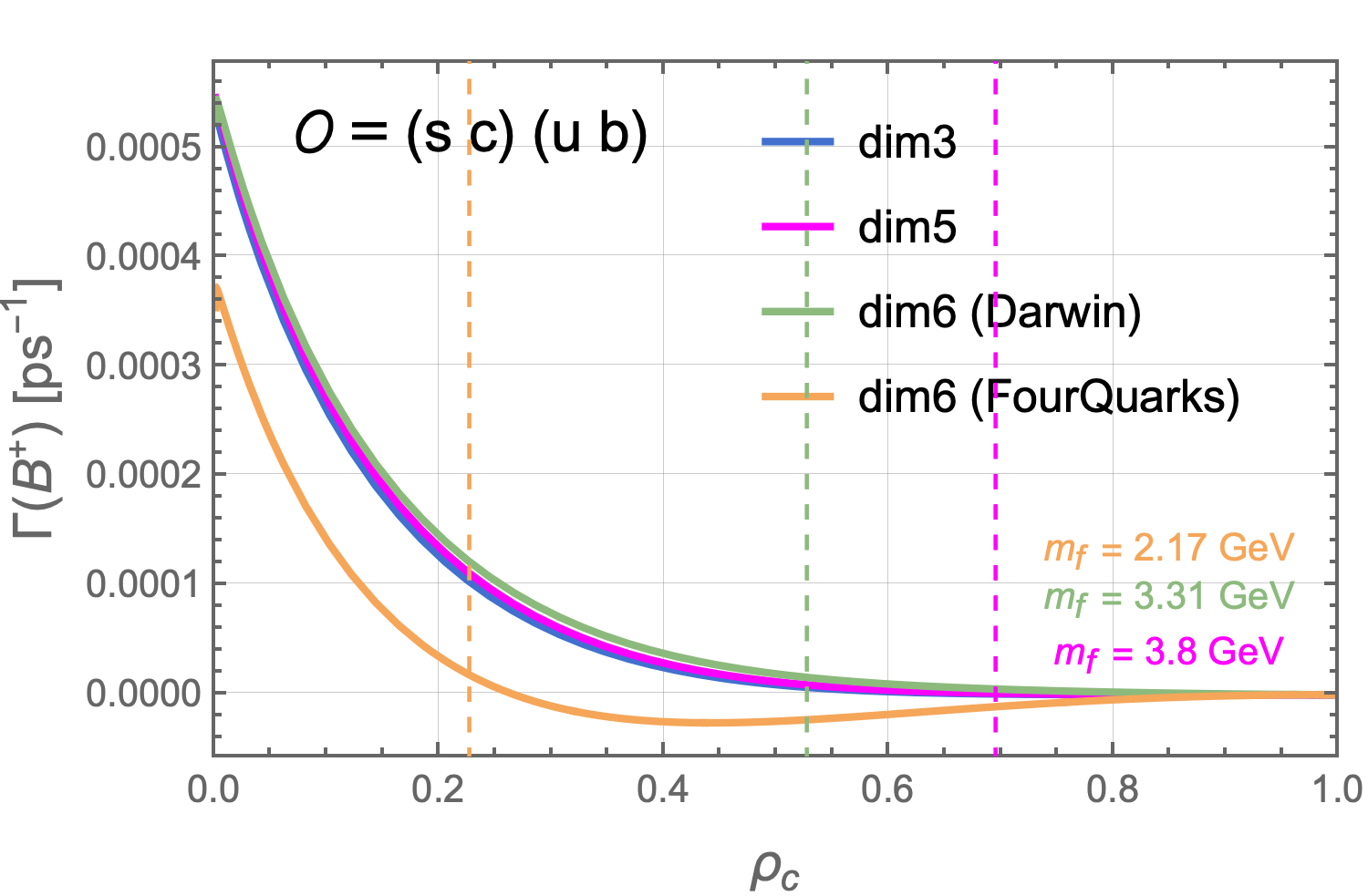}
    \end{subfigure}
        \caption{(Part 1): Total decay width of $\Gamma(B^+)$ arising purely from the Mesogenesis (BSM) operators as a function of $\rho_f= (m_f/m_b)^2$. The figure displays the cumulative sum of all terms up to the indicated dimension. The vertical lines indicate the value of $\rho_f$, where the contribution exceeds the leading dimension-three $\Gamma_3$ term (the corresponding $m_f$ values are also indicated).}\label{fig:DecayRatesTotal2}
\end{figure}
\begin{figure}[ht]
    \centering
    \hfill
    \begin{subfigure}{0.40\textwidth}
        \centering
        \includegraphics[width=\textwidth]{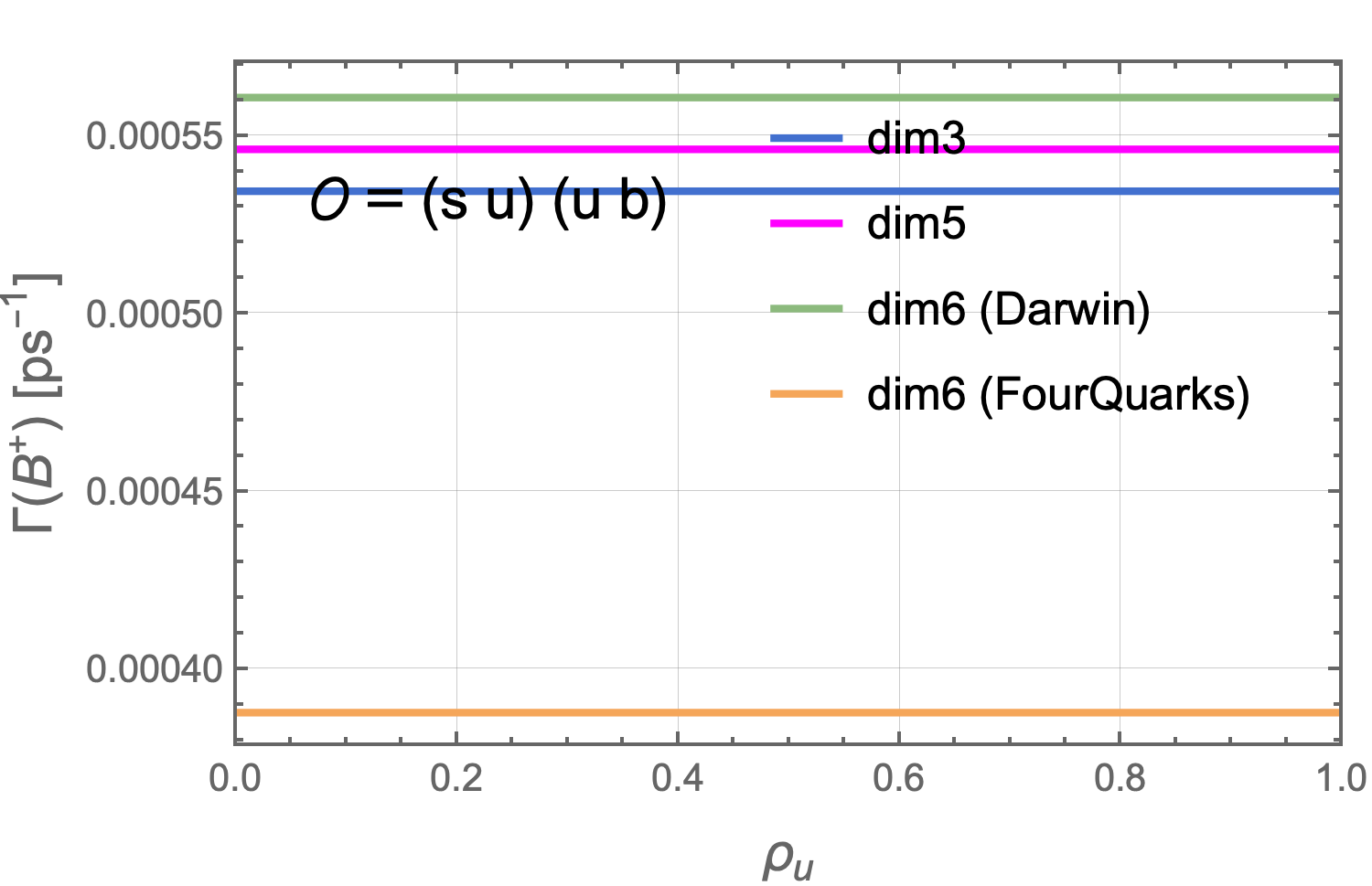}
    \end{subfigure}
     \hfill
    \begin{subfigure}{0.40\textwidth}
        \centering
        \includegraphics[width=\textwidth]{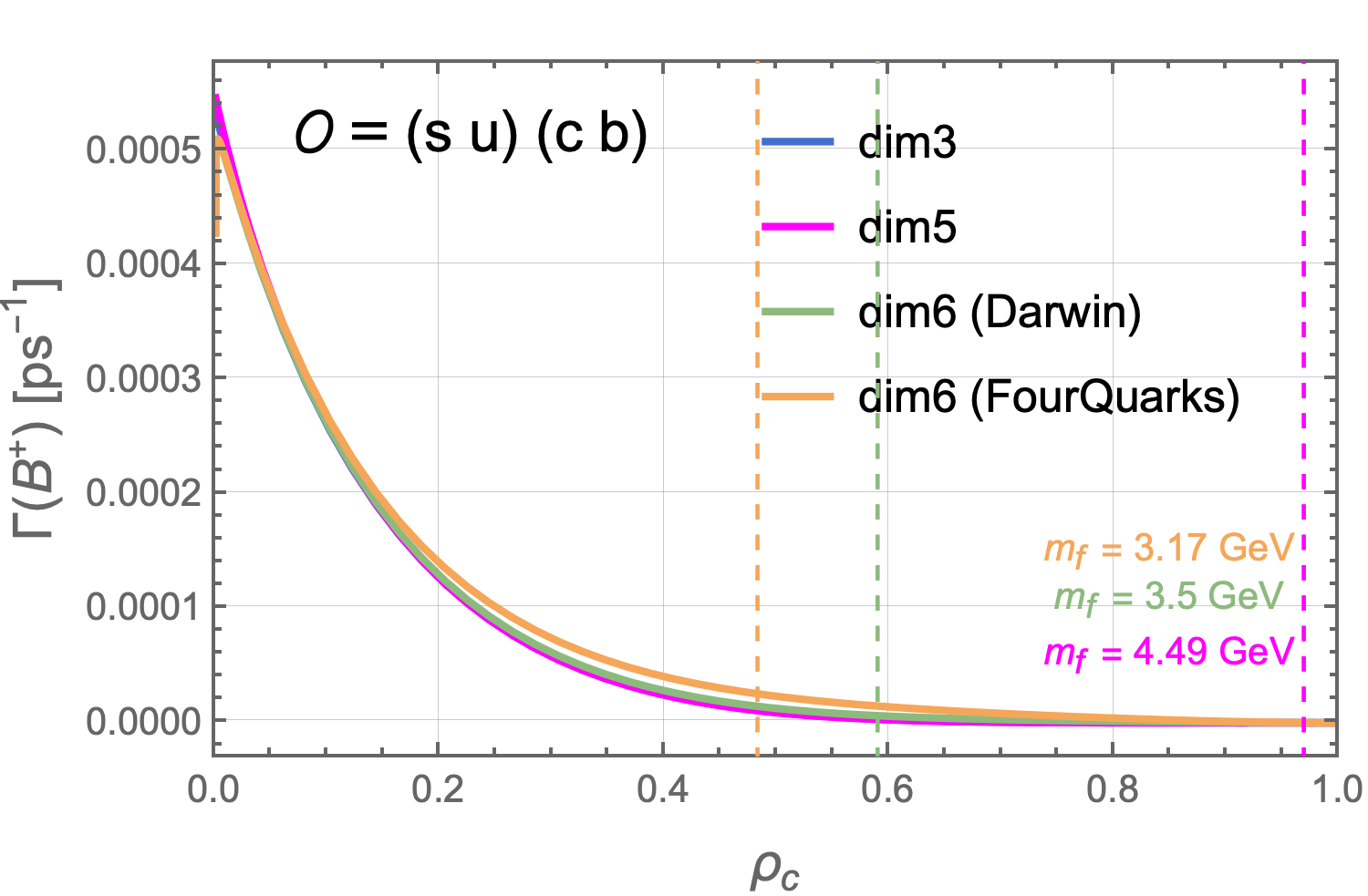}
    \end{subfigure}

  \medskip % Adds a small vertical space between the rows
    
    \begin{subfigure}{0.43\textwidth}
   %     \centering
        \includegraphics[width=\textwidth]{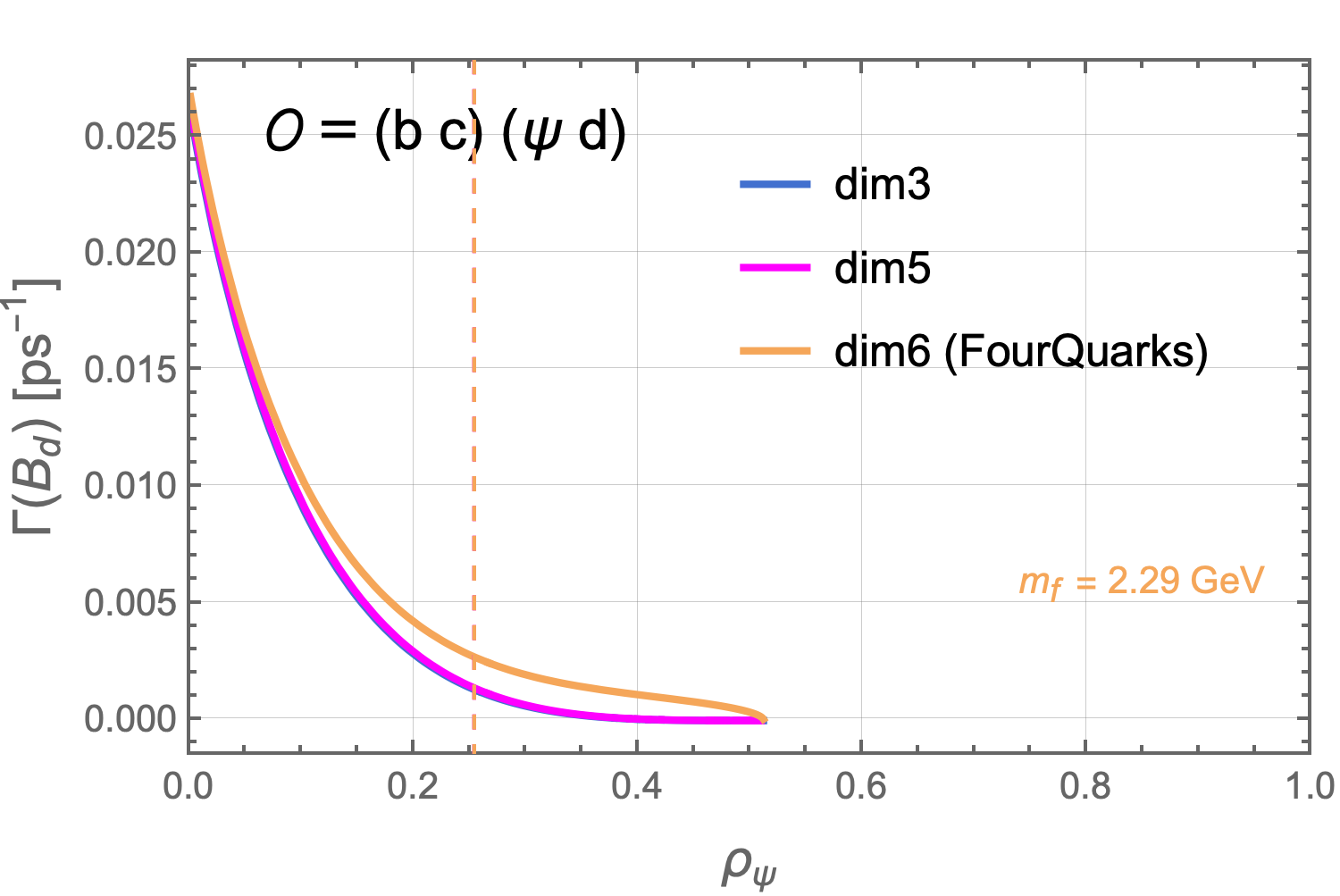}
    \end{subfigure}
        \caption{(Part 2): Total decay width of $\Gamma(B^+)$ arising purely from the Mesogenesis (BSM) operators as a function of $\rho_f= (m_f/m_b)^2$, with the same colour coding as in the previous figure. }\label{fig:DecayRatesTotal3}
\end{figure}

\begin{figure}[ht]
    \centering
    \begin{subfigure}{0.40\textwidth}
        \centering
        \includegraphics[width=\textwidth]{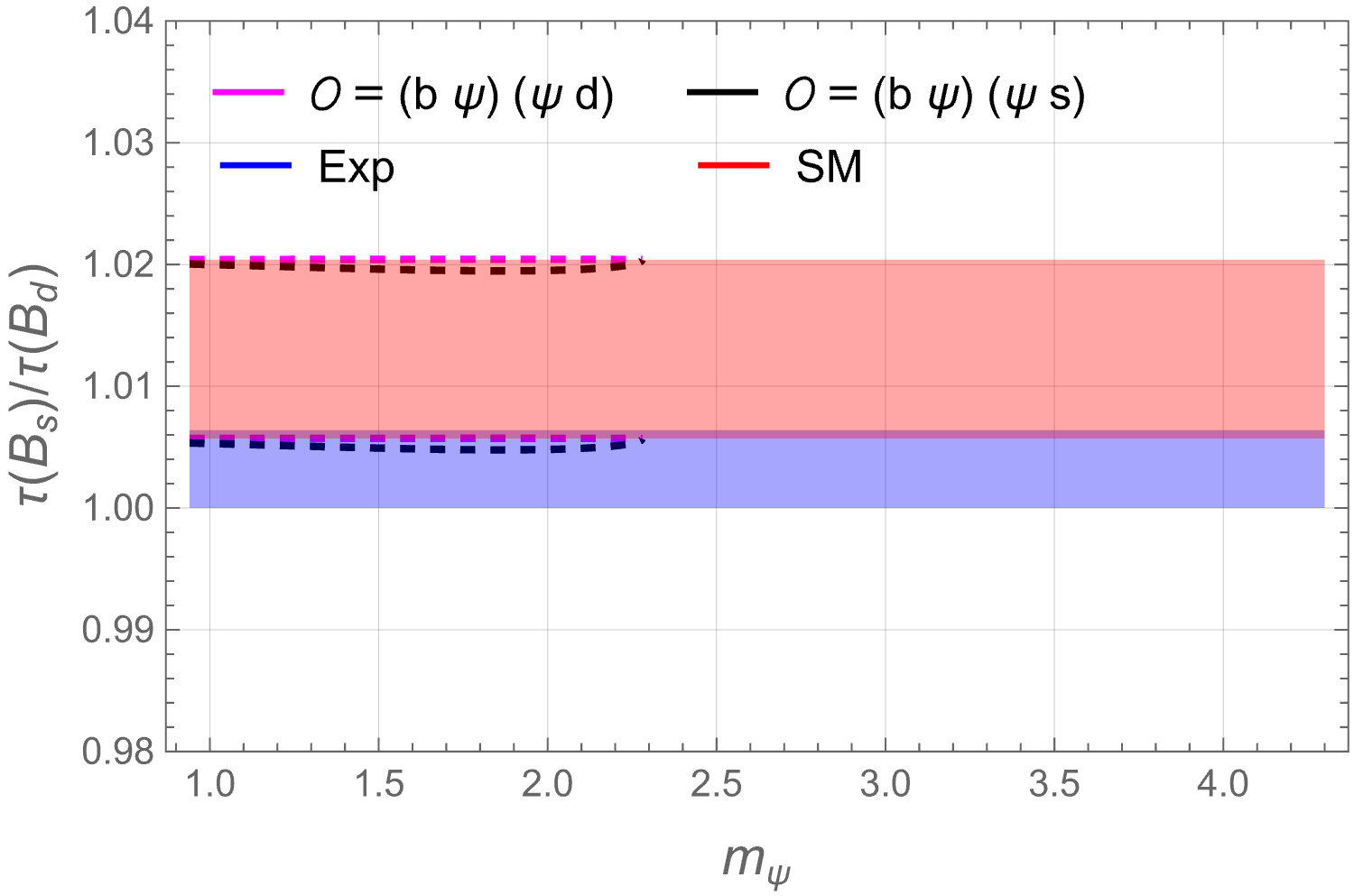}
    \end{subfigure}
    \hfill
    \begin{subfigure}{0.40\textwidth}
        \centering
        \includegraphics[width=\textwidth]{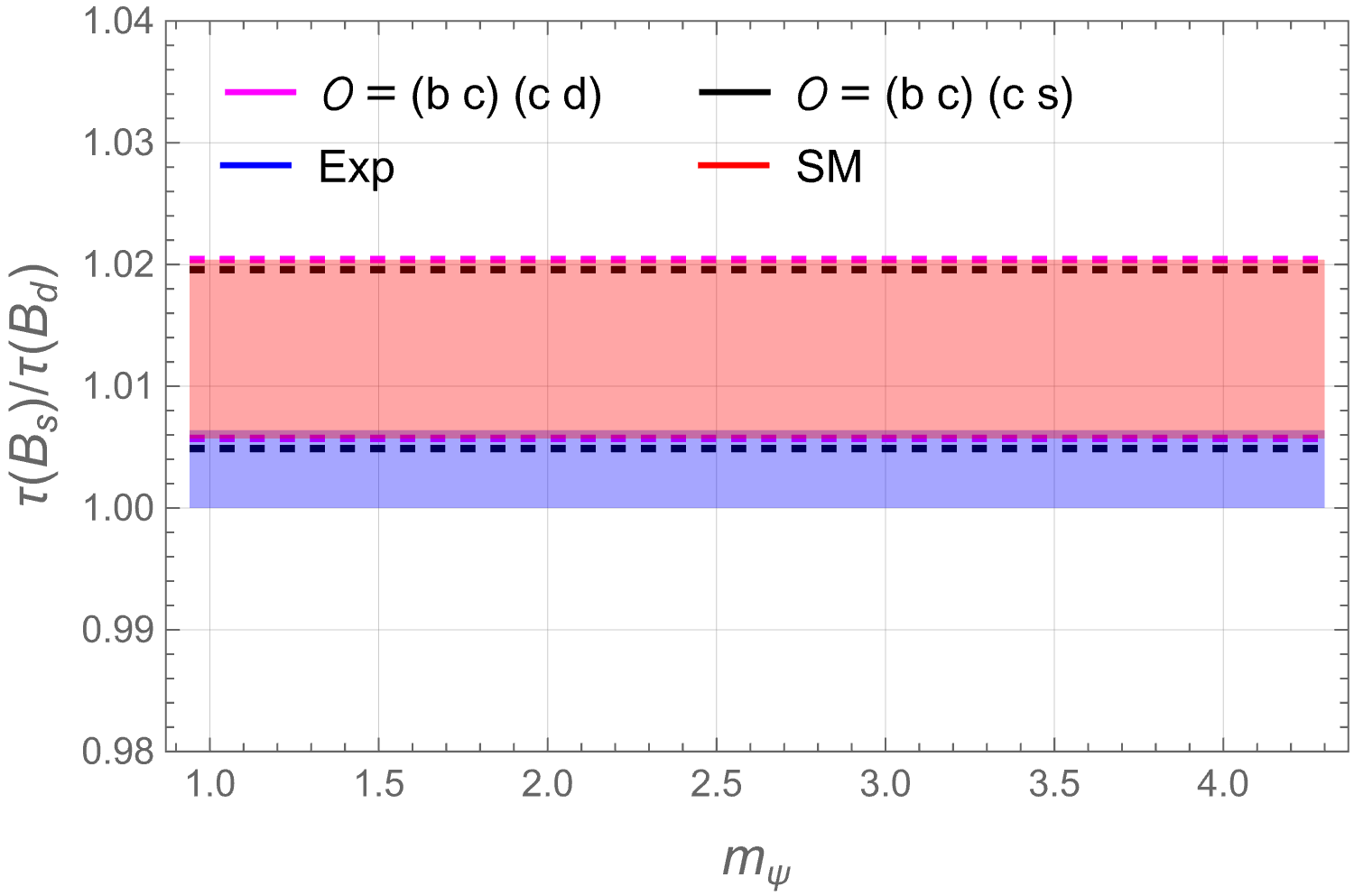}
    \end{subfigure}
     \hfill
    \begin{subfigure}{0.40\textwidth}
        \centering
        \includegraphics[width=\textwidth]{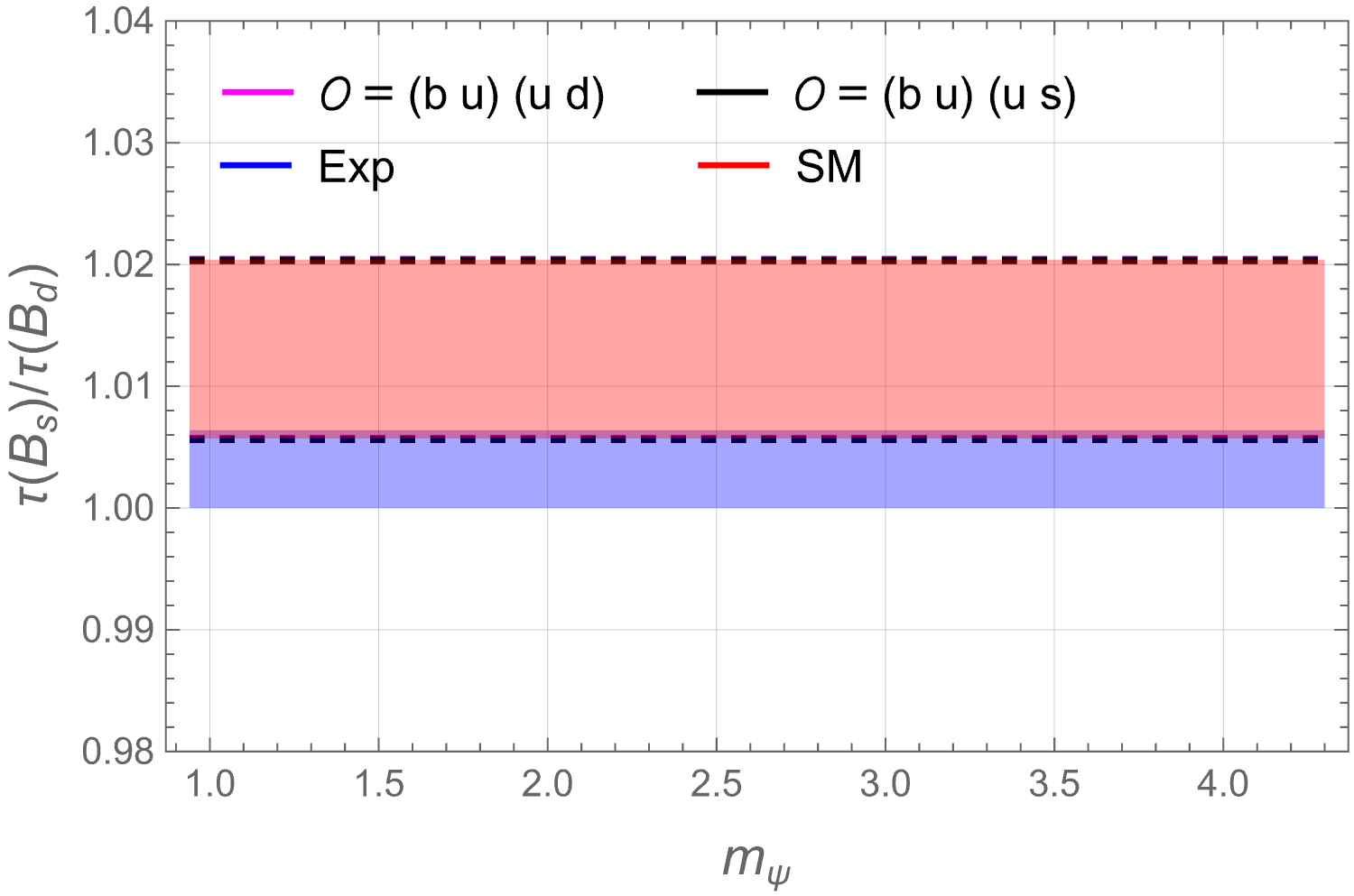}
    \end{subfigure}
     \hfill
    \begin{subfigure}{0.40\textwidth}
        \centering
        \includegraphics[width=\textwidth]{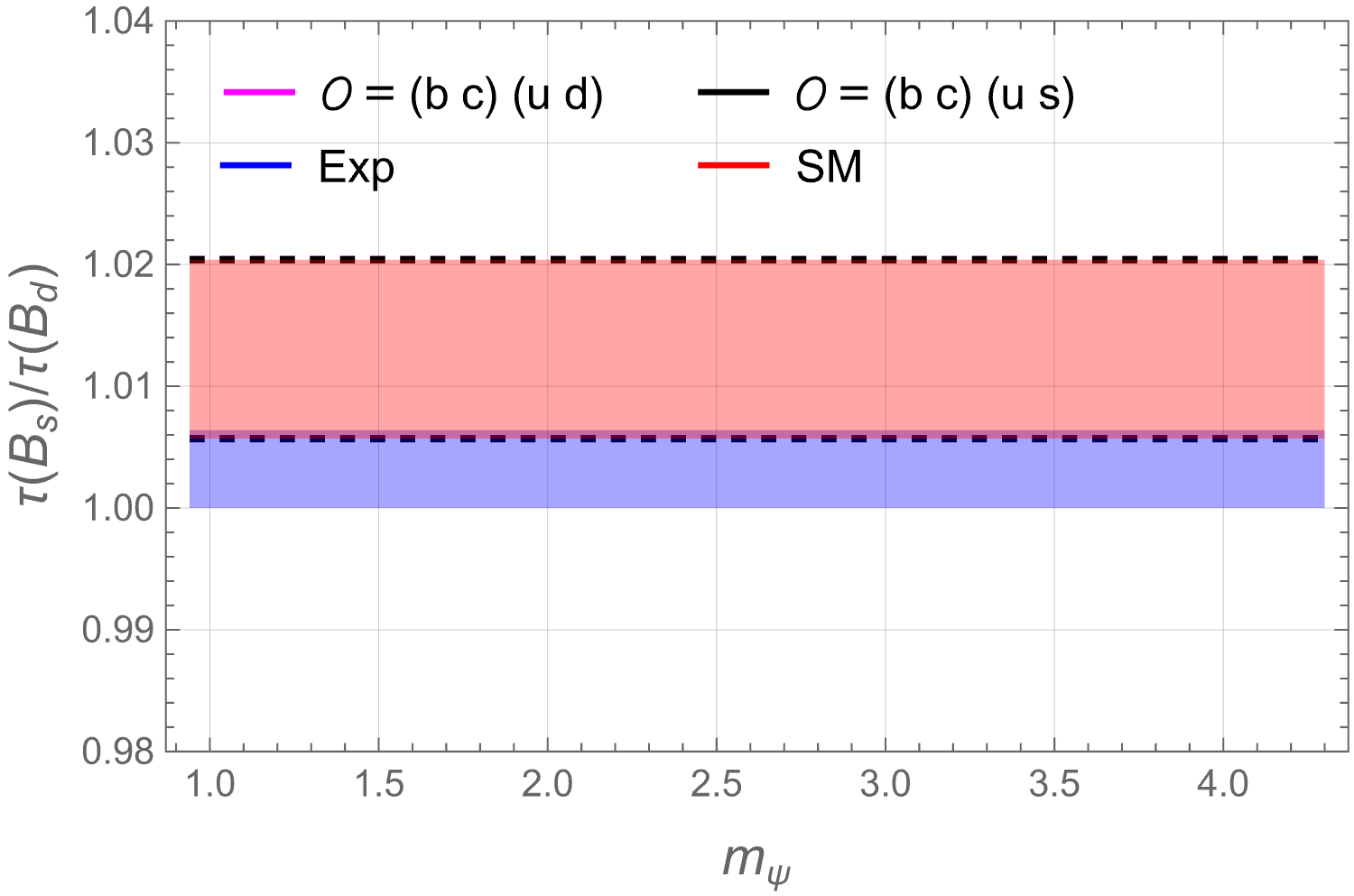}
    \end{subfigure}
     \hfill
    \begin{subfigure}{0.40\textwidth}
        \centering
        \includegraphics[width=\textwidth]{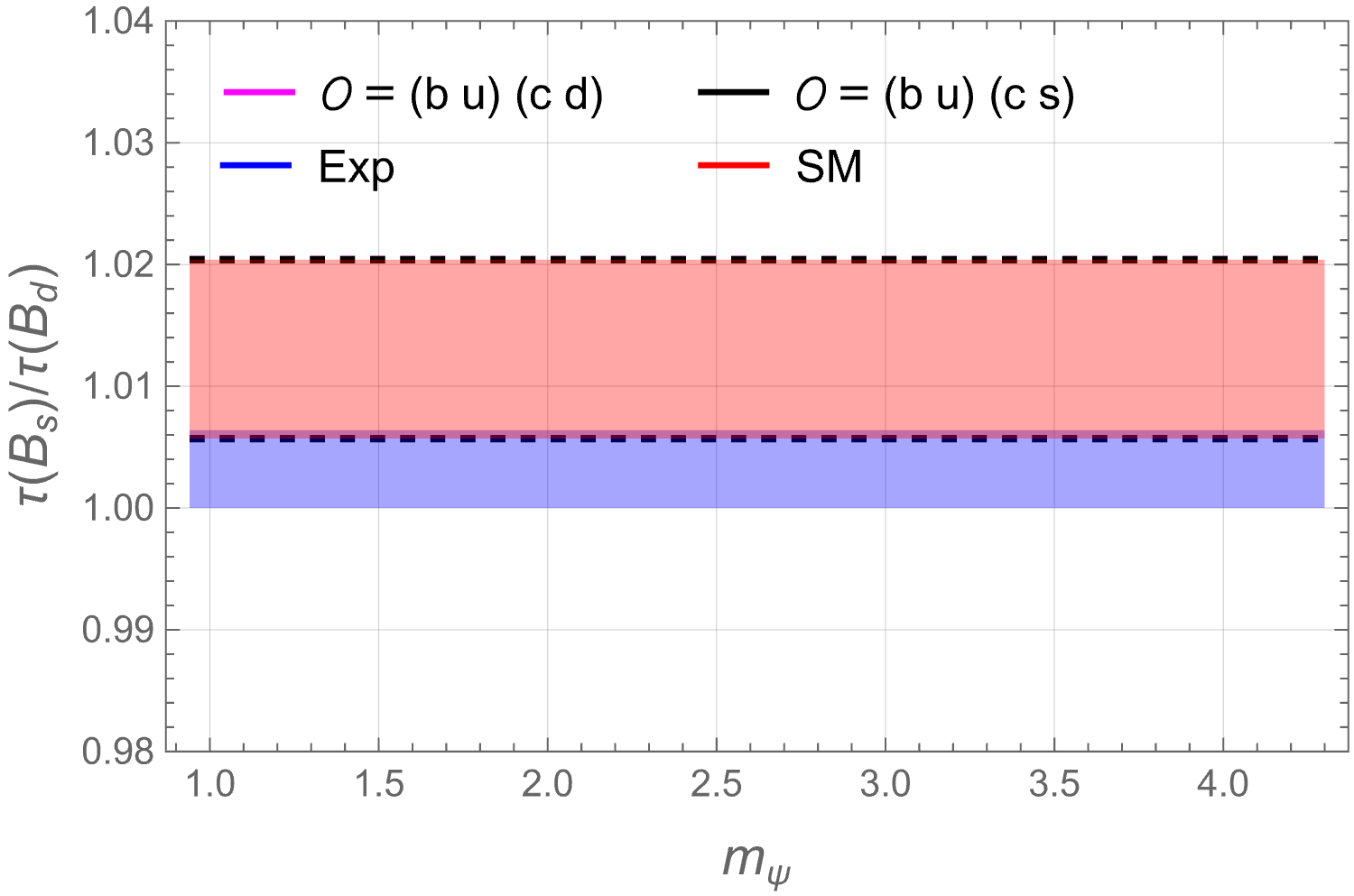}
    \end{subfigure}
       \caption{Lifetime ratio $\tau(B_s)/\tau(B_d)$ as a function of the dark fermion mass $m_\psi$, 
including both SM and SM+NP predictions for representative operators.  
The NP contribution is found to be negligible compared to present theoretical uncertainties. (Operators involving two distinct massive final states are not shown, as the corresponding Darwin-term contribution has not yet been calculated.)}
        \label{fig:BsBdAll}
\end{figure}

\begin{figure}[ht]
    \centering
    \begin{subfigure}{0.40\textwidth}
        \centering
        \includegraphics[width=\textwidth]{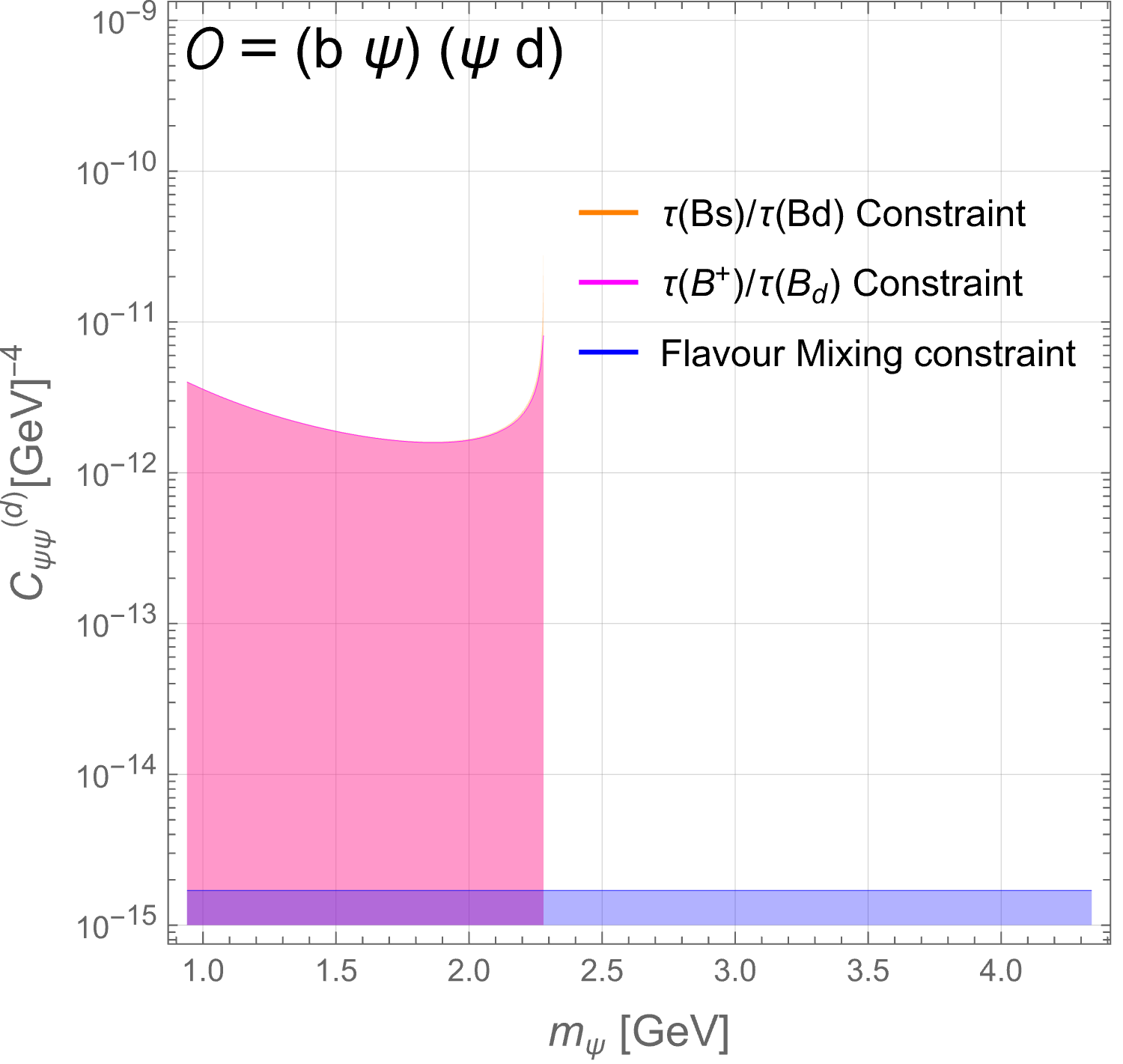}
    \end{subfigure}
    \hfill
    \begin{subfigure}{0.40\textwidth}
        \centering
        \includegraphics[width=\textwidth]{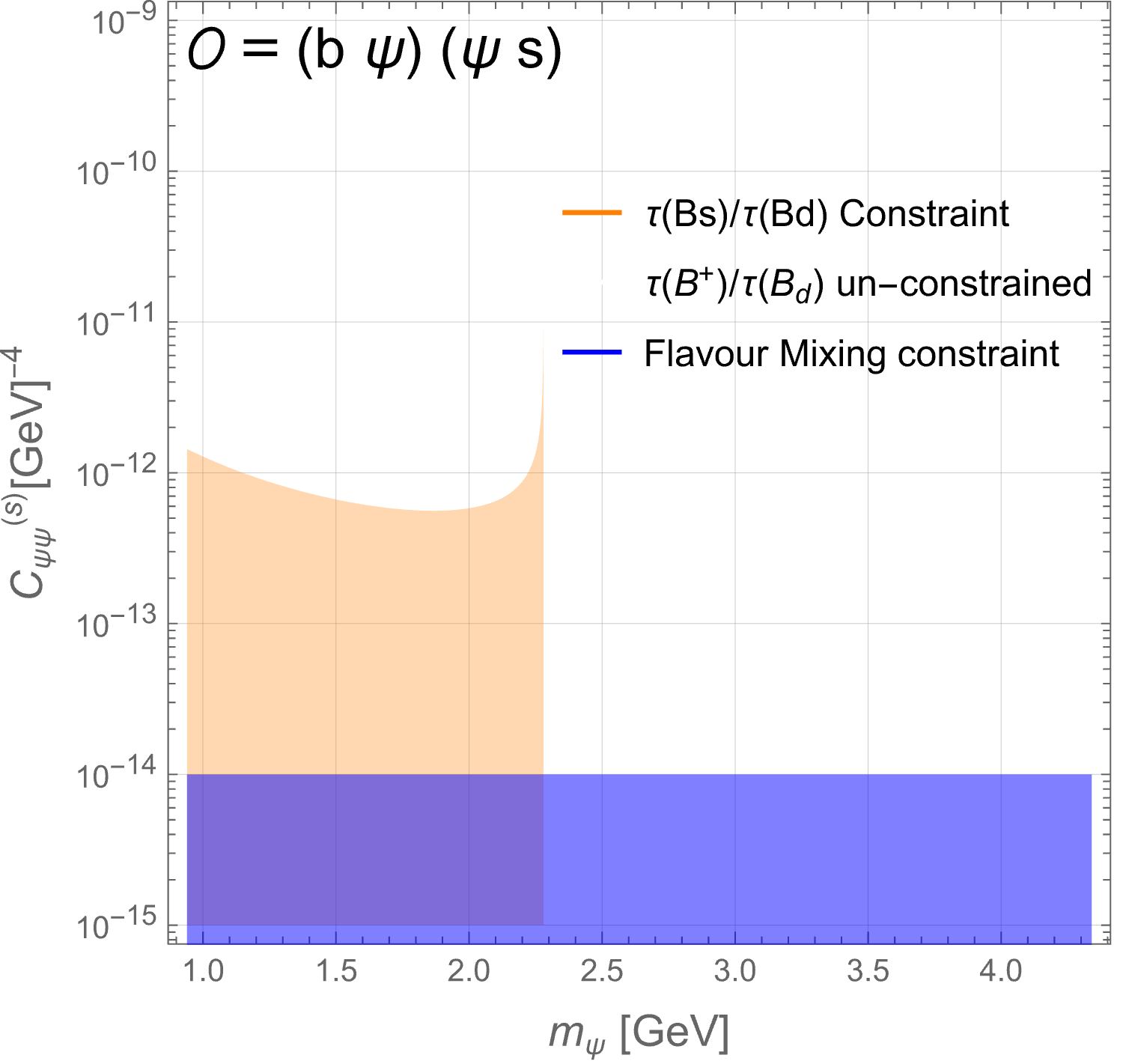}
    \end{subfigure}
    \begin{subfigure}{0.40\textwidth}
        \centering
        \includegraphics[width=\textwidth]{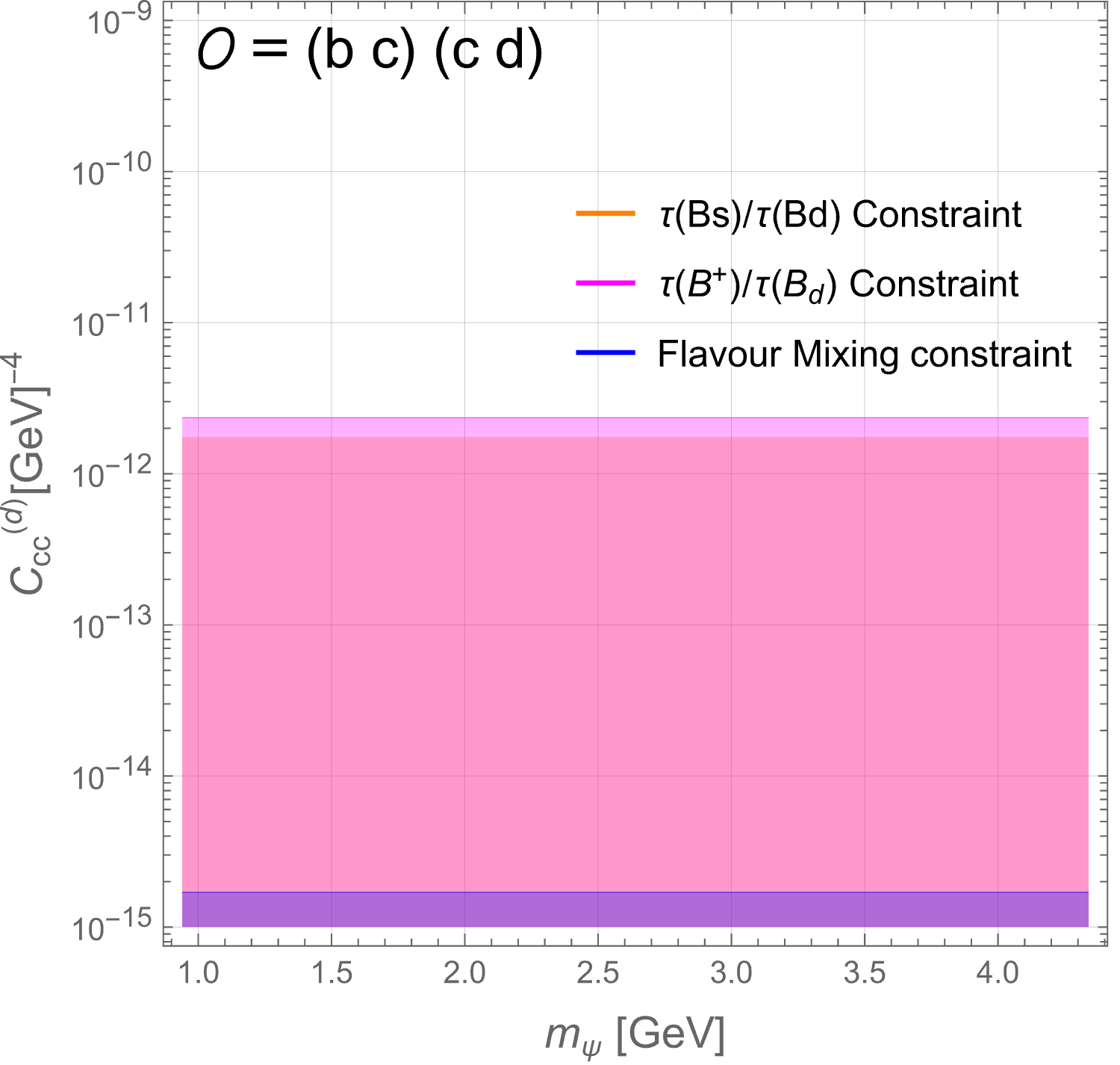}
    \end{subfigure}
      \hfill
    \begin{subfigure}{0.40\textwidth}
        \centering
        \includegraphics[width=\textwidth]{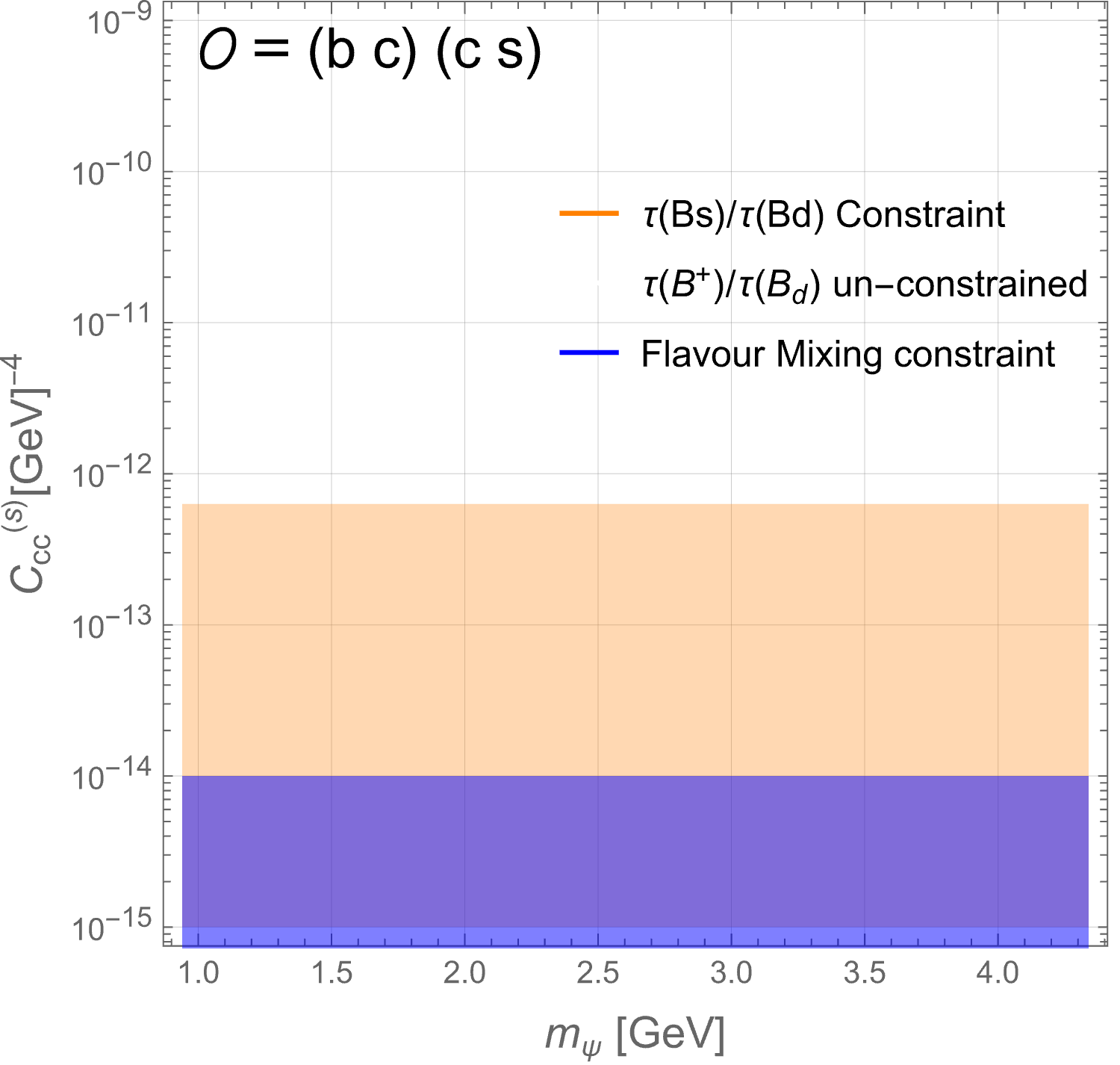}
    \end{subfigure}

     \begin{subfigure}{0.40\textwidth}
        \centering
        \includegraphics[width=\textwidth]{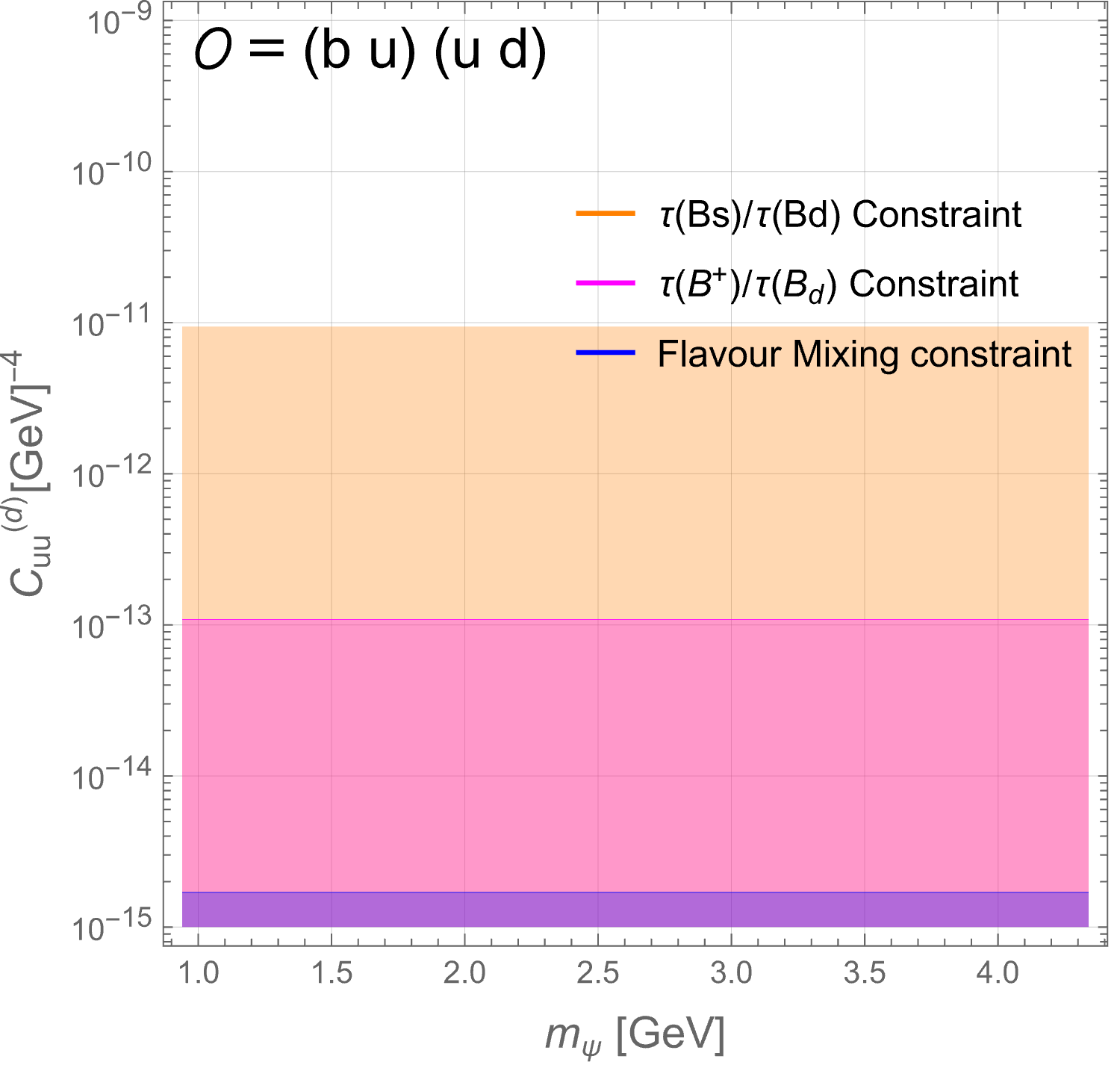}
    \end{subfigure}
      \hfill
    \begin{subfigure}{0.40\textwidth}
        \centering
        \includegraphics[width=\textwidth]{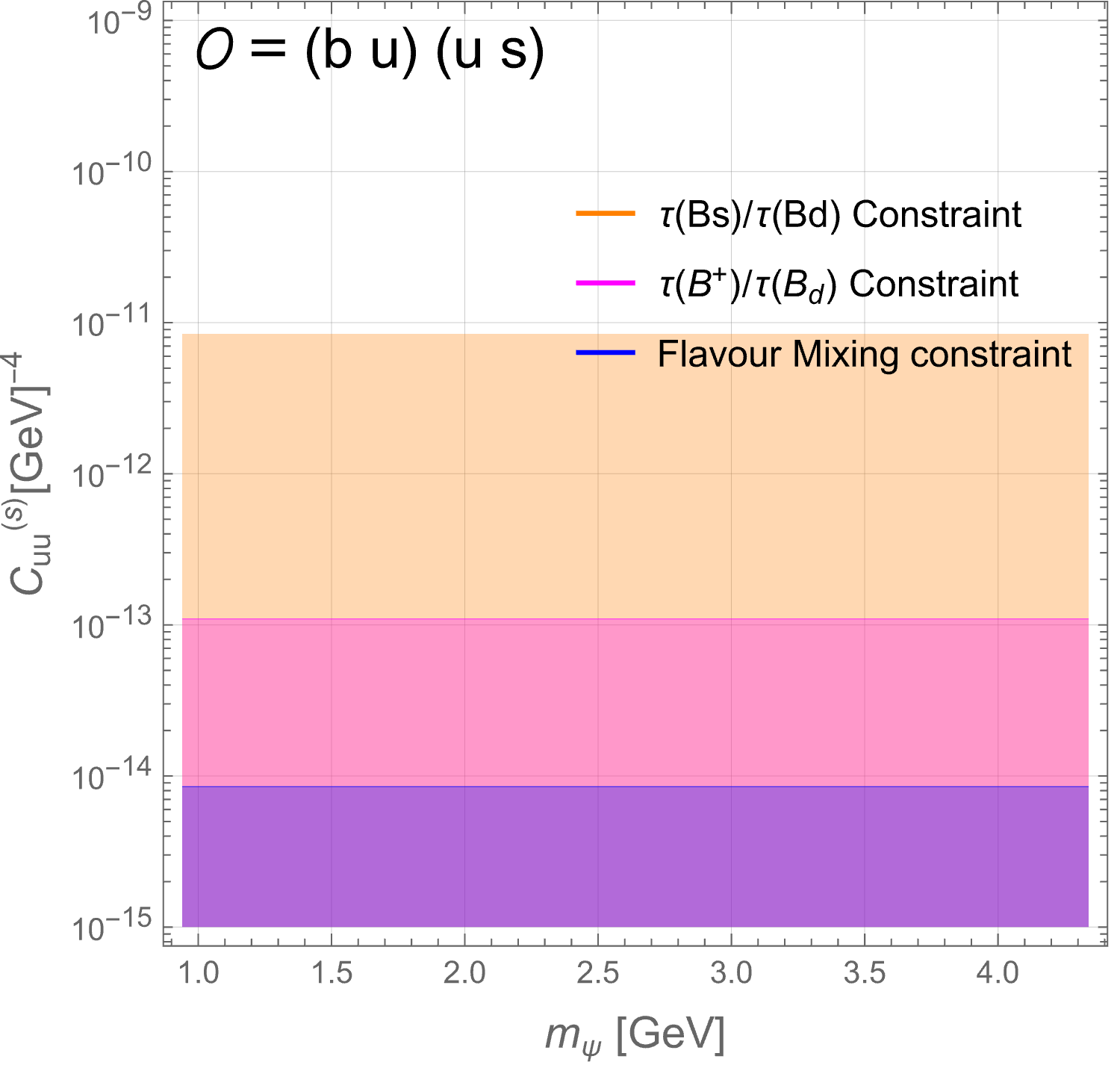}
    \end{subfigure}
       \caption{(Part 1): Constraints on the Wilson coefficients as a function of the dark fermion mass $m_\psi$, derived from the lifetime ratios $\tau(B_s)/\tau(B_d)$ (orange) and $\tau(B^+)/\tau(B_d)$ (magenta), compared to collider constraints (blue).}
        \label{fig:BsBdConstraints2}
\end{figure}

\begin{figure}[ht]
    \centering
    \begin{subfigure}{0.40\textwidth}
        \centering
        \includegraphics[width=\textwidth]{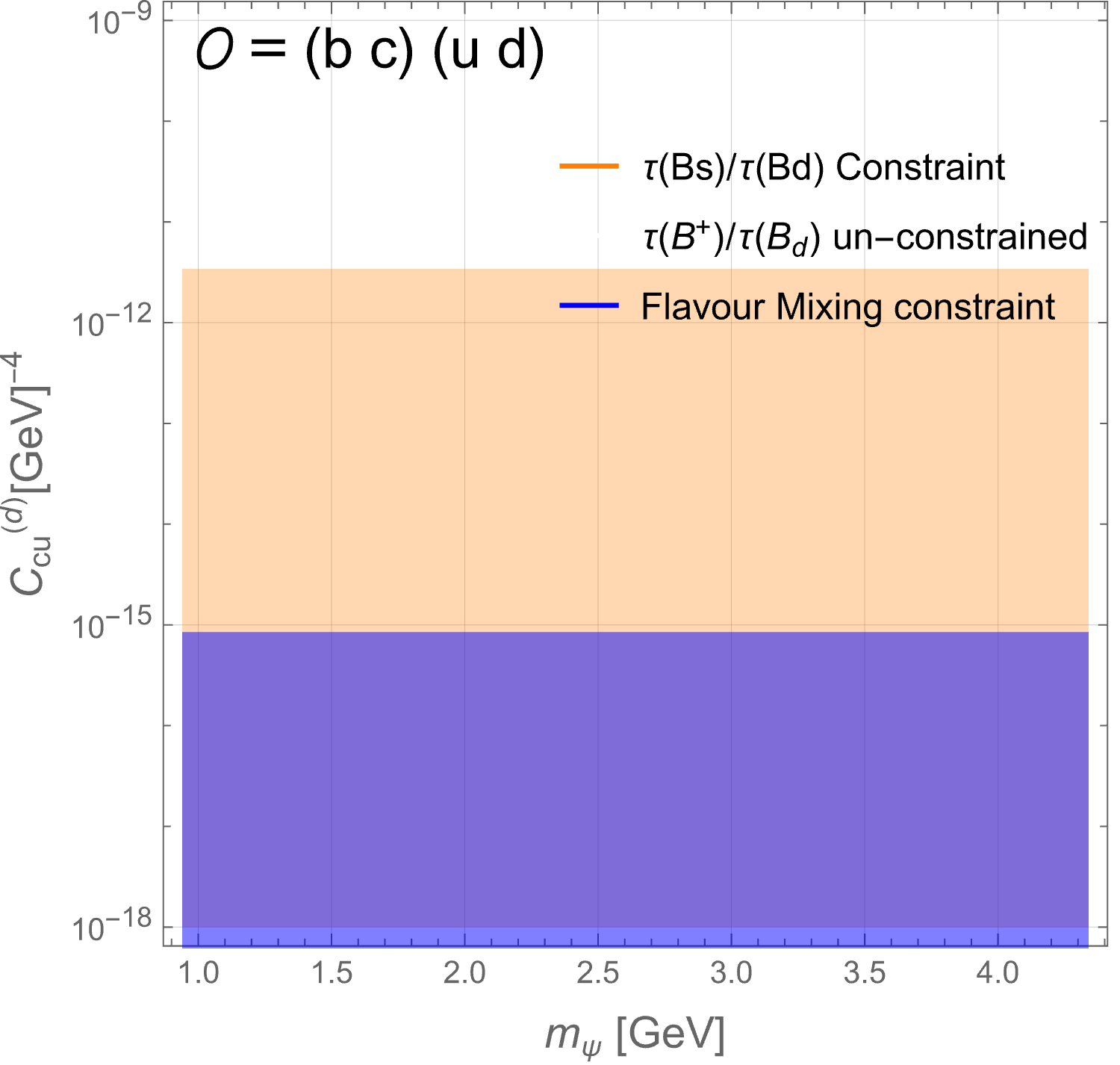}
    \end{subfigure}
    \hfill
    \begin{subfigure}{0.40\textwidth}
        \centering
        \includegraphics[width=\textwidth]{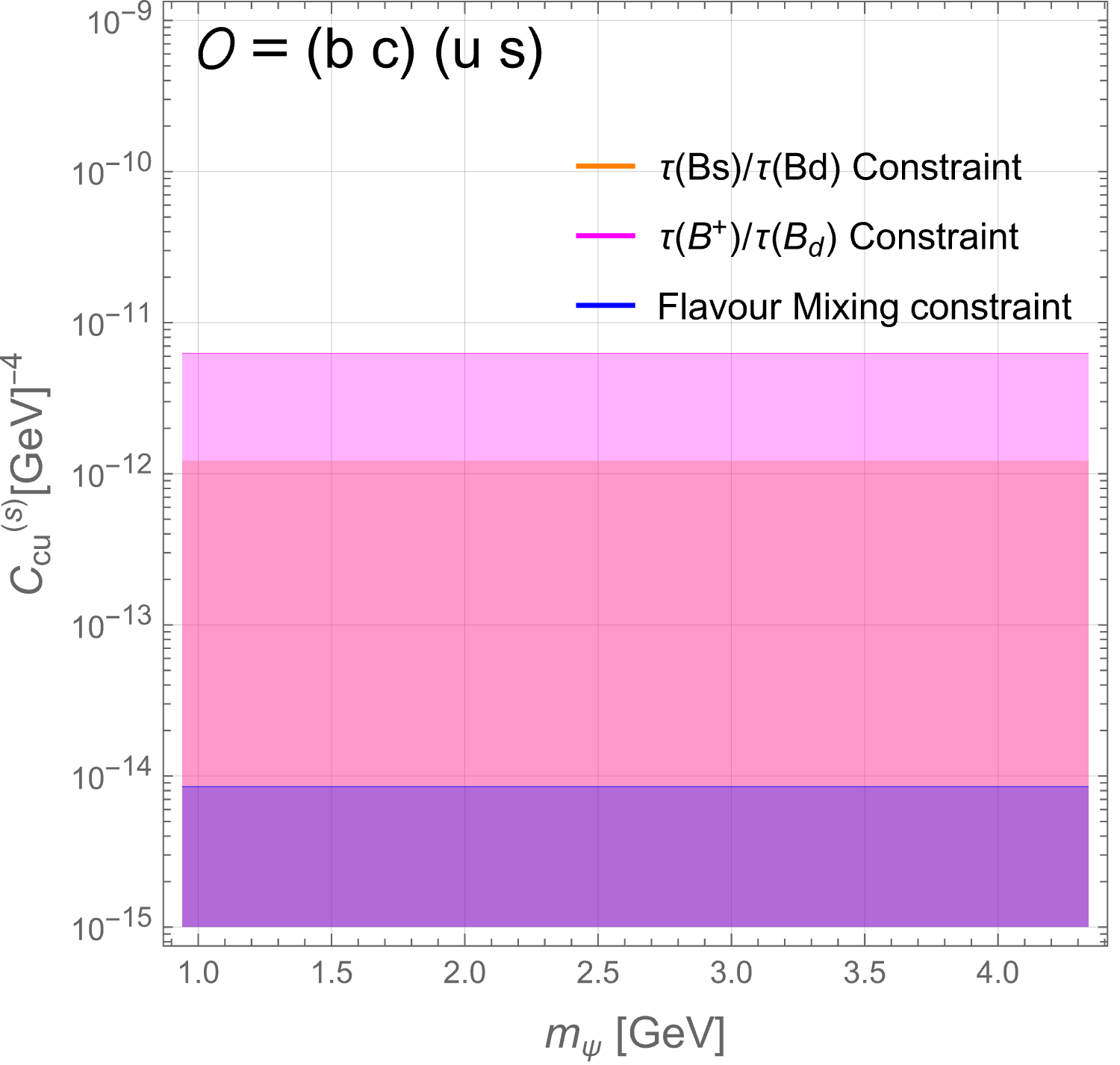}
    \end{subfigure}
     \begin{subfigure}{0.40\textwidth}
        \centering
        \includegraphics[width=\textwidth]{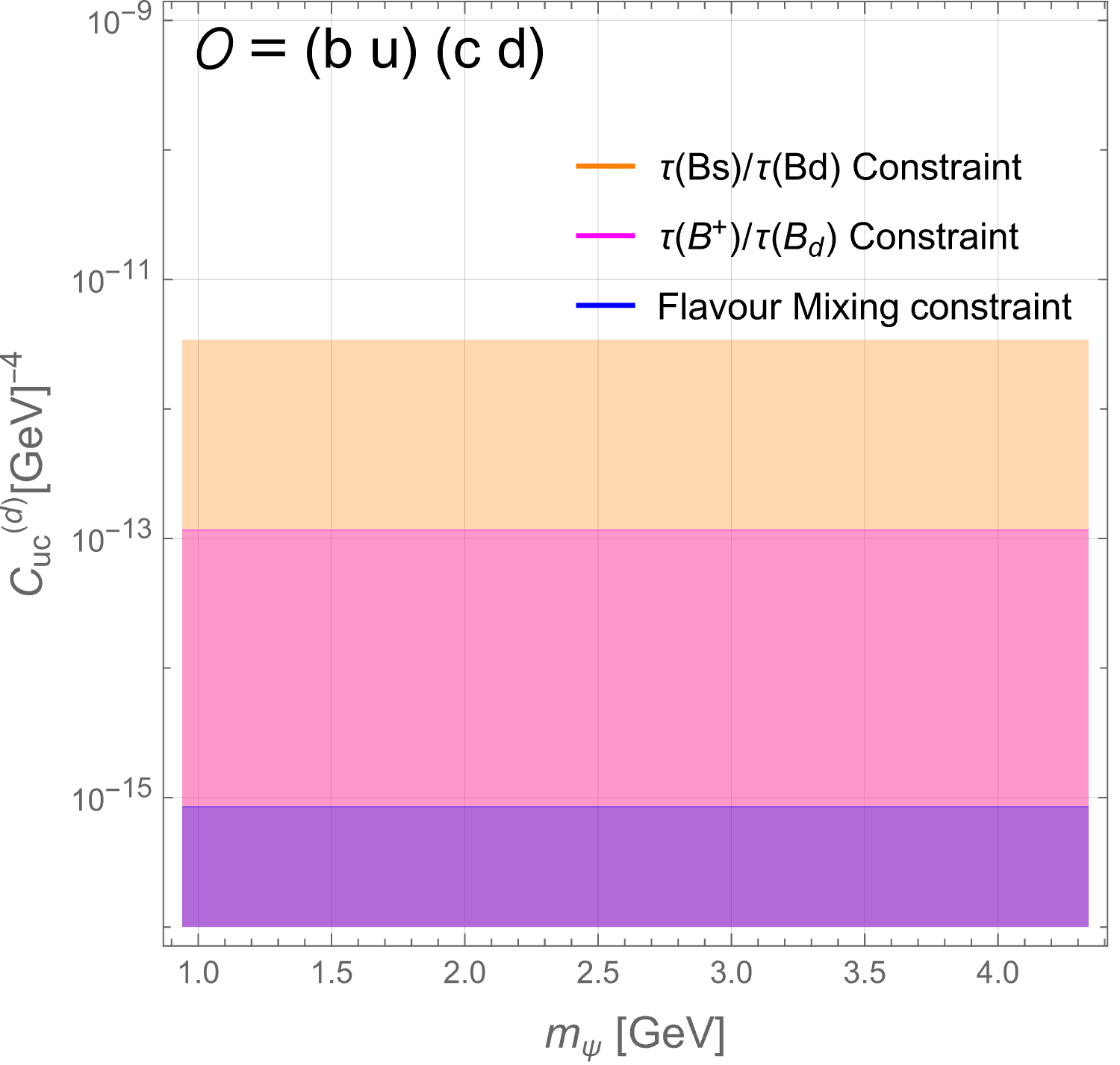}
    \end{subfigure}
    \hfill
    \begin{subfigure}{0.40\textwidth}
        \centering
        \includegraphics[width=\textwidth]{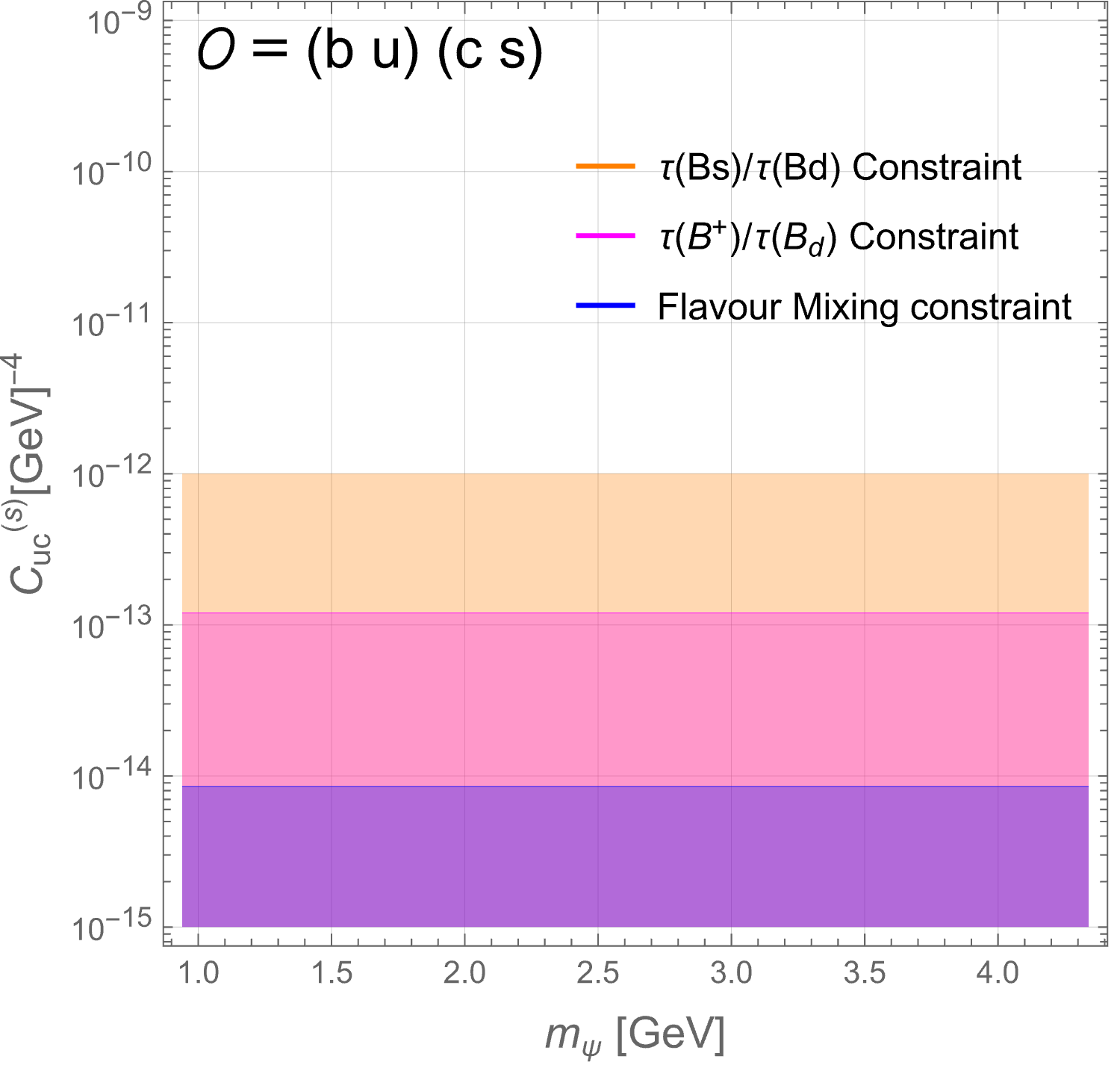}
    \end{subfigure}
       \caption{(Part 2): Constraints on the Wilson coefficients as a function of the dark fermion mass $m_\psi$, with the same colour coding as in the previous figure. (Operators involving two distinct massive final states are not shown, as the corresponding Darwin-term contribution has not yet been calculated.)}
        \label{fig:BsBdConstraints3}
\end{figure}

\begin{figure}[ht]
    \includegraphics[scale=0.8]{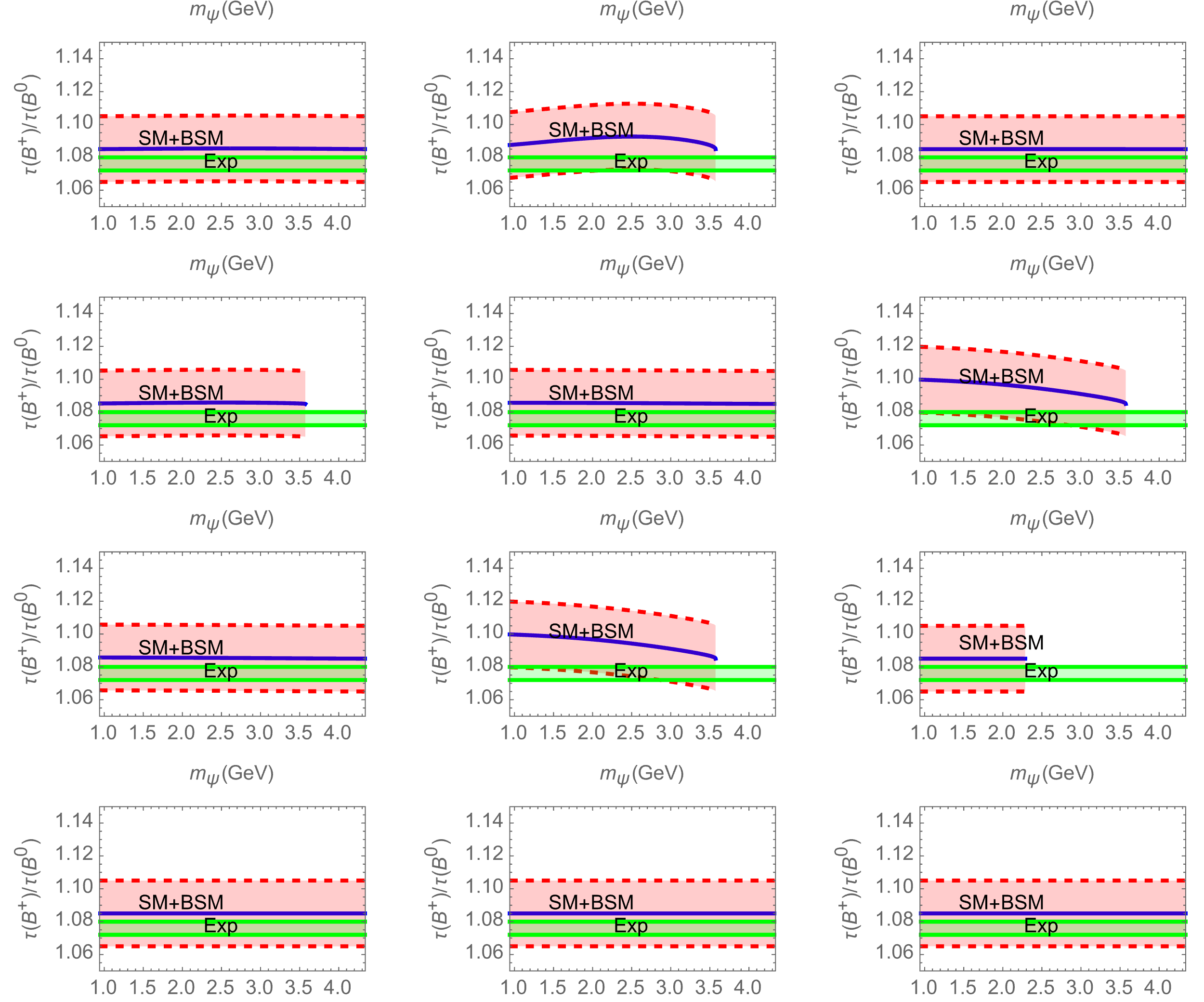}
    \caption{The lifetime ratio ${\tau (B^+)}/{\tau (B_d)}$ for $\mathcal{H}_{-1/3}$ as a function of $m_{\psi}$, using the upper limits of one of the constants $G^{q1q2}_{X}$ and setting the others to zero. The red region shows the SM error, whereas the experimental value is shown in green. The non-vanishing constants  from up to down and left to right are $G^{\psi u}_{III}, G^{\psi c}_{III},\, G^{u \psi}_{III}, \, G^{c \psi }_{III}, G^{\psi u}_{II}, \, G^{\psi c}_{II}, \, G^{u \psi }_{II},\, G^{c \psi}_{II},\, G^{\psi \psi}_{I},\, G^{u u}_{I} = G^{c c}_{I},\, G^{c u}_{I},\, G^{u c}_{I} $ }
    \label{fig:Lifetime1/3-BdAllPlots}
\end{figure}
\begin{figure}[ht]
    \includegraphics[scale=0.8]{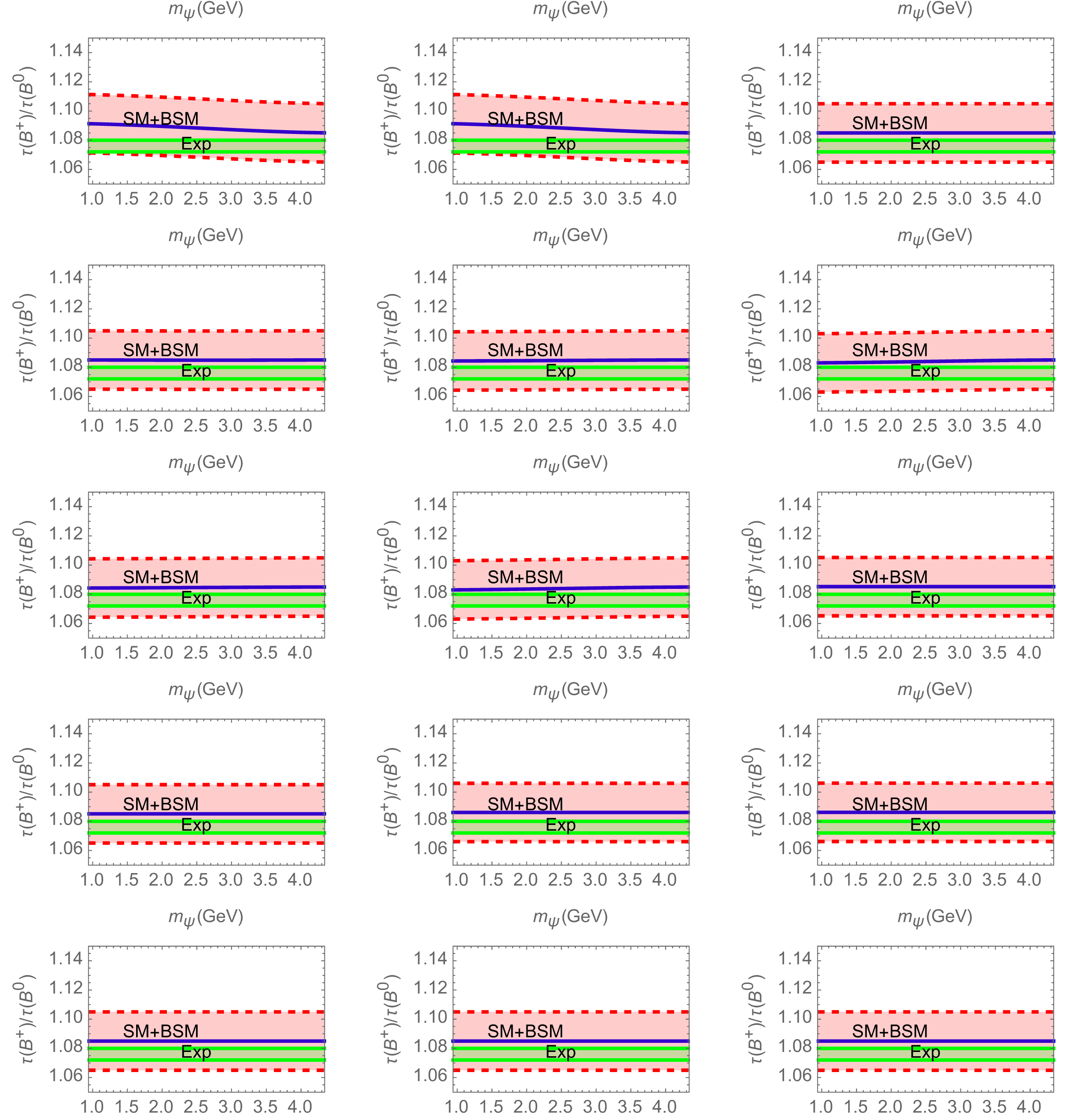}
    \caption{The lifetime ratio ${\tau (B^+)}/{\tau (B_d)}$ for $\mathcal{H}_{-1/3}$ as a function of $m_{\psi}$, using the upper limits of one of the constants $G^{f_1f_2}_{X}$ and setting the others to zero. The red region shows the SM error, whereas the experimental value is shown in green. The non-vanishing constants  from up to down and left to right are $G^{\psi d}_{IV}, G^{\psi s}_{IV},\, G^{\psi d}_{III}, \, G^{\psi s}_{III}, G^{\psi d}_{V}, \, G^{\psi s}_{V}, \, G^{\psi d}_{VI},\, G^{\psi s}_{VI}, \, G^{u d}_{IV}, \, G^{c d}_{IV}, \, G^{s c}_{I}, \, G^{s u}_{IV}, \, G^{d u}_{I}, \, G^{d c}_{I}, \, G^{s c}_{I}, \, G^{s u}_{I}.$ }
    \label{fig:Lifetime1/3-BPlusAllPlots}
\end{figure}
\begin{figure}[ht]
    \includegraphics[scale=0.8]{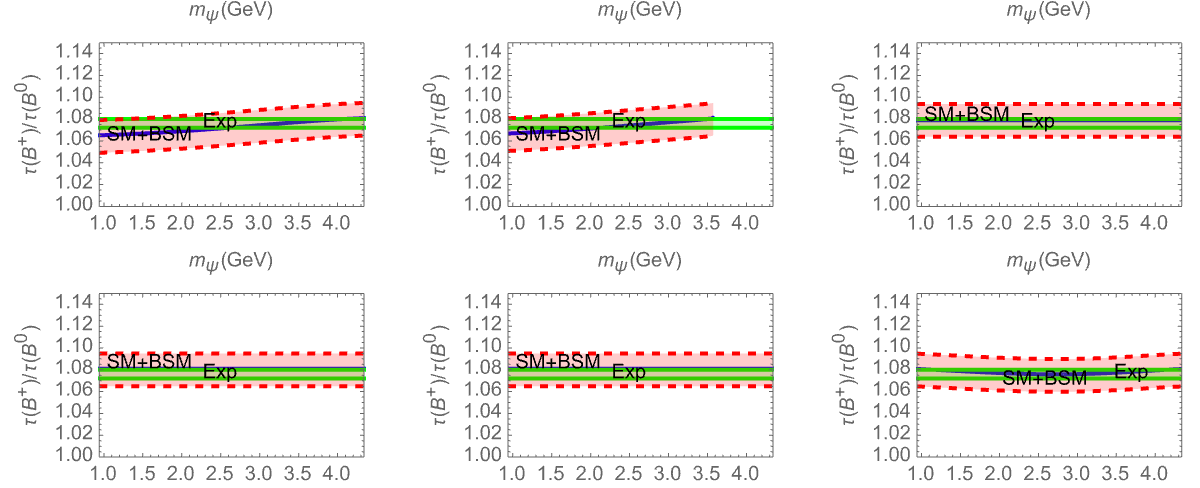}
    \caption{The lifetime ratio ${\tau (B^+)}/{\tau (B_d)}$ for $\mathcal{H}_{2/3}$ as a function of $m_{\psi}$, using the upper limits of one of  the constants $G^{f_1f_2}_{X}$ and setting the others to zero. The red region shows the SM error, whereas the experimental value is shown in green. The non-vanishing constants  from up to down and left to right are $G^{u \psi}_{IV}, G^{c \psi}_{IV},\, G^{d s}_{IV}, \, G^{d s}_{I}, G^{s s}_{I}= \, G^{d \psi}_{III}, \, G^{s \psi}_{III},$}
    \label{fig:Lifetime2/3AllPlots}
\end{figure}

\begin{figure}[ht]
    \centering
    \begin{subfigure}[t]{0.45\textwidth}
        \centering
        \includegraphics[width=\linewidth]{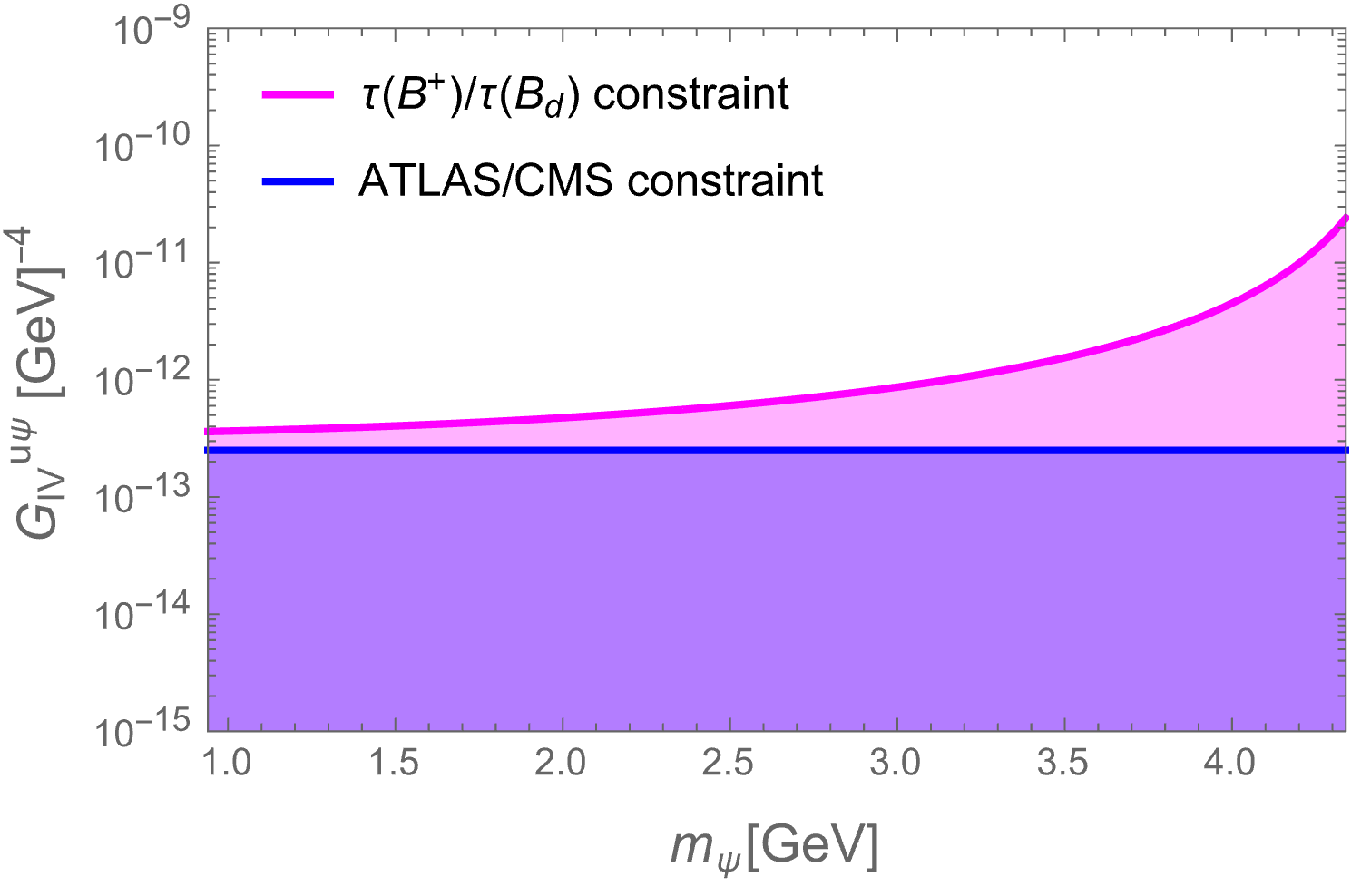}
    \end{subfigure}
    \hfill
    \begin{subfigure}[t]{0.45\textwidth}
        \centering
        \includegraphics[width=\linewidth]{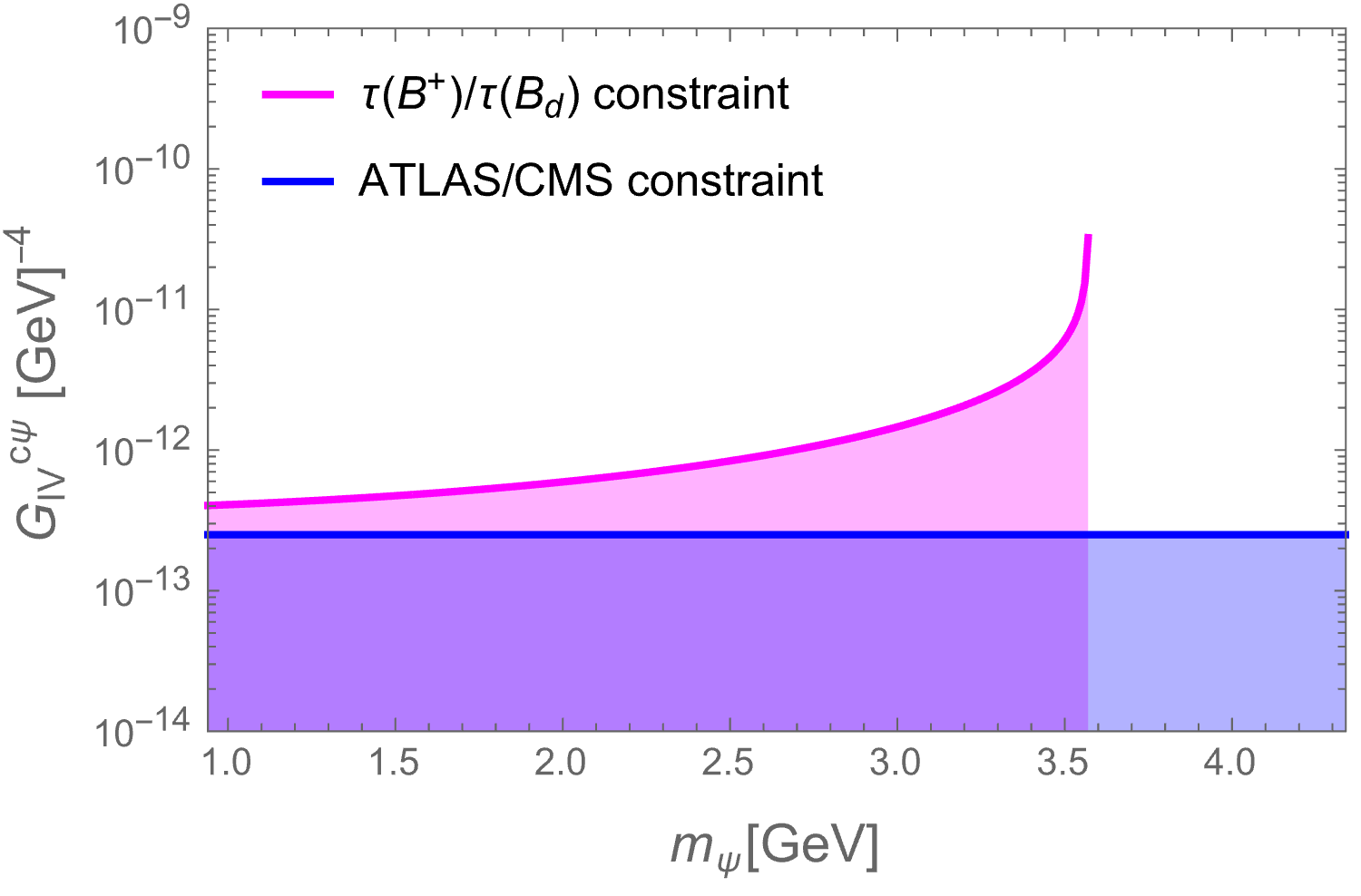}
    \end{subfigure}

    \vspace{0.1cm}

    \begin{subfigure}[t]{0.45\textwidth}
        \centering
        \includegraphics[width=\linewidth]{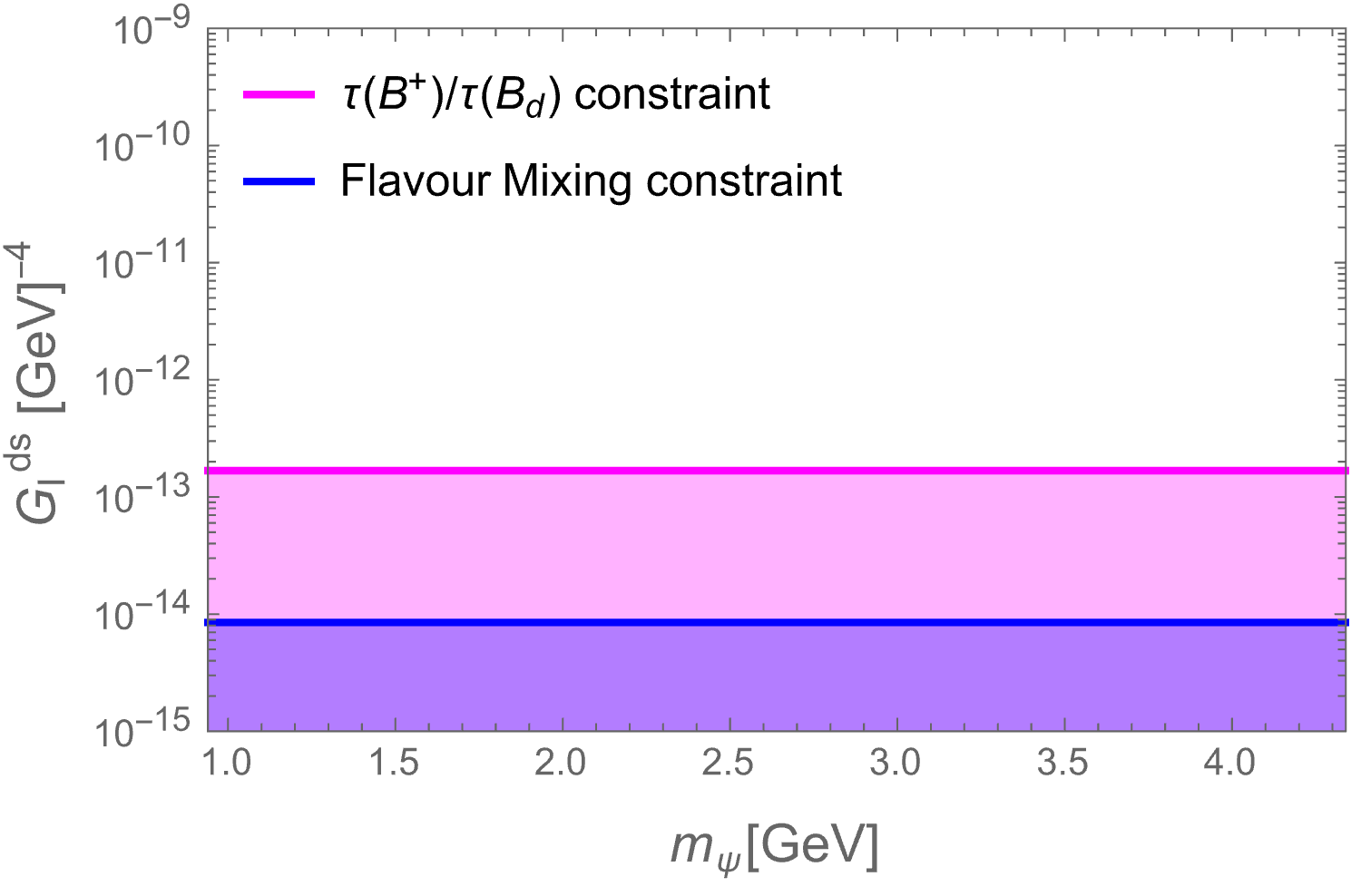}
    \end{subfigure}
    \hfill
    \begin{subfigure}[t]{0.45\textwidth}
        \centering
        \includegraphics[width=\linewidth]{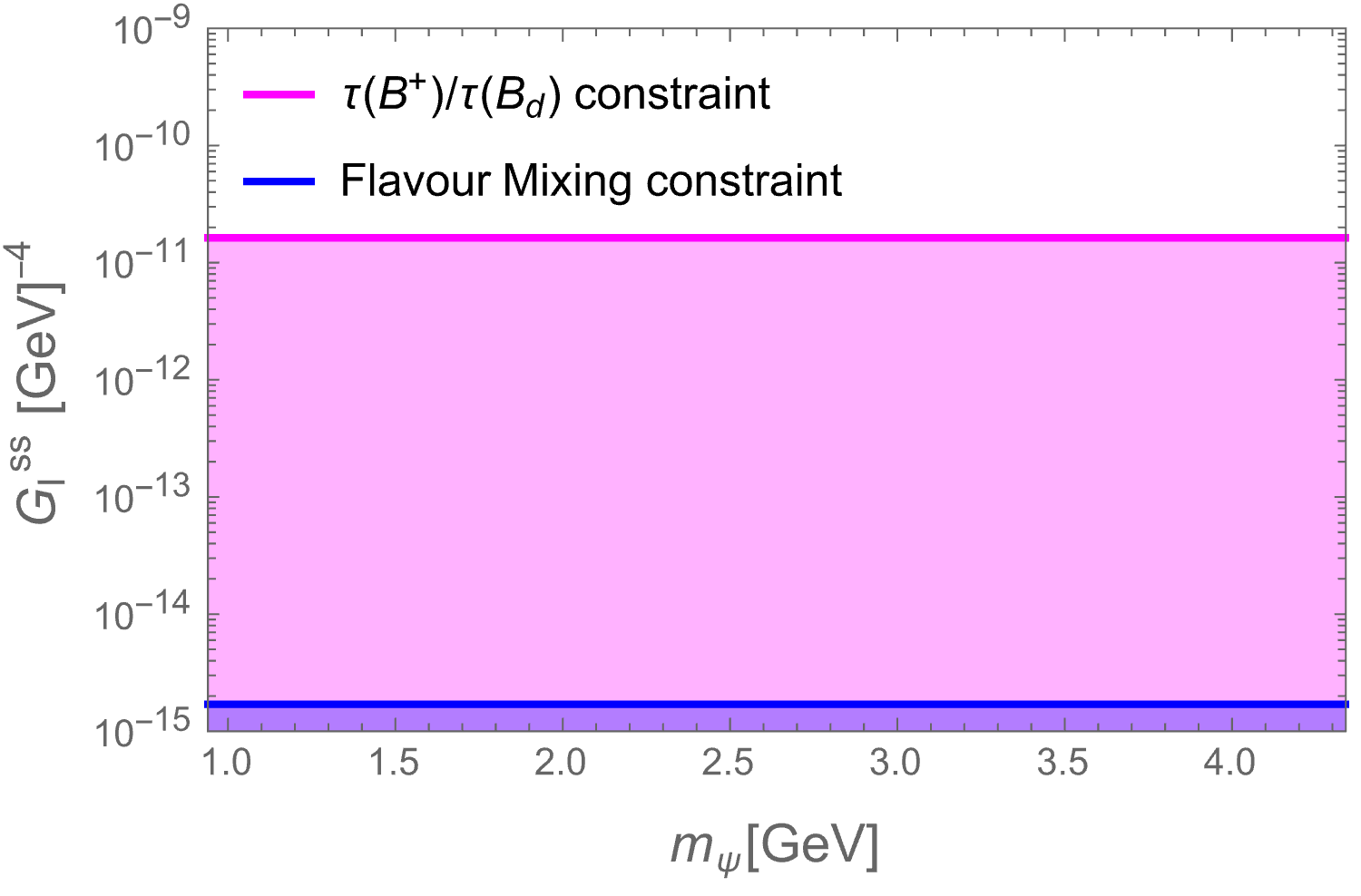}
    \end{subfigure}

    \vspace{0.1cm}

    \begin{subfigure}[t]{0.45\textwidth}
        \centering
        \includegraphics[width=\linewidth]{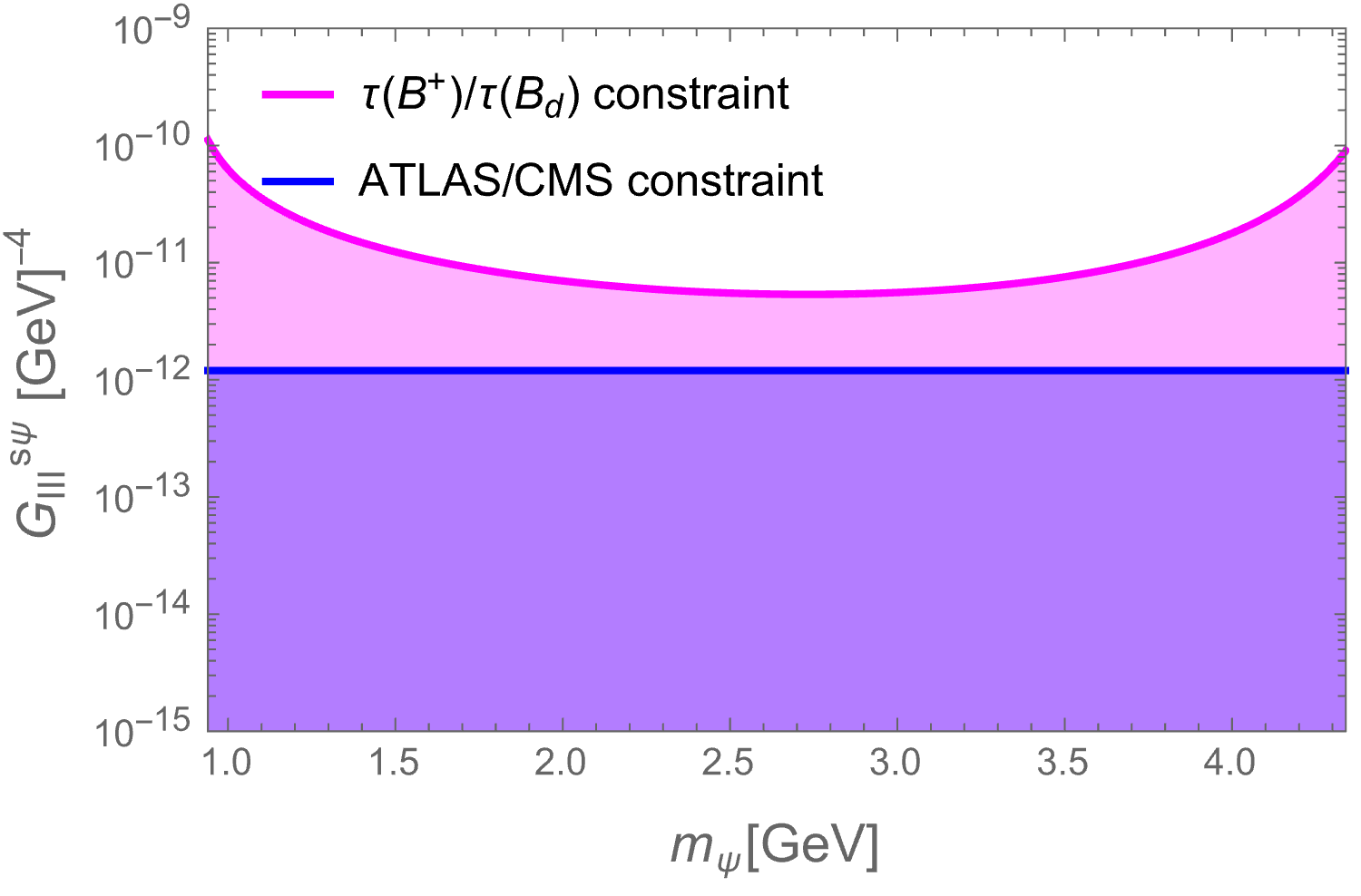}
    \end{subfigure}
    
   \caption{{$\mathcal{H}_{2/3}$: Constraints on the couplings $G^{f_1 f_2}_{X}$ derived from various sources. The magenta regions indicate the values allowed by the lifetime ratio $\tau(B^+)/\tau(B_d)$, while the blue region corresponds to the constraints obtained from collider signals or flavour mixing, as studied in ~\cite{Alonso-Alvarez:2021qfd}.}}
\label{fig:CouplingsConstraints23}
\end{figure}

\begin{figure}[ht]
    \centering
    \begin{subfigure}[t]{0.45\textwidth}
        \centering
        
        \subcaptionbox{}[0.99\textwidth]{\includegraphics[width=\linewidth]{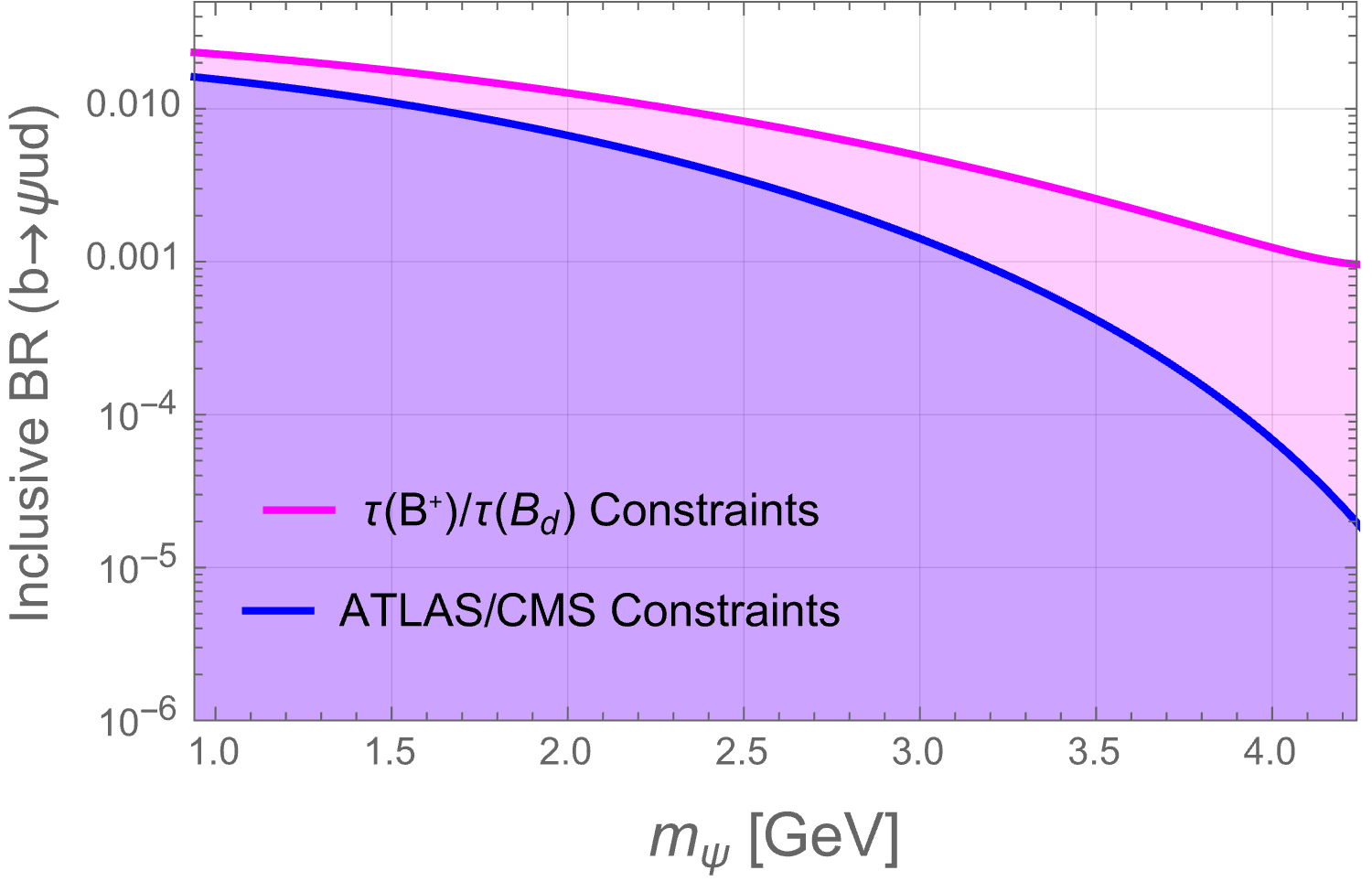}
        }
    \end{subfigure}
    \hfill
    \begin{subfigure}[t]{0.45\textwidth}
        \centering
        \subcaptionbox{}[0.99\textwidth]{\includegraphics[width=\linewidth]{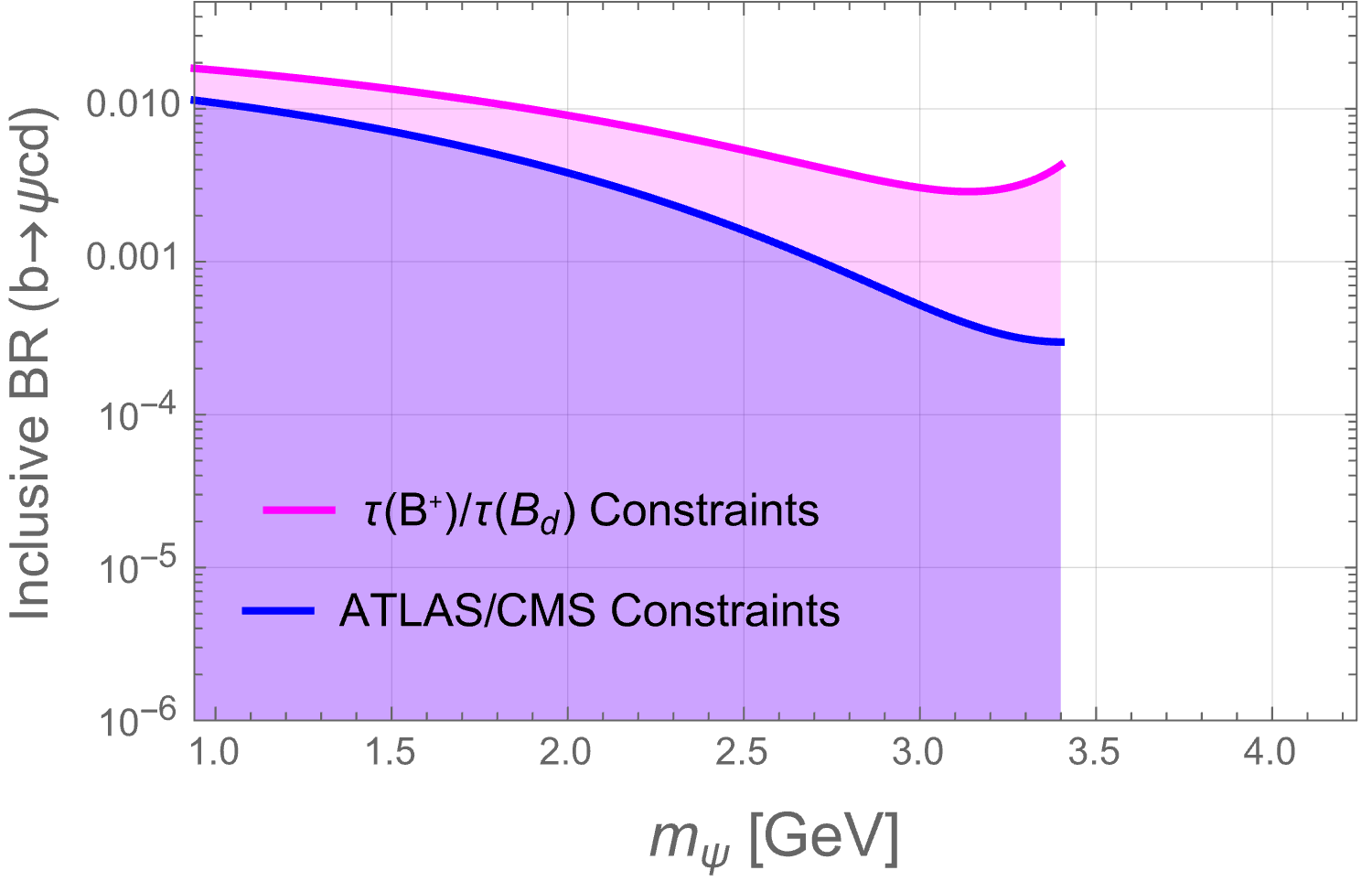}
        }
    \end{subfigure}

    \vspace{0.1cm}

    \begin{subfigure}[t]{0.45\textwidth}
        \centering
        \subcaptionbox{}[0.99\textwidth]{\includegraphics[width=\linewidth]{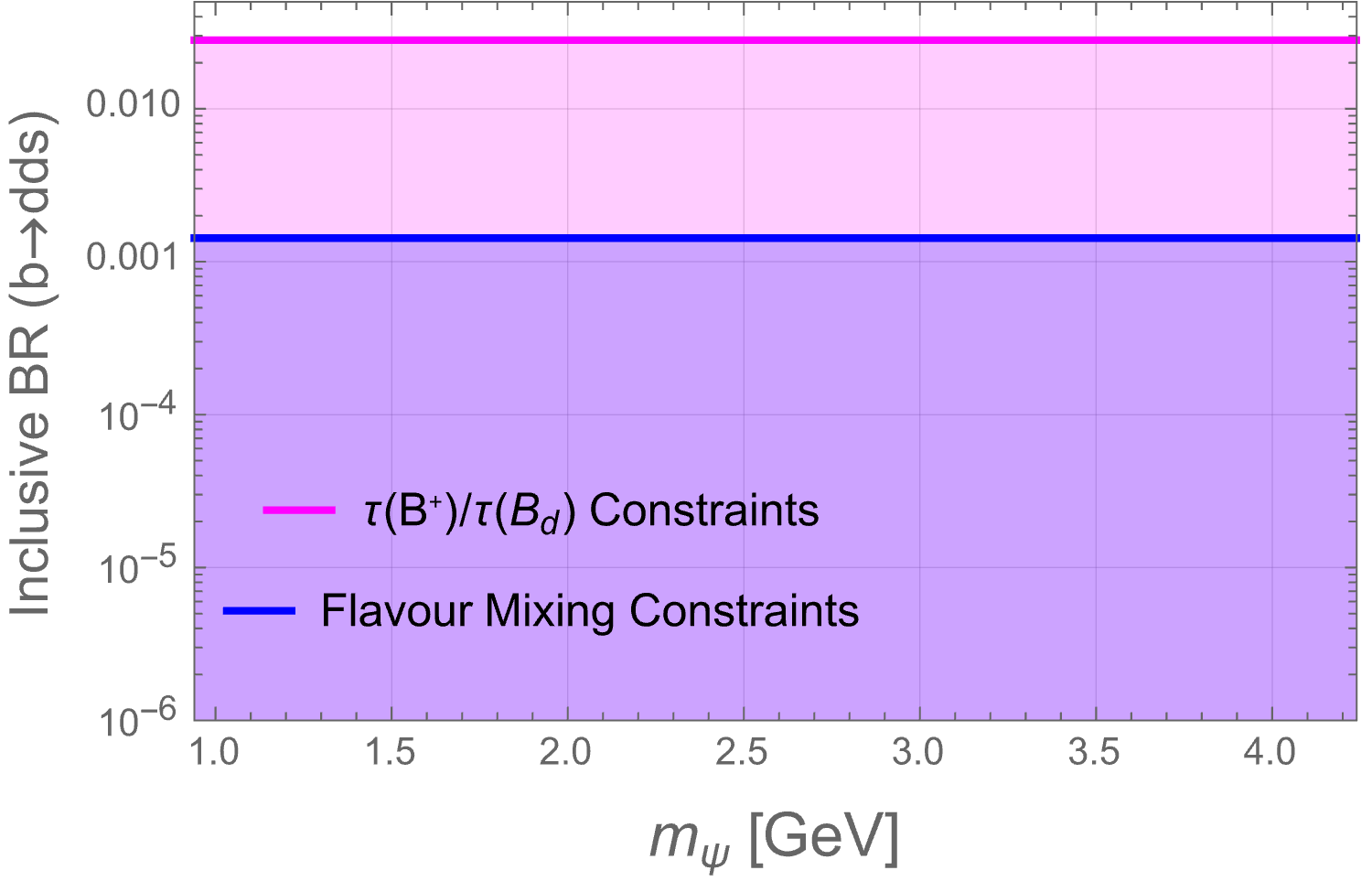}
        }
    \end{subfigure}
    \hfill
    \begin{subfigure}[t]{0.45\textwidth}
        \centering
        
        \subcaptionbox{}[0.99\textwidth]{\includegraphics[width=\linewidth]{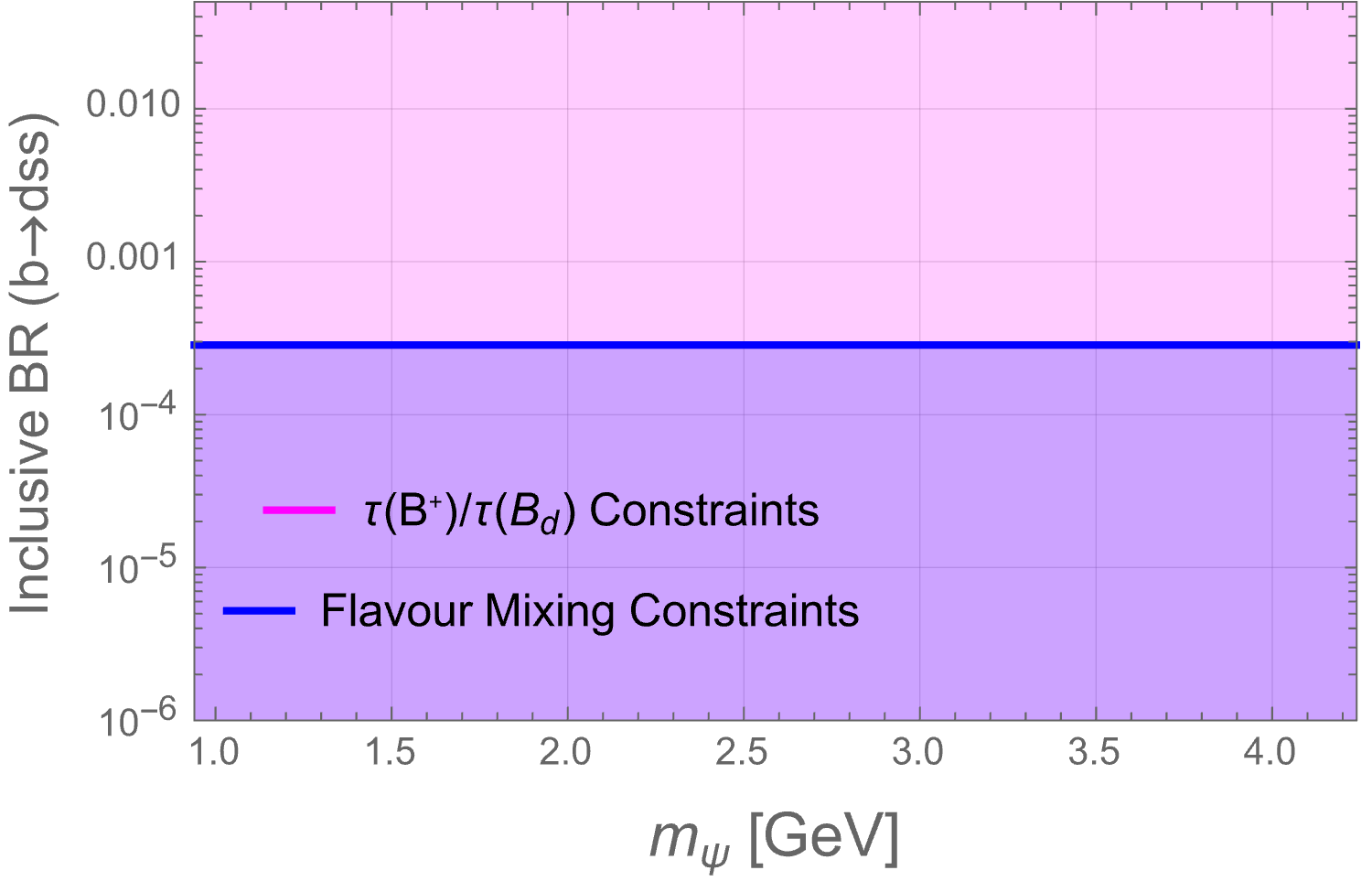}
        }
        
    \end{subfigure}

    \vspace{0.1cm}

    \begin{subfigure}[t]{0.45\textwidth}
        \centering
        \subcaptionbox{}[0.99\textwidth]{\includegraphics[width=\linewidth]{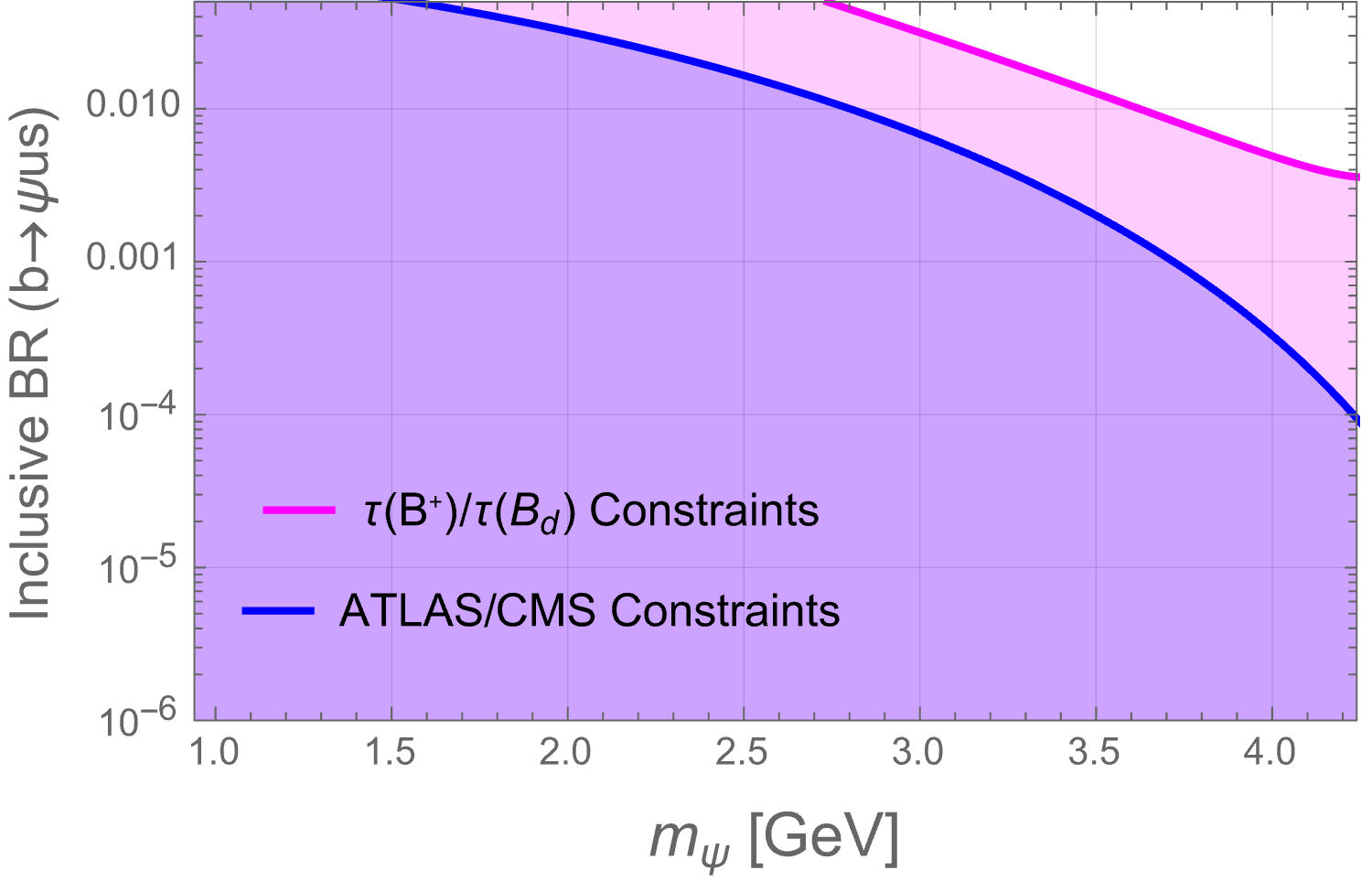}
        }
    \end{subfigure}
    
   \caption{{$\mathcal{H}_{2/3}$: Allowed regions of the inclusive branching ratio of the different operators proportional to the couplings $G^{f_1 f_2}_{X}$. Values of the inclusive branching fraction above $5\%$ are excluded by the bounds  from the total decay rate of $B$ mesons, therefore  we do not show regions above 0.05 in the plots. The figure shows the difference between using the values allowed by the lifetime ratio $\tau(B^+)/\tau(B_d)$ (magenta) and those allowed by collider signals or flavour mixing (blue), as studied in~\cite{Alonso-Alvarez:2021qfd}, to compute the inclusive branching ratio. The plots correspond to the following couplings: (a) $G_{IV}^{u\psi} + G_{III}^{d\psi}$ (b) $G_{IV}^{c\psi}$ (c) $G_{I}^{ds}+G_{IV}^{ds}$
   (d) $G_I^{ss}$ (e) $G^{s\psi}_{III}$.}}
   \label{fig:model23-branching}
\end{figure}

\begin{figure}[ht]
    \centering
    \begin{subfigure}[t]{0.45\textwidth}
        \includegraphics[width=\linewidth]{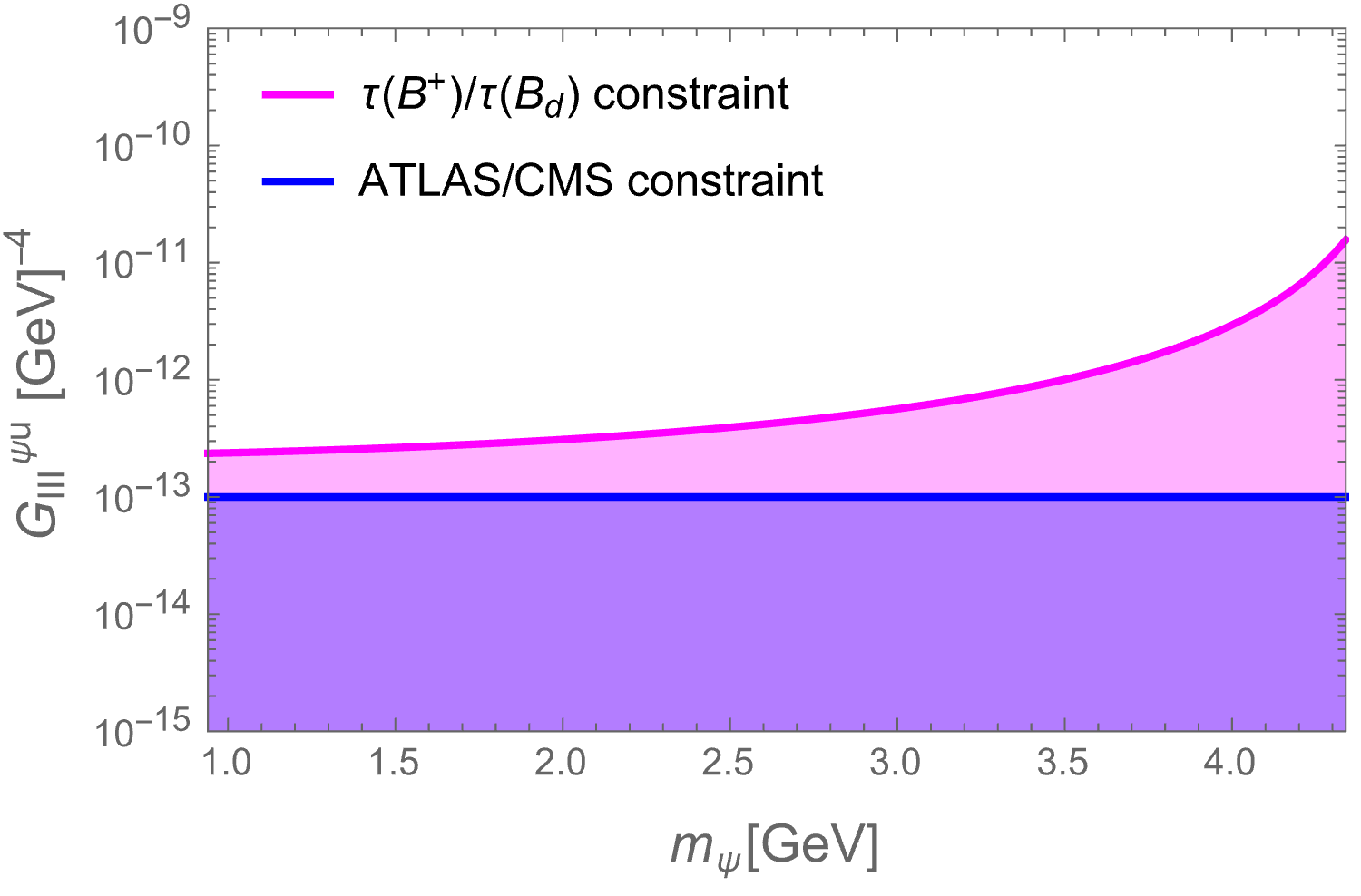}
    \end{subfigure}
    \hfill
    \begin{subfigure}[t]{0.45\textwidth}
        \includegraphics[width=\linewidth]{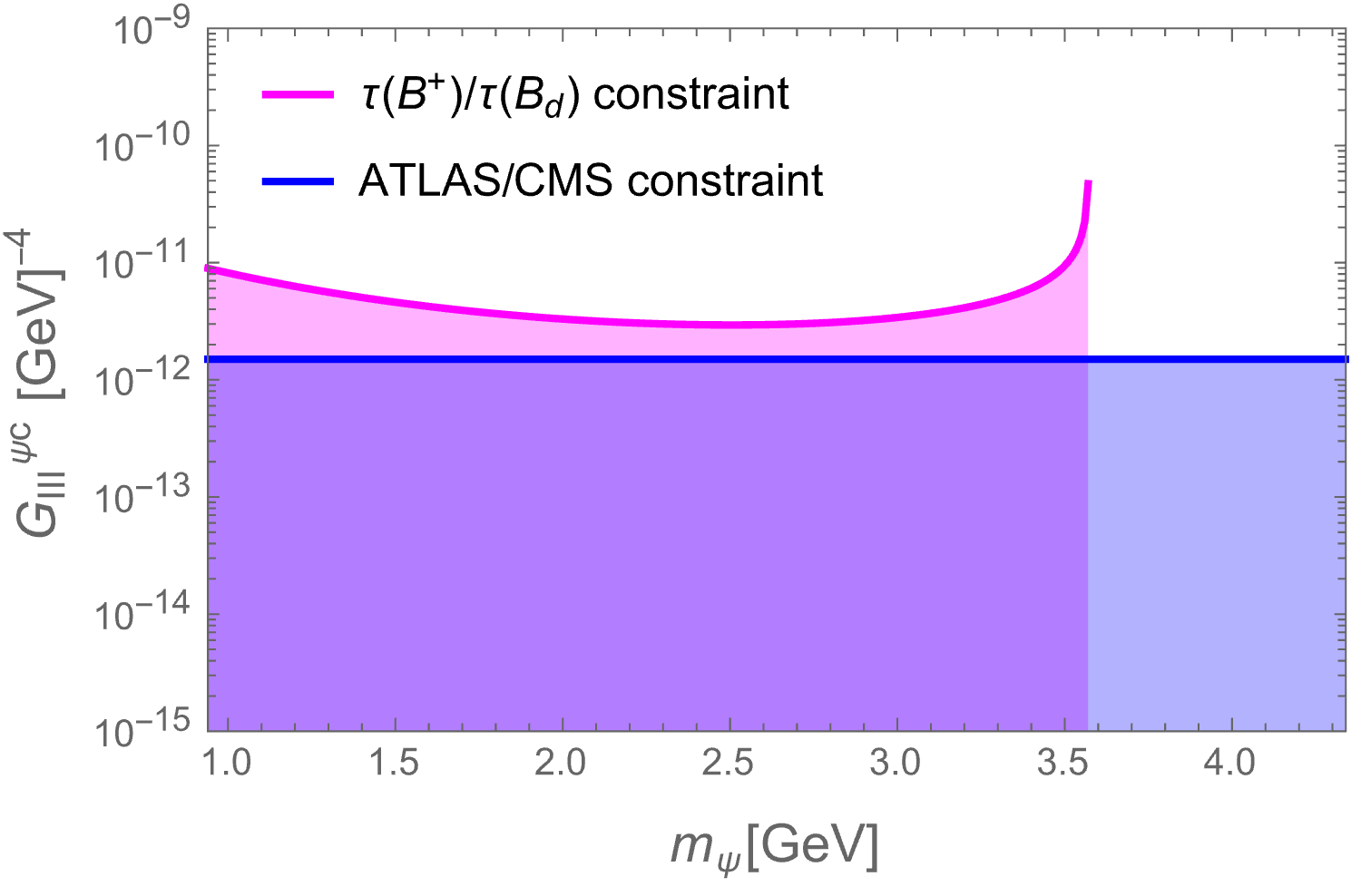}
    \end{subfigure}
    \vspace{0.1cm}

    \begin{subfigure}[t]{0.45\textwidth}
        \includegraphics[width=\linewidth]{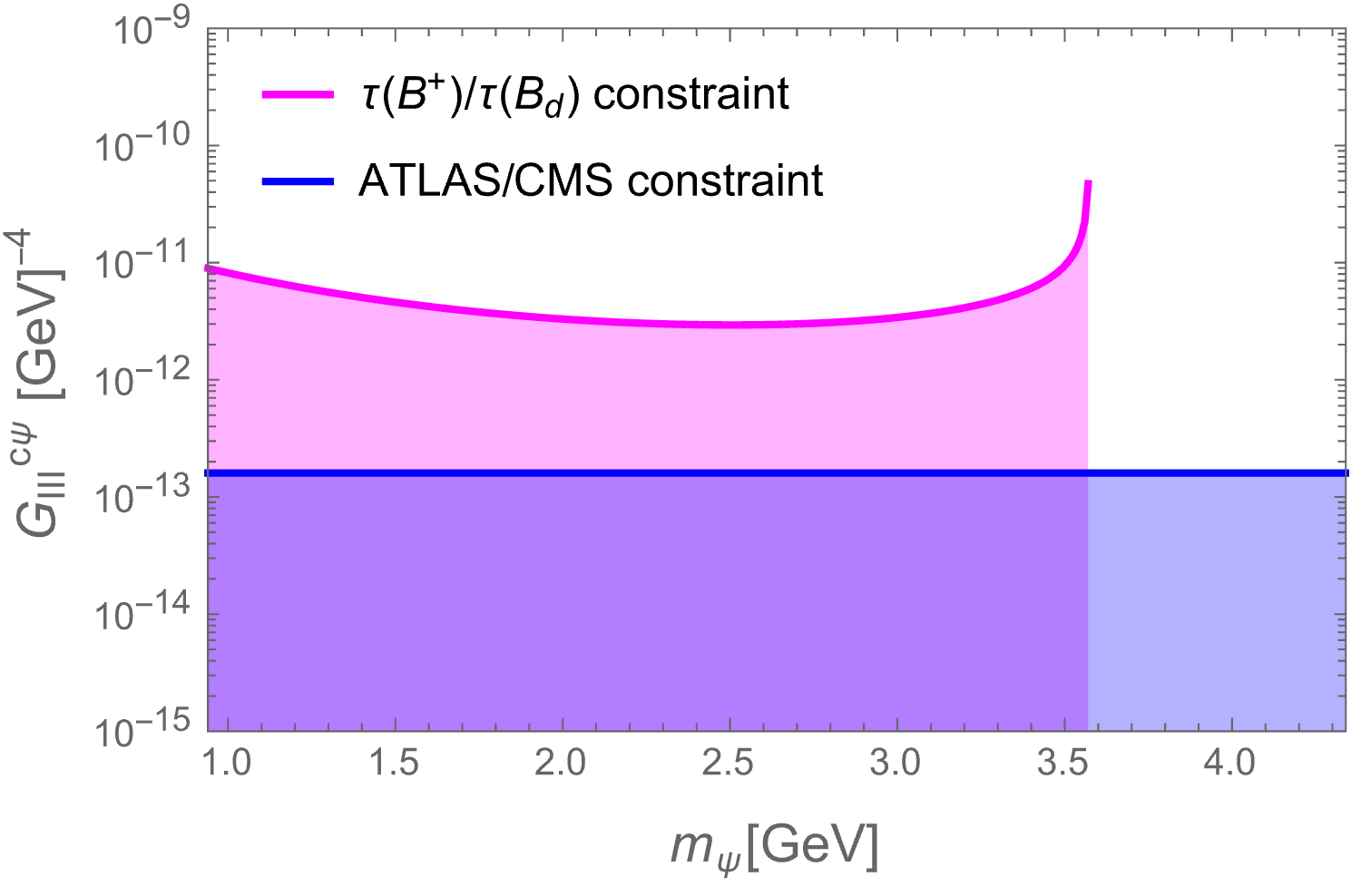}
    \end{subfigure}
    \hfill
    \begin{subfigure}[t]{0.45\textwidth}
        \includegraphics[width=\linewidth]{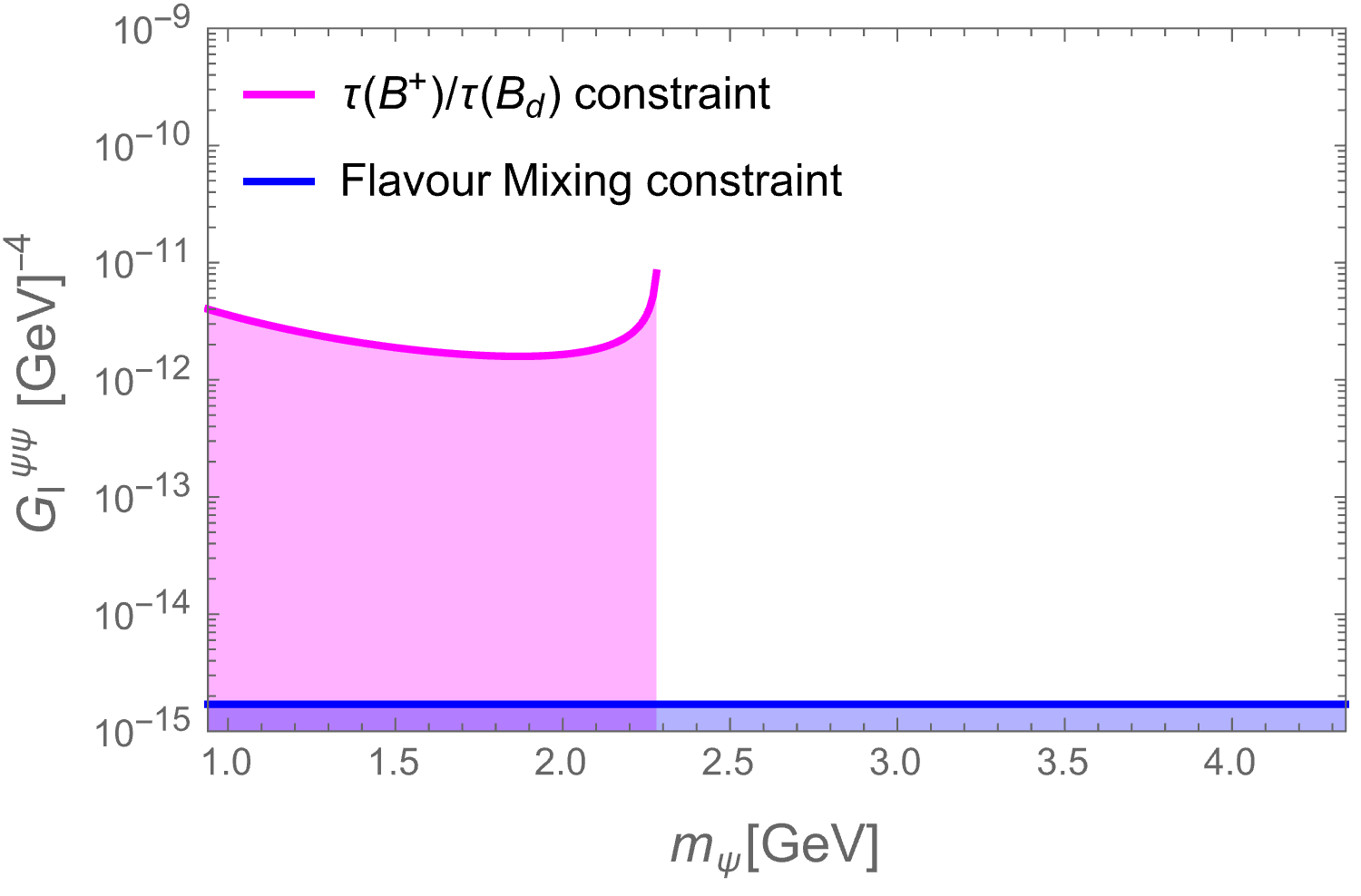}
    \end{subfigure}
    \vspace{0.1cm}

    \begin{subfigure}[t]{0.45\textwidth}
        \includegraphics[width=\linewidth]{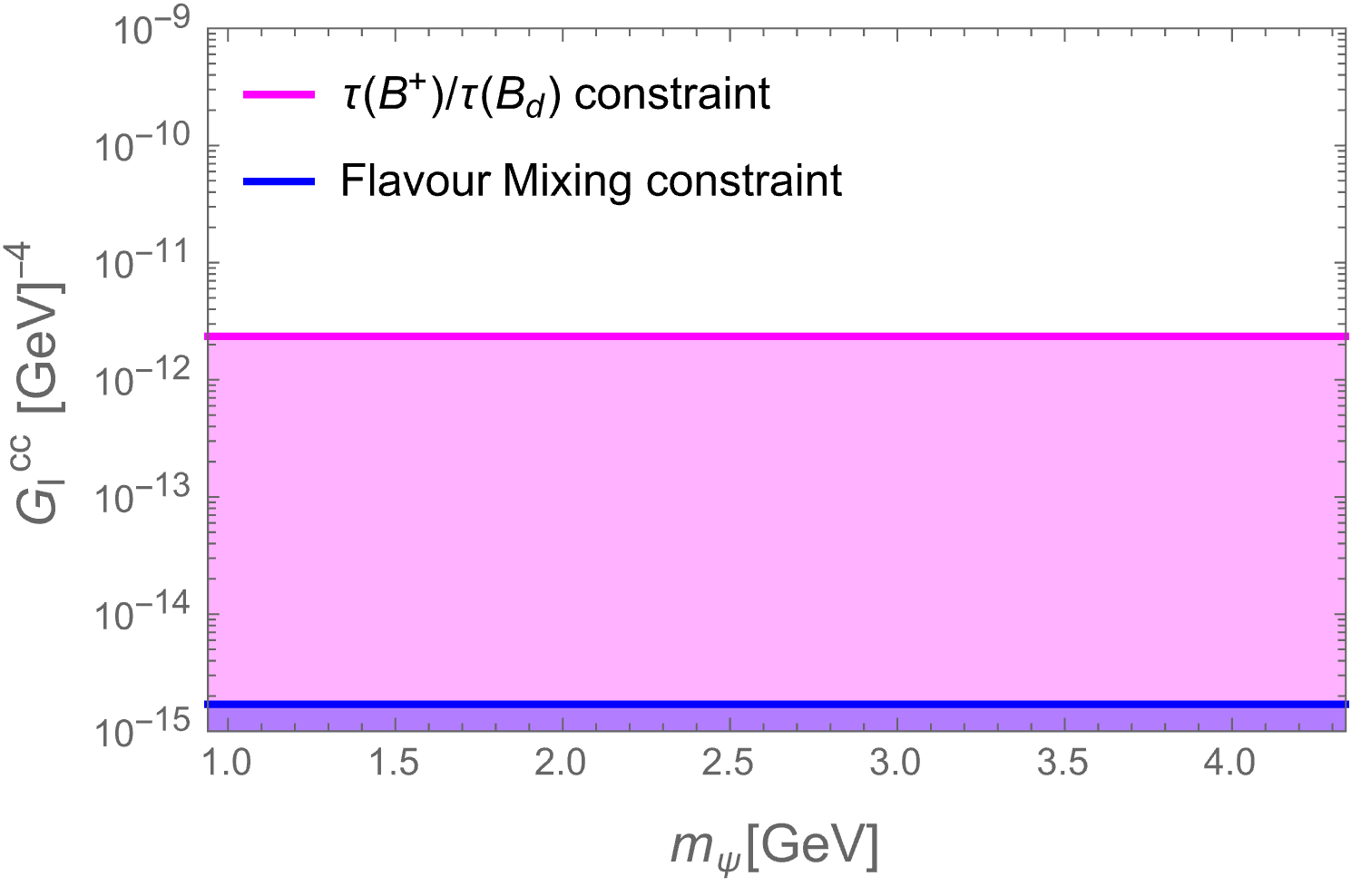}
    \end{subfigure}
    \hfill
    \begin{subfigure}[t]{0.45\textwidth}
        \includegraphics[width=\linewidth]{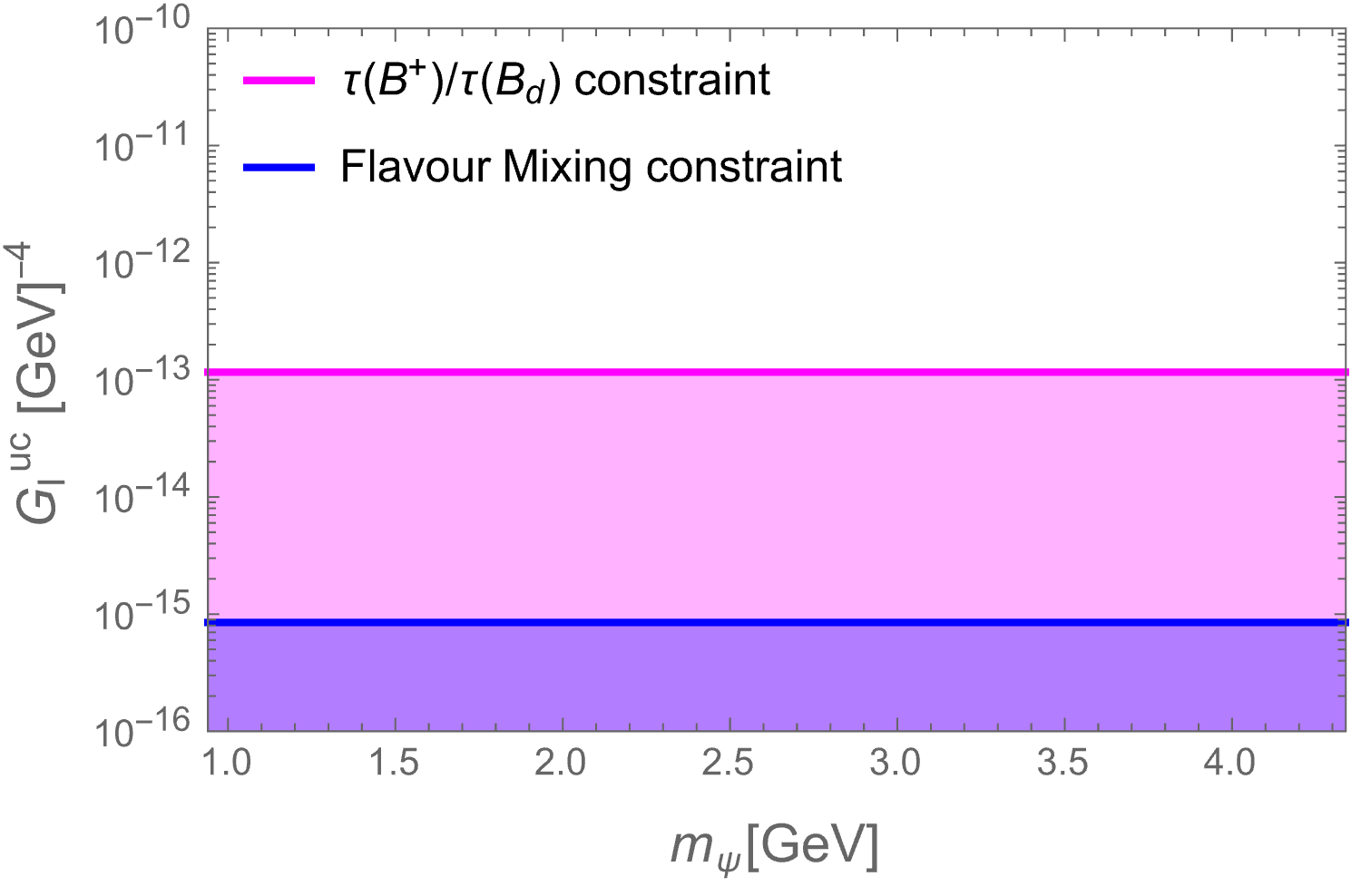}
    \end{subfigure}

    \caption{{$\mathcal{H}_{-1/3}$ (Part 1): Constraints on the couplings $G^{f_1 f_2}_{X}$ derived from various sources. The magenta regions indicate the values allowed by the lifetime ratio $\tau(B^+)/\tau(B_d)$, while the blue region corresponds to the constraints obtained from collider signals or flavour mixing, as studied in~\cite{Alonso-Alvarez:2021qfd}.}}
\label{fig:CouplingsConstraints131}
\end{figure}

\begin{figure}[ht]
    \centering
    \begin{subfigure}[t]{0.45\textwidth}
        \includegraphics[width=\linewidth]{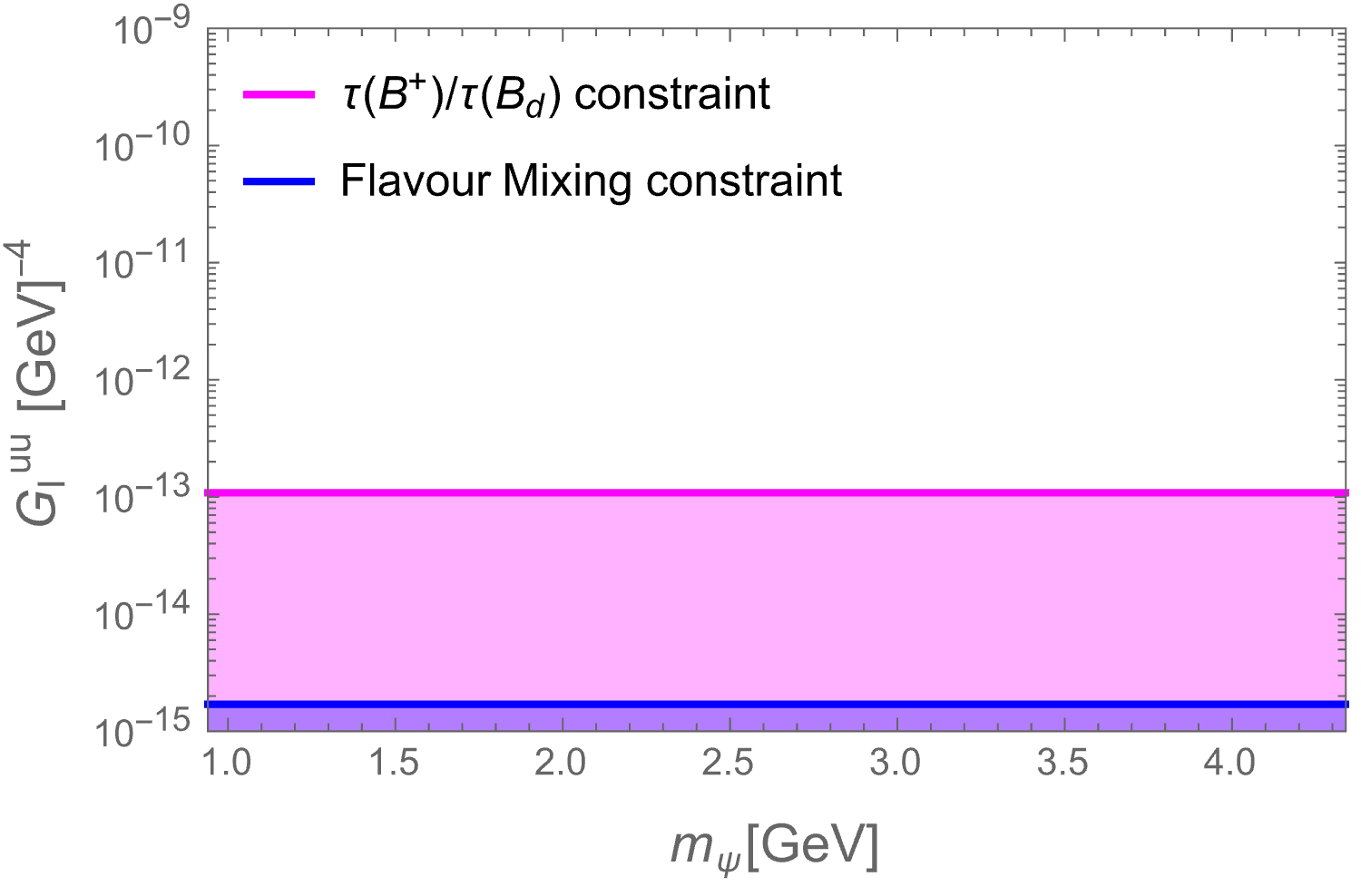}
    \end{subfigure}
    \hfill
    \begin{subfigure}[t]{0.45\textwidth}
        \includegraphics[width=\linewidth]{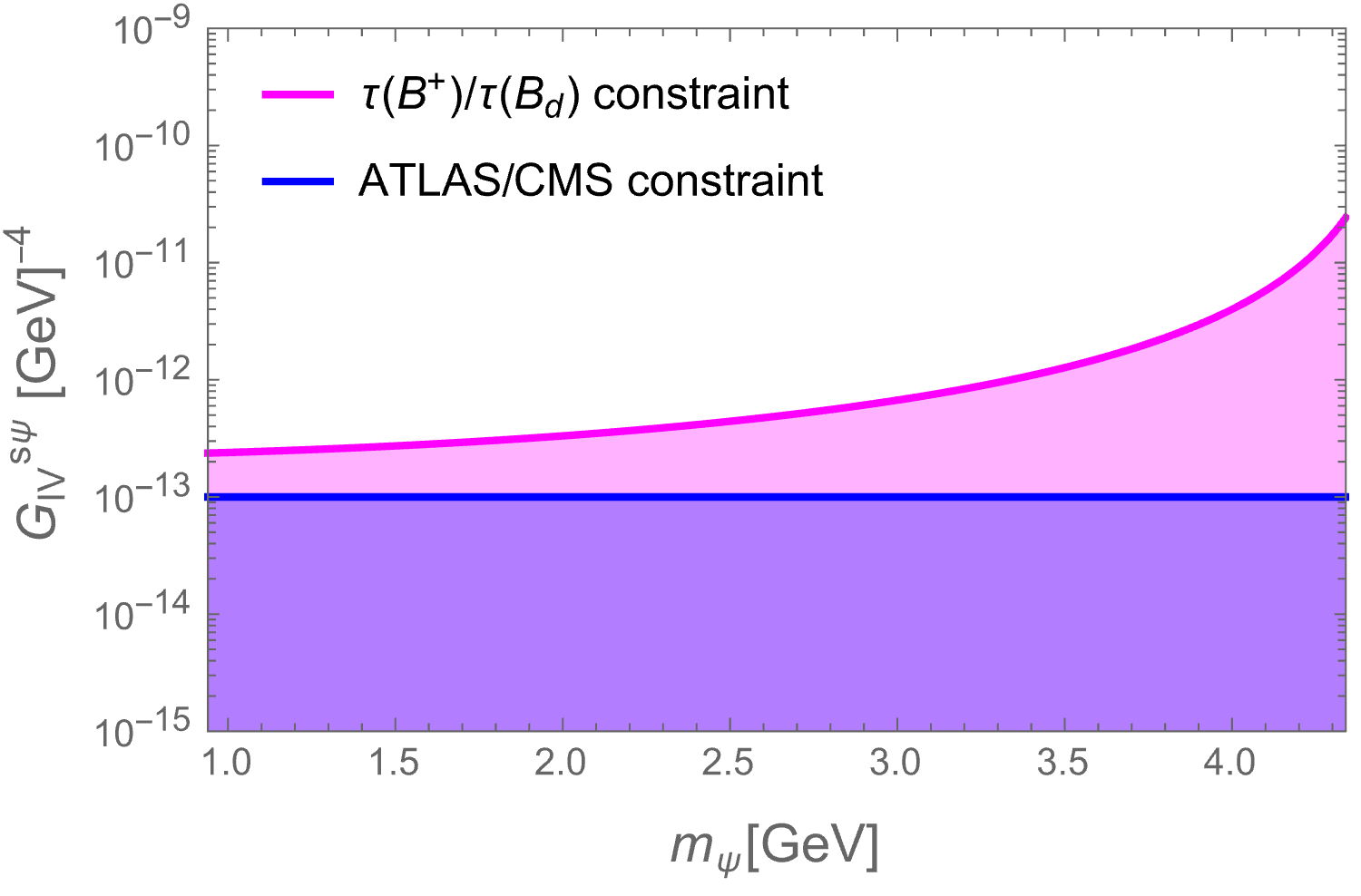}
    \end{subfigure}
    \vspace{0.1cm}

    \begin{subfigure}[t]{0.45\textwidth}
        \includegraphics[width=\linewidth]{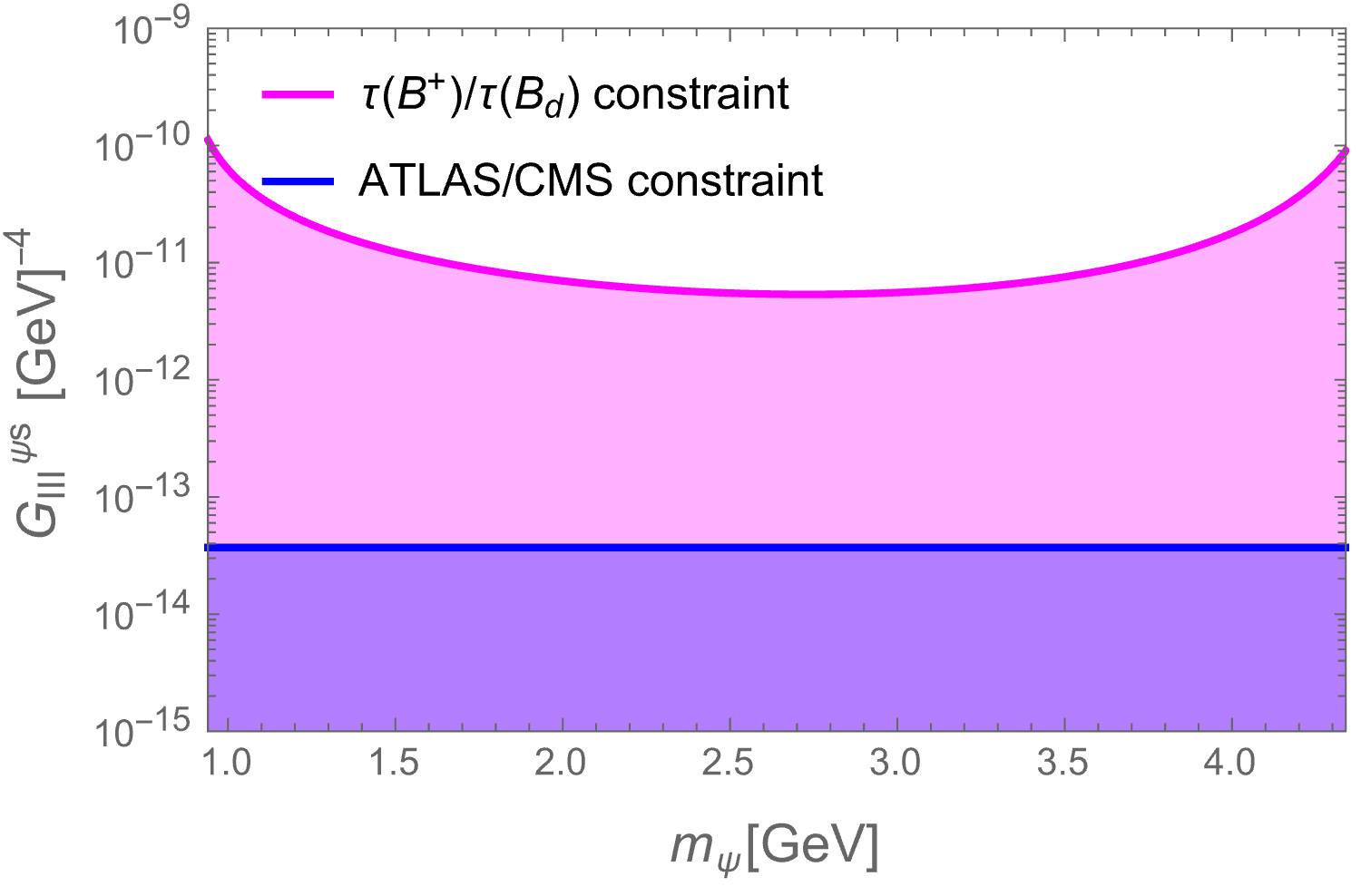}
    \end{subfigure}
    \hfill
    \begin{subfigure}[t]{0.45\textwidth}
        \includegraphics[width=\linewidth]{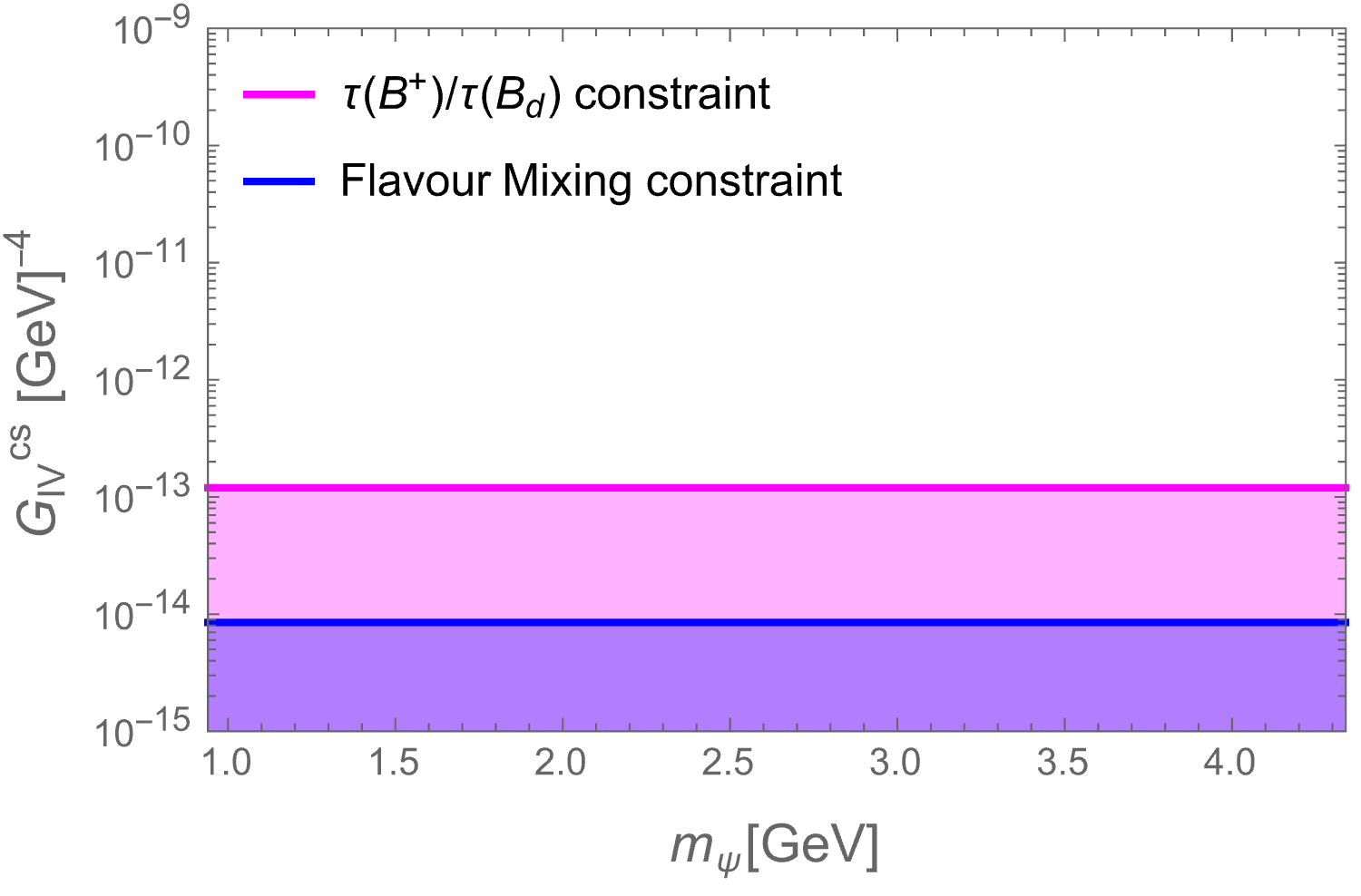}
    \end{subfigure}
    \vspace{0.1cm}

    \begin{subfigure}[t]{0.45\textwidth}
        \includegraphics[width=\linewidth]{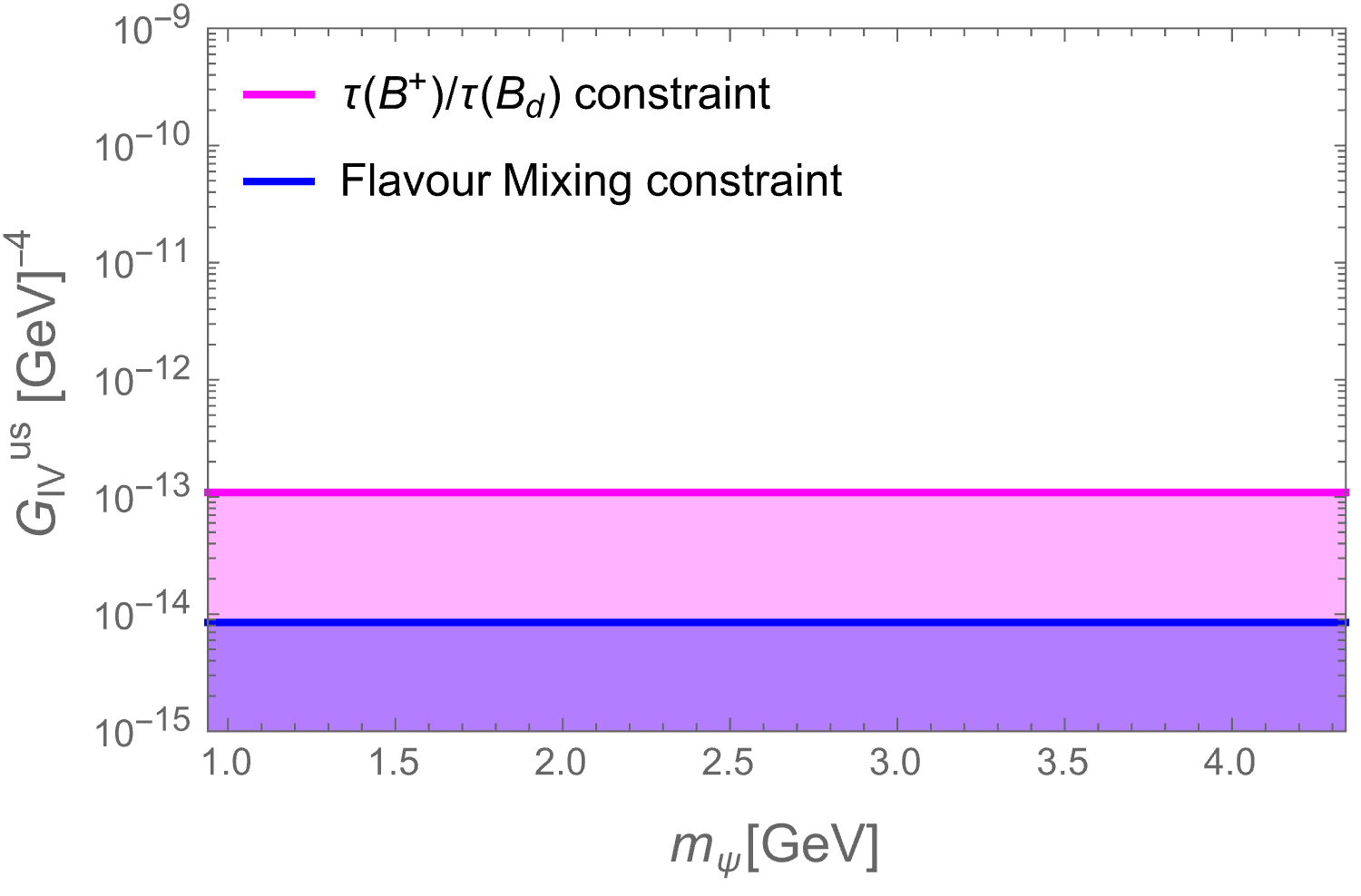}
    \end{subfigure}
    \hfill
    \begin{subfigure}[t]{0.45\textwidth}
        \includegraphics[width=\linewidth]{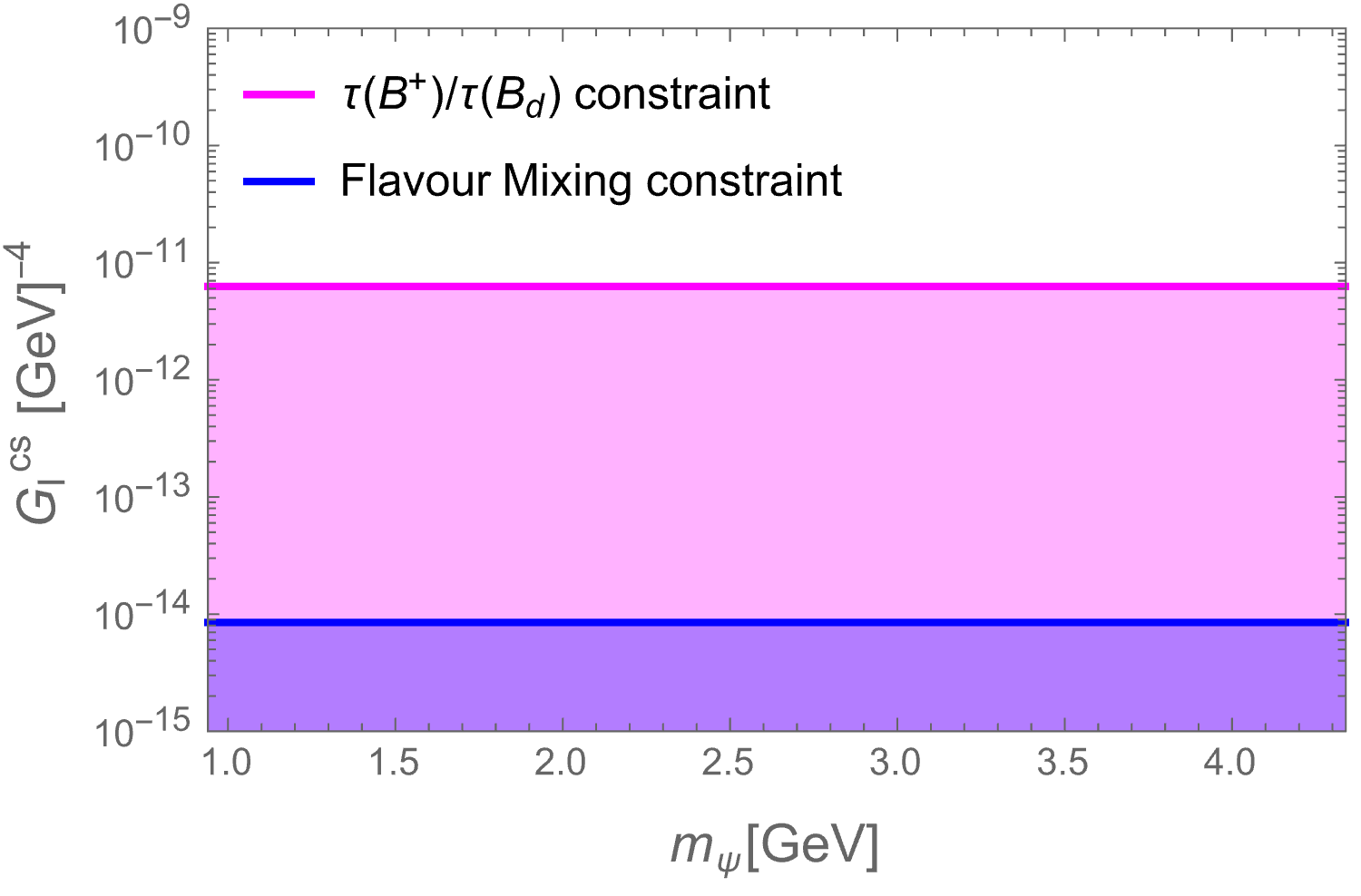}
    \end{subfigure}

    \caption{{$\mathcal{H}_{-1/3}$ (Part 2): Additional constraints on the couplings $G^{f_1 f_2}_{X}$ with the same colour coding as in the previous Figure.}}
\label{fig:CouplingsConstraints132}
\end{figure}

\begin{figure}[ht]
    \centering
    \begin{subfigure}[t]{0.45\textwidth}
         \subcaptionbox{}[0.99\textwidth]{\includegraphics[width=\linewidth]{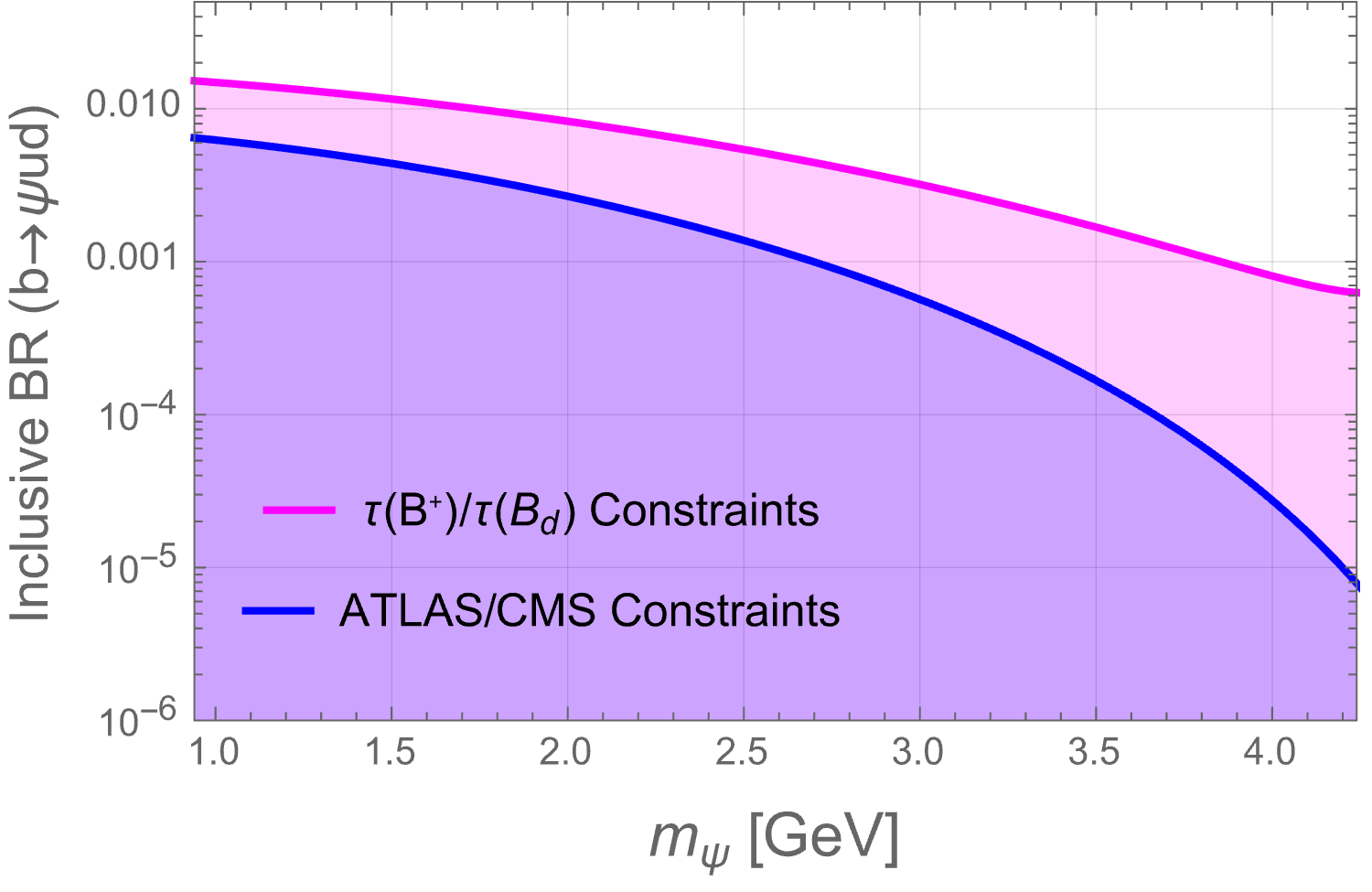}
         }
    \end{subfigure}
    \hfill
    \begin{subfigure}[t]{0.45\textwidth}
         \subcaptionbox{}[0.99\textwidth]{\includegraphics[width=\linewidth]{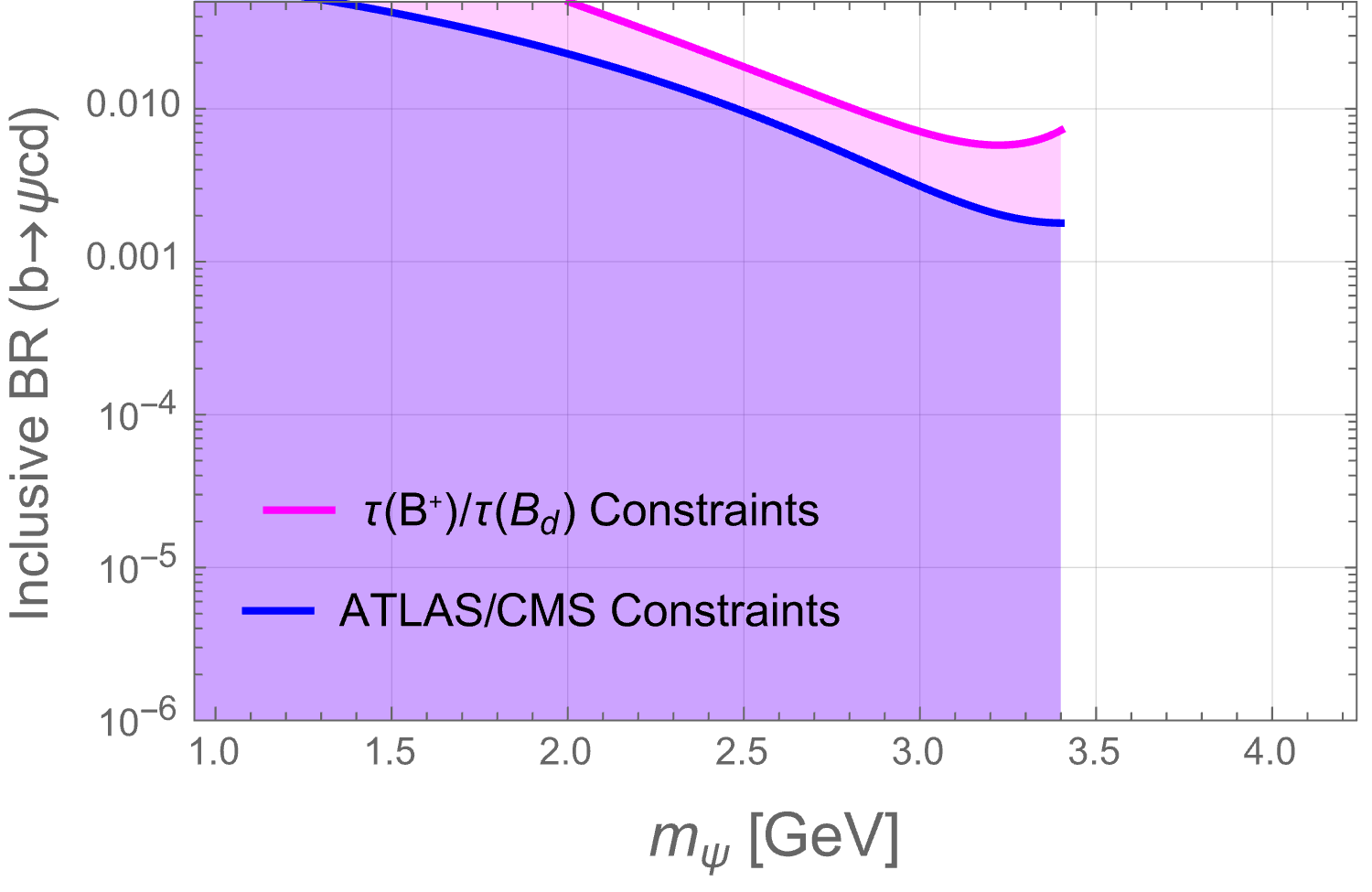}
         }
    \end{subfigure}
    \vspace{0.1cm}

    \begin{subfigure}[t]{0.45\textwidth}
         \subcaptionbox{}[0.99\textwidth]{\includegraphics[width=\linewidth]{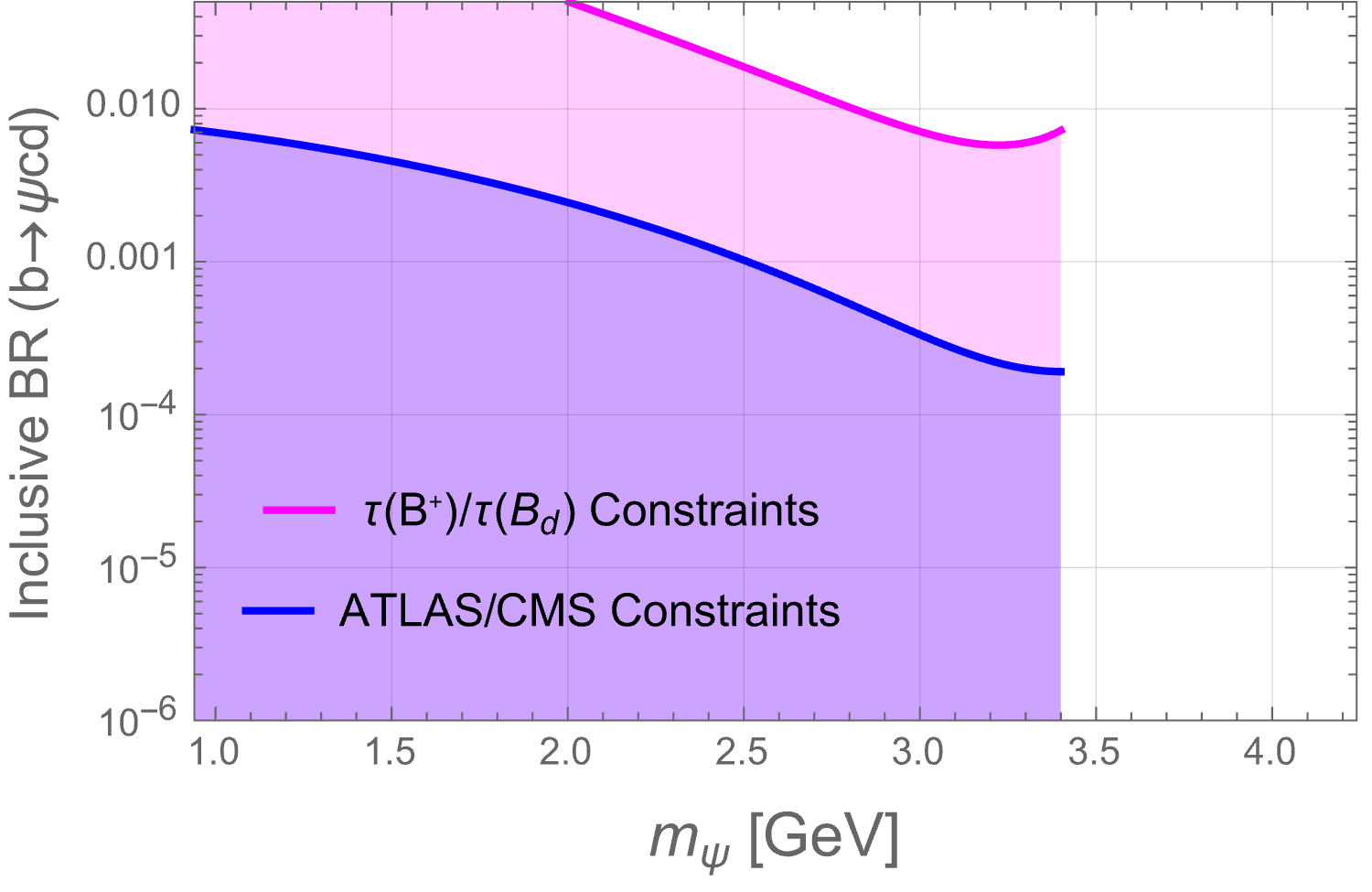}
         }
    \end{subfigure}
    \hfill
    \begin{subfigure}[t]{0.45\textwidth}
         \subcaptionbox{}[0.99\textwidth]{\includegraphics[width=\linewidth]{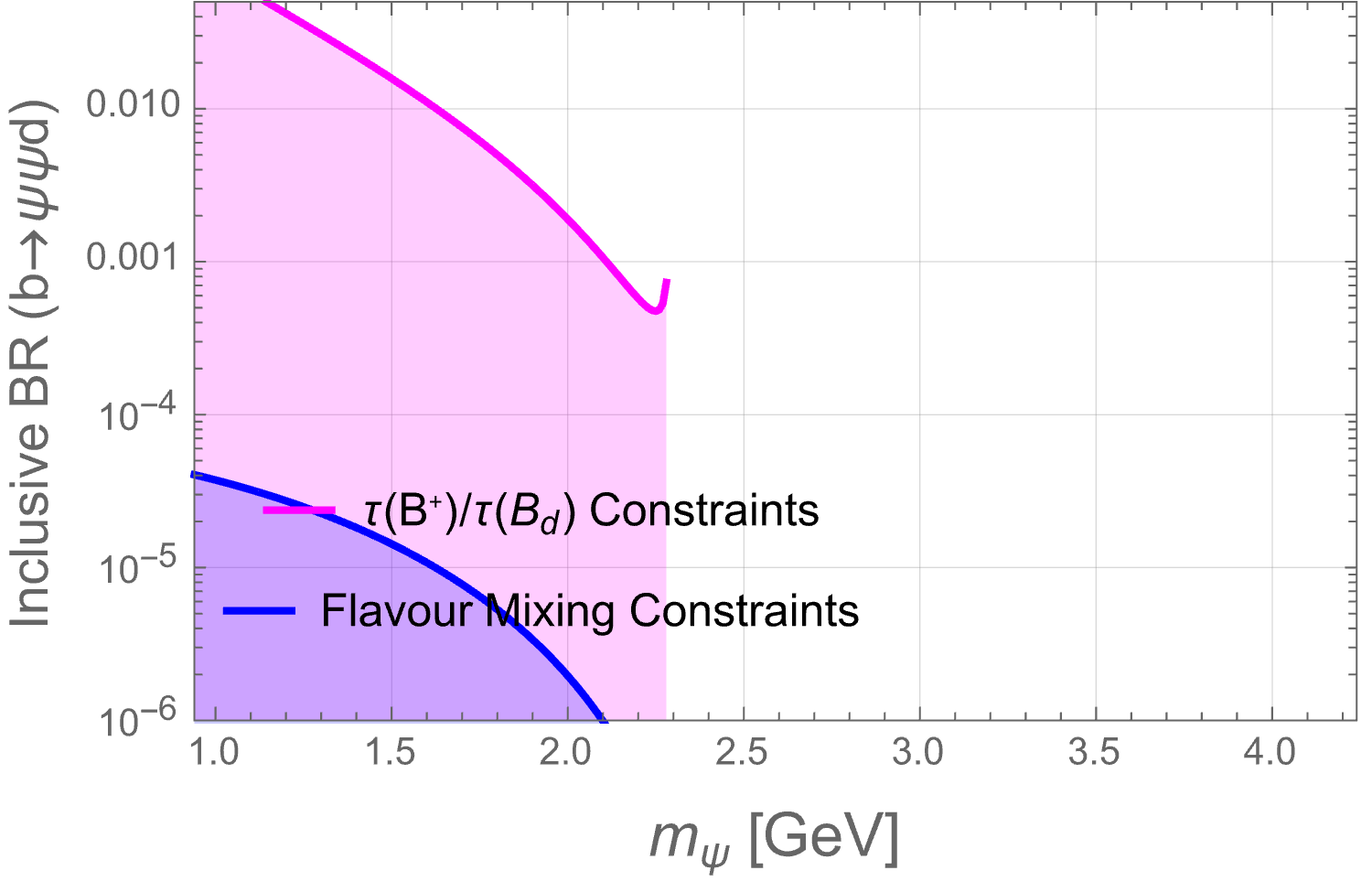}
         }
    \end{subfigure}
    \vspace{0.1cm}

    \begin{subfigure}[t]{0.45\textwidth}
         \subcaptionbox{}[0.99\textwidth]{\includegraphics[width=\linewidth]{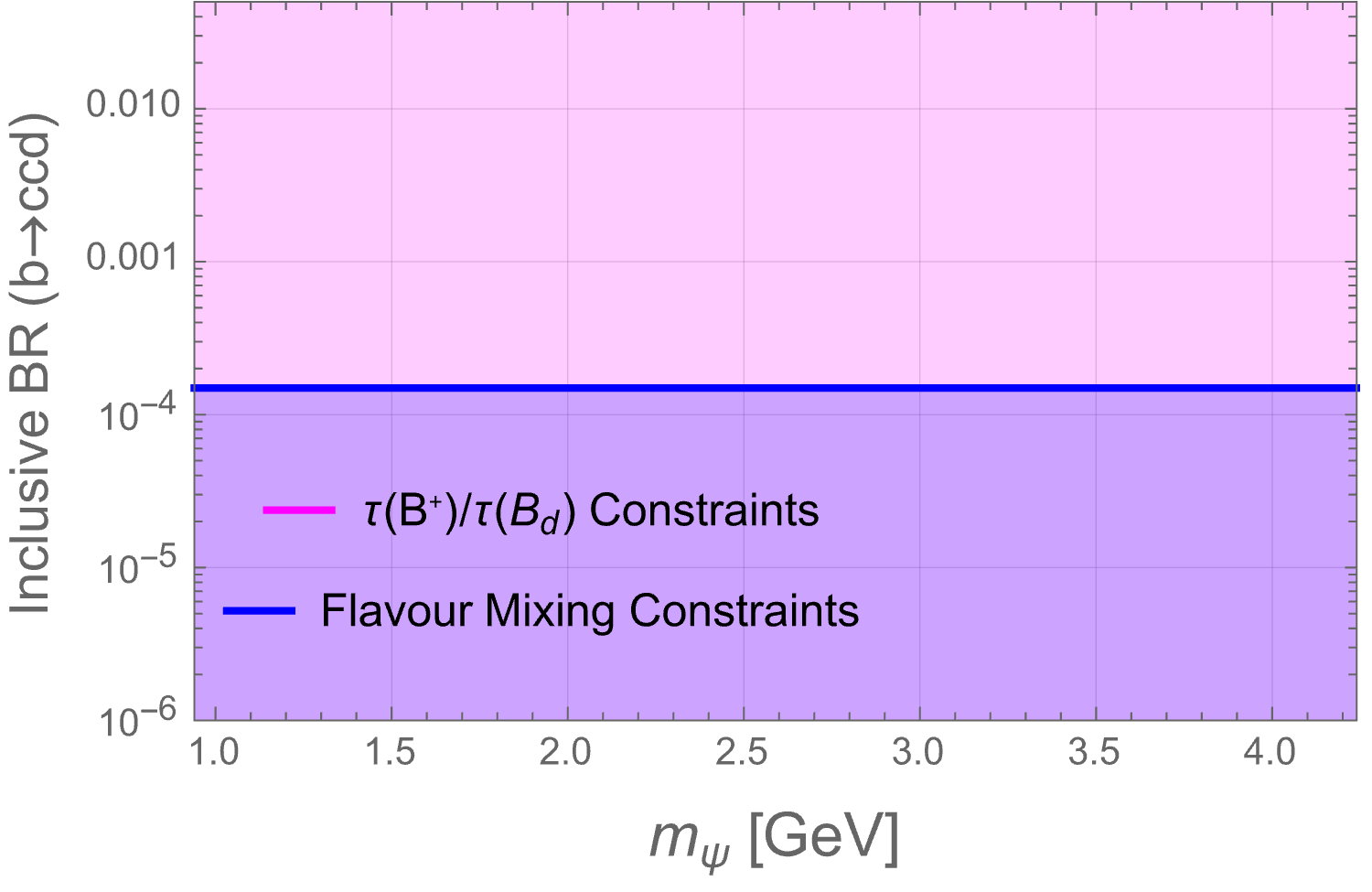}
         }
    \end{subfigure}
    \hfill
    \begin{subfigure}[t]{0.45\textwidth}
         \subcaptionbox{}[0.99\textwidth]{\includegraphics[width=\linewidth]{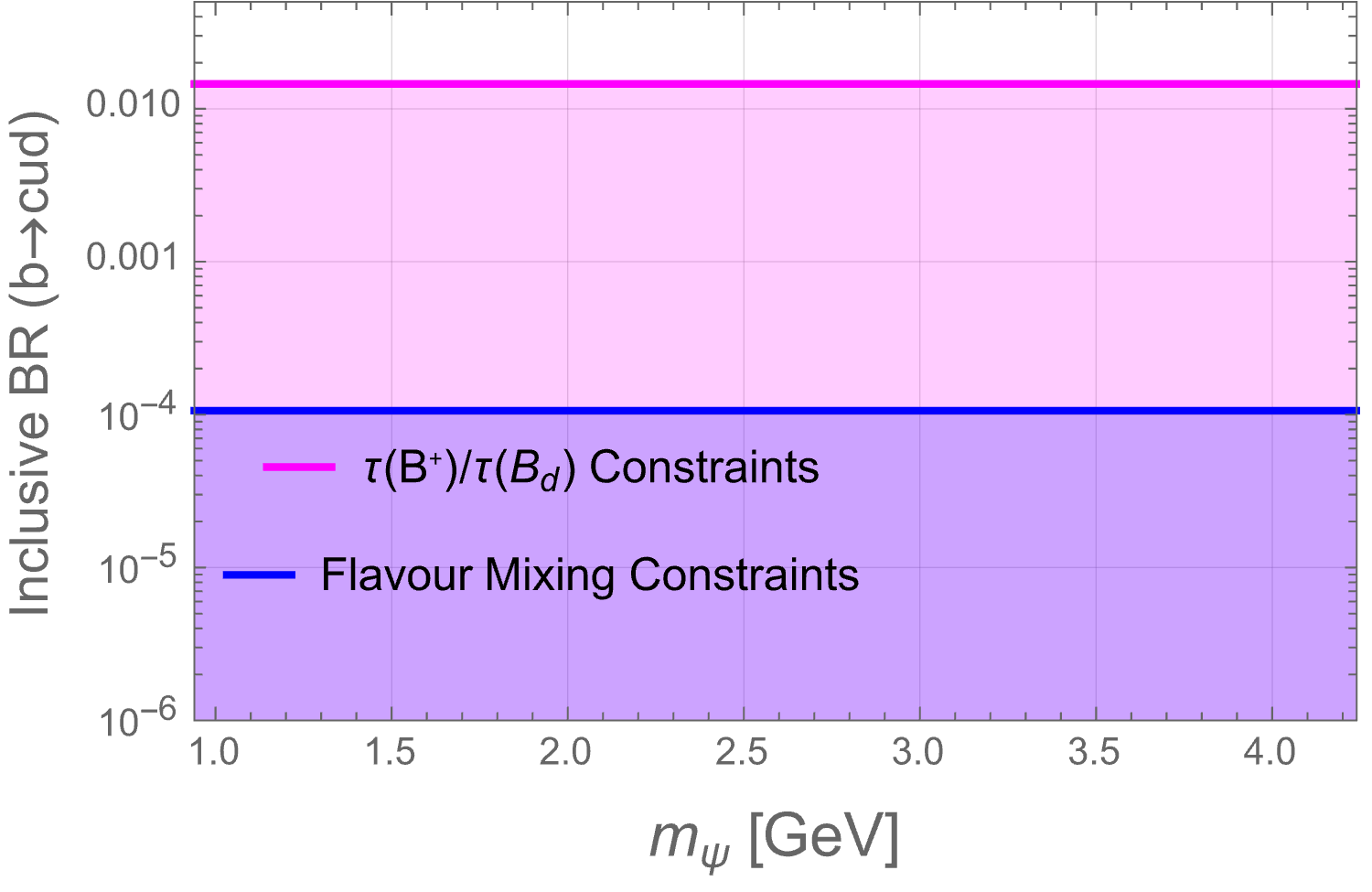}
         }
    \end{subfigure}

    \caption{{$\mathcal{H}_{-1/3}$ (Part 1): The inclusive branching ratio of the different operators proportional to the couplings $G^{f_1 f_2}_{X}$. Values of the inclusive branching fraction above $5\%$ are excluded by the bounds  from the total decay rate of $B$ mesons, therefore  we do not show regions above 0.05 in the plots. The figure shows the difference between using the values allowed by the lifetime ratio $\tau(B^+)/\tau(B_d)$ and those allowed by collider signals or flavour mixing, as studied in~\cite{Alonso-Alvarez:2021qfd}, to compute the inclusive branching ratio. The plots correspond to the following couplings: (a) $G_{III}^{\psi u}+G_{IV}^{d \psi}$ (b) $G_{III}^{\psi c}$ (c) $G_{III}^{c \psi}$ (d) $G_I^{\psi \psi}$ (e) $G_{IV}^{u d}$ (f) $G_I^{cc}$. }
    }
      \label{fig:model13-branching1}
\end{figure}

\begin{figure}[ht]
    \centering
    \begin{subfigure}[t]{0.45\textwidth}
         \subcaptionbox{}[0.99\textwidth]{\includegraphics[width=\linewidth]{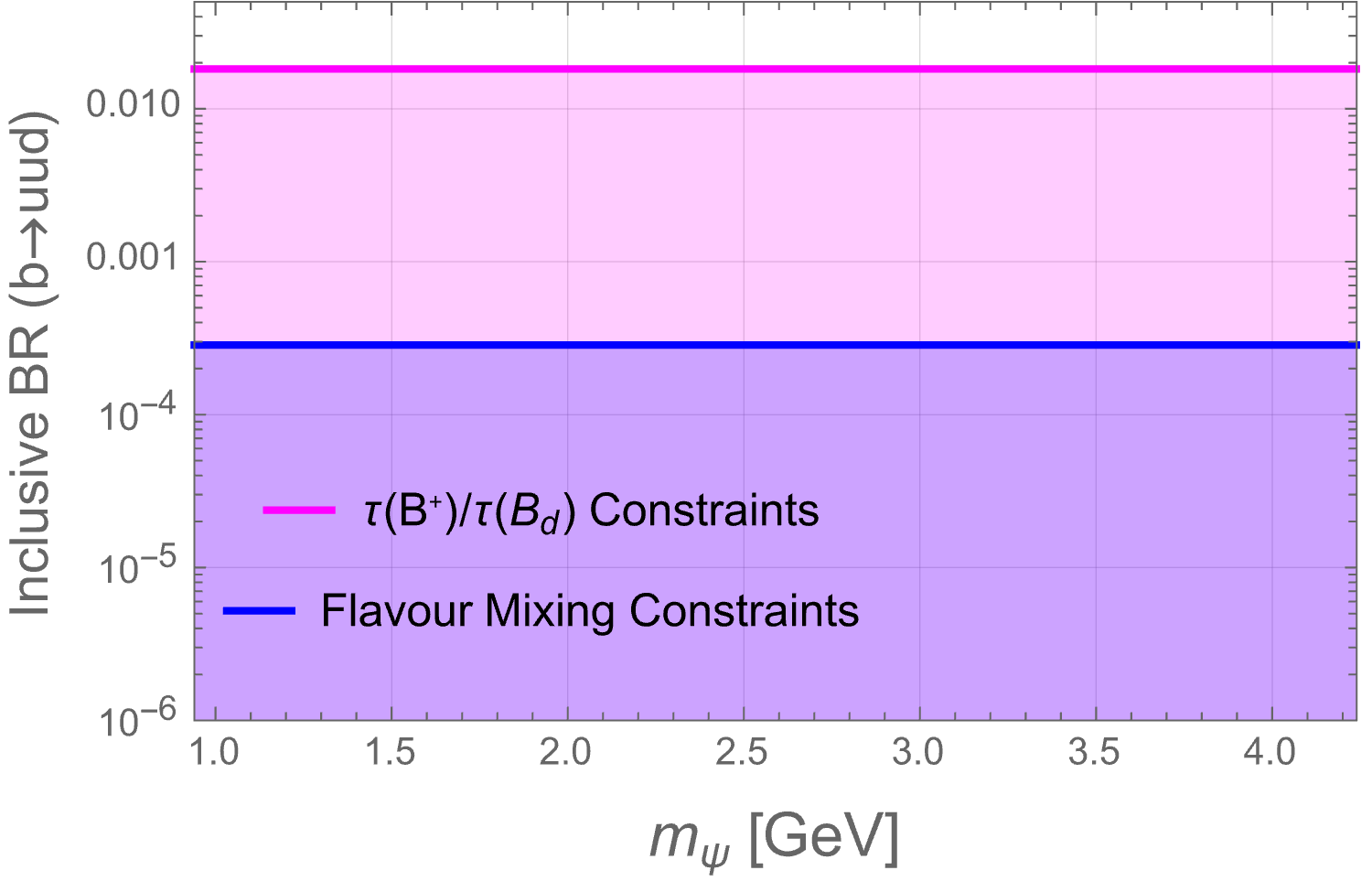}
         }
    \end{subfigure}
    \hfill
    \begin{subfigure}[t]{0.45\textwidth}
         \subcaptionbox{}[0.99\textwidth]{\includegraphics[width=\linewidth]{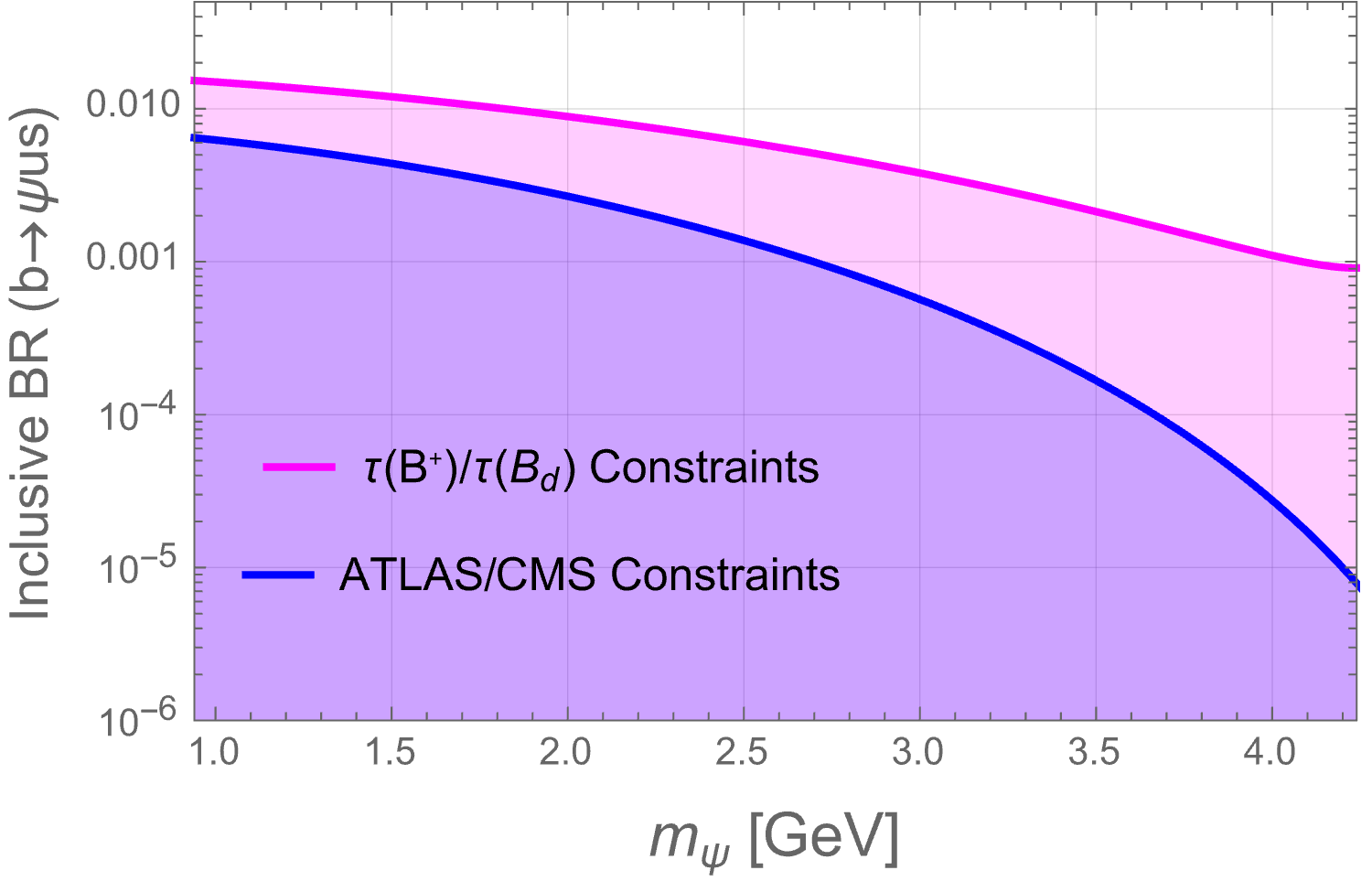}
         }
    \end{subfigure}
    \vspace{0.1cm}

    \begin{subfigure}[t]{0.45\textwidth}
         \subcaptionbox{}[0.99\textwidth]{\includegraphics[width=\linewidth]{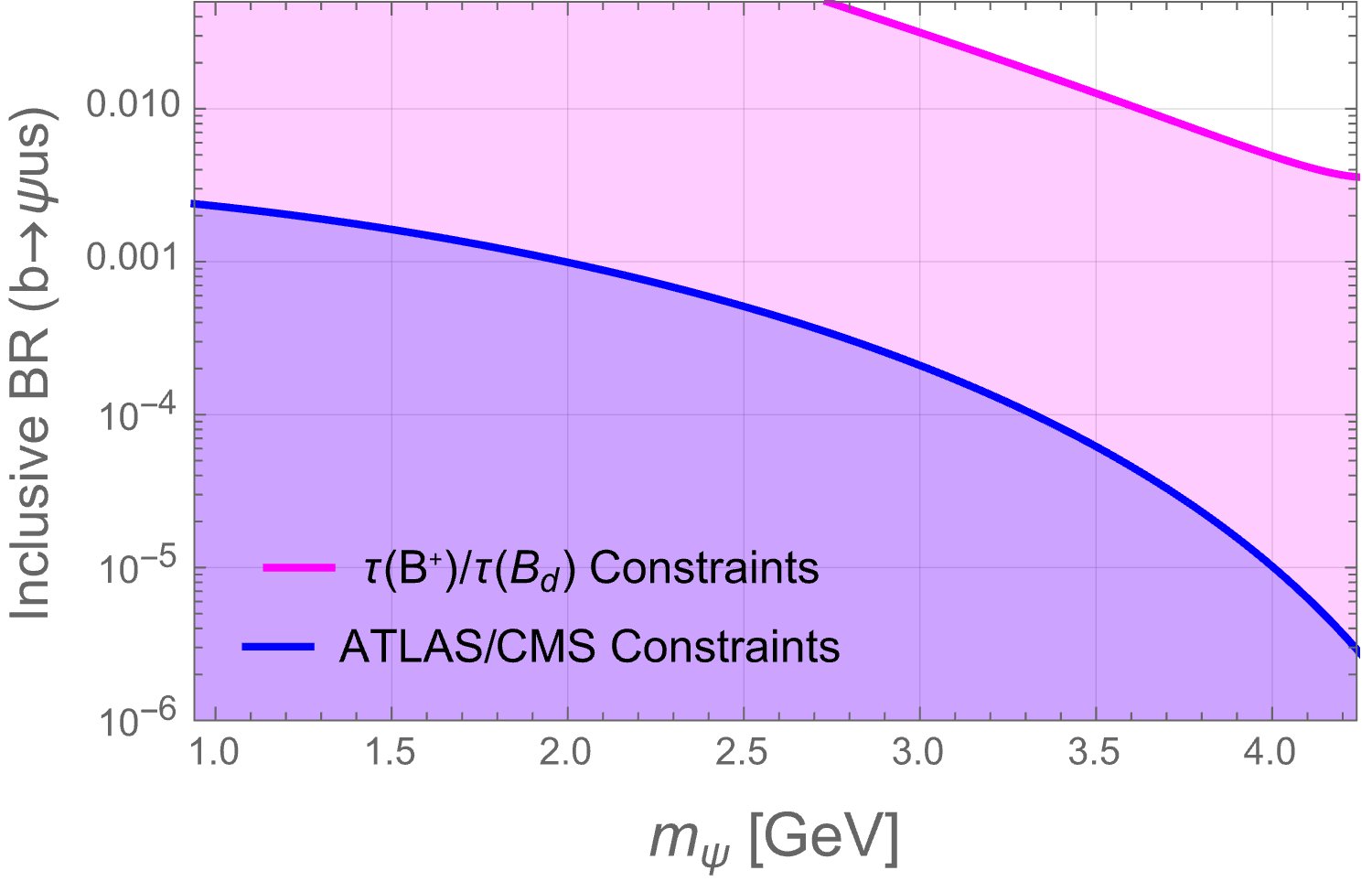}
         }
    \end{subfigure}
    \hfill
    \begin{subfigure}[t]{0.45\textwidth}
         \subcaptionbox{}[0.99\textwidth]{\includegraphics[width=\linewidth]{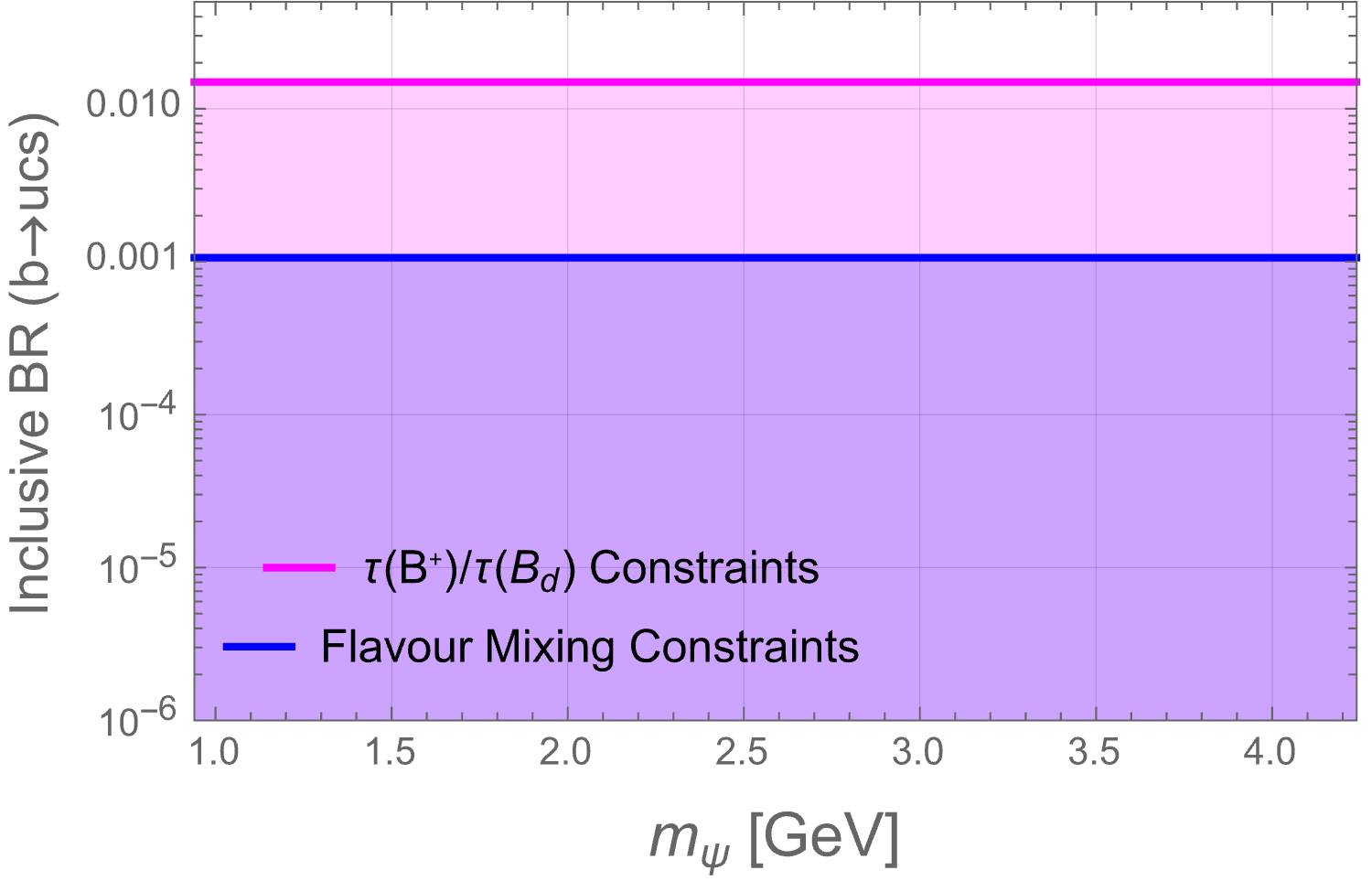}
         }
    \end{subfigure}
    \vspace{0.1cm}

    \begin{subfigure}[t]{0.45\textwidth}
         \subcaptionbox{}[0.99\textwidth]{\includegraphics[width=\linewidth]{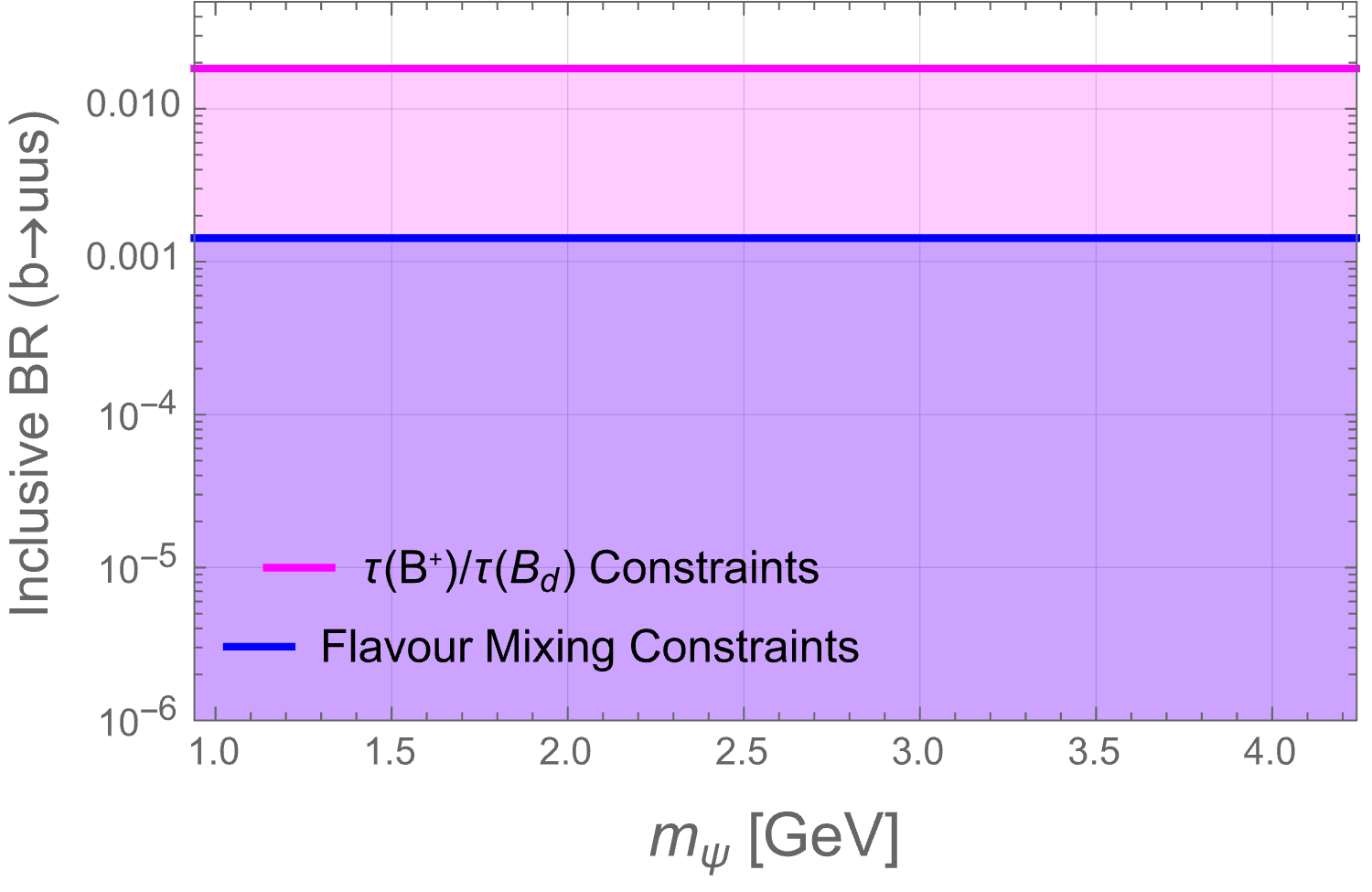}
         }
    \end{subfigure}
    \hfill
    \begin{subfigure}[t]{0.45\textwidth}
         \subcaptionbox{}[0.99\textwidth]{\includegraphics[width=\linewidth]{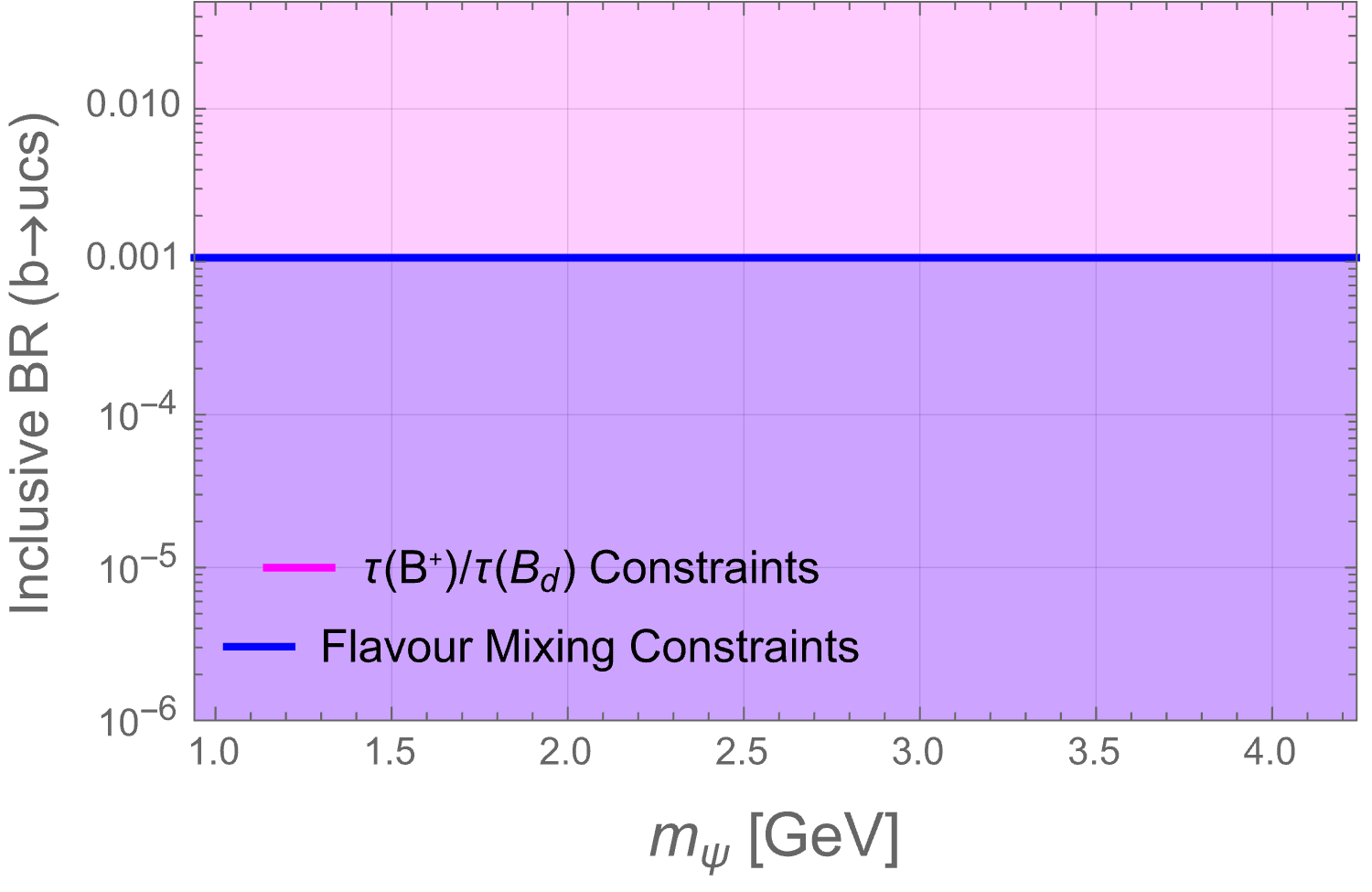}
         }
    \end{subfigure}

    \caption{{$\mathcal{H}_{-1/3}$ (Part 2): Additional branching ratio plots with the same colour coding. Values of the inclusive branching fraction above $5\%$ are excluded by the bounds  from the total decay rate of $B$ mesons, therefore  we do not show regions above 0.05 in the plots. The plots correspond to the following couplings: (a) $G_{I}^{u c}+G_{IV}^{c d}$ (b) $G_{IV}^{s \psi}$ (c) $G_{III}^{\psi c}$ (d) $G_{IV}^{u s}+_{I}^{u s}$ (e) $G_{IV}^{c s}$ (f) $G_I^{cs}$. }
    }
    \label{fig:model13-branching2}
\end{figure}

\chapter*{Acknowledgments}
\addcontentsline{toc}{chapter}{Acknowledgments}
One of the most important aspects of a doctoral journey is the supervisor. In this respect, I consider myself very fortunate. I first met my supervisor, Alexander Lenz, during a course he taught while I was a master’s student in Siegen. It did not take long for me to decide that I wanted to pursue my PhD under his supervision, and I was delighted when he accepted me as his student.
\\Throughout my doctorate and master studies, he had an exceptional sense for choosing projects that challenged me without becoming overwhelming, allowing me to develop steadily as a researcher. At the same time, he gave me the freedom to grow independently. He trusted me to explore ideas on my own, to make mistakes, and to learn from them. Looking back, I realize that this trust was one of the greatest gifts a supervisor could have given me.
\\Beyond research, I was always grateful for his understanding of my personality. He never expected me to participate in every social event or group activity, but instead respected the way I preferred to work and interact with others. That level of trust and acceptance is something I believe is rare, and I deeply appreciate it.
\\Beyond his own scientific contributions, I have always admired Alex's commitment to strengthening the flavour physics community. He has devoted an enormous amount of time and energy to bringing physics to the public, inspiring young people to pursue the subject, and building the flavour physics group in Siegen into one of the leading groups in the field. Although many of these efforts naturally benefit the University of Siegen, their impact extends far beyond it.
Investing in a strong and visible flavour physics community is, in the long run, an investment in the future of the field itself.
\\Much of what exists today had to be built from the ground up, and I can only imagine how demanding that must have been. Despite these many responsibilities, he always found time to guide and collaborate with his students. 
Even when I pursued projects independently, he never hesitated to carefully read my manuscripts, provide detailed comments, and offer valuable advice. 
Through his example, he also taught me that being a good physicist extends beyond doing good research. Equally important is learning how to communicate one’s work effectively, engage with the broader scientific community, and contribute to the growth of the field. For his trust, generosity with his time, and unwavering support, I will always be deeply grateful.
\\On a personal level, I have always admired his attitude toward both success and setbacks. Whatever the circumstances, he would meet them with optimism, a smile, and simply keep moving forward. In that respect, I think we are quite similar—perhaps with the exception of the smile.
\\I am deeply grateful to Maria Laura Piscopo and Alexey Rusov. I had the pleasure of interacting closely with both of them during my master’s thesis and throughout my PhD. For some reason I still do not fully understand, they always took me seriously. Every discussion with them left me with a much clearer idea of how to solve the problem I am having. In many respects, I learned the practical aspects of the Heavy Quark Expansion from them. Their guidance played a crucial role in my transition from being a student to becoming a researcher.
\\My sincere thanks also go to Alexander Khodjamirian. It has been a privilege to collaborate with one of the pioneers of sum rules and someone whose work I had admired long before we began working together. Learning the method of light-cone sum rules from Alex has been one of the highlights of my PhD.
What has always impressed me is his approach to research. Even today, he carries out calculations by hand whenever possible and is never satisfied with merely obtaining the correct result. Whenever something unexpected appears, his first instinct is to understand its origin. Watching him work has taught me the importance of developing physical intuition alongside technical calculations.
Beyond research, Alex has always been incredibly generous with his time. Whether I needed advice on a calculation, comments on a manuscript, or a recommendation letter, he was always willing to help. In many ways, it felt as though I had a second supervisor.
My only regret is that our collaboration began just as he was approaching retirement. I sincerely hope that it will continue for many years to come.
\\I would also like to thank Thomas Mannel. Although I have not yet had the pleasure of collaborating with one of the leading figures of the field, our discussions have had a profound influence on the way I think about physics. It is often the case that one can spend hours discussing a problem without making real progress. With Thomas, however, the opposite is true. Within minutes, he identifies the essence of the problem and immediately suggests promising ways of approaching it. 
What I admire most is that, despite the fact that my calculations are often carried out using methods different from his own, he always makes the effort to understand the way I think. Rather than insisting on his own perspective, he explains the physics in terms that I can readily relate to. This ability to adapt one’s explanation to the person in front of them is, in my experience, remarkably rare and something I have learned a great deal from.
\\I would also like to acknowledge my collaborators. 
I am especially grateful to Martin Lang, whose remarkable efficiency and dedication made him one of the most effective collaborators I have had the pleasure of working with.
I also sincerely thank Aritra Biswas and Zachary Wüthrich
for their careful cross-checks of many of the calculations presented in this thesis.
\\I also owe a great deal to the professors of the particle physics group in Siegen, with whom I had the opportunity to interact throughout my master’s studies and my PhD. Each of them contributed, in their own way, to my development as a physicist. I am particularly grateful to Tobias Huber, from whom I learned the techniques of loop calculations through one of the best courses I have ever attended. Guido Bell introduced me to the elegance of effective field theories, Thorsten Feldmann taught me the principles of QCD and hadron physics, Wolfgang Kilian deepened my understanding of the formal aspects of quantum field theory, and Oliver Witzel introduced me to the foundations of lattice QCD. I also benefited greatly from many discussions with Alexey Pivovarov, particularly during my work on the dimension-seven HQE project. He repeatedly challenged me to view the same calculation from different perspectives. 
\\
Looking back, I realize that the broad education I received in Siegen was the result of learning from all of them. What makes the particle physics group in Siegen truly exceptional is not only its scientific excellence, but also its welcoming and collaborative atmosphere. Every office door was always open, and I never hesitated to knock on one for a discussion, advice, or simply a different perspective on a problem. Although I was fortunate to have Alex as my supervisor, I always felt supported by the entire group. In that sense, I benefited from the guidance of far more people than one could normally expect during a PhD. As Damir Bečirević once told me,
``You are in a good group, of good people, in good hands.'' I could not agree more.
\\
I would also like to thank Gudrun Hiller and Miguel Escudero for the many fruitful discussions and the continuous exchange of ideas. Together with Maria Laura, they made my visit to CERN possible and created an exceptionally welcoming and inspiring environment. Thanks to the three of them, my time at CERN became one of the most memorable and rewarding experiences of my PhD.
\\Finally, I would like to express my deepest gratitude to my parents. 
Everything I have achieved rests upon the foundation they built for me, and I would not be where I am today without them.
\\
Throughout my life, I have been inspired by the achievements of many remarkable scientists. Yet, with time, I came to realize that the achievements which have shaped my own life the most are not those found in textbooks or scientific discoveries, but the countless sacrifices made by my parents.
Through their own example, they taught me how much one is willing to do for the happiness and future of their children. They also taught me that no matter what one achieves in life, there is always room to grow, to learn, and to become a better person.
\\
Of all the countless gifts they gave me, the greatest was, in my opinion, freedom. They gave me the freedom to discover who I wanted to become, to make my own choices, and to follow my own path. More importantly, they created an environment in which I could exercise that freedom with confidence, knowing that I would always have their unconditional love and support.
When I decided to pursue physics, they supported me wholeheartedly, even though the details of my work are far removed from their own experiences. 
They never asked me to become what they wanted me to be; instead, they encouraged me to become who I wanted to be. 
That is the greatest gift I have ever received, and one I will cherish for the rest of my life.

\cleardoublepage
\bibliographystyle{JHEP}
\bibliography{References}

\end{document}